\documentclass[%
aip,
jcp,
10pt, 
amsmath,
amssymb,
floatfix,
nofootinbib,
]{revtex4-2}
\usepackage{titlesec}
\usepackage{latexsym}
\usepackage{tabularx} 
\usepackage{array} 

\titlespacing*{\section}
{0pt}{1.5ex}{1.5ex}
\titlespacing*{\subsection}
{0pt}{1ex}{1ex}
\titlespacing*{\subsubsection}
{0pt}{1ex}{1ex}
\usepackage{setspace}

\usepackage[version=3]{mhchem} 
\usepackage[dvipsnames]{xcolor}
\usepackage{amsfonts}

\usepackage{graphicx}
\usepackage[colorlinks=true,citecolor=red]{hyperref}

\usepackage[dvipsnames]{xcolor}
\usepackage{url}
\usepackage{color}
\usepackage{dcolumn}
\usepackage{bm}
\usepackage[mathlines]{lineno}
\usepackage{comment}
\usepackage[version=3]{mhchem}
\usepackage[top=1cm, bottom=1.8cm, left=2.2cm, right=1.8cm, heightrounded]{geometry}

\usepackage[capitalise]{cleveref} 

\usepackage{graphicx} 
\usepackage{amsmath}
\usepackage{longtable}
\usepackage{wrapfig,floatflt}
\usepackage{etoolbox}

\usepackage{setspace}

\usepackage{mlmodern}

\usepackage{dsfont}
\usepackage{multirow}
\renewcommand{\bm}[1]{\mathbf{#1}}
\newcommand*{\myfont}{\fontfamily{phv}\selectfont}
\DeclareTextFontCommand{\textmyfont}{\myfont}

\DeclareMathOperator*{\argmin}{argmin}
\DeclareMathOperator*{\argmax}{argmax}

\newcommand*\pol[1]{\textcolor{black}{#1}}

\let\oldbibliography\thebibliography
\renewcommand{\thebibliography}[1]{\oldbibliography{#1}
\setlength{\itemsep}{0pt}} 
\allowdisplaybreaks

\makeatletter

\renewcommand{\thesubsection}{\arabic{section}.\arabic{subsection}}
\renewcommand{\p@subsection}{}
\renewcommand{\thesubsubsection}{\arabic{section}.\arabic{subsection}.\arabic{subsubsection}}
\renewcommand{\p@subsubsection}{}
\makeatother

\counterwithin*{equation}{section}

\usepackage[most]{tcolorbox}
\newtcolorbox{myquote}[1][]{%
    colback=black!5,
    colframe=black!5,
    notitle,
    sharp corners,
    borderline west={2pt}{0pt}{red!80!black},
    enhanced,
    breakable,
}
\usepackage{capt-of}

\begin{document}

\singlespacing


\title{A Practitioner's Guide to Materials Theory and Computation in the Age of Big Data and Artificial Intelligence}

\author{Jack Yang}
\affiliation{School of Materials Science and Engineering, University of New South Wales, Kensington, New South Wales 2052, Australia}
\email{jianliang.yang1@unsw.edu.au}

\date{\today}

\begin{abstract}
Computational modelling has become a cornerstone of materials science, with advances in atomistic simulations, high-throughput computation, and artificial intelligence transforming how materials are understood, discovered, and designed. These developments build upon generations of theoretical innovations that have translated complex mathematical and physical principles into practical predictive tools. However, the growing diversity of computational methods presents challenges for practitioners seeking a coherent understanding of their theoretical foundations. Drawing upon the author's research experience, this monograph provides a practitioner-oriented, pedagogical overview of computational materials science, unified through the concepts of potential-energy landscapes and statistical sampling. It explores how these concepts underpin structural stability, phase behaviour, and electronic and magnetic properties, through a systematic treatment of empirical force fields, density-functional theory, many-body electronic-structure methods, statistical sampling, and property prediction. It further examines high-throughput workflows, materials databases, and the probabilistic foundations of artificial intelligence. Particular emphasis is placed on atomistic representations, graph neural networks, and generative models, highlighting the roles of symmetry, group theory, differential geometry, and stochastic processes in modern materials modelling. By connecting established physical theories with emerging data-driven approaches, this monograph offers a unified perspective on computational materials science and identifies opportunities and challenges toward physics-informed, efficient, and reliable materials discovery.

\end{abstract}	
\maketitle


\tableofcontents

%
%
\section{Introduction}
Materials innovation is one of the main driving force behind the modern socioeconomic advances. It encompasses multiple disciplinaries that impact directly upon our daily life. For examples, the development of microelectronics with low-power consumption and fast processing speeds,\cite{lin2019two} electrocaloric materials\cite{valant2012electrocaloric} that enables the effective harvesting of waste heats in microelectronics, next-generation materials that can effectively convert renewable energy sources into electricity,\cite{zhang2022renewed,han2022ferroelectric} catalysts for the large-scale synthesis of industrially relevant chemicals,\cite{chang2021emerging} new drug molecules for the therapeutic treatments of existing and emerging life-threatening diseases,\cite{tanoli2025computational} as well as materials that could sustain and perform under extreme environments (such as inside nuclear reactors or at the outer space).\cite{mondal2021thermal,tadjer2022toward} There has been a long quest to achieve the goal of \emph{dial-a-material} to combat the increasing socioeconomic demands, where one could design the chemistry and structure, followed by the successful synthesis, of new materials with pre-defined chemical/physical properties for specific application, but the  destination for such a quest is still seems to on a far reaching horizon of an uncharted land. The pursuit for a room-temperature superconductor for over a century,\cite{pickett2023colloquium} is probably one of the most compelling and representative examples.  

The challenges behind materials discoveries are multilevelled. 
At the first level, there is a combinatorially large space to explore, in which different chemical elements and the atomic configurations can be combined together to give new materials,\cite{reymond2015chemical,groom2016cambridge,draxl2019nomad} even one builds these structures by following some very simple chemical and structural rules, such as the octet rules and known crystal symmetries.
At the second level, even with a given chemical composition, material can exhibit multiple different structures (polymorphs) with widely varying thermodynamic stabilities.\cite{doye1999evolution,engel2018mapping,senocrate2019thermochemical,yu2010polymorphism,bauer2001ritonavir} To include the next level of complexity, the thermodynamic  stability for materials itself is a complex, multivariable metric to determine both experimentally and theoretically, which is tightly coupled with the occurrence of defects and disorders that heavily depends on the synthesis condition. Whilst basic chemical intuitions based on elementary bonding theory is often very useful for us to rationalise the stability of a material, compelling evidences do suggest that more exotic quantum phenomena can also play an important role in materials' structural stabilities, as in the case of ferroelectric (in)stability in perovskites.\cite{bersuker2013pseudo} This naturally leads to the final, fourth level of challenge in our ranking, in which many material properties are the manifests of the responses from complex many-body quantum interactions in the material as a response to the external stimuli.\cite{slack1973nonmetallic,allen1993thermal,Migdal1958,Eliashberg1960,giustino2017electron,donchev2006many,ramasubramaniam2012large} It means that many of the exotic phenomena that drives the current wave of materials innovations may not be understood with simple physical laws and equations such as the classical Newtonian mechanics or Maxwell's theory of electromagnetism.
Certainly, one may rank these challenges in a different order, depending on the perspective one tends to approach this problem.

Whilst our technological advancements are largely driven by experimental breakthroughs in materials discoveries of new materials and phenomena that are not existing in the current material space, the interwinding complexities discussed above are often bringing us tremendous challenges to understand the experimental results, derive useful unbiased generalisations, from which predictive decisions for materials discoveries can be made. As such, computational materials, particularly atomistic modellings, have been playing a pivotal rule in dissecting these complex factor, to deepen our understanding behind the fundamental materials physics and chemistry from the atomistic level. The main benefit of computational modelling, is to provide an pragmatic approach, which we can draw analogy to Taylor expansions in mathematics or perturbation theory in quantum mechanics, that enables us to isolate and examine how different (sets of) physical effect(s) contribute, either individually or corporately, to the new behaviours observed in experiments. Such a fine-grained control, that separate and isolate different physical contributions towards a specific observed phenomena may not always be achievable experimentally. 

At the most fundamental level, quantum mechanics outline the principles that govern the interactions among the nuclei and electrons, that lead to the  structures, stabilities and the properties of materials at the atomic scale. Nevertheless, the many-body quantum problems\cite{mahan2000many} are inherently complex, and often impossible to be solved, to truly reflect the behaviours of real materials in both the time and length scales that we are commonly dealing with. On the other hand, to link microscopic phenomena with macroscopic observables, we also need to apply tools from statistical mechanics to compute the thermodynamically relevant behaviours of materials. As a small divergence at this point, although the later merely compounded the challenges that we are facing, we must point out that both theoretical frameworks have a strong linkage with probabilistic theory, in which \emph{sampling} played a critical role in helping to establish the computational approaches to tackle problems in their respective fields, and tight strongly with modern machine learning (ML) and artificial intelligence (AI),\cite{murphy2012machine} which we shall elaborate further in this monograph.

To overcome the many-body challenges, theoreticians have long been working on the developments of plausible \emph{approximations} \cite{harrison2018review,kohn1999nobel,golze2019gw} to solve these problems, to which, now materials modelling practitioners have the access to a rich repository of practically feasible simulation methodologies to compute the structures, energetics and the associated properties of materials with predictive accuracies, enabling many new materials to be predicted before more expensive laboratory synthesis and characterisation efforts are devoted to minimise the costs in materials discoveries.\cite{Lv2015Weyl,pulido2017functional,Mounet2018Exfoliation,Sun2019TernaryNitrides} Some of these methods, which has profoundly shaped the current landscape of the materials modelling practice, and shall be discussed in more details in this monograph, include: (1) Empirical atom-atom potentials or forcefields,\cite{harrison2018review} that offered simple parameterised analytical expressions to compute the interaction energies for atomistic systems, the many fundamental concepts in which underpinned the development of mathematical representations of atomistic structures for modern ML.\cite{Musil2021StructuralRepresentations} (2) Density Functional Theory (DFT),\cite{kohn1999nobel} which unarguably one of the most successful approximations developed for treating the many-body electronic structure problem that is the main workhorse for computational chemistry and materials science. (c) Many-body methods, such as GW approximations\cite{golze2019gw} for materials, which provided us to compute materials' properties at even better accuracies. These are ultimately integrated with (d) statistical sampling methods that enable us to   sample  the potential energy surface (PES) to predict materials' behaviour that are thermodynamically relevant to experimental observations. On the other hand, the availability of these methods, require materials modelling practitioners\footnote{which, for this monograph, the name refer to those who are not directly involved in the development of new theories or methods, but primarily focusing on the applications of the established methods in materials simulations}  to have a good solid understanding behind the fundamental principles and assumptions adopted behind these methodologies, as well as the trade-off between their accuracies, computational costs and system sizes capable of dealing with, in order to appropriately apply them in real problem solving.

From the physics point of view, the uniting concept behind all these developed methodologies is the energy of the atomistic system, that enters either as the  output of the model (such as empirical potentials), one of the key quantity that many others (electron densities, response properties) are associated with or can be derived from (such as in DFT), or the primary quantity to be sampled in statistical samplings. 

Therefore, from both the realistic materials application and theoretical method development perspectives, surveying the PES (approximated at a chosen theory level) of a given material system, as a function of its atomistic structures $E(\bm{R})$, under some prescribed thermodynamic conditions, can be considered as a central theme of computational materials science. The determinations of other associated physical properties, such as the anharmonic lattice dynamics and thermal conductivities,\cite{knoop2024tdep,tadano2015self,phono3py} dielectric responses,\cite{souza2002first} magnetic ground states,\cite{fender2025altermagnetism} as well as electron-phonon interactions,\cite{giustino2017electron} can all be cast into some form of sampling problems in more complex manifolds, such as on a PES in $\mathbb{R}^{3N}$ (with $N$ being the number of atoms) that are approximated beyond a harmonic truncation of the Taylor expansion of $E(\bm{R})$ in real space or alternatively as the combinatorial couplings among different harmonic phonons in the reciprocal space, the inclusion of the spin degree-of-freedoms in the PES, or the coupling between electron and phonon degrees of freedoms in which the PES for both the electronic and phononic subsystems need to be sampled jointly. A more concrete example lies in the study of heterogeneous electrocatalysts in explicit solvents,\cite{Andrade2020ProtonTransfer,Qin2023CO2Electroreduction,Sakong2018ElectricDoubleLayer} which has become increasingly popular with the advance in computational power, that one finds many of the aforementioned problems meet in a single system at the intersection between chemistry and physics.

The increase in the computational power, that is backed by the developments of high performance computers (HPC), and more recently, Graphical Processing Units (GPU), not only enriched the individual's capabilities in tackling computational materials problems with greater complexities, but also the communal developments of automated workflow management research software\cite{ganose2025atomate2,ong2013python} to facilitate the high-throughput, multi-staged computational modelling over families of materials systems. This ultimately leads to the establishment and continuous growths of several community-driven large-scale computational materials databases\cite{jain_apl_mater,draxl2019nomad,kirklin2015open,tran2023open} over the past decade, that not only has become useful centralised repositories for exploring materials' stabilities, structures and various physical properties (such as electronic structures, phonon properties, and magnetic ground states), but most importantly, provided the much needed community standard for materials modelling practices to enhance the consistency, reproducibility and credibility of the new research outputs in similar area.

It is probably not a surprise that this community achievement comes at the same time when ML and AI become increasingly popular, and start to reshape the landscape of computational research in chemistry, physics and materials science. Fundamentally, ML and AI are data-driven and statistical approaches, that enable us to extract complex hidden patterns (knowledges) from the dataset,\cite{glielmo2021unsupervised} that may be further harnessed to build various different models\cite{deringer2021gaussian,behler2021four,duval2023hitchhiker} to accelerate the computation and predictions of materials' stabilities and properties. Lately,  the use of generative AI\cite{zeni2025generative} has opened up the real hope for us to predict new chemistries and structures for materials that are not yet known, which it shall be demonstrated in this monograph, that this new frontier has offered us a new paradigm to reformulate transitional statistical samplings methods, making them to become better guided by the (equilibrium) probabilistic distributions of materials.

Whilst the developments in ML and  AI have brought many excitements into the field, a key challenge behind the adaptation of these breakthroughs in computer science into molecular and materials domain, is to appropriately incorporate the fundamental principles in atomistic modellings as discussed above, perhaps most critically, \emph{symmetries}, into these models mathematically.\cite{bartok2013representing,musil2021physics,grisafi2018symmetry,duval2023hitchhiker,nigam2022unified} This has shown to be important for enhancing the models' data efficiency and extrapolability, regardless whether a simple ML surrogate or more complex AI model is to be used. There has been a long criticism that DFT has often been used blindly as `black-box' methods by the practitioners,\cite{kruse2012standard} and the shear mathematical complexities in the newly developed AI models across the community can impose similar risks, with models being chosen by the practitioners based on popularity (or more bluntly put, number of citations), without properly understand the `nitty-gritty' behind the models, especially how and why two models differ from each other.

To help overcoming this `black-box' perceptions, and facilitate the future computational materials modelling practitioners to understand the multiple connected layers in the physical models, their workable approximations, probability theory in statistical thermodynamics and modern AI model developments, as well as the vibrant field of group and representation theory that played a pivotal role in atomistic ML and AI, I believe it will be valuable in writing a monograph that is based on a centralised concept of statistical sampling, to provide the readers with an overview on the topics discussed above in a pedagogically approachable way. This contents and organisations behind this monograph is building on my own research and teaching experiences. In particular, key mathematical equations will be presented to retain the necessary theoretical rigorous, but I also tend to provide a more intuitive, hitch-hiker style interpretations behind physical meanings behind these equations. I found that such an approach will be particularly useful in understanding the latest generative AI models for molecules and materials, which have extensively borrowed  mathematical tools from differential geometries, group theory and stochastic processes, which some readers may found intimidated to comprehend.

Hence, the primary objectives behind the present monograph, is to facilitate the practitioners in atomistic materials modelling, to: Firstly, establish the necessary physical and mathematical foundations for atomistic modelling. Secondly, establish an appreciation behind the energy-structure-property relationships for materials, that underpinned the developments of physics-inspired (structural) material representations, that significantly shaped the current landscape in ML and AI-driven materials science.  Thirdly, establish the linkage among traditionally applied simulation method and modern data-driven approaches, with the former acting more than simply a data-generator for the latter, but more fundamentally and conceptually linked through statistical thermodynamics and stochastic sampling, which we hope can facilitate the reader to establish a more unified viewpoint in understand the methodologies that have been developing in the field. Ultimately, these knowledge will be essential in helping the practitioners to better identify the limitations behind the currently available modelling methods, and more importantly, a foundation to advance themselves into the more challenging endeavours of developing  new materials theory and computational methods to tackle emerging problems in materials physics and chemistry, as well as incorporating new AI methodologies in the tackling of these problems. 

Therefore, I hope that both the competent and newly entered practitioners in atomistic modelling can both find inspirations in this, somehow personal, account of the essential knowledges and developments that shaped the modern atomistic materials modelling.

This monograph is organised as following. \cref{chapter:energy_landscape_concept} introduces the concept of energy landscape and surveyed a variety of examples in which the concept is relevant to the problems in materials science. 
\cref{chapter:energy_models} reviews empirical forcefields (or atom-atom potentials), DFT and many-body GW methods, which we treat primarily as the `energy calculators', on the theoretical level of which the PES will be established.  
\cref{chapter:sampling_methods} discussed the traditional statistical sampling methods that includes Monte Carlo, molecular dynamics and other related enhanced sampling methods.  
\cref{chapter:physical_properties} examines approaches with which the structural, dynamic, electronic and magnetic properties can be determined. The modern approaches for studying the electron-phonon interactions is also briefly reviewed to showcase the capabilities for us to tackle many-body physics in materials science that give rise to a variety of exotic physical properties, including superconductivities, and thermoelectric effects. 
\cref{chapter:high-throughput} highlights the developments of high-throughput computational workflows and the large scape computational materials databases. 
\cref{sect:ML_AI_Mat_Sci} develops the foundations and applications of ML and AI models in molecular and materials simulations, in which we pay special focus on how symmetry is incorporated into the model by respecting the invariant and equivariant properties in the atomistic structures through the tools of group and representation theory.
The monograph concludes in \cref{sec:conclusion_outlook}, which some open challenges and future directions will be discussed.

\section{Energy Landscape as the Fundamental Concept in Material Science}
\label{chapter:energy_landscape_concept}

PES is an important concept that is universally applicable in physical, chemical, biomolecular, as well as materials science. Fundamentally, it determines whether a specific system, being either a pharmaceutical drug, a protein, or a mineral crystal is stable under a given thermodynamic condition, before it is even meaningful to discuss how it can be made, and where to apply its unique properties. 

\begin{figure*}[t]
    \centering
    \includegraphics[width=\textwidth]{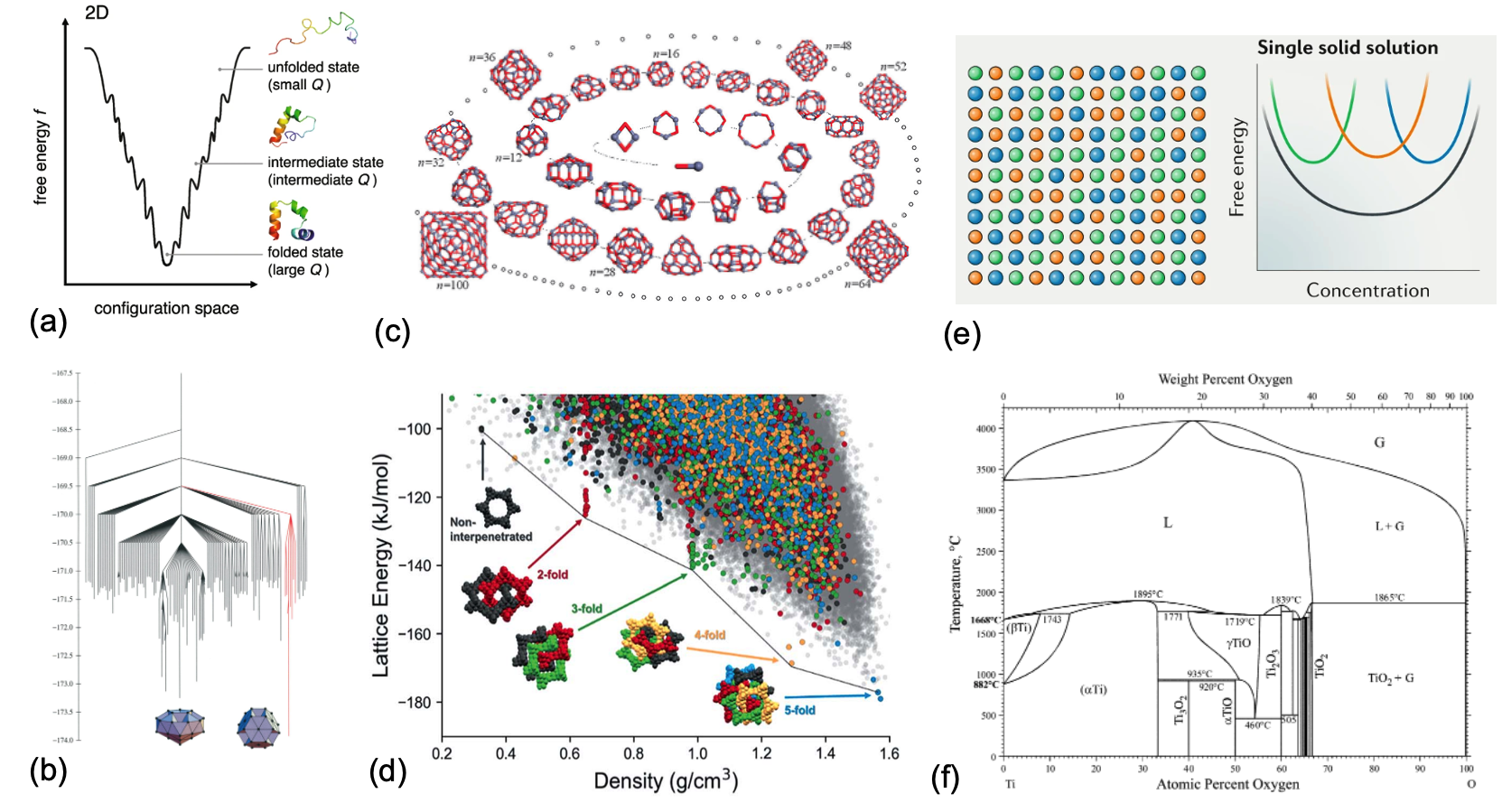}
    \caption{Demonstrative examples of energy landscapes in biological, chemical and material science. (a) Funnel-type energy landscape for protein folding,\cite{chong2019folding} showing that as the protein folds, the system free energy is decreased that is accompanied by the simultaneous decrease in the system entropy, where the latter is characterised by the increase in the number of native contacts $Q$. (b) Disconnectivity graph for the energy landscape of LJ$_{38}$ (Lennard--Jones cluster with 38 particles).\cite{doye1999evolution} The vertical scale is the total energy in the reduced unit ($\varepsilon=1$). The red line corresponds to the basin in which the global minimum cluster structure is located, which is separated from other wide energy basins with  large energy barriers. (c) Global minima for zinc oxide clusters (ZnO)$_n$ for $n=$1 to 32 and extended to larger sized bubble--shaped clusters, which are of distinctively different structure to bulk ZnO crystal.\cite{catlow2010modelling} (d) Predicted crystal structure landscape for adamantane--1,3,5,7--tetracarboxylic acid (\textbf{ADTA})\cite{cui2019mining}. Points on the landscape are color coded according to the crystal structure that contains interpenetrated \textbf{ADTA} molecules with different folds. The global minimum is the one with the highest density which follows the close--packing principle of molecular solids. (e) The formation of a homogeneous high--entropy alloy is driven by the free energy of a multicomponent solid solution (black curve) being always lower than that of a mixture with fewer components.\cite{george2019high} (f) Phase diagram for the binary Ti--O system.\cite{okamoto2011ti}}
    \label{fig:energy_landscape_examples}
\end{figure*}

In molecular biology, the folding of a long polypeptide chain into the tertiary structure of protein is believed to be governed by a `funnel--like' protein folding PES [\cref{fig:energy_landscape_examples}(a)].\cite{onuchic1995toward} As the protein folds, its energy decrease towards the \textit{global minimum}, which is the point with the lowest energy on the PES that represents the most stable structure of a protein. This is accompanied by the simultaneous decrease in the width of the PES funnel, reflecting the decrease in the system entropy by maximizing the intramolecular hydrogen bonds\cite{mu2005energy} that stabilises the complex three--dimensional structure of a protein. Given a typical structure stored in the protein databank  (PDB)\cite{berman2000protein} can contain anywhere between few hundred to more than 5000 amino acid residues,\cite{pdb2019protein} the number of possible intramolecular  hydrogen bonding patterns that can be formed by a peptide chain can be astronomically large. This is reflected by the presence of numerous \textit{local minima} on the PES with competing stabilities, which is one of the main reason why  \textit{de novo} prediction of protein structures has been a long--standing challenge.\cite{freddolino2010challenges} The recently developed AlphaFold\cite{jumper2021highly} model tackles this challenge by combining the power of deep learning and the rich structural information stored in the PDB to determine the best protein structure for a given peptide sequence, and has demonstrated a huge success and wide adaptation across the community.

In contrast, chemists study  systems that are generally much smaller in sizes. This significantly reduces the degrees--of--freedoms in the problems that are being dealt with, which, in many cases, it becomes not only possible but necessary to thoroughly enumerate the minima on, or even map out the topology of the entire PES, in order for us to obtain a more deeper understanding on the thermodynamics and kinetics of the atomistic structures that can be served to explain the phenomena observed in more complex systems, such as proteins and other macromolecules. For instance, the PES for small atomic clusters with up to 110 atoms  bounded by model Lennard--Jones potential have been mapped out,\cite{wales1997global} and it is clear that even the PES for such simple clusters can be more complex than a `single funnel' [\cref{fig:energy_landscape_examples}(b)].\cite{doye1999evolution} Extending this more realistic materials, such as \ce{MgO}\cite{roberts2001investigation}, \ce{TiO2}\cite{hamad2005structure} and \ce{ZnO}\cite{al2008structures} [\cref{fig:energy_landscape_examples}(c)], it is clear that when the materials are down to the size of a few atoms, they adopt an atomistic structure that is completely different from their bulk crystal structure, offering insights into nucleation and growth behaviours of the crystalline materials. For organic compounds, this phenomena is most commonly known as different isomers for a given compound. Molecular aggregates bounded by the intermolecular non--covalent interactions also exhibit rich potential energy landscapes, with the water\cite{wales1998global,engel2018mapping,monserrat2020liquid,reinhardt2021quantum} being one of the most intensively investigated system, not only for its physiological significance, but also for the occurrence of cooperative\cite{mahadevi2016cooperativity} and nuclear quantum effects\cite{ceriotti2016nuclear}  that continue to be a challenge for the developments of quantum chemical methods in treating this type of system.\cite{gillan2013energy,gillan2016perspective,yuan2017benchmark,manna2017conventional}  In the condensed phase, this leads to the  polymorphic behaviours of molecular crystals where the same molecule can crystallize into crystals of different structures but very similar thermodynamic stabilities.\cite{yu2010polymorphism,bauer2001ritonavir}  In this case, the ability to enumerate a large number of local minima on the lattice energy landscape of a given molecule [\cref{fig:energy_landscape_examples}(d)] (or a combination of a few molecules) is critical for screening  both pharmaceutical drugs and functional molecular materials,\cite{neumann2015combined,pulido2017functional} and is an active research field with the progress of which being tracked by the \emph{small molecule crystal structure prediction blind tests} to reflect the development of the theoretical methodologies in attacking such a challenge.\cite{reilly2016report,Hunnisett:aw5093,Hunnisett:aw5094} There is also a strong focus not just in enumerating the local minima on a single PES,  but also across multiple PES for the excited states, as well as tracing the pathways that connect all these local minima, which is particularly useful for determining the  energy barriers and kinetics for different chemical reactions.\cite{czako2011dynamics,motagamwala2020microkinetic,mai2020molecular}

\begin{figure*}[tb]
    \centering
    \includegraphics[width=\textwidth]{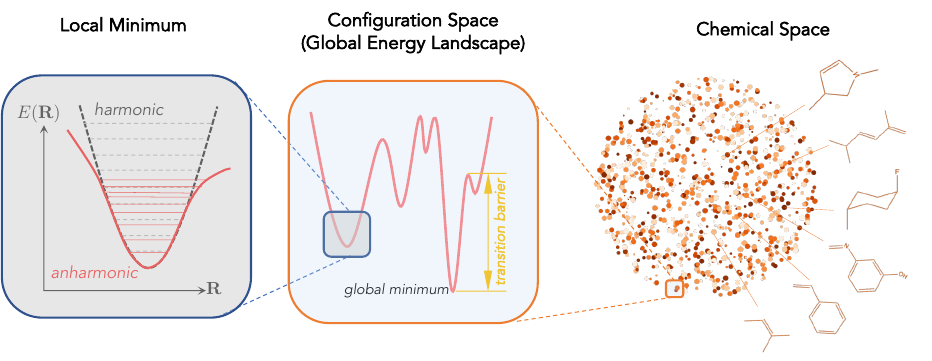}
    \caption{Hierarchical view of the energy landscapes. (Left) Local minimum for a stable structure that is represented by an anharmonic potential well. The shape of the well close to its bottom can generally be well approximated by a harmonic potential (see \cref{sect:phonon} for detailed discussions). The topology of the potential well determines the vibrational behavior of the material, in which the dashed and solid horizontal lines represent the vibrational eigenenergy levels for the harmonic and anharmonic potential, respectively. (Middle) The configuration space for materials with a given chemical composition [or also known as the global energy landscape as for the cases of (a), (b) and (d) shown in \cref{fig:energy_landscape_examples}], which consists of multiple local minima that are separated by transition barriers for a structure to inter--convert between two different local minima. The global minimum corresponds to the structure with the lowest energy on the entire potential energy surface. This energy landscape belongs to a specific compound that is part of the larger (Right) chemical space, in which here a schematic illustration for the chemical space representing small organic molecules is shown. }
    \label{fig:energy_landscapes_hierachical}
\end{figure*}

For material scientists, the concept of PES is most widely known as the \textit{phase diagram}. It provides the information on the compositions and structures that are stable under a specific thermodynamic condition (\emph{e.g.} temperatures and pressures) for a given combination of elements, such as the Fe--C\cite{okamoto1992c} system that is critical in the steel--making industry and  Ti--O\cite{wahlbeck1966reinvestigation} [\cref{fig:energy_landscape_examples}(f)] system that has attracted enormous attention in guiding the development of photocatalysts. A particular compound may also possess multiple different phases each with a different physical property. The ABO$_3$ perovskite is a particularly interesting system with a rich phase behaviour. For instances, as temperature drops, \ce{BaTiO3} and \ce{KNbO3} undergo a sequence of phase transformations from cubic paraelectric, to tetragonal, orthorhombic, and finally the ferroelectric rhombohedral  phase, whereas \ce{PbTiO3} will only undergo cubic to ferroelectric tetragonal phase.\cite{ravel1998local} These phase transformation behaviours are strongly affected by the presence of the imaginary phonon modes\cite{fleury1968soft,pallikara2022physical} in the system, which is fundamentally dictated by the topology of the local minimum on the PES. This has also a strong implication in, for example, the behaviours of ion diffusions in solid--state batteries and electrolytes.\cite{he2017origin,di2019superionic} Phases that are non--stable under a given thermodynamic condition are generally referred to as the `metastable' phases\cite{aykol2018thermodynamic,sun2016thermodynamic} (similar to the case of high energy polymorphs for organic crystals described above). These phases may be of interesting physical properties, attracting interests to explore strategies to stabilize them under standard thermodynamic conditions, for instances, through the applications of chemical\cite{eitel2002preparation,protesescu2015nanocrystals,tan2017chemical} or mechanical  pressure. The presence of foreign dopants as well as the native defects in the material can also perturb the chemical bonding states in the material, hence changing the total system energy. As such, investigating the defect formation energy and/or chemical doping limits in materials comprise a large discipline of `defect chemistry' that is particularly important to guide the synthesis of materials with targeted properties.\cite{freysoldt2014first,anand2022convergent,norby2004hydrogen,robertson2011limits,lany2007dopability} Last but not least, the ability for high--entropy metal\cite{george2019high} or ceramic\cite{oses2020high}
alloys, which contains at least five different principal metal elements intermixed at a type of lattice site to form a homogeneous solid solution, is dictated by the competitions between the materials' enthalpy (bonding strengths) and configurational entropy, whereby the latter measures whether a specific spatial arrangements of atoms is energetically more preferred or degenerate compared to many other possible arrangements [\cref{fig:energy_landscape_examples}(e)].

Regardless of the different application scenarios discussed above, the concept of the PES can be  mathematically abstracted as a function $E(\bm{R})$ that maps the total energy of the system $E$ to the positions of $N$ atoms in the high--dimensional space $\bm{R}\in \mathbb{R}^{3N}$.  This generally defines a \emph{configuration space} in which both the types and numbers of different chemical elements in the material system, which will be denoted as $\mathcal{C}=\{\mathcal{X}_i,n_{i}\}$ (with $\mathcal{X}_i$ denotes the $i$--th chemical element in the compound and $n_i$ being its stoichiometry with $\sum_i n_i=N$), are fixed. Isomers of an organic compounds, different conformers of an organic molecule, high--entropy alloys with a fixed chemical composition, as well as different phases of perovskite \ce{BaTiO3}, can be characterised by the PES $E(\bm{R})$, where the experimentally observed structure corresponds to one of the many local minima on the PES. One can extend this to a collection of PES with different chemical compositions as $\left\{E_\chi(\bm{R})\vert \chi\in  \left\{\mathcal{C}_1,\mathcal{C}_2,\ldots\mathcal{C}_{k} \right\} \right\}$. This essentially defines a \emph{chemical space}, which is particularly popular in organic chemistry for searching drug--like and other functional compounds,\cite{reymond2015chemical,gomez2018automatic} whereas the phase diagram that is commonly used by material scientists can be considered as a restricted chemical space with fixed $\{\mathcal{X}_i\}$ but variable $\{n_i\}$.  As such, one can start to appreciate PES as a key universal concept in fundamental science, and  \cref{fig:energy_landscapes_hierachical} provides an overview behind this critical concept.

\subsection{Energetic Measures for Material Stability}

\begin{figure*}[tbp]
    \centering
    \includegraphics[width=\textwidth]{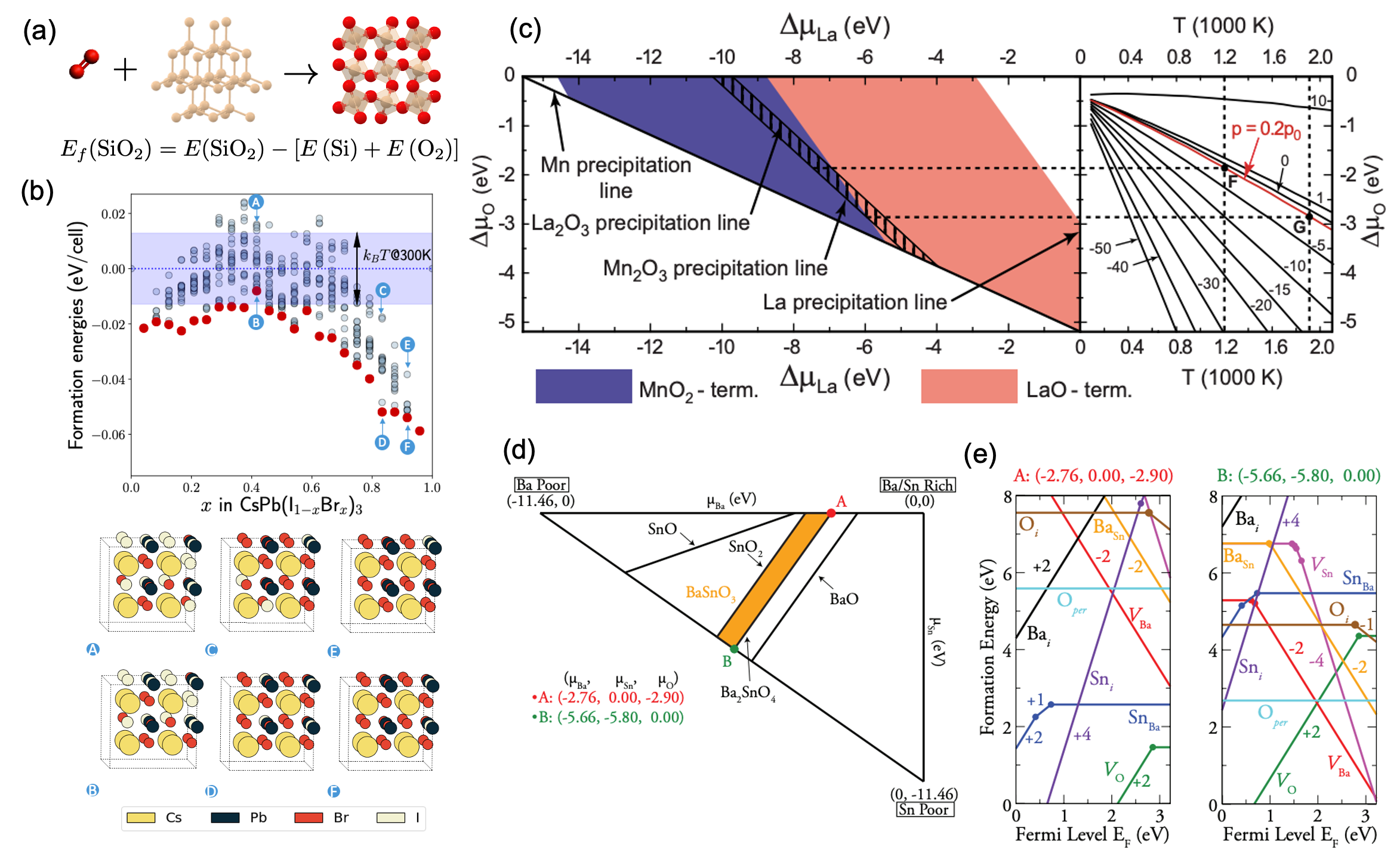}
    \caption{Illustrations of some common stability measurements in computational material science. (a) The formation energy $E_f$ of a compound is defined by referencing to the energies of its constituent elements in their standard form, which is demonstrated here for the chemical reaction of Si$(s)$+O$_{2}(g)\to$SiO$_{2}(s)$ that corresponds to  the formation of silicon dioxide. (b) (Top) Energy landscape for the intermixing between cubic halide perovskites of \ce{CsPbI3} and \ce{CsPbBr3} forming the solid solution of CsPb(I$_{1-x}$Br$_{x}$)$_3$.\cite{yang2020correlating} The red dots highlight the convex hull of the energy landscape. (Bottom) Exemplary atomistic structures of the  CsPb(I$_{1-x}$Br$_{x}$)$_3$ solid solutions with different concentrations ($x$) and stabilities. (c) Thermodynamic stability diagram as a function of O and La chemical potentials built for \ce{LaMnO3} (001) surfaces.\cite{piskunov2008electronic} It contains comparison of stability of both LaO-- and \ce{MnO2}--terminated (001) surfaces and accounts for precipitation conditions for La and Mn metals and their trivalent oxides \ce{La2O3} and \ce{Mn2O3}. Stable region is shown as the hatched area between \ce{La2O3} and \ce{Mn2O3} precipitation lines. The right side shows a family of oxygen chemical potentials under different conditions. The label $m$ indicates the \ce{O2} gas partial pressure: $10^{m}$ atm. Red gray line corresponds to oxygen partial pressure $p=0.2p_0$ as in the ambient atmosphere. (d) Range of the chemical potentials $\mu_{\mbox{\scriptsize{Sn}}}$ and $\mu_{\mbox{\scriptsize{Ba}}}$ (yellow shaded region)  in which the transparent oxide \ce{BaSnO3}  \cite{scanlon2013defect} is chemically stable. The verticies of the triangle are determined by the formation enthalpy $\Delta H_f$ of \ce{BaSnO3}. (e) Formation energies of the native defects in \ce{BaSnO3} as a function of the Fermi level at two diffent chemical potentials A and B as highlighted in (d).}
    \label{fig:stability_measures}
\end{figure*}
 
Practically, the potential energy of a material $E(\bm{R})$  can be computationally determined from its atomic configurations based on different level of approximations, which will be  discussed with details in \cref{chapter:energy_models}. However, regardless of what approximation is adopted,  the absolute value of the potential energy $E$ itself that is calculated for a given atomic configuration $\{\bm{R}\}$ bears little physical significance, and must be interpreted by comparing it to the energy of a reference state in order to provide a meaningful measure of the material stabilities. Conceptually, this is equivalent to the consideration of a  reaction $\mathcal{A}\to\mathcal{B}$, where the reaction energy is always measured as the difference in energy of the product(s) with respect to the reactant(s) as $\Delta E=E[\bm{R}(\mathcal{A})]-E[\bm{R}(\mathcal{B})]$. Although practically there may not be a chemical reaction happening, for example, in the case where $\mathcal{A}$  and  $\mathcal{B}$  corresponds to two different conformers of a molecules  or polymorphs of a crystal.  In this section, some important energetic  merits for material stability that frequently arises in computational modellings will be briefly reviewed, and we focus our discussions on the energetic measures at 0 K. 

The \textit{formation energy} ($E_f$) of a compound, for instance, A$_l$B$_m$C$_n$, is defined for the formation reaction from its constituent elements in their standard form, i.e.
\begin{equation}\label{eq:formation_chemical_reaction}
    l\mbox{A}+m\mbox{B}+n\mbox{C}\to \mbox{A}_l\mbox{B}_m\mbox{C}_n,
\end{equation}
with 
\begin{equation}\label{eq:formation_energy}
    E_{f}\left(\mbox{A}_l\mbox{B}_m\mbox{C}_n\right)=\frac{1}{N}\left[E\left(\mbox{A}_l\mbox{B}_m\mbox{C}_n\right)-lE(\mbox{A})-mE(\mbox{B})-nE(\mbox{C})\right],
\end{equation}
where $N$ is the total number of atoms in the (simulation) cell. This is demonstrated in \cref{fig:stability_measures}(a) for \ce{SiO2}.  Compounds that possess $E_f<0$ are generally considered to be stable (against decomposition to its constituent elements). For crystalline compounds, $E_f$ is sometimes also referred as the \textit{cohesive energy} or the \textit{lattice energy}. If all the reactants on the left--hand--side of the reactions are further constrained to the gas phase of individual element, then $E_f=-\Delta H_{sub}$ where $\Delta H_{sub}$ is the sublimation enthalpy of the solid. 

Note that although we have chosen to illustrate the concept of $E_f$ using a compound with strong chemical bond (\ce{SiO2} in this case), it is equally applicable to aggregates bonded together with weak bonding, such as van der Waals (vdW) interactions. Normally, this is referred to as the \emph{interaction} or \emph{binding energy} $E_{b}$, such as the binding energies for a graphene bilayer, water dimer, or the cohesive energy of molecular crystals. The vdW binding energy is typically an order or two weaker than a strong chemical bond.  For example, the bonding energy of a C--C single bond is around 90 kcal/mol compared to $\sim$2.5--3.0 kcal/mol for the binding energy of a benzene dimer.

A more generalised version of the reaction (\cref{eq:formation_chemical_reaction}) to be considered, that is better aligned with the concept of phase diagram [\cref{fig:energy_landscape_examples}(f)] in material science, is to rewrite reaction (\cref{eq:formation_chemical_reaction}) as 
\begin{equation}\label{eq:convex_hull_reaction}
n_{1}\mathcal{P}_{1}+n_{2}\mathcal{P}_2+\cdots\to \mathcal{C}^{(i)},
\end{equation}
which represents the formation of a new compound $\mathcal{C}$ in a particular phase $(i)$, or, the $i$--th configuration of a solid solution, with a fixed chemical composition, from the stable parent phases  $\{\mathcal{P}_{1},\mathcal{P}_{2},\ldots\}$ on the phase diagram. In this case, $\mathcal{P}$ could be either an element or a compound, and $\mathcal{C}$ could occur in multiple different phases or atomic configurations. For example, \cref{fig:stability_measures} (b) demonstrates the formation of  mixed halide solid solutions, where  $\mathcal{C}$ in (\cref{eq:convex_hull_reaction}) corresponds to CsPb(I$_{1-x}$Br$_{x}$)$_3$ with a fixed halide concentration $x$, while $\mathcal{P}_{1}$ and $\mathcal{P}_{2}$ are the parent \ce{CsPbBr3} and \ce{CsPbI3} perovskites. With this in mind, a negative (positive) reaction energy, or most commonly referred to as the \emph{mixing enthalpy} 
\begin{equation}\label{eq:mixing_enthalpy}
\Delta H_{mix}^{(i)}=E[\mathcal{C}^{(i)}]-\sum_k n_k E[\mathcal{P}_{k}]
\end{equation}
indicates whether the compound is less (more) likely to undergo chemical decompositions into its parent phases.   By further sampling the possible structures $\{\mathcal{C}^{(1)},\mathcal{C}^{(2)},\mathcal{C}^{(3)},\ldots\}$ at a particular chemical composition to obtain their corresponding mixing enthalpies $\{\Delta H_{mix}\}=\{\Delta H_{mix}^{(1)},\Delta H_{mix}^{(2)},\ldots\}$, the position for the \textit{convex hull} of stability can be identified as the point corresponding to $\min\{\Delta H_{mix}\}$ [highlighted with red dots in \cref{fig:stability_measures}(b)], whereas
\begin{equation}\label{eq:convex_hull_energy}
\Delta E_{hull}^{(i)}=\Delta H_{mix}^{(i)} - \min\{\Delta H_{mix}\}
\end{equation}
is the convex hull energy for a given material, which is typically applied as a measure of the material's metastability in its $i$--th phase.\cite{sun2016thermodynamic}

Beyond the phase stabilities mentioned above, for some materials that are particularly prone to chemical degredations under the ambient conditions, it is also necessary to further compare their phase stabilities with respect to the energies of the degradation reactions for these materials, in order to obtain a more complete assessment on their chemical stabilities. Considering this effect has proven to be critical for designing materials with outstanding physical properties that have been intensively studied in the recent years, including the oxidative degradation of halide perovskites as photovoltaic materials,\cite{senocrate2019thermochemical,senocrate2018interaction} black phosphorous,\cite{zhou2016light,zhang2018degradation} as well as thermoelectric SnSe.\cite{li2016high} 

Instead of fixing the energies of the  elemental reference states in \cref{eq:formation_energy}, we recognise that these energies are thermodynamic state variables, where the most commonly used one being the the \emph{chemical potential} $\mu$. Practically, this means that rather than being a single--valued quantity, the formation energy $E_f$ is a multi--dimensional function of the chemical potentials of its constituent elements, \emph{e.g.}:
\begin{equation}\label{eq:formation_energy_chem_pot_func}
E_f(\mbox{A}_l\mbox{B}_m\mbox{C}_n; \mu_A,\mu_B,\mu_C) =\frac{1}{N}\left[ E\left(\mbox{A}_l\mbox{B}_m\mbox{C}_n\right)  -l\mu_{A}-m\mu_B-n\mu_C\right],
\end{equation}
whereby the stability region for a specific structure is the range of $\mu$'s within which $E_f<0$. It is apparent from \cref{eq:formation_energy_chem_pot_func} that  different compounds/phases formed with a fixed set of elements  will have different energies $E$, as such the stability regions for these different compounds/phases will cover different ranges of accessible $\mu$'s on the phase diagram.  This further means that for a given structure to be \emph{the} most stable one, its $E_f$ must be more negative than those for all other competing structures at some $\mu$'s, which may further narrow down the stability region for a given structure. Therefore, the ability to map out the the stability regions for different competing structures on the phase diagram  spanned by the chemical potentials of the constituent elements can provide the important guideline in determining the suitable experimental conditions  for fabricating a specific material/structure. An example of this is a (computationally determined) stability diagram for the (001) surfaces of a complex oxide \ce{LaMnO3} [\cref{fig:stability_measures}(c)],\cite{piskunov2008electronic} showing two distinct stability regions corresponding to the  surfaces with either \ce{LaO}-- and \ce{MnO2}--terminations, respectively, which can be applied to guide the experimental synthesis of complex oxide thin films with vacuum depositions. In particular, as shown in \cref{fig:stability_measures}(c), the chemical potentials of the oxygen gas can be determined by its partial pressure and temperature. 

Finally, we briefly consider the \textit{formation energies of charged defects}  in crystalline solids. The presence of point defects in materials plays an important role in determining its physical properties. The formation of defects in the materials is generally unavoidable, but may be controlled with a good understanding in their formation thermodynamics. Given its profound importance, there already exist many prior reviews specifically devoted to the discussions of defect chemistry and physics in crystalline solids.\cite{freysoldt2014first,anand2022convergent} As such, we will only briefly highlight the key ideas behind the defect thermodynamics here. Following Annand \emph{et al.}\cite{anand2022convergent}, consider the formation reaction of an oxygen vacancy V$_{\mbox{\footnotesize{O}}}$ in a binary oxide that are both mass and charge conserved:
\begin{equation*}
    \mbox{MO}\to\mbox{MO}_{1-\delta}+\delta \mbox{O} + q\delta \mbox{e}^{-},
\end{equation*}
where $q$ is the charge transferred from the solid into the reservoir upon the formation of the charged defect, and in this case $q=2$ Now if one follows the formulation of formation energy as in \cref{eq:formation_energy_chem_pot_func}, the  formation energy of a V$_{\mbox{\footnotesize{O}}}$ in the binary oxide MO would be
\begin{equation}\label{eq:formation_defect_chem_pot}
E_f\left(\mbox{V}_{\mbox{\footnotesize{O}}}\right)=\left[E\left(\mbox{MO}_{1-\delta}\right)-E\left(\mbox{MO}\right)\right]-\delta \mu_{\mbox{\footnotesize{O}}}-q\delta\mu_{e}.
\end{equation}
Here, $E\left(\mbox{MO}_{1-\delta}\right)$ and $E\left(\mbox{MO}\right)$ are the total energies of the solid with and without the oxygen vacancy. Generally, the chemical potential of an electron in a semiconductor is defined as the Fermi level $(E_F)$ with respect to the energy position of the valence band maximum (VBM), hence $\mu_e=E_{F}+E_{\mbox{\footnotesize{VBM}}}$. Replacing this definition of $\mu_e$ into \cref{eq:formation_defect_chem_pot}, it is apparent that $E_f\left(\mbox{V}_{\mbox{\footnotesize{O}}}\right)$ will not only depend on the chemical potential of the oxygen that is being removed from the solid, but also the Fermi level and the charge state of the oxygen vacancy. This concept is illustrated in \cref{fig:stability_measures}(d--e) where the formation energy landscapes of the native defects (with respect to $E_{F}$ and $q$) in perovskite \ce{BaSnO3} are compared at two different sets of chemical potentials at which the \ce{BaSnO3} phase is stable. It demonstrates that the native defects in undoped  \ce{BaSnO3} are energetically stable when $E_{F}$ is close to the VBM, posing a challenge to achieve $n$--type electrical conductivity in undoped  \ce{BaSnO3}.

\subsection{The Needs for Sampling the Energy Landscape} 
\label{sect:sampling_examples}

In the traditional material  research, computational modelling had largely been applied for post-experimental rationalisation, which often starts from a very limited set of well defined chemical compositions and crystal structures. This was fundamentally limited by the high  costs associated with the computations, which are usually performed at the DFT-level of theory, and the computational power available at the time. With  more efficient implementations of the electronic structure theory codes, \emph{e.g.} on the Graphic Processing  Units (GPUs)\cite{hutchinson2012vasp,maintz2018strategies}, and increasing computational power on the  High Performance Computing (HPC) facilities, it has become increasingly popular and important to perform  samplings on the chemical and structural spaces for a specific material system to obtain its energy landscape as governed by some of the energetic merits discussed above, in order to facilitate the holistic determination and understanding of the materials' structures and properties, as well as proposing new material systems with unconventional properties. It also inspires the continuous development of cheaper yet accurate modelling methods, in particular, machine-learning models, for us to simulate materials in complex scales that were not accessible before. Here, we shall provide a brief account of the recent literature to showcase different areas of applications in which structural sampling in computational material studies is particularly important and useful.

\subsubsection{Structure Elucidations} Whilst XRD (X-Ray Diffraction) and TEM (Transmission Electron Microscopy) are the two mostly used and standard method for determining the structures of a given material, there exists scenarios when, a high quality crystalline sample cannot be obtained for detailed structural characterisations, or the material is in a form that only exists under extreme environments that cannot be characterised with the standard experimental techniques.

For examples, solid state nuclear magnetic resonance (NMR)\cite{hodgkinson2020nmr} has now become an important technique for determining the crystal structures of organic molecules that only exist in a power form either naturally or due to to the experimental limitations, which prevents crystals for diffraction being obtained. The NMR signals are sensitive to the local chemical environment around the NMR-active nuclei in the crystal, which is in term determined by the overall crystal packing. Therefore, the NMR signal can be considered as a \emph{fingerprint} of the crystal structure, where the later can be determined if this particular type of fingerprint can be decoded. By first applying organic crystal structure predictions\cite{price2014predicting} with cheap atom-atom potential to generate a  set of hypothetical crystal structures that are thermodynamically stable, and then subjecting them to periodic DFT calculations that enable the determinations of their corresponding (theoretical) NMR chemical shift signals,\cite{van1993first,ashbrook2016combining} one can reveal the crystal structure of the compound being the one with its theoretical NMR signal that best matches the experimental result.\cite{baias2013novo,selent2017clathrate} 

Similar approaches have also been adopted to facilitate the determination of the crystal structures for inorganic minerals and small gaseous molecules under thermodynamic conditions that are inaccessible under standard laboratory conditions, which is typically the case for geo- and planetary sciences.\cite{wang2014perspective} For instances, the crystal structures of the superionic phase of ice\cite{wang2011high} had been proposed under a high pressure ($>$14 Mbar) and zero temperature, which is believed to exist in the cores of planets such as Uranus and Naptune,\cite{guillot1999interiors} among the solid phases of many other small molecules.\cite{han2019predicting,cable2020properties,ennis2020mixed,maynard2022crystal} Using the complementary crystal structure prediction algorithms developed for inorganic materials,\cite{glass2006uspex} it has enabled geoscientists to gain more detailed structural insights into the phase behaviours of minerals that exists in the earth's mantle, such as the post-aragonite phase of \ce{CaCO3} at a pressure above 40 GPa,\cite{oganov2006high} the anisotropic post-perovskite structure of \ce{MgSiO3} at 120 GPa,\cite{oganov2004theoretical} as well as iron-carbides and iron-hydrides at 330-364 GPa.\cite{bazhanova2012fe} More recently, it has also been applied to provide structural models for understanding the mechanisms of high-temperature superconductivities in metal hydrides,\cite{salke2019synthesis,troyan2021anomalous,chen2021high} in which the anharmonic behaviour of lattice vibrations (see \cref{sec:anharmonic_vibs}) are also believed to be playing a critical contribution to the observed superconducting behaviours. 

\subsubsection{Phase Diagrams} Rather than being used for determining a particular structure of a material, more importantly, the enumeration of the global energy landscape for material with a specific composition is a critical steps towards constructing their phase diagram under different thermodynamic boundary conditions, which is important for assessing the (meta)-stabilities of different phases.\cite{nyman2015static,sun2016thermodynamic} These information may be further used to guide the discovery  of new materials,\cite{aykol2018thermodynamic} most importantly, to plan and design the pathways of solid state synthesis of complex multicomponent solids.\cite{hautier2012accuracy,mcdermott2021graph} 

For instance, the voltage that can be delivered by Li-battery may be correlated with the free-energies of  Li-intercalated compounds with different Li concentrations,\cite{van2013understanding} which is an important principle that can be applied in the computational designs of ion batteries, provided that one can map out the compositional dependent phase diagrams containing the thermodynamic stability information for all phases that may be formed by the target set of elements. \cite{doe2008first,albina2018finite} Elemental gallium possesses many intriguing physical properties, such as low melting point, density anomaly and the coexistence of covalency and metallicity. Recently, a full phase diagram has been constructed as facilitated by the neural network (NN) potential, showing excellent agreement with the experimental phase diagram, and the formation of the metastable $\beta$-Ga is s kinetically driven process over the formation of the thermodynamically stable $\alpha$-Ga.\cite{niu2020ab} Application of computational crystal structure prediction to the W-Se system\cite{zhu2020ab} reveals that the trigonal crystal structure (space group $P\bar{3}m1$) containing layers of van der Waals bonded \ce{WSe2} is the thermodynamic stable phase under ambient temperature and pressure. Increasing the pressures to 33 GPa, then subsequently 105 GPa, an orthorhombic $(Pmmn)$ and a cubic $(Pm\bar{3}m)$ phase can be stabilised with a notable change in their atomic coordination environments. 

\subsubsection{Phase Transitions and Lattice Dynamics}   Beyond the enumeration of different phases for a given material in its phase diagram, there is a strong desire to understand both the thermodynamic and atomistic origins behind the transitions between different phases for a given material, in order to gain better control in their phase behaviours and corresponding physical properties. For example, the inorganic halide perovskite \ce{CsPbI3}  has a suitable bandgap of 1.73 eV and has attracted immense intention to be developed as the next-generation solar cell, but it has a low phase stability, making it prone to transform into the wide-band-gap tetragonal phase at room temperature.\cite{swarnkar2016quantum} As such, computer simulation has become pivotal in providing detailed atomistic understanding behind the phase transition mechanisms in order to devise new strategies to stabilize the desired phase of technologically important materials under their standard operating condition. This generally requires the performance of long-time molecular dynamic (MD) simulations to achieve a sufficient sampling of the phase space in order to detect  the  `rare-events', such as the occurrence of a phase transition. These samplings could be particularly challenging when bond breakage and rearrangements, as well as anharmonic vibrations are involved,\cite{lahnsteiner2018finite,marronnier2017structural} making classical (harmonic) forcefield become incapable\cite{lahnsteiner2018finite} of producing the phase transition events, whereas DFT becomes too costly to achieve a thorough sampling of the relevant phase space. 

Such a challenge has gradually be overcome with the development of machine-learning potentials that is capable of reproducing the DFT-PES but at the cost of parameterised atom-atom forcefields. In the early years, these potentials are pre-trained specifically for a given compound (typically with a simple chemical composition) using high-dimensional NNs\cite{behler2011neural}, to examine, for example, the pressure-assisted  graphite-to-diamond phase transformation,\cite{khaliullin2011nucleation}  and the mechanical behaviour of the phase changing compound GeTe.\cite{sosso2013fast}.  More recently, with the development of better  kernel for atomistic machine learnings,\cite{bartok2013representing,de2016comparing} that is highly effective in interpolating the  PES,\cite{bartok2010gaussian} new scheme to train machine-learning potential on-the-fly from DFT-MD samplings have been developed. This method has since been successfully applied to study the entropy-driven temperature-dependent phase transformations in methylammonium lead iodide and its inorganic counterparts,\cite{jinnouchi2019phase} revealing systematically, that the cation radii are playing an important role in determining the phase transition temperatures of lead halide perovskites. The same method has also be applied to study the melting behavior of Si,\cite{jinnouchi2019fly} and \ce{ZrO2}.\cite{verdi2021thermal}

At a more fundamental level, the emergence of phase transition can be attributed to the presence of lattice instability that is characterised by the possession of imaginary phonon modes under the harmonic approximation,\cite{pallikara2022physical} reflecting the highly anharmonic nature of the lattice vibration. In particular, by continuously distorting the structure along the eigenvector of a given imaginary mode is another viable way to determine the phase transition pathway between a pair of crystal polymorphs.\cite{rahim2020polymorph,pallikara2021phase} Modelling the lattice vibrations, in particular, the anharmonic contributions to the lattice vibrations, also involves a sampling process, either in a brute-force enumeration or a Bayesian inference approach, to obtain a set of atomic displacements and their corresponding atomic forces, which are more elaborately discussed in \cref{sect:phonon_harmonic_and_anharmonic}. This often forms an initial setup to further sample the thermal induced fluctuations in the electronic structures of solids, which are important for characterizing the electron-phonon coupling behaviours that subsequently determine their properties such as the (photoexcited) carrier life-time and thermoelectric transports, particularly in mechanically soft materials.\cite{li2018spin,li2019anharmonicity,yang2020correlating,yang2022mapping,fratini2017map}  The modelling of electron-phonon couplings will be further discussed in \cref{sec:eph_coupling}.

\subsubsection{Intrinsic Defects and Chemical Dopants} Manipulating the physical properties of materials through controlling the defects and chemical dopants have been the central theme in traditional material science. As such, extensive efforts have been devoted to apply modelling as a key tool in facilitating the determinations and understanding of the atomistic structures and associated physical properties of defects and dopants in solids, which has been extensively discussed in many prior reviews.\cite{freysoldt2014first,anand2022convergent} Here, we focus on a few aspects for which the idea of sampling is particularly prevalent in the computational investigations of defects and dopants in solids.

Both point defects and dopants break the symmetry and long-range periodicity of a crystalline solids, and they are typically modelled with the `supercell' approach in DFT.\cite{makov1995periodic} For this, it is often necessary to minimize the artefacts induced by the unphysical Coulombic interactions among the periodic images of the defects/dopants by either sampling supercells with increasing sizes to extrapolate the asymptotic limit at infinite cell,\cite{kantorovich1999elimination} or applying correction to the results obtained from small supercell.\cite{freysoldt2009fully} The increase in the computational power has propelled the community to investigate the defect and dopant chemistry in materials with more complex chemical compositions and crystal structures, meaning defects/dopants across a wider range of the periodic table, as well as multiple different lattice occupancies, can be sampled in the host materials. The latter is important to investigate the possible configurational entropic stabilisation of defects/dopants in the host lattice, which counterbalances the enthalpic penalties in their formations. These need to be further accompanied with, as shown in \cref{eq:formation_defect_chem_pot}, the investigations of charge-dependent stabilities of defects/dopants, which consequently determines their electronic energy levels in the band structures. 

Some examples that best illustrate the above aspects includes: The modelling the defect chemistries in lithium  solid electrolytes (\ce{Li7La3Zr2O12}) with a complex garnet-type crystal structure,\cite{miara2015first,squires2019native} cobaltate thermoelectrics with incommensurate layered structures,\cite{rebola2012first,lei2014synthesis} as well as complex oxides (\emph{e.g.} \ce{SrCoO}$_{x}$\cite{jeen2013topotactic} and \ce{SrFeO}$_{x}$\cite{nguyen2020topotactic}) that can undergo  topotatic phase transformations which can be tuned by the oxygen vacancy concentration. In the last case, it implies that the defect formation energy for a given compound at a given concentration needs to be additionally sampled across a few different phases in order to identify the thermodynamically most likely phase to accommodate oxygen vacancies with a specific concentration. The occurrence of defect complexes in solids, particularly hydrogens,\cite{morris2008hydrogen,lange2010native}  further combinatorially increases the size of the search space to identify the lowest-energy lattice configurations for these complexes. In this regards, whilst most of the prior computational studies constructed supercells containing defects/dopants through ion replacements on lattice site(s) followed by an energy minimisation to the \emph{local} minimum, a recent investigation\cite{mosquera2022identifying} has highlighted the fact that this approach does not guarantee the reach of their energetic ground-state defect/dopant configurations, rendering further configuration samplings around the local minimum should be further incorporated into the protocol for modelling defects/dopants in crystalline solids.

The above discussions have ignored the temperature contributions to the thermodynamic and physical properties of dopants or defects, which may come from a few different aspects that will be briefly discussed here. Firstly, the equilibrium concentration $n_{D}^{q}$ for a given dopant/defect $D$ in the charge state $q$ is given by minimizing the free energy of defect formation with respect to $n_{D}^{q}$,
\begin{equation*}
    G(T,p)=\sum_{q}n_D^q \Delta G_D^{q}(T,p)-G^{0}(T,p)-T\Delta S_{conf},
\end{equation*}
in which $G^{0}(T,p)$ and $\Delta G_D^{q}(T,p)$ are the temperature-dependent Gibbs formation energies of solid without and with the charged dopant/defect that may or may not further include the vibrational  (see \cref{sect:phonon}) and electronic contributions to the system entropies. Note that the formation of point defects in solids is expected to change the bonding strengths in a solid, and hence leads to a change in its phonon eigenspectrum and the vibrational entropies determined therefrom. The reduction in the crystal symmetry due to the presence of point defects also increases the number of symmetry inequivalent atomic displacements that need to be sampled for obtaining the phonon eigenspectrum.

Nevertheless, assuming all the defects are non-interacting hence different defect configurations are energetically degenerate, then the configuration entropy $S_{conf}$ is only dependent on the available dopant/defect sites ($N_D$) in the crystal and their corresponding occupancy $(n_{D}^{q})$.  (This is generally not the case, see the following discussions on high-entropy alloys) In this case, the minimisation problem $\partial G(T,p)/\partial n_{D}^q=0$ can be solved analytically, leading to the following Boltzmann-like expression for the equilibrium point defect/dopant concentration:\cite{kasamatsu2011theoretical}
\begin{equation*}\label{eq:defect_concentration}
    \left[D^q\right]=\frac{N_{D}}{V}\frac{\exp[-\beta\Delta E_f(D^{q})]}{1+\sum_{q^\prime} \exp[-\beta\Delta E_f(D^{q^\prime})] },
\end{equation*}
where $V$ is the cell volume per formula unit, $\beta=1/k_{B}T$ with $k_{B}$ being the Boltzmann constant, $q^{\prime}$ denotes all possible charge state for the defect/dopant, and $\Delta E_{f}$ being its corresponding formation energy [see \cref{eq:formation_defect_chem_pot}]. 

Secondly, in (strongly) anharmonic solids with notable electron-phonon coupling, additional molecular dynamic simulations may also be needed in order to model the thermally induced fluctuations in the defect/dopant electrons energy levels, in order to gain a more thorough understanding on the materials' physical properties close to the normal operating conditions, which may deviate significantly from the results obtained from static calculations.\cite{cohen2019breakdown,li2019influence,qiao2020extending,chu2020low}

\subsubsection{From Binary to High-Entropy Alloys}\label{sect:HEA}
Whilst defects and dopants are important in material designs, they are often present in the host lattice in low concentrations that do not fundamentally change the physical properties of the host materials. By \emph{randomly} intermixing two more more principal elements (often with similar chemical/physical properties) in high or equimolar concentrations to form homogeneous solid solutions (or alloys), it offers material scientists a  strategy to obtain new, high-entropy materials\cite{george2019high,oses2020high} with unusual chemical bonding states between pairs of elements, which otherwise may not be  stable as a binary compound. In these type of materials, the atoms for the same element occupying different lattice sites are surrounded by atoms of a different set of elements, which `confuses' the atoms to find an enthalpically favored pathway to `phase-segregate' into multiple different phases made from subsets of the constituent atoms. 

Thermodynamically, the stability of a multicomponent alloy is determined by its temperature-dependent formation free energy $\Delta G_f(T)=\Delta H_f-T\Delta S_{conf}$. Due to the presence of unusual bonding state, generally we have $\Delta H_f>0$, implying the system can only be stabilised if the configuration entropy that is contributed by different energetically (similar) atomic configurations in a multicomponent alloy, $T\Delta S_{conf}$ is larger than $\Delta H_f$. As pointed out in the seminal work by \citet{yeh2004nanostructured} and the discussions above on defects/dopants, when intermixing $n$ different principal elements \textit{in equimolar ratio} that do not have mutual interactions amongst each other, $S_{conf}=R\ln n$ (with $R$ being the ideal gas constant), showing $TS_{conf}>\Delta H_f$ is a condition that may be met when $n$ is large, such that a high-entropy system is often regarded as those with $n\geq 5$, although  $S_{conf}$ is not non-zero even in a binary mixture. However, in real materials, the principal elements interacts with each others, such that every unique atomic configuration of the principal elements in the lattice may have a different total energy $E^{(i)}$. This must be taken into consideration when determining the configuration entropy in a multicomponent alloy, which can be obtained through the Boltzmann relationship $G_{conf}=-k_{B}T\ln Z_{conf}$, where the configurational partition function
\begin{equation}\label{eq:configuration_partiton_function}
\mathcal{Z}_{conf}=\sum_{i=1}^{\mathcal{N}}g_{i}\exp[-\beta E^{(i)}],
\end{equation}
in which $g_i$ is the degeneracy factor. If one wishes to accurately determine the configurational entropy, then
\begin{align}\label{eq:conf_entropy}
    S&_{conf}=-\left(\frac{\partial G_{conf}}{\partial T}\right)_V\nonumber\\
    &=\frac{\partial}{\partial T}(k_{B}T\ln Z_{conf})_V\nonumber\\
    &=k_B \ln \sum_{i=1}^{\mathcal{N}}\exp[-\beta E^{(i)}]+\frac{1}{T}\frac{\sum_{i=1}^{\mathcal{N}}E^{(i)} \exp[-\beta E^{(i)}]}{\sum_{i=1}^{\mathcal{N}}\exp[-\beta E^{(i)}]},
\end{align}
assuming $g_i=1$. Strictly speaking, at finite temperature, the energy per configuration $E^{(i)}$ which enters \cref{eq:conf_entropy} should really be the Helmholtz free energy [\cref{eq:helmholtz}] that takes into account the vibrational entropic contributions at finite temperature.\cite{van2002effect} However, computing the Helmholtz free energy individually for each configuration across a wide set of configurations for high entropy materials with a brute-force approach is generally not feasible. Some recent attempt\cite{yang2025accounting} along this line can be seen with the use of deep-learning potentials that provide significant acceleration in atomistic simulations for new materials.  
Before proceeding, we note that the configurational entropy from the rotational degree-of-freedom is also important contribution to the free energies of flexible molecules, the determination of which has been a long time focus in theoretical chemistry.\cite{karplus1981method,hnizdo2008efficient,pracht2021calculation}
 
In solid-state materials, obtaining a converged estimation of $Z_{conf}$ based on \cref{eq:configuration_partiton_function} is a genuine sampling challenging, particularly when DFT is used for determining $E^{(i)}$. Firstly,  large supercell containing $N$ sites that can be replaced by up to $K$ different elements is needed to accurately simulate \emph{each} atomistic model of a high-entropy alloy, which itself is computationally demanding. Note that although there exists method, such as the virtual crystal approximation (VCA),\cite{bellaiche2000virtual} that enables the simulation of alloys using small unit cells using the compositionally averaged crystal potential, it will not allows us to explicitly enumerate individual $E^{(i)}$ term that goes into \cref{eq:configuration_partiton_function} for the corresponding determination of $Z_{conf}$. Secondly, the total number of configurations $\mathcal{N}=N!/\sum_{j=1}^k N_j!$,\cite{he2017formation} where $N_j$ is the number of sites that can be occupied by element $j$ and $\sum_{j=1}^k N_j=N$, showing the number of symmetrically independent configurations that need to be sampled grows factorially as $N$ (or equivalently, the supercell size) increases. Finally, the number of possible combinations of elements that can be obtained by choosing them from a pool of total $m$ elements, forming alloys with $i=2,3,\dots, m$ principal elements is $\sum_{i=2}^{m} C_{i}^{m}$, where $C_{i}^{m}$ is the number of combinations of $i$ items from a total of $m$. While this means we have a large chemical space (e.g. 8165 combinations in total for $m=13$) to search for new alloys, it also imposes additional challenges in computational material discoveries.

The enormously large number of configurations that can exist in a high-entropy alloy means there is a large variety of the local (metastable) chemical bonding environments that may not be found in the conventional materials, which may lead to outstanding catalytic properties. It also means that the interstitial crystal sites will exhibit a distribution in their sizes, with each being surrounded by a different set of ions, providing a higher probability for the high-entropy alloy to have the interstitial sites that are just big enough to accommodate hydride ions, making them to be ideal candidates to be further developed as hydrogen storage materials.\cite{marques2021review} In this regard, that is additional to determining the degree of entropy stabilisation in high-entropy alloys, sampling is important  in the computational design of functional high-entropy alloys as their overall properties, being either the catalytic performances or the hydrogen storage capacities, will be a statistical average over the probabilities that different atomic coordination environments or the interstitial void spaces will occur in their crystal lattice. For instances, by combining DFT and ML model, \citet{pedersen2020high} sampled the adsorption energies of CO on the surfaces of two high-entropy alloys \ce{CoCuGaNiZn} and \ce{AgAuCuPdPt} over 1000+ different local chemical environments (as characterised by the neighbouring atoms) of the adsorption sites, demonstrating a broad distribution in the absorption energies from weak non-binding ($\sim 0.5$ eV) to strong binding ($\sim -2$ eV) could be observed in a single high-entropy alloy. Similar approach has been applied to the design of high-entropy alloy catalysts for the oxygen reduction reactions (ORR),\cite{batchelor2021complex,pedersen2021bayesian} \ce{NH3} decomposition\cite{xie2019highly,yao2020computationally} and  \ce{CO2} reduction.\cite{chen2019identifying} Lastly, both the atomic binding energies and averaged atomic volumes (which can be correlated with the volumes of the interstitial sites) have been incorporated in the screening approach recently proposed by \citet{witman2021data} to optimize the compositions of the high-entropy alloys for hydrogen storage.

Inspired by the concept of high-entropy metal alloys, there is an increasing focus in the design of entropy stabilised functional ceramics, that are facilitated by the high-throughput computational screening. For instances, the emergence of methylammonium lead iodide (MAPbI$_3$) photoabosrbers\cite{zhou2014interface} has stimulated a significant body of works being devoted to the theoretical investigations of mixing different cationic/anionic species in the host halide perovskites to achieve tunable photovoltaic properties, including band gaps and carrier lifetimes, in a configurational-entropy-stabilised random alloy  that is resistant to phase separations.\cite{brivio2016thermodynamic} Some  examples include: A-site cation mixed perovskite (MA$_{x}$FA$_{1-x}$)PbI$_{3}$ (FA=formamidinium),\cite{senno2021mixed,schelhas2019insights} B-site cation mixed perovskite Cs(Pb$_{x}$Sn$_{1-x}$)X$_{3}$ (X=Cl, Br, I),\cite{yang2020mapping} FA(Pb$_{x}$Sn$_{1-x}$)I$_{3}$,\cite{kahmann2022compositional} X-site anion mixed perovskites CsPb(I$_{x}$Br$_{1-x}$)$_3$,\cite{yang2020correlating}  {C}{H}$_3${N}{H}$_3${P}b({I}$_{1-x}${B}r$_{x}$)$_3$,\cite{brivio2016thermodynamic} hetero-valent-cation-doped CsPbIBr$_2$,\cite{hu2022valence}  as well as low dimensional antimony perovskites Cs$_3$Sb$_2$X$_{y}$I$_{9-y}$.\cite{yang2021composition}  Anion mixing has also been applied to enhance the photovoltaic properties of kesterites, such as  Cu$_2$ZnSn(S,Se)$_4$.\cite{shang2014cation}
Because of the significant electronic interactions amongst two types of different but chemically similar ions occupying the same lattice site, resulting in a non-linear behaviour in their thermodynamic stabilities and electronic properties that deviates from the linear Vergard's law for a binary mixture,\cite{dalpian2019formation} performing configurational samplings in the computational studies of these perovskite alloys thus become particularly important and interesting in understanding their fundamental chemical and physical properties.  Alloying has also been adapted as a strategy to tune the electronic and phononic properties of layered two-dimensional materials such as {S}n({S}$_{1-x}${S}e$_x$)\cite{gunn2019thermodynamics,skelton2020lattice}  MXenes,\cite{han2020tailoring,nemani2021high} and transition metal dichalcogenides\cite{cavin20212d} for optical, thermoelectric and catalytic applications. The charge states of the transition metal B-site cations in complex oxide perovskites are sensitive to the oxygen stoichiometries, for which the subtle interplays between these two factors can be used to design thermal storage materials by interconverting between heat and the chemical energies in the B-O bonds,\cite{lei2022redox} This bonding strengths can be further manipulated by changing the B-site coordination environment through A-site alloying.\cite{sai2020exploring} 

It is important to note that the configurational entropy reflects the topology of the PES that is `accessible' at a given temperature $T$. Due to the high cost for computing the configurational entropy,  the term $E^{(i)}$ that enters \cref{eq:conf_entropy} is usually the enthalpy of the configuration corresponding to the local energy minimum at 0 K. However, at finite $T$, a configuration is not frozen at its minimum energy  atomic arrangement taking a single energy state of $E^{(i)}$, but constantly vibrating around the minimum within the local energy basin. In other words, there is no reason why the phononic contribution (\cref{sect:phonon}) to the stability of individual configuration should not be accounted for when calculating the configurational entropy. As formulated by \citet{van2002effect}, this can be achieved by replacing $E^{(i)}$ by its corresponding Helmholtz free energy [\cref{eq:helmholtz}, which is effectively the expected energy for a  given configuration that is confined and vibrating in its energy basin at $T$] in \cref{eq:conf_entropy} to effectively include the vibrational entropy $S_{vib}$ in the configurational entropy. Such a calculation would be prohibitively expensive with DFT, and  solution to such a challenge using machine-learning atom-atom potential has emerged recently, \cite{tolborg2023low} which, for the solid solution of Na$_{1-x}$K$_{x}$Cl, including $S_{vib}$ has resulted in the theoretically calculated miscibility dome for this particular solid solution become more closely resembling the one that was determined experimentally.

\subsubsection{Amorphous Solids} The amorphous (glassy) state of solids constitutes a higher level up in the structural complexities of solids. Compared to the high-entropy alloys, not only is the long-range periodicity absent in the amorphous solids, the short-range atomic environments as characterised by the bond lengths and angles, particularly, the coordination numbers, can be completely different even between two neighbouring atoms. These diverse chemical bonding environments have contributed to unique strengths, corrosion resistance and chemical activeness that benefit the application of amorphous materials as coating layers,\cite{joshi2016amorphous} catalysts,\cite{anantharaj2020amorphous}  photovoltaics,\cite{stuckelberger2017progress} and many others. A detailed understanding of the atomic bonding environments in amorphous solids will inevitably play a critical role to facilitate the understanding of the origin behind their unique physical and chemical properties. However, because there exists limited (experimental) knowledge of the detailed atomistic structures for amorphous solids, some types of unbiased sampling will be necessary for us to obtain an ensemble of realistic atomistic models of the amorphous solids. 

Computationally, this can be achieved from running molecular dynamic (MD) simulations following the typical temperature profile of how amorphous solids are formed experimentally, that is superheating the material and equilibrating it in its liquid state, followed by a rapid quenching to freeze the material into a snapshot of its nonequilibrium structure. A few challenges exist in performing this type of samplings. Similar to the high-entropy alloys, large supercells containing $10^3$-$10^5$ atoms are needed for simulating amorphous solids that are lack of the long-range periodicity. Although systems of this size can typically be handled with classical forcefields, they are typically parameterised (biased) against a subset of prototypical structures, resulting them being incapable of accurately reproducing the diverse chemical bonding environments that are revealed from detailed experimental characterisations.\cite{de2019transferability} Whilst this parameterisation problem can be mitigated by running the simulation with DFT, the poor scaling behaviour of DFT has significantly limited the system sizes (up to a few hundred atoms) that can be modelled, which must be further accompanied by a fast quenching rate ($10^{14}$ K/s).\cite{cooper2000density} On the other hand, a cooling rate that is three orders of magnitudes slower is required, for example, to correctly reproduce the bonding environments in amorphous silicon\cite{deringer2018realistic} as revealed by the solid-state NMR measurements. 

Significant progress in overcoming the aforementioned challenges to obtain realistic models of amorphous solids comes with the developments of ML forcefields that are trained on the DFT-determined energies and forces for a particular material with a diverse range of coordination environments, particularly the works on the application of the GAP (Gaussian Approximation Potential) model\cite{bartok2010gaussian} to amorphous systems as led by Cs\'{a}nyi and Deringer. For example, in one of the earliest works,\cite{deringer2017machine} GAP is applied to simulate the structures of both the bulk amorphous carbon (a-C) and its surfaces. Unlike the forcefield model, GAP was successful in reproducing the experimentally determined contents of $sp^3$-coordinated C atoms across the spectrum from low (1.5 gcm$^{-3}$) to high (3.5 gcm$^{-3}$) density a-C. It also shows that the extent of surface graphitisation increases with the annealing temperature, that is also in compliance with the experimental observations. The robustness of the model was subsequently put to test on simulating the depositions of a-C films with incident ion energies from 5 to 100 eV.\cite{caro2018growth,caro2020machine} The simulations showed that low impact energies give rise to low density a-C that is dominated by graphite-like $sp^2$-C, whereas high impact energies lead to the growth of high density a-C with predominately diamond-like $sp^3$-C. The simulation clearly revealed that, contradicting to the proposed `subplantation' mode, the high-density a-C is grown via a `peening' mode where the high-energy carbon impacted onto the surface generates a pressure wave that induces densifications away from the impact site. Using a cut-off radius (1.9 $\mbox{\AA}$) based criteria, it was also found that a small but non-negligible number of C atoms was in a fivefold coordination environment resulting from high energy deposition. Detailed bonding analysis from quantum mechanical calculation reveals that the fifth C atom, that is the atom furthest from the center atom but residing within the cut-off radius, shows negligible bonding strength with the center C. This means that the fivefold coordinated C is an artifact from how the coordination sphere was algorithmically identified, rather than a defect of the GAP model. 

Subsequently, the GAP model is  applied to investigate the amorphous silicon (a-Si) system using supercell containing $10^5$ atoms.\cite{deringer2021origins} The  vitrification process of Si is simulated from a liquid state at 1,500 K to a lower temperature of 1,175 K over a 1 ns-long MD trajectory, which clearly reveals the gradual decrease in the Si coordination number from $>8$ to the normal 4-fold tetrahedral coordination, that is accompanied with the lowering of the atomic energy, indicating the stabilisation effect. For the first time, this work also provided detailed atomistic insight into the pressure induced structural transitions in a-Si, showing that, at 13 GPa, the low-coordination a-Si transforms into a transient VHDA (very high density amorphous) state with coordination number $\geq 10$. Further increasing the pressure to 15 GPa, crystallisation is initiated with the emergence of crystalline domains that are primarily made of 6-fold coordinated Si, where the system fully transforms into a polycrystalline state at 20 GPa. More uniquely, the pressure-induced crystallisation is accompanied with an electronic  transition to the  metallic phase at pressure $\geq 12$ GPa, which is revealed by applying an ML model that enables the prediction of the electronic density-of-states (DOS) over an ensemble of disordered structures.\cite{mahmoud2020learning}

\subsubsection{Surfaces, Interfaces and Heterogeneous Catalysts} The creation of a surface breaks both the translational symmetry, as well as the equilibrium atomic coordination environments in the bulk crystal structure. Consequentially, this brings the system to a higher energy state, and the magnitude of this energy increase determines the thermodynamic stabilities of surfaces, which can be affected by factors such as the surface orientations and termination chemistries or stoichiometries, meaning that these factors need to be thoroughly sampled in order to obtain a comprehensive surface stability diagram for a given material [\emph{c.f.} \cref{fig:energy_landscape_examples}(c-d)]. To minimize such an energetic penalty in the surface creations, the topmost atomic layer will typically undergo surface reconstructions \emph{in vacuo} to compensate the surface dangling bonds, and some of these reconstructions can lead to a large surface unit cell\cite{kruger2006generalized,battaglia2009structure} that imposes additional challenges in their computational investigations. By joining two surfaces, from either different materials to form a heterointerface, or the same material to form grain boundaries, new physical properties may emerge. Although there exist a large body of theoretical works being devoted to  epitaxial heterointerfaces, less is so for the studies of grain boundaries, as the (often unknown) atomistic structures at the grain boundary regions are expected to be significantly different from the bulk, that is imposed by the crystallographic orientations and the grain boundary angles.\cite{chua2010genetic} This requires significant computational power to determine the lowest energy grain boundary structure through sampling, especially when the supercell size for sampling is generally inversely proportional to the grain boundary angles. Finally, accurate surface models are the pre-requisite for theoretically  investigating the  catalytic properties of heterogeneous catalysts, in which samplings are further needed to determine the catalytic active sites on the surfaces, the possible number of choices for which increases with lowering surface symmetry and degree of chemical complexities (such as the high-entropy alloys discussed above).

We note that the sampling can become more challenging if the surface adsorbates contain additional internal degrees of conformational flexibility and different charge states, given that the simulation of hybrid organic/inorganic interface itself is already considered to be theoretically challenging.\cite{draxl2014organic} For example, as demonstrated recently by \citet{maksimov2021conformational}, in which they have sampled more than 1000 configurations of arginine on three different metal surfaces, from which the results revealed the accessible molecular surface area of surface adsorbed arginine is strongly depending on the protonation state of the arginine molecule. 

\subsubsection{Closing Remarks}  
The above examples show that sampling is an overarching theme in modern theoretical material science, which is enabled by the increasing computational power and the development of ML methodologies. As a result, rather than focusing on examining a specific physical property $\mathcal{A}$ for material with a given atomic configuration $\bm{R}$, as how computation was used to be applied in supplementing experimental observations, there is an increasing capability for material theoreticians to better predict the properties of a specific material system by taking an `ensemble averaged' to determine the materials' expected property $\langle \mathcal{A}\rangle_{T}$ at their working temperature $T$ as
\begin{equation}\label{eq:general_sampling_1}
\langle \mathcal{A}\rangle_{T} = \int_{\bm{R}} \mathcal{A}(\bm{R})\mathcal{P}\left[E(\bm{R}),T\right]\mbox{d}\bm{R} \quad \mbox{(continuous)},
\end{equation}
or
\begin{equation}\label{eq:general_sampling_2}
\langle \mathcal{A}\rangle_{T} = 
\sum_i \mathcal{A}(\bm{R}_i)\mathcal{P}\left[E(\bm{R}_i),T\right] \quad \mbox{(discrete)}, 
\end{equation}
in which $\mathcal{P}$ denotes the corresponding statistical distribution of atomic configuration $\bm{R}$ with an energy of $E(\bm{R})$. It is obvious that the determination of $\langle \mathcal{A}\rangle_{T}$  is nowadays the key challenge that one wishes to tackle in modern computational material science, which heavily relies on our modelling capabilities to: (1) generate a database of structures $\left \{\bm{R}\right\}$ for new materials, (2) evaluate their corresponding stabilities $E(\bm{R})$ and (3) the associated properties $\mathcal{A}(\bm{R})$. The following three sections of this review will therefore be devoted for discussing these three aspects in more detail.

\section{Theoretical Methods for Material Modellings at Atomistic Scale}\label{chapter:energy_models}
Having outlined seven major themes to demonstrate energy landscape being a universal concept in (computational) material science, in this section, we will review the theoretical methods to obtain the total energy of the system $E$ as the function of the atomic positions $\bm{R}$.  

\subsection{Empirical Forcefield}
\label{sect:empirical_forcefield}

The empirical forcefield (FF) provides us with the most trivial conceptual understanding of the PES, since it seeks  an analytical expression for the total energy, where, in the most general form, it can be written as a series expansion of the  coordinates ($\bm{r}$) for two, three, up to $n$ interacting atoms in the system:
\begin{align}\label{eq:total_energy_atom_pos_expand}
E(\bm{R})=V_{0}+&\sum_{i}V^{(1)}(\bm{r}_{i},\bm{\beta}^{(1)})
               +\frac{1}{2!}\sum_{i,j}V^{(2)}(\bm{r}_{i},\bm{r}_{j},\bm{\beta}^{(2)})+\cdots\nonumber\\
               &+\frac{1}{n!}\sum_{i,j,\ldots,n}V^{(n)}(\bm{r}_{i},\bm{r}_{j},\ldots,\bm{r}_{n},\bm{\beta}^{(n)}).
\end{align} 
Here, each term of $V^{(i)}$ takes a well-defined mathematical form that is formulated based on the known physics of (classical) mechanics and electrostatics. With this, both the total system energy and its derivates with respect to the atomic positions can be easily and efficiently calculated, enabling us to study the structures and properties of materials at the length  (more than a million atoms) and time (up to milliseconds\cite{pierce2012routine}) scales that are relevant to experimental observations. Examples that are specifically relevant to material science include: the determination of phonon density-of-states (see \cref{sect:phonon}) to supplement the inelastic neutron scattering experiments,\cite{cheng2019simulation} thermal transport,\cite{esfarjani2011heat} crystallisation and polymorphism in molecular crystals,\cite{raiteri2005exploring} phase transformation,\cite{pols2021atomistic} and many others. We would like to highlight that, while the current discussions  around \cref{eq:total_energy_atom_pos_expand} focus on atom-atom interaction, the idea of many-body expansion is general across many aspects of material modelling. For example, in determining the total interaction energy of non-covalently interacting molecular aggregates using the fragmented approach,\cite{mahadevi2016cooperativity,wen2012practical} the series will be expanded based on the positions of $n$--interacting molecules, in which the higher--order terms are particularly important for the interaction energies contributed by the long-ranged effect of charge polarisation. As will be further discussed in \cref{sec:anharmonic_vibs}, the anharmonic phonon vibrations in solids are primarily dictated by the non-negligible higher-order terms $(n>2)$ in   \cref{eq:total_energy_atom_pos_expand}. 

The key idea to review the constructions of major  forcefields used in material and molecular simulation is to be compared with the later review in the development of machine-learning forcefield, which learns the PES $E(\bm{R})$ solely from the training data without explicitly assuming the underlying functional forms that are necessary for describing these interactions.

\subsubsection{Parameterisation of the Forcefield}

FF methods are commonly referred to as being \emph{empirical} because they contain a set of parameters $\{\bm{\beta}^{(i)}\}$ [\cref{eq:total_energy_atom_pos_expand}] that is specifically fitted to a particular material system for a particular functional form of the FF, such that the materials' properties obtained from the FF  $\mathcal{P}^{FF}(\{\bm{\beta}^{(i)}\})$ reproduce the targeted properties $\mathcal{P}^{target}$ obtained from either experiments or higher level theoretical calculations. In another word, the values of $\{\bm{\beta}^{(i)}\}$  are determined by minimising the following error:
\begin{equation}\label{eq:FF_fitting}
\min_{\bm{\beta}}\left\Vert \mathcal{P}^{FF}(\{\bm{\beta}^{(i)}\})-\mathcal{P}^{target}\right\Vert.
\end{equation}
Before we move on to discuss some of the common FF types in materials modelling, we will provide a brief account of different choices $\mathcal{P}^{target}$ that have been used in fitting the FF parameters.

For the use of experimental data as $\mathcal{P}^{target}$, several common choices have been reported in the literature. For instances, the measured sublimation enthalpy of crystal can be directly correlated with the cohesive energy calculated from a model FF. These thermochemical data, which are often augmented with structural data such as lattice constants, have been applied to fit the FF parameters for crystalline materials, such as molecular crystals.\cite{williams1967nonbonded,warshel1970consistent,pyzer2016optimized,bunte2000molecular} 
For biological and aqueous systems, neutron scattering is capable of providing detailed structural and dynamic informations, particularly for its high sensitivity to the atomic position of the hydrogen atoms. In this regard, the energy and momentum dependent neutron scattering cross section $S(\bm{Q},E)$  has been applied as $\mathcal{P}^{target}$ in refining the FF  to obtain more accurate structures for biomolecules\cite{lynch2017automated,pluhackova2015extension} and the hydration environments of the physiologically important ions.\cite{martinek2018calcium,kohagen2016accounting,duboue2017hydration} The vibrational energy levels, which is determined by the Hessian of the PES function [\cref{eq:force_constant}], can be compared to the vibrational frequencies measured from the Raman spectroscopy to refine $\{\bm{\beta}^{(i)}\}$. Such strategy had been applied in developing the FF for $n$-butane,\cite{murphy1991harmonic} and $\beta$-\ce{Ga2O3}.\cite{dohy1982raman} The Raman intensities, which can be  calculated from the dipole-dipole autocorrelation functions extracted from molecular dynamic trajectories, have also been applied as $\mathcal{P}^{target}$ to refine the polarisable FF for water.\cite{sidler2018efficient} Finally, X-ray and electron diffraction measurements can provide direct and detailed information on the equilibrium bond lengths, bond angles and van der Waals (vdW) interaction distances, which are some of the important parameters needed in the FF.\cite{sun1998compass,zgarbova2013toward,aytenfisu2017revised} 

By parameterising the FF directly against experimental data provides a significant advantage that enables the predicted properties $\mathcal{P}^{FF}$ for new systems modelled with the FF to become directly comparable with the experiments for verifying the theoretical predictions. However, the qualities and the consistency of the experimental data, the account of thermal effects, as well as the difficulties in simultaneously fitting multiple inter-dependent parameters, make the parameterisation of the empirical FF a non-trivial, often  tedious process. Some of these challenges may be mitigated with the usage of training data that is generated from \emph{ab initio} calculations.

For examples, by simultaneously fitting both the intra- and inter-molecular PES to the one that is consistently derived from periodic DFT calculations, Neumann\cite{neumann2008tailor} developed a recipe for constructing tailored-made FF for crystals of flexible organic molecules, thus enabling the accurate ranking of the relative stabilities of different polymorphs of organic crystals which are typically separated by energy gaps in the order of $\sim 1$ kJ/mol. Achieving such a level of accuracy has important implications in   guiding the screening of crystallisations of drug molecules in the pharmaceutical industry.\cite{neumann2015combined} Note that Neumann adapted a simple Buckingham\cite{buckingham1938classical} potential of the form $E(r)=A\exp(Br)-C/r^6$ in his FF, which is a variation of the Lennard-Jones potential commonly used in many empirical forcefields (see Fig. 2 in \citet{heinz2013thermodynamically}),  to describe the vdW interaction.  One of the challenges here is to reach a balanced fitting for the sharp-rising exponential function at short-distance and slowly varying $1/r^6$-dependency at long range. The application of Symmetry-Adapted Perturbation Theory (SAPT)\cite{misquitta2005intermolecular}, which is an electronic structure theoretical method, enables the intermolecular interaction energies to be decomposed into  contributing terms, whereby each term can be individually fitted to a well-defined analytical function,\cite{misquitta2016ab} thus mitigating the problem of cross-coupling among the FF parameters. In an effort to make the FF parameterisation become more of a `black-box' method, Grimme proposed a quantum mechanically derived forcefield (QMDFF) for organic molecules that can be parameterised using only the results from quantum mechanical calculations,\cite{grimme2014general} particularly the atomic charges and the Hessian, albeit the resulting FF is only aimed to reproduce the PES close to a specific equilibrium structure obtained from quantum mechanical calculation. 

Having briefly reviewed the strategies for parameterising the empirical FFs, we moved on to give a high-level overview of the functional forms of the major FFs that are commonly used in atomistic modelings. The discussion here will mainly focus on how different types of  FFs are categorised according to the fundamental physics behind the atom-atom interactions that are aimed to be captured by the specific functional forms of the FFs, with some modifications following the work from \citet{heinz2013thermodynamically} More specifically, they are (a) common harmonic that describes interactions among the bonded (including both covalent/ionic and vdW) atoms, (b)  many-body/reactive FFs in which the intramolecularly bonded atoms are allowed to vary during the simulation, and finally, (c) polarisable FFs which are able to account for the change in the strong electrostatic interactions resulted from the deformations of the electron densities as the interacting atoms approach each others. 

\subsubsection{Common Harmonic Terms}
In real practice, most of the FFs applied in material modelling do not use a functional form  that is dependent directly on the Cartesian coordinates of the interacting atoms [\emph{c.f.} \cref{eq:total_energy_atom_pos_expand}], but the internal coordinates that represent the bond lengths ($r_{ij}$, between a pair of atoms), bond angles ($\theta_{ijk}$, among a triplet of atoms) and torsional angles ($\varphi_{ijkl}$, among a quartet of atoms). Such a choice is not only based on our intuitive understanding on the material chemistry, but also, as has been demonstrated in many prior works,\cite{baker1999geometry,buvcko2005geometry,echenique2006definition} that the molecular motions along the curvilinear space spanned by (an appropriate choice for the set of) internal coordinates tend to follow low-energy-barrier pathways along the PES. Consequently, this can improve the stabilities in gradient-based geometry optimisations, as well as the efficiencies in sampling the PES  for structure predictions (see \cref{sect:MonteCarlo}). 

With this, the lowest level of approximation to construct the FF is to assume all bonded interactions are \emph{harmonic}, whereas the electrostatic interactions between pairs of atoms are modelled in the form of simple Coulomb interactions between \textit{fixed charges}. For example, the CHARMM FF that is commonly used in modelling biomolecular systems takes the following specific form:\cite{mackerell1998all}
\begin{align}\label{eq:charmm_general}
    E=&\sum_{ij \mbox{\scriptsize{,bonded}}}V_{\mbox{\scriptsize{bond}}}\left(r_{ij}\right)
    +\sum_{ijk,\mbox{\scriptsize{bonded}}}V_{\mbox{\scriptsize{angle}}}\left(\theta_{ijk}\right)+\sum_{ijkl,\mbox{\scriptsize{bonded}}}V_{\mbox{\scriptsize{proper-tor}}}\left(\varphi_{ijkl}\right)\nonumber\\
&+\sum_{ijkl,\mbox{\scriptsize{bonded}}}V_{\mbox{\scriptsize{improper-tor}}}\left(\chi_{ijkl}\right)+\sum_{ij,\mbox{\scriptsize{nonbonded}}}V_{\mbox{\scriptsize{Coulomb}}}\left(r_{ij}\right)+\sum_{ij,\mbox{\scriptsize{nonbonded}}}V_{\mbox{\scriptsize{LJ}}}\left(r_{ij}\right),
\end{align}
where each term in the above expression is
\begin{subequations}
\begin{align}
V_{\mbox{\scriptsize{bond}}}\left(r_{ij}\right)&=K_{r,ij}\left(r_{ij}-r_{ij,\mbox{\scriptsize{eq}}}\right)^2\label{eq:bond}\\
V_{\mbox{\scriptsize{angle}}}\left(\theta_{ijk}\right)&=K_{\theta,ijk}\left(\theta_{ijk}-\theta_{ijk,\mbox{\scriptsize{eq}}}\right)^2\label{eq:angle}\\
V_{\mbox{\scriptsize{proper-tor}}}\left(\varphi_{ijkl}\right)&=\frac{1}{2}K_{\varphi,ijkl}\left[1+\cos\left(n\varphi_{ijkl}-\varphi_{ijkl,\mbox{\scriptsize{eq}}}\right)\right]\label{eq:proper_tor}\\
V_{\mbox{\scriptsize{improper-tor}}}\left(\chi_{ijkl}\right)&=K_{\chi,ijkl}\left(\chi_{ijkl}-\chi_{ijkl,\mbox{\scriptsize{eq}}}\right)^2\label{eq:improper-tor}\\
V_{\mbox{\scriptsize{Coulomb}}}\left(r_{ij}\right)&=\frac{1}{4\pi\varepsilon}\frac{Q_{i} Q_{j}}{r_{ij}}\label{eq:Coulomb}\\
V_{\mbox{\scriptsize{LJ}}}\left(r_{ij}\right)&=\varepsilon_{ij}\left[\left(\frac{r_{ij,\mbox{\scriptsize{eq}}}}{r_{ij}}\right)^{12}-2\left(\frac{r_{ij,\mbox{\scriptsize{eq}}}}{r_{ij}}\right)^{6}\right]\label{eq:LJ}.
\end{align}
\end{subequations}
FFs adapting this type of functional from are usually dubbed as the Class I FF, which, apart from the CHARMM FF, also include FFs such as AMBER,\cite{pearlman1995amber} OPLS-AA,\cite{jorgensen1996development} DREIDING\cite{mayo1990dreiding} and many others. The so-called Class II FF (\emph{e.g.} the COMPASS\cite{sun1998compass} FF) includes cubic and quartic terms that accounts for the anharmonic effects (see \cref{sec:anharmonic_vibs}) in molecular motions, as well as the cross-coupling terms, among the bonded atoms. The latter terms are included because  internal coordinates  are not necessarily orthogonal to each other, changing the torsional angles in a molecule, for example, may simultaneously requires certain bond angles to be changed at the same time in order to avoid the steric clashes of the bulky side groups. The inclusion of the cross-coupling term increases the flexibility of the FF to capture this important physical behaviour, which can be  important for dynamic simulations to guide the molecule to transform from one conformation to another by traversing through part of the PES with lower transition barriers. 

The simple functional forms adopted in both the Class I and II FF to describe the PES also imposes some limitations in modelling the physical and chemical behaviours of materials. Here, we focuses on the two major limitations.

Firstly, the FF assumes and maintains a fixed bond topology, meaning no chemical reactions (forming/breaking of the covalent bonds) can occur throughout the simulation. Fundamentally, this is limited by the functional form [\emph{c.f.} \cref{eq:bond}] chosen to describe the bond stretching, which does not include the long-range asymptotic behaviour $E(r\to\infty)=0$ that promotes bond breakage. In addition, \cref{eq:bond} only contains \emph{one} well-defined local minimum at the (parameterised) equilibrium bond distance between the two atoms $i$ and $j$ that are predefined to be bonded together, meaning this particular functional form of the FF is physically constraining the motion of the atoms in the vicinity around the local minimum at  $r_{ij,\mbox{\scriptsize{eq}}}$ on the PES. Secondly, both Class I and II FFs model the electrostatic interactions [\cref{eq:Coulomb}] between two atoms by using fixed point charges on atoms. Whilst this simple form of Coulomb interaction is extremely fast to evaluate in large-scale atomistic modelling, it ignore the important physics of the electronic charge densities around atoms, the shape of which can change when two atoms come close to each other, contributing to the important polarisation effects that consequently affect the strength of the Coulomb interactions between two interacting atomic moieties. For instances, ignoring the charge polarisation may lead to 40 \% error in the interaction energies between two water molecules when the intermolecular oxygen--oxygen distances fall under 3 $\mbox{\AA}$.\cite{ferenczy2001modeling} In order to model the PES  over a wider range and with better accuracy,  reactive and polarisable FFs have been proposed and developed to overcome these two limitations, respectively, which will be briefly reviewed in the following sections.

\subsubsection{Many-Body and Reactive Forcefields}

A useful approach to overcome the limitation imposed by the predefined bond topology/connectivity is to consider the total system energy as a sum of  the energies from individual atoms in the system, each interacting with multiple neighbouring atom that can be identified as those fall within a predefined cut-off distance $r_c$ to the central atom. This  constitutes a many-body FF as the number of atoms that can be considered as interacting with the central atom grows as $r_c$ increases, which is particularly useful for developing FFs for solids,  as the atoms in solids typically have  higher coordination numbers compared to the number of bonds that can form with the C atoms in organic molecules. It also allows the number of neighbouring atoms to changes, thus making it possible to model the change in the total system energy as a result of a chemical reaction. As shall be demonstrated in the latter sections [\cref{sec:physics_informed_representation}], the fundamental concept behind many-body FFs has played a particularly significant role in the design of atom-centred structure descriptors for modern machine-learning methods in material modelling. For this reason, and its importance in modelling inorganic crystalline solids, which is of the main interest for the present review, we will focus on three major many-body/reactive FFs in more details, namely the embedded atom method (EAM), the bond-order potential (BOP) and the charge optimised many-body (COMB) potential.
 
\paragraph{\textbf{The embedded atom method}} was first proposed by \citet{daw1984embedded} for modelling the impurities, defects and surfaces of metals, to overcome the unsatisfactory results in using pairwise FF in treating systems such as H in the metal lattice. The EAM is inspired from the Hohenberg-Kohn formulation of the DFT\cite{hohenberg1964inhomogeneous} which views each atom in the solid as an impurity that is embedded in the potential generated by all the other atoms in the solid. With this, the total energy in the EAM takes the following form:
\begin{equation}\label{eq:EAM_total_energy}
E=\sum_{i}^N F(\rho_{h,i})+\frac{1}{2}\sum_{i\neq j}\phi(r_{ij}).
\end{equation}
Here, $F(\rho_{h,i})$ is the energy required to embed atom at the site $i$ of the host lattice, which is a function of the electron density of all atoms in the host lattice at this site $\rho_{h,i}$. The latter can be built by summing a (model) electron densities at all sites except $i$, \emph{i.e.} $\rho_{h,i}=\sum_{j\neq i}\rho_j^a(R_{ij})$. The  core-core repulsive interaction is described by the term $\phi(r_{ij})$. Similar to the pair potentials discussed in the previous section, practical application of EAM potential will require a specific functional form for the embedding and core-core repulsion potentials,  where some of the popular choices are briefly discussed below which a specific focus on the physical motivations behind the chosen functional forms.

In the original formulation of the EAM for Ni and Pd,\cite{daw1984embedded}  the  charge densities $\rho_i(r)$ were obtained from the parameters for the atomic wavefunctions for the $s$ and $d$ orbitals determined from the \emph{ab initio} Hartree-Fock calculations.\cite{clementi1974roothaan} With this, the total charge density of an atom $\rho(r)=N_s\rho_s(r)+(N-N_s)\rho_d (r)$ where $N$ is the total number of outer electrons and $N_s$ is the number of $s$ electrons.  From here, $F(\rho_{h,i})$ was obtained as a cubic spline function satisfying the constraint of it being of a monotonically decreasing function from $F(\rho)\vert_{\rho=0}=0$, which is set to be the zero point energy. Because  $N_s$ cannot be determined from homonuclear data, it becomes a free-parameters in the potential which must be determined by fitting to the experimental data (such as lattice constants, elastic constants and cohesive energies of the equilibrium structure), in order to determine the exact shape of $F(\rho)$. For the core-core repulsion energy, a simple form of Coulombic interaction $\phi_{ij}(r)=Z_i(r)Z_j(r)/r$ is adopted. The distance-dependent effective charge $Z(r)$ is also numerically fitted, which must satisfy the boundary conditions where $Z(0)=Z^{*}$ (the atomic number) and must asymptotically decrease to 0 when $r\to a_0$, which is the atomic radius. In a subsequent work, \citet{foiles1986embedded} developed a consistent form of the embedding function and short-ranged repulsive pair interactions for Cu, Ag, Au, Ni, Pd, Pt and their binary alloys. In this work, the information about $F(\rho)$ was obtained from the knowledge of system energies away from $\rho_{eq}$, the charge densities at the equilibrium crystal structure, when the latter is expanded or shrinked. In particular, this takes the form of $E(a)=-E_{sub}(1+\Delta a)e^{-\Delta a}$, where $E_{sub}$ is the sublimation enthalpy of the metal, and $\Delta a=\left[\frac{a}{a_{0}}-1\right]/\sqrt{\frac{E_{sub}}{9B\Omega}}$ measures the deviation from the equilibrium lattice constant $a_{0}$ where $B$ is the Bulk modulus and $\Omega$is the equilibrium cell volume per atom. 

Later in the work by \citet{ackland2004development}, where EAM potential was developed for modelling P impurity embedded in Fe, the embedding potential was chosen to take the simple form of $F(x)=-\sqrt{x}+a_2 x^2+a_4 x^4$. The $\sqrt{x}$-dependency takes the inspiration from the tight-binding model\cite{finnis1984simple} with the quadratic and quartic terms accounts for the kinetic energy of electrons in the hopping and free-electron regime. The model parameters were obtained by fitting to the data derived from \emph{ab initio} calculations, including crystal structures, surface energies, lattice constants, as well as vacancy and defect formation energies. In the other works which aimed at modelling the Ni-H\cite{angelo1995trapping,baskes1997trapping} and W-H\cite{wang2017embedded} systems, the function $F(x)=c_6 x^2+c_7 x+\frac{c_8 x^{5/3}}{c_4+x}$ was proposed to be a suitable form for modelling H in the metal matrix. The spherical atomic densities was assumed which was further augmented with a exponentially decaying cut-off function $f_{cut}(r)=\exp\left[\frac{1}{r-r_{cut}}\right]$. The usage of cut-off function with similar form is quite common in the construction of atom-centred structural descriptors\cite{bartok2013representing,behler2011neural} for atomistic machine-learnings. Finally, \citet{sheng2011highly} developed a set of highly optimised EAM potentials for 14 FCC metals. To ensure the properties modelled by the parameterised EAM potentials, the \emph{ab initio} determined lattice constants and energies were further re-scaled to the experimental values during the parameterisation process. They adopted a quintic spline function for the embedding function, which enables them to obtain a smooth interpolation of the Hessian matrix, leading to the successful reproduction of the phonon dispersion relationship for Au that is closer to the 80 K experimental results from neutron diffraction, compared with that derived from pure \emph{ab initio} calculation.  By explicitly including the \emph{ab initio} molecular dynamics data for simulating the melting and liquid states of the metals, the EAM potential is capable of sampling the liquid structures and dynamics for the metals that have been parameterised. 

\paragraph{The bond-order potential} which can be represented by the Tersoff potential\cite{tersoff1986new} that is designed for covalent materials, such as silicon, to model its surface reconstruction behaviours that cannot be satisfactorily described by a simple two-body and three-body potential due to a fundamental change in the atomic coordination environments upon the formation of covalent surfaces. Similar to the two body potentials, the Tersoff potential model the total system energy as the sum of energies for all interacting atomic pairs $E=\sum_{j>i}V_{ij}$, where the pair interaction energy is the balance between the attractive and repulsive interaction energies that both take an exponential form of distance-dependency\cite{abell1985empirical}:
\begin{equation}\label{eq:tersoff}
V(r_{ij})=f_c(r_{ij})\left[A^{\mbox{\scriptsize{repul}}}\exp\left(-\lambda^{\mbox{\scriptsize{repul}}}r_{ij}\right)
-B^{\mbox{\scriptsize{attr}}}_{ij}\exp\left(-\lambda^{\mbox{\scriptsize{attr}}}r_{ij}\right)\right],
\end{equation}
in which $A$, $B_{ij}$ and $\lambda$'s are parameters and $f_c(r_{ij})$ is the smooth cut-off function, thus the summation $\sum_{j>i}$ operation will be performed for all atom pairs that fall within the cut-off radius $r_{cut}$ of any given central atom. Unlike the simple pair-potentials, the attractive parameter $B_{ij}$ takes into account the bond orders, as well as the change in the bonding strength due to the change in the local bonding environments (\emph{i.e.} the bond lengths and angles formed between atom $i$ and other atoms $k\neq j$). More explicitly, 
\begin{subequations}
    \begin{align}
    B_{ij}&=B_{0}\exp(z_{ij}/b)\label{eq:tersoff_b}\\
        z_{ij}&=\sum_{k\neq i,j}\left(\frac{w(r_{ik})}{w(r_{ij})}\right)^n \frac{1}{c+\exp(-d\cos\theta_{ijk})}\label{eq:tersoff_z}\\
        w(r_{ij})&=f_c(r_{ij})\exp\left(\lambda^{\mbox{\scriptsize{attr}}}r_{ij}\right)\label{eq:tersoff_w}
    \end{align}
\end{subequations}
 Physically, $z_{ij}$ measures the number of competing bonds with bond $ij$ and the parameter $b$ determines how rapidly the strength of bond $ij$ falls with the increase in the bond coordination number. $w(r_{ij})$  is the unscaled bond strength, and the first term in \cref{eq:tersoff_z} is the ratio in bond strengths between bond $ik$ and $ij$, where the power $n$ determines how significant bonds between closely neighbouring atoms are weighted against the distanced atoms. Finally, the second term in \cref{eq:tersoff_z} gives the weighting factor based on the bond angle $\theta_{ijk}$. Fundamentally, a change in the local coordination number changes the shape of the charge densities around the central atom, thus affects the overlap of charge densities between the neighbouring atoms which consequentially changes the bond strengths.\cite{abell1985empirical} In a later development, for example, \citet{brenner1990empirical} extended the bond order potential to account for the occurrence of chemical reactions by further refining the $B$ parameters depending on the nature of the chemical bonds (\emph{e.g.} the $\sigma$ and $\pi$ bonds). Similar idea of local coordination dependency is also reflected in Grimme's --D3\cite{grimme2010consistent}  scheme to correct the inaccuracy of DFT in dealing with the non-covalent interactions, where the elemental $C_6$ coefficient for the dispersion interaction can be attenuated depending on the specific coordination number of that element in a particular compound. Many structural fingerprints [\emph{c.f.} \cref{sec:feature_eng}] that are devised for atomistic machine learnings, such as SOAP\cite{bartok2013representing} and Behler's atom-centred symmetry functions,\cite{behler2011atom} can see their inspirations from the construction of bond-order potentials, such as the one of Tersoff.
 
Several material systems have been parameterised with their corresponding Tersoff potential, mostly against \emph{ab initio} training data. For instances, \citet{lindsay2010optimized} optimised the potential for graphene and carbon nanotubes against their experimental phononic properties, with a specific aim in applying the potential for modelling the phonon transport properties. More specifically, in order to model the bonding environments in the carbon nanotube, an additional term that depends on the  torsion angles coupling four atoms in the system is included for the $B$ parameter. \citet{munetoh2007interatomic} parameterised the potential for the Si-O system, which is capable of reproducing the properties of various crystalline silica polymorphs as well as its amorphous state. The parameters optimised by \citet{kumar2016optimised} for  B-N enable them to perform molecular dynamic studies for examining the fracture behaviours of two-dimensional BN nanosheets in atomistic details. \citet{billeter2006ab}  developed the Tersoff potential for the silicon oxynitride system, which can be applied for investigating its deposition behaviour on the silicon substrate, for its promising physical properties to be applied as gate dielectrics in the microelectronics. Similar potential has been developed for a series of semiconductor nitrides,\cite{karaaslan2020assessment} and In-doped GaAs systems.\cite{titantah2007bond}
 
\paragraph{\textbf{Charge optimised many-body potential} } In all types of many-body potentials discussed above, the electrostatic energy between two interacting atoms is assumed to possess a fixed charge that is independent from its chemical environment. This poses a fundamental limitation in using the potential to model systems that involves complex chemical bonding environments that may also be dynamically varying,\cite{liang2013classical} such as the heterogeneous interfaces formed between two different materials, as well as  the chemical reactions occurring at these interfaces where the charges on atoms changes as they migrate from one material from another. 

A remedy for this problem is to extend the QEq charge equilibration approach\cite{rappe1991charge} that is widely used in the \textit{ab initio} atomistic simulations into empirical potentials, which leads to the development of the charge optimised many-body (COMB) potentials. In this case, the functional form for the total energy of the system is not only just the function of the atomic position vector $\bm{r}$, but also the vector $\bm{q}$ of the point charges on individual atom. With this,  the COMB potential takes the following form:\cite{yu2007charge}
\begin{equation*}
    E(\bm{q},\bm{r})=V_{\mbox{\scriptsize{short}}}(\bm{q},\bm{r}) +V_{\mbox{\scriptsize{es}}}(\bm{q},\bm{r})+V_{\mbox{\scriptsize{vdw}}}(\bm{r})+V_{\mbox{\scriptsize{corr}}}(\bm{r}).
\end{equation*}
Here, the short-ranged interaction $V_{\mbox{\scriptsize{short}}}$ takes the form of the bond-order potential [\cref{eq:tersoff}], except now both the attractive and repulsive terms have extra term that is charge dependent.  $V_{\mbox{\scriptsize{vdw}}}$ describes the van der Waals interaction between two atoms and takes the form of a 12-6 potential. $V_{\mbox{\scriptsize{corr}}}$ is introduced in the latest version of the COMB potential to additionally improve the description of bond angle terms. Finally, the electrostatic interactions \begin{equation*}
V_{\mbox{\scriptsize{es}}}=V_{\mbox{\scriptsize{self}}}+V_{\mbox{\scriptsize{qq}}}+V_{\mbox{\scriptsize{qZ}}}+V_{\mbox{\scriptsize{polar}}}
\end{equation*} 
is the summed interaction energies contributed from: (a) the self energy $V_{\mbox{\scriptsize{self}}}(\bm{q},\bm{r})$ describes the energy cost associated with the formation of a charge ion, which generally takes the form of a Taylor series expansion based on the atomic charge, (b) $V_{\mbox{\scriptsize{qq}}}(\bm{q},\bm{r})=\frac{1}{2}\bm{q}\mathbf{J}^{qq}\bm{q}$ is the Coulomb interaction between two spherical $s$--type charge densities with the strength of the interaction given by the tensor $\mathbf{J}^{qq}$, similarly
(c) $V_{\mbox{\scriptsize{qZ}}}(\bm{q},\bm{r})=\frac{1}{2}\bm{q}\mathbf{J}^{qZ}\bm{Z}$ is the Coulomb interactions between electrons and the nuclei. Finally, to further improve the accuracy of the electrostatic interactions, the polarisability interactions (d) $V_{\mbox{\scriptsize{polar}}}=\sum_i \frac{\mu_i^2}{2P_i}+\sum_{i}\bm{\mu}_i\cdot \bm{E}^{q}_i+\sum_{i,j}\bm{\mu}_i\mathbf{T}_{ij}\bm{\mu}_j$ is included where $\bm{\mu}_i$ is the electronic dipole moment,  $\bm{E}^{q}_i$ is the electric field and $\mathbf{T}_{ij}$ is the dipole-dipole interaction tensor (c.f. \citet{stone2013theory}). 

Since the atomic charges are no longer  fixed parameters but variables in the COMB potential, in the dynamic simulation of material systems using this potential, the effect of charge flows amongst the atomic species are additionally included in the Lagrangian to optimised the charge distribution in the simulated system, that is further subjected to the condition of electronegativity equalisation.\cite{rappe1991charge,rick1994dynamical}  In this regard, the COMB potential can also be considered as a type of polarisable forcefield (discussed in detailed in the next section), except the key difference here is that the COMB potential is an reactive potential where the bonded atoms are allowed to change during the simulation.

Within this framework of modelling the atom-atom interactions, several technologically important materials, the interfacial behaviours in which are critical for microelectronic devices, have been successfully parameterised and modelled using the COMB potential. Some of the notable examples include Si/\ce{SiO2},\cite{yu2007charge} amorphous silica,\cite{shan2010second} Hf/\ce{HfO2},\cite{shan2010charge} Ti/\ce{TiO2},\cite{cheng2014charge} as well as \ce{Al2O3}.\cite{choudhary2015charge}

\subsubsection{Polarisable Forcefields}\label{sect:polarisable_FF}

\begin{figure*}[thb]
    \centering
    \includegraphics[width=\textwidth]{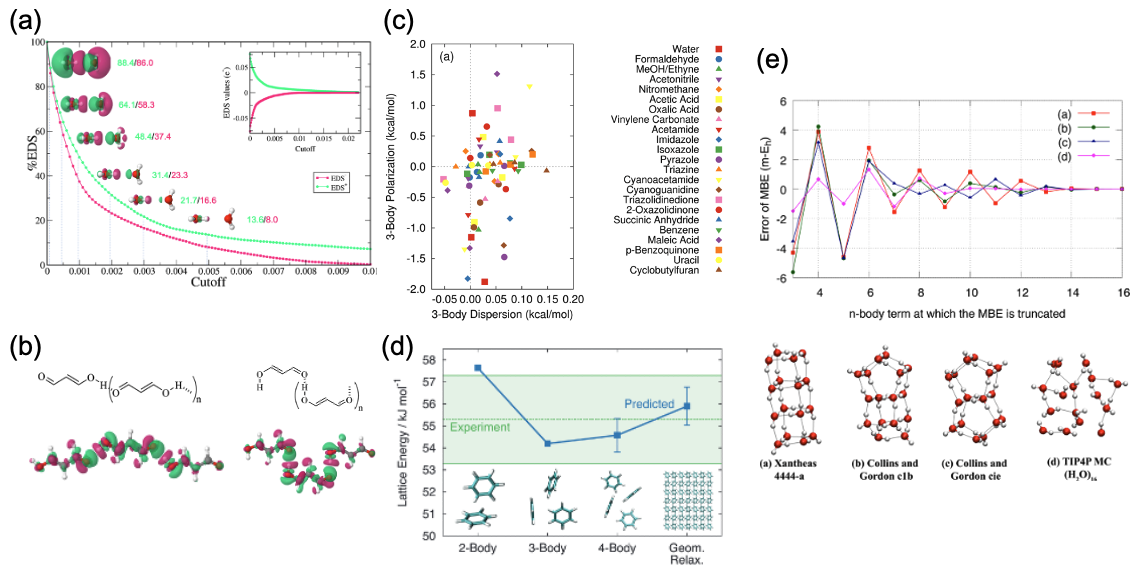}
    \caption{\pol{(a-b) Measurements of electron density shift (EDS) in pairs of non-covalently bounded molecules.\cite{iribarren2021evaluation} Here, the EDS is defined as $\mbox{EDS}=\rho(abc\ldots n)-\sum_{i=1}^{N}\rho(i)$, which is the difference in the electron density for a mutually interacting cluster of $N$ molecules, and the sum of electron densities for isolated monomers in their respective geometries exactly as in the cluster. Green and magenta represent positive and negative values for the EDS, respectively. (a) Evolution of the percentage of EDS inside the cutoff (a.u.) and EDS values (in-plots) with respect to the cutoff for \ce{H2O}$\cdots$\ce{H2O}. (b) EDS maps at the 0.001 a.u. cutoff for trimers of malonaldehyde arranged in linear and zigzag shapes. (c)  Benchmark results for 3-body energies $\Delta E_{3b}$ for 69 small molecular trimers bounded by non-covalent interactions, which were extracted from 23 different molecular crystals.\cite{rezac2015benchmark} Here the total $\Delta E_{3b}$ is broken down into the contributions from the electronic polarisations and dispersions to showcase the relationships between them. (d) Convergence of the lattice energy for the benzene crystal calculated at the highly accurate coupled cluster level of theory using many-body expansion approach,\cite{yang2014ab} which shows that including the contributions from up to the 4-body term is necessary to converge the lattice energy to the experimental value (adopted from \citet{beran2014new}). (e) Erratic convergence of the many-body energies (see \cref{eq:p-body-nonadd})  towards the full cluster energies for  four different (\ce{H2O})$_{16}$ clusters.\cite{ouyang2014trouble}}}
    \label{fig:cooperativity_illustration}
\end{figure*}

\paragraph{\textbf{Preliminary:  non-additive interactions in many-body systems}} For the majority of the forcefield models discussed above, the total interaction energy of the entire system is generally assumed to be a simple sum of the interaction energies amongst all subsets of bodies in the system [\emph{c.f.} \cref{eq:total_energy_atom_pos_expand}]. In other words, given an $N$ body system, once parameterised, the interaction energy $V^{(n)}(\{\bm{r}_{n}\},\bm{\beta}^{(n)})$ for a subset of $n\subseteq N$ bodies is independent of the presence of the rest of bodies $\{i\in N:i\notin n\}$ in the system. However, this is not necessarily true even for simple material systems. Fundamentally, as illustrated in \cref{fig:cooperativity_illustration}(a-b), this is originated from the polarisability of the electron density centred on each body, where upon mutual contact, the electron densities can polarise each other through space in a geometry-dependent way, causing changes in the overall electron density distributions in the entire system, even they are bounded with (weak) non-covalent interactions. Since the interactions among electron densities play significant contributions to the total interaction energy, it can be appreciated that the total system energy of a many-body system will no longer be a simple summation of the  energies for all interacting pairs, each consider in isolation from the other bodies. This is because the latter energy is modified when the pair is brought into the contact with the remaining bodies in the whole system.

To formalise this argument,\cite{kaplan1995non} consider the total energy $E(abc\cdots n)$ for a system with $N$ bodies. In a non-interacting system, the total energy is a sum of individual one-body energies, \emph{i.e.} $E(abc\cdots n)=E_{1b}(abc\cdots n)=\sum_{i=1}^{N}E(i)$. In an additive system, the total system energy needs to take into account only the contributions from all the two-body interactions in the system
$E(abc\cdots n)=E_{1b}(abc\cdots n)+\Delta E_{2b}(abc\cdots n)$, in which
\begin{equation}\label{eq:two-body-nonadd}
    \Delta E_{2b}(abc\cdots n)=\sum_{i<j}^{{N \choose 2}}E(ij)-a_{2N}^{1}E_{1b}(abc\cdots n),
\end{equation}
where $a_{mn}^{k}=\frac{(n-k)!}{(n-m)!(m-k)!}$ and $E(ij)$ is the total energy of a two-body subsystem in isolation. As discussed, in a non-additive system, the two-body terms $\Delta E_{2b}(abc\cdots n)$ must be further corrected by the presence of the other bodies that are also interacting with the two-bodies, such that the total system energy can be written as the following series:
\begin{equation}\label{eq:nonadd-total-energy}
E(abc\cdots n)=E_{1b}(abc\cdots n)+\sum_{p=2}^{N}\Delta E_{pb}(abc\cdots n),
\end{equation}
which can be considered as a variant of \cref{eq:total_energy_atom_pos_expand}, and it can be shown by induction that the non-additive $p$-body $(p\leq N)$ interaction energy
\begin{equation}\label{eq:p-body-nonadd}
    \Delta E_{pb}(abc\cdots n)=\sum_{i<j<\ldots<p}^{{N \choose p}}E(ij\cdots p)-\sum_{i=1}^{p-1} a_{pN}^{i}\Delta E_{pb}(abc\cdots n).
\end{equation}

It is clear from \cref{eq:p-body-nonadd} that,  calculating the non-additive $p$-body energies for an $N$-body system in general faces similar  sampling challenges in high-entropy systems (\cref{sect:HEA}), as the number of different $p$-body configurations grows factorially as a function of $p$ and $N$. Tackling this challenge, specifically in the field of non-covalent interactions in (supra)molecular systems.\cite{mahadevi2016cooperativity} Generally, clusters of polar molecules exhibit large non-additive many-body energies. For instance, benchmark calculations [\cref{fig:cooperativity_illustration}(c)] shows that the average $\Delta E_{3b}$ is $\sim 0.46$ kcal/mol for water trimer,\cite{rezac2015benchmark} whereby the largest $\Delta E_{3b}$  for water trimer was found to be $-2.42$ kcal/mol, which contributes significantly towards its stability (18.4$\%$ of the total interaction energy) that is comparable in magnitudes to the usual two-body terms. For non-polar  molecules that are predominantly bound together via dispersion interactions, such as benzene, in which the mutual interactions between benzene molecules will not significantly redistribute the charge density in the monomers, the corresponding $\Delta E_{3b}$'s in the benzene trimers are usually small. This has been exploited in an earlier study\cite{yang2014ab} to utilise the many-body expansion [\cref{eq:nonadd-total-energy}] for accurately determining the lattice energy of benzene crystal using quantum-chemistry method based on localised atomic orbitals, where sufficiently good agreement with experimental result is already reachable by truncating \cref{eq:nonadd-total-energy} at the 4-body term [\cref{fig:cooperativity_illustration}(d)]. This is not generalisable to polar molecules, however, as shown in \cref{fig:cooperativity_illustration}(e), where $\Delta E_{nb}$ shows strongly oscillating behaviours for three model examples of water clusters,\cite{ouyang2014trouble} making the expansion [\cref{eq:nonadd-total-energy}] difficult to converge. Various approaches have been developed to overcome such a challenge, and it is beyond the scope of discussion here, thus we direct interested readers to other reviews in this field.\cite{beran2016modeling}  

\paragraph{\textbf{Popular polarisable forcefields in material modellings}} The need to capture the atomic position dependent deformation in the electronic density  (\emph{i.e.} electron polarisations) for accurate modelling the non-additive interactions in atomistic simulation transcends into the more complex functional forms and parameterisations that needs to be added into the atom-atom potential, which leads to the (specialised) development of polarisable forcefields. The COMB potential discussed in the previous section is a variant of he polarisable forcefield, but this term is more commonly used for categorising the forcefields that have been better developed for (bio)molecular simulations. A possible reason behind is that the magnitude of the polarisation components can be accurately quantified by decomposing it from the total interaction energy between two non-covalently bounded molecular moieties using SAPT,\cite{misquitta2005intermolecular,misquitta2016ab} enabling it to be efficiently parameterised with a well-defined  functional form with solid physical origins. Therefore, the primary focus of this section is to provide a brief overview on the three major strategies for including polarisation effects in forcefields for molecular simulations.\cite{jing2019polarizable} 

In the forcefields such as AMOEBA\cite{ponder2010current} and AMBER,\cite{cieplak2001molecular,wang2012development} electronic polarisation is accounted with the \emph{induced dipole model}. Here, distributed multipole analysis\cite{stone2013theory,cisneros2012application} is first applied to expand the molecular charge densities into atom-centred multipoles up to  quadrupolar level for obtaining an accurate description of molecular electrostatics. On each polarisable site, an  induced dipole in response to the external field $\bm{\mu}_{\mbox{\scriptsize{ind}}}=\alpha \bm{E}_{\mbox{\scriptsize{ext}}}$ is introduced, in which $\alpha$ is the electronic polarisability. Within this framework, the additional contribution to the total electrostatic interaction energy in the system is modelled as $E_{\mbox{\scriptsize{ind}}}=\sum_{i}\frac{1}{2}\alpha_i^{-1}\bm{\mu}^2_{i,\mbox{\scriptsize{ind}}}$, where the summation is performed over all polarisable sites. The molecular polarisability is determined iteratively using Thole's damped interaction method,\cite{thole1981molecular} which considers the  all polarisable sites can mutually interact with each other, which changes the charge distribution in the system until it becomes stabilised.

On the other hand, the \emph{Drude oscillators model}\cite{lemkul2016empirical} takes a more classical  approach to account for the polarisation effects using point charges, which, has been implemented in forcefield such as CHARMM-Drude.\cite{baker2011development,jana2015charmm} In the Drude model, electronic polarisation is modelled as a point charge attached to the polarisble site with a parameterised force constant $k_i$, such that the induced electrostatic interaction takes the form of a usual harmonic potential $E_{\mbox{\scriptsize{ind}}}=\frac{1}{2}\sum_i  k_i\bm{d}_i^2$, where $d_i$ is the displacement of the point charge.

Lastly, the \emph{fluctuating charge model}, which is used in, for example, the AMBER-$\sigma\pi$ forcefield,\cite{chen2007qtpie,patel2004charmm} takes similar approach at the COMB potential discussed above to optimise the atomic charges $\{ q_i \}$ in the system, where the total system energy is minimised by allowing charges to flow from one atom to another while maintaining the overall charge balance in the entire system. From here, the induced electrostatic energy $E_{\mbox{\scriptsize{ind}}}=\sum_i\left(\chi_i q_i+\eta_i q_i^2\right)$. Using the atomic ionisation potential (IP) and electron affinity (EA), the Mulliken electronegativity is then $\chi_i=\frac{1}{2}(\mbox{IP}+\mbox{EA})=-\mu_i$, whereas the Parr-Pearson hardness can be determined as $\eta_i=\frac{1}{2}(\mbox{IP}-\mbox{EA})$.

\paragraph{\textbf{Exemplary applications of polarisable forcefields in condensed matter modellings}} Although less common, some investigations have been dedicated to parameterise and apply the polarisable forcefields in modelling technologically important materials in the condensed phases. \citet{rathnayake2020evaluation} parameterised the AMOEBA forcefield specifically for metal halide perovskites against the \emph{ab initio} reference data, although molecular dynamic simulations performed using the as-developed forcefields lead to an overestimation of the phase transition temperatures in these materials. It highlights the challenges in accurately modelling the electrostatic interactions between localised (molecular cations) and delocalised (inorganic frameworks) electronic states in a hybrid system, that is an over-lasting challenge in material modellings. Similarly, polarisation forcefield has been applied to study metal-organic frameworks (MOFs) and their gas uptaking capacities,\cite{mcdaniel2015evaluation,becker2017polarizable,becker2018polarizable} which highlighting the importance of including the polarisation effects in accurate determination of the thermodynamics of gas absorption in MOF. Although neglecting such effect may also lead to acceptable result, the outcome is most likely an error cancellation effect arises from the overestimation of the dispersion energies. Another directly relevant field of application is surface chemistry for sensors and catalysts. In this regard, for examples, \citet{schyman2013exploring} parameterised the OPLS\cite{jorgensen1995importance}-AAP forcefield for water adsorption on carbon surfaces, and achieved good agreement with the thermochemical data obtained from high-level quantum chemical calculations. Polarisbale CHARMM forcefields for gold\cite{wright2013golp} and silver\cite{hughes2013biomolecular} have also been developed, enabling the efficient investigations of the interactions between biomolecules and metal nanoparticles for sensing applications using atomistic modellings.

\citet{walz2018phase} developed polarisable forcefield to examine the phase transition and molten behaviours of alkaline halide crystals. Predicting the solubilities of molecular crystal is critically important for the development of new pharmaceutical drugs, \citet{schnieders2012structure} tested the applicability of the AMOEBA forcefield for modelling the structure, thermodynamics and the solubility of the hexanamide organic crystal, achieving an accuracy of 1.1 kcal/mol in predicting the solvation free energy. By parameterising against DFT results, \citet{corradini2015dft} developed forcefields that accurately predicted the stabilities for the polymorphs of \ce{TiO2} and \ce{SiO2}, and subsequently applied it to model their amorphous structure, achieving excellent agreement with the experimental neutron structural factors.

\subsection{Density Functional Theory}
\label{sec:DFT}
\subsubsection{The Many-Body Schr\"{o}dinger Equation}

From the viewpoint of scientific discovery, it comes quite so often that we would like to predict and study the behaviours of new materials, for which no parameterised energy function [\emph{c.f.} \cref{eq:total_energy_atom_pos_expand}] exists. Fortunately, quantum mechanics provides us with a fundamental theory which says, given a system with $m$ atoms containing $n$ electrons, its total energy can be, in principle, determined by the \emph{non-parameterised many-body} Schr\"{o}dinger equation that accounts for the interaction energies amongst all the nuclei and electrons in the material:
\begin{equation}\label{eq:full_Sch_eqn}
\left[\hat{\mathcal{K}}_N+\hat{\mathcal{K}}_e+\hat{\mathcal{V}}_{Ne}+\hat{\mathcal{V}}_{NN}+\hat{\mathcal{V}}_{ee}\right]\psi\left(\{\bm{R}\},\{\bm{r}\}\right)=E\psi\left(\{\bm{R}\},\{\bm{r}\}\right),
\end{equation}
the solution of which is the system \emph{wavefunction} $\psi\left(\{\bm{R}\},\{\bm{r}\}\right)$ that completely determines its quantum state with the given nuclei and electron coordinates, $\{\bm{R}\}$ and $\{\bm{r}\}$, respectively. The left-hand-side of \cref{eq:full_Sch_eqn} expresses the total system energy $E$ as sum of different energetic components, that are written in the form of \emph{operators}, including:
\begin{enumerate}
    \item The nuclear kinetic energy operator $\hat{\mathcal{K}}_N=\displaystyle{-\sum_{j=1}^{m}\frac{\hbar}{2M_j}\nabla_j^2}$, where $M_j$ is the nuclear mass for atom $j$, $\hbar$ is the Planck's constant and $m$ is the total number of atoms in the system.
    \item The electron kinetic energy operator $\hat{\mathcal{K}}_e=\displaystyle{-\sum_{i=1}^{n}\frac{\hbar}{2m_e}\nabla_i^2}$ with $m_e$ being the mass of electron and $n$ is the total number of electrons considered in the system.
    \item The  Coulomb attraction between nuclei and electrons $\hat{\mathcal{V}}_{Ne}=-\displaystyle{\sum_{i,j}\frac{e^2}{4\pi\epsilon_{0}}\frac{Z_j}{\left\vert \bm{r}_i-\bm{R}_{j}\right\vert}}$, here $\epsilon_{0}$ is the vacuum permittivity and $Z_j$ is the nuclear charge of the atom $j$.
    \item The Coulomb repulsion between nuclei  $\hat{\mathcal{V}}_{NN}=\displaystyle{\sum_{j<j^\prime}\frac{e^2}{4\pi\epsilon_{0}}\frac{Z_j Z_{j^\prime}}{\left\vert \bm{R}_{j^\prime}-\bm{R}_{j}\right\vert}}$, and
    \item The Coulomb repulsion between electrons $\hat{\mathcal{V}}_{ee}=\displaystyle{\sum_{i<i^\prime}\frac{e^2}{4\pi\epsilon_{0}}\frac{1}{\left\vert \bm{r}_{i^\prime}-\bm{r}_{i}\right\vert}}$.
\end{enumerate}

Since the nuclei are much heavier and move slower than the electrons, solving the Schr\"{o}dinger equation typically invokes the \emph{Born-Oppenheimer (BO) approximation} which drops the $\hat{\mathcal{K}}_N$ term, and treats $\hat{\mathcal{V}}_{NN}$ classically as a constant background electrostatic potential. However, exceptions do exist, such as the motion of light-weighted hydrogen atoms that plays a key role in hydrogen bonding, where the nuclei must also be treated quantum mechanically.\cite{ceriotti2016nuclear} which is beyond the focus of the present review. With this, \cref{eq:full_Sch_eqn} reduces to the following form for the electronic wavefunction $\psi(\left\{\bm{r}\right\})$ (note that the curly brackets are kept here to emphasise that this is a function of the collection of Cartesian coordinates for all the electrons in the system):
\begin{equation}\label{eq:e_Sch_eqn}
\left[\hat{\mathcal{K}}_e+\hat{\mathcal{V}}_{Ne}+\hat{\mathcal{V}}_{ee}\right]\psi\left(\{\bm{r}\}\right)=E\psi\left(\{\bm{r}\}\right).
\end{equation}

\begin{figure}[htb]    
    \centering
    \includegraphics[width=0.48\textwidth]{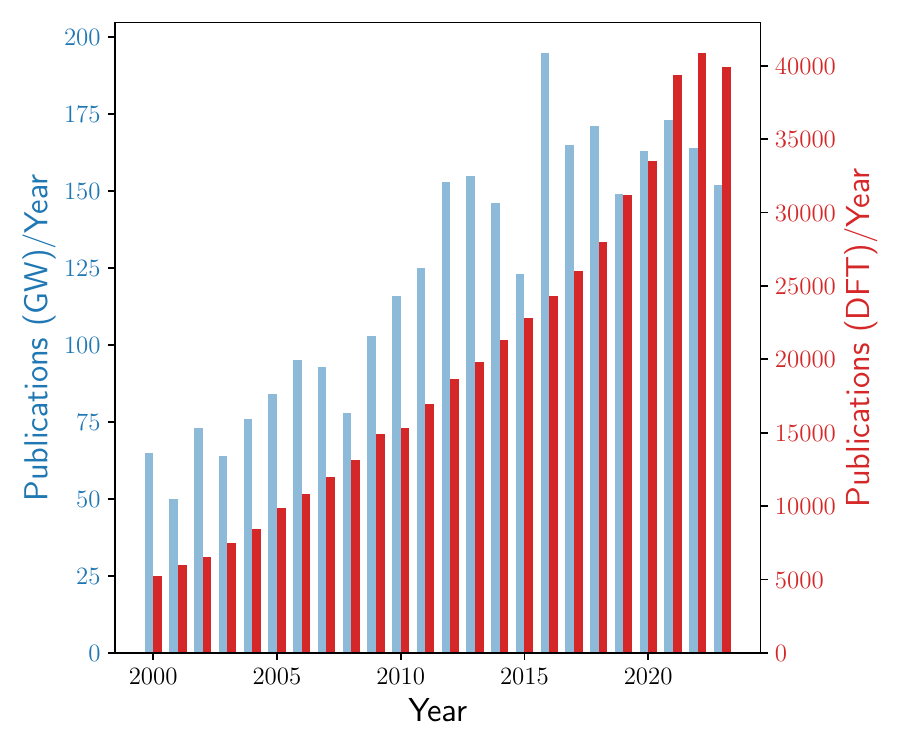}
    \caption{Number of publications per year since 2000 which involve `Density Functional' or `DFT' according to the Web of Science. In comparison, the same trend for publications including `GW approximations' is also shown.}
    \label{fig:DFT_statistics}
\end{figure}
However, even with the BO-approximation, \cref{eq:e_Sch_eqn} is still intractable to be solved for realistic materials, for instance, it is estimated that $\sim10^{46}$ complex numbers are necessary to fully specify the wavefunction  of the 10 valence electrons for crystalline silicon containing two atoms in its cubic unit cell with a cell length of 5.42 $\mbox{\AA}$.\cite{giustino2014materials} Fortunately, the formulation and development of the  Density Functional Theory (DFT) have provided a practical solution to overcome such a challenge and has become the backbone of modern material theory and discoveries [\cref{fig:DFT_statistics}]. Note that, although  Density Functional \emph{Theory} is the most commonly used term in the literature, what it really encompasses, are the formally exact Hohenberg-Kohn-Sham (HKS) theorems, and the Density Functional \textit{Approximations} (DFAs) that brings the theory into practicalities, and these will be followed and briefly reviewed in the following two sections.
\subsubsection{The Hohenberg-Kohn Theorems and the Kohn-Sham Equations }

Modern DFT is built upon the foundation of the pivotal theorems established by Hohenberg and Kohn,\cite{kohn1999nobel} which states that the  Hamiltonian for a many-electron system is uniquely determined by its ground-state electron density $\rho(\bm{r})$, from which one can derive the corresponding system energy $E$. Most importantly, Hohenberg and Kohn demonstrated that the exact ground state energy can be determined by minimising the \textit{total energy functional} $E[\rho]$ over all possible electronic density, \emph{i.e.}
\begin{equation}\label{eq:DFT_min_rho}
E^{\mbox{\scriptsize{gs}}}=\min_{\rho}E[\rho(\bm{r})]    
\end{equation} 
This unique theory has provided the much needed simplification to tackle the many-body quantum mechanical problems for real materials by mapping an electronic wavefunction with $3n$ variables into the electron density $\rho(\bm{r})$ which is a function of only  three Cartesian coordinates. 

In order to make the Hohenberg-Kohn Theorems practical for performing realistic calculations, it is  necessary to write down the expressions for $E[\rho]$, and this is achieved later\cite{kohn1965self} with the introduction of the famous Kohn-Sham equation for DFT, in which the total energy functional is divided into the sum of multiple terms:
\begin{equation}\label{eq:KS_eqn}
E_{\mbox{\scriptsize{KS}}}[\rho]=\sum_{i=1}^{occ}\int\vert \nabla\phi^{\mbox{\scriptsize{KS}}}_i (\bm{r})\vert^2 \mbox{d}\bm{r} + \int v(\bm{r})\rho(\bm{r})\mbox{d}\bm{r}
+\frac{1}{2}\int \mbox{d}\bm{r} \int \mbox{d}\bm{r}^\prime \frac{\rho(\bm{r})\rho(\bm{r}^\prime)}{\left\vert \bm{r}-\bm{r}^\prime\right\vert} +E_{NN}+E_{xc}[\rho].
\end{equation}
Here, the first three terms (electron kinetic energy, electron-nuclear Coulomb interactions and electron-electron Coulomb interactions, respectively) can be written analytically as the functionals of the electronic density, that can be self-consistently determined by the occupied Kohn-Sham orbitals $\phi^{\mbox{\scriptsize{KS}}}(\bm{r})$, where $\rho(\bm{r})=\sum_{i=\mbox{\scriptsize{occ}}}|\phi^{\mbox{\scriptsize{KS}}}_i(\bm{r})|^2$. $E_{NN}$ is the Coulomb repulsions between the nuclei, which is deterministic from the nuclei charges and positions. These four terms, collectively, could be understood from the classical electrostatics, particularly based on the functional forms they adopt. In contrast, the \emph{exchange-correlation functional} $E_{xc}[\rho]$ describes the pure quantum mechanical nature of electrons, and can generally be separated into the sum of exchange and correlation contributions as $E_{xc}[\rho]=E_{x}[\rho]+E_{c}[\rho]$. 
Physically, the exchange energy is the energy associated with exchanging the positions of two electrons, while maintaining the antisymmetry of the electronic wavefunction according to the Pauli's exclusion principle. With this definition and utilising the Kohn-Sham orbitals $\phi^{\mbox{\scriptsize{KS}}}(\bm{r})$, the exchange energy, which is a \emph{non-local} quantity in $\bm{r}$, in principle, can be calculated exactly  using its definition from the Hartree-Fock (HF) theory:\cite{staroverov2012density}
\begin{equation}\label{eq:exchange_HF_DFT}
E_{x}^{\mbox{\scriptsize{exact}}}=-\sum_{i,j=1}^{n/2}\int \mbox{d}\bm{r}\int \mbox{d}\bm{r}^\prime  \frac{\phi_{i}(\bm{r}) \phi_{i}^{*}(\bm{r}^\prime) \phi_{j}^{*}(\bm{r})\phi_{j}(\bm{r}^\prime)}{\left\vert \bm{r}-\bm{r}^\prime\right\vert}.
\end{equation}
The negative sign in the above expression shows that the exchange interactions always contribute to the stabilisation of the system. On the other hand, electron correlations, which also contributes to stabilising forces and describes the extent to which the movement of one electron is affected by the presence and motion of all other electrons in the system, lack of a closed mathematical form such as \cref{eq:exchange_HF_DFT}. 

It turns out that simply appending approximated $E_{c}[\rho]$ to $E_{x}^{\mbox{\scriptsize{exact}}}$ do not generally lead to good performing results, because of the opposite effects on how the exchange and correlation interactions can affect the distributions of two particle electron density $\rho(\bm{r}_{1},\bm{r}_{2})$ in space. As such, \emph{the general lack of an analytical  form for  $E_{xc}[\rho]$, particularly one that is local, meaning only dependent on the electron density $\rho(\bm{r})$ at a single point in space}, is the major challenge behind DFT, and is feuling the ongoing development of DFAs to accurately approximate the true $E_{xc}$ for a many-electron system without significantly scarifying the computational efficacy of DFT. In what follows, we will briefly review  the major categories of DFAs that have been developed and matured over the years.  
\subsubsection{The Jacob's Ladder for DFAs}

To categorise the increasing level of complexity of DFAs in terms of the ingredients that entering the DFAs, as well as the modelling accuracies that are achievable with these DFAs, John Perdew and his colleague first introduced the `Jacob's ladder for DFAs'.\cite{perdew2001jacob} 

In the bottom three rungs of the Jacob's ladder, the general idea behind the development of DFAs is to increase its accuracy by gradually including more topological information on the single-particle electron density $\rho(\bm{r})$. With this, the exchange-correlation functional can be written in the general form of 
\begin{equation}\label{eq:Exc_general}
E_{xc}[\rho(\bm{r})]=\int (e_x^{F}[\rho(\bm{r})] + e_c^{F}[\rho(\bm{r})]) \mbox{d}\bm{r},
\end{equation}
where, starting from the exchange $e_{x}[\rho]$ and correlation $e_{c}[\rho]$ energies for homogeneous electron densities that can be either exactly determined or numerically parameterised, the exchange and correlation energies for inhomogeneous electron densities could be modelled by including an additional inhomogeneous factor $F$ that incorporates the information of electron density topologies, thus
\begin{equation}\label{eq:Exc_inhomo}
    e^{F}_{x,c}[\rho(\bm{r})]=F_{x,c}[\rho(\bm{r})] e_{x,c}[\rho],
\end{equation}
where the inhomogeneous factor can be written in a general form of:\cite{mardirossian2017thirty}
\begin{equation}\label{eq:inhomo_factor_general_form}
F_{IJ}[\rho(\bm{r})]=\sum_{i=0}^{I}\sum_{j=0}^{J} s_{ij}u_\sigma^i w_\sigma^j, 
\end{equation}
in which $s_{ij}$ are parameters. Dropping the explicit dependency on $\bm{r}$ for clarity, in \cref{eq:inhomo_factor_general_form}, the spin $(\sigma)$ dependent parameter $u_\sigma=\beta x_\sigma^2/(1+\beta x_\sigma^2)$ where $x_\sigma=|\nabla\rho_\sigma|/\rho_\sigma^{4/3}$ is so-called the reduced density gradient, and $w_\sigma=(t_\sigma-1)/(t_\sigma+1)$ where $t_\sigma=\frac{3}{5}(6\pi^2)^{\frac{2}{3}}\frac{\rho^{5/3}_\sigma}{\tau_\sigma}$ is related to the kinetic energy of a Kohn-Sham system $\tau_\sigma=\frac{1}{2}\sum_{i=1}^{\mbox{\scriptsize{occ}}}|\nabla \phi_{i,\sigma}^{\mbox{\scriptsize{KS}}}|^2$, which can be considered equivalent to measuring the curvatures of the electron density distributions. 

Using this formalism, the first rung of the Jacob's ladder contains exchange-correlation functionals that are constructed using $F_{00}$, which are functionals that only depend on the density of a homogeneous electron gas. This leads to the developments of \emph{Local Density Approximations} (LDA). The LDA has seen the initial success of DFAs in modelling condensed matters in which the electron densities vary slowly across the space. It provides good  values of the lattice constants albeit overestimates the binding energies of common inorganic solids.\cite{stampfl1999density} On the other hand, LDA did not win the confidence over DFAs in the chemistry community because its fundamental assumption of homogeneous electron density does not compile with the realistic behaviours of electrons involved in chemical bonds. This problem was overcome with the introduction of the \emph{Generalised Gradient Approximations} (GGA), by adding the reduced density gradients $x_\sigma$ into the recipe for constructing the DFAs that corresponds to the inclusion of $F_{10}$ term in \cref{eq:Exc_general}.  This has led to, for examples, the  PBE (Perdew-Burke-Ernzerhof)\cite{perdew1996generalized} functional that remains to be the workhorse in the condensed matter community as exemplified by the Materials Project,\cite{jain2013commentary} and the BP86\cite{becke1988density,perdew1986density}(augmented with dispersion correction) being a good choice in quantum chemistry to obtain good starting geometries for  more detailed analysis. Moving up in the Jacob's ladder,  with the $F_{11}$  term in \cref{eq:Exc_general}, one reaches the \emph{meta-GGA} functionals that include additionally, the electron kinetic energy in its construction recipe, which further improves the binding energies of molecular systems compared to pure-GGA.\cite{perdew1999accurate} 

As discussed, although direct combination of $E_{x}^{\mbox{\scriptsize{exact}}}$  with a correlation functional does not lead to good performing DFAs, the existence of a closed expression for $E_{x}^{\mbox{\scriptsize{exact}}}$ remains to be a highly attractive features for quantum chemists to explore different approaches to combine it with approximated correlation term to construct the \emph{hybrid functionals} that sit on the fourth rung of the Jacob's ladder. Notably this leads to the  hybrid functional B3LYP\cite{10.1063/1.464913,lee1988development,vosko1980accurate,stephens1994ab} that is most widely used in computational chemistry, although its stand-alone usage without  additional corrections are known to be not recommended for obtaining accurate results.\cite{kruse2012standard} Formally, the B3LYP functional is constructed\cite{stephens1994ab} as
\begin{equation}\label{eq:b3lyp_func}
    E_{xc}^{\mbox{\scriptsize{B3LYP}}}=(1-a_0)E_{x}^{\mbox{\scriptsize{LDA}}}+a_0 E_{x}^{\mbox{\scriptsize{exact}}}+a_x \Delta E_{x}^{\mbox{\scriptsize{B88}}}
    +a_c E_{c}^{\mbox{\scriptsize{LYP}}}+(1-a_c)E_{c}^{\mbox{\scriptsize{VWN}}}.
\end{equation}
Here, part of the LDA exchange is replaced by $E_{x}^{\mbox{\scriptsize{exact}}}$  with additional inclusion of Becke's gradient correction\cite{becke1988density} to the correlation functional. For the correlation energy, the Lee-Yang-Parr (LYP)\cite{lee1988development} functional, which lacks  an easily
separable local component, is combined with the local correlation
functional formulated by Vosko, Wilk, and Nusair (VWN).\cite{vosko1980accurate} $a_0$, $a_x$ and $a_c$ are three adjustable mixing parameters, which, in the case of B3LYP functional, does not contain explicit dependency on interatomic distance, such that functionals with this type of constructions are generally referred to as the \emph{global hybrid}. However, the Hartree-Fock exact exchange  describes the electronic interactions better than semilocal DFAs at large separations $r$ between electrons, whereas DFAs works better at short $r$'s due to Hartree-Fock's intrinsic lack of correlations between electrons with the same spins, this inspired the development of \emph{range-separated hybrids}, as pioneered by the CAM-B3LYP\cite{yanai2004new} functional that attenuates the mixing parameters with $r$. This  basic construct was modified by Heyd, Scuseria, and
Ernzerhof to circumvent the challenges of computing the exact exchange for periodic systems, leading to the HSE functionals\cite{heyd2003hybrid} as being the popular choice of hybrid functional in the condensed matter community.

Lastly, the \emph{double hybrid  functionals}, which are at the top rung of the Jacob's ladder, were proposed to overcome the inability for all lower-rung functionals in modelling the long-ranged dynamic electron correlation energies that play important contribution towards the non-covalent dispersion interactions in molecular system. Similar to how $E_{x}^{\mbox{\scriptsize{exact}}}$  is acquired in a hybrid DFA, the exchange-correlation energy from the double hybrid functional can be achieved by mixing part of the energy from hybrid DFA with the correlation energy obtained from a  second-order M\o ller-Plesset perturbation theory (MP2)\cite{moller1934note} calculation using both the occupied and virtual Kohn-Sham orbitals  $\left\{ \phi_{i}^{\mbox{\scriptsize{KS}}}\right\}$. Unlike DFAs, second-order perturbation theory belongs to the family of \emph{wavefunction theoretical methods} that are \emph{systematically improvable in terms of treating the challenging electron correlation problems in quantum chemistry,} which is beyond the scope of the present review. Using this strategy, \citet{grimme2006semiempirical} proposed the first double hybrid  functional dubbed as B2PLYP, in which the total exchange correlation energy is given by 
\begin{equation}\label{eq:B2PLYP_xc}
E_{xc}^{\mbox{\scriptsize{B2PLYP}}}=(1-a)E_{x}^{\mbox{\scriptsize{GGA}}}+a E_{x}^{\mbox{\scriptsize{exact}}}+bE_{c}^{\mbox{\scriptsize{GGA}}}+cE_{c}^{\mbox{\scriptsize{MP2}}},
\end{equation}
with $a$, $b$ and $c$ are the mixing parameters. Following this foundational constructions of double-hybrid exchange-correlation functionals, a range of different ideas across the chemistry community to construct similar functionals\cite{goerigk2014double} to achieve better accuracy in determining the energetics of molecular systems. Direct application of double hybrid DFAs in condensed matter simulations remains rare until these days due to the challenge in developing implementations to efficiently calculate the MP2 correlation energy with the planewave basis. \cite{carter2008challenges,pisani2008periodic}

\paragraph{\textbf{Parameterisation of DFAs}}
Although the goal of density functional \emph{theory} is to provide a formalism to solve the many-body Schr\"{o}dinger equation with no empirical parameters, which will enable us to predict the ground state energies and associated physical properties for new material systems, it can be seen above [\emph{e.g.} \cref{eq:b3lyp_func} and \cref{eq:B2PLYP_xc}] that some parameters are still necessary to be introduced in order to approximate the unknown exchange correlation functionals to make DFAs practical for solving real material problems. Typically, particularly in the early stages of DFA developments, these parameters are fitted to enable the total Kohn-Sham energies to reproduce the experimental thermochemical data. 

For examples, the popular B3LYP functional [\cref{eq:b3lyp_func}] was derived through a linear square fit to a dataset containing 56 atomisation energies, 42 ionisation potentials, 8 proton affinities and 10 first-row total atomic energies.\cite{beeke1993density} The mixing parameters in the B2PLYP functionals are fitted to the G2 dataset containing the heat of formation for 148 neutral molecules.\cite{curtiss1997assessment} The same dataset is also used for determining the Coulomb attenuation parameters in the CAM-B3LYP\cite{yanai2004new} functional. The hope here is to construct a DFA that can be universally applied to most chemical systems. However, this is also a monumental endeavour as the effects of electron correlations manifest themselves differently in different chemical systems (\emph{e.g.} main group molecules, transition metal complexes, non-covalent interactions, and so on), requiring system-specific functional forms and parameterisations of DFAs in order to achieve good performance. This is the motivation behind the development of the Minnesota family of density functionals,\cite{peverati2014quest,mardirossian2016accurate} and the associated database\cite{yu2016mn15} to optimise the parameters in the DFAs, depending on the specific (sub)sets of chemical problem(s) that a DFA is constructed to solve. 

\subsubsection{Challenges and Pitfalls in Standard DFAs} \label{sect:DFA_challenge}

Because of the necessity to approximate the exchange-correlation energy $E_{xc}[\rho(\bm{r})]$ as a functional of single-electron density $\rho(\bm{r})$,  there are challenging cases where DFAs failed to determine the energetics of molecular and material systems quantitatively or even qualitatively.\cite{cohen2012challenges} In particular, the delocalisation error is regarded as the most outstanding challenge in the developments of DFAs,\cite{bryenton2023delocalization} which is most relevant to the \textit{band gap problem}\cite{perdew1985density} in the community of materials modelling. The semilocal nature of $E_{xc}[\rho(\bm{r})]$ also leads to the failure in determining the correct (equilibrium) structures and interaction energies of systems that are bounded by the (non-covalent) dispersion interactions. Fortunately, many robust and  efficient theoretical approaches have been developed to correct such a deficiency in standard DFAs.\cite{grimme2016dispersion} Here, we aim to provide some concise and explanatory accounts on these two challenges for readers  who are  interested in applying DFT solving problems in real material designs and discoveries to become aware of the limitations in the underlying methodologies applied in their research. For readers who are more interested in   different theoretical aspects and the latest developments behind DFT, we refer them to this interesting article\cite{DFT_exchange} that exchanges  viewpoints on DFT by the leading theoreticians in the research community.

\paragraph{\textbf{The delocalisation error}} The delocalisation error is a term that expands\cite{perdew1982density,ruzsinszky2006spurious,mori2006many} on the originally coined self-interaction error (SIE)\cite{perdew1981self}  to cover more general context in DFT. \citet{hait2018delocalization} have provided a good explanation on the relationship between these two terms, which will be quoted here (in \emph{italics}). The delocalisation error originates in defining the electrostatic repulsion term $E_{J}=\frac{1}{2}\iint \frac{\rho(\bm{r})\rho(\bm{r}^\prime)}{\left\vert \bm{r}-\bm{r}^\prime\right\vert}d\bm{r}d\bm{r}^\prime$ [\cref{eq:KS_eqn}] based on the one-particle electron density. \emph{However, $\rho(\bm{r})$ is made up of discrete number of electrons, which (despite being delocalised over space) should not be able to interact with themselves. $E_J$ therefore contains a spurious self-repulsion energy. The exchange-correlation energy }($E_{xc}[\rho(\bm{r})]$) \emph{in principle should fix this, but the correction is very approximate for most functionals on account of the $E_{xc}$ functional having substantial local components, while self-repulsion is explicitly non-local. The electron density subsequently delocalises (becomes less compact) in order to minimise the residual self-repulsion}, resulting in the delocalisation error.

\begin{figure}[tb]
    \centering
    \includegraphics[width=\linewidth]{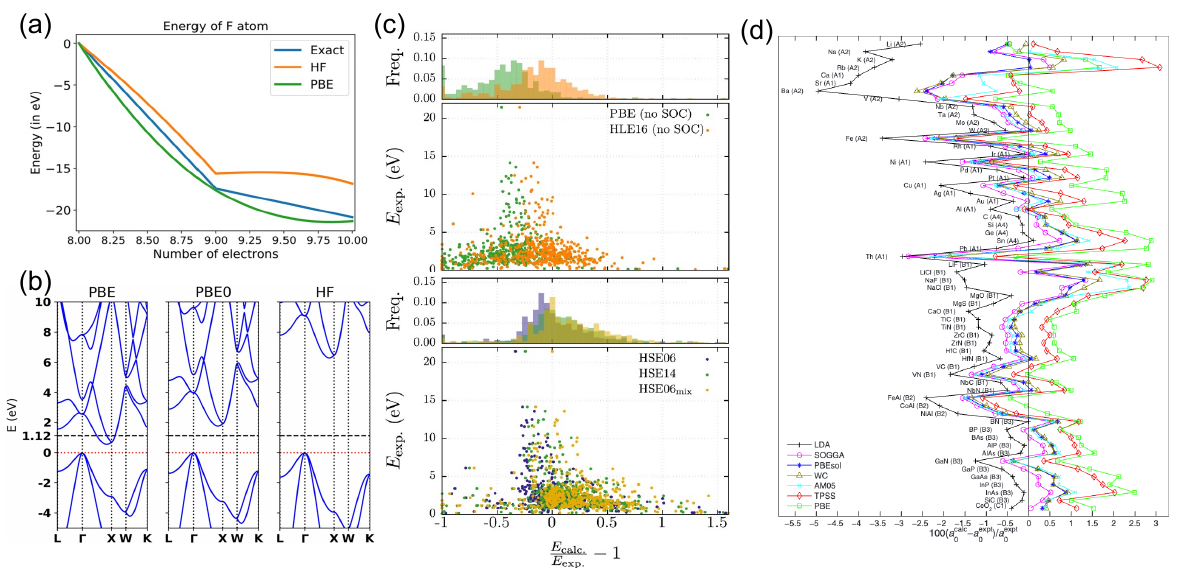}
    \caption{Fractional charges and the band gap problems in DFT. (a) Effect of delocalisation error on fractional charges, showing the overestimation of energy by HF and underestimation by a typical GGA functional (PBE) [from \citet{hait2018delocalization}]. (b) Variations in the computed electronic band structures for bulk silicon, showing strong dependence on the content of exact HF exchange (PBE: 0\%, PBE0: 25\% and HF: 100\%) ,ixing in the exchange-correlation functionals [from \citet{bryenton2023delocalization}
]. (c) Benchmarking the performances of different DFAs in predicting the band gaps for 472 non-magnetic materials. The figures shows the relative frequency and experimental band gap ($E_{\mbox{\scriptsize{exp.}}}$) as a function of the ratio between theoretical ($E_{\mbox{\scriptsize{calc.}}}$) and experimental band gaps for the different exchange-correlational functionals [from \citet{borlido2019large}].  To compare the performance of GGA functionals in reproducing the structural properties of solids, in (d), we also show the relative error  in the calculated lattice constants for selected inorganic solids with respect to the experimental data. Overall, the PBEsol functional gives the lowest mean absolute relative error of 9.5\% compared to all other functionals tested.  [from \citet{haas2009calculation}] }
    \label{fig:DFT_bandgap}
\end{figure}

In their review, \citet{bryenton2023delocalization} summarised a few different images that can be applied to interpret the origins of, and consequences resulting from the SIE in predicting the properties of molecular and material systems using DFT. Here, we focus on the \emph{fractional charge} picture, which is best suitable for explaining the well-known \emph{band gap problem} in DFT, particularly for strongly correlated systems and Mott insulators,\cite{cohen2012challenges} that is most relevant to the field of theory-driven material designs.

In this seminal work, \citet{perdew1982density} shows that the total energy of a multielectron system containing a fractional number of electrons $N+\omega$ [$N\in \mathbb{Z}$ and $\omega \in (0,1)$] should be $E(N+\omega)=(1-\omega)N(E)+\omega E(N+1)$, which is a straight line that interpolates between the total energies of two systems containing $N$ and $N+1$ electrons, respectively, with discontinuities at the integer numbers of electrons. However, as semilocal DFAs (such as PBE) tend to stabilises the delocalised electron densities, they generate states with fractional charges that are lower in energies than what is expected from the linear interpolation [\cref{fig:DFT_bandgap}(a)]. A more concrete example is to consider the H$_{2}^{+}$ ion at its dissociation limit, which, instead of producing the correct configuration of H$^{+}\cdots$H, with the charge localised predominantly on one hydrogen ion, most DFAs will stablise the complex  H$^{\delta +}\cdots$H$^{\delta +}$ with fractional charges equally distributed on both ions. It should also be noted that, the energy curve $E(N+\omega)$ corresponding to HF shows a concave-down behaviour [\cref{fig:DFT_bandgap}(a)], indicating its tendency to produce more localised electron densities.

The challenge behind standard DFAs in correctly predicting the electronic band gap could be understood from their molecular equivalent of the HOMO-LUMO gap, which \citet{janak1978proof} first showed that the orbital eigenenergies are the left- and right-derivatives with respect to the electron number,
\begin{equation}\label{eq:HOMO_LUMO_def}
\varepsilon^{\mbox{\scriptsize{HOMO}}}=\lim_{x\to N^{-}}\frac{E(x)-E(N)}{x-N},\qquad \varepsilon^{\mbox{\scriptsize{LUMO}}}=\lim_{x\to N^{+}}\frac{E(x)-E(N)}{x-N}.
\end{equation}
As shown in \cref{fig:DFT_bandgap}(a), the `concave-up' behaviour in the $E(N)$ curve for semilocal GGA functional, it leads to $\varepsilon^{\mbox{\scriptsize{HOMO}}}$ becomes too high whereas $\varepsilon^{\mbox{\scriptsize{LOMO}}}$
becomes too low, resulting in the underestimation of the HOMO-LUMO gap, or the band gap for solids [see \cref{fig:DFT_bandgap}(b) for the illustration of the band structures for silicon]. On the other hand, it can be seen that the opposite behaviour in the $E(N)$ curve for HF resulted in a drastic overestimation of the band gap. This can be used to understand why, by including 25 \% of HF exchange into the functional PBE (PBE0),\cite{adamo1999toward} the problem of band gap underestimation by pure-GGA can be alleviated by opening up the estimated band gap values.

Putting the arguments above into a broader context, \cref{fig:DFT_bandgap}(c) shows a large-scale benchmark of the exchange-correlation functionals in calculating band gaps for 472 non-magnetic solids against their experimental values. Using the metric $\eta=E_{\mbox{\scriptsize{calc.}}}/E_{\mbox{\scriptsize{exp.}}}-1$ to compare the magnitudes of the calculated ($E_{\mbox{\scriptsize{calc.}}}$) versus experimental $(E_{\mbox{\scriptsize{exp.}}})$, \citet{borlido2019large} shows that statistically LDAs underestimate $(E_{\mbox{\scriptsize{exp.}}})$ by 50 \% while GGAs underestimate $E_{\mbox{\scriptsize{exp.}}}$ by 40 \%. Meta-GGA functionals, such as SCAN\cite{sun2013semilocal} and mBJ (modified Becke-Johnson potential)\cite{becke2006simple,tran2009accurate} show improvements from LDA/GGA results with $\sim$37 \% underestimation of  $E_{\mbox{\scriptsize{exp.}}}$. The range-separated hybrid HSE06\cite{krukau2006influence} gives the smallest mean absolute error of 0.53 eV, particularly for small band gap materials, the mean percentage error in estimating the band gap is $\sim$ 31 \%. However, the high computational costs associated with hybrid functionals generally prohibit them from being applied in studying exotic electronic properties of materials where a large number of band gap calculations are required, such as electronic dephasing in photo-absorbing semiconductors originating from the strong electron-phonon coupling in the material.\cite{li2018time,li2019anharmonicity,yang2020correlating} 

Another important consequence of the delocalisation error in DFT is the challenge to correctly determine the energetics of the charge-transfer complexes when DFT is used for predicting the properties of materials. Here, we focus on the calculation of the charge transfer integral $J_{ij}$, an important parameter in molecular electronics that is the energy required for an electron/hole to be transferred from one monomer to another in a non-covalently bounded molecular dimer [\emph{c.f.} \cref{Markus}], which is equivalent to the reaction energy of D$^{-}$+A$\to$D+A$^{-}$ (for electron transfer where D stands for donor and A for acceptor). Due to the presence of delocalisation error, standard DFT cannot be applied to determine the reaction energy since it fails to localise an extra charge on either the donor or acceptor. Methods such as the constrained density functional theory (CDFT)\cite{wu2006extracting} and frozen density embedding (FDE)\cite{pavanello2011modelling} are some of the proposed solutions to this problem, however, it still requires the careful choice of appropriate DFA to reach the right answer. For example, \citet{kubas2014electronic} put forward a benchmark set of $J_{ij}$ for 11 small molecule homodimers, which shows that PBE+50\% HF exchange gives the best result ($\sim 0.8$ \% error), whereas overestimates $J_{ij}$ by 38.7 \% because of the overly diffused charge densities produced by the pure PBE functional. In the context of energy landscape, recent study from \citet{leblanc2018pervasive} showed that, standard dispersion-corrected DFT methods failed to correctly determine the relative energetic stabilities of the proton-transferred acid-base forms of multicomponent molecular crystals over their co-crystal forms, which poses challenges in applying organic crystal structure prediction methods\cite{reilly2016report,Hunnisett:aw5093,Hunnisett:aw5094} for polymorph screenings with computationally efficient methods based on electronic structure theory.

\begin{figure}[tb]
    \centering
    \includegraphics[width=0.7\linewidth]{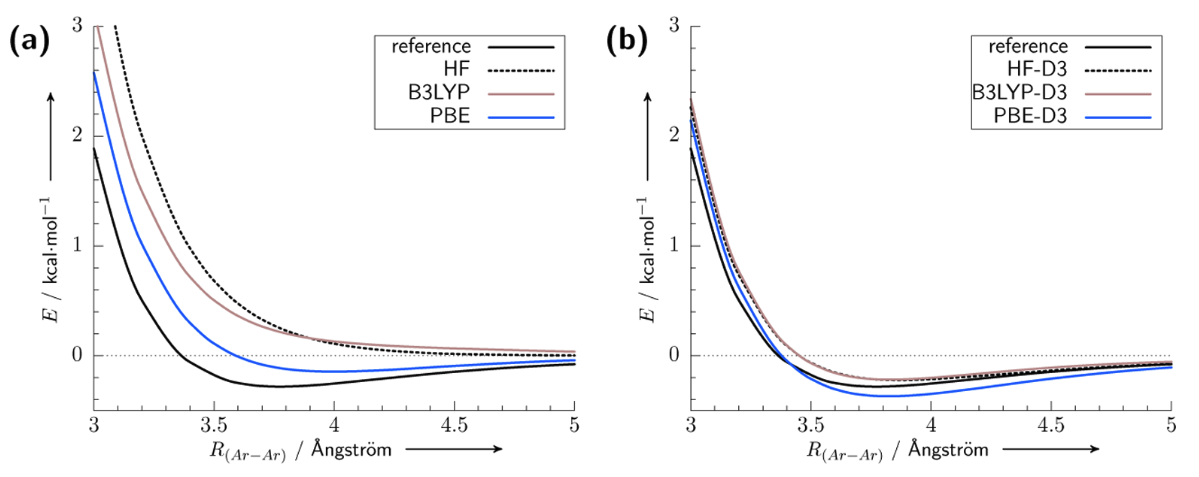}
    \caption{Computed potential energy curves for two interacting Ar atoms with and without dispersion correction with respect to an accurate reference [Adapted from \citet{grimme2016dispersion}].}
    \label{fig:Ar_dimer_PES}
\end{figure}

\paragraph{\textbf{Dispersion interactions in non-covalently bounded systems.}} Dispersion interaction is the attractive part of the vdW interaction in non-covalently bounded complexes. Being able to correctly determine the magnitudes of dispersion interactions is critical across a wide range of atomistic modellings with broad implications, including chemical reactions, catalysts, molecular recognitions, protein folding, gas adsorptions, and 2D materials. However, interaction energies determined from semilocal DFAs are generally too repulsive thus fail to give the correct binding curves of vdW complexes [\cref{fig:Ar_dimer_PES}]. This failure arises because the dispersion interaction is essentially a long-range dynamical electron correlation effect, whereby the electron densities on two interacting bodies A and B are constantly fluctuating, which generate instantaneous dipole moment on each body that attracts each other via Coulomb interaction. From the perspective of MP2 (second-order M\o{}ller–Plesset perturbation) theory,\cite{moller1934note} this correlation energy is related to the interaction energies of electrons transit from the ground to the excited states on the two interacting bodies:
\begin{equation}\label{MP2_corr}
    E_{\mbox{\scriptsize{corr}}}^{AB}=-\sum_{ia}\sum_{jb}\frac{(ia\vert jb)[(ia\vert jb)-(ib|ja)]}{\varepsilon_{a}+\varepsilon_{b}-\varepsilon_{i}-\varepsilon_{j}},
\end{equation}
in which $(ia\vert jb)=\iint \phi_i^\dagger (\bm{r}_{1})\phi_a(\bm{r}_1)\frac{1}{|\bm{r}_{1}-\bm{r}_2|} \phi_j^\dagger (\bm{r}_{2})\phi_b(\bm{r}_1)  d\bm{r}_1 d\bm{r}_2$ is the two electron integral for the particle-hole excitations involving the excitation of electron from occupied orbital $i$ (or $j$) to the unoccupied orbital $a$ (or $b$) that are localised on A (or B), with $\varepsilon$ being the corresponding orbital energy. It is therefore clear that, the involvement of excited states (contrary to DFAs which are build upon the ground state electron densities), as well as long-range electron-electron interactions (whereas DFAs are local in $\bm{r}$), make it challenge for standard DFAs to quantitatively, or even qualitatively, determines the strengths of the dispersion interactions in atomistic modelling.

In empirical atom-atom forcefields, the dispersion interaction can be modelled as $E_{\mbox{\scriptsize{disp}}}=\sum_{i<j}-C_6/R^6_{ij}$ (with the possibilities to include higher order terms such as $-C_8/R^8$ and $-C_{10}/R^{10}$) with $R_{ij}$ being the interatomic distance between atom $i$ and $j$. Modern \emph{dispersion-corrected DFAs}\cite{grimme2016dispersion} primarily builds upon this idea and correct the DFA interaction energy between two bodies A and B as $E_{\mbox{\scriptsize{tot}}}^{\mbox{\scriptsize{AB}}}=E_{\mbox{\scriptsize{DFA}}}^{\mbox{\scriptsize{AB}}}+E_{\mbox{\scriptsize{disp}}}^{\mbox{\scriptsize{AB}}}$ where $E_{\mbox{\scriptsize{disp}}}^{\mbox{\scriptsize{AB}}}$ takes the same atom pairwise form as in the empirical atom-atom forcefields, with atom-atom dispersion coefficients $C_{n}$ $(n=6,8,10\ldots)$ determined from high-level quantum chemical calculations, thus minimising the level of empiricism in the overall method. This is fundamentally possible, for example, by starting from the MP2 correlation energy [\cref{MP2_corr}], the $C_6$ coefficient is related to the frequency-dependent  polarisability $(\alpha)$ of the interacting atomic dipoles as the Casimir-Polder integral\cite{casimir1948influence} 
\begin{equation}\label{eq:casimir_eq}
    C_6^{\mbox{\scriptsize{AB}}}=\frac{3}{\pi}\int_{0}^{\infty} \alpha_{\mbox{\scriptsize{A}}}(i\omega)\alpha_{\mbox{\scriptsize{B}}}(i\omega)\mbox{d}\omega,
\end{equation}
where $\alpha$ can be calculated from time-dependent density functional theory. \cref{eq:casimir_eq} also implies that atom with lower coordination number (CN) in the structure will have a more diffused electron density, thus becomes more polarisable resulting in a larger $C_6$ coefficient. This geometry-dependent attenuation effect of dispersion coefficients is built into the major modern dispersion correction approaches to enhance the accuracy, where in the following, we will briefly review the key features of some dispersion correction schemes that are mostly used in the materials modelling community. 

\subsubsection{Modern Dispersion Corrected DFAs Commonly Applied in Materials Modelling}

It should be noted that, the meaning of `commonly applied in materials modelling', we implies scenarios where periodic boundary conditions and planewave (or equivalent) basis sets are usually applied in materials simulations, and the dispersion correction schemes discussed below have their equivalent versions being implemented for computations under periodic boundary conditions. However, here we mainly focus on how the dispersion correction schemes were originally developed for molecular systems, in which atom-centred basis sets are used, highlighting how modern dispersion correction scheme account for the chemical environments in an atomistic system. The extension of these from molecular to periodic systems are  technical details on how the total dispersion energies are summed across interacting atomic pairs (and multiplets), that, in our opinion, do not change how the dispersion coefficients are determined by taking into account the local chemical environments, which is the key theoretical aspect we would like to focus here.

\begin{figure}[tb]
    \centering
    \includegraphics[width=\linewidth]{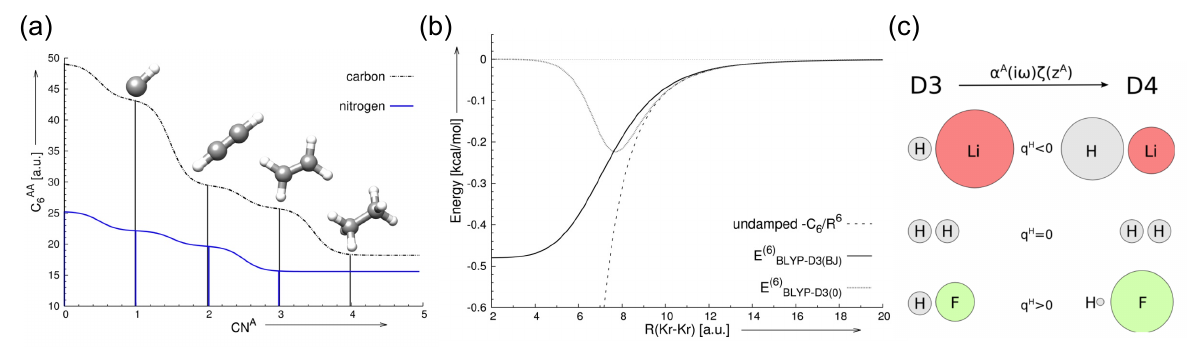}
    \caption{Key features in Grimme's $-$D3 and $-$D4 dispersion correction schemes. (a) CN-dependent atom pairwise dispersion coefficient for homoatomic pairs $C_{6}^{AA}$ as a function of the coordination number CN$^{A}$. The reference points $C_{6,ref}^{AA}$ are drawn as vertical sticks in the respective colours (black for carbon, blue for nitrogen). For carbon, the structures of CH, ethyne, ethene, and ethane (corresponding to the coordination numbers 1-4) are depicted as well. (b) Different D3 damping schemes for the $E_{\mbox{\scriptsize{disp}}}^{(0)}$ contribution of the Kr-Kr dispersion energy. For D3(BJ) (solid black) and D3(0) (dotted gray), the curves are given using the damping parameters for the BLYP functional. The undamped $-C_{6}/R^{6}$ contribution is given for comparison (dashed black). [Adapted from \citet{grimme2016dispersion}] (c) Schematic representation of the $\zeta$-functionality which rescales atomic polarisabilities based on input Mulliken charges $q$ in the $-$D4 model. The left side shows the standard partitioning of atomic polarisabilities in D3. On the right side, the scaled atomic polarisabilities are shown. The depicted radii of the atoms directly correspond to the calculated atomic $C_{6}^{AA}$ coefficients. [Adapted from \citet{caldeweyher2017extension}]}
    \label{fig:grimme_dispersions}  
\end{figure}

\paragraph{\textbf{Grimme's -D3 and -D4 schemes}} The popular -D3 correction was first proposed by \citet{grimme2010consistent} in 2010, where the total dispersion energy is given by the sum of an atom pairwise and triplewise (given in the Axilrod-Teller-Muto form\cite{axilrod1943interaction,muto}) terms:
\begin{equation}\label{eqn:D3}
E_{\mbox{\scriptsize{disp}}}^{{\mbox{\scriptsize{D3}}}}=-\sum_{AB}\sum_{n=6,8}s_n \frac{C_n^{\mbox{\scriptsize{AB}}}}{R^n}f_{\mbox{\scriptsize{damp}}}^{(n)}(R)-\sum_{ABC}s_9 \frac{C_{9}^{\mbox{\scriptsize{ABC}}}(3\cos\theta_a \cos\theta_b\cos\theta_c +1)}{(R_{AB}R_{AC}R_{BC})^3}f_{\mbox{\scriptsize{damp}}}^{(9)}(\bar{R}_{ABC}).
\end{equation}
Here, the leading $C_6^{\mbox{\scriptsize{AB}}}$ coefficients are calculated from its corresponding reference value $C_{6,\mbox{\scriptsize{ref}}}^{\mbox{\scriptsize{AB}}}$ determined from the polarisabilities of model hydrides A$_m$H$_n$ and B$_k$H$_l$ systems using a modified version of \cref{eq:casimir_eq}:
\begin{equation}\label{eq:d3-casimir-eq}
C_{6,\mbox{\scriptsize{ref}}}^{\mbox{\scriptsize{AB}}}(\mbox{CN}^A,\mbox{CN}^{B})=\frac{3}{\pi}\int_0^\infty\frac{1}{m}\left[\alpha^{A_m H_n}(i\omega)-\frac{n}{2}\alpha^{H_2}(i\omega)\right]\times\frac{1}{k}\left[\alpha^{B_k H_l}(i\omega)-\frac{l}{2}\alpha^{H_2}(i\omega)\right]\mbox{d}\omega,
\end{equation}
in which CN for atom A in the reference molecule is given by:
\begin{equation}\label{eq:D3_coord_nummer}
\mbox{CN}^A=\sum_{B\neq A}^{N_{atoms}}\left\{1+\exp\left[-16\frac{4(R_{A,\mbox{\scriptsize{cov}}}+R_{B,\mbox{\scriptsize{cov}}})}{3R_{AB}-1}\right]\right\}^{-1},
\end{equation}
where $R_{A,\mbox{\scriptsize{cov}}}$ is the covalent radius of element A. As illustrated in \cref{fig:grimme_dispersions}(a), the inclusion of the CN effectively reduces the magnitude of the atomic $C_6$ coefficients for bare atoms to those with a high CN (hence less exposed electron density). With the precomputed pairwise $C_{6,\mbox{\scriptsize{ref}}}^{\mbox{\scriptsize{AB}}}$, in the actual D3 calculation for the target system, the pairwise atom $C_{6}^{\mbox{\scriptsize{AB}}}$ is further calculated as the Gaussian average of the precomputed values:
\begin{equation}\label{eq:actual_c6_grimme}
C_{6}^{\mbox{\scriptsize{AB}}}(\mbox{CN}^A,\mbox{CN}^{B})=\frac{\sum_i^{N_A}\sum_j^{N_B}C_{6,\mbox{\scriptsize{ref}}}^{\mbox{\scriptsize{AB}}}(\mbox{CN}^A,\mbox{CN}^{B})L_{ij}}{\sum_i^{N_A}\sum_j^{N_B}L_{ij}},
\end{equation}
with $L_{ij}=\exp\left\{-4[(\mbox{CN}^A-\mbox{CN}_i^A)^2+(\mbox{CN}^B-\mbox{CN}_i^B)^2] \right\}$. The next order $C_8^{\mbox{\scriptsize{AB}}}$ take into account the dipole-quadrupole contribution to the dispersion interaction, which is computed recursively as $C_8^{\mbox{\scriptsize{AB}}}=3C_6^{\mbox{\scriptsize{AB}}}\sqrt{Q_A Q_B}$ with $Q_{A}=\sqrt{Z_A}\frac{\langle r_A^4\rangle}{\langle r_A^2\rangle}$ where $Z_A$ is the atomic number of element A and $\langle r_A^4\rangle$ and $\langle r_A^2\rangle$ are multipole-type expectation values
derived from atomic densities. The scaling factor $s_6$ is typically set to unity and $s_8$ term is used to absorb higher order multipolar terms into the $R^{-8}$ term. Finally, the three-body dispersion coefficient is approximated as $C_{9}^{\mbox{\scriptsize{ABC}}}\approx-\sqrt{C_{6}^{\mbox{\scriptsize{AB}}}C_{6}^{\mbox{\scriptsize{AC}}}C_{6}^{\mbox{\scriptsize{BC}}}}$.

Furthermore, as shown in \cref{fig:grimme_dispersions}(b), the undamped dispersion energy $-C_6/R^6$ decays to negative infinity at small interatomic distance, that can lead to the unphysical behaviours of atoms collapsing into each other when combined with standard DFT calculations. This problem can be overcome by including an additional damping function $f_{\mbox{\scriptsize{damp}}}^{(n)}(R)$. For the -D3 scheme, both zero damping\cite{chai2008long} (which brings the dispersion energy to zero at zero interatomic separation) and Becke-Johnson (BJ)\cite{johnson2005post,grimme2011effect} damping are available, with the later being the default choice for the -D3 scheme [dubbed as -D3(BJ)].

In 2017, the -D3 model is extended to its new version, termed as the -D4 model.\cite{caldeweyher2017extension,caldeweyher2019generally} The significant improvement in the newer model [\cref{fig:grimme_dispersions}(c)] is the scaling of the polarisabilities [\emph{c.f} \cref{eq:d3-casimir-eq}] based on the atomic charge $\alpha^{A}(i\omega)\to\alpha^{A}(i\omega)\zeta(Z^A)$ computed in the target molecular environment. Consequently, the dispersion coefficient can no longer be precomputed and the Casimir-Polder integral must be evaluated on-the-fly to obtain the atomic dispersion coefficients that can more accurately account for the electrostatic environment of each atom in specific molecules.

\paragraph{\textbf{The Tkatchenko-Scheffler (TS) models}} In 2009, \citet{tkatchenko2009accurate} put forward an atom pairwise dispersion correction scheme that is similar to the -D3 correction, except that the atom polarisability $\alpha_A$ and the $C_6^{AA}$ dispersion coefficients are density-dependent by scaling them from their free-atom values as:
\begin{equation}\label{eq:TS_pol_scaling}
    \alpha_A=v_A\alpha_{\mbox{\scriptsize{free}}}^A\qquad\mbox{and}\qquad C_6^{AA}=v_A^2C_{6,{\mbox{\scriptsize{free}}}}^{AA},
\end{equation}
in which the scaling coefficient is computed based on the Hirshfeld charge density partition scheme\cite{hirshfeld1977bonded} as
\begin{equation}\label{eq:hirshfeld}
v_A=\frac{\displaystyle{\int} r^3 w_A(\bm{r})\rho(\bm{r})\mbox{d}^3\bm{r}}{\displaystyle{\int} r^3 \rho^{\mbox{\scriptsize{free}}}_{A}(\bm{r})\mbox{d}^3\bm{r}}, \qquad \mbox{with}\qquad w_A(\bm{r})=\frac{\rho^{\mbox{\scriptsize{free}}}_{A}(\bm{r})}{\sum_B \rho^{\mbox{\scriptsize{free}}}_{B}(\bm{r})}.
\end{equation}

Subsequently, the TS dispersion model is extended beyond pairwise additivity to account for the dispersion interaction that arises from the collective fluctuation of electron densities in the interacting molecules, thus reflecting the intrinsic many-body effects behind dispersion interactions\cite{tkatchenko2012accurate,tkatchenko2013interatomic} [also see \cref{sect:polarisable_FF} for detailed discussion]. This means that the atomic polarisabilities and dispersion coefficients for an atom in the target molecular system will be affected by those of the other atoms present in the same system. To account for such an effect, the many-body TS model represents all atoms in the system as a collection of quantum harmonic oscillators, which are able to interact with each others and alter the dynamic electronic polarisability of every atom. As such, 
the imaginary frequency dipole polarisability of any atom $p$ needs to be computed self-consistently from the polarisabilities of all other atoms $q\neq p$ in the system as
\begin{equation}\label{eq:SCS_polarisation}
\alpha_{p}^{\mbox{\scriptsize{SCS}}}(i\omega)=\alpha_{p}^{\mbox{\scriptsize{TS}}}(i\omega)+\alpha_{p}^{\mbox{\scriptsize{TS}}}(i\omega)\sum_{p\neq q}\hat{\mathcal{T}}_{pq}\alpha_{q}^{\mbox{\scriptsize{SCS}}}(i\omega),
\end{equation}
 in which $\hat{\mathcal{T}}_{pq}$ is the dipole-dipole interaction tensor. From here, the many-body dispersion energy is then computed using the coupled fluctuating dipole model by diagonalising the coupling matrix of the fluctuating dipole moments.

An other extension to the many-body TS model is the fractionally ionic approach (FIA)\cite{gould2016fractionally} to properly account for the ionic charge and volumes in fully ionic systems that can differ significantly from molecular systems. Here, the frequency-dependent atomic dipole polarisability model employed for a system with non-integer electron number $N$ is a linear interpolation of the polarisabilities for system with $M=\lfloor N \rfloor$ and $M+1$ electrons. Compared to the standard -D3 and TS dispersion correction scheme (MAE $\sim 1.5-1.8$ meV/\AA) , the FIA (MAE $\sim 0.9$ meV/\AA) provides notable improvements in the binding energies of ionic 2D materials.\cite{tawfik2018evaluation} This clearly shows that practitioners of computational materials studies should be more cautious in using `generally applicable' methodologies, particularly be aware of the assumption(s) adapted behind the developed method.

\paragraph{\textbf{The exchange-dipole moment (XDM) model}}
The XDM model is the density-dependent dispersion correction model proposed by Becke and Johnson in 2005.\cite{johnson2005post,becke2005density,becke2005exchange,johnson2006post,becke2007exchange,otero2013many} The starting point of the XDM model is to reformulate the exact exchange interactions [\cref{eq:exchange_HF_DFT}] as (spin-dependent, labelled with $\sigma=\uparrow$ and $\downarrow$) a localised electron density interacting with a nonlocal exchange hole $h_{X\sigma}(\bm{r},\bm{r}^\prime)$ as
\begin{equation}\label{eq:exchange_interaction_hole_form}
    E_{x}=\frac{1}{2}\sum_{\sigma}\int \mbox{d}\bm{r}\int\mbox{d}\bm{r}^\prime \rho_\sigma (\bm{r})\frac{h_{X\sigma}(\bm{r},\bm{r}^\prime)}{|\bm{r}-\bm{r}^\prime|}.
\end{equation}
whereby in Kohn-Sham DFT, the exchange hole can be computed as
\begin{equation}\label{eq:exchange_hole}
h_{X\sigma}(\bm{r},\bm{r}^\prime)=-\frac{1}{\rho_\sigma(\bm{r})}\sum_{ij} \phi_{i\sigma}(\bm{r}) \phi_{j\sigma}(\bm{r}) \phi_{i\sigma}(\bm{r}^\prime) \phi_{j\sigma}(\bm{r}^\prime),
\end{equation}
which involves the summation over the occupied spin-orbitals ($\psi_{i\sigma}$). Typically, the exchange hole is not spherically symmetry around the electron, leading to a non-vanishing exchange (hole) dipole moment and analogously higher order exchange hole multipoles, thus one can imagine the dispersion interaction arises from the instantaneous interaction between the exchange and induced dipole (and multipole) moments of two fragments A and B (which, in the XDM model, are defined using the Hirshfeld partitioning scheme as in \cref{eq:hirshfeld}). The exchange hole dipole moment, in this case, is
\begin{equation}\label{eq:exchange_dipole_moment}
\bm{\mu}_{X\sigma}(\bm{r})=\int \bm{r}^\prime h_{X\sigma}(\bm{r},\bm{r}^\prime)\mbox{d}\bm{r}^\prime -\bm{r}.
\end{equation}
From here, the square of the $l$-pole operator can be calculated as:
\begin{equation}\label{eq:XDM_l_pole_moment}
\langle M_l^2\rangle_A=\sum_\sigma\int \rho_\sigma(\bm{r})[r_A^l -(r_A-\mu_{X\sigma}(\bm{r}))^l]^2\mbox{d}\bm{r},
\end{equation}
which allows the two-body dispersion coefficients to be obtained as:
\begin{align}
    C_6^{\mbox{\scriptsize{AB}}}&=\frac{\alpha_A\alpha_B\langle M_{1}^2\rangle_A\langle M_{1}^2\rangle_B}{\alpha_B \langle M_{1}^2\rangle_A+\alpha_A \langle M_{1}^2\rangle_B}, \label{eq:XDM_C6}\\
    C_8^{\mbox{\scriptsize{AB}}}&=\frac{3}{2}\frac{\alpha_A \alpha_B (\langle M_{1}^2\rangle_A \langle M_{2}^2\rangle_B+ \langle M_{2}^2\rangle_A\langle M_{1}^2\rangle_B)}{\alpha_B \langle M_{1}^2\rangle_A+\alpha_A \langle M_{1}^2\rangle_B}, \label{eq:XDM_C8}\\
    C_{10}^{\mbox{\scriptsize{AB}}}&=2\frac{\alpha_A \alpha_B (\langle M_{1}^2\rangle_A \langle M_{3}^2\rangle_B+ \langle M_{3}^2\rangle_A\langle M_{1}^2\rangle_B)}{\alpha_B \langle M_{1}^2\rangle_A+\alpha_A \langle M_{1}^2\rangle_B}+\frac{21}{5}\frac{\alpha_A\alpha_B\langle M_{2}^2\rangle_A\langle M_{2}^2\rangle_B}{\alpha_B \langle M_{1}^2\rangle_A+\alpha_A \langle M_{1}^2\rangle_B}, \label{eq:XDM_C10}
 \end{align}
in which $\alpha$ are the atomic polarisability determined as in \cref{eq:TS_pol_scaling}. With these, the dispersion correction to the DFT energy takes the form of 
\begin{equation}\label{eq:XDM_disp_energy}
E_{\mbox{\scriptsize{disp}}}^{\mbox{\scriptsize{XDM}}}=-\sum_{AB}\sum_{n=6,8,10}\frac{C_n^{\mbox{\scriptsize{AB}}}}{R^n}f_{\mbox{\scriptsize{damp}}}^{(n)}(R),
\end{equation}
where $f_{\mbox{\scriptsize{damp}}}^{(n)}(R)$ is the Becke-Johnson damping function. Extension of the XDM model to include the three-body Axilrod-Teller-Muto term had also been established,\cite{otero2013many} but it is not used by default in XDM as the dispersion energy vary significantly with the underlying choice of the damping function

\paragraph{V\textbf{ydrov and Van Voorhis functionals}} Instead of applying a \emph{post hoc} correction of the dispersion interaction energy to the DFT energy, alternative approach that seeks to construct nonlocal correlation functionals that use electron density as the direct input, and an integration kernel $\Phi(\bm{r},\bm{r}^\prime)$ that correlates the electron densities at two electron coordinates:
\begin{equation}\label{eq:nonlocal_disp_corr_kernel}
E_{c}^{\mbox{\scriptsize{NL}}}=\frac{1}{2}\int \mbox{d}\bm{r} \int \mbox{d}\bm{r}^\prime\rho(\bm{r})\Phi(\bm{r},\bm{r}^\prime) \rho(\bm{r}^\prime) .
\end{equation}
Along this venue of method development,\cite{grafenstein2009efficient,dion2004van,lee2010higher,vydrov2009improving,vydrov2009nonlocal,vydrov2010nonlocal} the VV10 functional constructed by \citet{vydrov2010nonlocal} in 2010 represents the one with the simplest functional form and the best accuracy for molecular systems. Here, the correlation kernel is introduced \emph{ad hoc} based on author's experience:
\begin{equation}\label{eq:vv10_kernel}
\Phi(\bm{r},\bm{r}^\prime)=-\frac{3}{2gg^\prime(g+g^\prime)}, 
\end{equation}
with $g=\omega_0(\bm{r})|\bm{r}-\bm{r}^\prime|^2+\kappa(\bm{r})$ and $g^\prime=\omega_0(\bm{r}^\prime)|\bm{r}-\bm{r}^\prime|^2+\kappa(\bm{r}^\prime)$. Here, $\omega_0(\bm{r})$ is related to the local plasma frequency  $\omega_p
(\bm{r})=\sqrt{4\pi\rho(\bm{r})}$ and local band gap $\omega_g(\bm{r})=\sqrt{C\left\vert\nabla \rho(\bm{r})/\rho(\bm{r})\right\vert}$ 
(with $C$ being the adjustable parameter) as $\omega_0(\bm{r})=\sqrt{\omega_g^2(\bm{r})+\omega_p^{2}(\bm{r})/3}$, and the other component in $g$ is $\kappa(\bm{r})=b[3\pi\rho(\bm{r})^{1/3}]/\omega_{p}(\bm{r})$, in which $b$ is a parameter can be adjusted\cite{hujo2011performance} to different standard density functionals that is similar to the DFT-D3 schemes. Finally, a constant multiplied by the number of electrons in the system is included additionally in the correlation kernel to ensure that the long-range correlation energy vanishes in the uniform electron gas limit. Another important aspect of this particular nonlocal density functional is that there exists a self-consistent implementation of the functional,\cite{vydrov2010implementation} which enables one to study the influence of dispersion interactions on the topology of  electron  densities, otherwise not possible from \emph{post hoc} dispersion-corrected DFT methods.
\subsubsection{Practical Aspects in DFT Calculations}\label{sect:practical_DFT}

Having reviewed the fundamental concepts and the construction of DFAs across different rungs of the Jacob's ladder, in this section, we shall briefly review some of the key concepts that are related to both the numerical implementations and practical applications of DFT calculations.

\paragraph{\textbf{Self-consistent field (SCF)}} From \cref{eq:DFT_min_rho} and \cref{eq:KS_eqn}, it can be seen that, the solution of the Kohn-Sham equation provides us with the ground state single electron density $\rho(\bm{r})$ for a system with a fixed nuclei position. However, the construction of the Kohn-Sham energy functional itself requires the knowledge of $\rho(\bm{r})$, which is not known \emph{apriori}. This is seemingly like a `\emph{chicken-and-egg}' problem, Mathematically, such a problem can be tackled by computing the SCF in DFT, which is typically formulated by rewriting the Kohn-Sham energy in \cref{eq:KS_eqn} as its corresponding Kohn-Sham potential $V_{\mathrm{KS}}(\bm{r})=[\delta E_{\mathrm{KS}}(\bm{r})/\delta \rho]\vert_{\rho(\bm{r})}$. 

\begin{figure}[t] 
    \centering
    \includegraphics[width=0.6\textwidth]{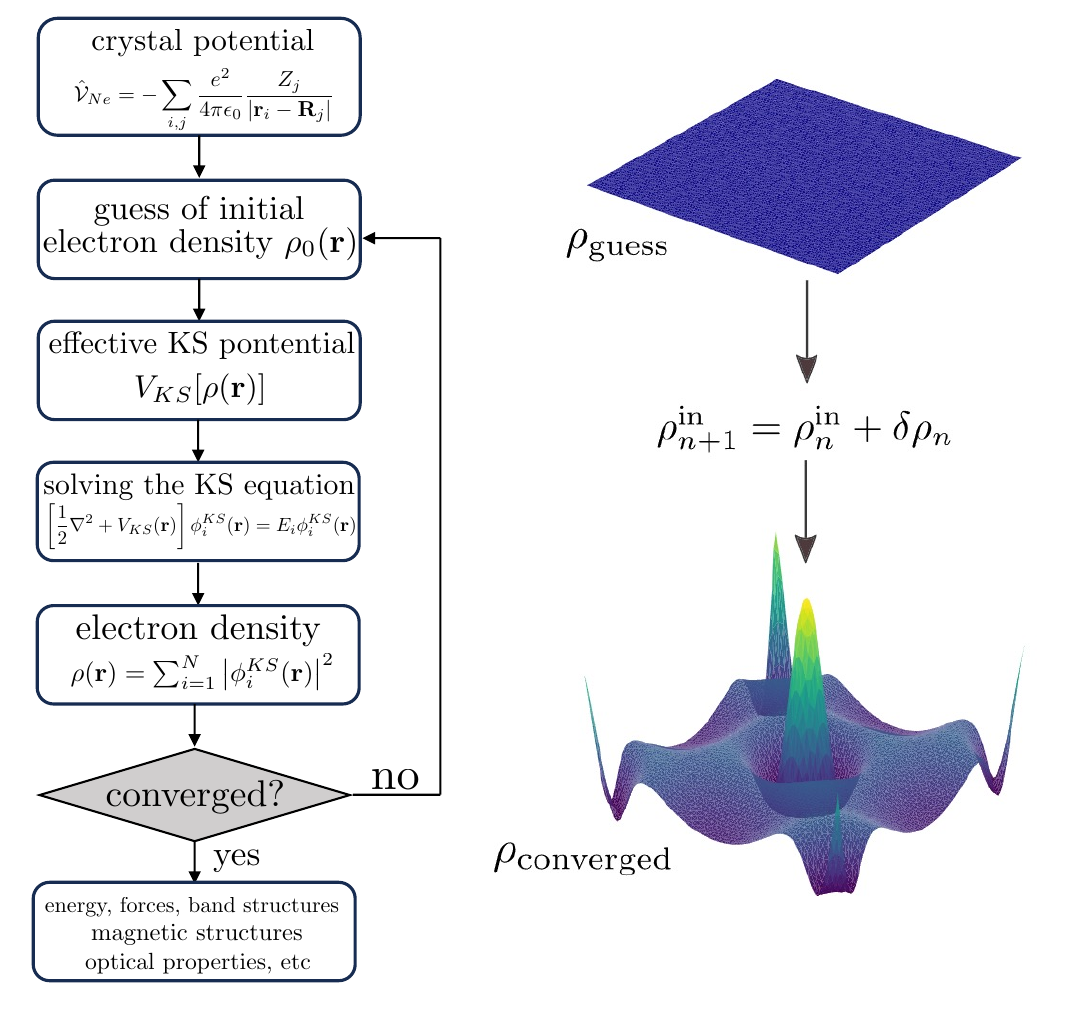}
    \caption{(left) Schematic flowchart for the self-consistent field procedure for solving the Kohn-Sham equation. (right) An iterative algorithm generates a series of perturbations to the density $\delta \rho_n$ in order to converge the initial guess density (top) to the fixed-point density $\rho_{*}$ (bottom). Example shown for an FCC four atom aluminium unit cell. [\citet{woods2019computing}]}
    \label{fig:KS-SCF}
\end{figure}

Practically, as shown in \cref{fig:KS-SCF}(left), the SCF starts with a initial guessed electron density $\rho_{0}(\bm{r})$ based on the crystal potential (here refer to a periodic calculation), that is determined by the positions of the nuclei. Then, the initial density $\rho_{0}$ can be used  as the input density $\rho^{\mathrm{in}}$ to construct the effective Kohn-Sham potential $V_{\mathrm{KS}}$ that can be plugged into the corresponding Kohn-Sham equation, solving which leads to an updated electron density $\rho^{\mathrm{out}}$. By comparing the difference $\| \rho^{\mathrm{out}}-\rho^{\mathrm{in}}\|$, one can gauge if the input density correctly reflects the underlying crystal potential, and if the discrepancy between the two is large, then   $\rho^{\mathrm{out}}$ will be refined (\emph{e.g.} with quasi-Newton method) and plugged back into the Kohn-Sham equation as the new $\rho^{\mathrm{in}}$ to solve for a new $\rho^{\mathrm{out}}$. This procedure will be carried out iteratively until $\| \rho^{\mathrm{out}}-\rho^{\mathrm{in}}\|$ is lower than a pre-defined convergence criteria. \cref{fig:KS-SCF}(right)\cite{woods2019computing} provides a schematic illustration on how the electron density evolves in an SCF procedure to reflect the underlying electrostatic potential. 
 
Because the SCF equation is solved iteratively to reach self-consistency, there are considerations that need to be accounted for, with corresponding approaches and methodologies being developed to address these issues to accelerate the convergence of electronic SCF in practical DFT calculations, which we will outline a few of them that occur commonly in the literature, whereas for more comprehensive discussion on SCF, we refer the readers to the review by \citet{woods2019computing}.

Secondly, it was found in the SCF procedure, that directly using $\rho^{\mathrm{out}}_n$ from the $n$-th iteration as the input $\rho^{\mathrm{in}}_{n+1}$ for the $(n+1)$-th interaction often led to numerical instability thus slow SCF convergence. A practical solution to this problem is to adapt a \emph{density-mixing} scheme, in which $\rho^{\mathrm{in}}_{n+1}$ is constructed by proportionally mixing $\rho^{\mathrm{out}}_{n}$ with the residues $\delta\rho_n=\|\rho^{\mathrm{out}}_{n}-\rho^{\mathrm{in}}_{n}\|$ from previous $n$ iterations. A similar approach in modern deep learning can be seen in the architecture of the residual multilayer perceptron (MLP) to accelerate model training.\cite{he2016deep} Some of the widely used density-mixing schemes in DFT include: (1) The Pulay scheme,\cite{pulay1980convergence} or otherwise known as Direct Inversion of the Iterative Subspace (DIIS), in which one seeks an ansatz $\rho_{n+1}^{\mathrm{in}}=\sum_{i=n-m+1}^{n}c_i \rho^{\mathrm{out}}_i$ as the linear combination of the output density $\rho^{\mathrm{out}}_i$ from previous $m$ $(m<n)$ SCF interactions, with the coefficient $c_i$ being determined from minimising the weighted residue of $\|\sum_{i=n-m+1}^{n}c_i \delta\rho_i\|^2$. (2) The Broyden scheme,\cite{johnson1988modified,broyden1965class} further generalise the Pulay scheme. Generally, in a gradient-based optimisation, the function can be updated by using the relationship $\rho_{n+1}=\rho_{n}-\bm{J}_n^{-1}\delta \rho_n$ with $\bm{J}_n^{-1}$  being the inverse Jacobian matrix. However, direct calculation and inversion of the Jacobian matrix at each iteration is costly, and instead, according to \citet{broyden1965class}, one could initiate $\bm{J}_{0}^{-1}$ with a simple diagonal matrix and iteratively update it along with $\rho$ in the SCF procedure. Note that these mixing procedures are proposed as the general accelerating schemes for local optimisation problems, and are similarly applied to other problems in atomistic materials modelling, such as geometry optimisations.

Finally, the electron density is constructed from the Kohn-Sham orbitals as 
\begin{equation}
    \label{eq:electron_density_KS_orbital}
\rho(\bm{r})=\sum_{i=1}^{N}c_i|\phi^{\mathrm{KS}}_i(\bm{r})|,
\end{equation}
in which $c_i=2$ and 0 for occupied and unoccupied non-spin-polarised orbitals at 0 K, respectively. Whilst this is physically correct, the discontinuity in the orbital occupations $c_i$ can lead to numerical instability in the SCF calculation. One solution to overcome this problem is to allow fractional occupancies by exciting electrons from the occupied to the unoccupied orbitals whilst satisfying the constraints of the exclusion principle, \emph{i.e.} $c_{i}\in [0,2]$, whilst $\sum_{i=1}^{N} c_i=N_e$, the total number of electrons in the system. For solid-state calculations, in particular, this may be achieved with Fermi smearing by introducing an electronic temperature $T_i$ such that $c_i=1/[\exp (\epsilon_i-\mu)/T+1]$ according to the Fermi-Dirac statistics, where $\mu$ is the Fermi energy.

\paragraph{\textbf{Basis functions}} One question that has not been addressed in the SCF procedure is how one should construct the Kohn-Sham orbitals $\{\varphi^{\mathrm{KS}}_i(\bm{r})\}$ that leads to the electron density [\cref{eq:electron_density_KS_orbital}], which also determines how the electron density is algorithmically optimised to the one corresponding to the Kohn-Sham ground state. 

We answer this question by first looking at the case of molecular systems in the field of quantum chemistry, whereby the finite spatial extent of a molecule renders the corresponding SCF calculations to be performed with orbitals represented in real space, which is conceptually easy to understand. For this case, to be in line with the molecular orbital theory in physical chemistry that takes the molecular orbitals as the \emph{linear combination of atomic orbitals} (LCAO), we construct the Kohn-Sham orbital using \emph{atom-centred basis functions} $\{\varphi_{\mu}(\bm{r})\}$ such that 
\begin{equation}
\label{eq:molecular_basis_func_expansion}
\phi_i(\bm{r})=\sum_{\mu}c_{i\mu}\varphi_{\mu}(\bm{r}).
\end{equation}
Under such a construction, the SCF is essentially a procedure that iteratively adjusts the expansion coefficients $\{c_i,c_{i\mu}\}$ to reach the optimum $\rho(\bm{r})$ for a given molecule. A trivial choice for $\{\phi_{\mu}(\bm{r})$ is to follow the solution of hydrogen atom from the Schr\"{o}dinger equations by using a functional form of 
\begin{equation}
\varphi_{\mu}^{\mathrm{Slater}}(r)\sim r^n e^{-\zeta r},
\end{equation}
which leads to the design of the Slater basis functions\cite{slater1930atomic} that gives the correct cusp and exponential decay behaviour with respect to $r$ as derived from fundamental quantum mechanics. However, the Slater basis functions are rarely used in modern quantum chemistry practices because the computation of the overlap integral [such as \cref{eq:exchange_HF_DFT}
 for the exact exchange] is difficult thus not numerically efficient. 

To overcome this problem, \citet{boys1950electronic} introduces the Gaussian basis set:
\begin{equation}
\varphi_{\mu}^{\mathrm{Gaussian}}(r)\sim r^n e^{-\zeta r^2},
\end{equation}
which is now the standard choice in molecular calculations, for which the molecular integrals between two atomic orbitals can be analytically calculated. This significantly enhances the computational efficiency, however, the Gaussian basis function do not have the correct cusp and long-ranged decay behaviours as in the Slater's basis function. Solving this problem gave rise to the famous Pople basis functions\cite{hehre1969self}, in which a contraction scheme was devised to optimise a combination of the Gaussian functions to achieve a good approximation on the shape of a particular Slater orbital, while the numerical efficacy of the Gaussian basis functions could be retained, achieving the desirable outcomes from both the Slater and Gaussian basis functions. 

Theoretically, one can anticipate from  \cref{eq:molecular_basis_func_expansion} that the most accurate system energy can be acquired if the expansion is complete $\mu\to\infty$, which leads to the so-called complete basis set limit. However, this can never be achieved practically whereby the basis function must be truncated up to a certain order $\mu_{\max}$. In quantum chemical calculations on the non-covalent interaction energy between two fragments $\mathcal{A}$ and $\mathcal{B}$, this gives rise to the basis set superposition error (BSSE)\cite{boys1970calculation} in the computed interaction energy. Computing the interaction energy is usually achieve by computing the total energies for the isolated systems $\mathcal{A}$ and $\mathcal{B}$, as well as the interacting system $\mathcal{AB}$ first in the basis functions of $a\in\mathcal{A}$, $b\in\mathcal{B}$ and $a\cup b$ for $\mathcal{AB}$, respectively. From here we have [\emph{c.f.} \cref{eq:formation_energy}]:
\begin{equation}
    \Delta E(\mathcal{AB})=E^{a\cup b}(\mathcal{AB})-\left(E^{a}(\mathcal{A})+E^{b}(\mathcal{B}) \right).
\end{equation}
However, it can be seen from this expression that, when computing $E^{a\cup b}(\mathcal{AB})$, the computed energy of fragment $\mathcal{A}$ ($\mathcal{B}$) will be `contaminated' by the basis functions $b$ ($a$) from its interacting counterpart, since the size of the basis functions $a\cup b$ is larger than both $a$ and $b$, which is not the case when computing the terms $E^{a}(\mathcal{A})$ or $E^{b}(\mathcal{B})$. This generally results in the overbinding of the non-covalent complexes, and can be corrected by applying the counterpoise correction scheme.

\begin{wrapfigure}[30]{r}{0.35\textwidth}
\includegraphics[width=0.35\textwidth]{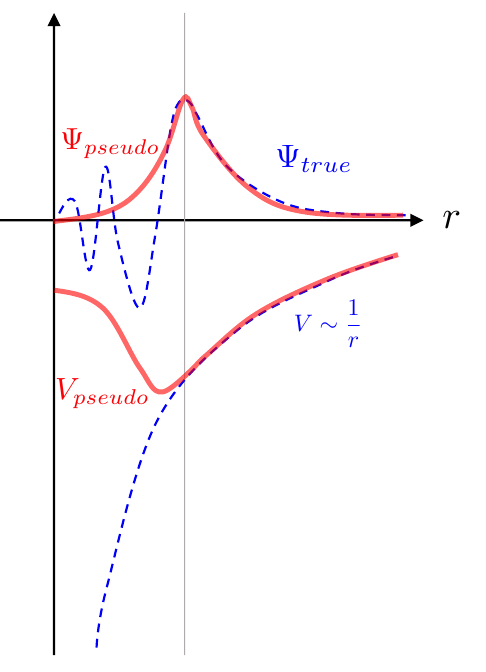}
\caption{Illustration of the construction of pseudopotential (red), which replaces the diverging Coulomb potential (blue) below the cut-off radius (grey solid line) with one  has finite value when $r\to 0$. This results in a pseudowavefunction that is more smooth and less oscillatory within the core region of an atom.}
   \label{fig:pseudopotential}
\end{wrapfigure}

For solid state calculations, on the other hand, one typically applies the planewave basis function in respect of the periodic boundary conditions. Based on the Bloch theory,  a suitable form of the planewave basis function can be written as 
\begin{equation}
    \label{eq:bloch_wave}
    \Psi_{n,\bm{k}}(\bm{r})=e^{i\bm{k}\cdot\bm{r}} u_{n,\bm{k}}(\bm{r})=\sum_{\bm{G}}c_{n,\bm{k}}(\bm{G}) e^{i(\bm{k+G})\cdot\bm{r}},
\end{equation}
in which $u_{n,\bm{k}}(\bm{r})$ is the wavefunction that reflects the local crystal potential and $\bm{G}$ is the reciprocal lattice vector. Similar to the situation in molecular calculations where the accuracy increases with the increasing sizes of basis set, in periodic calculations, the number of reciprocal lattice points included in $\{\bm{G}\}$ will dictate the accuracy of the simulation, which can be selected based on the cutoff of the electronic kinetic energy as $\frac{\hbar^2}{2m}|\bm{k}+\bm{G}|^2\leq E_{\mathrm{cutoff}}^{\mathrm{kin}}$. Unlike localised basis functions, because the planewave basis functions is fully delocalised in space, and the kinetic energy cutoff is also independent from the atomic positions, it does not suffer from BSSE for noncovalently bound systems, such as molecular crystals.


Nevertheless, direct application of \cref{eq:bloch_wave} in implementing periodic DFT is not numerically efficient. This is because, close to the nuclei, the core electronic wavefunction typically exhibits wild oscillations, which require superpositioning many planewaves to correctly model this behaviour because a large $\bm{G}$ is needed to model core electrons with high $E^{\mathrm{kin}}$.  This is not a problem when using the localised basis functions, in which both the core and valence electrons can be modelled with basis functions of the same form. In order to balance both the accuracy of all-electron calculations and the efficacy of planewave basis functions for periodic systems, the projector augmented wave (PAW)\cite{blochl1994projector} was proposed and has been a successful scheme that is adopted, for example, in the widely used planewave DFT code VASP (Vienna \emph{ab initio} Simulation Package).\cite{kresse1996efficient} In its essence, the PAW is constructed as follows
\begin{equation}
    \label{eq:PAW}
    \left|\Psi\right\rangle=\left(|\bar{\Psi}\rangle - \sum_i |\tilde{\phi}_i\rangle \langle\tilde{p}_i|\bar{\Psi}\rangle\right) + \sum_i |\phi_i\rangle \langle\tilde{p}_i|\bar{\Psi}\rangle,
\end{equation}
in which $|\bar{\Psi}\rangle$ represents the pseudo wavefunction that is expanded in the planewave basis, $|\phi_i\rangle$ and $|\tilde{\phi}\rangle$ are the pseudo wavefunction for valence and core electrons in a localised basis, respectively. Finally $\langle\tilde{p}_i|$ is a projector function. With this, the bracketed term in \cref{eq:PAW} removes the part in the planewave that is suppose to model the core electronic behaviour, that is subsequently replaced by [remaining term in \cref{eq:PAW}]  an oscillating core-like wavefunction that is projected onto the planewave basis.

\paragraph{\textbf{Pseudopotentials}}  To further overcome the problem of having to use a large number of $\bm{G}$ vectors in the reciprocal space for expanding the core electron wavefunctions with the planewave basis functions, we utilise the fact that, in comparison with the spatially delocalised valence electrons, the core electrons play a minimal role in determining the chemical bondings and electronic properties of materials. This means that in practical DFT calculations, one only needs to model the behaviours of core electrons approximately. This leads to the concept and developments of pseudopotentials that are often applied in conjunctions with the planewave basis functions in DFT calculations for periodic solids.

\cref{fig:pseudopotential} provides an illustration on the core idea behind the atomic pseudopotential $V_{\mathrm{pseudo}}(\bm{r})$, which is a smoothly varying potential, such that below the core/valence cut-off distance $r_c$, $V_{\mathrm{pseudo}}(\bm{r})$ varies slowly to a finite value as $\|r\|\to0$, instead of diverging to $-\infty$ as for the case of a true Coulomb electrostatic potential generated from the positively charged nuclei. On the other hand, when $\bm{r}>\bm{r}_c$, $V_{\mathrm{pseudo}}(\bm{r})$ resembles the original form of the Coulomb potential for a particular atom. Consequently, the pseudowavefunction $\Psi_{\mathrm{pseudo}}(\bm{r})$ that is solved from the Schr\"{o}dinger equation with $V_{\mathrm{pseudo}}(\bm{r})$ should become a slowly varying one when $\bm{r}<\bm{r}_c$, while resembling the correct tailing wavefunction in the valence region, as if the original Coulomb potential was used.  Here, we will briefly review the two major categories of the constructions for the pseudopotentials that are widely used in the community.

The \emph{norm-conserving pseudopotentials}\cite{hamann1979norm,vanderbilt1985optimally,hamann2013optimized} are constructed by directly applying the concept discussed above through imposing the constraint such that the norm of the pseudowavefunction $\Psi_{\mathrm{pseudo}}(\bm{r})$ within the cut-off distance should equal to that of an all-electron wavefunction $\phi_{\mathrm{all-electron}}(\bm{r})$,  \emph{i.e.}
\begin{equation}\label{eq:norm_conserve_constraint}
    \int_0^{\bm{r}_c}\left\|\Psi_{\mathrm{pseudo}}(\bm{r})\right\|^2 \mbox{d}\bm{r} = \int_0^{\bm{r}_c}\left\|\phi_{\mathrm{all-electron}}(\bm{r})\right\|^2 \mbox{d}\bm{r}.
\end{equation}
Additionally, $\Psi_{\mathrm{pseudo}}(\bm{r})$ needs to be a continuous and differentiable function at $\bm{r}_c$. Typically the all electron wavefunction for each orbital (based on the quantum number $\ell$) is first solved for a given atom with, for example, localised basis function, which allows one to fit a corresponding $\Psi_{\mathrm{pseudo}}^{\ell}(\bm{r})$ under the norm-conservation constraint. To obtain the final pseudopotential, we note that $\Psi_{\mathrm{pseudo}}^{\ell}(\bm{r})$ should satisfy the following Schr\"{o}dinger equation:
\begin{equation*}
    \left[-\frac{1}{2}\nabla^2_{\bm{r}} +\frac{\ell(\ell+1)}{2r^2}+V_{\mathrm{pseudo}}^\ell(\bm{r})\right]\Psi^{\ell}(\bm{r})=\varepsilon_\ell  \Psi^{\ell}(\bm{r}).
\end{equation*}
By harnessing the linearity of the Schr\"{o}dinger equation, we can compute the corresponding pseudopotential from $\Psi^{\ell}(r)$ and $\varepsilon_\ell$ directly as:
\begin{equation*}
    \label{eq:norm_conserved_PP}
    V_{pseudo}^\ell (\bm{r}) = \varepsilon_\ell+\frac{1}{2\Psi^\ell (\bm{r})}\nabla^2_{\bm{r}} \Psi^{\ell}(\bm{r})-\frac{\ell(\ell+1)}{2r^2}.
\end{equation*}
While this is conceptually simple, the pseudowavefunction $\Psi_{\mathrm{pseudo}}(\bm{r})$ that is produced under the norm-conserving constraint can still retain significant oscillatory behaviours, such that the problem of large $\bm{G}$ vector needed in constructing the  planewave basis function cannot be fully circumvented, particularly for transition metals with very localised valence states.

The \emph{ultrasoft pseudopotentials}\cite{laasonen1991implementation,kresse1999ultrasoft} were introduced to further overcome the problem in the norm-conserving pseudopotentials by lowering the energy cutoff to enhance the computational efficiencies. It achieves this by  relaxing the norm-conserving condition [\cref{eq:norm_conserve_constraint}] to achieve a smoother $\Psi_{\mathrm{pseudo}}(\bm{r})$ through allowing less charge density to be retained within the core region whilst preserving the total electron density of a given atom, which means now the constraint [\cref{eq:norm_conserve_constraint}] should be rewritten as:
\begin{equation}
    \label{eq:ultrasoft_PP_charge_constraint}
    \Delta \rho=\int_0^\infty \left(\Psi^* (\bm{r})\Psi(\bm{r})-\phi^*(\bm{r})\phi(\bm{r})\right)\mbox{d}\bm{r}\to 0.
\end{equation}
Because the pseudowavefunction $\Psi_{\mathrm{pseudo}}(\bm{r})$ is very smooth, it generally cannot preserve the total charge of an atom. The solution proposed for this problem is to reconstruct the total charge density by rewritting the total atomic charge density as:
\begin{equation*}
    \rho(\bm{r})=\sum_{i}f_i \|\phi_i (\bm{r})\|^2+\rho^{\mathrm{aug}}(\bm{r}),
\end{equation*}
in which $f_i$ is the atomic orbital occupation number. The augmented electron density $\rho^{\mathrm{aug}}(\bm{r})=\sum_{ij}q_{ij}Q_{ij}(\bm{r})$ effectively projects the differences in the total charge densities calculated from localised and planewave basis functions, \emph{i.e.} $Q_{ij}(\bm{r})=\phi_{i}^{*}(\bm{r})\phi_{j}(\bm{r})-\Psi_{i}^{*}(\bm{r})\Psi_{j}(\bm{r})$, onto the localised basis function with $q_{ij}=\sum_n f_n \langle \phi_n|\beta_i\rangle\langle\beta_j |\phi_n\rangle$, with $|\beta_i\rangle$ being a projector operator that depends on $\phi_i$. Essentially $\rho^{\mathrm{aug}}(\bm{r})$ measures how much charge density is missing when a particular atomic orbital is relaxed from $\phi$ to its corresponding $\Psi$. Unlike the norm-conserving pseudopotentials, the formulation for the ultrasoft pseudopotential does not give a potential directly, but rather the ingredients to reconstruct the Kohn-Sham equations to solve for the corresponding Kohn-Sham orbitals:
\begin{equation*}
    \left[-\frac{1}{2}\nabla^2+\hat{V}_{local}+\hat{V}_{NL}\right]\psi_n^{\mathrm{KS}}=\varepsilon_n  \hat{S} \psi_n^{\mathrm{KS}}.
\end{equation*}
Here, the local potential $\hat{V}_{local}$ contains the Hartree, exchange-correlation and ionic contributions that can be obtained from $\rho(\bm{r})$ by solving the Poisson equation. The non-local potential is given by the projector $\hat{V}_{NL}=\sum_{ij}D_{ij}|\beta_i\rangle\langle\beta_j|$, in which $D_{ij}$ is the non-local potential coefficient obtained by matching the eigenvalues of the pseudo atomic wavefunction to its true wavefunction. Finally, $\hat{S}=1+\sum_{ij}q_{ij}|\beta_i\rangle\langle\beta_j|$ is the orbital overlap operator. 

The development and wide adaption of pseudopotentials have undoubtedly become an inseparable factor for the success of DFT in material chemistry and physics. However, as in any approximations proposed to simplify the practical implementation of the theory, it also comes with its own limitations. Firstly,  by approximating the core electrons in pseudopotentials prevents accurate modelling of physical effects in which core electrons can also play a critical role, such as spin-orbit coupling,\cite{dolg2012relativistic} high-pressure physics,\cite{oganov2003all} and core-level spectroscopies. Secondly, because the targets for fitting the pseudopotentials are the  orbitals of individual isolated atoms, it cannot reflects the change in the atomic potential due to the neighbouring bonding environments, which could reduce their transferabilities. Thirdly, as one can see in the formulation of the ultrasoft pseudopotential, its practical implementation is directly tightened up with the corresponding implementation of the Kohn-Sham equation, which can further limit its  transferability across different DFT codes.   
 
\subsubsection{Performance Benchmarking for Density Functional Approximations}

The idea of using a single-electron density to approximate the intricate many-body interactions in multielectronic systems continues to be an attractive pursuit to develop better and computationally cheap functionals, in order to facilitate our understanding on the fundamental properties of materials thus driving the discovery of novel materials that do not exist in the current material space. It is therefore not surprising that a `zoo'\cite{goerigk2019trip} of more than 200 density functionals\cite{mardirossian2017thirty} have been developed to date, making it a profound challenge for DFT practitioners to choose the most appropriate functional to be applied for a specific modelling problem at hand, if not simply resolving to the most popular one(s).\cite{kruse2012standard}

In this regard, benchmarking the performance of DFAs in reproducing the reference material property data, derived either experimentally or from high-level theoretical calculations (such as coupled-cluster theory\cite{bartlett2007coupled}) becomes a critical component in the research community of DFT. There are a few key rationales behind the efforts in carefully collating and assembling comprehensive benchmark sets: (1) Each DFA was designed and developed under some specific assumptions and approximations to the multielectronic behaviours in real materials, a quantitative measure is needed in order for us to assess how good these approximations are. (2) It is no surprise that electrons behave differently in different material systems, such as metals and insulators. Therefore, it is important to calibrate DFAs that are designed to satisfy different `boundary conditions' on their capabilities in correctly predicting the behaviours of different materials.  On this regard, whilst the construction of a universal DFA may be a desirable pursuit, there is no compelling reason to believe such a functional, if it exists, would be easy to implement or numerically efficient.\cite{DFT_exchange} (3) By understanding the limitations in different functionals, for example, GGA being able to produce reliable structures for solids but failed for estimating the band gap energies [\cref{fig:DFT_bandgap} (c,d)], it can guide the future designs and choices of DFAs for future improvements in electronic structure theory and better results in materials modelling. This is particularly useful when we can quantify the uncertainties in reproducing the benchmark values with a specific DFA. Finally, (4) it helps establishing practical standards for the computational chemistry and materials science community, which encourages data sharing, as well as increases the trust and transferability of computational modelling as an essential tool for scientific discoveries.\cite{DFT_exchange}

\begin{figure}[tb]
    \centering
    \includegraphics[width=\textwidth]{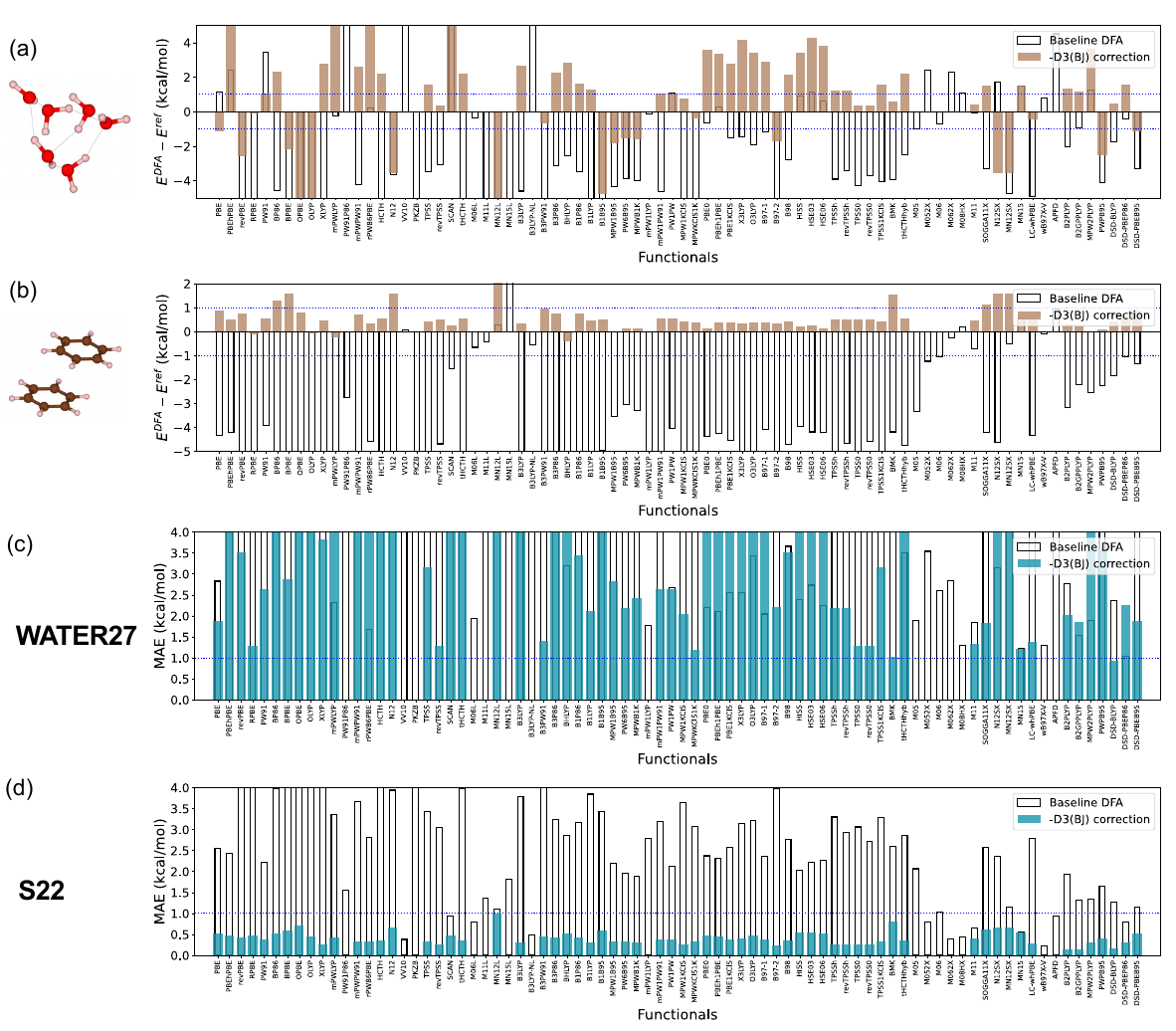}
    \caption{Benchmarking the performances of DFAs for determining the energetic stabilities  of model chemical systems for the purposes of assessing their applicability. (a) Binding energy of a small hydrogen bonded (\ce{H2O})$_6$ cluster as in the WATER27 (binding energies for small neutral, protonated and deprotonated water clusters) benchmark set,\cite{bryantsev2009evaluation} where the reference value $E^{ref}$ was reported in \citet{manna2017conventional} at the MP2/CBS+$\Delta$CCSD(T) level of theory. (b) Binding energy of a dispersion force bounded benzene dimer from the S22 (noncovalent interaction energies for 22 dimers of small molecules) benchmark set,\cite{jurevcka2006benchmark} with the reference value determined at the CCSD(T)/CBS level that is subsequently corrected with $\delta_{\mbox{\scriptsize{MP2}}}^{\mbox{\scriptsize{CCSD(T)}}}$ values.\cite{marshall2011basis} In (a-b), the performances of DFAs are measured based on the unsigned difference between the energies determined by a given DFA and the reference value, namely $\Delta E=E^{DFA}-E^{ref}$, which provides an indication on whether a specific DFA tends to under- or over-bind the molecular complex. The mean absolute errors (MAE) for DFAs in determining the binding energies of molecular complexes, which are determined as $\frac{1}{N}\sum_{i=1}^{N}|\Delta E_i|$ across all $N$ molecules in the WATER27 (c) and S22 (d) sets for each DFA examined. In all figures, the performance of the baseline DFAs (empty bars), and those that are dispersion-interaction-corrected (filled bars, if available) by Grimme's D3(BJ) scheme.\cite{grimme2010consistent,grimme2011effect} The blue dashed lines designate the position where $|\Delta E|$ or MAE is  1 kcal/mol, the limit within with the DFAs are considered as being able to predict the molecular binding energies with the so-called `chemical accuracy'. All data presented here are extracted from the new GMTKN55 quantum chemistry benchmark database.\cite{goerigk2017look}}
    \label{fig:DFA_benchmarks}
\end{figure}

Here, we shall provide   high-level overviews on the key features in some of the well-known benchmark sets in the community. 

In the chemistry community, the GMTKN55\cite{goerigk2017look} (\textbf{G}eneral \textbf{M}ain group \textbf{T}hermochemistry \textbf{K}inetic \textbf{N}on-covalent interactions) database, which is an extended version to its popular predecessor GMTKN30,\cite{goerigk2011efficient} represents one of the most comprehensive benchmark for main group chemistries. It tested the performance of 217 dispersion-corrected and -uncorrected DFAs in reproducing 1505 reference values of relative energies that were primarily determined from high-level wavefunction theory calculations. It is the \emph{relative energy}  because every system that has been benchmarked can be casted into the form of reactants$\to$products, with $\Delta E=E_{total}(\mbox{products})-E_{total}(\mbox{reactants})$. 

In a high-level overview, the GMTKN database is organised in such a way that the performances of individual DFA, measured in terms of the deviation of DFA-determined relative energies to the corresponding reference values, can be assessed on the levels of (1) A specific chemical system. For instance, \cref{fig:DFA_benchmarks}(a) and (b) show the relative unsigned errors in calculating the interaction energies of a small hydrogen-bonded water cluster and a $\pi$-stacked benzene dimers. It  can be seen that for purely dispersion-bounded benzene dimers, the inclusion of -D3(BJ) corrections consistently improves the results in determining the interaction energy, but less so for water clusters that are strongly bounded by dipole-dipole interactions. (2) A set of chemically similar systems, such as the WATER27 and S22 sets shown in \cref{fig:DFA_benchmarks}(c) and (d) whereby the molecular systems within a given set share some commonalities in the electronic natures that determine their interaction strengths, and thus the benchmark data reflects the capabilities of different functional designs in capturing those critical electronic behaviours in the systems. (3) Overall performances of DFAs (particularly within given rung of the Jacob's ladder) across the whole benchmark, providing an evidence-based and data-driven recommendation for the best DFAs across each rung of the Jacob's ladder, which, can often be used as a starting point for new research project. For example, in the context of high-throughput screening where high computational speed is required, GGA functional revPBE-D3\cite{zhang1998comment} would be the best choice, however, double-hybrid functionals are critical for accurate thermochemistry and non-covalent interactions. 

Of course, the use of double-hybrid functionals may not be always computationally feasible. In this case, a careful sift through the benchmark for alternatives lower-rung DFAs would be needed, which can be a costly exercise.  In this regard, \citet{gould2018diet} has put forward an interesting genetic-algorithm approach to construct a `diet-GMTKN55' benchmark consisting of only 100 representative chemical systems that are capable of closely reproducing error metrics from the full set. 

The capability for us to theoretically collate accurate reference data for molecular systems with wavefunction method can be largely attributed to the efficiency of localised basis functions in quantum chemistry calculations. This is in stark contrast to periodic solids, for which the use of planewave basis sets are the practical norm, making it become very expensive for post-Hartree-Fock correlation calculations. On the other hands, systems such as Mott insulators\cite{imada1998metal} and transition metal oxides\cite{tokura2003correlated} that exhibit strong electron correlation effects would require more specialised theoretical treatments\cite{georges2004strongly,imada2010electronic} that are beyond the scope of the present review. In this regard, most benchmark studies for solids would focus on comparing the DFT-derived properties with known experimental results [\cref{fig:DFT_bandgap}]. Here, we shall briefly outline a few exemplary studies/datasets that have been widely referred to in the past decade.

In solid-state DFT calculations, the choice of pseudopotential and the corresponding cutoff radius for each element plays a crucial role in balancing the computational accuracy and speed. For this, \citet{prandini2018precision} have compiled a benchmark of standard solid-state pseudopotential (SSSP). Through careful testing of the convergence behaviours in phonon frequencies, band structure, cohesive energy and pressure calculations, for each element in the periodic table, the authors have provided the optimum choice of pesudopotential type and the corresponding cut-off radius based on two criteria: SSSP-efficiency tailored for high-throughput materials screenings and SSSP-accuracy for high-accurate (\emph{e.g.} phonon) calculations. Different research groups often select different DFT codes as their default choice, in this regard, \citet{lejaeghere2016reproducibility} had compared the reproducibility of DFT calculations from 15 different solid-state DFT codes in 2016. Even though there is a lack of reference DFT implementation, the study has already found that, the continuous improvement in the software developments, the results obtained from mainstream DFT codes are essentially identical, only varying subtly due to differences in the numerical implementation. 

Molecular crystals represent a unique set of material systems which, on one hand, their long-ranged electronic structures need to be accounted for with periodic DFT, on the other hand, the atom-atom pairwise non-covalent interactions that plays an important role in holding the molecules together in their crystalline form, rely heavily on our knowledge in the development of dispersion-corrected DFT from quantum chemistry. With this, the X23 benchmark for molecular crystals\cite{otero2012benchmark,reilly2013understanding,dolgonos2019revised} has tabulated reference lattice constants and sublimation enthalpies for 23 crystals of small molecules, which are  corrected for thermal vibration and expansion effects from the corresponding experimental values. Benchmarking studies showed that dispersion-correction in DFT, which takes into account the many-body effect, tends to perform best in reproducing the lattice energies. It has also played a critical role in assessing the accuracy of atom-atom forcefields for fast, high-throughput organic crystal structure predictions,\cite{nyman2016accurate}  and validates the application of dispersion-corrected DFT as a useful post-prediction method for correcting the forcefield-derived lattice energies to provide better energetic ranking of the polymorph stabilities from structure predictions, which are reflected in the last two blind tests.\cite{reilly2016report,Hunnisett:aw5094}

Finally, in the past decade, we have seen the establishment and continuous evolution of large-scale high-throughput computational material databases, as represented by the Materials Project Database,\cite{jain_apl_mater} AFLOW Database,\cite{CURTAROLO2012218} Open Quantum Materials Database (OQMD),\cite{kirklin2015open} and 2D Materials Database.\cite{haastrup2018,gjerding2021recent} These databases, being assembled for $\sim$150 K up to $\sim 2$ M inorganic materials, contain results from DFT calculations with the underlying method carefully tested to ensure the results are sufficiently accurate and consistent throughout the database, have provided the community a set of benchmarks for standardising and sanity checking future DFT calculations, which has already significantly enhance our capabilities to cross-compare results from different literatures and discover new phenomena in functional materials based on existing knowledges.
\label{sec:DFT_benchmark}

\subsection{The GW Approximation}\label{sect:GW_approximation}
\subsubsection{Motivations}\label{sect:motivation_GW}

\begin{figure}[t]
    \includegraphics[width=1\linewidth]{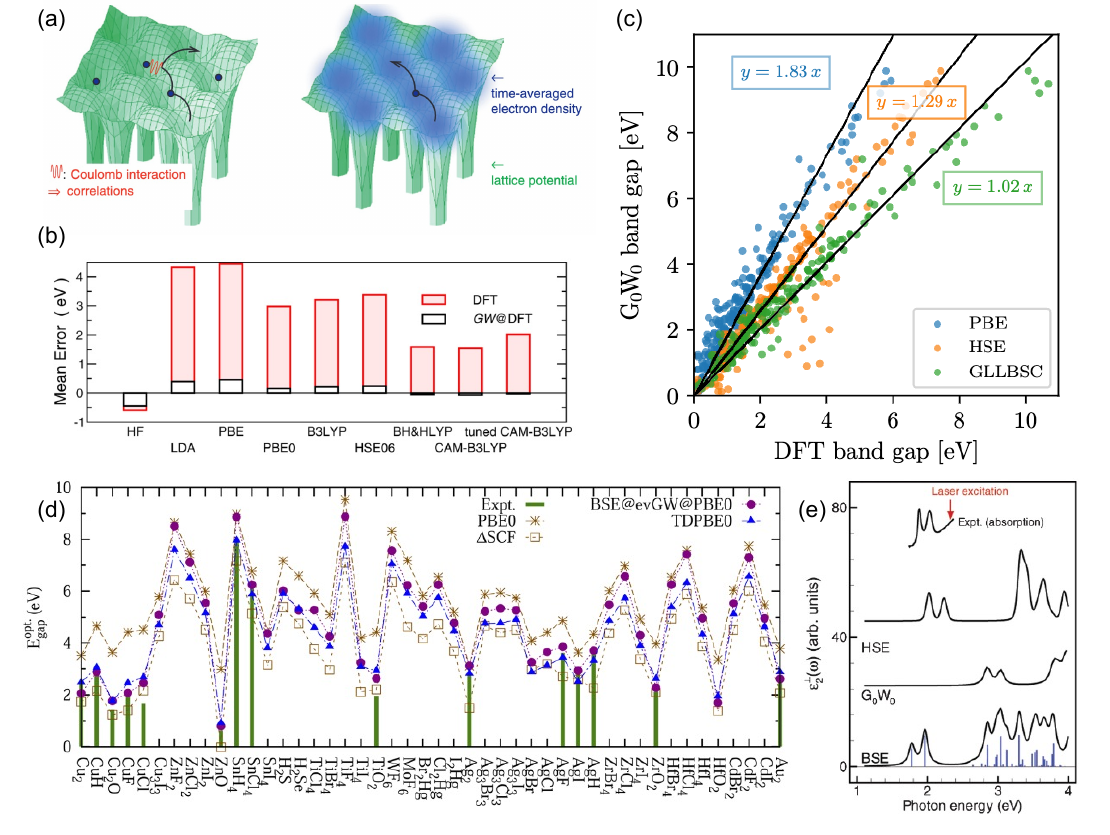}
    \caption{(a) In the true many-body system, the electronic correlation is both position- and time-dependent (left), in which the movement of electrons (black curly arrows) are affected by the Coulomb repulsions (red curly line) induced by other neighbouring electrons in the system. On the other hand, mean-field electronic structure theory (such as DFT) simplifies the many-body problem by describing the behaviour of a single electron as moving in a time-averaged potential generated from the other electrons (right) [Adopted from \citet{held2007electronic}]. (b) Deviation of the HOMO energies from the reference CCSD(T) values for a benchmark set of 34 small molecules,\cite{bruneval2013benchmarking} showing the results from different generalised Kohn-Sham (GKS) calculations, that are compared to GW calculations starting from the corresponding GKS orbitals. (c) Band gaps obtained with different DFT functionals plotted against the $G_{0}W_{0}$ results for a database of monolayer 2D materials. The band gaps from PBE (blue), HSE06 (orange), and GLLBSC\cite{kuisma2010kohn} (green) are fitted linearly to the $G_0 W_0$ results. [Adopted from \citet{haastrup2018}]  (d) Optical gaps from PBE0, $\Delta$SCF, the BSE, and TDPBE0, compared to experimental data. [Adopted from \citet{korbel2014benchmark}] (e) Imaginary part of transverse dielectric constant for monolayer \ce{MoS2}, $\varepsilon_{2}^{\perp}(\omega)$, as a function of photon energy $(\hbar\omega)$. Computed spectra are presented for three levels of theory: HSE functional, $G_{0}W_{0}$, and BSE. Vertical (blue) bars represent the relative oscillator strengths for the optical transitions. The two lowest-energy peaks in the spectrum (first two bars) correspond to the A and B excitons. As seen, the closest agreement with experiments in locating the excitionic peaks is obtained at the level of BSE calculations. [Adopted from \citet{ramasubramaniam2012large}]} 
    \label{fig:GW_approximation_diagram}
\end{figure}

Electron correlation arises as the result of the electron-electron Coulomb repulsive potential $\hat{\mathcal{V}}_{ee}$ that enters the many-body Schr\"{o}dinger equation \cref{eq:full_Sch_eqn}. In principle, such effect must be described by the many-body electron wavefunction that varies both in positions and time, \cref{fig:GW_approximation_diagram}(a, left), in which an electron travelling in a material is constantly experiencing and repelled from its original path by the Coulomb repulsions from other surrounding electrons, which are also in constant motions while responding to the Coulomb interactions from each other. This seemingly intractable collective electronic motion is greatly simplified by DFT, which is often referred to as a \textit{mean-field electronic structure theory} because the electron density, the basic variable in DFT, reflects the time-averaged distributions of electrons in the system [\cref{fig:GW_approximation_diagram}(a, right)]. This means that the Kohn-Sham equation effectively describes the behaviour of a single electron in the \textit{mean field} generated by the other electrons in the material. One could imagine this provides a rather crude description on the responses of other electrons.

This problem becomes particularly `magnified' when one considers interpreting the electron spectroscopies.\cite{golze2019gw} In the case of an electron being added to (removed from) a many-body system, the extra electron (hole) will interact with its surrounding electrons, which effectively form a polarisation cloud. This mutual interaction is happening in both ways: the movement of the  extra electron (hole)  is affected by the polarisation cloud, whereas the `shape' of this polarisation cloud will change in response to the Coulomb interaction with the  extra electron (hole)  that is added to the system. Effectively, the true behaviour of this many-body system should be described in terms of \textit{quasiparticle}, which is the extra electron (hole) that is \emph{screened} by its surrounding polarisation cloud. In the many-body theory, the modification to the single-particle energy due to its interaction with the surrounding electrons is given by the \textit{self-energy} $\Sigma$, and the quasiparticle energy is the energy required to remove (add) an electron from HOMO (to LUMO) of the  system, taking into account the electron correlation effect, which corresponds to the ionisation potential (electron affinity) that is measured by the (inverse) photoemission spectroscopy, \emph{i.e}
\begin{equation}
    \label{eq:ionisation_potential}
    \varepsilon_{\mbox{\scriptsize{HOMO}}}=-IP=-[E(N-1)-E(N)],
\end{equation}
and
\begin{equation}
    \label{eq:electron_affinity}
    \varepsilon_{\mbox{\scriptsize{LUMO}}}=-EA=-[E(N)-E(N+1)].
\end{equation}
Here, $E(N)$ represents the (ground-state) energy of a system with $N$ electrons, whereas $E(N+1)$ and $E(N-1)$ represent the (excited-state) energies of the same system with an extra electron added or removed, respectively. However, for DFT that accounts for the electronic correlation effects in a mean-field approach, only \cref{eq:ionisation_potential} holds if the exact exchange-correlation functional is used,\cite{kleinman1997significance} and $\varepsilon_{\mbox{\scriptsize{LUMO}}}\neq-EA$ due to the lack of treatment of the screening effects. This results in the problem such as the underestimation of fundamental gaps in solids and molecules.
\subsubsection{A Quick Primer on Many-Body Green's Function}


The many-body perturbation theory, which utilises the mathematical tool of Green's function\cite{economou2006green} provides the systematic framework to treat the electron-correlation effects in complex material systems. Here, we will provide a gentle introduction on many-body Green's function.

\begin{wrapfigure}[27]{r}{0.3\textwidth}
  \begin{center}
    \includegraphics[width=0.3\textwidth]{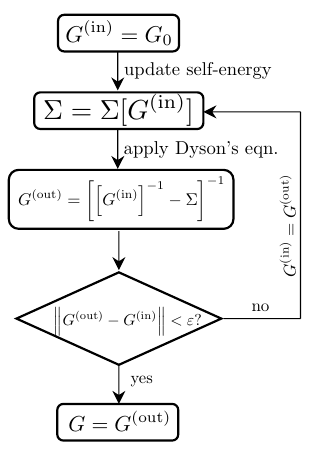}
    \caption{Flowchart illustrating the self-consistent procedure for solving the Dyson's equation in many-body physics.}
    \label{fig:dyson_solution}
  \end{center}
\end{wrapfigure}

In the time-dependent formalism, the Green's function $G(\bm{r},\bm{r}^\prime, t,t^\prime)$ is a function that describes the evolution of the amplitude of electron wavefunction from $(\bm{r}^\prime,t^\prime)$ to $(\bm{r},t)$ in the time-position space under the influence of the Coulomb interactions from all the other electrons. Here, we are interested in its Fourier transformation into the energy domain $G(\bm{r},\bm{r}^\prime,\omega)$, which is usually defined in the complex plane of energy. This can be seen from the Lehman representation of the Green's function,\cite{economou2006green}  for example, giving the quantum states of the system described by a set of basis functions $\{\phi_i(\bm{r})\}$ with $\{\varepsilon_i\}$ being the corresponding eigenenergies, we have,
\begin{equation}
    \label{eq:Lehman}
    G(\bm{r},\bm{r}^\prime,\omega)=\sum_{i}\langle\phi_i|\phi_i\rangle \left[\frac{\Theta(\varepsilon_i-E_F )}{\hbar\omega-(\varepsilon_i -i\eta)}+\frac{\Theta(E_F-\varepsilon_i )}{\hbar\omega-(\varepsilon_i +i\eta)}\right],
\end{equation}
in which $E_F$ is the Fermi energy,  $\Theta$ is the Heaviside step function and $\eta\to0^{+}$. The poles of the Green's function, which occur at $\{\varepsilon_i\}$, correspond to the quasiparticle energies, can be directly matched with the  experimentally measured EA and IP.

Apparently, the quasiparticle energies $\{\varepsilon_i\}$ are supposed to be obtained from solving the full many-body Schr\"{o}dinger equation, from which the corresponding Green's function can be defined according to \cref{eq:Lehman}, but we know this is generally not possible to achieve. This problem is solved by applying the \emph{Dyson's equation}.\cite{dyson1949} Starting from the non-interacting single-particle Green's function $G_0(\omega)$, which can be constructed using the Lehman's representation [\cref{eq:Lehman}] with, for example, the Kohn-Sham orbitals and energies, the corresponding many-body Green's function $G(\omega)$ can be determined with the knowledge of the self-energy $\Sigma(\omega)$ as:
\begin{equation}
    \label{eq:Dyson_equation}
    G(\omega)=G_0(\omega)+G_0(\omega)\Sigma(\omega)G(\omega).
\end{equation}

The Dyson's equation is usually solved iteratively until self-consistency is reached, for which \cref{fig:dyson_solution} illustrates the general procedure behind. It can be seen that the self-energy $\Sigma(\omega)$ itself is also a function of the Green's function. More specifically, as discussed later,  the self-energy $\Sigma(\omega)$ is the product of the Green's function $G(\omega)$ and the screened Coulomb interaction $W(\omega)$, thus taking into account the mutual interaction between an electron and its surrounding polarisation cloud, leading to the name of $GW$ approach.

\subsubsection{The $G_{0} W_{0}$ Approximation}

Practically, the computational complexity of the $GW$ approximation depends critically upon the number of iterations needed to solve  Dyson's equation [\cref{fig:dyson_solution}] in order to obtain the full interacting Green's function. Depending on the implementation, the computational costs associated with $GW$ calculations are typically in the order from $\mathcal{O}(N^4)$ to $\mathcal{O}(N^6)$.\cite{golze2019gw} Nevertheless,  a one-shot $GW$, or $G_0 W_0$ approximation, in which only one iteration of the Dyson's equation is performed, is most widely used for computational efficiencies, and is often sufficient for achieving accurate results compared with the experimental data.\cite{yan2010band,liu2011he,li2012impact} As such, we will focus the subsequent discussions on the $G_0 W_0$ approximation, by following the procedures on how it is computationally achieved to allow practitioners to develop an intuitive understanding of the physical meanings behind. For a more complete overview on $GW$ method beyond the $G_0 W_0$ approximation, we refer the readers to the recent review by \citet{golze2019gw}.

Typically, the converged Kohn-Sham orbitals and energies are often used as the starting point to construct $G_0$ in the $GW$ calculations. Even though the Kohn-Sham anzatz has already included some effects of electron correlations through the mean-field exchange-correlation potential, from the view of the perturbation theory, the Kohn-Sham energies, particularly those obtained from hybrid DFT calculations, can be considered as a good baseline to the quasiparticle energies, making them a reasonable choice as the starting point for the $GW$ calculations. 

For instances, \citet{bruneval2013benchmarking} tested the starting point dependence for the $GW$ calculations for a benchmark set containing 34 small molecules [\cref{fig:GW_approximation_diagram}(b)], which shows that the HOMO energies obtained from the tuned CAM-B3LYP orbitals with the highest percentage of exact exchange give the lowest mean absolute error of 0.1 eV.  In the 2D material database established by \citet{haastrup2018}, the authors had examined the correlations between the band gap energies from $G_0 W_0$ approximation and those from generalised Kohn-Sham calculations [\cref{fig:GW_approximation_diagram}(c)], which also showed good agreement between the $G_0 W_0$ band gaps and GLLBSC functionals with exact exchanges tailored for solid-state band gap calculations, rendering the later being a cheaper alternative to the $G_0 W_0$ method for accurate band gap predictions.  Similar results were found by \citet{korbel2014benchmark} when benchmarking the performance of $GW$ approximations on small binary transition metal molecules, in which the IPs and EAs calculated with $G_0 W_0$@PBE0 level of theory give MAEs of 0.2 and 0.3 eV, respectively. In terms of the quasiparticle gaps, the $G_0 W_0$@PBE0 yields an MAE of 0.1 eV. Subsequent calculation of the optical gap with the Bethe-Salpeter equation (BSE, see below) also leads to a reasonable agreement with the experimental data with MAE $\sim 0.3$ eV, for which time-dependent DFT (TDDFT) was found to give similar level of performance  [\cref{fig:GW_approximation_diagram}(d)].

The next key ingredient in solving the $GW$ equations is the self-energy term $\Sigma(\omega)$, which contributes to the change in the free-electron Green's function due to the screened Coulomb interaction with other electrons. Effectively, the electron is interacting with a polarisation cloud that is induced by all other moving electrons in the system, from which one can appreciate that calculating $\Sigma(\omega)$ will involve the computation of the electronic contribution to the frequency-dependent polarisability $\chi_{0}(\bm{r},\bm{r}^\prime,\omega)$, which modifies the bare electrostatic Coulomb potential $v(\bm{r},\bm{r}^\prime)$ to its screened counterpart $W_{0}(\bm{r},\bm{r}^\prime,\omega)$. With this in mind, given that the self-energy is given by the following convolution:
\begin{equation}
    \label{eq:GW_self_energy}
    \Sigma(\bm{r},\bm{r}^\prime,\omega)=\frac{i}{2\pi}\int G_{0}(\bm{r},\bm{r}^\prime,\omega+\omega^\prime) W_0 (\bm{r},\bm{r}^\prime,\omega^\prime)\mbox{d}\omega^\prime.
\end{equation}
To compute the self-energy, we have, first of all, the free-particle Green's function $G_{0}(\bm{r},\bm{r}^\prime,\omega)$ from the Kohn-Sham orbitals and energies as given in \cref{eq:Lehman}. Next, the independent particle polarisability is calculated under the random phase approximation (RPA)\cite{ren2012random} as
\begin{equation}
    \label{eq:rpa_pol}
    \chi_0 (\bm{r},\bm{r}^\prime,\omega)=-\frac{i}{2\pi}\int G_{0}(\bm{r},\bm{r}^\prime,\omega+\omega^\prime) G_0 (\bm{r},\bm{r}^\prime,\omega^\prime)\mbox{d}\omega^\prime.
\end{equation}
In brief, RPA considers the electronic polarisations coming predominantly from the Coulombic interactions in an excited electron-hole pair while ignoring the contributions from the exchange-correlation interactions.  It assumes that the phases of the wavefunctions of the excited electrons and holes are random enough that over time, they average out. Practically, \cref{eq:rpa_pol} can be computed using the corresponding Adler-Wiser expression based on the Kohn-Sham orbitals and energies:\cite{adler1962quantum,wiser1963dielectric}
\begin{equation}
    \label{eq:adler-wiser}
     \chi_0 (\bm{r},\bm{r}^\prime,\omega)=\sum_j^{\mbox{\scriptsize{occ.}}}\sum_k^{\mbox{\scriptsize{virt.}}}\left[ \frac{\phi_j^{*}(\bm{r})\phi_k(\bm{r})\phi_k^{*}(\bm{r}^\prime)\phi_j(\bm{r}^\prime)}{\hbar\omega-(\varepsilon_k -\varepsilon_j)+i\eta} +\frac{\phi_j (\bm{r})\phi_{k}^{*}(\bm{r})\phi_{k}(\bm{r}^\prime)\phi_i^* (\bm{r}^\prime)}{\hbar\omega+(\varepsilon_k -\varepsilon_j)-i\eta}\right].
\end{equation}
With this, the frequency-dependent dielectric constant 
\begin{equation}
    \label{eq:freq_dielectric_constant}
    \varepsilon(\bm{r},\bm{r}^\prime,\omega)=\delta(\bm{r},\bm{r}^\prime)-\int  v(\bm{r},\bm{r}^{\prime\prime})\chi_0  (\bm{r}^{\prime\prime},\bm{r}^\prime,\omega)\mbox{d}\bm{r}^{\prime\prime},
\end{equation}
from which the  screened Coulomb potential can be computed as:
\begin{equation}
    \label{eq:screened_Coulomb}
    W_0 (\bm{r},\bm{r}^{\prime},\omega)=\int v(\bm{r},\bm{r}^\prime)\varepsilon^{-1}(\bm{r},\bm{r}^{\prime\prime},\omega)\mbox{d} \bm{r}^{\prime\prime}. 
\end{equation}
Substituting \cref{eq:Lehman} and  \cref{eq:screened_Coulomb} into \cref{eq:GW_self_energy}, one obtains the self-energy for $G_0 W_0$ approximation. 

However, notice here that in both \cref{eq:Lehman} and \cref{eq:adler-wiser}, the orbital energies are the Kohn-Sham energies, whereas our targeted key quantities are the quasiparticle energies $\{\varepsilon_i^{\mbox{\scriptsize{QP}}}\}$.  A numerically efficient implementation in calculating the the quasiparticle energy is using the following linearised equation:\cite{liu2016cubic,wilhelm2016gw}
\begin{equation}
    \label{eq:linearised_QP_eqn}
    \varepsilon_{i}^{\mbox{\scriptsize{QP}}}= \varepsilon_{i}+Z_{i}\langle \phi_i(\bm{r})|\mbox{Re}(\Sigma(\varepsilon_{i}^{\mbox{\scriptsize{QP}}}))-V_{\mbox{xc}}|\phi_{i}(\bm{r}) \rangle,
\end{equation}
where $Z_i$ is the quasiparticle renormalisation constant
\begin{equation}
    \label{eq:QP_renormal_const}
    Z_i=\left[1-\frac{\mathrm{d}}{\mathrm{d}\omega}\langle \phi_i(\bm{r})|\Sigma (\omega)|\phi_i(\bm{r})\rangle\right]^{-1}.
\end{equation}
\cref{eq:linearised_QP_eqn} is solved self-consistently. It starts from computing $\Sigma$ by setting $\varepsilon_{i}^{\mbox{\scriptsize{QP}}}=\varepsilon_i$, the Kohn-Sham orbital energy, which the self energy is plugged back into  \cref{eq:linearised_QP_eqn} to update the quasiparticle energy $\varepsilon_{i}^{\mbox{\scriptsize{QP}}}$. The process iterates until $\varepsilon_{i}^{\mbox{\scriptsize{QP}}}$ converges, which usually can be achieved in a few iterations. 

\subsubsection{The Bethe-Salpeter Equation}

BSE\cite{salpeter1951relativistic} is a useful tool developed based on the many-body perturbation theory to describe the dynamic interactions of excited particles (\emph{e.g.} excitons) through the exchange of virtual particles such as photons, in solids and molecules. When applied in conjunction with the quaisparticle energies from $GW$ calculations, it gives excellent outcomes in reproducing the experimentally observed excitionic emission spectra. For instance, \cref{fig:GW_approximation_diagram}(e) compares the transverse dielectric constant for monolayer \ce{MoS2}, $\varepsilon_{2}^{\perp}(\omega)$.\cite{ramasubramaniam2012large} It can be seen that the hybrid HSE functional produces two emission peaks (just below 2 eV) that are slightly higher in photon energy compared to the experimental spectrum. $G_{0}W_{0}$-approximation, on the other hand, overestimates the HSE-gap, by $\sim 1$ eV. Further incorporating the electron-hole interactions  through BSE is found to be essential in correcting the HSE-gap and produces the excitionic peaks at photon enegies that match exactly with the experimental spectrum. Note that the two exciton peaks observed in  \ce{MoS2} are originated from the spin-orbit couplings induced by the heavy Mo atom, which is also important to be included in the simulation. 

The general form of BSE resembles that of a Dyson equation, which is a matrix eigenvalue equation to be solved for obtaining the two-particle wavefunction $\Psi$. Hence, it has the general form of $\Psi=G_0 G_0 K\Psi$, in which $G_0$ is the free-particle wavefunction and $K$ is the interaction kernel that takes into account both the Coulomb attraction and exchange repulsion between two particles. For modelling the excitonic interactions in solids, BSE takes the following form
\begin{equation}
    \label{eq:BSE}
    \left( E_c - E_v \right) A_{vc\bm{k}} + \sum_{v'c'\bm{k}'} K^{eh}_{vc\bm{k},v'c'\bm{k}'} A_{v'c'\bm{k}'} = \Omega A_{vc\bm{k}},
\end{equation}
which $E_{c}$ and $E_{v}$ refer to the quasiparticle conduction and valence band energies from $GW$ calculations, whereas the interaction kernel $K$ further takes into account the electronic states at different reciprocal space vectors through the $\bm{k}$ indicies. $A$ contains the information of the  amplitudes of electron excitation from the valence to the conduction band at wavevector $k$. Solving the BSE thus gives the excitation energy $\Omega$ that can be directly matched with the optical spectra. Solving the BSE is computationally intensive, which quadratically scales with the product of  the number of valence and conduction bands, as well as the $k$ points involved. As such, in the practical application of the BSE, one often restricts the computation to only involve  the conduction band, the valence band and a few other bands directly above/below.

\section{Traditional Methods of Structure Predictions and Sampling}\label{chapter:sampling_methods}

\begin{figure}[t]
    \centering
    \includegraphics[width=0.9\textwidth]{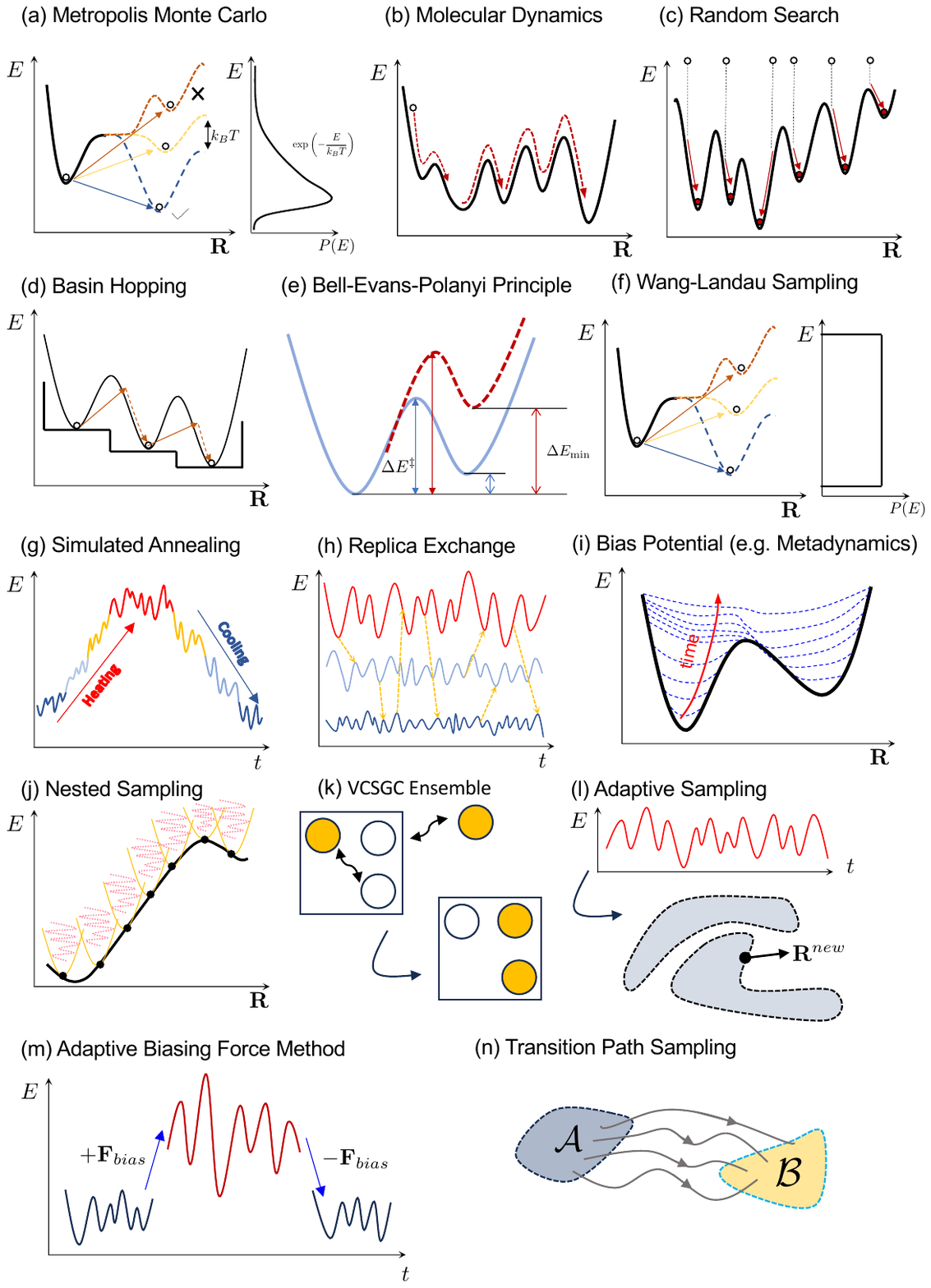}
    \caption{Illustration of different configuration sampling methods for exploring the potential energy landscapes in atomistic modelling.}
    \label{fig:sampling_methods}
\end{figure}
In \cref{eq:general_sampling_1} and \cref{eq:general_sampling_2},  we have established the statistical viewpoint on the physical/chemical properties for a given material system as the ensemble average of the corresponding property for individual configuration on the PES, which is drawn from seven different scenarios highlighted in \cref{sect:sampling_examples}  where sampling the PES are relevant in the fundamental materials science. In order to perform the computation as in \cref{eq:general_sampling_1} or \cref{eq:general_sampling_2}, the  first piece of recipe is to have a suitable energy model with which we can compute the corresponding energy of a given material with known atomistic structure $\bm{R}$, that we have  reviewed comprehensively in \cref{chapter:energy_models}. However, very often, we may not have a full dataset of possible structures for a given material readily available, in which case, we need a method to pragmatically sample the PES to generate a distribution of stable structures $\mathcal{P}(E(\bm{R}))$ at a given thermodynamic ensemble $(P,V,T,\mu)$ specified by the thermodynamic variables of pressure $T$, volume $V$, temperature $T$ and chemical potential $\mu$. Having access to these structures is the pre-requisite before subsequent property calculations can be conducted, thus PES sampling constitutes a critical component in modern computational materials science, which will be the focus for this section. More specifically, we will be focusing on the \textit{traditional} PES sampling methods as illustrated in \cref{fig:sampling_methods}, which exclude those driven by the recently popular generative AI models, that will be reserved for the later sections.

However, we do like to highlight here, that both the PES sampling and generative AI methods are targeted towards conquering the probability distribution $P(E(\bm{R}))$, which is  the Boltzmann distribution 
\begin{equation}
    \label{eq:Boltzmann_distr}
    P(E)=e^{-\beta E}/\mathcal{Z},
\end{equation}
in which $\beta=1/k_B T$, and $\mathcal{Z}$ is the normalisation constant as given in \cref{eq:configuration_partiton_function}. This seemingly simple distribution is generally unknown for a given material system because the normalisation constant $\mathcal{Z}$ is determined based on the energy of all configurations on the PES. As the potential energy function $E(\bm{R})$ lies in a high-dimensional space with complexity that grows with the dimension of $\bm{R}$, the number of potential local minima becomes  analytically intractable, leading to the difficulty of obtaining $\mathcal{Z}$ deterministically. In this sense, methods for sampling the PES are developed, primarily, in the hope to find:
\begin{enumerate}
    \item the global minimum, which corresponds to the most stable configuration on the PES, and
    \item  a set of low-energy metastable configurations that contribute significantly towards the Boltzmann weight $e^{-\beta E}$.
\end{enumerate}

Broadly speaking, methods for PES sampling may be classified into two classes of approaches, namely, the ones with energy guidance and the ones without. The well-known Monte Carlo [MC, \cref{fig:sampling_methods}(a)] and molecular dynamics [MD, \cref{fig:sampling_methods}(b)] are typical examples of energy-guided sampling approaches, whereby the evolution of the configurations and/or the acceptance of new configurations are tightly coupled with the underlying energy model employed. On the one hand, this allows us to directly generate the corresponding Boltzmann distribution at the sampling temperature [\cref{eq:Boltzmann_distr}]. On the other hand, it can encounter difficulties or even impossibility in reaching deep minima separated by high energy barriers, such that ergodicity can never be reached.\cite{wales1997global,doye1999evolution} The energy-guided samplings will be the major focus of this section.

In contrary, random search [\cref{fig:sampling_methods}(c)] first generates a set of candidate structures according to a specific set of system-dependent geometric rules, such as space group symmetries and molecular orientations in molecular crystals,\cite{case2016convergence}, after which energy minimisations are performed on each candidate structure to their respective local minimum. It provides us with the possibility to sample the PES globally, and the possibility to benchmark the performances of different energy models on an identical set of structures.\cite{Hunnisett:aw5094} However, one caveat behind the suitability of using random search for PES sampling is that the degrees of freedom being sampled should be as independent from each other as possible.\cite{echenique2006definition} In the case when multiple different degrees of freedom are coupled together, such as the torsions and bond angles in a molecule, they must be changed cooperatively in order to make a stable molecular conformation by avoiding steric clashes, the chances of locating low-energy conformers are significantly lowered when these degrees of freedom are searched independently, rending the necessity to guide the search with the topologies of the underlying PES, such as the low-mode search.\cite{kolossvary1996low}
\subsection{Monte Carlo Sampling}\label{sect:MonteCarlo}

Sampling atomistic configurations from the Boltzmann distribution with MC is trivial to understand and implement numerically. Begin with a starting structure $\bm{R}^{\mbox{\scriptsize{start}}}$, one take a step along the PES by randomly perturbing the atomic positions of $\bm{R}^{\mbox{\scriptsize{start}}}$ to generate a new structure $\bm{R}^{\mbox{\scriptsize{new}}}$, which will be either accepted (in this case, set $\bm{R}^{\mbox{\scriptsize{start}}}=\bm{R}^{\mbox{\scriptsize{new}}}$), or rejected (in which case we revert to $\bm{R}^{\mbox{\scriptsize{start}}}$) based on a pre-defined \emph{acceptance criteria}. The most common acceptance criterion is the Metropolis acceptance\cite{metropolis1953equation} [\cref{fig:sampling_methods}(a)], in which the probability of accepting $\bm{R}^{\mbox{\scriptsize{new}}}$ at temperature $T$ is given by:
\begin{equation}
    \label{eq:metropolis}
    p(\bm{R}^{\mathrm{new}}|\bm{R})=\min\left[1,\frac{\pi(\bm{R}^{\mathrm{new}})}{\pi(\bm{R})}\right],
\end{equation}
which is written in the form of transition probability to be in line with the subsequent discussions on the probability theories for machine learning [\emph{c.f.} \cref{sec:prob_theory_fundamentals}], $\pi(\bm{R})$ is the probability of observing a structure $\bm{R}$ at a given temperature following \cref{eq:Boltzmann_distr},
and the new structure will be accepted if such a transition probability $p(\bm{R^{\mbox{\scriptsize{new}}}}|\bm{R})>\mbox{rand}(0,1)$. 
This procedure will be repeated for a large number of $N_{\max}$ steps to reach adequate sampling. It guarantees that the Markov chain is sampled from the equilibrium Boltzmann distribution, which corresponds to a canonical ensemble and satisfies the condition of detailed balance. To test whether MC sampling converges, one could run multiple MC samplings from different starting structures and compare the similarities of the final probability distributions acquired from these samplings, for instance, using the Kullback–Leibler divergence.\cite{kullback1951information} 

 The choice of the MC  moves plays an important role in enhancing the acceptance of new structures, thus increasing the efficiency of MC samplings for low energy structures. The basin-hopping\cite{wales1997global,wales1998global} [\cref{fig:sampling_methods}(d)] is  a widely adopted MC sampling approach in atomistic modelling, in which each MC move consists of a random perturbation of the atomic positions followed by a structural optimisation to the nearest local minimum on the PES. The local optimisation leads to a much lower $E(\bm{R}^{\mbox{\scriptsize{new}}})$, which correspondingly increases the acceptance probability. In this sense, instead of the full PES, basin hopping samples the reduced energy surface [straight lines in \cref{fig:sampling_methods}(d)]. Nevertheless, the design of the random perturbation in basin-hopping, which is typically system-specific, is still critical in ensuring we can achieve a high acceptance probability in MC sampling.  For this, the general guidance to design efficient MC moves is to follow the Bell-Evans-Polanyi principle\cite{bell1936theory,evans1938inertia} [\cref{fig:sampling_methods}(e)] which states that a lower (higher) $E(\bm{R}^{\mbox{\scriptsize{new}}})$ is typically separated from $E(\bm{R}^{\mbox{\scriptsize{start}}})$ with a lower (higher) energy barrier, hence taking a MC move that crosses the low energy barrier will tend to result in a higher acceptance probability. 

Here, we shall briefly review some examples of MC moves that have been applied in sampling the PES for chemical and material systems. The simplest move is the random perturbation on the Cartesian coordinates that translates the particle positions, which is effective for sampling the PES of systems bonded by a simple energy function, such as the Lennard-Jones clusters.\cite{wales1997global,wales1998global}  For flexible molecules, such as peptides, one could focus on perturbing the rotatable bonds independently while preserving the atomic connectivity.\cite{echenique2006definition} However, this will not work effectively for systems such as cyclic molecules in which the rotation of 4-body torsion angles is highly coupled with 3-body bond angle change, for which low-mode search\cite{kolossvary1996low} could be a better alternative. In this case, the low-energy vibrational eigenmodes  point to a set of orthogonal directions for atomic movements, along which the internal coordinates of a molecule could change cooperatively and optimally into a new conformation by avoiding steric clashes, following the Bell-Evans-Polanyi principle. However, deliberately biasing the MC move to cross high energy barriers could also be a desirable feature for exploring chemical reaction pathways. \cite{shang2013stochastic} 

In the context of solid state materials, the eigenvectors for the imaginary phonon modes point to the direction of phase transitions, which can be used to guide the exploration of the underlying PES.\cite{togo2013evolution,da2015phase}  In the case where the cell volume $V$ is allowed to change, the MC simulation will be performed under the $NPT$ ensemble with a fixed pressure $P$ (see later), in which the acceptance probability [\cref{eq:metropolis}] needs to be changed accordingly into:
\begin{equation*}
    P(\mbox{new})=\min\left\{(1,\exp\left[-\beta\left(\Delta E+P\Delta V-Nk_B T\ln \frac{V^{\mbox{\scriptsize{new}}}}{V^{\mbox{\scriptsize{start}}}} \right)\right]\right\}.
\end{equation*}
For multicomponent alloys, one could construct phase diagrams by performing samplings under the Variance-Constrained Semi-Grand Canonical (VCSGC) ensemble [\cref{fig:sampling_methods}(k)],\cite{aangqvist2019icet,sadigh2012scalable} which enables both the atomic positions and the chemical composition of the system to be `swapped' under certain constraints of the chemical potentials. In this case, the acceptance probability is formulated as:
\begin{equation*}
P(\mbox{new})=  \min\left\{1, \exp\left[-\beta \left( \Delta E + N \bar{\phi} \Delta c + N \bar{\kappa} (\Delta c)^2 \right) \right] \right\},
\end{equation*}
in which $\bar{\phi}$ and $\bar{\kappa}$ are dimensionless parameters controlling the average and variance of the composition, respectively. $\Delta c$ is the change in concentration resulting from the trial move. Finally, $N$ denotes the total number of particles in a system.

\subsection{Molecular Dynamics}\label{sect:molecular_dynamics}

In contrast to MC, MD, which is often described analogously as `rolling a ball over a rugged carpet'  [\cref{fig:sampling_methods}(b)], every move in the sampling process is completely guided by the topology of the underlying PES by integrating the Newtonian equation of motion
\begin{equation}
    \label{eq:Newtonian}
    M\ddot{\bm{R}}=-\nabla_{\bm{R}}E(\bm{R})
\end{equation}
to determine the velocity of individual atom for it to propagate through the space at every time step. Unlike MC in which the simulation temperature is introduced through the acceptance criterion, MD controls the simulation temperature by adjusting the particle velocity through thermal noises with stochastic thermostats and/or velocity rescaling. To facilitate practitioners in understanding the machinary behind MD simulations, in this part, we will be discussing the three key concepts in MD simulations: (1) Equilibration process, which typically use a velocity scaling algorithm, to bring the system to the desired temperature for production runs. (2) The integration schemes to solve the Newton's equation for propagating the particles, and (3) The main mechanisms for controlling the thermodynamic ensembles in MD simulations.

\subsubsection{The Initial Velocities}\label{sect:initial_velocity}
MD is typically carried out at a desired temperature $T_{\mbox{\scriptsize{desired}}}$, however, the input structure for the MD simulation typically corresponds to an idealised equilibrium structure at 0 K, which one must assigned initial velocity $v_0$ to each atom as the boundary condition for integrating the Newton's equation. The most straightforward approach to achieve this is by realising the distribution of atomic kinetic energy at temperature $T$ is given by
\begin{equation}
    \label{eq:particle_velocity_distribution}
    P(v)=\sqrt{\frac{M\beta}{2\pi}}\exp\left(-\frac{\beta Mv^2}{2}\right)\sim\mathcal{N}\left(0;\sqrt{\frac{1}{M\beta}}\right),
\end{equation}
which is a Gaussian distribution centred at 0 with a standard deviation of $\sqrt{1/M\beta}$, with $\beta=1/k_B T$. In this case, the initial atomic velocities can be drawn randomly from this normal distribution. To ensure the materials' centre-of-mass is stationary throughout the simulation, the centre-of-mass velocity will be subtracted from the velocity of each atom. This may cause the actual system temperature $T_{\mbox{\scriptsize{actual}}}$ to deviate from $T_{\mbox{\scriptsize{desired}}}$, in which case, one can determine $T_{\mbox{\scriptsize{actual}}}$ from the total kinetic energy of the system
\begin{equation*}
    \frac{1}{2}\sum_{i=1}^{N}M_i v_i^2 = \frac{3}{2}Nk_B T_{\mbox{\scriptsize{actual}}},
\end{equation*}
in which $N$ is the total number of atoms in the system. From here, one can rescale the atomic velocity $v_i\leftarrow\lambda v_i $ by applying the scaling parameter
\begin{equation*}
\lambda=\sqrt{\frac{T_{\mbox{\scriptsize{desired}}}}{T_{\mbox{\scriptsize{actual}}}}}.
\end{equation*}

Practically, especially when $T_{\mbox{\scriptsize{desired}}}$ is high, one does not assign the initial velocity in an one-shot manner, as sudden temperature increase (which is equivalent to a wild MC move) can cause the atoms to move excessively, resulting in unphysical configurations with high degree of structural instabilities. Instead, one performs an \textit{equilibration MD run} over a time period $t_{\max}$ that brings up the system temperature slowly with a small time increment $t$, for example, with a linear time-dependent temperature scheduler:
\begin{equation*}
T(t)=T_{\mbox{\scriptsize{initial}}}+\left(\frac{T_{\mbox{\scriptsize{desired}}}-T_{\mbox{\scriptsize{initial}}}}{t_{\max}}\right)t.
\end{equation*}
Equilibrating structures to $T_{\mbox{\scriptsize{desired}}}$  in this way also allows the dynamical evolution of the system to closely follows the topology of the underlying PES, enabling the total system energy to be properly distributed between the potential and kinetic components, leading to a more realistic representation of the high energy structure. Conceptually, the equilibration procedure similarity with the \emph{forward diffusion} process in the diffusion generative AI model,\cite{ho2020denoising} in which  noises are  added to destroy the original samples into pure Gaussian noises. These noises are added progressively to ensure the process is stable, and the noises are learnable for the reverse process to reconstruct meaningful samples.

\subsubsection{The Integration Schemes}
To propagate the atomic positions along the PES according to the Newton's equation [\cref{eq:Newtonian}], it must be integrated and converted to a finite-difference form with respect to the time step $\Delta t$, in order to determine the change in the atomic positions and velocity from $t$ to $t+\Delta t$. This is typically achieved with a symplectic algorithm to better preserve the integrity of the atomic structures over time. The Verlet algorithm is one of the most commonly used and the simplest form of the integration scheme for solving the Newton's equation of motion,\cite{verlet1967computer,swope1982computer} which consists of following four substeps in each time step from $t$ to $t+\Delta t$:
\begin{subequations}
    \begin{align}
        &(1) \quad \bm{v}_{i}\left(t+\frac{\Delta t}{2}\right) =\bm{v}_{i}(t)+\left(\frac{\bm{F}_i (t)}{2M_i}\right)\Delta t\quad \mbox{velocity update over a  half time-step} \label{eq:velocity_update_1}\\
        &(2) \quad \bm{R}_i (t+\Delta t)=\bm{R}_i (t)+\bm{v}_{i}\left(t+\frac{\Delta t}{2}\right) \quad \mbox{postion update over a full time-step}\label{eq:position_update}\\
        &(3) \quad \mbox{Compute the atomic force } \bm{F}_i \mbox{ at the new } \bm{R}_i  \mbox{ 
 from the energy model} \nonumber\\
        &(4) \quad \bm{v}_{i}(t+\Delta t)=\bm{v}_{i}\left(t+\frac{\Delta t}{2}\right) + \left(\frac{\bm{F}_{i}(t+\Delta t)}{2M_i}\right)\Delta t. \quad \mbox{velocity update over the full time-step} \label{eq:velocity_update_2}
    \end{align}
\end{subequations}
From here, it is apparent that the choice of time step $\Delta t$ plays an important role in MD simulations that balances efficiency and stability, in which a small $\Delta t$ will require a long MD simulation to reach ergodicity, whereas a large $\Delta t$ forces atoms to take a large step forward which may results in unrealistic thus unstable structures. A variant of the Verlet integration is the Leap-Frog algorithm\cite{allen2017computersimulationofliquid} which asynchronously updates the velocity and positions in two steps, and is more commonly used with thermostats that controls the system temperature through velocity scaling, and its procedure is as follow:
\begin{subequations}
    \begin{align}
        &(1)\quad \bm{v}_{i}\left(t+\frac{\Delta t}{2}\right)=\bm{v}_{i}\left(t-\frac{\Delta t}{2}\right)+\left(\frac{\bm{F}_i (t)}{2M_i} \right)\Delta t\label{eq:LP_velocity_update}\\
        &(2)\quad   \bm{R}_i (t+\Delta t) =\bm{R}_i (t)+\bm{v}_{i}\left(t+\frac{\Delta t}{2}\right) \Delta t.
    \end{align}
\end{subequations}

Before proceeding, we note that since MD simulation applies the above integration schemes to propagate systems along the PES, the underlying energy models $E(\bm{R})$ that are deployed to model the atom-atom interactions in the systems will have a strong influence of the topology of the PES that is being sampled by MD. For examples, if a harmonic forcefield is employed, then the MD simulation will only sample the harmonic portion of the PES. In contrary, when the MD simulation is coupled with DFT-based energy models, which leads to the \emph{ab initio} molecular dynamics (AIMD), it will be able to sample parts of the PES that possesses an anharmonic topologies. The later can provide intriguing information in multiphonon [\cref{sec:anharmonic_vibs}] and electron-phonon [\cref{sec:eph_coupling}] interactions in materials, which have become more widely adapted in modern computational materials science with the increasing computational powers.

\subsubsection{Controlling the Thermodynamic Ensembles}

Choosing and controlling the correct thermodynamic ensembles play the critical role in MD simulations for determining the (dynamic) properties of a material that closely reflect the corresponding experimental conditions under which these properties are determined in the laboratory. Theoretically, a physical property of a given material is, in additional to its intrinsic bonding characteristics, a function of the thermodynamic states determined by the macroscopic thermodynamic variables that include temperature $T$, pressure $P$, volume $V$, internal energy $E$ and chemical potential $\mu$. Some of these variables are kept fixed at the measurements, such that similar conditions should also be met in the MD simulation through, for examples, applying the appropriate thermostats (to control the temperature) and/or barostats (to control the pressure and volume), so the simulation leads to the physical observables under the correct \emph{thermodynamic ensemble}. In \cref{tab:thermodynamic_ensembles} we listed the four most commonly used thermodynamic ensembles applied in MD simulations with the corresponding mechanisms for maintaining the ensemble conditions during the simulation, and common scenarios in which these ensembles will be applied.

\begin{table}[tbh]
\footnotesize
   \centering
    \caption{Common thermodynamic ensembles in molecular dynamics simulations}
\label{tab:thermodynamic_ensembles}
\begin{tabular}{lp{2.2cm}p{2.8cm}p{2cm}p{3cm}p{4cm}}

    \hline\hline
       \textbf{Ensemble}  & \textbf{Name} & \textbf{Fixed Quantities} & \textbf{Fluctuating Quantities} & \textbf{Control Mechanisms} &  \textbf{Applications} \\\hline
        $NVE$ & Microcanonical & Particle numbers, Volume, Energy & Temperature, Pressure & Not available & Phonon density-of-states, otherwise not widely used  \\\hline
        $NVT$ & Canonical & Particle numbers, Volume, Temperature & Energy, Pressure  & Langevin, Nos\'{e}-Hoover, Berendsen, Andersen thermostats & protein folding, chemical reactions and solid-state experiments at constant temperature  \\ \hline

        $NPT$ & Isothermal-Isobaric & Particle Numbers, Temperature, Pressure & Energy, Volume  &  Parrinello-Rahmann, Berendsen, Andersen barostats and thermostats & Liquid materials under pressure, high-pressure polymorph predictions for solids  \\ \hline
        $NPH$ & Isoenthalpic-Isobaric & Particle numbers, Pressure, Enthalpy & Volume, Temperature & Barostat, conserve total enthalpy $H=E+PV$. For this, the sampled distribution [\cref{eq:Boltzmann_distr}] needs to be modified into $P(H)=e^{-\beta H}/\mathcal{Z}.$ &    Adiabatic processes at constant pressure\\

         \hline\hline
\end{tabular}
\end{table}

In this section, we will first briefly outline the key features in each thermostat and barostat listed in \cref{tab:thermodynamic_ensembles} to help the practitioners in understanding the fundamental principles behind.

\paragraph{\textbf{Langevin thermostat}} is a stochastic thermostat that was first applied by \citet{schneider1978molecular} to control the system temperature in the MD simulation. It is established based on the Langevin equation that couples the Newtonian dynamics of the system with the interaction with a heat bath plus some thermal noise:
\begin{equation}
    \label{eq:Langevin_dynamic}
    M\ddot{\bm{R}}=-\nabla E(\bm{R})-\gamma M\dot{\bm{R}}+\sigma (t),
\end{equation}
in which $\gamma$ is the frictional term that determines the coupling strength to the heat bath, and $\sigma(t)\sim\mathcal{N}(0,1)$ is the time-dependent noise term that is drawn from a normal distribution. Controlling the simulation temperature is realised by invoking the fluctuation-dissipation theory\cite{kubo1966fluctuation} (see later) that couples $\gamma$ and $\sigma$ together by satisfying the condition of:
\begin{equation}\label{eq:langevin_fluctuation}
\left\langle\sigma(t)\sigma(t^\prime)\right\rangle=2\gamma M k_B T\delta(t-t^\prime).
\end{equation}
The Langevin dynamics is not only useful for simulating the dynamic responses of physical systems, but  also plays a central role in the sampling process of the diffusion generative AI models\cite{ho2020denoising} with applications such as image generation due to its stochastic nature.

\paragraph{\textbf{Andersen thermostat}} is also a stochastic thermostat that was proposed by \citet{andersen1980molecular}. It is conceptually easy to understand, whereby at each fixed time interval, or based on a predefined probability, the algorithm randomly reassigned the particles' velocities in accordance to the probability distribution as given by \cref{eq:particle_velocity_distribution}.

\paragraph{\textbf{Nos\'{e}-Hoover thermostat}} Unlike the previous two thermostats, the Nos\'{e}-Hoover thermostat\cite{nose1984unified,hoover1985canonical} is a deterministic approach to control the system temperature in the MD simulations. This is achieved through applying the following Lagrangian to determine the system dynamics:
\begin{equation}
    \label{eq:NH_lagrangian}
    \hat{\mathcal{L}}=\sum_{i=1}^{N}\frac{1}{2}M_is^2\dot{\bm{R}}_i^2-E(\bm{R})+\frac{1}{2}Q\dot{s}^2-3N k_B T\ln s,
\end{equation}
in which $Q$ is a fictitious thermostat mass. The kinetic energies of the atoms are controlled through a scaling variable $s$, which is determined through the following Newtonian-like equation of motion:
\begin{equation}
    \label{eq:NH_KE_control}
Q\ddot{s}=\sum_{i=1}^{N}M_i\dot{\bm{R}}_i^2-\frac{3Nk_B T}{s},
\end{equation}
which essentially tries to match the system's kinetic energy with the thermal energy of the heat bath at the targeted temperature $T$.

\paragraph{\textbf{Berendsen thermostat}} Proposed by \citet{berendsen1984molecular}, it applies a similar approach as the velocity scaling discussed in \cref{sect:initial_velocity}, where the particle velocities are adjusted through a time-dependent factor $\lambda(t)$ as $\bm{v}_{i}(t+\Delta t)=\lambda(t)\bm{v}_i(t)$, in which 
\begin{equation}
    \label{eq:berendsen_lambda}
    \lambda(t)=\sqrt{1+\frac{\Delta t}{\tau_{th}}\left(\frac{T_{\mbox{\scriptsize{target}}}}{T_{\mbox{\scriptsize{current}}}(t)}-1\right)},
\end{equation}
with $\tau_{th}$ is the relaxation time dictating how fast the system temperature should relax to the targeted temperature.

\paragraph{\textbf{Parrinello-Rahman barostat}} is designed for applying MD simulations to study the polymorphic transitions of crystalline materials that approximately generates an $NPT$ ensemble.\cite{parrinello1980crystal,parrinello1981polymorphic} It aims to simulate the response of materials that are subjected to a constant external pressure by allowing its cell matrix $\mathbf{h}=[\bm{a},\bm{b},\bm{c}]^{T}$ to deform, meaning that both the shape and the volume of the cell are allowed to be changed during the simulation. The Cartesian positions of the constituent atoms can be determined from the cell matrix $\textbf{h}$ and their fractional coordinates $\bm{s}$ with the dot product $\bm{R}=\textbf{h}\bm{s}$. With this, the Lagrangian that controls the Parrinello-Rahman dynamics is of the following form
\begin{equation}
    \label{eq:parrinello-rahman-dynamic}
    \hat{\mathcal{L}}=\sum_{i=1}^{N}\frac{1}{2}M_i\left\vert \textbf{h}\dot{\bm{s}_i}\right\vert^2+\frac{1}{2}w\mbox{Tr}\left[\dot{\textbf{h}}^{T}\dot{\textbf{h}}\right]-E(\bm{R})+\textbf{P}_{ext}V,
\end{equation}
in which $w$ is a fictitious mass associated with the cell degree-of-freedom, $\dot{\textbf{h}}$ is the time-dependent change in the cell matrix,  $\textbf{P}_{ext}$ is the matrix of external pressure and $V=\mbox{det}[\textbf{h}]$ is the cell volume that can be directly computed by taking the determinant of the cell matrix. This leads to the following coupled differential equations for the positions and cell matrix to be solved to propagate the system on the PES:
\begin{equation}
    \label{eq:PR_barostat_coupled_ODE}
    M_{i}\ddot{\bm{s}}_i=-\nabla_{\bm{s}}E(\bm{R})+f(\dot{\textbf{h}},\ddot{\textbf{h}}),\qquad  w\ddot{\textbf{h}}=V(\bm{\sigma}-\textbf{P}_{ext}\textbf{I}).
\end{equation}

\paragraph{\textbf{Berendsen barostat}} that is proposed in the same publication as the Berendsen thermostat,\cite{berendsen1984molecular} the Berendsen barostat allows the cell volumes to be changed over time in order to match  the external pressure under the control of the rate constant $\tau_p$:
\begin{equation}
    \label{eq:berendsen_barostat}
    \frac{\mathrm{d}V}{\mathrm{d}t}=\frac{1}{\tau_p}\left(P_{\mbox{\scriptsize{target}}}-P_{\mbox{\scriptsize{current}}} \right)V.
\end{equation}

\paragraph{\textbf{Andersen barostat}} that is proposed in the same publication as the Andersen thermostat,\cite{andersen1980molecular} which, unlike the Berendsen barostat, allows sampling from an $NPT$ ensemble to be undertaken in a statistically correct way. Similar to the Parrinello-Rahman barostat, this is achieved by modifying the Lagrangian for the system dynamics through introducing a fictitious degree of freedom $\mathcal{M}$ associated with the cell volume:
\begin{equation}
    \label{eq:andersen_lagrangian}
    \hat{\mathcal{L}}=\sum_{i=1}^{N}\frac{1}{2}M_i \left(\dot{\bm{R}}_i \cdot\dot{\bm{R}}_i\right)-E(\bm{R})+\frac{1}{2}\mathcal{M}\dot{V}^2-P_{ext}V,
\end{equation}
for which the atomic positions and cell volume are updated through solving the following coupled differential equations:
\begin{equation}
    \label{eq:andersen_coupled_ODE}
    M_i\ddot{\bm{R}}_i=-\nabla_{\bm{R}}E(\bm{R})+f(V), \qquad \mathcal{M}\ddot{V}=P_{int}-P_{ext},
\end{equation}
where the second differential equation changes the cell volume to match the difference between the internal and external pressure.

With MD simulations performed under the suitable thermodynamic ensemble, one can extract materials' corresponding thermodynamic properties in a physically meaningful way. The underlying physical principles behind this practice is the \emph{fluctuation-dissipation theory}\cite{kubo1966fluctuation} which states that \emph{the linear response of a given system to an external perturbation is expressed in terms of fluctuation properties of system in equilibrium}, and the system responses to this perturbation by dissipating its energy to its surrounding. For a given physical observable $\mathcal{A}$, its response to the external perturbation is a time-dependent phenomenon, which depends on how long the system can remember its past states. This can be described by the autocorrelation function
\begin{equation}
    \label{eq:autocorrelations}
    c_\mathcal{A}(t)=\int_{0}^{\infty}\langle A(t+\tau)A(t)\rangle \mathrm{d}\tau,
\end{equation}
where $\langle\cdot\rangle$ represents the ensemble average. At time zero, the autocorrelation function takes the value of $c_{\mathcal{A}}(0)=\langle\mathcal{A}^2(0)\rangle$, which gives the estimation on the amplitude of fluctuation. Subtracting the squared mean from the amplitude, $\langle\mathcal{A}^2\rangle-\langle\mathcal{A}\rangle^2$, gives the actual fluctuation. Based on this formation, \cref{tab:properties_from_MD} has listed a range of thermodynamic properties that can be extracted from MD simulations under a specific ensemble.

\begin{table}[tbh]
    \centering
    \footnotesize
\caption{Physical properties extractable from MD simulations based on fluctuation-dissipation theory.}
\label{tab:properties_from_MD}
    \begin{tabularx}{\textwidth}{p{2.8cm}p{2.5cm}llp{4cm}}
    \hline\hline
      \textbf{Symbol and Definition}   & \textbf{Name} & \textbf{Computation Formula} & \textbf{Ensemble}& \textbf{Note} \\ \hline
        $\displaystyle{C_{V}=\left(\frac{\partial U}{\partial T}\right)_V}$ & Heat capacity at constant volume & $\displaystyle{\frac{\langle E^2\rangle-\langle E\rangle^2}{k_B T^2}}$ & $NVT$ & Fluctuation of the total energy over time.\\
        $\displaystyle{C_{p}=\left(\frac{\partial U}{\partial T}\right)_p}$ & Heat capacity at constant pressure & $\displaystyle{\frac{\langle H^2\rangle-\langle H\rangle^2}{k_B T^2}}$ & $NPT$ & Fluctuation of the system enthalpy $(H=E-PV)$ over time.\\
        $\displaystyle{\kappa_T=-\frac{1}{V}\left( \frac{\partial V}{\partial P}\right)_T}$ & Isothermal compressibility & $\displaystyle{\frac{\langle V^2\rangle - \langle V\rangle^2}{k_B T \langle V\rangle} }$ & $NPT$& Fluctuation of the system volume over time, from which $K=1/\kappa_L$ gives the system's bulk modulus.\\
        $\displaystyle{\alpha=\frac{1}{V}\left(\frac{\partial V}{\partial T} \right)_p}$ & Thermal expansion coefficient & $\displaystyle{\frac{1}{k_B T^2}\frac{\langle VH\rangle-\langle V\rangle\langle H\rangle}{\langle V\rangle}}$ & $NPT$ & \\
        $\kappa_{\alpha\beta}$ & Lattice thermal conductivity tensor & $\displaystyle{\frac{1}{k_B T^2 V}}\int_0^\infty \langle \mathcal{J}_{\alpha}(t)\mathcal{J}_{\beta}(0)\rangle \mbox{d}t$ &$NVT$& Autocorrelation from the heat current operator, based on the Green-Kubo formula.\\
        $\varepsilon$ & Lattice dielectric constant & $\displaystyle{1+\frac{4\pi}{3k_B TV}\left(\langle \bm{M}^2\rangle-\langle\bm{M}\rangle^2\right) }$ & $NVT$& Fluctuation in the total dipole moment $\bm{M}$.\\
        $\varepsilon(\omega)$ & & $\displaystyle{1+\frac{4\pi}{3k_B TV}\int^\infty_0 e^{i\omega t}}\langle\bm{M}(t)\bm{M}(0)\rangle \mbox{d}t$ & & Frequency-dependent dielectric constant based on the Fourier transformation of the autocorrelation function that describes the relaxation of the dipole moment. \\
        $D$ & Atomic diffusion coefficient & $\displaystyle{\lim_{t\to\infty}\frac{1}{6t}\langle |\bm{R}(t)-\bm{R}(0)|^2\rangle }$ & $NVT$ or $NPT$ & \\
            & &$\displaystyle{\frac{1}{3}}\int_0^\infty \langle \bm{v}(t)\bm{v}(0)\rangle \mbox{d}t$ & & Alternative formulation based on the Green-Kubo formalism with the velocity-autocorrelation function.\\
        $g(\omega)$ & Phonon density-of-states & $\displaystyle{\sum_{i}^{N}\int_0^\infty e^{i\omega t} \frac{\langle\bm{v}_i (t)\bm{v}_i(0)\rangle}{\langle\bm{v}_i (0)\bm{v}_i(0)\rangle} \mbox{d}t}$ & $NVE$ & Fourier transformation of the velocity autocorrelation function, should not use thermostat to avoid modification of particle velocity through collision with heat bath.\\
     \hline\hline
    \end{tabularx}

\end{table}

\subsection{Enhanced Samplings for Rare Events}\label{sect:enhanced_smapling}
A common term used to access the sampling quality is the so-called \emph{ergodic principle}. Referring to \cref{eq:general_sampling_2}, the thermodynamic averaging of the property of a material is supposed to be exact when the MD sampling time $t$ or the number of MC steps $N$ goes to infinity:
\begin{equation}
    \label{eq:ergody_condition}
    \langle\mathcal{A}\rangle_T=\lim_{t\to\infty}\int_0^t\mathcal{A}(\bm{R}(t))\mbox{d}t =\int \mathcal{A}(\bm{R}) P(\bm{R})\mbox{d}\bm{R}.
\end{equation}
However, ergodicity can only be met if the underlying PES being sampled is relatively flat with shallow minima separated by low energy barriers. In realistic materials systems, very often, one sees deep local minima/global minimum that are separated from the main part of the energy landscape with high energy barriers [see, for example, \cref{fig:energy_landscape_examples}(b)]. On one hand, the low energy of these configurations means they contribute significantly to the Boltzmann weighting, and are important `landmarks' of the PES. On the other hand, the high-energy barriers translate to a low probability or even the impossibility of visiting these deep minima in an equilibrium MC/MD simulation. Visiting these deep minima thus becomes the `rare events' in equilibrium samplings, and overcoming this limit is the key driving force behind the development of the enhanced sampling methods in atomistic modelling, which we will review some of the major approaches that have been developed in the field.
\subsubsection{Wang-Landau Sampling} 
This approach is also referred to as the `flat histogram sampling'. In Wang-Landau sampling [\cref{fig:sampling_methods}(f)],\cite{wang2001efficient,wang2001determining} a density-of-states histogram is being built over a targeted range of energies $[E_{\min},E_{\max} ]$ as the sampling progresses. When a new sample with energy of $E_{new}$ is generated, the histogram at this energy state will be updated and the `flatness' of the histogram will be assessed. The new sample will be accepted if it leads to a more flattened histogram and rejected otherwise. Provided that $E_{\max}  - E_{\min}$ is (much) larger than the thermal energy $k_B T$, for which a canonical MC sampling will be performed otherwise, the Wang-Landau acceptance criteria will, therefore, encourage high-energy configurations to be accepted as the intermediate steps (anchoring points) for crossing the high energy barriers in the hope of reaching other deep local minima. Essentially, the flat histogram can be regarded as the result of combining the Boltzmann distributions from multiple temperatures, therefore, the Wang-Landau sampling leads to a \textit{multicannonical ensemble} of configurations for a given material.
\subsubsection{Simulated Annealing}
Different to MC, in MD simulations where temperature is directly embedded into the Lagrangian that determines the atomic movements, one could attenuate the kinetic energies of the atoms  through dynamically controlling the simulation temperature, in order to promote the system to cross high energy barriers. As named, this enhanced sampling approach [\cref{fig:sampling_methods}(g)] simulates the annealing process in real material processing by first slowly increasing the temperature in the MD simulations with a time-dependent temperature profile, assisting the system to cross high energy barriers through increasing its kinetic energies.  After holding the system at a high temperature for some time, it will be cooled down  to the starting temperature of the simulation, which facilitates the recovery of the ground-state configuration at the new local minimum.
\subsubsection{Replica exchange (or Parallel Tempering)}

Instead of running a single MD trajectory over a range of temperatures, replica exchange [\cref{fig:sampling_methods}(h)] seeks to concurrently run $n$ independent MD simulations at multiple temperatures $[T_1,T_2,\ldots,T_n]$, in which the PES at different windows of energetic ranges are explored simultaneously. To promote energy barrier crossing, at each given time interval, the algorithm exchanges two structures (replicas) from trajectories sampling at $T_i$ and $T_j$ respectively, with a probability determined by
\begin{equation}
    \label{eq:replica_exchange}
    \min\left\{1, \exp\left((\beta_i-\beta_j)(E_i-E_j)\right)\right\},
\end{equation}
which may be considered  analogously as an MC move that suddenly excites or relaxes a structure without introducing unphysical distortions that may destabilise the structure. 
\subsubsection{Metadynamics} 
\label{section:metaD}
An important class of enhanced sampling methods is to modify the topology of the underlying PES to ease the barrier crossing, as represented by metadynamics pioneered by \citet{laio2002escaping} [\cref{fig:sampling_methods}(i)].  In order to cross the energy barrier that separates the two energy basins in MD, a time and position bias potential $V(\bm{R},t)$ is progressively added to the underlying energy landscape to `fill up' the basins, until some point $t_{final}$ where the energy basins become completely `flat' and the system can move without any energy barrier between the two basins.  Because of the dynamic nature of the MD simulations, they not only provide information on the potential energy but also allow direct access to the system's free energy that accounts for both the vibrational and configurational degrees of freedom. This leads to the unique advantage of applying metadynamics in PES sampling whereby, upon convergence, the negative of the total bias potential added during the simulation gives the system's free energy landscape, $i.e.$ $F(\bm{R})=-V(\bm{R},t_{final})$.

Practically, the bias potential in  metadynamics  is added to the so-called \emph{collective variables} (CVs), which are  (or derived from) a subset of all the atomic coordinates in the system. They are a `focused' set of degrees of freedom which usually define the slowest parts of the system dynamics that are of interest. This implies that when the system is traversing the space spanned by the CVs, it needs to cross an energy barrier that is significantly larger than $K_B T$. 

Here, we will provide some examples of the typical choices of CVs for applying metadynamics in molecular and materials simulations.  For chemical reactions involving small molecules, using interatomic distances as CVs is the simplest choice and usually leads to meaningful chemical insights, as seen in the examples of studying the autoionisation of water \cite{liu2023mechanistic} as well as the kinetics of oxygen reduction reactions at the air/water interface for catalyst designs.\cite{yang2023neural} Bond angles consistute a set of CVs for examining the conformational dynamics of small molecules such as peptitdes,\cite{mccarty2017variational}  whereas the radius of gyration is typically selected for simulating the folding dynamics of polypeptides and proteins.\cite{vymetal2011gyration} For solid-state systems, the Steinhardt order parameters\cite{steinhardt1983bond} are often used as indicators for the emergence of cubic and hexagonal orderings from disordered structures by measuring the angular distributions of particles around a given structural centre. Both the standard form of the order parameters and their extension have been applied for simulating the nucleation of ice\cite{quigley2008metadynamics} and molecular crystals\cite{giberti2015metadynamics} with metadynamics.  A closely related type of CV is the coordination numbers\cite{iacovella2007efficient} which can be used for monitoring the changes of local coordination environments during the simulation. The lattice vectors\cite{martovnak2003predicting} are an effective set of CVs for simulating the phase transformations of crystalline materials, which comes naturally from the ideas behind the Parrinello-Rahman barostat.\cite{parrinello1980crystal} 

Identifying a set of suitable CVs is critical in metadynamics simulations. Traditionally, CVs are chosen based on chemical intuitions to identify the slowest degrees of freedom along the pathways of chemical reactions or phase transitions, which may not always be a trivial task for complex systems such as proteins and solids with more than a few atoms in the unit cells. The dimensionality of the CVs also needs to be controlled to lower the computational complexities and costs while ensuring all the relevant slow modes in the system have been sufficiently sampled. In this sense, the CVs identified for a particular system are usually not transferable to another. Finally, as the usage of CVs essentially reduces the dimensionality of the sampling problem, it leads to difficulties in judging whether hidden barriers or other important dynamics have been missed in the sampling, particularly for complex material systems. In this regard, the wide adaptation of machine-learning in atomistic simulations has opened new pathway to identify CVs that is solely based on the intrinsic dynamics of the systems itself, for example, using dimensional reduction techniques such as principle component analysis\cite{sittel2018perspective} to identify the slowest degree of freedom from existing MD trajectories. 

\subsubsection{Nested Sampling}\label{sec:nested_sampling}

With the basic understanding behind metadynamics, it becomes relatively trivial to rationalise the idea behind  nested sampling. It is  usually achieved through a constrained molecular dynamics simulation with a slow-growth approach\cite{woo1997combined,jarzynski1997nonequilibrium} to simulate the free-energy required for driving the system from its initial ($\zeta_i$) to the final $(\zeta_f)$ state  [defined by the CV $\zeta$], which can be obtained from the thermodynamic integration as:\cite{sprik1998free}
\begin{equation}
    \label{eq:thermodynamic_integration}
    W_{i\to f}=\int_{\zeta_i}^{\zeta_f}\left\langle \frac{\partial \mathcal{H}}{\partial \zeta}\right\rangle_{\zeta^\prime}^{\mbox{\scriptsize{cond.}}} \mbox{d}\zeta^\prime ,
\end{equation}
in which $\mathcal{H}$ is the Hamiltonian that determines the system dynamics and the conditional average $\langle\cdot\rangle^{\mbox{\scriptsize{cond.}}}$ are typically computed through performing MD simulations by constraining the CVs at  fixed $\zeta^\prime$, which are different points that lie on an interpolated path between $\zeta_i$ and $\zeta_f$, while allowing other degrees of freedom in the system to freely evolve at the simulating temperature [\cref{fig:sampling_methods}(j)]. 

\subsubsection{Adaptive Biasing Force (ABF) Method}
The nested sampling scheme assumes there exists some prior knowledge on the underlying (free) energy surface to be sampled. The adaptive biasing force method [\cref{fig:sampling_methods}(m)]\cite{darve2008adaptive,rodriguez2004assessing,darve2008adaptive} aims to overcome this problem by utilising the information gathered on-the-fly in an MD run to determine a bias force $F_{bias}$ that should be added to the system, facilitating it to overcome the free energy barriers. Similar to the nested sampling, in ABF, the CVs are first discretised into multiple points $\bm{\zeta}=\{\zeta_1,\zeta_2,\ldots,\zeta_n \}$ along the reaction pathway, where constrained MDs are performed for some time $t$ at each $\zeta_i\in\bm{\zeta}$. For each MD run, the instantaneous force $\bm{F}(\zeta,t)=\langle \partial_{\zeta}\mathcal{H}\rangle_{\zeta^\prime}^{\mbox{\scriptsize{cond.}}}$ is computed, whereby taking the time average over the MD trajectory gives the mean force $\langle \bm{F}(\zeta)\rangle$ that drives the system dynamics. If now one applies a biasing force that is opposite to the mean force, $i.e.$ $\bm{F}_{\mbox{\scriptsize{bias}}}(\zeta)=-\langle \bm{F}(\zeta)\rangle$ for the subsequent MD simulation, it will help to drive the system to cross the energy barriers by overcoming the mean forces. In this case, \cref{eq:thermodynamic_integration} for free energy computation is modified to 
\begin{equation*}
    W=-\int \bm{F}_{\mbox{\scriptsize{bias}}}(\zeta)\mbox{d}\zeta.
\end{equation*}
\subsubsection{Adaptive Sampling}
Similar to the ADF scheme, adaptive sampling also tries to utilise the information gathered throughout the MD sampling process  to guide the subsequent MD moves, but differently, it applies data analytic tools such as dimensionality reductions, clustering analysis and Markov state model to identify poorly sampled configuration spaces, towards which the subsequent samplings will be biased. For instance, by building up  Markov state model  from  the collected MD frames and analysing the transition matrix or using cluster analysis, \cite{zimmerman2015fast,doerr2016htmd} one could identify microstates that have not been frequently visited in the MD runs, which can be used as the starting point for new MD simulations to enhance the exploitation on a particular part of the energy landscape. For large biomolecular simulations, a contact matrix\cite{weber2011characterization,doerr2014fly} that counts the number of times states $i$ and $j$ were adjacent to each others in  the trajectory can be built to identify states that have the minimum contacts, which are selected as new starting points for MD simulations. The idea of structural-based adaptive sampling has important implication that underscores, for example, the development of on-the-fly machine learning forcefields\cite{jinnouchi2019fly}  to accelerate \emph{ab initio} MD simulations, which uses an agnostic structural descriptor and similarity kernel\cite{bartok2013representing} to identify representative structures on the PES for forcefield training. As the final example, \citet{perez2020adaptivebandit} developed the AdaptiveBandit framework for biomolecular simulations to tackle the exploitation-exploration dilemma, in which the decision to take an action $a$ to restart the sampling from the starting point at a given time $\bm{R}(t)$ is cast into a maximisation problem over all possible actions $\mathcal{A}$:
\begin{equation*}
    a(t)=\underset{a\in\mathcal{A}}{\mbox{argmax}} \left[\mathcal{Q}_t(a)+c\sqrt{\frac{\ln t}{N_{t}(a)}}\right],
\end{equation*}
in which $\mathcal{Q}_t(a)$ is the estimated reward for taking the action $a$, $e.g.$ the likelihood that a new structure will be generated, $c$ is a parameter aim to balance  exploitation-exploration, and $N_t (a)$ is the number of time that the action $a$ has been taken up to time $t$.

\section{Material Properties from Atomistic Modelling}\label{chapter:physical_properties}

\subsection{Structural Dynamics}\label{sect:phonon_harmonic_and_anharmonic}

\subsubsection{Phonons and Thermal Properties Under (Quasi)--Harmonic Approximations}\label{sect:phonon}

The local minimum on the zero-Kelvin PES $E(\bm{R})$ provides us with the information of a static equilibrium structure for a solid. However, under thermal excitation, atoms are constantly vibrating around their equilibrium positions $\bm{R}_{eq}$, and the thermal vibrations, or \emph{phonons}, in a solid has a profound influence on its corresponding physical properties. In terms of the theory of energy landscape, the vibrational behaviour of a structure ($\bm{R}_{eq}$) is fundamentally determined by the topology of the PES around the local minimum corresponding to this particular structure $\bm{R}_{eq}$.

Considering  the vibration of the $i$--th atom along the $\alpha$--Cartesian direction in a crystal unit cell with $N$ atoms, the restoring force $\bm{F}_{i}^{\alpha}$ experienced by this atom is given by the first derivative of the PES with respect to the atomic displacement $\bm{R}_{i}^{\alpha}$ as 
\begin{equation}\label{eq:vib_restor_force}
    \bm{F}_{i}^{\alpha}=-\frac{\partial E(\bm{R})}{\partial \bm{R}_{i}^{\alpha}}.
\end{equation}
For small amplitude atomic vibration, it can be described under the harmonic approximation according to the Hooke's law, where $\bm{F}_{i}^{\alpha}$ is linearly correlated with the atomic displacements of all other atoms in the system along the three orthogonal Cartesian directions:
\begin{equation}\label{eq:harmonic_forces}
     \bm{F}_{i}^{\alpha}=\sum_{j=1}^{N}\sum_{\beta=x,y,z}\bm{\Phi}_{ij}^{\alpha\beta}\bm{R}_{j}^{\beta},
\end{equation}
in which $\bm{\Phi}$ is the so--called \emph{force constants} matrix that give the curvature of the local minimum on the high--dimensional PES, where its matrix element:\cite{chaput2011phonon}
\begin{equation}\label{eq:force_constant}
    \bm{\Phi}_{ij}^{\alpha\beta}=\frac{\partial^{2}E(\bm{R})}{\partial \bm{R}^{\alpha}_{i}\partial \bm{R}^{\beta}_{j}}=\frac{\partial}{\partial \bm{R}_{j}^{\beta}}\left(\frac{\partial E}{\partial \bm{R}_{i}^{\alpha}}\right)=-\frac{\partial \bm{F}_{i}^{\alpha}}{\partial \bm{R}_{j}^{\beta}},
\end{equation}
in which the last expression is obtained with the use of \cref{eq:vib_restor_force}, and with the use of finite difference, this first derivative can be calculated as [\emph{c.f.} \cref{fig:anharmonic_phonon_illustration}(d)]:
\begin{equation}\label{eq:FC_finite_difference}
\bm{\Phi}_{ij}^{\alpha\beta}\approx -\frac{\bm{F}_{i}^{\alpha}(\Delta\bm{R}_{j}^{\beta})}{\Delta\bm{R}_{j}^{\beta}},
\end{equation}
meaning the matrix element $\bm{\Phi}_{ij}^{\alpha\beta}$ can be calculated based on the force experienced by $i$--th atom along the $\alpha$--Cartesian direction when the $j$--th atom in the unit cell is displaced from its equilibrium position along the $\beta$--Cartesian direction by  $\Delta\bm{R}_{j}^{\beta}$. \cref{eq:FC_finite_difference} is the working equation for determining the force constants in real space through the finite--displacement or `frozen--phonon' method.\cite{phonopy}  It is apparent that this procedure must be repeated for all $3N$ degrees of vibrational freedoms in a solid in order to build the full force constant matrix, indicating that the calculation of $\bm{\Phi}$ is also a problem of configuration sampling on the PES.  However, the number of configurations need to be sampled can be significantly reduced by harnessing the crystal symmetry if two displacements $\Delta \bm{R}_{j}^{\beta}$ and $\Delta \bm{R}_{j^\prime}^{\beta^\prime}$ can be transformed into each other by a crystal symmetry. Furthermore, in this real space approach, a supercell of the crystal lattice is often required to account for the contributions from the long--ranged atom--atom interactions to the force constants, or equivalently, allows the long wavelength phonons to be thoroughly sampled. This can significantly increase the computational costs associated with phonon calculations. The major advantages behind such an approach are (a) The generation of atom--displaced configurations is decoupled from the force evaluation. This allows the forces of a fixed set of atom--displaced configurations to be independently evaluated with different energy calculation methods (provided the gradients are available), and benchmark the performance of different theoretical methods in determining the dynamic stabilities of crystals. (b) The force evaluations for different configurations can be carried out concurrently to maximally harness the benefit of parallel computing in modern HPCs.

By making a mass--reduced Fourier transform of $\bm{\Phi}$, one can obtain the dynamical matrix $\bm{D}(\bm{q})$. Upon solving its corresponding eigenvalue equation $\bm{D}(\bm{q})\bm{\mathcal{Q}}_{\bm{q},n}=\omega^2_{\bm{q},n}\bm{\mathcal{Q}}_{\bm{q},n}$ provides a set of (up to $3N$ in case of no degeneracy) vibrational eigenmodes and eigenfrequencies $\{\bm{\mathcal{Q}}_{\bm{q},n},\omega_{\bm{q},n}\}$ for a given crystal structure.

Different from the real-space approach discussed above,  the Density Functional Perturbation Theory (DFPT)\cite{baroni2001phonons} is another approach to determine $\bm{D}(\bm{q})$ directly in the reciprocal space. This approach gives the phonon eigenmodes and eigenfrequencies at the $\Gamma$ point of the reciprocal space corresponding to phonon vibrations with infinite wavelength, as such, it does not require a supercell to be used in the practical calculation but the phononic properties other than the  $\Gamma$ point one cannot be obtained. In brief, we treat the small displacement of  positions for $i$-th atom, $\bm{u}_i(\bm{q})$, that is expressed in the reciprocal space, as a small perturbation to the total system (Kohn-Sham) Hamiltonian. It is related to the real-space atomic displacement via the Fourier transformation:
\begin{equation*}
    \bm{R}_{I}(\bm{u}_i(\bm{q}))=\bm{R}_\ell+\tau_{s}+\bm{u}_i(\bm{q})\exp(i\bm{q}\cdot\bm{R}_\ell),
\end{equation*}
where $\tau_s$ takes into account the translation across different unit cells.

\begin{wrapfigure}[33]{r}{0.3\textwidth}
  \begin{center}
    \includegraphics[width=0.3\textwidth]{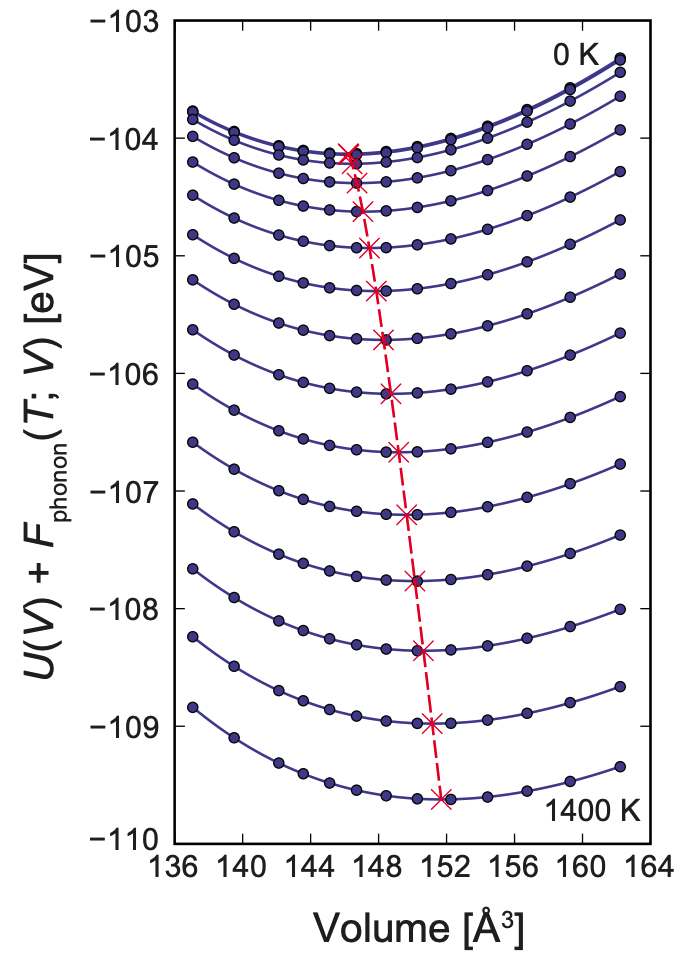}
    \caption{Gibbs free energy $G(T,V)$ (at zero pressure) for \ce{Ti3SiC2} as a function of the unit cell volumes at temperatures between 0 and 1400 K. The solid curves show the fitted $G(V)$ curves at different $T$. The minimum values of the $G(V)$ curves at temperatures are depicted by the crosses. The dashed curve passing through the crosses is for guide only. [Adapted from  \citet{togo2010first}]}
    \label{fig:free_energy_QHA}
  \end{center}
\end{wrapfigure}
By sampling the $\bm{q}$-points on a set of pre-defined reciprocal space grids $(n_{q_x},n_{q_{y}},n_{q_z})$ enables us to directly compute $\bm{D}(\bm{q})=\partial^2 E/\partial \bm{u}_i^{*,\alpha}(\bm{q})\partial \bm{u}_j^{\beta}(\bm{q})$ in the reciprocal space. It is immediately clear that computing the long-wavelength $(\bm{q}\to 0)$ limit of $\bm{D}(\bm{q})$ using this approach is trivial, which bypass the necessity of using large supercell as needed in the real-space approach. By treating $\bm{u}$ as a small perturbation to the system KS Hamiltonian, DFPT considers how such a  perturbation leads to the changes in the corresponding Kohn-Sham potential $\Delta V^{\mbox{\scriptsize{KS}}}(\bm{r})$, which translates to the distortion of electron density $n(\bm{r})$ that ultimately determines the changes in the KS eigenstates $\Delta\psi_{n,\bm{k}}^{\mbox{\scriptsize{KS}}}$. Consider only the first-order term in the perturbation theory, these quantities can be computed self-consistently from the Sternheimer equation:\cite{sternheimer1954electronic}
\begin{equation}
    \label{eq:sternheimer}
    (\hat{H}^{\mbox{\scriptsize{KS}}}-\varepsilon_{n,\bm{k}})\Delta\psi_{n,\bm{k}}^{\mbox{\scriptsize{KS}}}=-\Delta V_{n,\bm{k}}^{\mbox{\scriptsize{KS}}}\psi_{n,\bm{k}}^{\mbox{\scriptsize{KS}}},
\end{equation}
which gives the value of $\partial E/\partial \bm{u}_{i}^{\alpha}({\bm{q}})$ that is needed for computing the Hessian matrix.


The spectrum of phonon eigenfrequencies $\{\omega_{\bm{q},n}^2\}$ is an important material property indicator which is commonly used in computational material screenings. Systems with $\omega_{\bm{q},n}^2\geq 0$ for all $(\bm{q},n)$ are said to be dynamically (or mechanically) stable, which can be observed for materials with strong chemical bondings (\emph{e.g.} silicon). It implies that the topology of the local energy minimum is harmonic, such that $E(\bm{R})$ can be effectively expressed in terms of the othorgonal eigenmodes as
\begin{equation}\label{eq:PES_normal_mode_displaced}
    E(\mathcal{Q})=\frac{1}{2}\omega^2\mathcal{Q}^2.
\end{equation} Systems which possess  $\omega_{\bm{q},n}^2<0$, or commonly known as the \emph{imaginary phonons}, are said to be \emph{anharmonic}, which will be further discussed in \cref{sec:anharmonic_vibs}.

Access to the knowledge of $\{\omega_{\bm{q},n}^2\}$ (or the \emph{phonon dispersion relationship}) allows us to further calculate other fundamental thermodynamic properties of solids,\cite{phonopy} which can complement their determinations through the corresponding experiments to enhance our understanding on the fundamental properties of the materials. For examples, the phonon density--of--states (DOS) is given by:
\begin{equation}\label{eq:ph_DOS}
    g(\omega)=\frac{1}{N}\delta\left(\omega-\omega_{\bm{q},n}\right),
\end{equation}
whereby such quantity can be  measured with the inelastic neutron scatterings.\cite{osborn2001phonon,cheng2019simulation} The energy of the phonon subsystem at a given temperature $T$ can be calculated as:
\begin{equation}\label{eq:total_E_ph}
    \langle E_{ph}\rangle=\sum_{\bm{q},n}\hbar\omega_{\bm{q},n}\left[\frac{1}{2}+\frac{1}{\exp(\hbar\omega_{\bm{q},n}/k_{B}T)-1}\right],
\end{equation}
from which the constant volume heat capacity is given by
\begin{equation}\label{eq:Cv}
    C_v (T)=\left(\frac{\partial \langle E_{ph}\rangle}{\partial T}\right)_V
    =\sum_{\bm{q},n}k_{B}\left(\frac{\hbar\omega_{\bm{q},n}}{k_{B}T}\right)^2\left[\frac{\exp(\hbar\omega_{\bm{q},n}/k_{B}T)}{(\exp(\hbar\omega_{\bm{q},n}/k_{B}T)-1)^2}\right],
\end{equation}
which is the major contribution to the heat capacity in semiconducting and insulating materials, and  can typically be measured with differential scanning calorimetry.

Thermodynamically, phonon vibrations contributes towards the Helmholtz free energy $F_{vib}(T)$ of a crystalline solid at a constant volume $V$. According to statistical mechanics, 
\begin{equation}\label{eq:free_energy_statistical_mech}
    F_{vib}(T)=-k_{B}T\ln Z_{vib},
\end{equation}
where $Z_{vib}(T)=\prod_{\bm{q},n}\frac{\exp(-\hbar\omega_{\bm{q},n}/2k_{B}T)}{1-\exp(-\hbar\omega_{\bm{q},n}/k_{B}T)}$ is the phonon partition function, such that:\cite{phonopy}
\begin{equation}\label{eq:helmholtz}
F_{vib}(T)=\frac{1}{2}\sum_{\bm{q},n}\hbar\omega_{\bm{q},n}+k_{B}T\sum_{\bm{q},n}\ln\left[1-\exp(-\hbar\omega_{\bm{q},n}/k_{B}T)\right].
\end{equation}
Correspondingly, the vibrational contribution to the entropy is
\begin{equation}\label{eq:vib_entropy}
S_{vib}(T)=-\left(\frac{\partial F_{vib}}{\partial T}\right)_V=\frac{1}{2T}\sum_{\bm{q},n}\hbar\omega_{\bm{q},n}\coth\left(\frac{\hbar\omega_{\bm{q},n}}{k_{B}T}\right) - k_{B}\sum_{\bm{q},n}\ln\left[2\sinh\left(\frac{\hbar\omega_{\bm{q},n}}{k_{B}T}\right)\right].
\end{equation}

Alternatively, the vibrational contribution to the Helmholtz free energy can also be calculated from the phonon DOS $g(\omega)$ extracted from the equilibrium MD calculations under canonical ensemble from the velocity autocorrelation function [\emph{c.f.} \cref{tab:properties_from_MD}], from which the corresponding vibrational contribution to the free energy can be calculated by invoking the relationship of \cref{eq:free_energy_statistical_mech}. Under the harmonic approximation, it can be shown that the free energy takes the following from (for a system containing $N$ atoms):
\begin{equation}
    \label{eq:phonon_FE_from_density}
    F_{vib}=-\frac{k_{B}T}{N}\int_{0}^{\omega_{\max}}g(\omega)\ln\left[2\sinh\left(\frac{\hbar\omega}{2k_{B}T}\right)\right]\mbox{d}\omega.
\end{equation}

Ultimately, the thermodynamic stability of a solid is governed by its Gibbs free energy that is a function of the state variables [temperature ($T$), volume ($V$) and pressure ($p$)]. Knowing the Gibbs free energy also provides the information on the thermal expansion coefficients for a solid. This is because under the \emph{quasiharmonic approximation}, the equilibrium volume at a given temperature is the one that minimizes the Gibbs free energy at a given $T$:\cite{togo2010first}
\begin{equation}\label{eq:Gibbs_energy_minimisation}
    G(T,p)=\min_{V}\left[U(V)+F_{vib}(T,V)+pV\right],
\end{equation}
in which $U(V)$ is the internal energy (\emph{e.g.} total energy from electronic structural calculation) for the system at a given volume $V$. \cref{eq:Gibbs_energy_minimisation} shows that the phonon vibration play a major contribution to the Gibbs free energy of a solid. Practically, in order to determine $G(T)$ (at a given pressure, $e.g.$ $p=0$), phonon and Helmholtz free energy calculations need to be performed for a given solid at its equilibrium volume, as well as a few extra  expanded and shrinked volumes in order to identify the free energy minimum at a given temperature [\cref{fig:free_energy_QHA}].  This further forms the basis for the \emph{ab initio} determination of the phase diagrams and critical phase transition temperature $T_{c}$ for polymorphic inorganic\cite{da2015phase} and organic\cite{nyman2015static} solids: for two competing phases $P_1$ and $P_2$, the stable polymorph at a given temperature is the one with a lower Gibbs free energy, whereby $T_{c}$ is the solution for $G_{P_1}(T)=G_{P_2}(T)$, if exists.

\subsubsection{Anharmonic Vibrations in Solids}\label{sec:anharmonic_vibs}
Fundamentally, the existence of non--zero temperature--dependent volume expansion coefficients for solids arise from the fact that atomic vibrations are not strictly harmonic. The anharmonic nature of the lattice vibration is often reflected in the change of the vibrational frequencies with respect to the cell volume, a quantity that is captured by the mode--dependent Gr\"{u}neisen parameter,\cite{dolling1966thermodynamic} defined as:
\begin{equation}\label{eq:gruneisen}
    \gamma_{\bm{q},n}=-\frac{\partial \ln \omega_{\bm{q},n}}{\partial \ln V}.
\end{equation}
It can be shown that the lattice expansion coefficient $\alpha=(\partial V/\partial T)_p$ can be derived from the change in the Helmholtz free energy [\cref{eq:helmholtz}] with respect to volume, which can only be non--zero if some phonon modes of the material possess  non--zero $\gamma_{\bm{q},n}$. 

Returning back to the real space view of atomic vibrations, the vibrational anharmonicity arises when the PES $E(\bm{R})$ traced by an vibrating atom in the solid, particularly at large vibrational amplitudes, can only be satisfactorily described by taking the  Taylor expansion of $E(\bm{R})$ that involves the contributions from higher--order force--constants:
\begin{align}\label{eq:higher_order_vib_Hamiltonian}
E(\bm{R})=&\mathcal{U}(\bm{R}_{0})+\bm{\Phi}_{i}^{\alpha}\Delta\bm{R}_{i}^{\alpha}+\frac{1}{2!}\bm{\Phi}_{ij}^{\alpha\beta}\Delta\bm{R}_{i}^{\alpha}\Delta\bm{R}_{j}^{\beta}\nonumber \\&+\frac{1}{3!}\bm{\Phi}_{ijk}^{\alpha\beta\gamma}\Delta\bm{R}_{i}^{\alpha}\Delta\bm{R}_{j}^{\beta}\Delta\bm{R}_{k}^{\gamma}+\frac{1}{4!}\bm{\Phi}_{ijkl}^{\alpha\beta\gamma\delta}\Delta\bm{R}_{i}^{\alpha}\Delta\bm{R}_{j}^{\beta}\bm{R}_{k}^{\gamma}\Delta\bm{R}_{l}^{\delta}+\mathcal{O}(\Delta\bm{R}^{5}),
\end{align}
meaning the topology of $E(\bm{R})$ deviates significantly from a parabolic shape. In highly anharmonic materials, such as cubic perovskites \ce{SrTiO3}, significant contributions to the above expansion can even be found from the sixth order force constant.\cite{tadano2015self} This also implies that, in practical computation of anharmonic lattice dynamics of materials, one should use an energy model, such as DFT [\cref{sec:DFT}], rather that empirical atom-atom potentials [\cref{sect:empirical_forcefield}] that typically truncate the above series at the second order, as it is usually easier to parameterise.

\paragraph{\textbf{Weakly anharmonic systems}} For systems with weak vibrational anharmonicity, the expansion beyond the third order in \cref{eq:higher_order_vib_Hamiltonian} can be satisfactorily disregarded. Extending the finite--displacement method for calculating the harmonic force constant [\cref{eq:FC_finite_difference}], the third order force constant can be calculated as:
\begin{equation}\label{eq:FC3_finite_difference}
    \bm{\Phi}_{ijk}^{\alpha\beta\gamma}\approx-\frac{\bm{F}_{i}^{\alpha}(\Delta\bm{R}_{j}^{\beta},\Delta\bm{R}_{k}^{\gamma})-\bm{F}_{i}^{\alpha}(\Delta\bm{R}_{j}^{\beta})}{\Delta\bm{R}_{j}^{\beta}\Delta\bm{R}_{k}^{\gamma}}. 
\end{equation}
Calculating $\bm{\Phi}_{ijk}^{\alpha\beta\gamma}$ via this approach is in principle only feasible for crystals with small unit cell and high symmetry. This is because \cref{eq:FC3_finite_difference} involves the force contributed by displacing a combination of two atoms along two different Cartesian directions, and the total number of unique combinations in a crystal structure grows factorially as the total number of atoms per cell increases and the crystal symmetry reduces. Although in rare cases, it has been applied in the investigations of selected complex materials, such as hybrid halide perovskites.\cite{zhu2019mixed}

Taking the Fourier transform of $\bm{\Phi}_{ijk}^{\alpha\beta\gamma}$ leads to third--order force constant $\bm{\Phi}_{\lambda\lambda^{\prime}\lambda^{\prime\prime}}$ in the reciprocal space, which contains the information on the strength of interactions among three phonons,\cite{phono3py} and $\lambda=(\bm{q},n)$ is an abbreviation for a specific phonon mode. It means the lowest order contribution to the anharmonic lattice vibration comes from the three--phonon scattering process,\cite{pathak1965theory} which further leads to the finite lattice thermal conductivity ($\kappa_L$) in solids. Otherwise, $\kappa_L$ becomes infinite if the lattice vibration is strictly harmonic in a defect--free crystal.\cite{allen1993thermal} Furthermore, since the crystal is weakly anharmonic, the three--phonon scatterings can be treated perturbatively as a correction to the harmonic phonon, which gives rise to the phonon self--energy $\Sigma_\lambda (\omega_\lambda)$. The imaginary part of $\Sigma_\lambda (\omega_\lambda)$, dubbed as $\Gamma_{\lambda}(\omega_\lambda)$,  represents the finite width  of phonon peak at $\omega_\lambda$, which is proportional to $|\bm{\Phi}_{\lambda\lambda^{\prime}\lambda^{\prime\prime}}|^2$ and can be determined based on the Fermi Golden rule.\cite{maradudin1962scattering} Consequently, this results in the finite lifetime for a given phonon mode $\tau_\lambda=[2\Gamma_{\lambda}(\omega_\lambda)]^{-1}$ according to the Heisenberg's uncertainty principle. This reciprocal relationship means that a phonon mode that is strongly scattered by the others will have a particularly short lifetime. Together with the phonon group velocity $\bm{v}_{\lambda}=\partial \omega_\lambda/\partial \bm{q}$, the lattice thermal conductivity tensor may be obtained by solving the linearised phonon Boltzmann equation under the relaxation time approximation as:\cite{chaput2013direct}
\begin{equation}\label{eq:ph_boltzmann}
    \bm{\kappa}_{L} = \frac{1}{NV}\sum_\lambda C_\lambda \bm{v}_{\lambda}\otimes \bm{v}_{\lambda} \tau_\lambda,
\end{equation}
in which $C_\lambda$ is the mode contribution to the heat capacity [\emph{c.f.} \cref{eq:Cv}]. The above discussion highlights the most commonly adopted approach for computationally determining the lattice thermal conductivities for weakly anharmonic materials.\cite{phono3py,li2014shengbte}  

\paragraph{\textbf{Quantifying the degree of vibrational anharmonicity}} 

Given that the phonon life time $\tau_\lambda$ in \cref{eq:ph_boltzmann} is determined by treating the three--phonon interactions perturbatively, a nature question to ask is how the degree of vibrational anharmonicity can be quantified in order to facilitate the choice of appropriate theoretical approach to model and understand both the structural dynamics and thermal properties of functional materials.  

Theoretically, AIMD simulation is capable to provide us with a set of atomic displacements and forces $\{\Delta\bm{R},\bm{F}\}$, that is governed by  the structural dynamics of a given material in which  all degrees of anharmonicity are accounted for [\cref{eq:higher_order_vib_Hamiltonian}]. If the corresponding harmonic force constant $\bm{\Phi}_{ij}^{\alpha\beta}$ is also known, the corresponding \emph{harmonic force} $\bm{F}^{(2)}$ at any given $\Delta\bm{R}$ can also be determined based on \cref{eq:harmonic_forces}, whereby the difference $\bm{F}^{A}=\bm{F}-\bm{F}^{(2)}$ is naturally the remainder of the atomic force that is contributed by the force--constants beyond the harmonic terms. Based on this fundamental understanding, \citet{knoop2020anharmonicity} proposed an \emph{anharmonic score} to access the degree of vibrational anharmonicity by comparing the standard deviations in $\bm{F}^{A}$ and $\bm{F}$ as sampled from AIMD simulations:
\begin{equation}\label{eq:anharmonic_score}
    \sigma^{(2)}(T)=\sqrt{\sum_{i,\alpha}\left\langle \left(F^{\alpha,A}_{i}\right)^2 \right\rangle_{T}\Big/ \sum_{i,\alpha}\left\langle \left(F_{i}^{\alpha}\right)^2 \right\rangle_{T}},
\end{equation} 
where the summation is performed over the  atoms ($i$) in the simulation cell and the Cartesian direction ($\alpha$). Generally, systems with $\sigma^{(2)}<0.5$ are considered to be weakly anharmonic in which the expansion \cref{eq:higher_order_vib_Hamiltonian} can be safely truncated at the third--order term, whereas those with $\sigma^{(2)}>1$ are structurally not stable at a given $T$. This  criteria is supported by the results obtained from screening the room--temperature stabilities across a large set of 3819 cubic perovskite.\cite{yang2021altas}

\paragraph{\textbf{Strongly anharmonic systems}} 

\begin{figure*}[tb]
    \centering
    \includegraphics[width=\textwidth]{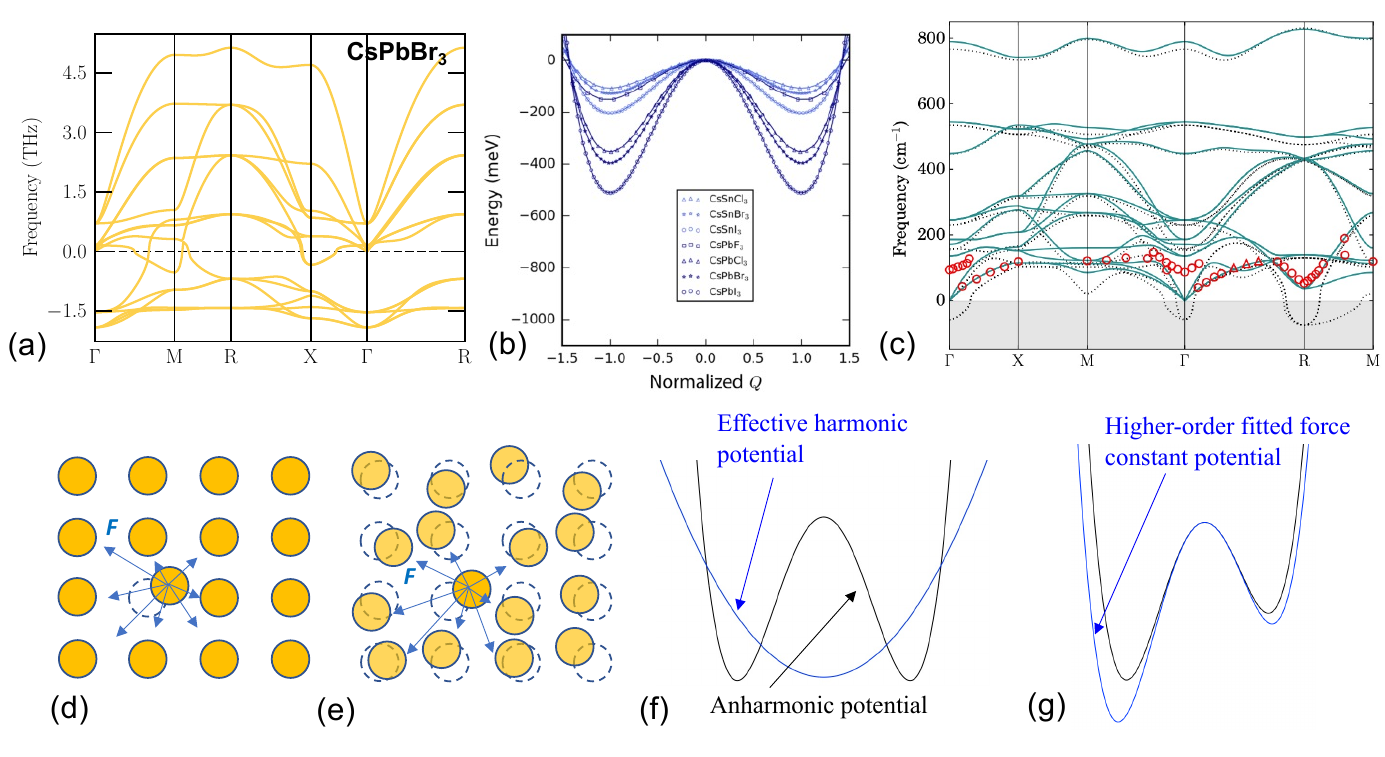}
    \caption{Strongly anharmonic phonons in cubic perovskites. (a) Phonon dispersion relationship for cubic \ce{CsPbBr3}\cite{yang2021altas} showing the presence of imaginary phonon modes in the system. (b) Double--well potential energy surface of phonon instabilities  in various cubic cesium halide perovskites. The potential energy surface is plotted against the normalised atomic displacements along the normal mode $\bm{\mathcal{Q}}$ that corresponds to the $M_3^{+}$ octahedral tilting.\cite{yang2020assessment}  (c) Comparison of the anharmonic renormalised (solid line) and harmonic (dashed line) phonon dispersion relationship for cubic \ce{SrTiO3},\cite{tadano2015self} showing the dynamic stabilisation of \ce{SrTiO3} at 300 K with the disappearance of the imaginary modes, which is consistent with the experimental results.\cite{cowley1969relationship,stirling1972neutron} (d--e) Schematic illustration of the difference between phonon calculations performed with frozen phonon (d) and  temperature--dependent effective potential (TDEP)\cite{hellman2011lattice,hellman2013temperature} (e). In the frozen phonon approach, the force on a displaced atom is evaluated in an environment of its stationary neighbours, whereas in TDEP, the same force is evaluated in neighbouring atoms that are also moving under thermal excitation. (f) In the original TDEP formalism, an anharmonic double--well--type potential is refitted to an effective harmonic potential. (g) Extension of the TDEP concept to extract higher--order force constants enabling the reconstruction of the anharmonic potential energy surface.\cite{eriksson2019hiphive}}
    \label{fig:anharmonic_phonon_illustration}
\end{figure*}

Materials that possess imaginary phonon frequencies ($\omega_{\bm{q},n}^2<0$) are generally considered to be highly anharmonic, which is also referred to as the phonon instability. This is widely observed in material systems, such as the high--symmetry cubic ABX$_{3}$ perovskites [\cref{fig:anharmonic_phonon_illustration}(a)]. Geometrically, it implies that a given structure is located on a local maximum of the PES with negative curvature, rather than a minimum. Hence, according to \cref{eq:PES_normal_mode_displaced},  the atomic vibrations along the imaginary phonon mode provides additional energetic contributions to stabilise the system. For example, the octahedral tiltings [\cref{fig:anharmonic_phonon_illustration}(b)] in cubic perovskites stabilize the structure by minimising the effect of ionic size mismatches for the constituent ions in adopting a perfect cubic crystal structure in the space group of $Pm\bar{3}m$.\cite{aschauer2014competition} This also implies that, in the higher--order expansion of the PES [\cref{eq:higher_order_vib_Hamiltonian}],   it is critical to include the quartic order term in order to correctly model the lattice dynamics of perovskites that are strongly affected by the double--well potentials, such as the ones shown in  \cref{fig:anharmonic_phonon_illustration}(b). However, as discussed in the previous section, calculating the  force constants via the finite--displacement method is generally not computationally feasible due to the factorial growth in the number of atom combinations.

On the other hand, many perovskites of significant technological importance, such as \ce{SrTiO3} [\cref{fig:anharmonic_phonon_illustration}(c)]\cite{stirling1972neutron,cowley1969relationship}, \ce{CaSiO3}\cite{sun2014dynamic}, and \ce{CsPbI3}\cite{an2021structural} are known to be stable at elevated temperature, which is reflected by the disappearance of the imaginary phonon modes in the phonon dispersion relationships measured at these temperatures. This means that the high temperature lattice vibrations can be effectively described by a harmonic model.

Given the high relevance of strongly anharmonic phonons in these important materials, a range of theoretical and computational approaches have been developed to obtain deeper physical insights behind imaginary phonons,\cite{pallikara2022physical} particularly with the increasing adaption of (supervised) machine learning in material science. Here, we focus our discussions on the Temperature--Dependent Effective Potential (TDEP)\cite{hellman2011lattice,hellman2013temperature} and its extension,\cite{tadano2014anharmonic,eriksson2019hiphive} that has become one of the most widely used method in treating highly anharmonic systems.\cite{romero2015thermal,zhou2018electron,zhao2020quartic,ding2021soft} Under the  framework of TDEP, the vibration of a single ion in the potential field of its stationary neighbours [\cref{fig:anharmonic_phonon_illustration}(d)] is energetically unfavored, which destabilises the structure as reflected by the emergence of the imaginary phonon modes. In contrast, the same atomic vibration in the potential induced by its non--stationary neighbours that are also vibrating under thermal excitation [\cref{fig:anharmonic_phonon_illustration}(e)], is favored. It implies that the thermal vibrations renormalise the zero--Kelvin PES into an effective harmonic potential at finite temperature [\cref{fig:anharmonic_phonon_illustration}(f)], along which the lattice vibrations do not exhibit instability [\cref{fig:anharmonic_phonon_illustration}(c)]. By sampling the finite--temperature atomic vibrational amplitudes ($\Delta \bm{R}$) and forces ($\bm{F}$) with AIMD, this effective \emph{harmonic} potential ($\tilde{\bm{\Phi}}_{ij}^{\alpha\beta}$) can be obtained by minimising the following loss function\cite{hellman2011lattice}:
\begin{equation}\label{TDEP_fitting}
    \min_{\tilde{\bm{\Phi}}}\left\Vert \sum_{i,\alpha}\left[\sum_{j,\beta}\tilde{\bm{\Phi}}_{ij}^{\alpha\beta}\Delta \bm{R}_{j}^{\beta} -\bm{F}_{j}^{\beta}\right]\right\Vert.
\end{equation}

Since $\bm{F}$ contains the information of PES including the anharmonic contributions from all orders [\cref{eq:higher_order_vib_Hamiltonian}], a natural extension of \cref{TDEP_fitting} is to obtain higher--order force constants $\bm{\Phi}^{(n)}$ up to the $n$--th order by fitting the AIMD displacement and force data,\cite{tadano2014anharmonic,eriksson2019hiphive}  by the usage of compressive sensing algorithm.\cite{zhou2014lattice} In such case, the fitting procedure is aiming at closely reconstructing the anharmonic PES [\cref{fig:anharmonic_phonon_illustration}(g)] rather than renormalising it, effectively resulted in a \emph{force constant potential} (FCP) that may be further applied to perform dynamic simulations.\cite{eriksson2019hiphive,tadano2015self} The major advantage of FCP is enabling us to obtain higher--order force constants,  otherwise unfeasible with the finite--displacement approach. The access to the higher--order force constants will further allow us to determine the temperature--renormalised phonon frequencies at a given temperature non--perturbatively, thus taking into account the quantum mechanical nature of the lattice vibrations which is ignored in the quasiharmonic approximation. This is important, even for the prediction of the temperature--dependent  phonon frequencies in materials such as silicon.\cite{kim2018nuclear} As later shown by \citet{masuki2022anharmonic}, the definition of thermal expansions via the Gr\"{u}neisen parameter [\cref{eq:gruneisen}] is only valid when the ratio $\langle \bm{\Phi}^{(4)}\rangle/\langle \bm{\Phi}^{(2)} \rangle$ is small, otherwise, properties such as the thermal expansion coefficients, phonon specific heat, as well as the lattice thermal conductivity\cite{zhao2021lattice,tadano2015self} and the Helmholtz free energy\cite{oba2019first} should be determined based on the temperature--renormalised phonon frequencies. One approach to achieve this is through the application of the self--consistent phonon (SCPH) approach which treats the problem with many--body Green's function that naturally accounts for the quantum nature of anharmonic vibrations. Here, the free--particle (harmonic) phonon propagator is `dressed' by the self--energy contributed from the quartic,\cite{werthamer1970self,tadano2015self} and additionally cubic\cite{goldman1968improved,oba2019first} force constants via the Dyson's equation, which needs to be solved self--consistently to obtain the effective harmonic frequencies.

\subsubsection{Ion Diffusions}
Ionic conductors are a unique class of solid-state materials which contain mobile ions, that are not spatially confined to particular lattice sites, but act as mobile ions that move across the lattices, conducting both masses and charges through the systems. Determining the ionic conductivities plays an important role in understanding the physical properties, and helps in the design of functional materials, including, for instances, (a) the search for fast ion conductors for safe and compact solid-state electrolytes,\cite{bachman2016inorganic} (b) enhancing the hydrogen storage capabilities in the solid-state hydrogen storage materials,\cite{berube2007size} (c) the stabilities of mixed halide perovskite photovoltaic materials against photo-induced phase segregations,\cite{hu2022valence} (d) ionic contribution to the Seebeck coefficients in  ionic thermoelectrics,\cite{song2022ionic} and (e) the formation and stabilities of oxygen vacancies, which dictate the ferroelectric properties of perovskites.\cite{yang2013strain}

\begin{figure}[htb]
    \centering
    \includegraphics[width=0.7\linewidth]{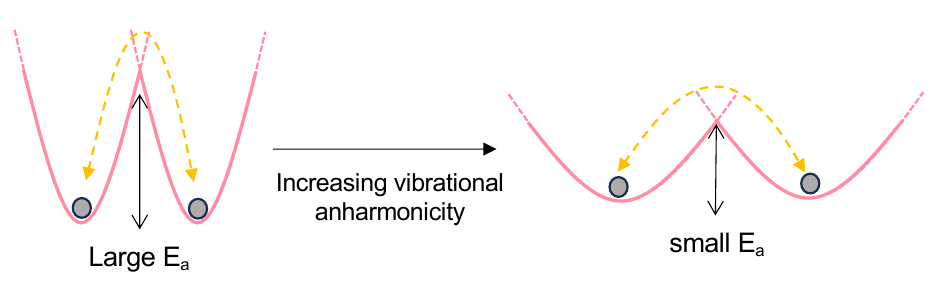}
    \caption{Schematic illustration of how softening of the lattice, which enhances the vibrational anharmonicity, reduces the activation energy $E_a$ for ion diffusions, leading to an enhancement in the corresponding ion diffusivity.}
    \label{fig:anharmonic_diffusion}
\end{figure}

Fundamentally, ions diffuse through a periodic lattice by hopping between two adjacent local minima on the PES [\cref{fig:anharmonic_diffusion}(left)] that are separated by an energy barrier, which is commonly referred to as the activation energy $E_{a}$. Typically, when the mobile ions diffuse through the lattice, they need to `push' away their surrounding atoms by overcoming the repulsive forces with these surrounding atoms to open up the diffusion pathway, such that $E_{a}$ is strongly correlated with the strengths of these repulsive interactions. If one assumes the host lattice is frozen, then the thermal energy is primarily contributing for the mobile ions to overcome the activation energy in achieving the hopping between two adjacent local minima, such that the temperature-dependent ion diffusivity  $D(T)$ can be determined  by the Arrhenius relationship:
\begin{equation}
    \label{eq:arrhenius_eq}
    D(T)=D_0\exp\left(-\frac{E_a}{k_B T}\right).
\end{equation}
However, in reality, the diffusion of the mobile ions are strongly coupled with the vibration of the host lattice to achieve fast ion conduction, in which many different material-specific mechanisms behind such coupling have been proposed.\cite{jun2024diffusion} More specifically, with respect to the above discussions on the anharmonic lattice vibrations, it is commonly understood that anharmonicity flattens the PES, thus reducing the $E_{a}$ to promote fast ion conduction in solids [\cref{fig:anharmonic_diffusion}(right)]. 

Practically, the ion conductivity ($\sigma$) is often the measured experimental quantity that characterises the performance of ionic conductors, which relates to the ion diffusivity via the Einstein relationship $\sigma=nq^2D/k_B T$, in which $n$ is the concentration of the charge carriers and $q$ is the associated charge. From here, it is not difficult to appreciate that, being able to determine the ion diffusivity $D$ will be the key in applying computational methodologies to supplement experimental results in understanding the mechanisms behind fast ion conduction in solids. Computationally, $D(T)$ can be computed from MD simulations at a constant temperature $T$ via the mean-squared-displacements (MSD) of ions over the simulation time:
\begin{equation}
    \label{eq:ion_MSD}
    D(T)=\lim_{t\to \infty}\frac{1}{6Nt}\sum_{i=1}^{N}\left\langle \left\vert \bm{R}_i(t)-\bm{R}_i(0) \right\vert^2\right\rangle.
\end{equation}
In addition, MD simulations are capable of providing a rich set of structural information which is particularly useful for identifying the most likely ionic conduction pathway in materials. This information can be further utilised to conduct further theoretical modelling, for example, with the Nudged Elastic Band (NEB) method, to map out the detailed PES along the ion diffusion pathway,\cite{he2017origin} from which not only the corresponding $E_a$ can be determined, but also obtain more detailed atomistic insights on the atomic interactions that lead to a high/low $E_a$ along a particular ion conduction pathway. Further supplementing MD simulations with harmonic phonon calculations could help identify vibrational modes that couple strongly with the conduction of mobile ions, of which anharmonic vibrations may play a significant role in contributing towards the fast ion conduction in the system.

\subsection{Electronic and Magnetic Properties}

\subsubsection{Electronic Band Structures and Density--of--States}\label{sect:band_structure}

Determining the electronic structures is probably the most important rationale behind the application of the electronic structure methods (DFT and GW) in materials modelling. Fundamentally, it enables the characterisations of materials into metals, semiconductors and insulators as their basic electronic properties, which dictate their practical applications and many other associated physical and chemical properties. For examples, by projecting the electronic eigenfunctions to atomic orbital basis functions, one can obtain further fundamental insights into the origins of electronic states near the Fermi level from the perspective of orbital hybridisations,\cite{dronskowski1993crystal} which also play important roles in dictating the materials' stabilities. For instance, in lead halide perovskites, the distortion of atomic positions along particular phonon modes can lead to redistribution of the electronic charge densities to increase its structural stability,\cite{yang2020composition} which can be viewed as a manifestation of the pseudo-Jahn-Teller effect.\cite{bersuker2013pseudo} The topological information of the electronic band structure can be utilised directly to compute the electronic conductivity tensors for a material subjected to external fields (electric, magnetic, thermal gradients, and so on). The Kohn-Sham eigenenergies and eigenfunctions serve as the critical ingredient for computing the frequency-dependent dielectric function  [\cref{eq:adler-wiser}], which is not only an important step in GW calculation, but also useful for computing the optical spectra of semiconductors.\cite{ramasubramaniam2012large} Finally, the band structure is also an important component in facilitating the understanding of more complex many-body phenomena, such as band-mediated magnetisms,\cite{walsh2008theoretical} in solid state materials.

Formally, the electronic band gap energy $E_g$ of a system should be taken as the energy difference between the LUMO and HOMO states, i.e. $E_g=\varepsilon_{\mbox{\scriptsize{LUMO}}}-\varepsilon_{\mbox{\scriptsize{HOMO}}}$, which, as can be seen in \cref{eq:ionisation_potential} and \cref{eq:electron_affinity}, that, fundamentally, $E_g$ is determined not only by the ground, but also the excited state energies of a many-electron system. However, DFT calculation only gives the ground-state energy of an $N$ electron system with fixed orbital occupation under the mean-field approximation. What enables us to overcome this limitation, and extract the band gap energies from a ground-state DFT calculation is the Janak theorem,\cite{janak1978proof} which proved that $\partial E/\partial c_i=\varepsilon_i$, $i.e.$ the derivative of the Kohn-Sham total energy with respect to the occupation of the $i$-th eigenorbital is the corresponding orbital eigenenergy. The theory was established based on a generalisation of the Kohn-Sham theory, whereby in the electron density $\rho(\bm{r})=\sum_i c_i|\varphi_i(\bm{r})|^2$ [\cref{eq:electron_density_KS_orbital}], one considered the case where the orbital occupation $c_i$ was allowed to be a continuous (fractional) number between 0 and 2 (constraint of the Pauli exclusion principle). With this, the orbital eigenenergy can be written as a function of the orbital occupancy, or $\varepsilon(n_i)$, and $E(N+1)-E(N)=\int_0^1 \varepsilon(n_i)dn_i$. This serves as the fundamental basis to compute the electronic band structures of materials with DFT. Nevertheless, one should note that the problem of mean-field DFT in underestimating $E_{g}$ (\cref{sect:DFA_challenge}) does not invalidate the Janak's theorem, but a problem of approximating the many-body exchange-correlation potential with an one-electron density matrix. The application of post-DFT method, such as the GW approximation, in which the quasiparticle energy now follows the definition of  \cref{eq:linearised_QP_eqn}, should (partially) overcome the limitations in mean-field DFT, leading to more accurate estimation of the electronic energy levels.

Practically, when using the planewave basis functions [\cref{eq:bloch_wave}] to solve the Kohn-Sham equation for solids, one first seeks to construct the basis function on a series of discrete $k$-points in the reciprocal space, either along high-symmetry $k$-point paths for calculating the band structures or evenly spaced within the first Brillouin zone (BZ). At each selected $k$-point, solving the Kohn-Sham equation will give rise to a spectrum of energy eigenvalues $\{\varepsilon_{i\bm{k}}\}$, in which $i$ is often referred to as the band index. The electronic band structure can thus be constructed by connecting $\varepsilon_{i\bm{k}}$ that belong to the same band index with different $k$ by following the  $k$-point paths that were pre-defined based on the crystal symmetry, whereas the electronic density-states (DOS) as a function of electronic energy levels are computed by integrating across all $k$-points in the first BZ as:
\begin{equation}
    \label{eq:electronic_DOS}
    \rho(\varepsilon)=\sum_i\int_{\mbox{\scriptsize{BZ}}} \frac{\mbox{d}\bm{k}}{\Omega_{\mbox{\scriptsize{BZ}}}}\delta(\varepsilon-\varepsilon_{i\bm{k}}).
\end{equation}

\subsubsection{Transport Properties}
The transport of electrons, phonon\cite{chen2018manipulation} and spins\cite{zutic2004spintronics,hirohata2020review} with/without the couplings amongst these three subsystems,  is a critical material property to be understood for developing new (nano)electronic technology through material innovations. The transport phenomena are associated with the dynamics of the carriers under external stimulus (\emph{e.g.} electric and/or magnetic field, temperature gradient) that induces an imbalance in the statistical distributions of the carrier across the sample.\cite{onsager1931reciprocal,onsager1931reciprocal2} It is therefore unsurprising that a plethora of theoretical approaches have been established to model the transport phenomena in different materials across a range of scales. 

In the context of atomistic modellings, for instances, in molecular and nano electronics with a device at the atomistic length scale, the quantum mechanical nature of the carrier plays an important role in determining the corresponding transport behaviours such that the transport phenomena is often interpreted as the time--evolution of the particle wavefunction.\cite{economou2006green} In this regime, the transport phenomena is often modeled with the Green's function approach under either the Landauer--B\"{u}ttiker\cite{buttiker1986four}  or the Kelydsh\cite{jauho1994time} formalism. The former is usually applied for modeling the phase coherent transport.\cite{xue2001charge,datta2005quantum} In contrast, the latter approach is more versatile to include the contributions from many--body  effects in modeling the inelastic transports.\cite{lu2007coupled,galperin2006resonant,liu2006inelastic} It has been combined with DFT for modelling the transport behaviors in realistic atomistic transport junctions,\cite{frederiksen2007inelastic,stradi2016general} but this method is highly specialised and its wide adaption, for example, in high--throughput computational screening is not common.

Charge transports in (crystalline) organic semiconductors, on the other hand, is most often interpreted as the hopping of electrons from one molecule to a neighboring one that are bounded together by the weak van der Waals interactions.\cite{coropceanu2007charge} The probability for the hopping to occur is characterised by the transfer integral $J_{ij}$ between the frontier orbitals of the two neighboring molecules, which itself is strongly correlated with the orbital overlaps $S_{ij}$.\cite{gajdos2014ultrafast} The latter provides a workable heuristic rule that is often applied, for instance, in guiding the design of organic semiconductors with high charge mobilities.\cite{li1998highly,takimiya2011thienoacene,gomez2017hydrogen,campbell2017predicted,yang2018large} Furthermore, the acceptance/donation of a charge carriers by a molecule will change its vibrational energy landscape $E(\bm{R})$ as a function of its atomic coordinates $\bm{R}$, resulted in the additional free energy cost associated with the charge hopping (most commonly referred to as the reorganisation energy $\lambda$). This `localised polaron' model is mathematically formulated in the Markus theory of charge transfer,\cite{marcus1993electron} in which the charge transfer rate between two molecule is given by
\begin{equation}\label{Markus}
\mathcal{W}_{ij}=\frac{|J_{ij}|^2}{\hbar}\sqrt{\frac{\pi}{\lambda k_B T}}\exp\left(-\frac{\lambda}{4k_{B}T}\right).
\end{equation}
The carrier mobility can then be obtained by using \cref{Markus} as inputs for  kinetic Monte Carlo simulations to model the diffusion pathways of charge carriers.\cite{coropceanu2007charge} Despite many known limitations,\cite{shuai2020applying} this model has widely be used in modelling the (upper bound limits of)  carrier mobilities in soft organic materials due to its conceptual simplicity. 

For inorganic solids, where the neighboring atoms are bounded by strong ionic/covalent interactions, their intrinsic  transport properties can be derived from their electronic band structures and/or densities--of--states,\cite{madsen2006boltztrap} which are readily accessible from most electronic structure calculations for periodic solids (for phonon transport, see discussions in  \cref{sec:anharmonic_vibs}). According to the Boltzmann transport theory, the electric current under an external gradient of temperature, electric and magnetic field is:
\begin{equation}\label{eq:boltzmanneq}
    j_i=\sigma_{ij}\bm{\mathcal{E}}_{j}+\sigma_{ijk}\bm{\mathcal{E}}_{j}\bm{\mathcal{B}}_{k}+v_{ij}\nabla_j T+\cdots
\end{equation}
From the topologies of the electronic bands, the group velocities and the tensor of inverse effective masses can be determined:
\begin{equation*}
    \bm{v}_\alpha(i,\bm{k})=\frac{1}{\hbar}\frac{\partial \varepsilon(i,\bm{k})}{\partial \bm{k}_{\alpha}}, \qquad \bm{M}^{-1}_{\alpha\beta}(i,\bm{k})=\frac{1}{\hbar^2}\frac{\partial^2 \varepsilon(i,\bm{k})}{\partial \bm{k}_{\alpha}\partial \bm{k}_\beta}.
\end{equation*}
Under the relaxation time approximation, the band--dependent conductivity tensors are then given by: 
\begin{equation}
\label{eq:e_cond_tensor}
    \bm{\sigma}_{\alpha\beta}(i,\bm{k})=e^2\tau_{i,\bm{k}}\bm{v}_\alpha(i,\bm{k})v_\beta(i,\bm{k}),
\end{equation}
\begin{equation}
\label{eq:hall_cond_tensor}
    \bm{\sigma}_{\alpha\beta\gamma}(i,\bm{k})=e^3\tau_{i,\bm{k}}^2\varepsilon_{\gamma uv}\bm{v}_\alpha(i,\bm{k})v_v(i,\bm{k})\bm{M}^{-1}_{\beta u},
\end{equation}
in which $\varepsilon_{ijk}$ is  the Levi--Civita symbol, and the band--dependent relaxation time $\tau_{i,\bm{k}}$ is usually assumed to be constant in most modelling cases. By further integrating the conductivity tensors over $\bm{k}$--points with a given energy, the density--of--states projected conductivity tensors $\bm{\sigma}_{\alpha\beta}(\varepsilon)$ and $\bm{\sigma}_{\alpha\beta\gamma}(\varepsilon)$ can be determined, from which the temperature and chemical potential dependent conductivity tensors [c.f. \cref{eq:boltzmanneq}] can be calculated as
\begin{subequations}
\begin{align}
&\mbox{electronic conductivity:}\nonumber  \\  &\sigma_{\alpha\beta}(T,\mu)=\frac{1}{\Omega}\int\bm{\sigma}_{\alpha\beta}(\varepsilon)\left[-\frac{\partial f_{\mu}(T,\varepsilon)}{\partial \varepsilon}\right]\mbox{d}\varepsilon, \label{electronic_cond}\\
&\mbox{electronic energy conductivity:}\nonumber\\     &\nu_{\alpha\beta}(T,\mu)=\frac{1}{eT\Omega}\int\bm{\sigma}_{\alpha\beta}(\varepsilon)(\varepsilon-\mu)\left[-\frac{\partial f_{\mu}(T,\varepsilon)}{\partial \varepsilon}\right]\mbox{d}\varepsilon, \label{electronic_energy_cond}\\
&\mbox{electronic thermal conductivity:}\nonumber\\    &\kappa_{\alpha\beta}^{e}(T,\mu)=\frac{1}{e^2T\Omega}\int\bm{\sigma}_{\alpha\beta}(\varepsilon)(\varepsilon-\mu)^2\left[-\frac{\partial f_{\mu}(T,\varepsilon)}{\partial \varepsilon}\right]\mbox{d}\varepsilon,\label{electronic_th_cond}\\
&\mbox{Hall conductivity:}\nonumber\\    &\sigma_{\alpha\beta\gamma}(T,\mu)=\frac{1}{\Omega}\int\bm{\sigma}_{\alpha\beta\gamma}(\varepsilon)\left[-\frac{\partial f_{\mu}(T,\varepsilon)}{\partial \varepsilon}\right]\mbox{d}\varepsilon, \label{Hall_cond}
\end{align}
\end{subequations}
where $f_{\mu}(T,\varepsilon)$ is the Fermi--Dirac distribution function. In particular, the total thermal conductivity of a given material can be obtained as a sum of its electronic thermal conductivity [\cref{electronic_th_cond}] and lattice thermal conductivity, providing a theoretical basis for the computational discoveries of thermoelectric materials.\cite{yang2008evaluation,hong2016full,zhang2017titanium,wang2017exceptional}

It can be seen that the above approach based on the Boltzmann transport equations establishes a nice link to bridge atomistic electronic structure simulation and the transport properties at the macroscopic limit. It avoids the complexity in considering the `split--over' of the electronic wavefunctions from the contact into the channel, which requires partitioning the systems into contact and channel regions.\cite{datta2005quantum} Although in its lowest level of approximations, detailed carrier scattering mechanisms (such as electron--phonon and impurity scatterings) are ignored, good agreements between the theoretically and experimentally determined transport characters had been demonstrated for materials such as  PbTe,\cite{singh2010doping} SnSe,\cite{zhao2016ultrahigh} Mg$_3$Sb$_{1.5}$Bi$_{0.5}$,\cite{zhang2017high} as well as a selected set of elemental and binary compound semiconductors.\cite{ricci2017ab}  Recent advances in material theory and availability in computational resources have seen more elaborated methods being developed and applied for determining the carrier relaxation times contributed by different scattering from first principles, enabling a more accurate prediction in the transport properties of exotic materials.\cite{krishnaswamy2017first,ganose2021efficient,jung2021prediction}

\subsubsection{Polarisations and Dielectric Properties}
The dielectric response in a semiconducting or insulating material arises when the electric charges in the material cannot be effectively drained away when an external electric field $\bm{E}$ is applied across the material. Instead, the electronic and ionic charges in the material are redistributed, inducing an internal polarisation to counter--balanced the external field. The strength of the (frequency--dependent) dielectric response $\bm{D}(\omega)$ is correlated with the electric field  as:
\begin{equation}\label{dielectric_reponse}
    \bm{D}(\omega)=\varepsilon(\omega)\bm{E}(\omega),
\end{equation}
where  $\varepsilon(\omega)$ is the tensor of dielectric permittivity that is an intrinsic property of the material. Generally for a complex material, its dielectric response is contributed by multiple different mechanisms, including (in the order of decreasing magnitudes) space--charge ($\varepsilon_{sc}$), dipole reorientation ($\varepsilon_{dip}$), electronic (or optical, $\varepsilon_{elec}$) and ionic (or phonon, $\varepsilon_{ion}$) contributions. Based on the \emph{modern theory of polarisation},\cite{spaldin2012beginner} algorithms for calculating the latter two quantities in condensed matters have been implemented in many modern electronic structure codes. Therefore, the discussions in this section will be focusing on the electronic and ionic contributions to the polarisations.

\paragraph{\textbf{Electronic polarisation}}

In molecular systems, the dipole moment for an solated molecule is trivial to define and understand.\cite{stone2013theory} For example, a simple linear molecule of carbon monoxide (CO) is said to carry a \textit{permanent dipole} moment because of the difference in the electronegativity between the C and O atom, leading to a partial point charge of $+\delta$ and $-\delta$ to be possessed by each atom, respectively. The corresponding dipole moment for CO is therefore $\bm{\mu}=\delta\mathbf{r}/|\mathbf{r}|$, where $\mathbf{r}$ is the vector joining the two nuclei in the molecule. Within this point charge model, molecules such as O$_2$ is said to be nonpolar for its homoatomic nature. If we further take  the quantum mechanical nature of the electrons into account, then we need to think of the electrons in the molecules as a spatial distribution of charge densities around the nuclei, regardless of whether the molecule is polar or nonpolar. Firstly, when an electric field is applied to the molecule, an \textit{induced dipole} on the molecule is generated resulted from the distortion in the shape of the charge density as a response to the external field. Secondly, a molecule always possess an \textit{instantaneous dipole} due to the non--static nature of the electron. The constant spatial movements of electrons causes a fluctuation in the spatial distribution of the electronic charge density with respect to the positions of the nuclei, meaning at any time instance $t$, there will be some finite net charge of $\delta_{t,i}$ associated with the $i$--th atom in the molecule, thus gives rise to a dipole moment in the molecule. Further considering the electronic correlation effects, the fluctuations of the electron densities between two nearby molecules\cite{dobson2001prediction} or nanostructures\cite{tawfik2018evaluation} are correlated, which contributes to the important London dispersion interactions\cite{stone2013theory} between two non--covalently bounded entities.

However, such a simple exercise in counting the ionic charges and interatomic distances to determine the dipole polarisations fails for extended solids,\citep{spaldin2012beginner} which leads to the electronic polarisation become multi--valued that is dependent on the choice of the lattice vector. This problem can be overcome by making a connection to the experiments, in which the \emph{change in the polarisation} $\Delta\mathbf{P}_{elec}$ between the initial  ($\lambda_i$) and final ($\lambda_f$) state in a sample, defined as\cite{resta1992theory}
\begin{equation}\label{polarisation_change}
\Delta\mathbf{P}_{elec}=\int_{\lambda_i}^{\lambda_f}\frac{\partial \mathbf{P}_{elec}}{\partial \lambda}d\lambda,
\end{equation}
is measured as an electric current flowing through a closed circuit connecting the two ends of the sample. Here, $\lambda$ is the coordinate in the phase space that can be spanned by time, ionic coordinates, external electric and/or magnetic fields, \emph{etc}, that drive the change in $\mathbf{P}_{elec}$ along a path that goes from $\lambda_i$ to $\lambda_f$ in the phase space. For simplicity here, we consider $\lambda$ is represented by the time variable, such that $\frac{\partial \mathbf{P}_{elec}}{\partial \lambda}=\frac{\partial \mathbf{P}_{elec}}{\partial t}$. This means now, in order to obtain the change of polarisation over time, we need to take into account the time--dependent evolution of a wavefunction $\psi(\bm{r}(t))$ under a time--dependent Hamiltonian $\mathcal{H}(\bm{r}(t))$, in which $\partial \bm{P}/\partial t$ can be correlated with the change in the electronic wavefunction over $\Delta t$, and the latter can be obtained from the perturbation theory as an expansion involving the operator $\partial \mathcal{H}(\bm{r}(t))/\partial t$. However, note that in the ground state electronic structure theory, \emph{e.g.} DFT, the time--dependency is not included. This problem is overcome under the theory of Berry phase with the recognition of the chain rule for differentiation $\partial \mathcal{H}(\bm{r}(t))/\partial t=\bm{\nabla}_{\bm{r}}\mathcal{H}\cdot\dot{\bm{r}}$, such that \cref{polarisation_change} can be cast into a form that is  only dependent on the atomic coordinates $\bm{r}$ but not time. For periodic DFT, we further take into account the KS orbitals in the form of the Bloch wave $\psi_{n,k}(\bm{r})=e^{i\bm{k}\bm{r}}u_{n,k}(\bm{r})$, we arrive at the  following workable definition of electronic polarisation that depends only on the occupied electronic states:
\begin{equation}\label{electron_pol_DFT}
\bm{P}_{elec}=\frac{e}{(2\pi)^{3}}\int_{\bm{k}\in BZ}\mathcal{A}(\bm{k})\mbox{d}\bm{k},
\end{equation}
in which $\mathcal{A}(\bm{k})=-i\langle u_{n,k}|\bm{\nabla}_{\bm{k}}|u_{n,k}\rangle$ is the so--called Berry curvature.  Based on \cref{dielectric_reponse}, the corresponding electronic dielectric tensor can be obtained from:\cite{souza2002first}
\begin{equation}\label{eq:electronic_dielectric_tensor}
    \varepsilon^{\alpha\beta}_{elec}(\omega)=\delta_{\alpha\beta}+4\pi\frac{\partial \bm{P}_{\alpha}}{\partial \bm{E}_{\beta}},
\end{equation}
implying that to determine $\varepsilon_{elec}(\omega)$ algorithmically, the polarisation tensor needs to be evaluated under the application of different electric field, from which the partial derivative in \cref{eq:electronic_dielectric_tensor} can be evaluated. 

\paragraph{\textbf{Ionic polarisation}} The above discussions has highlighted the approach to calculate the electronic polarisation in a material with fixed ionic positions $\bm{R}_{0}$, such as the equilibrium structure that corresponds a local minimum on the PES. On the other hand, the ionic polarisation $\bm{P}_{ion}$ reflects the change in the system polarisation resulted from the variation in the  charge density distribution when the ions change their positions from $\bm{R}_{0}$ to $\bm{R}$ along a particular normal mode $\bm{\mathcal{Q}}$ of lattice vibrations. This is the case because under the Born--Oppenheimer approximation,  the electron density is assumed to change its shape instantaneously with respect to the change in the ionic positions. It also implies that evaluating $\bm{P}_{ion}$ requires the knowledge of the vibrational eigenmodes and eigenfrequencies $\{\bm{\mathcal{Q}}_n,\omega_n\}$, together with  $\bm{P}_{elec}$ to be determined for different $\bm{R}$. 

To map the above interpretation back to a more classical picture of dipole moments (\textit{c.f.} the discussions about molecular dipoles), the Born effective charge tensor on ion $i$ in a solid is defined\cite{ghosez1998dynamical} as the change in the polarisation along the $\alpha$--Cartesian direction when the ion is displaced along the normal mode in the $\beta$--direction at zero field:
\begin{equation}\label{eq:born_charge}
    Z_{i,\alpha\beta}^{*}=\left.\frac{\Omega}{|e|}\frac{\partial \bm{P}_{\alpha}}{\partial \bm{\mathcal{Q}}_\beta^{(i)}}\right\vert_{E=0},
\end{equation}
where $\Omega$ is the unit cell volume. With this, the ionic contribution to the polarisation in a crystal unit cell can be calculated as:
\begin{equation}\label{eq:ionic_pol}
    \Delta\bm{P}_{ion}=\frac{|e|}{\Omega}\sum_{i}\bm{Z}_{i}^{*}\bm{\mathcal{Q}}^{(i)},
\end{equation}
where the summation is performed all ions in the unit cell. The ionic contribution to the tensor of dielectric permittivity is then given by:\cite{gonze1997dynamical}
\begin{equation}\label{eq:ionic_epsilon}
     \varepsilon_{ion}^{\alpha\beta}(\omega)=\frac{4\pi}{\Omega}\sum_{n=1}^N\frac{\left[\sum_{s\alpha^\prime}Z_{s,\alpha\alpha^\prime}^{*}\bm{\mathcal{Q}_n^{*}}\right] \left[\sum_{s^\prime\beta^\prime}Z_{s^\prime,\beta\beta^\prime}^{*}\bm{\mathcal{Q}}_n\right]}{\omega_n^2-\omega^2},
\end{equation}
and the index $n$ runs over all phonon eigenstates. The concept of the Born effective charge has been a particularly useful property indicator to explain and screen for the emergence of spontaneous polarisation in materials, such as $d^0$ transition metal perovskites,\cite{matthias1949new} in which the $d^0$ B--site cation exhibits an anomalously large Born effective charge compared to its formal charge.\cite{armiento2011screening}

\subsubsection{Electron--Phonon Interactions}\label{sec:eph_coupling}
Apart from the electron-electron [\cref{sect:GW_approximation}] and phonon-phonon [\cref{sec:anharmonic_vibs}] interactions, the electron-phonon interactions (EPI) give rise to another many-body quantum effect that plays an important role in dictating the physical properties of materials. The fundamental concept of electronic band structures [\cref{sect:band_structure}] is formulated based on the Bloch's theorem which considers the electronic wavefunction in a solid is modulated by the ionic potential $V(\bm{R})$ that is perfectly periodic in space, such that electrons moving in this potential should experience no scattering from the lattice. It is apparent that such an idealisation breaks down at finite temperature, as the phonon vibrations in the crystal will deform the crystal potential, as $\Delta V(\bm{R}_i)\sim \Delta \bm{R}_i\nabla_{\bm{R}}V(\bm{R}_{i})$ to the first-order approximation, which contributes towards the coupling of electrons with acoustic phonons via the deformation potential.\cite{khan1984deformation} Particularly, in noncentrosymmetric crystals where the long-range cooperative acoustic phonon vibrations can also generate electric fields due to the piezoelectric effect that couples with the electrons, leading to the piezoelectric EPIs.\cite{zook1964piezoelectric}   Finally, the oscillating electric dipole moments induced by the optical phonon vibrations in polar crystals can also couple with electrons, leading to the so-called Fr\"{o}hlich interaction\cite{frohlich1954electrons} being another important mechanism behind EPI in materials. 

Formally, the Hedin-Baym equations,\cite{hedin1965new,baym1961field} which are a set of integro-differential equations derived from the many-body quantum field theory, provided the most sophisticated description of the EPI effect in solids to-date, with harmonic phonons being the only assumption.\cite{giustino2017electron} In a high-level view, the Hedin-Baym equations take into account the  mutual effects of EPIs which alter the dynamics of both the electronic and phononic subsystems via the electronic ($\Sigma_{\mbox{\scriptsize{eph}}}$) and phononic ($\Pi_{\mbox{\scriptsize{eph}}}$) self-energies, whereby the fluctuations in the lattice potentials introduce additional Coulomb screening on the electrons, which in terms also modifies the phonon  frequencies and lifetimes by changing the nuclear charge densities and nuclear-nuclear interactions. 

The complexity of the coupled integro-differential equations prevents the Hedin-Baym equations from being fully solved to obtain the self-consistent solutions for the energy shifts and lifetimes of the electronic and phononic subsystems with the presence of EPI. Nevertheless, the key physical quantity that determines the self-energies $\Sigma_{\mbox{\scriptsize{eph}}}$ and $\Pi_{\mbox{\scriptsize{eph}}}$ is the frequency- and momentum-dependent EPI strength $g_{mn\nu}(\bm{k},\bm{q})$, which couples the electronic Bloch states $|\psi_{m}(\bm{k}+\bm{q})\rangle$ and $|\psi_{n}(\bm{k})\rangle$ through a phonon vibration mode $\bm{\mathcal{Q}}_{\bm{q},\nu}$, where $m$, $n$ and $\nu$ are the band indicies and $\bm{k}$, $\bm{q}$ are the electronic and phonic wavevectors, respectively. The increasing computational power provided by the modern HPC facilities has made the computing $g_{mn\nu}(\bm{k},\bm{q})$ become readily accessible with major solid-state DFT codes,\cite{noffsinger2010epw,ponce2016epw,karsai2018electron} enabling us to gain critical insights into this many-body quantum effects through computational modellings. Here, we shall briefly account for the principles behind the two major state-of-the-art approaches to compute the EPI strengths in solids, which  follow directly  the real and reciprocal space formalism for harmonic phonon calculations as outlined in \cref{sect:phonon}. 

In the \emph{perturbative approach} based on DFPT,\cite{giustino2017electron} one works directly with the definition of the EPI matrix element
\begin{equation}
    \label{eq:epi_matrix_element_dfpt}
    g_{mn\nu}(\bm{k},\bm{q})=\left\langle\psi_{m}^{\mbox{\scriptsize{KS}}}(\bm{k}+\bm{q})\Bigg\vert \frac{\partial V^{\mbox{\scriptsize{KS}}}}{\partial \bm u_{\nu,\bm{q}}} \Bigg\vert  \psi_{n}^{\mbox{\scriptsize{KS}}}(\bm{k})\right\rangle,
\end{equation}
in which the computation of the perturbed KS potential $\partial V^{\mbox{\scriptsize{KS}}}$ follows the discussions in \cref{sect:phonon}. As discussed previously, DFPT would enable the direct computation of EPI strengths in which the contributions from long-wavelength ($\bm{q}\to0$) phonons are significant, without the need to use a large simulation supercell, as in the real-space \textit{finite difference} approach.\cite{monserrat2018electron} However, the later provides the additional flexibility whereby the lattice dynamics and electronic structures may be computed with two different levels of theories. In its essence, one recast \cref{eq:epi_matrix_element_dfpt} into a finite-difference form 
\begin{equation}
    \label{eq:epi_matrix_element_finite_diff}
    g_{mn\nu}(\bm{k},\bm{q})=\left\langle\psi_{m}^{\mbox{\scriptsize{KS}}}(\bm{k}+\bm{q})\Bigg\vert \frac{ V^{\mbox{\scriptsize{KS}}}(\Delta \bm u_{\nu,\bm{q}}) -V^{\mbox{\scriptsize{KS}}} (0) }{\Delta \bm u_{\nu,\bm{q}}} \Bigg\vert  \psi_{n}^{\mbox{\scriptsize{KS}}}(\bm{k})\right\rangle,
\end{equation}
which suggests that, under this formulation, one is required to perform separated calculations of the KS eigenstates at both the equilibrium structure (labelled as 0) and a series of  configurations that are generated by perturbing the equilibrium structure along different vibrational eigenstates $\{\bm{u}_{\nu,\bm{q}}\}$.

This approach tights up naturally with the core concept of PES sampling that underpins the current review. As pointed out by \citet{monserrat2018electron}, if one was interested in the change of some electronic property $\mathcal{A}$ (such as the band energy levels $\varepsilon_{n,\bm{k}}$ and band gap $E_g$) induced by the phonon vibration at temperature $T$ in general, without  consideration of the contributions from the specific vibrational modes or electronic states, we can first express $\mathcal{A}$ as a Taylor expansion of the vibrational coordinates, in the spirit of \cref{eq:higher_order_vib_Hamiltonian}:
\begin{equation}
    \label{eq:property_expand_vib_Hamiltonian}
    \mathcal{A}(\bm{R})=\mathcal{A}(0)+\sum_{\bm{q},\nu}\frac{\partial \mathcal{A}(0)}{\partial\bm{u}_{\bm{q},\nu}}\bm{u}_{\bm{q},\nu}+\sum_{\bm{q},\nu}\sum_{\bm{q}^\prime,\nu^\prime}\frac{\partial^2\mathcal{A}(0)}{\partial\bm{u}_{\bm{q},\nu}\partial\bm{u}_{\bm{q}^\prime,\nu^\prime}}\bm{u}_{\bm{q},\nu}\bm{u}_{\bm{q}^\prime,\nu^\prime}+\mathcal{O}(\bm{u}^3),
\end{equation}
which reflects how EPI arises from both harmonic and anharmonic phonons change the electronic behaviours of a material. If one truncates \cref{eq:property_expand_vib_Hamiltonian} at the harmonic term and taking the thermal average [in line with \cref{eq:general_sampling_2}], then, for example, the temperature-dependent shifts in the electronic state eigenenergies due to EPI at temperature $T$ will take the following form of Bose-Einstein averaging, leading to the famous Allen-Heine-Cardona (AHC) theory:\cite{allen1976theory,allen1981theory}
\begin{equation}
    \label{eq:electronic_state_epi_thermal_av}
    \varepsilon_{n\bm{k}}(T)=\varepsilon_{n\bm{k}}(0)+\sum_{\bm{q},\nu}\frac{1}{\omega_{\bm{q},\nu}}\frac{\partial^2 \varepsilon_{n\bm{k}}}{\partial \bm{u}^2_{\bm{q},\nu}}\left[\frac{1}{2}+\frac{1}{\exp(\hbar\omega_{\bm{q},\nu}/k_B T)-1} \right]
\end{equation}
in which $\varepsilon_{n\bm{k}}(0)$ is the electronic energy levels computed at 0 K. This provides an intuitive and semi-quantitative measure of the strength of EPI in semiconductors, as the shift of $\varepsilon_{n\bm{k}}$ (or equivalently, $E_g$) is directly proportional to $\frac{\partial^2 \varepsilon_{n\bm{k}}}{\partial \bm{u}^2_{\bm{q},\nu}}$. In addition to the physical origins of EPI as discussed above, the temperature-dependent variations in the electronic energy levels and band gaps can be understood as the results of modulating the orbital hybridisations between the bonding atoms in the solids due to the changes in the interatomic distances and angles from phonon vibrations. 

Another advantages of adopting the real-space approach is the capability to compute the anharmonic contributions to the EPIs by directly working with \cref{eq:property_expand_vib_Hamiltonian}. For example, \citet{saidi2016temperature} shows that the AHC theory based on harmonic phonon is insufficient to correctly capture the temperature-dependency of $E_{g}$ for  methylammonium lead iodide perovskite due to its strong vibrational anharmonicities. This can be achieved by performing AIMD simulations under the $NPT$ ensemble, one can access to a collection of time-dependent MD trajectories (frames) $\{\bm{R}(t)\}_T$ that correspond to the  configurations accessible on the fully anharmonic PES at a the target temperature $T$. By further computing the corresponding band gap energies on these MD frames, it becomes possible to retrieve the time-dependent evolutions of band gap energies $\{E_{g}(t)\}_T$ at the target temperature. From here, the expectation value of the band gap energies $\langle E_g(t)\rangle_T$ can be considered as the anharmonic equivalent to \cref{eq:electronic_state_epi_thermal_av}, providing information that is analogous to the real part of $\Sigma_{\mbox{\scriptsize{eph}}}$. At the same time, by computing the band gap dephasing function from $\{E_{g}(t)\}_T$ would further enable us to extract the band gap dephasing time due to the presence of EPI, which can be considered as a proxy for the imaginary part of $\Sigma_{\mbox{\scriptsize{eph}}}$.\cite{yang2020correlating,li2018time,li2019anharmonicity,hwang2004analysis}  

Apart from contributing to the temperature-dependent variation in $E_g$, EPI manifests itself in a wide range of other physical properties of materials, for which computational approaches is playing a significant role in facilitating the investigations of these phenomena. One of the most well-known effects is the phonon-assisted interband optical absorptions in the indirect band gap semiconductors, with Si being the most representative case,\cite{noffsinger2012phonon} which forms the basis of modern microelectronics and photovoltaics. In polar semiconductors, such as ZnO,\cite{teke2004excitonic} the photoexcited electron-hole pair can interact strongly via Coulomb interactions, forming quasiparticle called exciton, which can couple strongly to the longitudinal optical phonons that menifest in the appearance of multiple phonon sidebands in the optical spectra.

By solving Boltzmann transport equation [\cref{eq:e_cond_tensor}], taking into account the modification of the electron relaxation time due to EPI ($1/\tau\sim |g|^2)$, in which the EPI matrix elements $g$ was determined based on the Fermi-Golden rule, \citet{li2015electrical} demonstrated good agreement between the theoretically calculated and experimentally determined electron mobilities for semiconducting Si, metallic Al and 2D MoS$_2$, which can be considered a seminal work that providing a universal framework to include EPI in transport modellings for materials.  \citet{borysenko2010first} applied the DFPT scheme to compute the EPI matrix elements for graphene, which was further combined with the Boltzmann transport equation to demonstrate that both the acoustic and optical phonons contributed to the scattering of electrons at low temperature, whereas at the elevated temperature, the scatterings become dominated by the high-energy optical phonons. \citet{himmetoglu2014first} examined the strong contributions from optical phonon scattering to the electrical conductivities in \ce{SrTiO3}. Lately, \citet{mirjolet2021electron} provided the theoretical analysis on the $\rho(T)\sim  T^2$ dependency of resistivity for the strongly correlated perovskite of \ce{SrVO3}. Their investigation revealed that, in addition to the strong electron-electron coupling that renormalises the electron effective mass, the electronic band structure exhibits a cylindrical Fermi surface, in which a large portion of the Fermi surface can couple to a single phonon mode. This gives rise to polaron transport in the material due to strong EPI. 

EPI also plays the pivotal role in understanding the origins and predicting future candidate materials of high-temperature,  phonon-mediated superconductivity. It represents a type of superconductivity in materials (such as \ce{MgB2}\cite{joas2002theory} and FeSe/\ce{SrTiO3} interface\cite{yang2024phonon}), whereby the formation of electron pairs is mediated by the attractive polarisation fields that are generated by the lattice vibrations. The Migdal-Eliashberg theory,\cite{Migdal1958,Eliashberg1960} which extended the Bardeen-Cooper-Schrieffer (BCS) theory of super-conductivity\cite{Bardeen1957a,Bardeen1957b} that accounts for strong EPI. Upon the formation of electron pairing that contributes to the superconductivity in metals, its electronic DOS opens up a gap $(\Delta_{\mbox{\scriptsize{SC}}}$, commonly known as the superconducting gap. As the temperature rises beyond $T_c$, it closes the $\Delta_{\mbox{\scriptsize{SC}}}$, which destroys the electron pairing and the material returns back to its normal non-superconducting state. Based on this theory, McMilliam and Allan\cite{mcmillan1968transition,allen1975transition} established the following formula:
\begin{equation}
    \label{eq:mcmillan_Tc}
    k_B T_c=\frac{\hbar\omega_{\mbox{\scriptsize{log}}}}{1.2}\exp\left(-\frac{1.04(1+\lambda)}{\lambda-\mu^* (1+0.62\lambda)}\right),
\end{equation}
which provided the basic computational framework for determining the superconducting critical temperature $T_c$ for phonon-mediated superconductivity. Here, $\lambda$ is a material-specific parameter that is dependent on the frequency and momentum-dependent electron-phonon coupling strength, which can be determined from the isotropic Eliashberg function:\cite{allen1972neutron}
\begin{equation}
    \label{eq:eliashberg_function}
    \alpha^2 F(\omega)=\frac{1}{N_F}\int\frac{\mbox{d}\bm{k}\mbox{d}\bm{q}}{\Omega_{\mbox{\scriptsize{BZ}}}^2}\sum_{mn\nu}|g_{mn\nu}(\bm{k},\bm{q})|^2\delta(\varepsilon_{n\bm{k}}-\varepsilon_F)\delta(\varepsilon_{m,\bm{k}+\bm{q}}-\varepsilon_F)\delta(\omega-\omega_{q\nu}).
\end{equation}
It describes the modification of the electron and phonon density-of-states due to the presence of strong EPI in the materials system. From here, $\lambda=2\int_0^\infty\frac{\alpha^2 F(\omega)}{\omega}d\omega$, and the weighted logarithmic phonon frequency $\omega_{\mbox{\scriptsize{log}}}=\exp\left[\frac{2}{\lambda} \int_0^\infty\frac{\alpha^2 F(\omega)}{\omega}\log\omega d\omega\right]$. It should be immediately clear from \cref{eq:eliashberg_function} why strong EPI is prone to be observed in systems with cylindrical Fermi surfaces. Theoretically speaking, all ingredients to compute the spectral function with \cref{eq:eliashberg_function} are accessible with modern electronic structure theory. Practically, however, performing the integration would require sampling dense sets of electron and phonon wavevectors in the Brillouin zone, which can be seen as the main bottleneck in applying this method for studying materials with large and complex crystal structures.

Finally, the effect of EPI on the behaviour of the phonon subsystem should not be overlooked.\cite{giustino2017electron} Fundamentally, EPI contributes to the (first-order) phonon self-energy as
\begin{equation}
    \label{eq:phonon_self_energy}
    \Pi_{\bm{q}\nu}(\omega)=\frac{2}{N_{k}}\sum_{nm\bm{k}}\frac{|g_{nm\nu}(\bm{k},\bm{q})|^2[f(\varepsilon_{n\bm{k}})-f(\varepsilon_{m,\bm{k}+\bm{q}})]}{\hbar\omega+\varepsilon_{n\bm{k}}-\varepsilon_{m,\bm{k}+\bm{q}}+i\eta},
\end{equation}
in which $f(\cdot)$ is the Fermi-Dirac distribution function and $\eta\to 0^+$. From here, the shift in the phonon frequency from EPI can be computed as $\tilde{\omega}_{\bm{q}\nu}^2=\omega_{\bm{q}\nu}^2+2\omega_{\bm{q}\nu}\mbox{Re}\Pi_{\bm{q}\nu}(\tilde{\omega}_{\bm{q}\nu})$, and finite phonon lifetime $\gamma_{\bm{q}\nu}=-2\mbox{Im}\Pi_{\bm{q}\nu}(\tilde{\omega}_{\bm{q}\nu})$. It can be appreciated that when the effect of EPI on the phonon subsystem is significant, it can also lead to notable modifications on the lattice thermal conductivities of materials. A notable example of the effect of EPI on the phonon subsystem is the Kohn anomaly,\cite{kohn1959image} reflected as a sharp softening of phonon frequency. This effect is considered to be strong in 2D materials which exhibit a cylindrical Fermi surface and strong divergence of dielectric screening at the nesting wavevectors.\cite{zhu2015classification} For example, \citet{piscanec2004kohn} provided the theoretical analysis on the Kohn anomalies that are observed in graphite, in which they computed the phonon frequencies with DFPT at different levels of Fermi smearing, which affects the vertical transition probabilities between the occupied and virtual states, that can be considered as a manifestation of the EPI effect on the electronic subsystem.

\subsubsection{Magnetic Properties}
\label{sec:mag_prop}
Magnetic materials can be found across a wide range of modern technologically important applications, such as data storage and spintronic devices,\cite{zutic2004spintronics} quantum computing,\cite{de2021materials} energy conversions,\cite{kitanovski2020energy} and sensors.\cite{gloag2019advances}  Fundamentally, the ground-state magnetic structures of materials play a deterministic role in how the materials may be applied in functional magnetic devices, thus determining the magnetic ground states of solids, and how the interactions amongst spins, electrons and orbitals stabilise these magnetic states, are the key goal for applying electronic structure theory, particularly DFT, to study the magnetic properties of materials. From the perspective of energy landscape, this introduces additional degrees of freedom in the PES $E(\{\bm{R}_i\})\to E(\{\bm{R}_i\},\{\bm{s}_i\})$, in which now the energy of the atomic structure is not only dependent upon its ionic positions $\{\bm{R}_i\}$, but also the corresponding configurations of the net electron spin on the individual ion $\{\bm{s}_i\}$ in the structure. This implies that, for a fixed set of $\{\bm{R}_i\}$,  we will also need to sample different spin configurations $\{\bm{s}_i\}$, in account for this extra degree of freedom, to determine the magnetic ground state of a given material.

\begin{figure}[tb]
    \centering
    \includegraphics[width=\linewidth]{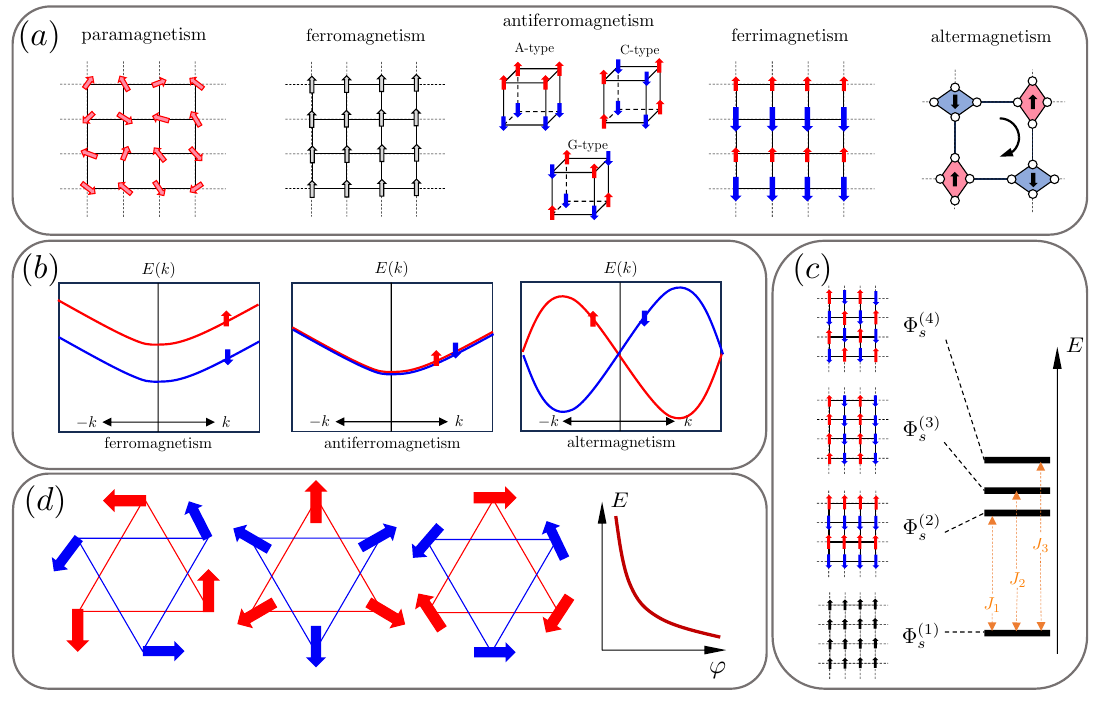}
    \caption{Landscape of magnetic states in solids. (a) Five types of commonly known colinear magnetic states, illustrating the spatial spin configurations in a simple square/cubic lattice. (b) Reciprocal-space (band-structural) interpretations of ferromagnetism, antiferromagnetism and altermagnetism in solids. (c) The energy landscape of different magnetic states in a given solid, in which the energetic differences between spin configurations can be used to infer the exchange coupling constants $J$. (d) Illustration of non-colinear spin configurations in a hexagonal lattice, in which the total energy of the system now varies as a function of the canting angle between the two neighbouring interacting spins. }
    \label{fig:mangetic_landscape}
\end{figure}

As such, to better understand the role of modern DFT in studying the magnetic properties of materials, it is important to first briefly account for the common types of magnetic states in materials. Broadly speaking, there are two categories of magnetic states in condensed matters, which, as to be discussed later, also leads to two different approaches on how spin densities are treated in DFT: (a) 
\textit{colinear magnetism} [\cref{fig:mangetic_landscape}(a)], in which the spin moments are all aligned along a fixed global axis (typically the $z$-axis), and (b) \textit{non-colinear magnetism} [\cref{fig:mangetic_landscape}(d)], in which spins are not restricted to be aligned with a given axis, but still adopt a distinct spiral/helical texture.

In the case of colinear magnetism, three different types of spin configurations are most well known from basic solid state physics:\cite{kittel2018introduction} (i) \textit{Ferromagnetism}, in which all neighbouring spins are parallel aligned with each other, giving rise to a net magnetic moment in the material. (ii) \textit{Antiferromagnetism}, in which the system exhibits no net magnetisation, since the total magnetic moment from the up-spin sublattice is perfectly compensated by that from the down-spin sublattice. Furthermore, depending on the relative spatial relationship between the up- and down-spin sublattices, the antiferromagnetic state can be further categorised as A-, C- or G-type, which can be energetically distinguished using computational techniques.\cite{horton2019high,chang2021electronic,kassimi2023electronic} (iii) \textit{Ferrimagnetism}, which is similar to antiferromagnetism except that the oppositely aligned magnetic moments are of different magnitudes, leading to an overall net non-zero magnetisation in the material. In  these three types of colinear magnetism, the spin-spin interactions can all be described by the Heisenberg exchange Hamiltonian $\hat{\mathcal{H}}=\sum_{ij}J_{ij}\bm{s}_i\cdot\bm{s}_j$. In terms of the electronic band structure theory [\cref{sect:band_structure} and \cref{fig:mangetic_landscape}(b)], ferromagnetic materials are typically characterised by non-degenerate spin-up and spin-down dispersion curves in the reciprocal space, whereas for antiferromangetic materials, the spin-up and spin-down energy bands are degenerated across the entire reciprocal space. Finally, (iv) \textit{altermagnetism} has recently attracted\cite{vsmejkal2022beyond,fender2025altermagnetism} many research interests as an unconventional colinear magnetism, although the phenomenon was first generally identified some thirty years ago.\cite{solovyev1997magneto} It typically appears in non-centrosymmetric crystals whereby two ferromagnetic sublattices are structurally related to each other by a rotational operation and become antiferromagnetically coupled to each other, hence the materials appear to be antiferromagnetic overall. In terms of the band structure theory, altermagnet manifests itself with the reversal of band energy ordering upon the inversion of the linear momentum, \textit{i.e.} $E(\bm{k},\bm{s}_i)=-E(-\bm{k},\bm{s}_i)$ [\cref{fig:mangetic_landscape}(b)], which necessitates the introduction of a linear-momentum-dependent exchange term in the Heisenberg Hamiltonian to describe altermagnetism, \textit{i.e.}, $\hat{\mathcal{H}}=\sum_{ij}J_{ij}(\bm{k})\bm{s}_i\cdot\bm{s}_j$.

Without considering the contributions from crystal structures, spin-orbit coupling in heavy elements is regarded as one of the prominent sources of non-colinear magnetism in materials. The spin-orbit coupling is described by the Hamiltonian $\hat{\mathcal{H}}=\lambda(\bm{L}\cdot\bm{s})$, which couples the orbital angular momentum $\bm{L}$ with the spin magnetic moment $\bm{s}$. This makes the spin orientation become tied to the direction of electronic motion around the nucleus, resulting in spin canting relative to each other. Non-linear magnetic states are also observed in 120$^\circ$-oriented triangular lattices,\cite{sakuma2002magnetic,wang2008spin,yin2022topological} which favour the spin configurations in which neighbouring spins pointing 120$^\circ$ away from each other and is an example of non-colinear magnetism that is induced by geometric frustration. As shown in \cref{fig:mangetic_landscape}(d), the energy of this system becomes a function of the canting angle between two neighbouring spin,\cite{baumsteiger2025exploring} which can also be physically described by the Heisenberg exchange Hamiltonian. Finally, the Dzyaloshinskii-Moriya\cite{dzyaloshinsky1958thermodynamic,moriya1960anisotropic} interactions, which is predominant in systems with strong spin-orbit interaction that is coupled with crystal structures lacking the inversion symmetry. The Dzyaloshinskii-Moriya interaction can be described by the Hamiltonian $\hat{\mathcal{H}}=\sum_{ij}\bm{D}_{ij}\cdot(\bm{s}_i\times\bm{s}_j)$, in which the cross-product antisymmetrised the spin-spin interactions under the exchange of $i$ and $j$, where as $\bm{D}_{ij}$ is the so-called Dzyaloshinskii-Moriya vector whose magnitude and direction are determined by the crystal symmetry and the strength of spin–orbit coupling. The outcome of this special type of spin interaction is that it favours the configuration of chiral spin canting, in which spins tend to twist around a given axis with a fixed chirality.   
 
We note that, strictly speaking, paramagnetism, in which spins across lattice sites are randomly aligned with respect to each other, is strictly speaking, a non-colinear magnetic state. It is the magnetic state that occurs above the magnetic phase transition temperatures that both colinear and non-colinear magnetic alignments become thermally randomised. It is only grouped into \cref{fig:mangetic_landscape}(a) because the concept of paramagnetism is traditionally introduced in the solid-state textbooks along with the colinear magnetism.

Building on the knowledge of these different spin configurations that can occur in solids, it enables us to construct a physically sensible workflow that combines with DFT to calculate and compare the relative energetic stabilities of different magnetic states in a material. This allows us to identify the most favourable ground state and to probe how competing interactions shape the magnetic landscape. In particular, the energy difference between the ground and next metastable magnetic state can be used to compute the magnitude of the exchange coupling constant [\cref{fig:mangetic_landscape}(c) as illustrated for the case of colinear magnetism], which can be further applied to estimate macroscopic properties, such as magnetic transition temperatures. For the case of non-colinear magnetisms, such as those occurring in the frustrated triangular lattices [\cref{fig:mangetic_landscape}], one would need to perform additional sampling of the energy landscape by varying the canting angle between a pair of neighbouring spins $\bm{s}_i$ and $\bm{s}_j$ while keeping the spin orientations on the other lattice sites fixed.\cite{baumsteiger2025exploring}  

Finally, before the end of this section, we shall briefly discuss the theoretical foundations of DFT to account for the spin degree of freedom in modelling the physical properties of materials. For colinear magnetism,\cite{von1972local,gunnarsson1976exchange} extending the KS formalism to account for spin polarisation is straightforward, in which the total electron density is simply the sum of the spin-up and spin-down densities over all the occupied states, \emph{i.e.} $\rho(\bm{r})=\sum\limits_{i=\mbox{\scriptsize{occ.}}}\sum\limits_{\sigma=\uparrow,\downarrow}\left\vert \varphi_{i,\sigma}(\bm{r}) \right\vert^2$. Differentiating the contributions from spin-up and spin-down electron densities to the total system energy can be achieved by making the KS potential become explicitly spin-dependent as $E_{xc}^{\sigma}(\rho_\sigma (\bm{r}))$, such that electrons with opposite spin orientations shall experience different spin-dependent potentials. In LDA, for instance, this can be achieved with the following functional for the exchange energy:
\begin{equation*}
    E_{x}[\rho_{\uparrow}(\bm{r}),\rho_{\downarrow}(\bm{r})]=-C_{x}\int\left(\rho_{\uparrow}(\bm{r})^{\frac{4}{3}}+\rho_{\downarrow}(\bm{r})^{\frac{4}{3}}\right) \mbox{d}\bm{r},
\end{equation*}
or equivalently expressed with the spin polarisation factor $\zeta=(\rho_{\uparrow}-\rho_{\downarrow})/\rho$, with $\rho=\rho_{\uparrow}+\rho_{\downarrow}$ as:
\begin{equation*}
    E_{x}(\rho,\zeta)=E_{x}(\rho,0)\frac{(1+\zeta)^{\frac{4}{3}}+(1-\zeta)^{\frac{4}{3}}}{2}.
\end{equation*}
On the other hand, to treat non-colinear magnetic states,\cite{kubler1988density,kurz2004ab,peralta2007noncollinear} one can no longer treat the electron density as a scalar quantity, but the spin-density matrix, which is constructed as the Kronecker product between two two-component spin-density vectors as: 
\begin{equation}
    \label{eq:density_matrix}
    \bm{\rho}(\bm{r})=\sum_{n\bm{k}}f_{n\bm{k}}\left(\varphi^{*}_{n\bm{k},\uparrow}(\bm{r}),\varphi^{*}_{n\bm{k},\downarrow}(\bm{r})\right)\left(\begin{array}{c}
        \varphi_{n\bm{k},\uparrow}(\bm{r})  \\
         \varphi_{n\bm{k},\downarrow}(\bm{r})
    \end{array}\right)=\left(\begin{array}{cc}
      \rho_{\uparrow\uparrow}(\bm{r})   &  \rho_{\uparrow\downarrow}(\bm{r})\\
      \rho_{\downarrow\uparrow}(\bm{r})   & \rho_{\downarrow\downarrow}(\bm{r})
    \end{array} \right),
\end{equation}
in which the off-diagonal elements $\rho_{\uparrow\downarrow}(\bm{r})$ and $\rho_{\downarrow\uparrow}(\bm{r})$ can map the electron densities to the magnetisations along the $x$ and $y$-directions $(m_x,m_y)$ in addition to the $z$-component $m_z$. This is made obvious by rewriting \cref{eq:density_matrix} in the following form
\begin{equation}
    \rho(\bm{r})=\frac{1}{2}\bigg(\rho(\bm{r})\bm{I}+\bm{m}(\bm{r})\cdot\vec{\sigma})\bigg),
\end{equation}
in which $\bm{m}(\bm{r})=(m_x,m_y,m_z)$ and $\vec{\sigma}$ is the vector of Pauli spin matricies. This thus concludes our short account for studying the magnetic properties of materials with computational approaches.

\section{High-Throughput Computation and Materials Informatics}\label{chapter:high-throughput}

\subsection{Motivations behind the Paradigm Switch to Materials Informatics}
In the previous three sections, we have discussed extensively the theoretical methodologies to compute the energy of atomistic structures of materials, to sample the materials' energy landscapes, and to determine the physical properties of materials. Although these methodologies appear to have covered a wide range of applicative scenarios, for a long period of time before the adaptation of the big-data-driven, as well as the open-sourced, research software and data practices, computational materials science has primarily been the endeavour of individual research groups, each performing calculations on a specific (or a small set of) material(s) by applying the methodologies discussed above. There are pros and cons to this traditional approach. Fundamentally, this was limited by how many of the computational resources were accessible to the individual research group. In the lack of community-based best practices for performing materials simulations, each researcher was required to focus on optimising the simulation parameters for a single structure in their own interests, that is often benchmarked against some known experimental data. While this promotes each researcher to exploit the scientific depth to the material system he/she is interested in, it also leads to significant limitations in the data thus generated. 

For instances, firstly, the generalisability and validity of the simulation methodologies adopted in each isolated study become questionable, meaning it is difficult to judge whether the same method will work equally well to other materials system alike, or whether the agreement with (mostly experimental) observation is incidental, that could have led to biased or even incorrect conclusions in the underlying mechanisms behind the physical or chemical properties investigated in each study. As mentioned before, in the lack of systematic benchmarks, selecting computational methods based on `popularity' not appropriateness or validity, such as PBE for solid-state and B3LYP\cite{kruse2012standard} for molecular systems thus had been, and to some extent in the present days, remains to be the `default' approach of density functional selection(s) in DFT-based materials studies. Secondly, research groups tend to develop their own set of computer scripts/programs to derive further physical properties by post-processing and analysing the DFT results, often leads to low reproducibility of similar  works across the research community. This heterogeneity in research methodologies together with the lack of open-sourcing practice in computational materials research naturally lead to challenges such as to perform statistical analysis on computational results to identify systematic structure-property relationships, as well as outliers amongst a group of materials, which significantly limited our capabilities to generate, not just qualitatively correct but also quantitatively useful insights, to guide the materials discoveries. 

On the experimental side, traditional trial-and-error approaches often lead to slow and labour-intensive cycles in identifying materials with desired properties. Over the past decades, advances in computational power, especially the emergence of GPU-accelerated materials simulations softwares, have fundamentally transformed this process. Large-scale computational pre-screenings based on either experimetally or computationally generated data, that are increasingly enhanced by ML/AI algorithms, can now efficiently narrow down candidate materials and substantially raise the success rate of experimental synthesis and discovery. These capabilities have become a major driving force behind the widespread adoption of high-throughput computational materials research across the community. 

For examples, \citet{raccuglia2016machine} applied ML algorithms to learn from a database containing 3,955 reactions, including failed ones, to guide the synthesis of new compounds that showed a significant increase of success rate to 89 \%. To facilitate the designs of molecular structures, \citet{ahneman2018predicting} applied ML algorithms and the Pd-catalysed Buchwald-Hartwig cross-coupling of aryl halides with 4-methylaniline under different inhibitory additives as examples, showcased how such theory-guided experimental approach was successful in minimising the side reactions, and enhance the stability of the catalyst towards the successful synthesis of the desired product.  Measuring the liftime of lithium-ion batteries is important for their commercialisation but also a time-consuming process. To accelerate the process of life-cycle measurements, \citet{severson2019data} devised ML approaches to extract features in the discharge voltage curves from early cycles, from which they were able to predict and classify the cells by life cycles. Similarly, \citet{zahrt2019prediction} applied ML-guided molecular design of organocatalysts to show how the reaction yields and stereoselectivities could be significantly enhanced. Despite the limitless choices in the structures of organic molecules, only less than 2000 organic photovoltaic donor molecules have been synthesised and tested in the photovoltaic cells, since organo-photovoltaics were first demonstrated in 1973,\cite{ghosh1973rectification} this reflects the long research cycles that involves molecular design, delicate control of the chemical synthesis conditions, laborious purifications and crystallisation processes as well as the challenges behind device fabrications. To overcome these challenges, \citet{sun2019machine} applied ML approach to extract the structure-property relationships from over 1700 experimentally reported chemicals, which led to the successful discovery of 10 newly designed donor molecules. 

The successes behind these works, which nowadays are commonly referred to as the \emph{data-driven} approach in materials discoveries or \emph{materials informatics}, are largely backed by the unique advantages for it to algorithmically extract hidden relationships among the chemical compositions, structures and the properties behind materials over a large materials dataset in a non-biased way, overcoming the limitations behind human intuitions. This has significantly enhanced the efficiency for us the navigate the materials spaces to find better candidates, identify potential pitfalls in the experiments, thus allowing us to better prioritise the compositions and conditions to synthesise new materials, ultimately shorten the materials development lifecycles. 

The paradigm shift from individuals performing isolated computational studies towards the global and communal effort in assembling large-scale material databases via high-throughput computations,\cite{jain2013commentary,saal2013materials,choudhary2020high,draxl2019nomad,haas2009calculation,gjerding2021recent,chanussot2021open,tran2023open} was not only to harness the continuously increasing available computational powers, but most importantly, to address and overcome the limitations mentioned above on the shortfalls behind isolated research efforts, as well as to further capitalise the success of data-driven material discoveries. Such a paradigm shift, together with the widely applied methods of structure predictions/energy landscape sampling as discussed in \cref{chapter:sampling_methods} have significantly transformed the field of materials science by allowing us to establish a holistic view of materials across both of its composition and structural spaces [\cref{fig:energy_landscapes_hierachical}].  In this section, we shall provide a brief account on  the development of materials databases that is enabled by high-throughput computations backed by efficient automated workflow management.

\subsection{Automated Workflows for High-Throughput Computational Materials Science}
\begin{figure}[htb]
    \centering
    \includegraphics[width=\linewidth]{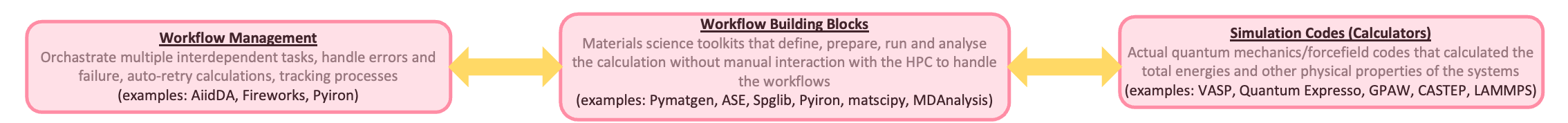}
    \caption{Three essential components of the high-throughput computation ecosystems.}
    \label{fig:ecosystem}
\end{figure}

It is clear that, performing high-throughput computation can no longer rely on individuals to manually set up, submit and execute, analyse, collect the final data, tear down the folders, handle errors and possibly the needs to resubmit failed jobs for every single calculation, as thousands of DFT or MD jobs need to be executed concurrently on the HPC systems.  Within each calculation, there could also exist complex dependent steps to obtain the final results, such as a procedure\cite{yang2021altas} that starts from optimising the input crystal structures, to the computation of its phonon properties and finite-temperature molecular dynamics for determining the dynamic stabilities of the structures, which eventually may lead to the computation of more complex physical behaviours of the materials, for example, the time-dependent fluctuation of the electronic band gap energies to investigate the electron-phonon couplings in the material.\cite{yang2022mapping} (Also, see \cref{tab:workflow_python} for a more generalised illustration of a typical computational materials workflow.) Furthermore, it is also essential to trace the metadata associated with each calculation, such as the version of the DFT codes, input parameters as well as the relationships between the input and output data in order to facilitate the data dissemination and enhance the reproducibility of the research. Finally, for computations that run on shared HPC facilities, it is often necessary to interact with their job scheduling systems to effectively handle inter-dependent jobs, be aware of the queue status and computational resource limitation, ensure efficient parallelisations, in order to optimise the strategies of chunking large computational workloads into smaller parts, minimising the queuing time on the HPC systems, thus increase the overall efficiency of the workflow. Manually handling these tasks thus becomes highly error-prone and inefficient, which has spawned the development of workflow management research software packages, constituting the highest level ``control centre'' for the high-throughput computation ecosystem [\cref{fig:ecosystem}] enabling a majority of the above tasks to be automated, significantly reducing the human labour and errors in  high-throughput computational materials science. These have become the driving forces for the development of Python-based computational materials for high-level workflow management packages such as atomate,\cite{mathew2017atomate} AFLOW,\cite{CURTAROLO2012218} AiiDA,\cite{uhrin2021workflows,huber2020aiida} Atomic Simulation Environment (ASE),\cite{larsen2017atomic} 
pymatgen,\cite{ong2013python} Fireworks,\cite{jain2015fireworks}  and pyiron.\cite{janssen2019pyiron}

\begin{table}[t]
\caption{An exemplary workflow in high-throughput computational materials science and common helper Python packages that are used to facilitate the tasks. }
    \label{tab:workflow_python}
    \centering
    \footnotesize
    \begin{tabularx}{\textwidth}{p{3.5cm}p{8cm}p{0.3cm}p{4.5cm}}
    \hline\hline
       \textbf{Stage}  & \textbf{Task Description}  && \textbf{Python Packages}\\
    \hline
       Structure generation  & Generating standard structures of bulk materials, surfaces, heterostructures, defects and random alloys  && ASE\cite{larsen2017atomic}, Pymatgen\cite{ong2013python}, spglib\cite{togo2024spglib} \\
       Job execution  & Interface with external electronic structure calculators (see \cref{tab:dft_codes}) to execute DFT/AIMD calculations. Interaction with the filesystem for generating and tidy up the calculation folder. Handling errors and restarting/continuing calculations. && ASE, Pymatgen\\
       Data parsing  & Preparing input files to the external codes, extract computational results from output files generated by the external codes (\emph{e.g.} energy, forces, structures, electronic properties, \emph{etc.}) && ASE, Pymatgen\\
       Analysis  & Compute thermodynamic and kinetic properties, data visualisations. && Pymatgen, sumo\cite{ganose2018sumo} \\
       Machine learning  & Computing structural descriptors, supervised and unsupervised learning. && matminer\cite{ward2018matminer}, Dscribe\cite{himanen2020dscribe}, scikit-learn\cite{scikit-learn}, scikit-matter\cite{goscinski2023scikit} \\
    \hline\hline
    \end{tabularx}
    
\end{table}

The next level in the high-throughput computational materials study ecosystem is the workflow building block. This is where the logic for coordinating, executing multistep and sequentially dependent computation procedure to determine a particular physical property of material with a specific atomistic structure, by following the published and rigorously validated theoretical frameworks for computing that physical property (for example, as those discussed in \cref{chapter:physical_properties}), is implemented. The rationale for implementing these workflow building blocks are as following. Calculating these physical properties generally requires some degree of pre-processing the input structures and selecting the suitable calculation parameters that maximally balance the computational speed and accuracy. This information is then passed to an external DFT simulation code (\cref{fig:ecosystem}) that does the heavy-lifting of computing the energies, electronic and other physical properties for the given structure(s). 

\cref{tab:dft_codes} lists some of the widely used electronic structure theory codes for materials and molecular system simulations. We note that whilst most of these codes are designed with a mindset of broad-range applicability in terms of the range of theoretical methods being implemented, there are unique strengths in each code on how some specific methodologies are theoretically and numerically implemented in that code, which have been briefly discussed in \cref{tab:dft_codes} that maybe helpful to guide the practitioners in choosing the most appropriate one for their problems at hand.   As these DFT computations are generally expensive and time-consuming, they are usually executed on the HPC queuing system asynchronously to the actual workflow, which may be managed by the workflow manager. 

For the same reason, these DFT codes are historically written in programming languages that are numerically more efficient than modern object-oriented Python, such as Fortran, thus writing the input files and parsing the program output files for post-processing is how these DFT codes can be interacted with inside a computational workflow. As such, the workflow building blocks also provide standardised I/O routines to interact with these DFT codes by writing and parsing the text-based files, extracting useful information for further processing, computing the relevant physical properties according to the known physics, and finally packaging the information in modernised Pythonic objects for data storage, sharing and future processing. As some of these core scientific logics for computing physical properties can be highly specialised, such as the phononic, transport and superconductive properties of condensed matter, there are specialised Python physics engines (see \cref{tab:python_physics_engines} for examples) developed by domain experts, which can be harnessed to assemble the workflow building blocks.

\begingroup 
\renewcommand{\arraystretch}{1.2}  
\LTcapwidth=\textwidth
\begin{longtable}{p{3cm}p{3cm}p{11cm}}

    \caption{List of (Python) packages that facilitate the computation of specific physical properties of materials in computational materials science.  }
    \label{tab:python_physics_engines}\\
    \hline\hline
     \textbf{\footnotesize{Domain}}    & \textbf{\footnotesize{Python package}}  & \textbf{\footnotesize{Core features}}\\
     \hline
     \endfirsthead
     
\caption*{Table \ref{tab:python_physics_engines} (Continued): List of specialised (Python) engines for physical property calculations.}\\\toprule
     \textbf{\footnotesize{Domain}}    & \textbf{\footnotesize{Python package}}  & \textbf{\footnotesize{Key features}}\\
     \hline
\endhead
\hline\hline
\endfoot
\hline\hline
\endlastfoot

\textbf{\footnotesize{Lattice dynamics}} \footnotesize{(incl. harmonic and anharmonic phonons) } 
& \footnotesize{Phonopy\cite{togo2015first,phonopy-phono3py-JPCM,phonopy-phono3py-JPSJ}} & \footnotesize{Harmonic phonons based on the finite-displacement method. Enables the computations of phonon dispersion relationship, density-of-states, vibrational thermodynamic functions with energy and forces from periodic DFT codes such as VASP and Quantum ESPRESSO. Anharmonic phonons, such as Gr\"{u}neisen parameters, thermal expansion coefficients, from quasi-harmonic approximation.}\\

& \footnotesize{Phono3py\cite{phonopy-phono3py-JPCM,phonopy-phono3py-JPSJ,phono3py-pssb-2015}} 
& \footnotesize{Anharmonic phonons; computation of the 3rd-order FCs to model the phonon-phonon interactions based on the finite-displacement approach; mode-dependent Gr\"{u}neisen parameters and phonon lifetime; combined with the Boltzmann transport equations for computing the lattice thermal conductivity of crystalline materials. } \\

& 
\footnotesize{TDEP\cite{knoop2024tdep}}
&\footnotesize{Anharmonic phonon properties; temperature-normalised phonon frequency and effective potential from AIMD through an efficient fitting procedure.}\\

& \footnotesize{ALAMODE\cite{phono3py-pssb-2015,tadano2015self}} 
& \footnotesize{Harmonic and anharmonic lattice dynamics; fitting temperature-dependent effective potential; temperature-dependent phonon frequencies through self-consistent phonon approach, enabling further computation of thermodynamic functions such as vibrational free energies.} \\

& \footnotesize{Hiphive\cite{eriksson2019hiphive}} 
& \footnotesize{Anharmonic phonon properties; efficient regression-based approach to extract higher-order interatomic FCs with space group symmetry enforced from AIMD atomic positions and forces; effective forcefield for dynamic simulations.}\\

& 
\footnotesize{DynaPhoPy\cite{carreras2017dynaphopy}}
&\footnotesize{Extracting temperature-dependent phonon frequency shifts from AIMD simulations, facilitating the examinations of anharmonic effects in lattice vibrations}\\

\footnotesize{\textbf{Electron-phonon interactions and carrier transports}}
& \footnotesize{EPW\cite{noffsinger2010epw,ponce2016epw}} 
& \footnotesize{Using Wannier interpolation to compute the EPI matrix elements; computing EPI self-energies; electron and phonon lifetime limited by EPI; anisotropic supercomputing gap from the Eliashberg theory; temperature-dependent electron/hole mobilities from the Boltzmann transport formalism.}\\

& \footnotesize{Perturbo\cite{zhou2021perturbo}}
& \footnotesize{Electron and Phonon transport capturing the phonon-limited transport properties, incl. conductivity, carrier mobility, and Seebeck coefficient \emph{etc.}; employing Wannier interpolation to reduced the computational cost.}\\

\footnotesize{\textbf{Thermal transports and thermoelectric properties}}
&  \footnotesize{BoltzTraP\cite{madsen2006boltztrap,madsen2018boltztrap2}}
& \footnotesize{Based on a smoothed interpolation of the electronic band structures to extract band curvature information to calculate electronic conductivity, electronic thermal conductivity \emph{etc.}, which can be subsequently used for computing the Seebeck coefficient, Hall conductivities, \emph{etc.}}\\

& 
\footnotesize{AMSET\cite{Ganose2021_AMSET}}
&\footnotesize{Calculate electron lifetime and transport properties in solids from first-principles; EPI scattering rate that is modelled within the Born approximation based on material properties such as phonon frequencies and dielectric constants. Transport properties computed based on the Boltzmann transport equation.}\\

& 
\footnotesize{ShengBTE\cite{li2014shengbte}}
&\footnotesize{3rd-order FCs for phonon-phonon scatterings; lattice thermal conductivity based on the Boltzmann transport equation. }\\

& 
\footnotesize{Phoebe\cite{cepellotti2022phoebe}}
&\footnotesize{Electron-phonon and phonon-electron scattering rates through Wannier interpolation from which phonon and electron transport coefficients may be determined; 3-phonon scattering rates are extracted from either phono3py or ShengBTE; lattice thermal conductivities taking into account the contributions from electron-phonon scatterings. }\\
\end{longtable}
\endgroup


\begingroup 
\renewcommand{\arraystretch}{1.2}  
\LTcapwidth=\textwidth

\begin{longtable}{p{2cm}p{3cm}p{12.5cm}}
    \caption{List of major electronic structure theory codes for materials and molecular science. For the periodic systems, we treat  VASP as the default go-to electronic structure theory package, and highlight the core features in other packages that are not implemented in the VASP package.}
    \label{tab:dft_codes}\\
    \hline\hline
     \textbf{\footnotesize{Code}}    & \textbf{\footnotesize{Basis set type}}  & \textbf{\footnotesize{Core features}}\\
     \hline
     \endfirsthead
\caption*{Table \ref{tab:dft_codes} (Continued): List of major electronic structure theory codes.}\\\toprule
     \textbf{\footnotesize{Code}}    & \textbf{\footnotesize{Basis set type}}  & \textbf{\footnotesize{Core features}}\\
     \hline
\endhead
\hline\hline
\endfoot
\hline\hline
\endlastfoot
       \multicolumn{2}{l}{\textbf{Periodic Systems}} &  \\
       \footnotesize{VASP\cite{Kresse1996CMS,Kresse1996PRB,Kresse1999PAW}}
            & \footnotesize{planewaves+PAW} 
            & \footnotesize{Highly optimised commercial  package; most globally dominant; backbone calculator for large-scale databased such as the Materials Project\cite{jain_apl_mater}; support DFT incl. hybrid GGA, GW+BSE, AIMD, and on-the-fly machine learning potential.\cite{jinnouchi2019phase}}\\
            
      \footnotesize{CASTEP\cite{Clark2005CASTEP}} 
            & \footnotesize{planewaves} 
            & \footnotesize{Similar to VASP in terms of core capability. Commercial package distributed via Materials Studios or standalone academic license (free for UK-based users).} \\
            
      \footnotesize{ONETEP\cite{Skylaris2005ONETEP}}
            & \footnotesize{localised psinc basis}
            & \footnotesize{Linear-scaling making  accessible to model large periodic structures, surfaces and macromolecules. Sharing core methodologies with the CASTEP code.}\\
            
      \footnotesize{Quantum ESPRESSO (QE)\cite{giannozzi2009quantum}}
            & \footnotesize{planewaves}
            & \footnotesize{Widely used open-sourced  package for solid; unique features such as  XDM-based dispersion correction model; support packages such as PHonon, EPW (for determining properties such as EPI strength and Eliashberg spectral function) and TDDFT (time-dependent DFT) \emph{etc.}}\\
            
      \footnotesize{ABINIT\cite{gonze2009abinit}}  
            & \footnotesize{planewaves+PAW}  
            & \footnotesize{Open-sourced research-grade electronic structure theory package with a strong  focus in advancing the modelling of many-particle physics in materials, with methods such as many-body perturbation theory [(self-consistent) GW approximation and Bethe-Salpeter   equation], dynamical mean-field theory (DMFT)\cite{kotliar2006electronic} for strongly correlated systems and temperature-dependent effective potential (TDEP).}\\
            
      \footnotesize{CP2K\cite{cp2k_package}} 
            & \footnotesize{planewaves+Gaussians} 
            & \footnotesize{Open-sourced  packages with the unique combination of localised Gaussian and delocalised planewave basis that is tailor designed to achieve linear-scaling behaviour for DFT, enable modelling systems with up to tens of thousands of atoms, particularly for AIMD; naturally handle both periodic and molecular systems; improved efficiency for HF, hybrid DFT and post-HF methods; constrained DFT for modelling charge transfer; hybrid QM/MM and density functional embedding methods.}\\
            
      \footnotesize{WIEN2k\cite{blaha2020wien2k}}
            & \footnotesize{augmented planewaves+local orbitals} 
            & \footnotesize{Licensed DFT packages; mixed basis sets providing good speed up for condensed matter, particularly AIMD simulations.}\\

      \footnotesize{FHI-aims\cite{blum2009ab,abbott2025roadmap}}
            &\footnotesize{Numerical atom-centred orbitals} & \footnotesize{Free/licensed package with small fee; utilisation of the localised basis function enabling efficient scaling for both periodic and molecular systems with good accuracy; excited state calculation at GW and coupled cluster level; dedicated developmental efforts towards lattice dynamics, anharmonic phonons, and electron-phonon coupling; transport properties (vibrational heat transport, electron transport and molecular transport); AI workflow integration.} \\
      
      \footnotesize{SIESTA}\cite{garcia2020siesta}
            &\footnotesize{Numerical atom-centred orbitals}  & \footnotesize{Free academic license; TDDFT; unique capability to model the electronic transport properties for nanoelectronics with the DFT+NEGF approach through the TRANSIESTA package.\cite{stokbro2003transiesta,papior2017improvements,wittemeier2025quantum}}\\
            
      \footnotesize{GPAW\cite{Enkovaara_2010,10.1063/5.0182685}} 
            & \footnotesize{Grid-based projector-augmented planewaves}
            & \footnotesize{Open-sourced \& Python-interfaced  package; dielectric screening for low-dimensional materials; magnetic responses within the TDDFT framework; computation of the electron-phonon coupling strengths and Raman spectra; non-linear optical response properties; circular dichoism; Ehrenfest dynamics; constrained DFT for charge transfer; solvated Jellium model for solvated electrochamical interfaces.} \\
            
      
      \footnotesize{CRYSTAL\cite{erba2017large}} 
            & \footnotesize{Atom-centred Gaussians}
            & \footnotesize{Commercial licensed  package; utilisation of all-electron Gaussian basis sets allows one to achieve quantum chemical accuracies in periodic systems; enhanced speed and efficiency for HF, hybrid DFT, MP2 and localised MP2 calculations.} \\
            
      \footnotesize{JDFTx\cite{sundararaman2017jdftx}}  
            &\footnotesize{Planewaves} 
            &\footnotesize{Open-sourced  package; grand canonical (fixed potential) for electrochemistry;  application of external electric field; charged-defect correction scheme; solvation models. } \\
\hline
      \multicolumn{2}{l}{\textbf{Molecular Systems}} &  \\
      \footnotesize{Gaussian}\cite{g16} 
            &\footnotesize{Atom-centred Gaussians} 
            &\footnotesize{Commercial  package; all-purposed, and often considered as the entry-level molecular quantum chemistry packages covering a broad range of semiempirical, DFT, WFT and QM/MM methods.} \\

      \footnotesize{Turbomole\cite{turbomole}} 
            &\footnotesize{Atom-centred Gaussians} 
            & \footnotesize{Commercial  package, employing the resolution-of-identity (RI)\cite{weigend2002fully} approximation to speed up the evaluation of four-electron Coulomb integrals in both DFT and WFT calculations. }\\

    \footnotesize{Amsterdam Density Functional (ADF)\cite{te2001chemistry}} 
            & \footnotesize{Atom-centred Slater-type orbitals} 
            & \footnotesize{Commercial package; the only quantum chemistry package that extensively uses the Slater-type orbitals; zero-order regular approximation (ZORA)\cite{van1999geometry} for treating relativistic effects; frozen-density-embedding method\cite{wesolowski2015frozen}; spectroscopic properties; periodic DFT with Slater-type orbitals through BAND; solvation effect through the COSMO-RS model\cite{klamt2018cosmo}.}\\
      
      \footnotesize{Orca\cite{neese2020orca}} 
            & \footnotesize{Atom-centred Gaussians}
            & \footnotesize{Free for academics; unique strength in computing spectroscopic properties, electron paramagnetic resonance properties, magnetically active transition metal complexes, bioinorganic systems; MP2 and CCSD(T) methods implemented under the domain-based local pair nature orbital (DLNPO) framework, offering CCSD(T) level-of-accuracy for treating large molecules ($\sim 10^3$ atoms) with linear-/square-scaling; accounting for the relativistic effects for $f$- and $d$-block chemistries.  }\\
     
      \footnotesize{Q-Chem\cite{epifanovsky2021software}}
            & \footnotesize{Atom-centred Gaussians} & \footnotesize{Commercial package; thermally-assisted-occupation DFT for treating large electronic systems with strong correlation; TDDFT, $\Delta$SCF and ROKS for excited states; MP$n$ method; CC/EOM-CC for open-shell and electronically excited species; configurational interaction (CI) for strongly correlated systems; nuclear-electronic orbital (NEO) framework for treating the nuclear quantum effects; solvation models; QM/MM and QM/MM embeddings; fragment-based approach (\emph{e.g.} for treating charge-transfer problems); symmetry-adapted perturbation theory (SAPT) for energy decomposition analysis of non-covalent interactions.}\\
      \footnotesize{NWChem\cite{nwchem}} 
            & \footnotesize{Gaussians/planewaves}
            & \footnotesize{Open-sourced package; DFT/WFT/forcefield; QM/MM}\\
            
      \footnotesize{Psi4\cite{WIRE_Psi4,psi_4}} 
            &  \footnotesize{Atom-centred Gaussians}
            & \footnotesize{Fully open-sourced quantum chemistry package, promote community-based developmental efforts; covering the full spectrum of quantum chemistry methods from HF, DFT, to coupled clusters, configuration interactions, multireference methods, as well as TDDFT and SAPT; input is naturally built upon the Python scripting language facilitating the integration with the high-throughput workflow. }\\

    \footnotesize{PySCF\cite{pyscf_1,pyscf_2}} 
            & \footnotesize{Atom-centred Gaussians}
            & \footnotesize{Similar design philosophy and capability as Psi4; support periodic systems. }\\
            
      \footnotesize{GFN2-xTB\cite{bannwarth2019gfn2}} 
            & \footnotesize{Atom-centred Gaussians}
            & \footnotesize{Open-sourced quantum chemistry package; semi-empirical, parameterised tight-binding method designed to handle molecular systems with $\sim 10^3$ atoms.}\\

      \footnotesize{MRCC\cite{mrcc_1,mrcc_2}} 
        & \footnotesize{Atom-centred Gaussians}
        & \footnotesize{Open-source licensed package; mostly focused on wavefuntion theory for quantum chemistry (HF, MP2, CC to multireference CI, excited states through linear response theory).}\\
      
      \footnotesize{Dalton\cite{dalton_package_2,dalton_package}} 
            & \footnotesize{Atom-centred Gaussians}
            &  \footnotesize{Open-sources quantum-chemistry package; special capabilities in wavefunction-based multireference methods, higher-order response theory, relativistic effects and spectroscopic properties from CC calculations.}\\

\end{longtable}
\endgroup


\begin{table}[htb]
    \centering
    \caption{Workflows and their corresponding supported Calculators as implemented in the \texttt{atomate2} workflow management package. The MLIP column represents all models that are supported by the atomate2 including CHGNet,\cite{deng2023chgnet} M3GNet,\cite{chen2022universal} MACE,\cite{batatia2023foundation} GAP\cite{bartok2010gaussian,bartok2013representing} and NequIP.\cite{batzner20223} $\bigcirc$ indicates supporting this workflow-engine combination is on \texttt{atomate2}'s near-term development roadmap. $\triangle$ is used for combos that are not currently planned but are considered a small implementation effort that could be added by adopting similar existing Makers.
    [Reproduced from \citet{ganose2025atomate2}]}
    \label{tab:automate2}
    \footnotesize
    \begin{tabular}{p{6cm}llllllllll}
        \hline\hline
        & & \multicolumn{9}{l}{\textbf{Supported calculators}} \\ \cline{3-11}
        \textbf{Workflow} & \textbf{System} 
        & ASE
        & FHI-AIMS
        & ABINIT
        & CP2K
        & JDFTx
        & MLIPs 
        & VASP 
        & OpenMM
        & Q-Chem
        \\
        \hline
        Geometry optimisation & Both & \checkmark & \checkmark & \checkmark & \checkmark & \checkmark & \checkmark & \checkmark & \checkmark & \checkmark \\
        Static & Both & \checkmark & \checkmark & \checkmark & \checkmark & \checkmark & \checkmark & \checkmark & \checkmark & \checkmark \\
        Dielectric & Periodic & -- & --& --& --& --& -- & \checkmark & -- & -- \\
        Polarisation & Periodic & -- & --& --& --& --& -- & \checkmark & -- & -- \\
        Electronic band structure & Periodic & -- &\checkmark &\checkmark &\checkmark  &-- &-- &\checkmark &-- &--\\
        Bond analysis workflow with LOBSTER\cite{nelson2020lobster}  & Periodic & --&--&$\triangle$&--&--&--&\checkmark &--&--\\
        Excited states (GW-BSE) & Periodic & $\triangle$ & \checkmark & \checkmark & \checkmark &-- &-- & \checkmark & -- & --\\
        \emph{Ab initio} molecular dynamics & Periodic & \checkmark & $\triangle$ & -- & -- & -- & -- &\checkmark &-- & --\\
        Force field molecular dynamics & Periodic &\checkmark &\checkmark &-- &-- &-- &\checkmark &\checkmark &-- &-- \\
        Elastic constants & Periodic & $\triangle$ &\checkmark & $\triangle$ & $\triangle$  &-- &\checkmark &\checkmark &-- &--\\
        Harmonic phonons & Periodic & $\triangle$ &\checkmark & $\triangle$ & $\triangle$ & -- &\checkmark &\checkmark &-- &--\\
        Equation of state & Periodic & $\triangle$ &\checkmark &$\triangle$ &$\triangle$ & -- & \checkmark & \checkmark & -- & -- \\
        Electron-phonon & Periodic &--&--&--&--&--&-- &\checkmark &--&--\\
        Gr\"{u}neisen & Periodic &-- &$\triangle$&--&--&--&\checkmark&\checkmark&--&--\\
        Matpes & Periodic &-- &-- &-- &-- &-- &-- &\checkmark  &-- &--\\
        Quasiharmonic approximation for phonons & Periodic &-- &$\triangle$  &-- &-- &--&\checkmark&\checkmark  &-- &--\\
        Anharmonic phonons & Periodic & $\triangle$ & $\bigcirc$ &$\triangle$ &$\triangle$ & -- &\checkmark &\checkmark & -- & --\\
        MPMorph & Periodic &$\triangle$ &$\triangle$ &$\triangle$ &$\triangle$ &-- &\checkmark &\checkmark &-- &--\\
        Magnetic ordering & Periodic & $\triangle$ & \checkmark & $\triangle$ &$\triangle$ & -- & -- &\checkmark &-- &--\\
        Adsorption & Periodic & $\triangle$ & $\triangle$ & $\triangle$ & $\triangle$ &$\bigcirc$ & $\triangle$ & \checkmark &-- &--\\
        Point defect & Periodic & $\triangle$ & $\bigcirc$ & -- & -- & -- & -- &\checkmark & -- & -- \\
        Anharmonicity quantification & Periodic & $\triangle$ & \checkmark & $\triangle$ &$\triangle$ & -- &$\triangle$ & $\triangle$ & --& --\\
        Electrode discovery & Periodic & $\triangle$ & -- & -- & -- & -- & -- &\checkmark & -- & -- \\
        Ferroelectric & Periodic & $\triangle$ & -- & -- & -- & -- & -- &\checkmark & -- & -- \\
        Materials project & Periodic & -- & -- & -- & -- & -- & -- & \checkmark & -- & -- \\
        Amset & Periodic & -- & -- & -- & -- & -- & -- & \checkmark & -- & -- \\
        Frequency flatterning optimiser workflow & Molecular  & -- & -- & -- & -- & -- & -- & -- & -- & \checkmark \\
        Classical molecular dynamics workflow & Molecular & \checkmark & -- & -- & -- & -- & -- & -- &\checkmark & -- \\
        \hline\hline
    \end{tabular}
\end{table}

Implementing these workflow building blocks has significantly accelerated the computation of physical properties of materials by reducing the human efforts in setting up, executing and analysing these calculations, thus reducing the likelihood of producing errors along the multistep workflow. This naturally increases the reproducibility and robustness of the published results. To provide more concrete examples of the workflow building blocks, we focus the following discussions using the example of the  \texttt{atomate2} workflow management framework.\cite{ganose2025atomate2} 

As listed in \cref{tab:automate2}, \texttt{atomate2} implemented 28 different workflows that are most commonly applied for studying the properties of materials, with a primary focus on periodic systems. It provides the interfaces to various different external calculators for heavy-lifting, including 7 commonly used (periodic and non-periodic) DFT codes, to ASE (which not only can be used as an interface to other external codes but also contains its own optimisers/calculators for tasks such as structural optimisation, molecular dynamics and so on), as well as to classical forcefields (through OpenMM\cite{eastman2023openmm}) and emerging machine-learning interatomic potentials (MLIPs). Since most of these external codes share a set of core capabilities, such as providing the system energies and forces for the atomistic models, the workflow implementations essentially abstracted this heavy-lifting part out as the `calculators' that can easily be swapped with different external codes to accomplish the same workflow with minimum efforts.  

\begin{figure}[tb]
    \centering
\includegraphics[width=\linewidth]{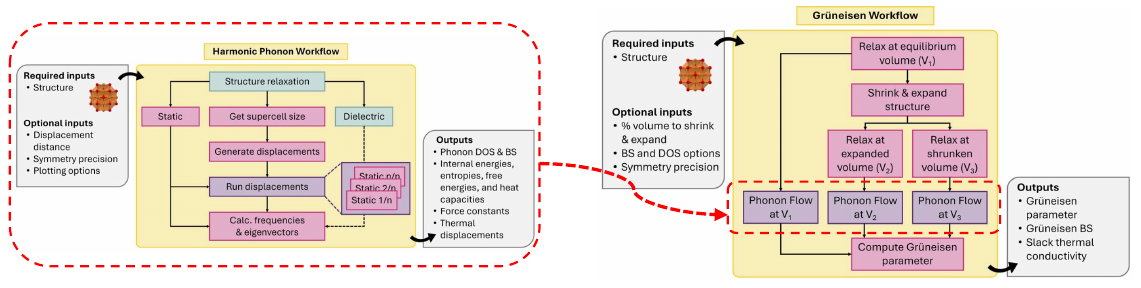}
\caption{Schematic illustrations of the harmonic phonon (left) and the Gr\"{u}neisen parameter (right) calculations workflows as implemented in  Atomate2.\cite{ganose2025atomate2} The composite nature of workflow building in computational materials science is demonstrated by the red dashed error, whereby the more complex Gr\"{u}neisen workflow is built upon the more basic workflow of harmonic phonon calculations.}
    \label{fig:complex_workflow}
\end{figure}

As concrete examples, \cref{fig:complex_workflow}(left) illustrate the \texttt{atomate2} implementation of the workflow for computing the harmonic phonon properties of solids, using the finite-displacement approach as discussed in \cref{sect:phonon_harmonic_and_anharmonic}. In its core, starting from a crystal structure as input, it utilises the \texttt{phonopy}\cite{phonopy} as the underlying framework to generate crystal structures with perturbed atomic positions, which are then passed to external DFT codes to compute the atomic forces for each one of the generated structures, before all these information are re-collected and fed back to  \texttt{phonopy} to construct and diagonalise the dynamical matrix to obtain the phonon eigenfrequencies and eigenvectors. In the language of software engineering, the design of workflow essentially `compartmentalises' the scientific logic, making it become an atomic and reusable part for assembling more complex workflows, which is illustrated on the right hand side of \cref{fig:complex_workflow}, whereby computing the Gr\"{u}neisen parameter for a given solid following its definition [\cref{eq:gruneisen}] would require three separated phonon calculations at the equilibrium, expanded and shrunken volumes of the solid in order to extract the variation of phonon eigenfrequencies with respect to the change in the cell volume.

In summary, automated workflows played a critical role as the enabler for high-throughput computational materials science. By orchestrating complex sequences of calculations, that often span across structure generation, electronic structure calculations, data collections, post-processing, error correction, and data management on HPC environments, utilising these frameworks enhances the  reproducibility, scalability, and robustness of the high-throughput computational materials researches. The flexibilities provided by these frameworks accelerates the exploration of the complex composition-structure-property relationship of materials, increase the flexibilities and adaptabilities of the research work for developing more complex and advanced computational methodologies for exploring more exotic properties of materials, and thus played a key role in the developments and curation of many generic and specialised computational materials databases, which will be discussed in the following section.

\subsection{Computational Materials Databases}
Through the implementations and the utilisation of standardised computational protocols through orchestrated workflows, the past decade has witnessed the rapid emergence of numerous large-scale computational materials databases that systematically and reproducibly curate vast collections of high-quality computational data for materials at unprecedented scale.  

For example, the \textit{Materials Project}\cite{jain_apl_mater} database is one of the earliest of its kind, that emerged in early 2010s as a transformative initiative under the US Materials Genome Initiative (MGI). Founded at Lawrence Berkeley National Laboratory, the project pioneered the vision of creating a freely accessible, community-driven repository of high-quality, standardised and consistent DFT calculations for crystalline and molecular materials. As discussed in the previous section, \texttt{pymatgen}\cite{ong2013python} is at the centre to the architecture of Materials Project, which provides a comprehensive suite of Python toolkit for preparing, executing, analysing, as well as  infrastructure for integrating with automated workflow management packages such as Fireworks\cite{jain2015fireworks} for performing calculations over large-scale HPC facilities. This synergy between standardised workflows and centralised data dissemination have enabled the Materials Project to systematically compute and validate properties for over 200,000 crystalline materials and 570,000+ molecules (as of late 2025), typically at the PBE+U level of theory. The \texttt{pymatgen} package has also  become widely adopted by invidual research groups to facilitate the building of their own customised workflows and data analysis softwares. Beyond the Materials Project, numerous other computational materials databases have been established over the years, and \cref{tab:materials_databases} provided a non-exhaustive list of some of the major computational materials databases that are widely used by the community. These databases typically cover a wide range of materials classes, including crystalline solids, 2D materials, and molecules whilst providing data on a generic set of materials properties such as structural information, thermodyanmic stabilities, electronic, dielectric/optical, magnetic, mechanical properties and more. 

\begingroup 
\renewcommand{\arraystretch}{1.2}  
\LTcapwidth=\textwidth

\begin{longtable}{p{3cm}p{3cm}p{11cm}}
\caption{List of major computational materials databases.}
    \label{tab:materials_databases}\\
    
    \hline\hline
     \textbf{\footnotesize{Databases}}    & \textbf{\footnotesize{Reference}}  & 
     \textbf{\footnotesize{Description}}\\
     \hline
     \endfirsthead
\caption*{Table \ref{tab:materials_databases} (Continued): List of major computational materials databases.}\\\toprule
     \textbf{\footnotesize{Databases}}    & \textbf{\footnotesize{Reference}}  &  
     \textbf{\footnotesize{Description}}\\
     \hline
\endhead
\hline\hline
\endfoot
\hline\hline
\endlastfoot

\footnotesize{Materials Project}
&\footnotesize{\citet{jain_apl_mater}}
&\footnotesize{Established under the US Materials Genome Initiative. This database contains the DFT calculation results at the PBE+U level for 200K+ crystalline materials and 570K+ molecules (as in Nov. 2025). The database holds the information of structures, DFT energies, magnetic structures, electronic band structures and DOS, elastic tensors \emph{etc.} The database can be accessed through the \texttt{pymatgen}\cite{ong2013python} Restful API (Application Programming Interface).}\\

\footnotesize{MP-ALOE}
&\footnotesize{\citet{kuner2025mp}}
&{\footnotesize{\textbf{M}aterials \textbf{P}roject-\textbf{A}ctive \textbf{L}earning of \textbf{O}ff
\textbf{E}quilibrium structures} database containing nearly 1 M DFT calculations for solid-state materials in and off equilibrium structures at the meta-GGA (r$^2$SCAN)\cite{furness2020accurate} level, generating higher accuracy data for AI model training.}\\

\footnotesize{OQMD (Open Quantum Materials Database)}
&\footnotesize{\citet{saal2013materials}}
&\footnotesize{This database contains the DFT computed thermodynamics and structural properties for 1.3M+ inorganic solids. It was curated by Chris Wolverton's research group from  Northwestern University.  It can be accessed through the OPTIMADE\cite{optimade} API.}\\

\footnotesize{JRVIS-DFT (\textbf{J}oint \textbf{A}utomated \textbf{R}epository for \textbf{V}arious \textbf{I}ntegrated \textbf{S}imulations)}
&\footnotesize{\citet{choudhary2020joint}}
&\footnotesize{This database contains the DFT calculation results for 80K+ materials, not only entailing the structures and their corresponding energy stabilities, but also databased containing specific materials properties including (a) optoelectronic properties computed with optB88vdw  and TBmBJ formalism\cite{choudhary2018computational}, (b) elastic properties for bulk and 2D materials with vdW density functionals\cite{choudhary2018elastic}, (c) nontrivial topological materials\cite{choudhary2019high}, (d) electric field gradients\cite{choudhary2020density}, and (e) computational scanning tunnelling microscopic images\cite{choudhary2021computational}. }\\

\footnotesize{Novel Materials Discovery (NOMAD) Repository}
&\footnotesize{\citet{draxl2019nomad}}
&\footnotesize{This is a community-driven database that hosts the computational materials data generated from various different DFT codes, it contains 19.2M+ entries for over 4.3 M represented materials.}\\

\footnotesize{Computational 2D Materials Database (C2DB)}
&\footnotesize{\citet{Haastrup_2018,Gjerding_2021}}
&\footnotesize{This database contains the computational results for  $\sim$4000 2D materials with more than 40 different crystal structures. It reported the corresponding structures, energetics, magnetic ground states, mechanical properties, electronic properties at PBE, HSE and G$_0$W$_0$ levels of theory, as well as RPA electronic polarisability and BSE optical absorbance. }\\

\footnotesize{Open Catalyst Project}
&\footnotesize{\citet{chanussot2021open} (OC20); \citet{tran2023open} (OC22)}
&\footnotesize{This database reported the structures and energetics for surface adsorbates at the revised-PBE level of theory\cite{Hammer1999RPBE}. The OC20 contains 1,281,040 DFT relaxation trajectories ($\sim$265M single point calculations), focusing adsorbate chemistries on nitrogen-, oxygen- and carbon-containing molecules. The OC22 focuses on oxide catalysts for the oxygen evolution reactions (OERs), which contains 62,311 DFT relaxation trajectories ($\sim$9.8 M single point calculations)}\\

\end{longtable}
\endgroup

While large-scale, general-purposed repositories such as the Materials Project, OQMD and NOMAD provide broad coverage of structures and a standard set of properties computed with consistent protocols, which excel at discovering trends in materials properties and providing wide-coverage training sets for  general ML models, many scientific and engineering questions demand datasets that are computed with workflows and parameters tuned specifically for a more targeted set of materials and properties. Therefore, many property-focused databases have also been curated by different research groups across the globe to fill in the gap left by these general-purposed databases. These represent dedicated efforts in the community to riguously test the convergence criteria, improve the data accuracies, develop specialised computational algortithms as well as post-processing and data analysis methodologies for targeted materials properties, such as the GW quasiparticle energies and excited-state BSE spectra,  electron--phonon coupling and superconducting spectral functions, anharmonic phonon scattering rates and lattice thermal conductivities, magnetic and ferroelectrics, for which \cref{tab:property_databases} provided more detailed account for these dedicated databases.

This specialsation is driven by three practical needs. Firstly, many properties require property-specific convergence and finite-size corrections, for example, determining the electron-phonon coupling constants requires reciprocal space sampling to be performed on densed $\bm{k}$- and $\bm{q}$-point grids, mouting to an $\sim N_{\bm{k}}\times N_{\bm{q}}$--fold increase in the computational cost per structure. Secondly, accurate evaluation typically needs additional post-processings, suuch as Wannierisation, interpolation of matrix elements, many-body self-energy workflows, phonon interpolation, and so on, that are most robust when implemented by more specialised domain expertises in the automated computational pipeline. Thirdly, focused datasets enable targeted validation against experiment or higher-level theory, producing reliable benchmarks for screening and ML training. As such, these specialised databases have critically complemented the large-scale generic computational materials databases to provide high-quality data for more complex materials properties that are not typically available in the general-purposed databases, which are essential for generalising robust knowledges of structure-property relationships for functional materials, as well as training physics-specific ML models for accelerating materials discoveries.

\begingroup 
\renewcommand{\arraystretch}{1.2}  
\LTcapwidth=\textwidth

\begin{longtable}{p{2cm}p{3cm}p{0.9cm}p{11cm}}
\caption{List of exemplary high-throughput computational studies that applied workflows as tabulated in \cref{tab:automate2} (or similar) to establish targeted databases of specific materials properties.}
    \label{tab:property_databases}\\
    
    \hline\hline
     \textbf{\footnotesize{Workflow}}    & \textbf{\footnotesize{Reference}}  & 
     \textbf{\footnotesize{Year}} &
     \textbf{\footnotesize{Description}}\\
     \hline
     \endfirsthead
\caption*{Table \ref{tab:property_databases} (Continued): List of notable computational materials properties databases.}\\\toprule
     \textbf{\footnotesize{Workflow}}    & \textbf{\footnotesize{Reference}}  &  
     \textbf{\footnotesize{Year}} &
     \textbf{\footnotesize{Description}}\\
     \hline
\endhead
\hline\hline
\endfoot
\hline\hline
\endlastfoot

\footnotesize{Dielectric and polarisation} 

& \footnotesize{\citet{yim2015novel}}
& \footnotesize{2015}
& \footnotesize{Established a library of electronic dielectric constants for $>$1800 binary and ternary oxides to facilitate the identification of high-$\kappa$ gate dielectric materials for microelectronics. }
\\
& \footnotesize{\citet{petousis2017high}}
& \footnotesize{2017}
& \footnotesize{Largest database (at the time of publishing) of dielectric constants and refractive index of materials computed with DFPT, containing 1,056 compounds, which is integrated into the Materials Project Database.\cite{jain_apl_mater}}\\

& \footnotesize{\citet{lee2018high}}
& \footnotesize{2018}
& \footnotesize{Extending the previous work\cite{yim2015novel} to include non-oxide materials, exploring the trends between $E_{g}$ and $\kappa$ for inorganic solids.}\\

& \footnotesize{\citet{choudhary2020high}}
& \footnotesize{2020}
& \footnotesize{Computation of the $\Gamma$-point phonons, infrared intensities, Born-effective charges, piezoelectric, and dielectric tensors for 5015 non-metallic materials in the JARVIS-DFT database.\cite{choudhary2020joint}}\\

\footnotesize{Electronic band structure}
&\footnotesize{\citet{setyawan2011high}}
&\footnotesize{2011}
&\footnotesize{One of the early adaptations of high-throughput workflow in computational material science, computing the band structures for 7,439 compounds at the GGA+$U$ level for the research of scintillator materials for $\gamma$-ray radiation detection.}\\

&\footnotesize{\citet{kim2020band}} 
&\footnotesize{2020}
&\footnotesize{Band gap energies for 10,481 materials at the hybrid-GGA level of theory, improving the underestimated GGA-band gap values.}\\

&\footnotesize{\citet{kawaguchi2024band}}
&\footnotesize{2024}
&\footnotesize{Electronic band structure database for 9,004 layered intercalation
compounds.}\\

\footnotesize{Bond analysis}
&\footnotesize{\citet{naik2023quantum}}
&\footnotesize{2023}
&\footnotesize{Database of chemical bonding analysis performed on 1,520 insulators and semiconductors.} \\

\footnotesize{Excited states (GW-BSE)}

&\footnotesize{\citet{rasmussen2021towards}}
&\footnotesize{2021}
&\footnotesize{$G_{0}W_{0}$-level electronic band structures for 370 two-dimensional materials covering 14 different crystal structures
and 52 chemical elements. Seminal work that lays the foundation towards automatised workflow for high-throughput $G_{0}W_{0}$ band structure calculations for solids.}\\

&\footnotesize{\citet{varrassi2024automated}}
&\footnotesize{2024} 
&\footnotesize{Development of a fully automated workflow for $G_0W_0$ calculation that is interfaced with VASP, with which a database of QP energies for 320 bulk structures was established.} \\

&\footnotesize{\citet{gao2025pah101}}
&\footnotesize{2025}
&\footnotesize{Dataset of $GW$+BSE for 101 polyaromatic hydrocarbon molecular crystals with $\sim 500$ atoms in the unit cell.}\\

\footnotesize{Elastic constants}
&\footnotesize{\citet{deJong2015_elastic_MP}}
&\footnotesize{2015}
&\footnotesize{Database of full elastic properties for 1,181 (at the time of paper publication) inorganic crystals, that are fully integrated into the Materials Project database.}\\

&
\footnotesize{\citet{Toher2017_AFLOW_elastic}}
&\footnotesize{2017}
&\footnotesize{Based on the elastic property library that was further combined with the thermomechanical property workflow to compute the thermoelastic properties for over 3,500 materials. }\\

&\footnotesize{\citet{choudhary2018elastic}}
&\footnotesize{2018}
&\footnotesize{Database collating the elastic properties for 11,067 bulk and 257 monolayer materials.}\\

&\footnotesize{\citet{nath2021high}}
&\footnotesize{2021}
&\footnotesize{High-throughput workflow that incorporated anharmonic effects to determine the (high-temperature) thermoelastic properties of ceramics.}\\

\footnotesize{Harmonic phonons}
&\footnotesize{\citet{petretto2018high}}
&\footnotesize{2018}
&\footnotesize{Phononic properties for 1,512 semiconductors computed based on density-functional perturbation theory}
\\
&\footnotesize{\citet{Togo2018_PhononDB}}
&\footnotesize{-}
&\footnotesize{Repository of computed phonon force constants and dispersion relationships collated by the original author of the \texttt{phonopy} code.}\\

\footnotesize{Electron-phonon}
&\footnotesize{\citet{Wang2022_diboride_EPC}}
&\footnotesize{2022}
&\footnotesize{Computed electron-phonon coupling constants for a small set of ternary metal diborides, from which 43 compounds were found to show electron-phonon coupling strengths that are stronger than \ce{MgB2}.}\\

&\footnotesize{\citet{Wines2023_2D_SC_EPC}}
&\footnotesize{2023}
&\footnotesize{Screened over 1000 2D materials, electron-phonon coupling strengths and the superconducting transition temperatures were computed based on the McMillan-Allen-Dynes formula for 165 of them, from which 34 dynamically stable structures were found with transition temperatures above 5K.}\\

&\footnotesize{\citet{Gibson2025_alpha2F_database}}
&\footnotesize{2025}
&\footnotesize{Including the computation of  superconducting spectral function $\alpha^2F(\omega)$ for 818 dynamically stable materials.}\\

&\footnotesize{\citet{daSilva2025_2D_ML_EPC}}
&\footnotesize{2025}
&\footnotesize{High-throughput screening of 4,383 2D materials, from which the critical temperature for $\sim 900$ dynamically stable 2D materials were computed based on the McMillan-Allen-Dynes formula.}\\

\footnotesize{Anharmonic phonons and related}
& \footnotesize{\citet{nath2016high}}
& \footnotesize{2016}
& \footnotesize{A benchmark set containing the specific heat capacities, thermal expansion coefficients, Gr\"{u}neisen parameters and bulk moduli for 130 compounds, which is aimed to serve as a benchmark set for the computational methods for predicting these properties.}\\

&\footnotesize{\citet{van2016high}}
&\footnotesize{2016}
&\footnotesize{High-throughput screening of $\sim 300$ cubic oxide and fluoro-perovskites with lattice thermal conductivities computed based on the 2nd- and 3rd-order force constants.}\\

& \footnotesize{\citet{xia2020high}}
& \footnotesize{2020}
& \footnotesize{A systematic study of the lattice thermal conductivities for 37 binary rocksalt and zincblend compounds with the explicit inclusion of three- and four-phonon scattering processes.}\\

& \footnotesize{\citet{pal2021accelerated}}
& \footnotesize{2021}
& \footnotesize{Computational screening performed on a set of 628 quaternary chalcogenides, with low $\kappa_L$ values for  $\sim10$ candidates verified through the Boltzmann transport equation calculations.}\\

& \footnotesize{\citet{yang2021altas}, \citet{yang2022mapping}}
& \footnotesize{2022}
& \footnotesize{Mapping the room-temperature stabilities and vibrational anharmonicity landscape for $\sim 4000$ cubic single perovskites and $\sim 2000$ cubic double perovskites.}\\

& \footnotesize{\citet{li2022high}}
& \footnotesize{2022}
& \footnotesize{Developed an efficient descriptor to facilitate the screening of 1,171 candidates from Materials Projects with strong contributions from rattling modes towards low $\kappa_L$, which was computed using the empirical equation for $\kappa_L$ proposed by \citet{yan2015material} for high-throughput computation.}\\

\footnotesize{magnetic ordering}
&\footnotesize{\citet{torelli2019high}}
&\footnotesize{2019}
&\footnotesize{Identified 11 ferromagnetic insulators from Computational 2D Materials Database (C2DB)\cite{Haastrup_2018,Gjerding_2021} containing $>3,700$ entries.}\\

&\footnotesize{\citet{horton2019high}}
&\footnotesize{2019}
&\footnotesize{Reporting the magnetic ordering workflow in the original atomate\cite{mathew2017atomate} framework that is used for screening the magnetic ground states for inorganic solids in the Materials Project database.}\\

&\footnotesize{\citet{frey2020high}}
&\footnotesize{2020}
&\footnotesize{More than 27,000 magnetic orderings for over 3,000 transition metal oxides in the Materials Project database were computed to determine their magnetic ground states and estimate their effective exchange parameters and transition temperatures. A thorough band topology analysis was performed, from which 18 new candidates of ferromagnetic topological semimetals, axion insulators, and antiferromagnetic topological insulators were identified.}\\

&\footnotesize{\citet{torelli2020high}}
&\footnotesize{2020}
&\footnotesize{This study screened 4,264 potential 2D materials from which 85 and 61 materials with ferromagnetic and antiferromagnetic ground states were identified, respectively. }\\

&\footnotesize{\citet{torelli2020high}}
&\footnotesize{2020}
&\footnotesize{From screening a database containing 167,767 bulk inorganic solids, this study first identified 4,264 potentially exfoliable 2D materials, from which further screening the colinear magnetic ground states led to the identification of 85 FM and 61 AFM 2D materials. }\\

&\footnotesize{\citet{su2022high}}
&\footnotesize{2022}
&\footnotesize{This work screened 1049 transition metal compounds without oxygens to identify potential magnetic topological materials. }\\

&\footnotesize{\citet{sodequist2024magnetic}}
&\footnotesize{2024}
&\footnotesize{High-throughput screening for the magnetic ground states in 2D materials reported in the C2DB, from which 58 ferromagnets, 21 collinear anti-ferromagnets, and 85 non-collinear ground states were identified, and 15 are chiral spin spirals driven by  Dzyaloshinskii-Moriya interactions.}\\

ferroelectrics
&\footnotesize{\citet{kruse2023two}}
&\footnotesize{2023}
&\footnotesize{This work screened the C2DB from which 63 potential 2D ferroelectrics were identified. }\\

&\footnotesize{\citet{ricci2024candidate}}
&\footnotesize{2024}
&\footnotesize{This work screened 1978 polar structures from the Materials Project database, with a new workflow that compared the energy difference between the polar and nonpolar phases of a given material, this work identified a subset of 182 materials that can be potentially ferroelectric}\\

\end{longtable}
\endgroup

These computational materials databases provide transformative contributions to the research community through several complementary mechanisms, which we shall briefly summarise before ending this section. These contributions include: (a) \textit{Standardisation and reproducibility}: By curating data at consistent theory levels with identical numerical settings, these databases establish a shared baseline for validating and comparing future computational work, directly enhancing reproducibility. (b) \textit{Reducing redundancy and accelerating model developments}: Centralised database eliminates unnecessary recalculation of existing data, thus improves the research efficiency and reduces the time to generate new insights to advance scientific discoveries. This is particularly valuable  for the development of new ML/AI materials models, where prior data generation is often prohibitively expensive. The Matbench\cite{dunn2020benchmarking} initiative further exemplifies this benefit: by providing standardised datasets, evaluation metrics, and public leader-boards for model comparison, it has substantially accelerated the development of ML/AI models in materials science and enabled users to select the most appropriate tools from the growing ecosystem. (c) \textit{Consistent structural starting points}: Databased structures serve as reliable starting points for further targeted studies of defects, interfaces, heterostructures, disordered and other more complex materials systems. (d) \textit{Pattern discovery via unsupervised learning}: The chemical and structural diversity in large databases enables unsupervised methods to uncover hidden structure--property relationships,\cite{glielmo2021unsupervised,nunez2019exploring,jia2022unsupervised} generating unexpected insights that are extremely valuable for guiding the future materials design. (e) \textit{Quantitative property prediction}: Supervised ML trained on database structure-property pairs yields quantitative models that enable fast, predictive, property-driven materials discovery.\cite{bartel2018physical,mazhnik2020application} (f) \textit{Infrastructure and FAIR principles}: Database curation establishes standardised computational workflows [\cref{tab:automate2}] and formats for data storage and exchange, naturally implementing the FAIR (Findable, Accessible, Interoperable, Reusable) data principle\cite{scheffler2022fair} and opening new research horizons.

\section{Machine Learning and Artificial Intelligence for Materials Science}
\label{sect:ML_AI_Mat_Sci} 

\subsection{Overview}\label{sec:ml_overview}
The electronic structure theory method discussed in \cref{chapter:energy_models} have been the workhorse for modern materials discoveries, enabling a wide range of materials properties to be computed from first-principles [\cref{chapter:physical_properties}], that ultimately led to the establishment of various large-scale computational materials databases [\cref{chapter:high-throughput}] that revolutionised how we conduct theoretical materials research in the past decade. However, the power scaling laws of these theoretical methods (\cref{tab:scaling_low}) have fundamentally limited the system sizes that can be practically simulated, typically to a few hundred atoms for DFT and only tens of atoms for more accurate many-body perturbation theory methods such as GW. This has imposed a limit on the range of chemical and configurational spaces that can be explored. It also constrained the thermodynamic conditions $(T, P, V)$ under which we can perform statistical samplings to obtain physical observables close to experimental conditions. 

\begin{table}[htb]
    \centering
    \caption{Scaling behaviours of the theoretical methods for materials modelling discussed in this review. The system sizes are measured in terms of the number of atoms $N$ in the systems being simulated.}
    \label{tab:scaling_low}
    \begin{tabular}{llll}
    \hline\hline 
    \textbf{Method} & \textbf{Formal Scaling} & \textbf{Limiting Step} & \textbf{Practical System Sizes} \\
    \hline
    Classical forcefields & $\mathcal{O}(N)$ & for short-range interaction defined by cut-off & up to 10$^6$ \\
                         & $\mathcal{O}(N\log N)$ & for long-range Ewald summation & \\
    Pure DFT  & $\mathcal{O}(N^3)$ & matrix diagonalisation for KS eigenvalues & $\sim 10^{2}-10^{3}$\\
    Hybrid DFT  & $\mathcal{O}(N^4)$ (Na\"{i}ve) & exact exchange integral evaluation & $\sim 10^{2}$ \\
                & $\mathcal{O}(N^3)$ (Practical) & screened long-range exchange & \\
    $G_{0}W_{0}$ & $\mathcal{O}(N^4)$ & computation of the polarisation $\chi_0(\bm{r},\bm{r}^\prime)$ & $\sim 10-10^{2}$ \\
                                   && Inversion of the dielectric matrix &\\
    self-consistent $GW$ & $\mathcal{O}(N^4\sim N^5)$ & repeated calculations of $G$ and $W$ & $<50$ \\
    \hline\hline
    \end{tabular}
\end{table}

Even with the rapid advancement in HPC, particularly in GPU (Graphic Processing Unit)-accelerated electronic structure theories, there remain many challenging areas in which calculations using currently available first-principle methods are far from routine practices. These include, but are not limited to: (a) the determination of accurate thermodynamic properties for alloys, defects and disordered systems with the explicit account of the lattice vibrational contributions,\cite{van2002effect} (b) many-body physics problems such as electron-phonon interactions (\cref{sec:eph_coupling})\cite{giustino2017electron} in polar semiconductors and high-temperature superconductors, (c) free energy sampling for solvated surfaces and heterogeneous catalysis under explicit solvent at finite temperatures,\cite{gono2019effect,sakong2015density,gross2022ab} (d) vdW heterostructures of 2D materials with incommensurate lattice parameters  and Moir{\'e} pattern superlattices, which often require the use of large simulation supercells,\cite{tang2020moire} and finally, (f)  determining the properties, such as the lattice thermal conductivities\cite{allen1993thermal} and catalytic activities\cite{sun2021high} of high-entropy materials, particularly taking into account the configurational entropic contributions through ensemble averaging [\cref{eq:general_sampling_2}]. 

From a practicality perspective, although empirical atom–atom forcefields (\cref{sect:empirical_forcefield}) can efficiently simulate systems with millions of atoms, their limited transferability and accuracy often hinder predictive materials modelling across diverse chemical and structural spaces. In particular, fitting such forcefields to DFT data is generally a non-trivial process. This is because standard DFT does not decompose the total energy into physically interpretable contributions from different physical terms (such as electrostatic, dispersions and exchange-repulsions) as well as their corresponding atom-atom contributions. As a result, parameters from different forcefield terms become strongly coupled. The rigid functional forms, combined with limited training data, also yield narrow and stiff optimisation landscapes, which usually result in poor numerical stability during the parameter fitting procedures. These turn forcefield fitting into a craft rather than a robust, reproducible procedure. One remedy, for example, when fitting intermolecular interaction potentials, is to use SAPT to decompose the interaction energy into individual physical components prior to fitting [see \citet{misquitta2005intermolecular,misquitta2016ab}], although the overall workflow is still far from trivial. In this regard, the rise of big-data in materials science and the progressive wide adoption of the ML/AI techniques, which are established on a probabilistic framework that naturally encourages the introduction of noise during the learning process,  have provided new opportunities to address these challenges.

From a fundamental perspective, uncovering and understanding the complex chemistry-structure-property relationships in materials to guide their future discoveries and development is an ongoing and prominent challenge in computational materials science, particularly with the continuously growing amount of data available from energy landscape sampling (\cref{chapter:sampling_methods}) and modern high-throughput computations (\cref{chapter:high-throughput}). This is because the  chemistry-structure-property relationship is often a high-dimensional function that is difficult for human to comprehend: the spatial arrangement of $N$ atoms in a molecule/materials leads to $3N$ degrees-of-freedoms, and from the viewpoint of DFT, the energy and physical properties of an atomistic system depends non-linearly on its ground-state electron density $\rho(\bm{r})$, which is also a geometric object with complicated shape in the 3D space. Consequently, the mapping from the atomic coordinates $\{\bm{R}\}$ to the electron density $\rho(\bm{r})$ and further to the total energy $E[\rho(\bm{r})]$ and other physical properties $\mathcal{P}[\rho(\bm{r})]$ are highly non-linear and high-dimensional functions that are difficult to be expressed in closed-form equations. Traditionally, this problem was overcome by the application of heuristic rules that were established based on careful human inspections on available materials data. Examples of these heuristic rules include (a) tolerance factors for rationalising the phase stabilities of ABX$_3$-type of perovskites,\cite{Goldschmidt1926_Krystallochemie,Shannon1976_IonicRadii,Bartel2019_NewToleranceFactor} (b) Pauling's rule for rationalising the close-packing and stabilities of inorganic crystals, (c) hydrogen bonding patterns and the stabilities of molecular crystals,\cite{gavezzotti2007molecular}   (d) $d$-band theory for rationalising the performance of transition-metal-based catalysts,\cite{Hammer1995_Nature_dband} and (e) suppression of lattice thermal conductivities due to mass and structural disorders in random alloys.\cite{Abeles1963_AlloyThermalConductivity} These heuristic rules are either quantitative or semi-qualitative, simple and intuitive for post-hoc analysis, but they are often limited in their accuracy and transferability, particularly when applied to predictive investigations on material systems that are outside the original domains in which they were devised. 

In this regard, the adaptation of ML/AI techniques in materials science during the past decade has brought a paradigm shift that enriched our capabilities in performing computational modelling on a more diverse range of materials. This lies critically on the capabilities of ML/AI models to learn and generate workable mathematical relationships among the chemistry, structure, energetic stabilities and physical properties for materials curated from structure predictions and high-throughput computations with little or no human bias. Direct utilisation of these mathematical relationships learned by the models provides orders-of-magnitude acceleration in computational materials modelling while retaining the accuracy of high-level electronic structure theories, thus circumventing the limitations imposed by the scaling laws. Augmented with advanced analysis and visualisation techniques further enables us to explore these complex structure-property relationships in an intuitive and visually appealing way to enhance our fundamental understanding on the behaviours of materials at the atomistic level. Finally, recent advancements in generative AI have also opened new opportunities for inverse materials design by holistically harnessing the complex chemistry-structure-property relationships from a diverse range of materials, where the desired properties are specified as inputs to generate candidate materials structures that can meet these property requirements. These advancements are inevitably fueled by the wide range of computational materials databases available, which provide consistent baselines for model training and validations.

Before diving into the details, we shall highlight some general conceptual points regarding ML and AI, where some of them are more specifically applicable to materials science.

Firstly, it is worth noting that ML and AI are often used interchangeably in the literature, and the techniques developed for ML and AI often complement each other in practice. In this review, particularly with respect to the research in materials science, we distinguish ML and AI based on the following general characteristics. For ML, we consider it as an approach that generally relies on some hand-crafted mathematical descriptors constructed from physical insights for representing materials' structures\cite{musil2021physics} and properties. Consequently, ML algorithms typically work on small to medium datasets to generate simple (often of linear characters) models with strong physical interpretability. On the other hand, AI or deep learning (DL, which we shall be using interchangeably from now on) from raw or minimally processed data to generate latent features (analogous to descriptors from ML) that link the input with the output.  DL models, through the implementations of NNs that were designed to mimic the way information is passed and processed through the interconnected neurons in the human brains,  usually possess complex model architectures with thousands or even millions of model parameters that need to be trained on a large amount of data. Whilst the increased complexity in the DL models often leads to better generalisability and higher accuracy, their physical interpretability can be significantly compromised, and very often the trained model can only be considered as a valid \emph{ansatz} within the domain of the training data and the chosen hyperparameters for the specific model architecture.   

Secondly, parallel to ML and AI are the concepts of supervised and unsupervised learning. Built upon some material representation $\mathcal{X}$ (which can be hand-crafted for ML or learned by the DL models), \textbf{\textit{supervised learning}} tries to learn a parameterised function $y=f_{\boldsymbol{\theta}}(\mathcal{X})$ that maps the input $\mathcal{X}$ to a known output $y$ (e.g., material properties) based on a set of training data $\{(\mathcal{X}_i, y_i)\}$, where $\boldsymbol{\theta}$ are the model parameters to be optimised during the training process. On the other hand, \textbf{\textit{unsupervised learning}} aims to discover the relationships amongst the materials (especially their structures) based on the similarities between materials, which are typically measured as some distance functions based on $\mathcal{X}$. It is important to emphasise that supervised and unsupervised learning do not necessarily work in isolation, but can be  applied in conjunction with each other, for example, in refining the material representation $\mathcal{X}$ to deliver the best learning outcome for both scenarios.\cite{helfrecht2020structure} 

Finally, different from the tasks such as image recognition and language modelling in AI, there are many existing physical and chemical rules which dictate the behaviours of molecules and materials at the atomistic scales. Rather than letting the models to discover these physical rules themselves, which could exponentially increase the amount of training data required, modern ML/AI seeks to incorporate these rules into the model as an \emph{inductive bias} or \emph{physical constraints} to reduce the training cost, improve the model's stability and generalisability. However, this sacrifices the model complexities, and implementing these architectures often requires deep and specialised mathematical knowledge in areas such as group and representation theories, as well as tensor algebra, which are not in the knowledge bases of common practitioners of computational materials.  This is usually coined as \emph{physically informed} ML/AI for materials. Examples of these physical constraints include (a) translational invariance and rotational equivariance of the physical observables with respect to the atomic positions,\cite{Bartok2013_RepresentingChemEnv,grisafi2018symmetry,duval2023hitchhiker} (b) conservation of energy and momentum in molecular dynamics simulations,\cite{chmiela2017machine} (c) crystal point-group and space-group symmetry for generating stable crystal structures,\cite{jiao2024space,zhu2024wycryst} (d) long-ranged atom-atom interactions in chemical structures.\cite{grisafi2019incorporating,frank2022so3krates} These will be discussed in more detail later. 

This part of the review is intended to provide an account for the core and generic concepts behind modern ML/AI that are highly relevant to the research in materials science to facilitate the development of a fundamental understanding of the field. 
When possible, we will try to establish some insightful connections among ML/AI methodologies with traditional atomistic modellings (e.g. forcefield methods) and energy landscape samplings. More specifically, this part of the review is organised as follows. 

\cref{sec:prob_theory_fundamentals} provides a brief overview of the fundamentals of probability theory that underpins modern ML/AI techniques.\cite{murphy2012machine,goodfellow2016deep} Such a viewpoint shares many conceptual similarities to the ensemble-averaging view of the structure-property landscapes for materials [\cref{eq:general_sampling_1} and \cref{eq:general_sampling_2}]. 

\cref{sec:ml_fundamentals} introduces the core concepts behind,  and techniques commonly applied for ML, including those for supervised and unsupervised learning. 

\cref{sec:feature_eng} discusses the construction of hand-crafted descriptors for representing materials' structures. We devote this section to discussing this important aspect in ML for materials not only because it plays a pivotal role behind the success of ML, but it also provides a strong foundation for understanding the incorporation of physical constraints in DL models to enhance the model stabilities and generalisabilities. 

\cref{sec:dl_fundamentals} will be focusing on the key concepts in modern DL architectures and model training, in particular, how the model architectures mimic our conventional understanding of atom-atom interactions in molecular and materials systems. We shall also discuss the importance of latent space in DL to enhance the model interpretability for facilitating the development of fundamental understanding. In this part of discussions around DL, we aim to avoid elaborate discussion on the architectural and implementation choices for specific models, as some of most essential mathematical frameworks behind the model designs are rather consistent amongst different models being designed. 

\cref{sec:generative_AI} discusses the recent advancement in generative AI for materials design with special attention being paid to the incorporation of physical constraints in materials generation. Here, we focus specifically on the development of variational autoencoders, diffusion and normalising-flow generative frameworks. Unlike many other sections in this monograph, which aims to provide a general overview on the theoretical methodologies, we have deliberately chosen  to provide more detailed account on specific models that have been developed under each framework. This is because we feel that the model design space, encompassing interdisciplinary knowledges in structural representation, symmetry constraints, statistical sampling, stochastic dynamics, differential geometries, group theory and beyond, is enormously large, that each one of the as-developed model is a unique and intriguing combination of these aspects, with much of the development being driven by computer scientists, rather than traditional practitioners of atomistic modellings. Given the recency of generative AI models, we thus believe a more centralised detailed account on the architectural designs behind these model will facilitate the atomistic modelling community to gain a better understanding of this rapidly developing field.


\subsection{Fundamentals of Probability Theory}
\label{sec:prob_theory_fundamentals}

\subsubsection{Motivations}

Although establishing `structure-energy-property' relationships has been the overarching goal for materials modellings, most practices have been long focusing on performing calculations on a small sets of well-defined input atomistic models, to extract the corresponding energy and physical properties for each one of them at 0 K and generate some fundamental understandings therefrom. Throughout this review, based on \cref{eq:general_sampling_1} or \cref{eq:general_sampling_2}, it is one of our main intention to encourage the readers to establish a probabilistic or ensemble view and adopt the corresponding approach in the practice of materials modelling, as the probability theory forms the mathematical foundation for the development of modern ML/AI methodologies.

As we have discussed in previous sections, the probabilistic viewpoint is highly relevant to materials modelling. Fundamentally, realistic materials function at finite temperatures, and may be intrinsically disordered (such as alloys) or contains defects (such as oxygen vacancies) that are randomly distributed in the system. Performing PES sampling on material systems of interest helps to determine an ensemble of most likely accessible low-energy and stable structures under the thermodynamic conditions at which the materials will be applied. Hence computational modelling can be viewed as providing some probabilistic guidance on what the most likely structural states and corresponding physical properties are for a given material under certain thermodynamic conditions. In this regard, some of the physical properties and the theoretical approaches for computing them already are already consistent with this probabilistic view, including, for instances, (a) the configurational [\cref{eq:conf_entropy}] and vibrational [\cref{eq:vib_entropy}] entropies that are computed based on the Boltzmann distribution of microstates at given temperature $T$, (b) physical properties extracted from MD sampling [\cref{tab:properties_from_MD}], (c) stabilities and transition probabilities between different magnetic states [\cref{fig:mangetic_landscape}] and (d) the determinations of the strength of electron-phonon couplings and temperature-dependent renormalisation of electronic band energies, which needs to average over the thermally excited phononic and electronic states [\cref{sec:eph_coupling}]. At a more practical level, the numerical implementations of the theoretical methodologies can introduce noises and uncertainties, even though the underlying theories are formulated in a deterministic way, such as the random initialisations in gradient-based optimisers and the choices of different grid sizes when performing numerical integrations. These factors, sometimes, can lead to notable differences in the calculation outcomes, particually when performing statistical samplings on some very flat landscapes. 

Whilst most traditional materials modelling approaches were guided by strong physical priors or principles to determine the most likely observable(s) for a given structure, taking the above examples into considerations, a more probabilistics-based approach should be encouraged in future modelling practices, particularly to make it more compatible with the ML/AI-facilitated materials studies. This is because the latter takes an opposite approach to traditional materials modelling, as ML/AI techniques generally rely on large amounts of data to learn the relationships between the input structures and the output properties based on the probability distributions of both, and thus naturally takes into account the presences of noises in the data for making decisions with minimum or no human bias.  To facilitate the understanding of the probabilistic viewpoint in materials modelling and its connections to ML/AI, we provide a brief overview of the fundamentals of probability theory that are highly relevant to the development of modern ML/AI methodologies. Interested readers are referred to \citet{murphy2012machine} and \citet{goodfellow2016deep} for more detailed discussions.

\subsubsection{Random Variables and Common Probability Distributions} 
We begin our discussions here from the idea of random variables, which are variables that can take different values based on some underlying probability distributions that naturally encapsulate uncertainties in the data arises from incomplete knowledge and intrinsic variabilities on the system as well as noises and approximation errors introduced by the (numerical) modelling approaches that are used to generate these data. More specifically, in the context of atomistic simulations, the randomness in the data can arise from various sources, including: (a) Intrinsic variability of the system due to thermal fluctuations in the atomic positions and velocities in MD simulations through their initialisations as well as the random collisions with the heat bath, resulting in the motions of atoms being governed not only by Newtonian mechanics but also stochastic dynamics, the latter also plays an essential role in modern diffusion generative models.\cite{song2020score} (b) Incomplete knowledge arising from the finite sampling of the true physical distributions, for example, when taking the discrete approximation [\cref{eq:general_sampling_2}] to the continuous ensemble average [\cref{eq:general_sampling_1}], the result can fluctuate based on the number of samples $M$ and the specific configurations $\{\bm{R}\}_i$ that are included in the sampling. (c) Errors arise from the approximations in the theoretical methods and numerical implementations, for example, the choice of exchange-correlation functionals in DFT calculations, the basis set incompleteness errors in quantum chemistry calculations, and how charge polarisation effects are introduced in the empirical forcefield models.

The next core idea in the probability theory is the actual probability distribution function that the random variables follow. In atomistic molecular and material modellings, the most important distribution is the Boltzmann distributions. However, such distribution is fundamental only at the microstate level and we can hardly work with this full microstate distribution due to its high dimensionality and possible uncertainties in energy estimation. On the other hand, many important physical observables can be projected onto a smaller degrees-of-freedoms that can be described by simpler distributions that are much easier to sample from. For example, taking the harmonic approximation of the local energy minimum $E(\bm{R})\sim\Phi\bm{R}^2$ leads to a multimodal normal distribution $p(\bm{R})\sim \exp(-\beta\Phi\bm{R}^2)$ that is generally much easier to sample. 

For this reason, in \cref{tab:discrete_prob_distributions} and \cref{tab:continuous_prob_distributions}, we summarise some of the most commonly used discrete and continuous probability distribution functions that are widely applied in ML/AI,  with specific examples on how each distribution may be relevant to the physical observables in molecular and materials science. These distributions form the mathematical foundation for many ML/AI techniques, including those for model training and sampling. Two widely used quantities for describing the shapes of the probability distribution functions, expressed with the notations of the probability theory, are the mean (or the expectation value):
\begin{equation}
    \label{eq:expectation_value}
    \mathbb{E}_{\bm{x}\sim p(\bm{x})}[\bm{x}]=\int \bm{x}p(\bm{x})\mbox{d}\bm{x},
\end{equation}
(with which \cref{eq:general_sampling_1} can be re-written as $\langle\mathcal{A}\rangle=\mathbb{E}_{\bm{x}\sim p(\bm{x})} [\mathcal{A}(\bm{x})]$), and the variance
\begin{equation}
    \label{eq:variance}
    \mbox{Var}(\bm{x})=\mathbb{E}_{\bm{x}\sim p(\bm{x})}\Big[\big(\bm{x}-\mathbb{E}_{\bm{x}\sim p(\bm{x})}[\bm{x}]\big)^2\Big],
\end{equation}
which are usually sufficient for characterising distributions that are symmetric in shape (\emph{e.g.} normal distributions), whereas higher order moments may be used for asymmetric distributions. 

Nevertheless, normal or Gaussian distribution is arguably the most important distribution in both materials modelling and ML/AI. In the probability theory, such a distribution is often denoted as $\mathcal{N}(\mu,\sigma^2)$, with $\mu$ being the mean and $\sigma^2$ being the variance. This is because of the \emph{central limit theorem}, which states that the sum of a large number of independent and identically distributed (i.i.d.) random variables will tend to follow a normal distribution, regardless of the original distributions of these variables. This theorem justifies the wide applicability of the normal distribution in describing various physical observables that arise from the collective contributions of many microscopic factors, such as thermal fluctuations in atomic positions and velocities, electronic density variations, and so on. In ML/AI, this theorem also underpins many algorithms that assume Gaussian noise in the data or model parameters, such as Gaussian process regression,\cite{bartok2010gaussian} variational autoencoders and diffusion generative models.\cite{ho2020denoising,song2020denoising}

 \begin{longtable}{p{2cm}
    >{\raggedright\arraybackslash}p{3cm}
    >{\raggedright\arraybackslash}p{1.5cm}
    >{\raggedright\arraybackslash}p{1.5cm}
    >{\raggedright\arraybackslash}p{3cm}
    p{6cm}}
\caption{Common discrete probability distribution functions used in ML/AI. PMF stands  for the probability mass function $\mbox{Pr}(x)$ for discrete probability distributions, under which we also presents the domains over which the random variables $x$ and parameter $\theta$ are defined. (Although the parameter in some PMFs has its unique symbol, here, we denote them all with the Greek letter $\theta$ to be consistent with subsequent discussions.) We also present the method for sampling the random variables for each distribution under the Sampling column, whereas under the Notes column we provide some reference examples on which each probability distribution is commonly used for or relevant to molecular and materials modelling.} 
\label{tab:discrete_prob_distributions}\\
  
\hline\hline
\textbf{\footnotesize{Distribution}} &  \textbf{\footnotesize{PMF } } & \textbf{\footnotesize{Mean}} & \textbf{\footnotesize{Variance}} & \textbf{\footnotesize{Sampling}} & \textbf{\footnotesize{Notes}} \\
\hline
\endfirsthead 

\caption*{Table \ref{tab:discrete_prob_distributions} (Continued): Common discrete probability distribution functions used in ML/AI.}\\
\hline\hline
\textbf{\footnotesize{Distribution}} &  \textbf{\footnotesize{PMF }} & \textbf{\footnotesize{Mean}} & \textbf{\footnotesize{Variance}} & \textbf{\footnotesize{Sampling}} & \textbf{\footnotesize{Notes}} \\
\hline
\endhead
\hline\hline
\endfoot
\hline\hline
\endlastfoot

\textbf{\footnotesize{Bernoulli}} 
& \footnotesize{$\mbox{Pr}(X=x)=\theta^{x}(1-\theta)^{1-x}$

$x \in \{0,1\}$, $\theta\in[0,1]$} 
& \footnotesize{$\theta$} 
& \footnotesize{$\theta(1-\theta)$} 
& \footnotesize{Draw $u\sim$Uniform$(0,1)$, return $1$ if $u<\theta$, otherwise, return $0$} 
& \footnotesize{Probability distribution for binary events (\emph{e.g.} coin tossing): occurrence of two possible states on a lattice site in an Ising model (\emph{e.g.} magnetic ordering or the presence of vacancies); connectivities of verticies in a stochastic network that describes the charge transport paths in molecular solids;\cite{baumeier2012stochastic} polymer modelling that describe the probability of adding one of the two possible monomeric units onto the growing polymer chain;\cite{sasanuma2014configurational} kinetics of chemical reactions, describing the probability of a given reaction might proceed through different reaction channels.\cite{tian2004binomial}} \\

\textbf{\footnotesize{Binomial}}  
& \footnotesize{$\mbox{Pr}(X=x)=\binom{n}{x}\theta^{x}(1-\theta)^{n-x}$

$x \in \{0,1,\ldots,n\}$, $\theta\in[0,1]$} 
& \footnotesize{$n\theta$} 
& \footnotesize{$n\theta(1-\theta)$} 
& \footnotesize{Draw $n$ times from Bernoulli$(\theta)$ and sum the results.} 
&\footnotesize{Probability couting the success of a given event in $n$ trials: site occupancies of specific elements in a random alloy\cite{robouch2014complex} or topological defects in solids;\cite{gomez2020full} probability distribution of products of ambimodal chemical reactions, which have post-transition-state bifurcation that leads to two different products via the same transition state.\cite{tyukina2024binomial} } \\

\textbf{\footnotesize{Categorical}} 
&\footnotesize{$\mbox{Pr}(X=i)=\theta_i$

$x \in \{1,\ldots,k\}$, $\theta = \{\theta_1,\ldots,\theta_k\}$, $\sum_{i=1}^{k} \theta_i = 1$}
& \footnotesize{$\theta_i$ for one-hot, or depend on the encoding}
& \footnotesize{$\theta_i (1-\theta_i)$ per one-hot component.}
& \footnotesize{Draw $u\sim$Uniform$(0,1)$, return the smallest $i$ such that $\sum_{j=1}^{i} \theta_j \geq u$}
& \footnotesize{Probablity distributions for finding each chemical element in molecules and crystals to be applied in generative models.\cite{le2023navigating,cao2025space}} \\

\textbf{\footnotesize{Multinomial}}
& \footnotesize{$\mbox{Pr}(\bm{x})=\frac{n!}{\prod_{i=1}^{k} x_i!} \prod_{i=1}^{k} \theta_i^{x_i}$

$\bm{x} \in \mathbb{N}^{k}$, $\sum_{i=1}^{k} x_i = n$, $\theta = \{\theta_1,\ldots,\theta_k\}$, $\sum_{i=1}^{k} \theta_i = 1$}
&   \footnotesize{$n \theta_i$ 

per component}
&   \footnotesize{$n \theta_i (1-\theta_i)$ per component}
&   \footnotesize{Draw $n$ times from Categorial$(\theta)$ and count the occurrences for each category.}
& \footnotesize{In the task of molecular generation, the multinomial distribution describes the probability of locating a substructure $k\in\{1,2,\ldots,K\}$ at the $\nu$-th $(\nu\in \{1,2,\ldots,N\})$ position in a molecule.\cite{fu2021mimosa}} \\

\textbf{\footnotesize{Poisson}}
& \footnotesize{$\mbox{Pr}(X=x)=\frac{\theta^{x} e^{-\theta}}{x!}$

$x\in \{0,1,2,\ldots\}$}
& \footnotesize{$\theta$}
& \footnotesize{$\theta$}
& \footnotesize{Standard Poisson sampler.}
& \footnotesize{Describes the probability of occurrence of random events within a fixed time interval or objects within a fixed region: occurrence of a number of random cluster-defects in an arbitrary area $(A)$ in the electron microscopic image;\cite{chandran2005duality} probability distributions of rare events in MD sampling;\cite{kroger2017assessing} dynamics of chemical reaction networks, describing the probability distributions for a series of random observations occur in time;\cite{anderson2011continuous} nucleation process of crystal, describing the distribution of crystal nuclei within any given time interval.\cite{jiang2011crystal} }
\end{longtable}

\begin{longtable}{p{2cm}
    >{\raggedright\arraybackslash}p{3.5cm}
    >{\raggedright\arraybackslash}p{1.0cm}
    >{\raggedright\arraybackslash}p{2cm}
    >{\raggedright\arraybackslash}p{2.5cm}
    p{6cm}}
\caption{Common continuous probability distribution functions used in ML/AI. PDF stands for probability density function $f(x)$ as a function of the random variable $x$.} 
\label{tab:continuous_prob_distributions}\\
  
\hline\hline
\textbf{\footnotesize{Distribution}} &  \textbf{\footnotesize{PDF } } & \textbf{\footnotesize{Mean}} & \textbf{\footnotesize{Variance}} & \textbf{\footnotesize{Sampling}} & \textbf{\footnotesize{Notes}} \\
\hline
\endfirsthead 

\caption*{Table \ref{tab:continuous_prob_distributions} (Continued): Common continuous probability distribution functions used in ML/AI.}\\
\hline\hline
\textbf{\footnotesize{Distribution}} &  \textbf{\footnotesize{PDF }} & \textbf{\footnotesize{Mean}} & \textbf{\footnotesize{Variance}} & \textbf{\footnotesize{Sampling}} & \textbf{\footnotesize{Notes}} \\
\hline
\endhead
\hline\hline
\endfoot
\hline\hline
\endlastfoot

\footnotesize{\textbf{Uniform}} 
& \footnotesize{$f(x|a,b)=\frac{1}{b-a}$, $x\in[a,b]$}
& \footnotesize{$\frac{a-b}{2}$}
& \footnotesize{$\frac{(b-a)^2}{12}$}
& \footnotesize{Draw $u\sim$Unifrom(0,1) and set $x=a+(b-a)u$.}
& \footnotesize{Wang-Landau or `flat-histogram' MC sampling of multicanonical ensemble of a physical system;\cite{wang2001efficient} initial distributions of positions and orientations in structure predictions;\cite{case2016convergence} initial particle positions and velocities for simulating nucleation in MD;\cite{diemand2013large} enumerating the configurations of site-disordered solids;\cite{holland2023workflow} uniform sampling of the torsional angles for molecular conformation generation;\cite{volokhova2024towards} combined with wrapped-normal distributions for sampling the fractional coordinates in generative models for crystal structures.\cite{luo2025crystalflow} }\\

\footnotesize{\textbf{Normal (Gaussian)}} 
& \footnotesize{$\mathcal{N}(x|\mu,\sigma^2)=\frac{1}{\sigma\sqrt{2\pi}}\exp\left[-\frac{(x-\mu)^2}{2\sigma^2} \right]$, $x\in\mathbb{R}$}
& \footnotesize{$\mu$}
& \footnotesize{$\sigma^2$}
& \footnotesize{Sample standard normal $z$ via Box-Muller method and set $x=\mu+\sigma z$.}
& \footnotesize{Most widely used distributions; basis function for molecular DFT calculations (\cref{sect:practical_DFT}); sampling initial particle velocities for MD simulations at a given temperature (\cref{sect:initial_velocity}); representing atomic densities in structural descriptors for atomistic ML;\cite{bartok2013representing} prior distributions in the Gaussian Process for, \emph{e.g.} constructing ML-based interatomic potentials;\cite{bartok2010gaussian} sampling noises in diffusion-based generative models for materials.\cite{zeni2025generative} Fundamentally, the log-likelihood constructed based on normal distribution leads to the L2-loss function (in the form of $|\bm{y}-\hat{\bm{y}}|^2$) in model training. }\\

\footnotesize{\textbf{Wrapped Normal}} 
& \footnotesize{$p(x|x_0)\propto\sum_{d\in\mathbb{Z}}\exp\left[-\frac{(x-x_0+p d)^2}{2\sigma^2}\right]$}
& \footnotesize{$x_0$}
& \footnotesize{$\sigma^2$}
& \footnotesize{Draw a sample from standard normal distribution and mapped it back to the domain in which $x$ is defined by removing the appropriate amount of  periodicity $p$.}
&\footnotesize{Fractional coordinates in crystals,\cite{zhu2024wycryst} and torsional angles of flexible molecules.\cite{jing2022torsional}}\\

\footnotesize{\textbf{Log-Normal}} 
& \footnotesize{$f(x|\mu,\sigma^2)=\frac{1}{x\sigma\sqrt{2\pi}} \exp\left[ -\frac{(\ln x-\mu)^2}{2\sigma^2}\right]$, $x>0$}
& \footnotesize{$e^{\mu+\frac{\sigma^2}{2}}$}
& \footnotesize{$e^{2\mu+\sigma^2}(e^{\sigma^2}-1)$}
& \footnotesize{Sample $z\sim\mathcal{N}(\mu,\sigma^2)$ and set $x=e^z$.}
& \footnotesize{Distribution of grain sizes in polycrystalline materials;\cite{abraham2023crystal} distribution of electronic conductance in single-molecular junctions.\cite{mejia2023microscopic}}\\

\footnotesize{\textbf{Exponential}} 
& \footnotesize{$f(x|\lambda)=\lambda e^{-\lambda x}$, $x\geq 0$}
& \footnotesize{$\frac{1}{\lambda}$}
& \footnotesize{$\frac{1}{\lambda^2}$}
& \footnotesize{Draw $u\sim$Uniform(0,1) and set $x=-\frac{1}{\lambda}\ln(1-u)$.}
& \footnotesize{Commonly used for describing the dynamics of systems. In surface diffusions, it describes the probability of finding an end state after some time $t$;\cite{andersen2019practical} in crystal nucleation, it describes the tail of the distribution for the first passage time to reach the critical nucleus size\;cite{nicholson2016analysis} in the Markov state models for molecular conformation dynamics,the dwell time is approximately exponentially distributed;\cite{chodera2014markov} distribution of the residence time of molecules adsorbed on surfaces.\cite{turchi2025segmentation}}\\

\footnotesize{\textbf{Gamma}} 
& \footnotesize{$f(x|k,\theta)=\frac{1}{\Gamma(k)\theta^k}x^{k-1}e^{-x/\theta}$, $x>0$}
& \footnotesize{$k\theta$}
& \footnotesize{$k\theta^2$}
& \footnotesize{Summed exponentials or use special library.}
& \footnotesize{In heterogeneous diffusions, the distribution of the squared displacements follows the Gamma distribution.\cite{matsuda2018estimation}}\\

\footnotesize{\textbf{Laplace}} 
& \footnotesize{$f(x|\mu,b)=\frac{1}{2b}\exp\left(-\frac{|x-\mu|}{b}\right)$, $x\in\mathbb{R}$}
& \footnotesize{$\mu$}
& \footnotesize{$2b^2$}
& \footnotesize{Draw $u\sim$Uniform(0,1), if $u<0.5$, then $x=\mu+b\ln (2u)$ else $x=\mu-b\ln(2(1-u))$.}
& \footnotesize{The log-likelihood built on this distribution leads to the L1 loss function (in the form of $|\bm{y}-\hat{\bm{y}}|$), which is often used (as the mean-absolute-error) in training ML-potentials,\cite{morrow2023validate} compress-sensing for lattice dynamics\cite{zhou2019compressive} and cluster expansion model.\cite{nelson2013compressive}}\\

\footnotesize{\textbf{Logistic}} 
& \footnotesize{$f(x|\mu,s)=\frac{\exp\left(-\frac{x-\mu}{s}\right)}{s\left(1+\exp\left(-\frac{x-\mu}{s}\right)\right)^2}$, $x\in\mathbb{R}$}
& \footnotesize{$\mu$}
& \footnotesize{$\frac{(\pi s)^2}{3}$}
& \footnotesize{Draw $u\sim$Uniform(0,1) and set $x=\mu+s\log\frac{u}{1-u}$.}
& \footnotesize{The PDF for Laplace distribution has the same functional form for the derivative of Fermi-Dirac distribution, which gives the weight of electron energies to the electron conductivity;\cite{madsen2006boltztrap} in chemical kinetic analysis, it describes the distribution of product population over time.\cite{burnham2017use}}\\

\footnotesize{\textbf{Dirichlet}} 
& \footnotesize{$f(\bm{x}|\bm{\alpha})=\frac{1}{B(\bm{\alpha})}\prod_{j=1}^{K} \bm{x}_i^{\alpha_i -1}$, $\bm{x}\in\Delta^{K-1}$ (simplex), $\bm{\alpha}\in (\alpha_1,\ldots,\alpha_K)$}
& \footnotesize{$\mathbb{E}[x_i]=\frac{\alpha_i}{\alpha_0}$}
& \footnotesize{$\frac{\alpha_i(\alpha_0-\alpha_i)}{\alpha_0^2(\alpha_0+1)}$, $\alpha_0=\sum_{j=1}^{K} \alpha_j$}
& \footnotesize{Sample $y_i\sim\mbox{Gamma}(\alpha_i,1)$ and set $x_i=\frac{y_i}{\sum_{j=1}^{K} y_j}$.}
& \footnotesize{Distribution of chemical composition in multicomponent alloys,\cite{clausen2023ab} can be used for facilitating the sampling of chemical elements in generative models.}\\ 
\end{longtable}
 
\subsubsection{Dealing with Multidimensional Joint Probability Distributions}
\begin{figure}[tbh]
    \centering
    \includegraphics[width=0.8\linewidth]{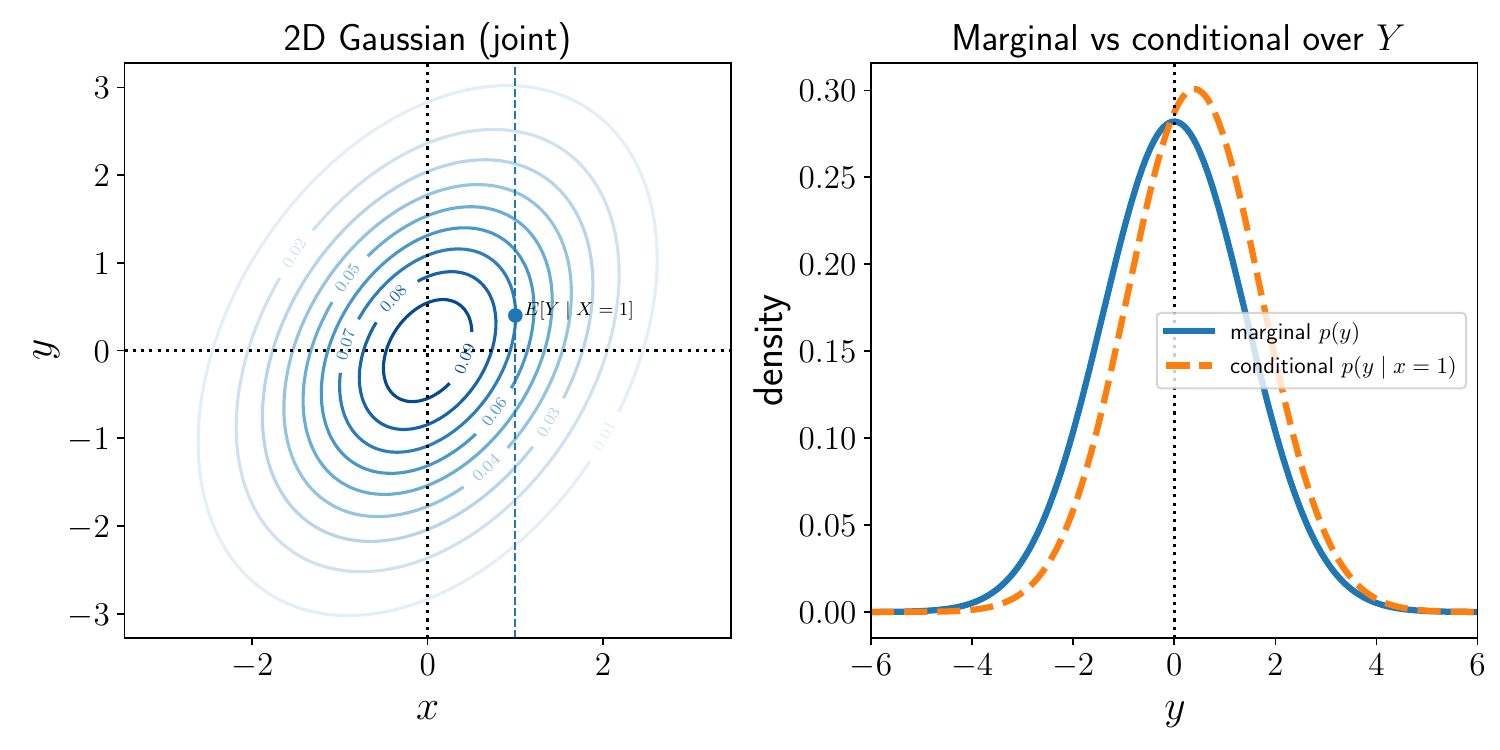}
    \caption{(Left) Contour plot illustrating the shape of the joint probability distribution function of a two-dimensional Gaussian $p(x,y)=\mathcal{N}(\bm{\mu},\bm{\Sigma})$ with mean at the origin and a covariance matrix of non-zero off-diagonal elements.  (Right) Showing the corresponding one-dimensional conditional distribution $p(y|x=1)$ (blue dashed line on the left figure, the blue dot on which indicates the conditional mean $\mathbb{E}[y|x=1]$) and the marginal distribution $p(y)$ that is marginalised over $x$ which is centred at $y=0$. }
    \label{fig:2D_Gaussian_demo}
\end{figure}

In reality, the physical behaviours of a system often depend on multiple (multidimensional) random variables $\{\bm{x}_1,\bm{x}_2,\ldots,\bm{x}_n\}$, such as the total energy of an atomistic system expressed in terms of the three-dimensional Cartesian coordinates of each constituent atoms $E(\bm{R}_1,\bm{R}_2,\ldots,\bm{R}_n)$. In this case, the probability that a system will take a particular state is described by the \textit{joint probability distribution} $p(\bm{x}_1,\bm{x}_2,\cdots\bm{x}_n)$, which is the probability that all the random variables will simultaneously fall into a set of given values [see \cref{fig:2D_Gaussian_demo}(left) for an illustration for a simple two-dimensional joint Gaussian distribution].  For atomistic systems, local minima on the PES correspond to local maxima of the joint probability distribution. It is also a common case, that for a system described by the joint probability distribution   $p(\bm{x}_1,\bm{x}_2,\cdots,\bm{x}_n)$, the value that each variable $\bm{x}_i$ will take is no longer independent by become highly correlated with the values of other variables $\{\bm{x}_j|j\neq i\}$. For instance, the distributions of dihedral angles in a polypeptide is highly correlated in order to avoid stereo-hinderance, which is commonly demonstrated by the Ramachandran plot. \cref{fig:2D_Gaussian_demo}(left) shows this behaviour with a 2D Gaussian that is not symmetric with respect to either the $x$- or $y$-axis, for which the variance is described by a matrix $\bm{\Sigma}$ where its diagonal element corresponds to the variance in the $i$-th dimension $\mbox{Var}[\bm{x}_i]$, and the off-diagonal element corresponds to the \emph{covariance}  $\mbox{Cov}[\bm{x}_i,\bm{x}_j]=\mathbb{E}\Big[[\bm{x}_i-\mathbb{E}[\bm{x}_i]][\bm{x}_j-\mathbb{E}[\bm{x}_j]]\Big]$ that measures the degree of correlation between the $i$- and $j$-th random variables. Physically, it is conceptually similar to quantities such as the off-diagonal terms in the force-constant matrix [\cref{eq:FC_finite_difference}] which reflects how strongly the simultaneous perturbation on two atomic positions will shift the system energy.\footnote{To see this, we take the harmonic approximation to the local PES as $E(\bm{x})=\frac{1}{2}\bm{x}^\top \bm{\Phi}\bm{x}$, such that under the canonical ensemble, the corresponding probability distribution function for the atomic displacements follows $p(\bm{x})=\exp(-\beta \bm{x}^\top \bm{\Phi}\bm{x}/2)$, which is a normal distribution with its covariance matrix being $\bm{\Sigma}=\bm{\Phi}^{-1}/\beta$. This shows that the covariance matrix of fluctuations in the atomic displacements is proportional to the inverse of the Hessian. More specifically, we have $\bm{\Sigma}(i,j)=\bm{\Phi}_{ij}^{-1}/\beta$, thus demonstrating the off-diagonal elements for the force-constant matrix is related to the covariance of the probability distribution.}

Directly working with high-dimensional joint probability distributions is often intractable. This can be solved by dealing with subsets of the joint probability distributions, for which the two concepts that are most relevant here are the conditional and marginal probability distributions [\cref{fig:2D_Gaussian_demo}(right)]. The conditional probability  $p(\bm{x}|\bm{\theta})$ describes the probability of occurrence for a particular event $\bm{x}$ when some other variables or parameters that might affect the shape of the probability distribution function is fixed at a specific choice $\bm{\theta}$. This gives us the language to formulate the probabilistic equivalent to forcefield training [\cref{eq:FF_fitting}] for training the ML/AI model, which is a process of optimising the parameters $\bm{\theta}$ until the conditional probability distribution $p_{\mbox{\scriptsize{tr}}}(\bm{x}|\bm{\theta})$ described by the model reaches the best match to the target distribution $p_{\mbox{\scriptsize{target}}}(\bm{x})$, and we shall further consolidate this idea in the subsequent discussions. Fundamentally, the joint  and conditional distributions are related as:
\begin{equation}
    \label{eq:joint_cond_prob}
p(\bm{x}_1,\bm{x}_2,\ldots,\bm{x}_n)=p(\bm{x}_1)\prod_{i=2}^n p(\bm{x}_i|\bm{x}_{i-1}).
\end{equation}
The marginal probability, on the other hand, is the probability distribution over a subset of random variables with the rest of the random variables being integrated out, \emph{e.g.}
\begin{equation}
    \label{eq:marginal_distribution}
    p_{X}(\bm{x})=\int p_{X,Y}(\bm{x},\bm{y})\mbox{d}\bm{y}.
\end{equation}
This follows that
\begin{equation*}
    p_{Y|X}(\bm{y}|\bm{x})=\frac{p_{X,Y}(\bm{x}|\bm{y})}{p(\bm{x})},
\end{equation*}
which is an expression that is most commonly used in the \textbf{\textit{Bayes' rule}} underpinning the modern probabilistic interpretations of ML/AI:
\begin{equation}
    \label{eq:bayes_rule}
    p(\bm{y}|\bm{x})=\frac{p(\bm{x}|\bm{y})p(\bm{y})}{p(\bm{x})}.
\end{equation}
For example, during training of an ML/AI model, one typically starts by initialising the model with a set of parameters $\mathbf{\theta}$ that are drawn from some prior belief on their values $p(\mathbf{\theta})$, and the likelihood that that training data $\bm{x}$ will be observed under these parameters is given by the conditional distribution $p(\bm{x}|\theta)$. After the model observed the training data given, the posterior probability distributions for the parameters will be updated according to \cref{eq:bayes_rule} as $p(\theta|\bm{x})=p(\bm{x}|\theta)p(\theta)/p(\bm{x})$, whereby parameters drawn from this updated distributions will lead to a model with better performance of predicting the outcome on the training data $\bm{x}$.

\subsubsection{Connection to Information Theory}

To build up the relationship among the probability theory, statistical thermodynamics and ML/AI model training, we need to introduce some of the basic concepts of the information theory, which provides us with the mathematical language to measure the information content, its associated uncertainties and how to reliably communicate information under the presence of noises, as one of the primary goal for developing ML techniques is to reduce and filter out the uncertainties in the information through model training. 

In the language of information theory, the \emph{information entropy} (or the \emph{Shannon entropy)} provides the mathematical measurement of the information uncertainty associated with the probability distribution of the information. For a continuous distribution, the information entropy is defined as
\begin{equation}
    \label{eq:information_entropy}
    S=-\int p(\bm{x})\log p(\bm{x})\mbox{d}\bm{x}=\mathbb{E}_{\bm{x}\sim p(\bm{x})}\left[-\log p(\bm{x})\right].
\end{equation}

\begin{figure}[tbh]
\centering
\includegraphics[width=0.5\linewidth]{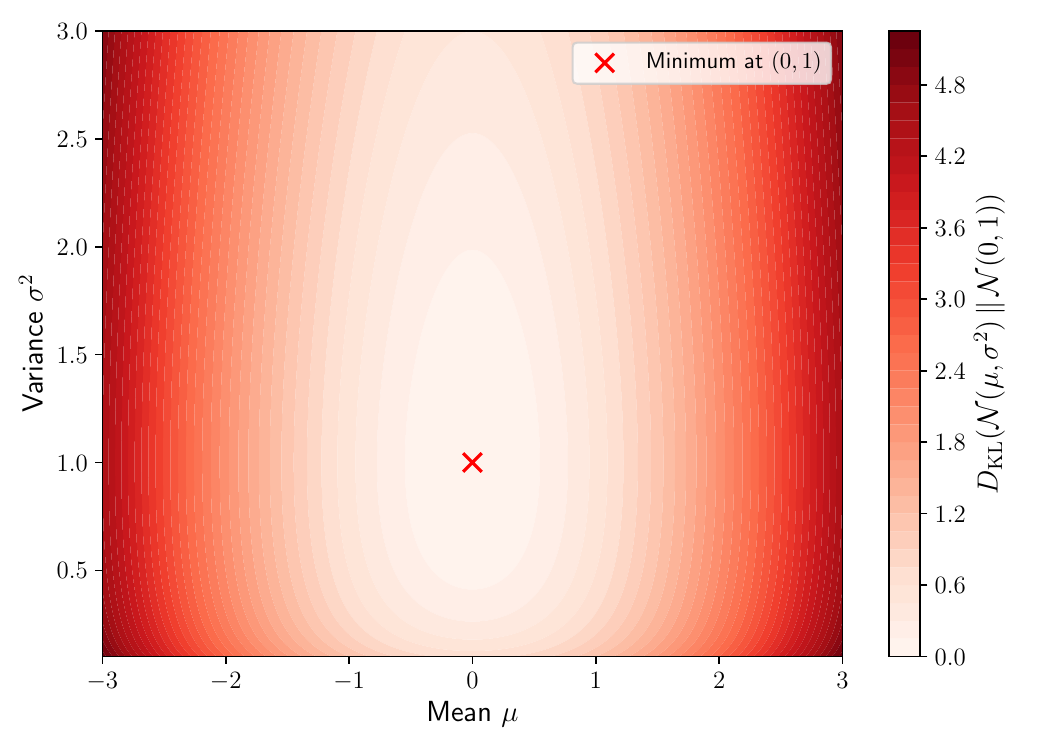}
\caption{Landscape of the KL divergence between a parameterised one-dimensional normal distribution $\mathcal{N}(\mu,\sigma^2)$ and a standard normal distribution $\mathcal{N}(0,1)$ as a function of the mean ($\mu$) and variance ($\sigma^2$).} 
\label{fig:KL_div_normal} 
\end{figure}

With the idea of information content, training a ML/AI model is a \textit{process that maximises the information content overlap between the model and the training data}, and this information overlap is measured in terms of the dissimilarity between two probability distributions $p(x)$ and $q(x)$, that can be quantified using the \textit{Kullback–Leibler (KL) divergence}:
\begin{equation}
    \label{eq:KL_divergence}
    \mathcal{D}_{KL}(p||q)=\sum_{i=1}^{N}p_i\log\frac{p_i}{q_i}\qquad\mbox{or}\qquad \mathcal{D}_{KL}(p||q)=\int p(x)\log\frac{p(x)}{q(x)}\mbox{d}x.
\end{equation}
If one of the distribution is modelled by a ML/AI model with parameter(s) $\boldsymbol{\theta}$, then the model training is essentially a process to minimise the KL-divergence between the target and model distributions with respect to $\boldsymbol{\theta}$:
\begin{equation}
    \label{eq:training_KL_minimise}
    \hat{\boldsymbol{\theta}}_{\mathrm{opt}} = \argmin_{\boldsymbol{\theta}}\mathcal{D}_{KL} \left(p_{\mathrm{target}}|q_{\boldsymbol{\theta}}\right),
\end{equation}
and to simplify the training, many model tend to choose simple distributions $q_{\boldsymbol{\theta}}$ such as those listed in \cref{tab:discrete_prob_distributions} and \cref{tab:continuous_prob_distributions} to derive analytical expressions for $\mathcal{D}_{KL}$ that are easy to compute. We shall provide more concrete examples of applying this in the following sections.  A simple example is shown here in \cref{fig:KL_div_normal} which shows how KL-divergence changes between parameterised one-dimensional normal distribution $\mathcal{N}(\mu,\sigma^2)$ with respect to the standard normal distribution $\mathcal{N}(0,1)$ as a function of the mean and variance.

Alternatively, in many ML/AI literatures, instead of minimising the KL-divergences between two probability distributions, the problem is often reformulated into an opposite problem of \emph{maximising the log-likelihood} between two distributions. By expanding
\begin{equation}
    \label{eq:DKL_expand}
    \mathcal{D}_{KL}(p||q)=\sum_i p_i\log p_i-\sum_i p_i\log q_{i,\boldsymbol{\theta}},
\end{equation}
and realising that the first term on the right hand side is essentially a constant, the optimal model parameter $\boldsymbol{\theta}$ can therefore be obtained equivalently by maximising the log-likelihood function $\sum_i p_i\log q_{i,\boldsymbol{\theta}}$.


\subsubsection{PES Sampling in the Context of Probability Theory}
\label{sec:ML_AI_PES_sampling_info_theory}

With the fundamentals of probabilistic theory introduced above, we shall briefly apply this probabilistic viewpoint to establish the connection with the PES samplings [\cref{chapter:sampling_methods}] to deepen our understanding of the abstract theory of probability. 

\begin{figure}[tbh]
    \centering
    \includegraphics[width=\linewidth]{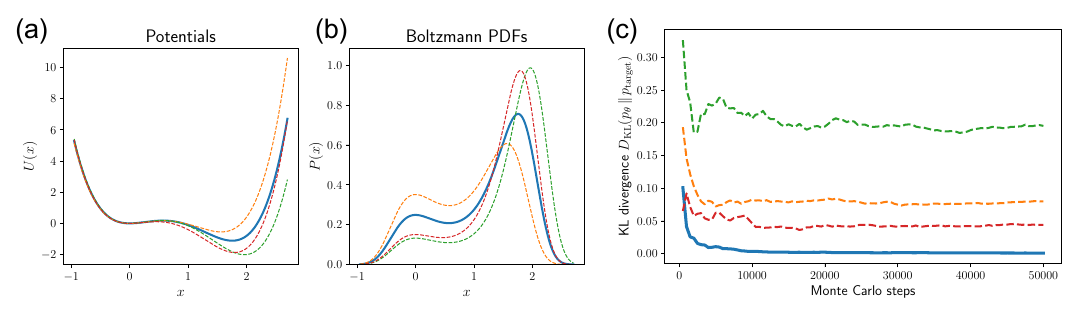}
    \caption{Probabilistic view on potential energy landscape sampling using 1D toy examples. (a) Shapes for the ground-truth potential energy surface $U(x)$ (blue solid line) and its three parameterised approximations $U_\theta (x)$, mimicking either traditional or ML/AI atom-atom potentials. (b) Corresponding probability distribution functions (PDFs) for each potential as defined by the Boltzmann distribution $P(x)=e^{-\beta U(x)}$, with $\beta=1$ a.u. (c) Building the underlying PDFs for $U(x)$ and $U_\theta (x)$ with Monte Carlo (MC) sampling, which will be dubbed as $P(x)$ and $P_\theta(x)$, respectively. This plot shows the evolution of the KL divergence ($D_{KL}$ of $P(x)$ or $P_\theta(x)$ with respect to the ground-truth PDF of  $P(x)$ as a function of the MC steps. If the sampling process follows $U(x)$, then $D_{KL}$ will eventually converge to zero, whereas for samplings performed on $U_\theta(x)$, the converged $D_{KL}$ will always be greater than one, where its absolute value depends on how close $U_\theta (x)$ is to $U(x)$ in its shape.}
    \label{fig:dummy_MC_sampling}
\end{figure}

\paragraph{\textbf{Energy model training}} We begin by examining the idea of training the energy models for atomistic simulations The basic idea is schematically illustrated with a one-dimensional toy model in \cref{fig:dummy_MC_sampling}. In an idealised scenario, we would like to compute the ground-truth potential energy function $U(x)$ [solid line in \cref{fig:dummy_MC_sampling}(a)] for a given material by solving the full many-body Schr\"{o}dinger equation, taking into account all the intricate physical interactions in that material. From here, we can easily compute the ground-truth PDF for this potential energy function as $p(x)=\exp[-\beta U(x)]$ analytically as the Boltzmann distribution[\cref{fig:dummy_MC_sampling}(b)], or through MC sampling, in which as the MC progress, we gradually build up our estimated density distribution. By computing the KL divergence of accumulated density at each MC step with respect to  $p(x)$, \cref{fig:dummy_MC_sampling}(c) shows that the KL divergence will eventually converge to zero, means the PES has been thoroughly sampled.

However, practically, we have to resort to some parameterised approximation $U_\theta (x)$ [dashed lines in \cref{fig:dummy_MC_sampling}(a)] to mimic the ground-truth PES $U(x)$, for example, using empirical forcefields, DFT, or ML/AI-based interatomic potentials. Correspondingly, we can also compute the approximate PDF as $p_\theta (x)=\exp[-\beta U_\theta (x)]$ [\cref{fig:dummy_MC_sampling}(b)], and perform MC sampling on this approximate PES to build up the estimated density distribution. In this case, as shown in \cref{fig:dummy_MC_sampling}(c), the KL divergence of the accumulated density with respect to the ground-truth PDF $p(x)$ will converge to a finite value greater than zero, which depends on how optimised the parameters $\theta$ are in terms of reconstructing the shape of $U(x)$ with $U_\theta (x)$. In realistic atomistic modelling, the problem of $U(x)$ being unknown is usually circumvented by using a high-level theory (such as hybrid-DFT and wavefunction method) to provide baseline or reference energies $U_{\theta}^{\mbox{\scriptsize{ref}}}(x)$ for optimising the parameters $\theta$ in $U_\theta (x)$ for lower-level approximations (such as forcefield and ML/AI models).  This is usually achieved by minimising some loss functions constructed based on $p_{\theta}^{\mbox{\scriptsize{ref}}}(x)$ and $p_\theta(x)$. In this regard, \textit{$U_{\theta}^{\mbox{\scriptsize{ref}}}(x)$ sets the `ceiling' for the accuracy that could be achieved by $U_\theta(x)$, and to what extent $x$ covers representative configurations on the PES (i.e. the size of the training set), will determine the generalisability of $U_\theta(x)$.} This is an important aspect to be mindful of when assessing the capabilities of ML/AI models.

In terms of the statistical sampling of configurations $\bm{R}$ from the PES (in which we use $\bm{R}$ here to denote the collection of all nuclear positions in the system), the probabilistic viewpoint also provides useful insights. In the language of probabilistic theory, what we would like to achieve from structural sampling is to directly draw a random variable $\bm{R}$ from some abstract space $\Omega$ that contains all possible configurations that a system might possess. The probability distribution $p(E(\bm{R}))$ of drawing $\bm{R}$ from $\Omega$ depends on the underlying potential energy model $E(\bm{R})$ and the statistical ensemble [\cref{tab:thermodynamic_ensembles}] that we are working with. 
However, since we do not have direct access to $\Omega$, $p(E(\bm{R}))$ and more importantly, the knowledge of how to inversely map the total system energy $U$ back to $\bm{R}$, we have to utilise the fundamental physics that governs the atom-atom interactions to sample the underlying PES to build up $p(E(\bm{R}))$ in traditional approach of PES sampling. On the other hand, as well shall see in \cref{sec:generative_AI}, modern generative AI models try to learn the mapping between some simple prior distributions (such as normal and uniform distributions) to $p(E(\bm{R}))$ based on large amounts of training data, and then directly draw $\bm{R}$ from $p(E(\bm{R}))$ without explicitly sampling the underlying PES. 

\paragraph{\textbf{Nested sampling and constrained MD}}
Next, we shall revisit the concepts of nested sampling and constrained MD discussed in \cref{sec:nested_sampling} in order to provide some further physical context behind the definitions of conditional and marginal probability distributions. Under the canonical ensemble, the purpose of performing PES sampling, via, \emph{e.g.} MD simulations, is to obtain the probability distribution 
\begin{equation}
    \label{eq:prob_full_sampling}
p(E(\bm{R}))=\pi(\bm{R})= \exp[-\beta E(\bm{R})]/\mathcal{Z},
\end{equation} 
from which the system entropy and free energy can be computed. To describe a chemical reaction, we shift our focus onto a subset of coordinates $\zeta\in\bm{R}$ that characterises the reaction progress, such as a bond length or dihedral angle. Further denoting $\bar{\zeta}=\bm{R}-\zeta$, the \cref{eq:prob_full_sampling} can be recasted into a form of joint probability distribution 
\begin{equation}
    \label{eq:joint_prob_full_sampling}
\pi(\zeta,\bar{\zeta})= \exp[-\beta E(\zeta,\bar{\zeta})]/\mathcal{Z}.
\end{equation} 
From here, the free energy profile along the reaction coordinate $\zeta$ can be computed via the marginal probability distribution
\begin{equation}    
\label{eq:marginal_prob_sampling}
\pi(\zeta)=\int \pi(\zeta,\bar{\zeta}) \mbox{d}\bar{\zeta}
\end{equation}
Determining this integrand can be achieved through constrained molecular dynamics, which effectively samples the conditional probability distribution $\pi(\zeta|\bar{\zeta})$ that leads to the integrand $\langle \partial\mathcal{H}/\partial {\zeta}\rangle_{\zeta^\prime}^{\mbox{\scriptsize{cond.}}}$ in \cref{eq:thermodynamic_integration}.\footnote{To show this, we start from the marginal probability distribution
$
    \pi(\zeta)=\frac{1}{\mathcal{Z}}\int\exp\left[-\beta E(\zeta,\bar{\zeta})\right]\mbox{d}\bar{\zeta},
$
differentiate both sides with respect to $\zeta$ gives
$
    \frac{\partial\pi(\zeta)}{\partial\zeta}=\frac{1}{\mathcal{Z}}\int\left(-\beta\frac{\partial E(\zeta,\bar{\zeta})}{\partial \zeta}\right)\exp\left[-\beta E(\zeta,\bar{\zeta})\right]\mbox{d}\bar{\zeta},
$
and 
$
    \frac{1}{\pi(\zeta)}\frac{\partial\pi(\zeta)}{\partial\zeta}=-\beta\int\frac{1}{\pi(\zeta)}\frac{e^{-\beta E(\zeta,\bar{\zeta})}}{\mathcal{Z}}\left(-\frac{\partial E(\zeta,\bar{\zeta})}{\partial \zeta}\right)\mbox{d}\bar{\zeta}.
$
Now using the relationship 
$
\pi(\bar{\zeta}|\zeta)=\frac{\pi(\zeta,\bar{\zeta})}{\pi(\zeta)}=\frac{1}{\pi(\zeta)}\frac{\exp\left[-\beta E(\zeta,\bar{\zeta})\right]}{\mathcal{Z}}
$
such that 
$
    \frac{1}{\pi(\zeta)}\frac{\partial\pi(\zeta)}{\partial\zeta}=-\beta\int\pi(\bar{\zeta}|\zeta) \frac{\partial E(\zeta,\zeta^\prime)}{\partial \zeta}\mbox{d}\zeta^\prime=-\beta\left\langle \frac{\partial E}{\partial\zeta}\right\rangle_{\zeta^\prime},
$
by using the definition of expectation value.
}

\paragraph{\textbf{(Information) entropy for a simple quantum harmonic oscillator}}
As the third example, we examine the definition of information entropy [\cref{eq:information_entropy}]. To put such definition into a more physical context, and establish its linkage to the concept of entropy in statistical physics, we consider the problem of vibrational entropy of a simple quantum harmonic oscillator. For this simple physical system, its energy eigenspectrum is $E_n=\left(n+\frac{1}{2}\right)\hbar\omega$, with $\omega$ being the characteristic frequency and $n=0,1,2,\ldots$. This gives the probability distribution function $p_n=\exp\left[-\beta\hbar\omega \left(n+\frac{1}{2}\right)\right]/\mathcal{Z}$ for finding an oscillator in its $n$-th quantum state, with the partition function 
\begin{equation*}
    \mathcal{Z}=\sum_{n=0}^{\infty}\exp\left[-\beta\hbar\omega \left(n+\frac{1}{2}\right)\right]=e^{-\beta\hbar\omega/2}\left(1-e^{-\beta\hbar\omega}\right),
\end{equation*}
that can be easily determined as the infinite sum of a converged geometric series. This leads to
\begin{equation*}
    p_n =\left(1-e^{-\beta\hbar\omega}\right)e^{-n\beta\hbar\omega}.
\end{equation*}
In statistical mechanics, the entropy for a system with microstate $i$ and probability $p_i$ [equivalent to \cref{eq:information_entropy}] is 
\begin{equation}
    \label{eq:entropy_stat_mech}
    S=-k_B\sum_i p_i\ln p_i.
\end{equation}
To simplify the notations, by letting $q=e^{-\beta\hbar\omega}$, we have $p_n=(1-q)q^n$, and $\ln p_n=\ln(1-q)+n\ln q$, we obtain the following alternative form for \cref{eq:entropy_stat_mech}:
\begin{equation}
    \label{eq:entropy_stat_mech_har_oscillator}
    S=-k_B\sum_n \left(p_n\ln (1-q)+np_n\ln q\right)=-k_B[\ln(1-q)+\langle n\rangle q],
\end{equation}
in which we have used the fact that $\sum_n p_n=1$ (the normalisation condition) and $\sum_n np_n=\langle n\rangle =\left[e^{\beta\hbar\omega}-1\right]^{-1}=q/(1-q)$ (expected number of excited oscillator at $T$). The latter expression gives $q=\langle n\rangle/(\langle n\rangle +1)$, which, upon substituting into \cref{eq:entropy_stat_mech_har_oscillator}, leads to the final expression for the entropy:
\begin{equation}
    \label{eq:entropy_stat_mech_har_oscillator_n}
    S=k_B\left[ (\langle n\rangle+1)\ln(\langle n\rangle+1) -\langle n\rangle \ln\langle n\rangle\right].
\end{equation}
It is not difficult to prove that $S$ is a monotonically increasing function of the expectation value $\langle n\rangle$ as given by \cref{eq:entropy_stat_mech_har_oscillator_n}, which means that at large $\langle n\rangle$ (or high temperature), more vibrational modes will get excited. Consequentially, it become less uncertain on exactly which mode(s) is excited at $T$, resulting in a higher vibrational (or information) entropy in the system.

\paragraph{\textbf{Probabilistic interpretation of vibrational anharmonicity measure}}
Finally, we provide the derivation here to showcase how  the vibrational anharmonicity score,\cite{knoop2020anharmonicity} originally defined based on the ratio between the standard deviations of anharmonic and total atomic forces [\cref{eq:anharmonic_score}], can be equivalently defined in terms of KL-diverges [\cref{eq:KL_divergence}] between two probability distributions and interpreted accordingly. For simplicity, we derive the analytical expressions for the anharmonic scores based on a simple form of one-dimensional anharmonic potential in the form of $U(x)=kx^2+k^{\prime}x^3$. 

We begin by first following the original definition of the anharmonicity score based on the standard deviation of the forces, in which we decompose the total atomic force $F_{\mbox{\scriptsize{tot}}}(x)$ into the sum of the harmonic $F_{\mbox{\scriptsize{harm}}}(x)=-2kx$ and anharmonic $F_{\mbox{\scriptsize{anh}}}(x)=-3k^{\prime}x^2$ components. Furthermore, we assume  $k\gg k^{\prime}$ such that we can treat the anharmonic force as a perturbation to the harmonic component. For the harmonic reference potential $U_{\mbox{\scriptsize{harm}}}(x)=kx^2$, the corresponding probability distribution is a Gaussian $p_0 (x)=\exp\left(-\beta kx^2\right)$. We know that for a standard Gaussian distribution $\mathcal{N}(x|0,\sigma^2)$, the expectation values for the moments of $x$ follows $\left\langle x^{2n}\right\rangle=\frac{(2n)!}{2^n\cdot n!}\sigma^{2n}$ and $\left\langle x^{2n+1}\right\rangle=0$ ($n\in\mathbb{Z}$ and $n\geq 0$), so for the probability distribution $p_0(x)$, we have the following:
\begin{equation*}
    \langle x\rangle_0=0,\quad \left\langle x^2\right\rangle_0=\frac{1}{2k\beta},\qquad \left\langle x^3\right\rangle_0=0,\qquad\left\langle x^4\right\rangle_0=\frac{3}{4\beta^2k^2}.
\end{equation*}
From here, one has the following expression for the variance for the harmonic force
\begin{equation} 
\label{eq:variance_F_harm}
\mbox{Var}\left(F_{\mbox{\scriptsize{harm}}}\right)=\left\langle F_{\mbox{\scriptsize{harm}}}^2\right\rangle-\left\langle F_{\mbox{\scriptsize{harm}}}\right\rangle^2=\left\langle F_{\mbox{\scriptsize{harm}}}^2\right\rangle=4k^2\left\langle x^2\right\rangle=\frac{2k}{\beta},
\end{equation}
using the fact that $\left\langle F_{\mbox{\scriptsize{harm}}}\right\rangle\sim\langle x\rangle_0=0$. Similarly, for the anharmonic force, its variance is 
\begin{equation}
    \label{eq:variance_F_anh}
    \mbox{Var}\left(F_{\mbox{\scriptsize{anh}}}\right)=9k^{\prime 2}\left[\langle x^4\rangle-\left(\langle x^2\rangle\right)^2\right]=\frac{9k^{\prime 2}}{2\beta^2k^2}=\frac{9}{4}\lambda\cdot\mbox{Var}\left(F_{\mbox{\scriptsize{harm}}}\right),
\end{equation}
in which we have defined $\lambda=k^{\prime 2}/(\beta k^2)$. Using the fact that $\mbox{Cov}(F_{\mbox{\scriptsize{harm}}},F_{\mbox{\scriptsize{anh}}})=0$, we can combine \cref{eq:variance_F_harm} and \cref{eq:variance_F_anh} to get the variance in the total atomic force being
\begin{equation}
    \label{eq:variance_F_total}
\mbox{Var}\left(F_{\mbox{\scriptsize{tot}}}\right)=\mbox{Var}\left(F_{\mbox{\scriptsize{harm}}}\right)+\mbox{Var}\left(F_{\mbox{\scriptsize{anh}}}\right)=\left(1+\frac{9}{4}\lambda\right)\mbox{Var}\left(F_{\mbox{\scriptsize{harm}}}\right),
\end{equation}
which by combining with \cref{eq:variance_F_anh}, we arrive at the following analytical expression for the anharmonicity score for this simple one-dimensional anharmonic potential:
\begin{equation}
    \label{eq:anh_score_1D_from_variance}
    \sigma^{(2)}=\sqrt{\frac{\mbox{Var}\left(F_{\mbox{\scriptsize{anh}}}\right)}{\mbox{Var}\left(F_{\mbox{\scriptsize{tot}}}\right)}} =\sqrt{\frac{\frac{9}{4}\lambda}{\left(1+\frac{9}{4}\lambda\right)}}\approx\frac{3}{2}\sqrt{\lambda},
\end{equation}
at the weak anharmonic limit ($k^\prime\ll k)$. To derive the similar result based on the definition of KL-divergence, we can re-use the results above and approximate the distributions of anharmonic and total forces as Gaussian distributions, \emph{i.e.}, $p_{\mbox{\scriptsize{anh}}}(F)\approx\mathcal{N}(0,\sigma_{\mbox{\scriptsize{anh}}}^2)$ and $p_{\mbox{\scriptsize{tot}}}(F)\approx\mathcal{N}(0,\sigma_{\mbox{\scriptsize{tot}}}^2)$, with $\sigma_{\mbox{\scriptsize{anh}}}^2/\sigma_{\mbox{\scriptsize{tot}}}^2=\left[\sigma^{(2)}\right]^2=u(\lambda)$. By using the analytical expression of the KL-divergence between two Gaussian distributions, we then have
\begin{equation}
    \label{eq:KL_div_anh_tot}
    \mathcal{D}_{KL}\left(p_{\mbox{\scriptsize{anh}}}||p_{\mbox{\scriptsize{tot}}}\right)=\frac{1}{2}\left[\frac{\sigma_{\mbox{\scriptsize{anh}}}^2}{\sigma_{\mbox{\scriptsize{tot}}}^2}-\ln\left(\frac{\sigma_{\mbox{\scriptsize{anh}}}^2}{\sigma_{\mbox{\scriptsize{tot}}}^2}\right)-1\right]
    =\frac{1}{2}\left[u(\lambda) -\ln u(\lambda)-1\right],
\end{equation}
which shows that $\mathcal{D}_{KL}$ is a monotonically decreasing function of the anharmonicity score in the region of $\sigma^{(2)}\in[0,1]$. This shows that, for example, in the limit of $k^\prime\to0$ (or small $\sigma^{(2)}$, the potential $U(x)$ is dominated by the harmonic part, such that $p(F_{\mbox{\scriptsize{anh}}})$ differs significantly from $p(F_{\mbox{\scriptsize{tot}}})$, leading to a large  $\mathcal{D}_{KL}$, and \emph{vice versa}. This completes our derivation of anharmonic scores from the perspective of KL-divergence between two probability distributions. 

\subsection{Fundamentals of Machine Learning}\label{sec:ml_fundamentals}

\subsubsection{Motivations}
Different from the limited amount of data being produced from individual experiments, the large amount of materials data that are generated from PES sampling, high-throughput computations and stored in large-scale public materials database requires fundamentally new approaches to extract the underlying relationships amongst the data, in order to deepen our scientific understanding on materials' behaviours and facilitate new scientific discoveries. 

From the perspective of information theory, the process of data collection in computational materials science can be considered as a process of information/data compression. A typical calculation $\bm{R}\xrightarrow{\hat{\mathcal{H}}}\mathcal{A}$ that maps the three-dimensional input structure $\bm{R}$ into its corresponding properties $\mathcal{A}$ (\emph{e.g.} energy, charge density, electronic structures \emph{etc.}) via a Hamiltonian $\hat{\mathcal{H}}$ (\emph{e.g.} DFT), where the final data to be collected is typically a tuple $(\bm{R},\mathcal{A})$, will inevitably lose many detailed information about $\hat{\mathcal{H}}$, which encodes the intriguing knowledge on how many-particle interactions in a system that leads to the physical observable $\mathcal{A}$. To overcome this problem of information loss, building on the theory of probability discussed in \cref{sec:prob_theory_fundamentals}, ML provides a set of (light-weighted) statistical tools to help us infer the relationship of the data through discovering recurring patterns in the data. In the context of atomistic materials modelling, it learns an approximate mapping between the \emph{mathematical representation} $\mathcal{X}$ of the materials and their corresponding properties, through capturing the dominant (most probable) features in $\mathcal{X}$ that contributed towards $\mathcal{A}$. 

Formally, on the highest level, many ML models aim to learn the conditional distribution $p(y|\bm{R})$ that captures the likelihood of a structure $\bm{R}$ will possess some property or label $y$. Fundamentally, the underlying physical principle that enables the application of probabilistic (or statistical) models to describe materials is the belief that chemically and structurally similar materials should have similar properties, such that data sharing many commonalities should naturally aggregate together and occur with high probability. The most prominent example behind this belief is the establishment of the periodic table of elements that categorised elements into periods and groups, long before the birth of ML, or even quantum mechanics. In this regard, previous discussions on the theory of probability (\cref{sec:prob_theory_fundamentals}) will not only help to provides the foundation for understanding the following discussions on the mechanistic details of different ML/AI models, but as we hope, to promote the establishment of linkage between statistical mechanics, energy landscape sampling, and other aspects of materials modellings that have been discussed in this monograph. 

This section will focus on discussing the tasks achievable and models that are commonly applied in supervised and unsupervised machine learnings, as well as how these models can be trained and interpreted. For the time being, we shall assume that a proper mathematical \textit{descriptor} (also known as the \textit{feature vector}, which we shall use interchangeably in the following discussions) $\mathcal{X}$ that represent the underlying data (predominantly the atomistic structures) as the input to the ML models is already available, and shall return to the problem of how to construct $\mathcal{X}$ in \cref{sec:feature_eng}.

\subsubsection{Supervised Machine Learning: General Framework and Applications}
\label{sec:supervised_ML}
\paragraph{\textbf{General framework of supervised ML}}
In the most general from, supervised ML models $\bm{y}=f_{\boldsymbol{\theta}}(\bm{x})$ are constructed to achieve two types of tasks, \textbf{\textit{regression}} and \textbf{\textit{classification}}, which depends on the type of the target variable $\bm{y}$. Generally, $\forall y_i\in\bm{y}$, $y_i\in\mathbb{R}$ takes the form of a continuous variable in regression, while $y_i\in\mathbb{Z}^{+}$ takes discrete values in the classification task, where each class is represented by a positive integer. Without loss of generality, we will focus on regression in the following discussion, and the same principles can be applied to classification by simply changing the target distributions from a continuous one into one for discrete random variables [see \cref{appendix:logistic_regression}].

At the core of the design of ML/AI models is how to construct the loss function $\mathcal{L}(\boldsymbol{\theta})$ that is used to train the model. Owing to the simplicity of the supervised ML framework, that is fundamentally identical to linear regressions, most works on training the supervised ML models adapted a $L2$ loss function without in-depth discussions, which minimises the mean squared error (MSE) between the model predictions and the target values to obtain a single set of optimum parameters (that is also widely used in training traditional atom-atom forcefields, see \cref{sect:empirical_forcefield}):
\begin{equation}
    \label{eq:simple_ML_optimisation}
    \hat{\boldsymbol{\theta}}=\argmin_{\boldsymbol{\theta}}\mathcal{L}(\boldsymbol{\theta})=\argmin_{\boldsymbol{\theta}}\left\Vert \bm {y}-f_{\boldsymbol{\theta}}(\bm{x})\right\Vert^2_2=\argmin_{\boldsymbol{\theta}}\sum_{i=1}^N (y_i-f_{\boldsymbol{\theta}}(x_i))^2.
\end{equation}

\begin{figure}[tb]
    \centering
    \includegraphics[width=\linewidth]{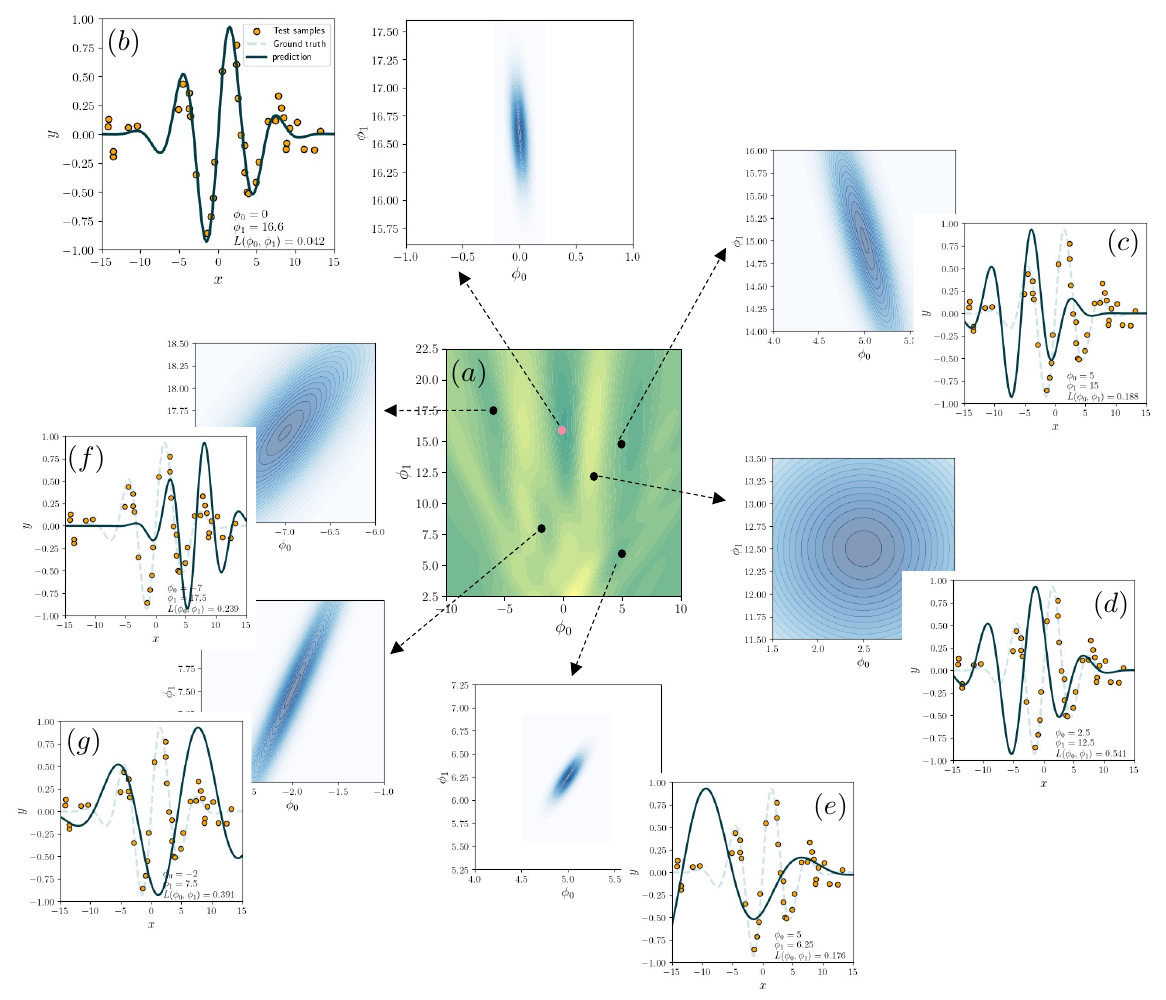}
    \caption{Probabilistic view of hyperparameter optimisation in rigde regression. (a) The loss function landscape as the function of the two hyperparameters $\phi_1$ and $\phi_2$, where the high (low) loss values are shown in yellow (green). (b) The global minimum of the loss function, which on the left panel shows the ground truth (dashed line) and predicted (solid line) Gabor function $y=f(x|\phi_0,\phi_1)$ perfectly overlapping with each other, and the training data (yellow dots) that are sampled from the ground truth function with noise. The right panel shows the probability distribution of the hyperparameters $\phi_0$ and $\phi_1$, which, at the global minimum, shows a narrow distribution with almost zero off-diagonal covariance. Panels (c-f) show the same information as (b) but computed at different points in the hyperparameter space, which are marked correspondingly in (a). To facilitate cross comparison, all probability distributions are plotted with $x$ and $y$ axis in the range of $[\hat{\phi}-1,\hat{\phi}+1]$ where $\hat{\phi}$ is the medium of the normal distribution. It can be seen that the probability distributions of the hyperparameters become broader and strongly correlated at the points where the loss function has high values. For details, see \cref{appendix:details_ridge_regression_loss_landscape}. [The underlying example is taken from Fig. 6.4 in \citet{prince2023understanding}]} 
    \label{fig:ridge_regression_prob_dist_loss_landscape}
\end{figure} 

However, as we have thoroughly discussed with more details in \cref{appendix:linear_regression}, the loss function can be derived from a probabilistic perspective. This perspective is established based on two equivalent viewpoints for training the supervised ML models, whereby model training can be viewed as a process to reach a probability distribution: (i) of the observables that maximises the likelihood for observing the training observables conditioned on corresponding model inputs and parameters [see \cref{eq:bayes_linear_variational_inf_expectation_variational_dist}, which in \cref{sec:ML_AI_PES_sampling_info_theory} we have also discussed extensively how this is related to statistical samplings on PES.] or, (ii) of the model hyperparameters to one that maximises the evidence for observing the training data conditioned on the model inputs [see \cref{eq:bayes_linear_variational_inf}, in which the integrand in the marginal distribution can be essentially bought down to a delta function $\delta(\boldsymbol{\theta}-\hat{\boldsymbol{\theta}})$ or alternatively a narrow Gaussian distribution around $\hat{\boldsymbol{\theta}}$]. The concept behind the second perspective is further illustrated in \cref{fig:ridge_regression_prob_dist_loss_landscape} by showcasing how the probability distributions of the model hyperparameters evolve across different parts of the landscape of the training loss function. 

As elaborated in \cref{appendix:ridge_regression}, one of the key mathematical tools for us to establish the second viewpoint for deriving the concrete form of the loss function to enable model training is the utilisation of \emph{variational Bayesian inference}, which is a powerful framework for approximating complex probability distributions through the introduction of a variational distribution $q(\boldsymbol{\theta})$ to approximate the true posterior distribution $p(\boldsymbol{\theta}|\bm{y},\bm{x})$ of the model parameters. This variational distribution also provides us a systematic way to encode our prior beliefs about the model parameters, which can be crucial for regularising the model training process to enhance the model stabilities and generalisation capabilities through minimising the tendency of overfitting the training data, particularly the outliers. Introducing a distribution over the model parameters also allows us to capture the uncertainty in the model predictions, which is essential for making informed decisions based on the model outputs. This can be either achieved in linear ridge regression [\cref{appendix:ridge_regression}], or more complex nonlinear models such as the Kernel Ridge Regression (KRR) [\cref{appendix:krr}] and the Gaussian process  [\cref{appendix:GP}] models. Finally, the variational Bayesian inference framework, particualy, the training process through maximising the evidence for observing the training data $p(\bm{y}|\bm{X})$ that can be casted into a problem of maximising the evidence lower bound (ELBO) [see \cref{eq:bayes_linear_variational_inf_ELBO}], provides us with the unified foundations for understanding the training process of a wide range of ML/AI models,\cite{prince2023understanding} particularly for the generative models, such as variational autoencoders (VAEs)\cite{diederik2019introduction} and diffusion models.\cite{sohl2015deep,song2019generative} For this reason, we believe that introducing the variational Bayesian inference framework to the readers within the context of simple linear regression is ideal for building the foundations for understanding the training process of more complex ML/AI models.

Based on the above conceptual framework, without diverting the readers into in-depth mathematical details, we provided readers who are interested these details in the \cref{appendix:linear_regression} on how to derive the loss function from the perspective of variational Bayesian inference for a few promoinent linear regression models in the context of materials and molecular modelling, and provides some general highlights on the key features of the models that have been discussed in \cref{appendix:linear_regression}. 

We begin from the simplest linear regression model, which is the ordinary least squares (OLS) regression, that aims to fit a linear relationship $y=\bm{x}^\top\boldsymbol{\beta}$ \footnote{Following the convention in statistics, we use $\boldsymbol{\beta}$ to denote the model parameters for linear regression, which is equivalent to $\boldsymbol{\theta}$ used in the general framework of supervised ML.} between the input features $\bm{x}$ and the target variable $y$, where by modelling both the target and predicted outcomes as random variables that follows Gaussian distributions, we cast the model optimisation problem into one that minimises the KL divergence between these two distributions, which naturally leads to the analytical form shown in \cref{eq:simple_ML_optimisation}. 

To avoid the model from overly weighting a few components $\beta_i\in\boldsymbol{\beta}$, which can be particularly problematic when the input features are highly correlated, ridge regression [\cref{appendix:ridge_regression}] introduces an additional regularisation term $\lambda\left\Vert \boldsymbol{\beta}\right\Vert^2_2$ to the loss function to stabilise the model training, where $\lambda$ is a hyperparameter that controls the strength of the regularisation. As shown in \cref{appendix:ridge_regression}, by adapting the variational Bayesian inference framework, we can formally establish the ridge regression process as one that tries to stabilise the OLS through encoding our prior belief on the probability distributions $q(\boldsymbol{\beta})$ that the model parameter should follow into the model training, such that the optimisation will focus on finding the optimal parameters in the region close to $\boldsymbol{\beta}$ provided by our prior belief. 

With the loss function for ridge regression derived based on this fundamental framework [\cref{eq:bayes_linear_variational_inf_loss}], it is easy to modify it to the corresponding loss function for (i) a classification task, such as logistic regression [\cref{appendix:logistic_regression}], by changing the conditional distribution of the target variable $p(y|\bm{x},\boldsymbol{\beta})$ from a Gaussian distribution to one that supports discrete variables, such as the Logistic distribution, and (ii) to adopt other forms of regularisation, such as the LASSO  regression, which as shown in \cref{appendix:lasso_regression}, is fundamentally driven by the inspiration of changing the prior distribution of the model parameters from a Gaussian distribution to a Laplace distribution, which has a sharper peak at zero and heavier tails than the Gaussian distribution, thus encouraging sparsity in the model parameters and enhancing the model interpretability by effectively performing feature selection with this type of regularisation. 

To keep the discussion general, in \cref{appendix:multimodal_bayesian_regression}, we also provided a brief account how this framework can be generalised to the case where the model parameters are expected to follow a complex multimodal distribution, such as one that is described by the Gaussian Mixture Model (GMM). Conceptually, we can draw some similarity with the training of classical atom-atom forcefields [\cref{sect:empirical_forcefield}] in which the forcefield parameters are simultaneously optimised against a set of heterogeneous targets $\bm{y}=(E,\mathbf{F},\boldsymbol{\Phi})^\top$ that comprised of energy, atomic forces and the force constant matrix.\cite{grimme2014general,rohskopf2017empirical} In this case, the model output can be described by a mixture of Gaussian distributions each for a different target: 
\begin{equation*}
p(\bm{y}\vert \bm{X},\sigma_E^2,\sigma_{\bm{F}}^2,\sigma_{\boldsymbol{\Phi}}^2)\sim\mathcal{N}\left(E|\bm{X}^\top\boldsymbol{\beta}_E,\sigma_E^2\bm{I}\right)\mathcal{N}\left(\bm{F}|\bm{X}^\top\boldsymbol{\beta}_{\bm{F}},\sigma_\bm{F}^2\bm{I}\right)\mathcal{N}\left(\bm{F}|\bm{X}^\top\boldsymbol{\beta}_{\boldsymbol{\Phi}},\sigma_{\boldsymbol{\Phi}}^2\bm{I}\right),
\end{equation*}
which, upon taking the log-likelihood, automatically leads to a loss function that is a weighted sum of the mean-squared-error for each target for model training:
\begin{equation*}
    \frac{1}{\sigma_E^2}\left\Vert E-\bm{X}^\top\boldsymbol{\beta}_E\right\Vert^2_2+\frac{1}{\sigma_\bm{F}^2}\left\Vert \bm{F}-\bm{X}^\top\boldsymbol{\beta}_{\bm{F}}\right\Vert^2_2+\frac{1}{\sigma_{\boldsymbol{\Phi}}^2}\left\Vert \boldsymbol{\Phi}-\bm{X}^\top\boldsymbol{\beta}_{\boldsymbol{\Phi}}\right\Vert^2_2.
\end{equation*}
However, we are not aware of the \emph{direct} wide practical applications of this framework in which the \emph{model parameters are directly drawn and optimised from a multivariant Gaussian distribution}, particularly in the context of materials and molecular modelling, because of the high computational cost associated with sampling from complex multimodal distributions that is required for the model training. 

To enhance the capabilities to handle nonlinear relationships between the input features and the target variable, the linear regression models can be further extended to their nonlinear variants, which are discussed in \cref{appendix:nonlinear_regression}. More specifically, we begin with the Kernel Ridge Regression (KRR) model [\cref{appendix:krr}] which extends the simple ridge regression by capturing the nonlinear relationships between the input features and the target variable through the utilisation of kernel functions, that implicitly map the input features into a higher-dimensional feature space and accounts for the nonlinear relationships in the feature space through the utilisation of nonlinear kernel functions. The so-called similarity kernel is a central concept in materials and molecular modelling in ML/AI, which we shall further elaborated in later sections.  To avoid the model being biased by a few outliers, the support vector machine (SVM) model [\cref{appendix:SVM}] stabilises the KRR model by introducing additional filters in the loss function to remove outlying points during the model training. Finally, in \cref{appendix:GP}, we provided detailed discussion on the Gaussian Process (GP) regression model from the perspective of variational Bayesian inference that optimises the probability distributions over the model function space rather than the model parameter space, which provides the greatest flexibilities for capturing complex nonlinear relationships between the input features and the target variable, and is one of the most popular supervised ML models in the atomistic modelling community.\cite{deringer2021gaussian}

\paragraph{\textbf{Supervised ML for atomistic materials and molecular modelling}}

Here, we shall briefly review some of the notable applications of supervised ML in atomistic materials modelling that are particularly related to the sampling of PES and physical properties of materials that have been discussed in the previous sections. 

\begin{figure}[tb]
    \centering
    \includegraphics[width=\textwidth]{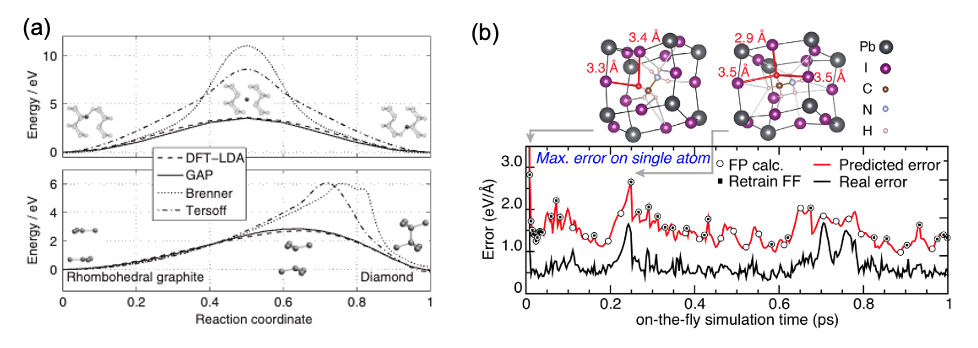}
    \caption{(a) Comparison of the performances for modelling the energy landscapes of a migrating vacancy in GaN (top) and phase transformation of rhombohedral graphite to diamond (bottom) as computed from DFT, the GAP model, and the empirical Tersoff and Brenner forcefields [adopted from \citet{bartok2010gaussian}]. (b) Time-dependent evolution of energy for MAPbI$_3$ perovskite computed with on-the-fly MLFF trained from the underlying DFT model using GP regression. [adopted from \citet{jinnouchi2019phase}], showing that the (re)-training of the MLFF can be triggered when the model uncertainty exceeds a certain threshold.}
    \label{fig:GAP_forcefield}
\end{figure}

\textbf{\textit{Energy Models}} One of the primary goal for applying supervised ML in atomistic modelling is to construct machine-learning force fields (MLFF) that can expedite the statistical samplings of PES with computational costs that is comparable to empirical forcefields, but with the accuracy that is in par with the first-principles calculations. \citet{bartok2010gaussian} pioneered the development of the Gaussian approximation potential (GAP) model, that is combined with the bispectrum form of the density-based representation atomic environment.\cite{bartok2013representing,willatt2019atom} As shown in \cref{fig:GAP_forcefield}(a), the GAP model can accurately capture the variation in the system energy across the defect migration and phase transition processes, which typically involve bond breaking and formation, which failed to be captured by the empirical Tersoff and Brenner forcefields, and would traditionally require the application of more advanced reactive forcefields. This provides a good example to showcase the capabilities of GP as a single framework to capture complex nonlinear relationships across regions of the PES that are fundamentally described by different physics.  \citet{thompson2015spectral} adopted  the spectral neighbour analysis potential (SNAP) model, to learn a linear relationship between energy, forces and stress tensor with the bispectrum representations of the local atomic environments to circumvent the speed bottleneck of nonlinear GP process to accelerate the modelling of large molecules. \citet{glielmo2017accurate} developed covariant kernel that respects the SO(3) rotational equvariance of the atomic forces to enable the direct learning of atomic forces using GP from the underlying DFT data. Finally, \citet{jinnouchi2019phase} developed an on-the-fly PES sampling scheme that is based on the GP regression model, in which the training of the MLFF is performed simultaneously with the sampling process, and the retraining of the MLFF can be triggered when the model uncertainty exceeds a certain threshold, which allows the model to achieve a good balance between accuracy and computational cost during the sampling process [see \cref{fig:GAP_forcefield}(b)]. The incorporation of this approach into the mainstream VASP code\cite{Kresse1996CMS,Kresse1996PRB,Kresse1999PAW} represents a significant milestone for the applications of MLFF in atomistic modelling.

In addition to the application of GP in developing MLFF, many other supervised ML models have also been explored for this purpose. For example, \citet{rupp2012fast} combined the KRR model with the Coulomb matrix representation of the molecular structure to develop a MLFF for small organic molecules, which can be used to predict the atomisation energies of these molecules with high accuracy. This represents one of the earliest applications of supervised ML for predicting the properties of molecules. Fundamentally, this simple model works because the Coulomb matrix representations, albeit its simple mathematical form, can capture the nonlinear atom-atom Coulomb interactions that are the dominant contribution to the total energy of the system, which provided a strong physical prior that is further enhanced by the capabilities to handle nonlinear relationships in the KRR model to achieve accurate predictions. \citet{xie2023ultra} went back to the drawing board to start from the many-body expansion of the total energy [\cref{eq:total_energy_atom_pos_expand}], which uses the B-spline basis functions to construct the feature vectors that captures the 2- and 3-body atom-atom interactions to be fed into the KRR model to develop a MLFF for small organic molecules. \citet{podryabinkin2017active} developed an active learning scheme to train a moment tensor potential (MTP) with linear regressions. Finally, as discussed in \cref{sec:anharmonic_vibs}, compressive sensing has been applied to develop TDEP\cite{hellman2013temperature,zhou2019compressive} for modelling the anharmonic vibrational behaviours of solids at finite temperatures. This is essentially developing a forcefield in which the energy is polynomially expanded with respect to the atomic displacements, and the ML model is trained to learn the coefficients of this expansion, which are the high-order force constant matricies that are used to capture the anharmonicity in the vibrational behaviours of solids.

For training the MLFF, the input $\bm{x}$ is typically some feature vector $\mathcal{X}$ that is constructed solely based on the atomic structure $\bm{R}$. As will be demonstrated in the subsequent discussions, by including other elemental physical properties into the feature vector that is to be combined with supervised ML approach will enables us to predict a wider range of physical properties. 
Instead of the actual ML model being employed, the focus of the following discussion will be on the key physical information that is used to construct the descriptor input $\mathcal{X}$, that are believed to provide dominant contribution to the target properties being predicted, which are the key enabler that allows complex physical properties to be learned with simple ML models. The rationale behind this approach can be understood by drawing parallel comparison to how classical physics laws were established. For examples, the Newton's second law for kinematics and Ohm's law for electrical circuits, in which there exists a small number of interdependent variables whereby the relationships among them were trivial to be discovered from a small set of experimental data. As the materials' properties that we are interested in become more complex, which are cooperatively determined by a large set of possible factors, some algorithmic approaches, such as ML, becomes indispensable in facilitating up to uncover the complex relationships amongst the physical and chemical properties of materials.

\textbf{\textit{Phononic Properties}} \citet{zhuo2018identifying} applied SVM to predicted the Debye temperatures for phosphors, which is a key proxy properties that can be correlated with the photoluminescence quantum yield of the phosphors, in which the input features are 150 dimensional that comprises a ranges of physical properties, such as the  radii and electronic configurations for the consistituent elements, as well as space group informations, which collectively determines the bonding strengths and the vibrational characteristics of the solids. \citet{juneja2019coupling} incorporated fundamental phonon properties of solids, such as the maximum harmonic phonon frequency, integrated Gr\"{u}neisen parameters (up to 3 THz), that are combined with other elemental and structural information, such as the atomic masses and unit cell volumes, that are subsequently fed into a GP model to accelerate the prediction of lattice thermal conductivities ($\kappa_L$) of solids. This provides an example on how ML models can be used to overcome the sampling challenges to compute the anharmonic phonon properties of solids from first principles that contribute towards $\kappa_L$ [\cref{sec:anharmonic_vibs}]. Similar approaches have been adapted in other works to expedite the determination of $\kappa_L$ for inorganic crystals.\cite{chen2019machine,jaafreh2021lattice}

\textbf{\textit{Electronic Properties}}  \citet{lee2016prediction} applied a set of supervised ML models, including ordinary least square, LASSO and SVR, to predict the $G_0 W_0$ band gaps for 270 inorganic compounds from KS band gaps, cohesive energy, crystalline volume per atom, and other fundamental information of constituent elements, achiving an accuracy of 0.24 eV in RMSE. \citet{zhuo2018predicting} applied a SVM classifier to first separate metals from non-metals for inorganic solids, then used a SVR algorithm to predict the band gaps for the non-metallic solids, in which an 136-dimensional feature vector that was constructed purely from the elemental properties was used as the input for the ML models. \citet{pilania2017multi} presented a multi-fidelity co-kriging statistical learning framework to predict the high-fidelity hybrid-DFT electronic band gap from  the low-fidelity GGA band gaps for 600 elpasolite compounds, in which the co-kriging model is essentially a GP regression model that is trained to learn the correlation between the low-fidelity and high-fidelity data to achieve accurate predictions for the high-fidelity data with reduced computational cost, based on features that include the chemical identities at the A, B, B$'$ and X sites, the electronegativity, first ionisation potentials, emprical atomic raidii and the Pettifor's Mendeleev numbers for the constituent elements. \citet{li2022predicting} applied similar approach to construct feature vectors that are combined with linear and kernel ridge regressions to predict the variations in the band gap energies for ABO$_3$ perovskites that were computed from PBEsol and HSE06 density functionals. Importantly, the authors applied the gradient boosted decision tree (GBDT) model to analyse the importance of different features for predicting the band gap variations, which is particularly useful for enhancing the interpretability of the model, thus facilitating the understanding of the important physical properties that contribute to the band gap variations in these perovskites. Finally, \citet{weston2018machine} developed SVR models to predict the band gap energies for 1568 kesterite compounds for photovoltaic applications using features that were constructed with the SVM model. 

\textbf{\textit{Electron-Phonon Interactions and Superconductivities}} In \cref{sec:eph_coupling}, we have discussed the fundamentals of EPI that plays a critical role in determining the temperature-dependent electronic and superconduncting properties of solids, in which the challenge behind the determinations of the EPI strengths in solids is, in the language of probability theory, the sampling of joint probability distributions of the electronic and phononic densities of states under the mutual interactions of the two subsystems. Hence, the purpose of developing simple ML surrogate models is to directly overcoming this sampling challenge to facilitate the direct learning of EPI properties from more elementary physical properties of materials. For examples, \citet{haldar2024machine} tried to predict the zero-point renormalisation of the electronic bandgaps of 2D semiconductors due to EPI using the Extra Random Tree model that was applied on 9 different features, including the zero-Kelvin electronic bandgap energies, the dielectric constants, as well as fundamental phononic and atomic properties. For similar purpose, \citet{nandi2024machine} applied GP model,  employing more than 37 features that chracterise the physical properties of materials to predict the parameters that is used in the O'Donnell model to predict the temperature-dependent bandgap renormalisation of metal oxides. \citet{tran2023machine} generated materials features using MATMINER\cite{ward2018matminer} to train a GP model to predict the EPI strengths and $\omega_{\mathrm{log}}$  to facilitate the discoveries of conventional superconductors based the McMillan model [c.f. \cref{eq:mcmillan_Tc}]. An interesting approach was put forward by \citet{xie2019functional}, which applied data-driven method using the SISSO\cite{ouyang2018sisso} algorithm to discover alternative functional form to the McMillan model for predicting the superconducting transition temperatures from the fundamental parameters of $\lambda$, $\mu^{*}$ and $\omega_{\mathrm{log}}$. Finally, EPI also plays an important role in attenuating the electronic transport properties of thermoelectric materials. For this, \citet{wee2019quantification} applied GP models to predict the EPI strengths $\left\vert g_{nm\nu}(\bm{k},\bm{q}) \right\vert^2\sim g_\nu^2(\varepsilon_{n\bm{k}},\varepsilon_{m\bm{k}+\bm{q}})$ as a function of the electronic energies $\varepsilon_{n\bm{k}}$ and $\varepsilon_{m\bm{k}+\bm{q}}$ for a range of materials.

\textbf{\textit{Electronic Charge Densities}} Charge density is a key fundamental property of materials and molecules, whereby in the DFT framework, having the access to the charge density is the key for solving the KS equation, from which other auxiliary properties can be computed. In this regards, the ability of predicting charge density using ML approach has fundamental interests for atomistic modellings. Using the symmetry-adapted GP regression (SA-GPR), 
\citet{grisafi2018transferable}  predicted the expansion coefficients to reconstruct the molecular charge densities from atom-centred basis functions, using the SOAP-type\cite{bartok2013representing} structure descriptors to represents the covalent bonding environments in molecules. 
The same approach was further extended to predict the charge densities in non-covalently bounded molecular systems,\cite{fabrizio2019electron} solids,\cite{lewis2021learning} that can be ultimately applied to accelerate the QM/MM simulations of electrochemical interfaces.\cite{grisafi2024accelerating} Ultimately, the goal behind the development of ML model is to try bypassing the need of solving the KS eigen-equations. 
In this regard, \citet{brockherde2017bypassing} developed a KRR model to directly learn the Hohenberg-Kohn map from the external potential to the ground state charge density, which can be used to predict the energies and atomic forces for MD simulations of molecules.
 
\textbf{\textit{Magnetic Properties}} By applying supervised ML models to directly predict the spin configurations and associated magnetic properties of materials can be considered as a promising route to overcoming the high computational costs associated with sampling the spin configurations in solids to determine their likely spin ground states, and derived associated magnetic properties therefrom [\cref{sec:mag_prop}]. 
For instances, \citet{dam2018important} employed the KRR model to predict the Curie temperatures of binary transition metal alloys, utilising elemental descriptors including information such as the $3d$ and $4f$ orbital and spin magnetic moments which play a significant role towards determining the ground state magnetic state of the binary alloys.
\citet{long2021accelerating} combined the ground-state magnetic configurations, chemical and structural information (via the SOAP\cite{bartok2013representing} descriptor), that were fed into the Random Forest model to predict the ground state magnetic configuration and Curie temperatures to accelerate the classifications of ferromagnetic and antiferromagnetic compounds.
\citet{ghosh2020machine} also employed the Random Forest model to predict the spin and orbital magnetic moments of uranium-based compounds. The descriptors that were used for their model learning include information such as structural information, cation $5f$-subshell occupation, which were further augmented with descriptors generated from these primitive information through mapping from 10 prototypical mathematical functions.
Lastly, the Random Forest model was also constructed in the study by \citet{khatri2023magnetization} to predict the magnetic moment per ion in Fe-based compounds, in which the orbital field matrix was used for representing the structures of the compounds. To enhance the model interpretability, the authors also performed SHAP (SHapley Additive exPlanations) feature importance analysis\cite{lundberg2017unified} to understand the contributions of orbital interactions in the feature vectors towards the accuracies of model predictions.

\textbf{\textit{Polarisations, Ferro- and Piezo-electric Properties}} \citet{xue2016accelerated} presented one of the earliest works in this area to help the design of \ce{BaTiO3}-based solid solutions as piezoelectric materials by constructing the phase diagram based on the Landau-Devonshire theory. In particular, Bayesian regression based on material-specific feature vectors was used to predict the unknown coefficients along the phase-boundary lines as a function of the compositions. Similar approach was employed in the design of \ce{BaTiO3}-based electrocaloric ceramics.\cite{yuan2022optimizing}
\citet{balachandran2018experimental} trained the SVR model to predict the $T_C$ values, thus guiding the synthesis of high-temperature ferroelectric perovskites, more specifically, the materials system of the type Ba(MM$^\prime$)O$_3$-\ce{PbTiO3}, using feature vectors including information such as the tolerance factor, atomic valency, electronegativity, bond length and the Mendeleev number. 
\citet{yuan2022machine} employed the KRR model to predict the $d_{33}$ values to facilitate the search of \ce{BaTiO3}-based ceramics with large piezoelectric constants, where 95 different features that may affect the  $d_{33}$ values, including electronegativity, ionic radius, unit cell volume, ionic displacements and polarisation strengths, were used.\cite{yuan2018accelerated,yuan2019accelerated}
Finally, \citet{he2021machine} applied regression model to predict the permanent polarisations in ferroelectrics, using features including the atomic sizes and numbers, electrochemical factors, as well as the bond strength factors.

\textbf{\textit{Structure Classifications}}  Noncentrosymmetric crystals are proven to be difficult to discover experimentally, but are of great importance for applications such as nonlinear optics and piezoelectrics. \citet{song2020machine} applied the Random Forest classifier to predict the noncentrosymmetricity of inorganic solids, which helps to dicsover the mode relevant chemical features that can be correlated with the tendency of a material to be noncentrosymmetric. For this model, the author applied elemental features that were constructed using MATMINER\cite{ward2018matminer} and Magpie\cite{ward2016general} libraries. \citet{gleason2024random} developed an architecture of hierarchical random forest classifier to predict the crystal system, space group and lattice parameters from 2D electron diffraction patterns. Similar approach was adopted by \citet{corriero2023crystalmela} to learn the crystal structural information from the X-ray diffraction patterns. Finally, \citet{zhang2020phase} applied genetic algorithms to select the best ML model and the descriptors to distinguish between homogeneous solid solutions from two-phased mictures in high-entropy alloys, with descriptors that are composed of a wide range of elemental properties as well as the compositional information of the alloys. The fundamental intuition here is that the intermixing tendencies of the alloys can be determined by the compatibility in the chemical properties of the consituent elements. 

\textbf{\textit{Classification of Electronic Phases}} Based on a dataset that consists of thermally-driven metal-insulator transition (MIT) compounds, as well as metals and insulators with similar compositions and stoichiometries, \citet{georgescu2021database} applied the XGBoost\cite{chen2016xgboost} classifier model to distinguish compounds with these three distinguishable electronic phases, using features composes of the compositional, strutural, electronic and energetic properties of the compounds, with the feature importance further analysed with the SHAP method.\cite{lundberg2017unified} \citet{stanev2018machine} developed ML model to analyse the superconductor dadabase containing the $T_c$ values for 12,000 superconducting materials, in which the classifier model was trained to predict whether a material is a superconductor with $T_c$ above 10 K, using features that are constructed from stoichiometric information, elemental properties, and electronic structure properties.

\paragraph{\textbf{Significance of Bayesian inference viewpoints for atomistic materials and molecular modelling}}

Before concluding this section, we would like to briefly discuss the significance of the Bayesian inference viewpoints for atomistic modelling, particularly for those that are backed by ML/AI models. Most importantly, we would like to stress that Bayesian inference is not just a mathematical trick for deriving the loss function, but deeply connected to the fundamental principles of statistical mechanics and thermodynamics, which are the cornerstones for understanding the behaviour of materials and molecules at the atomic level. 

First, we recite here the optimum model parameter for the ridge regression model derived in \cref{appendix:ridge_regression}:
\begin{equation}
    \label{eq:ridge_regression_optimum_parameter_bayesian_inference}
    \hat{\boldsymbol{\beta}}=\argmin_{\boldsymbol{\beta}}\left\{\mathbb{E}_{q}[\log p(\bm{y}|\boldsymbol{\beta})]-\mathcal{D}_{\mathrm{KL}}(q(\boldsymbol{\beta})|p(\boldsymbol{\beta}))\right\},
\end{equation}  
which essentially resembles the form of free energy in statistical mechanics. Here, the first term $\mathbb{E}_{q}[\log p(\bm{y}|\boldsymbol{\beta})]$ can be viewed as the negative of the internal energy, which quantifies how well the model predictions match the training data. Conceptually, the model is stable when the internal energy of the system is low, which is similar to the case when the ground truth can be well described by the model parameters. The second term $\mathcal{D}_{\mathrm{KL}}(q(\boldsymbol{\beta})|p(\boldsymbol{\beta}))$ can be viewed as the entropy of the system, which quantifies the uncertainty in the model parameters. Conceptually, the model is stable when the entropy of the system is high, which is similar to the case when the model parameters are not overly constrained by the training data and can capture a wide range of possible solutions to enhance the generalisability of the model. Therefore, training a regression model is a process that essentially balance the fidelity of the model to the training data and the complexity or the generalisability of the model, which is equivalent to the balance between the internal energy and the entropy in statistical mechanics.

Secondly, the Bayesian inference framework provides us with a systematic way to quantify the uncertainty in the model predictions, which, as we have seen for the development of on-the-fly PES sampling,\cite{jinnouchi2019phase} that maximises the trade-off between the accuracy and the computational cost in the computational process. 

Thirdly, the appropriate choice regularisation strategy in ML model training, as demonstrated in the LASSO regression [\cref{appendix:lasso_regression}], can be crucial for enhancing the interpretability of the model. This is an important aspect for atomistic modelling, in which enhancing our fundamental understanding of the underlying physical and chemical processes is often as important as making accurate predictions. It is particularly useful to reduce the model complexity thus enhance the model stabilities by focusing on a few key features that are most relevant to the target properties, as can be seen in the case of modelling the anharmonic vibrational behaviours of solids through the TDEP formalism\cite{hellman2013temperature} [\cref{sec:anharmonic_vibs}].

Finally, the feature vectors $\bm{\mathcal{X}}$ that are used as the input for ML models are  constructed based on the input structure $\bm{R}$ through some feature mapping functions $\bm{\mathcal{X}}=\bm{\mathcal{F}}_{\boldsymbol{\varphi}}(\bm{R})$, which are also parameterised by a set of hyperparameters $\boldsymbol{\varphi}$. In most ML models, these features are usually precomputed with a reasonable choice of $\boldsymbol{\varphi}$ due to the heaviness in computing these features, and the model training is only performed to optimise the model parameters $\boldsymbol{\theta}$. However, in principle, there is no barrier to further extend the loss function for the regression models to include the hyperparameters $\boldsymbol{\varphi}$, leading to the joint optimisation of $\mathcal{L}(\boldsymbol{\theta},\boldsymbol{\varphi})$ that simultaneously optimises the model parameters and the feature construction. Fundamentally, we can think of the application of deep learning models, such as deep NNs, that are applied to atomistic modelling as one that performs this process of joint optimisation.\cite{schutt2017schnet,gomez2018automatic,duval2023hitchhiker,hou2024unsupervised}

\subsubsection{Unsupervised Machine Learning: General Framework and Applications}
\label{sec:unsupervised_ML}
\paragraph{\textbf{General framework of unsupervised ML}}

Different from supervised ML, given a dataset $\{\bm{x}_i\}_{i=1}^{M}$ that may or may not have the corresponding labels $\{y_i\}_{i=1}^{M}$, which is generated from some underlying probability distribution $p(\bm{x})$, the application of unsupervised ML is aimed at learning the underlying structure of the data, which may be used for enhancing our understanding on the data, or facilitating the development of training protocols for the supervised ML models. 

It should be emphasised here that the \emph{structure} of the data to be discovered with the unsupervised ML approach is strongly coupled with the specific ML algorithm being used. For examples, signal decomposition analysis [\cref{appendix:signal_decomposition_analysis}] is designed to reveal the global trends across the data by revealing the dominant directions along which the data shows the most pronounced variations. In contrary, multidimensional scaling (MDS) [\cref{appendix:mds}] is designed to preserve the local relationships among the data, which facilitates the discovery of highly similar data points that maybe clustered together, which is the main idea behind clustering analysis [\cref{appendix:clustering_analysis}] that helps to discover recurrent patterns in the data. 

From the probabilistic perspective, the dimensionality reduction analysis (including both the signal decomposition analysis and the MDS) can be viewed as a process that maps the original data $\{\bm{x}_i\}$ to a lower-dimensional latent representation $\{\bm{z}_i\}$, where $\bm{z}_i \in \mathbb{R}^d$ ($d<M$), such that the density of the data being reconstructed in the latent space follows the same distribution as the original data, \emph{i.e.}, $\argmin_{\boldsymbol{\theta}}\mathcal{D}_{\mathrm{KL}}(p_{\boldsymbol{\theta}}(\bm{z})||p(\bm{x}))$. On the other hand, density estimates [\cref{appendix:density_estimation}] tries to develop an ML model (via $\argmin_{\boldsymbol{\theta}}[-\log (p(\bm{x}))]$) that provides the best estimation of the data density in the original space. 

Despite the conceptual simplicity of the unsupervised ML framework, one should be aware of the potential pitfalls when applying unsupervised ML algorithms to analyse the data. One of the most important pitfalls for dimensionality reduction is that it may cause distortion in reflecting the true spatial relationships among the data when projecting the data from a high-dimensional space to a low-dimensional space. Using Principal Component Analysis (PCA) [\cref{appendix:pca}] as an example [see also \cref{fig:PCA_ISOMAP_compare} for an illustration], when analysing a set of data $\{\bm{x}_i\}$ ($\bm{x}_i \in \mathbb{R}^D$), it gives a set of eigenvalues $[\lambda_1,\lambda_2,\cdots,\lambda_D]\in \mathbb{R}^D$. Generally, we have $\lambda_i>0$ $\forall i$, meaning the data has non-zero variance along all $D$ orthogonal directions. However, when projecting all data onto a $d$-dimensional subspace ($d<D$), the data variations along all other $D-d$ directions were ignored, thus such information is lost. In another word, while the peojection along the first $d$ eigenvectors can capture important global trends across the whole dataset, it is not possible to avoid the situations in which some pairs of data $(\bm{x}_m,\bm{x}_n)$ may show more pronouced variations along the $k$-th eigenvector $\bm{v}_k$ ($k>d$), such that the PCA projection may falsely projecting $\bm{x}_m$ and $\bm{x}_n$ very close to each other in the $d$-dimensional subspace, while they are actually far apart in the original $D$-dimensional space. A possible remedy to this problem is to combine dimensionality reduction with clustering analysis [\cref{appendix:clustering_analysis}] that works in the original high-dimensional space,\cite{musil2018machine} which can help to identify the data points that are similar to each other but being wrongly projected to be far apart in the low-dimensional space, or \emph{vice versa}. 

\paragraph{\textbf{Unsupervised ML for atomistic materials and molecular modelling}}

Here, we shall provide a brief account on some of the prominent application of unsupervised ML methods in the field of atomistic materials and molecular modelling. Unlike the previous section on the supervised ML, where we categorised the applications based on the specific target properties to be predicted with the models, here we categorised based on the specific ML algorithms being used. A recent comprehensive review on the applications of unsupervised ML in the field of atomistic materials and molecular modelling can be found in Ref. \citenum{glielmo2021unsupervised}.

\textbf{\textit{Density Estimation}} Estimating the density (or the probability of occurrence) of the data $\rho(\bm{x})$ that is generated from the simulation has important thermodynamic implications in materials modelling, as it is directly related to the free energy of the system via the relationship $F=-k_{\mathrm{B}}T\log \rho(\bm{x})$. For example, the vibrational free energy of solids can be derived from the density of the phonon frequencies $\rho(\omega)$, and similarly, the electronic free energy can be derived from the density of the electronic states $\rho(\epsilon)$. 

One of the most interesting and critical structural motifs in molecular systems is the hydrogen bonds. However, in realistic molecular systems, such as solvents and biomolecules, the atomic environments in which the hydrogen bonds may form are highly dynamic and heterogeneous, making it difficult to adopt a rigid and universal definition of hydrogen banding patterns in these systems. To address this problem, \citet{gasparotto2014recognizing} proposed a fully data-driven agnostic approach to identify hydrogen bonds and other interesting motifs\cite{gasparotto2018recognizing} directly from the simulation data. Here, the underlying probability density of the hydrogen bonding motifs $\rho(\bm{x})$ is first estimated with the kernel density estimation (KDE) method [\cref{appendix:non_parametric_density_estimation}]. Then a Gaussian mixture model (GMM) [\cref{appendix:non_parametric_density_estimation}] is used to fit $\rho(\bm{x})$ thus decomposing $\rho(\bm{x})$ into a set of $k$ Gaussian components, where the optimum number of the Gaussian components $k$ provides a fully data-driven definition of the hydrogen bonding motifs, which also allows one to directly infer the free energy associated with the formation of each type of the hydrogen bonding motifs.

To further illustrate the linkage between density estimation and PES sampling, \citet{rodriguez2018computing} proposed a novel approach to estimate the free energies of biomolecules without the explicit usage of the collective variables, in which the $k$-nearest neighbour (KNN) method is used to estimate the density of the biomolecular conformations $\rho(\bm{x_i})$ in the original high-dimensional space at each sampled point $\bm{x}_i$, from which the corresponding free energy is computed as  $F(\bm{x}_i)=-k_{\mathrm{B}}T\log \rho(\bm{x_i})$. \citet{westerlund2019inflecs} put forward an approach to identify and distinguish the stable and metastable states of biomolecules from the MD trajectory, whereby the desntiy estimation is applied directly to the MD data to determine $\rho(\bm{x})$, from which the gradients and Hessians can be directly computed on $\rho(\bm{x})$ to provide an indication on whether $\bm{x}_i$ corresponds to an energy minimum (stable state) or a saddle point (metastable state). \citet{invernizzi2020rethinking} provided an alternative formulation of metadynamics [\cref{section:metaD}]. Instead of the conventional formulation of metadynamics, where the bias potentials are being added continuously on the collective variables as the simulation progresses, here it viewed the sampling process as one the continuously performs unbiased probability density estimation as a function of the collective variables, until the reweighed free energy from the estimated density converges. Finally, although the density estimate approach was not directly applied, similar concept is adopted to infer the free energy landscape for Li-ion diffusions in solid electrolytes,\cite{gustafsson2023predicting} based on the likelihood of the spatial distribution of the Li-ion positions from statistical sampling. 

Before concluding this part, we would like to point out that, estimating the underlying probability distribution  function $p(\bm{x})$ from the simulation data with density estimate also  (in principle) establishes a critical bridge that allows us to directly draw new samples $\bm{x}^\prime$ from $p(\bm{x})$ with  some targeted probability of occurrences (\textit{e.g.} energy or some other physical properties). This is particularly the case when $p(\bm{x})$ can be described by some basic distributions that are easy to sample from [\cref{tab:discrete_prob_distributions} and \cref{tab:continuous_prob_distributions}]. However, in most realistic situations, the probability distribution $p(\bm{x})$ is high-dimensional, and a many-to-one mapping such that directly inverting $p$ may not be able to lead to a physically feasible new sample $\bm{x}^\prime$. This is where generative modelling become relevant and will be discussed in the later sections.

\begin{figure}[p]
    \centering
    \includegraphics[width=\linewidth]{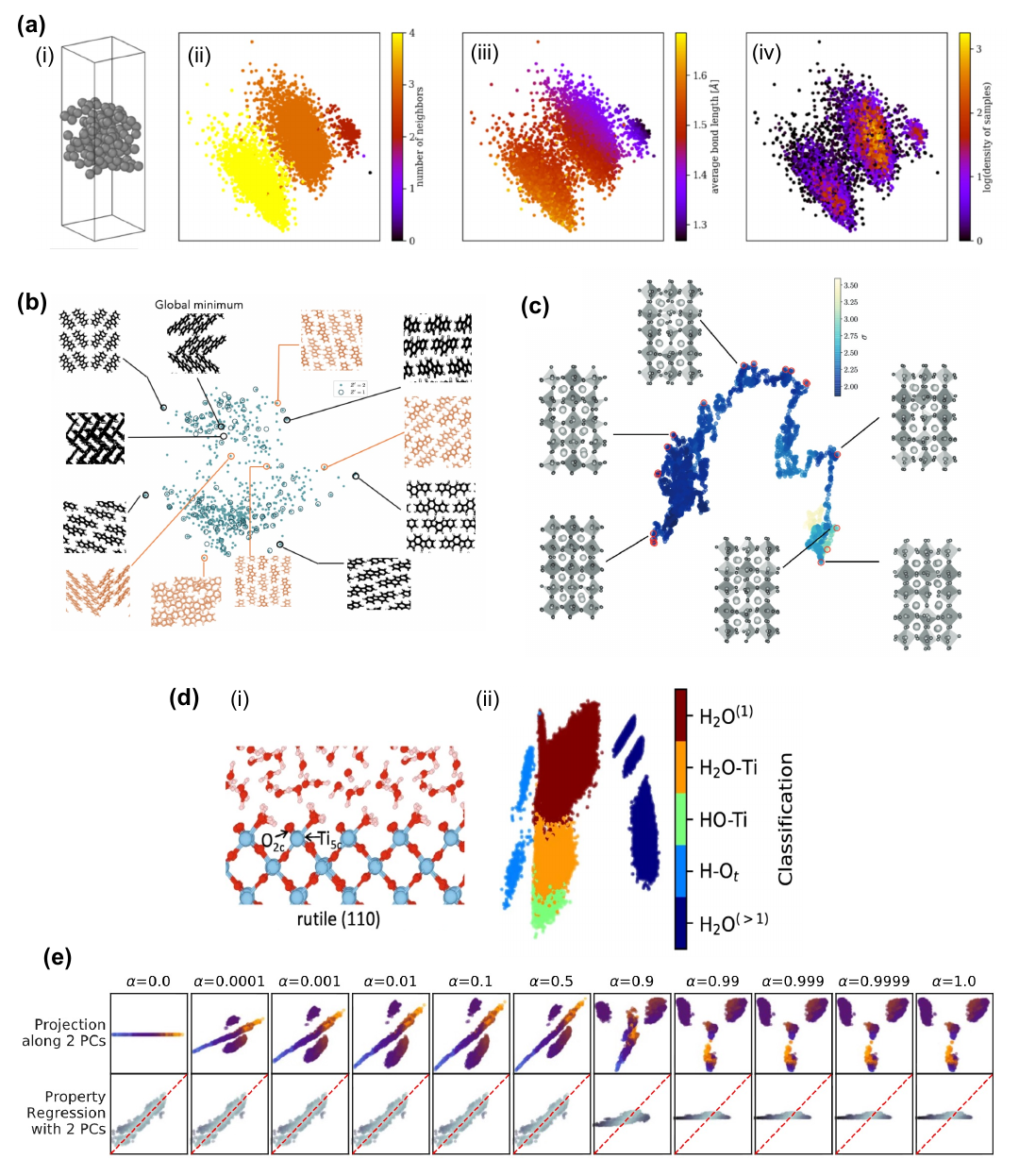}
    \caption{Examples of kernel principal component analysis applied in atomistic molecular and materials modellings. (a) Analysis of the structure-energy landscape of amorphous carbon thin film. (i) A snapshot of the structure\cite{deringer2018computational} (ii-iv) KPCA projection of the atomic environments in amorphous carbon films that are colour coded according to the coordination number, averaged bond length and the he logarithm of the relative probability of each atomic environment.\cite{cheng2020mapping} (b) KPCA map that compares the atomic environments in the predicted crystal structures for an pyrrole azaphenacene in both $Z^\prime=1$ and $Z^\prime=2$ settings.\cite{yang2019interplay}
    (c) KPCA map of the AIMD trajectory for a mixed halide perovskite with each point color-coded according to its vibrational anharmonicity score
    [\cref{eq:anharmonic_score}]. \cite{yang2020mapping}
    All structures in (a-c) are described by the SOAP-based\cite{bartok2013representing,de2016comparing} structural descriptor and similarity kernel. (d) (i) A snapshot of the structure of water on (110) rutile-\ce{TiO2} surface from AIMD and (ii) the corresponding mapping of the hydrogen environments on the same water-oxide interface from KPCA analysis.\cite{zeng2023mechanistic} (e) Principal Covariates Regression (PCovR) performed on the CSD-1000r dataset.\cite{paruzzo2018chemical} Here, a mixing parameter $\alpha$ is applied to toggle between best ridge regression quality (far left) and KPCA (far right) projection for the dataset.\cite{helfrecht2020structure} }
    \label{fig:kpca_examples}
\end{figure}

\textbf{\textit{Kernel Principal Component Analysis (kPCA)}} As a statistical analysis method, PCA has long been used in the field of (bio)molecular simulations to extract the essential dynamics, or equivalently, the dominant collective motions of biomolecules from MD simulations,\cite{amadei1993essential} which was much earlier before ML/AI become popular, particularly for atomistic materials simulations. This is most likely because the usage of bond lengths, angles, and dihedrals makes it trivial to define a set of frame-independent coordinates that capture the essential structural characteristics of molecules, which remains more challenging until the devise of  structural descriptors\cite{musil2021physics} that can equally applied to both materials and molecules. 

Here, we provide a few examples that combine one of the most agnostic structural descriptors, the smooth overlap of atomic positions (SOAP) descriptor\cite{bartok2013representing,de2016comparing} for  mathematically encoding the local chemical environments around each atom,  with the kernel PCA (kPCA) method [\cref{appendix:kpca}] for analysing the structure-property landscapes for molecules and materials, which allows the dominant trends across the whole dataset to be revealed. For examples, amorphous materials are characterised by the lack of long-range order that originates from a large variations in the random local coordination environments around each atom. With the kPCA technique that is applied to the amorphous carbon thin film structures generated from MD simulations,\cite{deringer2018computational} \citet{cheng2020mapping} [\cref{fig:kpca_examples}(a)] demonstrated the variations, diversities and populations of carbon atoms with different numbers of neighbouring atoms (coordination numbers) in amorphous carbon films, which can be applied to facilitate our understanding of how it correlates with the entropic contributions to the free energy of amorphous carbon films. The interplay between molecular structures and hydrogen bonding patterns has a long-lasting interest in the field of organic crystal engineering, and one of the central question is how such an interplay changes the preferences of molecules forming crystals with different numbers of molecules in the asymmetric unit ($Z^\prime$). Our own work on the crystal structure prediction for pyrrole azaphenacene molecules\cite{yang2019interplay} was the first of its kind that revealed the overall high similarity of the local hydrogen bonding environments in the predicted crystal structures for both $Z^\prime=1$ and $Z^\prime=2$ settings [\cref{fig:kpca_examples}(b)]. It also enabled us to propose a global similarity index that aims to quantify the overall similarity between  crystal structure landscapes of two structurally similar molecules. Later, similar techniques were applied to analyse the AIMD trajectories of perovskite photovoltaic materials,\cite{yang2020mapping} providing a visually appealing approach behind the strong correlation between the vibrational anharmonicity  and the changes in the local structural environments of the perovskite materials [\cref{fig:kpca_examples}(c)].

Studying the structures and dynamics of interfacial water on inorganic surfaces with MD has become increasingly popular in the recent years in the field of computational electrocatalysts to deepen our understanding behind the complex physical chemistry on these complex interfaces.\cite{sakong2018electric,kronberg2020revisiting,andrade2020free,kronberg2021reconciling,qin2023cation,chen2023accelerating,yang2023neural} Similar to other (bio)molecular simulations, the challenge here is to reach a quantitative understanding on the diverse structures and dynamics of (un)dissociated water molecules and associated hydrogen bonding patterns in the vicinity of the solid-water interfaces. For this, \citet{zeng2023mechanistic} proposed a set of easy-to-compute descriptors for the proton environments and applied to study the behaviour of water on the low-index \ce{TiO2} surfaces. To overcome the high computational cost associated with projecting a large number of proton environments onto its low-dimensional manifold, a sparsified kPCA method was used, in which a subset of most distinctive proton environments were identified through furtherest-point-sampling (FPS), on which kPCA was applied to compute the first two principal components. Subsequently, the entire dataset can be projected onto these two principal components, leading to a kPCA map that clearly reveals distinguishable clusters, each corresponding to a different type of proton observed in the dataset as shown in \cref{fig:kpca_examples}(d). This is the same as the `landmark' approach that is commonly used in the field of manifold learning, such as in the sketch-map metod,\cite{ceriotti2011simplifying} to reduce the computational cost.

Finally, to enhance the capability of KPCA method in producing meaningful structure-property relationships from atomistic modellings, \citet{helfrecht2020structure} showcased the working principles behind the Principal Covariates Regression (PCovR) approach. Here, the supervised regression and the unsupervised dimensionality reduction are combined together through minimising a loss function $\ell_{\mathrm{PCovR}}=\alpha\ell_{\mathrm{proj}}+(1-\alpha)\ell_{\mathrm{reg}}$, in which a weighting parameter $\alpha$ is applied to toggle between the best projection quality (when $\alpha=1$) and the best regression quality (when $\alpha=0$). With this approach, one can obtain a low-dimensional representation of the data that is not only able to capture the dominant trends across the whole dataset, but also able to reveal the structure-property relationships in the dataset. As shown in \cref{fig:kpca_examples}(e), the PCovR map of the CSD-1000r dataset\cite{paruzzo2018chemical} with $\alpha=0.5$ is able to reveal the overall structural trends across the dataset, whilst showing a clear correlation between the changes in the target properties and the first two principal components.

\begin{figure}[p]
    \centering
    \includegraphics[width=0.96\textwidth]{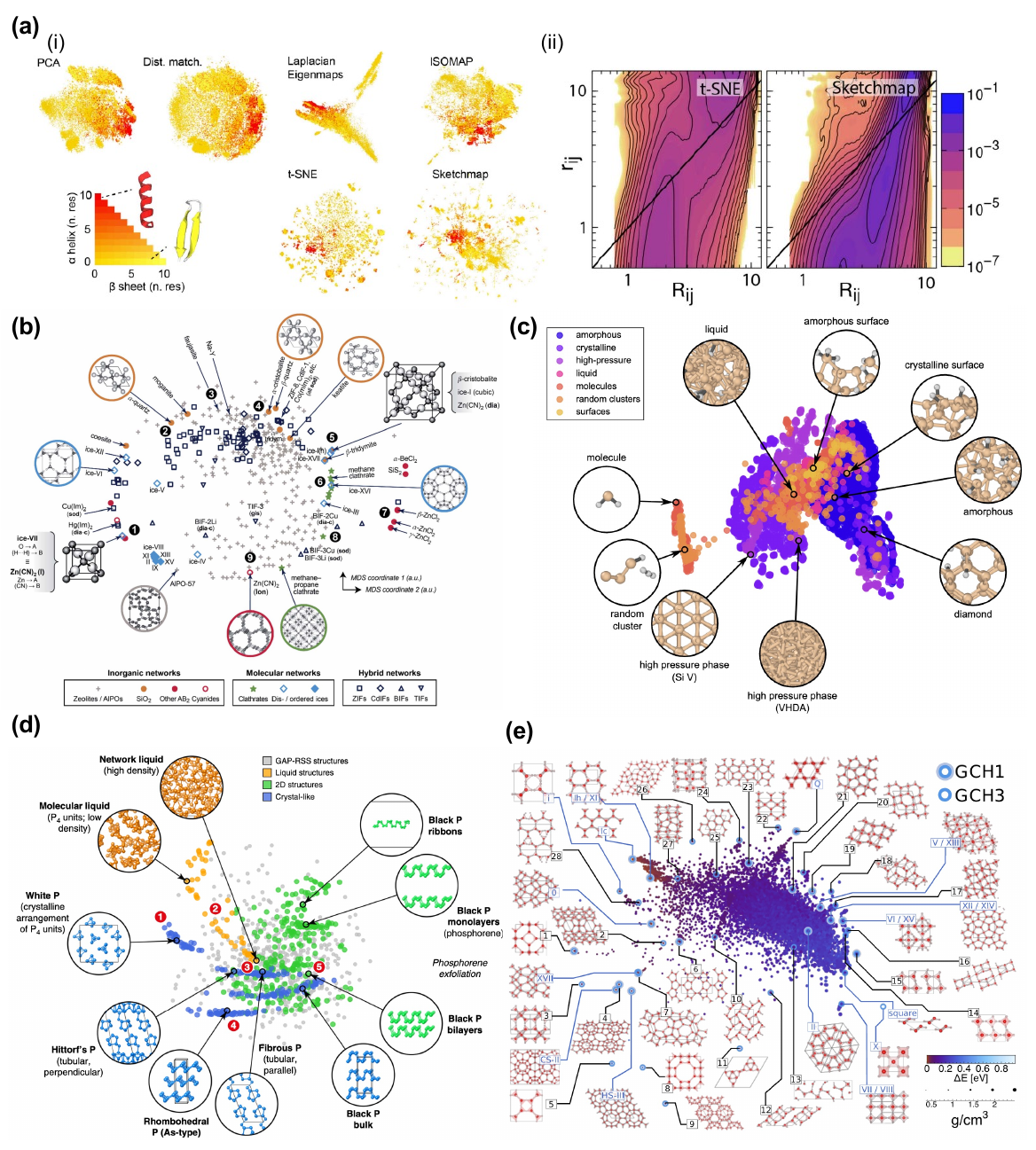}
    \caption{Examples of MDS maps applied in analysing the atomistic molecular and materials modelling data, where the closer the two points are on the maps, the more similar are the corresponding atomistic structures of the two molecules/materials, and \textit{vice versa}. (a) (i) Comparison of different dimensionality reduction techniques in mapping the MD trajectories for immunoglobulin binding domain B1 of protein G, where each point is colour-coded based on the number of residues found in the $\alpha$-helix and $\beta$-sheet structures.\cite{tribello2019using} (ii) Histograms illustrating the joint probability distribution functions for the distances between configurations in the trajectory as measured in the orginal high-dimensional space ($R_{ij}$) and projected 2D manifold ($r_{ij}$) with the t-SNE\cite{van2008visualizing} and sketch-map\cite{ceriotti2011simplifying} methods. (b) A two-dimensional map comparing the inorganic and hybrid tetrahedral crystal packing frameworks, demonstrating how two groups of distinctively different material systems may adopt very similar crystal structures.\cite{nicholas2020understanding} (c) A two-dimensional map showing the diversity of structures in the Si-H dataset, encompassing molecular, liquid, crystalline, amorphous and high-pressure phases, that are used for training the cluster expansion potentials for the Si-H system.\cite{rosset2025atomic} (d) Similar technique applied to analyse the bulk and nanostructured phosphorus for forcefield training.\cite{deringer2020general} (e) Sketch map of 15,896 distinct PBE-optimised structures of ice.\cite{engel2018mapping} The mapping shows strong correlation between the positions of the projected point and the corresponding configurational energy (coloured coded) and the density (not shown here) of each ice polymorph. GCH is the shorthand for Generalised Convex Hull,\cite{anelli2018generalized} with corresponding structures circled on the sketch map.}
    \label{fig:MDS_maps_examples}
\end{figure}

\textbf{\textit{Multidimensional Scaling (MDS)}} Unlike the PCA-based techniques that are designed to capture the global trends across the entire dataset, MDS put more focus on preserving the local relationships among the data points. Another important difference between PCA and MDS lies in the fact that PCA leads to a definitive mapping from the original high-dimensional space to the low-dimensional space, while MDS, in its most general form that [\cref{eq:MDS_minimisation}] aims to minimise the target function (\emph{e.g.} through gradient-based minimisation), which is the difference between the distances among the data points in the original high-dimensional space ($R_{ij}$) and the distances among the corresponding projected points in the low-dimensional space ($r_{ij}$), cannot always guarantee that: (a) the global optimal or the same local optimal solution can be reached with different initial conditions, and (b) $R_{ij}=r_{ij}$ to be satisfied for \textit{all} pairs of data points.\cite{tribello2019using} 

As further shown in \cref{appendix:mds},  variants of MDS differ from each other in the designs of the target function, through placing different weights on preserving the local relationships among the data points in different \textit{length scales} based on some basic understanding on the problem at hand, such that the general consensus is that no single MDS approach is unequivocally better than the others.\cite{duan2013evaluation} For instance, \citet{tribello2019using} systematically compared the performance of different dimensionality reduction techniques for analysing the MD trajectories of biomolecules, and for the specific example of the immunoglobulin binding domain B1 of protein G [\cref{fig:MDS_maps_examples}(a)(i)], \emph{all the algorithms do a reasonable job of separating the configurations that are predominantly $\alpha$-helix-like from those that have a structure that is predominantly composed of $\beta$-sheets}. A useful diagnostic tool to evaluate the performance of different MDS approaches is to compare the joint probability distribution functions $p(R_{ij},r_{ij})$ to understand how the high-dimensional spatial relationships among the data points may have been distorted in the process of projecting the data onto the low-dimensional manifold [\cref{fig:MDS_maps_examples}(a)(ii)]. Similar finding was reached in our own work that analysed the crystal packing landscape of a flat acene-type molecule.\cite{yang2019interplay} As such, the aim of the subsequent discussion is to provide a few notable examples on how MDS has been applied to reveal the complex structural landscapes for some seemingly simple material systems.

Tetrahedron is one of the most common form of connectivities observed in the inorganic crystals, with \ce{SiO2} being probably the most well-known example, which is also adopted by many crystals of hybrid organic-inorganic materials. To show case this fundamentally important phenomenon, by allowing inorganic and hybrid crystals to be compared in the same footing, \citet{nicholas2020understanding} devised a coarse-graining and standardisation procedure, which led to a unified two-dimensional structural map [\cref{fig:MDS_maps_examples}(b)] that exemplifies the structural heterogeneity of the tetrahedral crystal packing frameworks across inorganic crystals, clathrate hydrates and zeolitic imidazolate frameworks (ZIFs). Such map may be used to facilitate the understanding on how similar properties among different material systems may arise from the similar local tetrahedral connectivities in their crystal structures.

\cref{fig:MDS_maps_examples}(c) shows the map of Si-H structures that are used for training the cluster expansion potentials for the Si-H system.\cite{rosset2025atomic} Unsurprisingly, the map shows the molecular and cluster Si-H structures are significantly distinct from the crystalline and amorphous structures, forming a separate `island' on the two-dimensional map. However, despite the fact that crystalline, amorphous, and liquid-like structures differ fundamentally in terms of long-range ordering, they can be projected to be close to each other due to the strong similarities in the local tetrahedral bonding environments around the Si atoms that are preserved in different phases of the Si-H system. Similarly approach has also been applied to analyse the bulk- and nano-structure landscape of phosphorus [\cref{fig:MDS_maps_examples}(d)],\cite{deringer2020general} which is another material system that is well-known to be highly polymorphic. 

Last but not least, \citet{engel2018mapping} charted the structural landscape of 15,896 distinct ice structures that are both known experimentally and generated from random structure searching. As shown in \cref{fig:MDS_maps_examples}(e), the sketch-map projection shows a strong correlation between the structure and energetic stabilities of ice polymorphs. A particular interesting application of this map is its combination with the concept of the Generalised Convex Hull (GCH),\cite{anelli2018generalized} which is a data-driven generalisation of the conventional concept of convex hull based on thermodynamic constraints, that in the structural descriptor space, identify the structurally most distinctive on the global scale yet energetically most stable amongst a set of structural neighbourhood.  

\begin{figure}[tbh]
    \centering
    \includegraphics[width=0.85\textwidth]{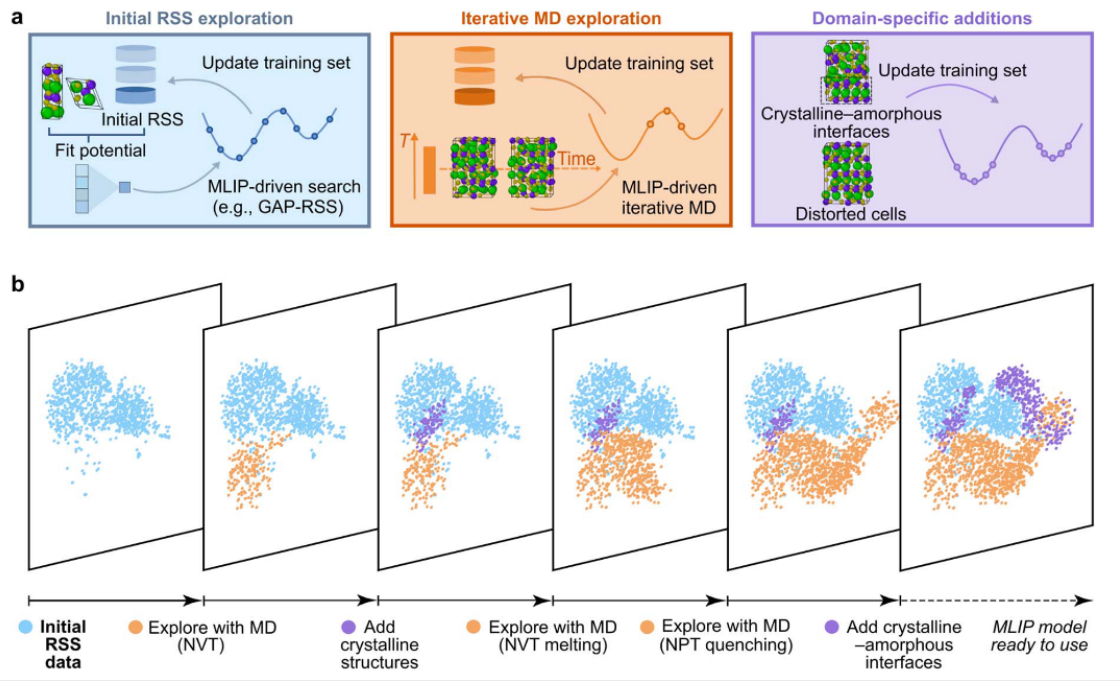}
    \caption{Synergistics between supervised and unsupervised ML in the development of ML energy models for an exemplary system of crystalline and amorphous \ce{BaZrS3}.\cite{pacsca2025machine} The UMAP\cite{mcinnes2018umap} projections based on the SOAP\cite{bartok2013representing} structural descriptor are applied to demonstrate how training data generated with three different sampling approaches (panel a) are progressively incorporated into the training dataset (panel b) to enhance the coverage of the training dataset in the configurational space.}
    \label{fig:MDS_sampling_model}
\end{figure}

Finally, the coverage of the structural space plays a critical role in the development and validation of ML/AI-based energy models,\cite{fu2022forces,yang2026mace} particularly in sustaining long-time dynamic simulation of molecules and materials. As shown in \cref{fig:MDS_sampling_model}, unsupervised ML is a powerful tool for us to monitor the coverage of the training dataset in the configurational space, that is use to guide the training process of energy models, particularly to understand whether (a) the portion of the configurational space in which the trained model will be subsequently applied, and (b) likely regions in which the models need to be able to extrapolate to in the subsequent applications, have been sufficiently covered in the training dataset.



\subsection{Feature Engineering for Machine Learning in Materials}\label{sec:feature_eng}
\subsubsection{Motivations}
\label{sec:descriptor_motivation}

As can be seen from the discussions in the previous section, both the supervised and unsupervised ML are statistical methods that can be applied to facilitate us in exploring the chemical-structural-property relationships for materials, therefore, the prerequisite that will enable us in applying these statistical methodologies, as opposed to the traditional approach of establishing qualitative understanding via known descriptive physical laws, is to \textit{construct physically meaningful mathematical representations} to describe the chemical, physical and/or structural attributes of the material systems to be studied. Very often, these mathematical representations are cast into the form of high-dimensional vectors $\boldsymbol{\chi}\in\mathbb{R}^N$, which shall be referred interchangeably as either the \textit{feature vectors} or \textit{descriptors}, such that we can enrich the information content that has been encoded for a given material, so it can be maximally differentiated from other materials within the set of materials to be studied.

More specifically, for supervised ML [\cref{sec:supervised_ML}], the feature vector $\boldsymbol{\chi}$ served as the \textit{independent variable}, which, will fundamentally allow a (often highly non-linear) mathematical relationship to be established, that correlates the target property with the independent variable, enabling the fast predictions of materials properties with the trained model. On the other hand, in unsupervised ML [\cref{sec:unsupervised_ML}], we often consider $\boldsymbol{\chi}$ as the \textit{coordinates to locate materials in the high-dimensional space}, providing a `geographic' interpretation of the materials space, upon which we can discover trends, patterns and outliers, similar to how we can identify the spatial relationships amongst objects on the geographical maps based on the altitudes and latitudes.  

Depending on the nature of the specific task that one is aiming to solve, here, we consider two general strategies with which  the feature vectors can be constructed to enable efficient learning, namely a \textbf{property-oriented} and a \textbf{structure-oriented} approach. 

The property-oriented approach has been covered in considerable details in our discussions [\cref{sec:supervised_ML}] on the applications of supervised ML for material property predictions, for which the purpose of learning is to establish a cheap surrogate relationship that allow exotic target properties, which are expensive to compute, becomes predictable from a set of physical/chemical properties that are relatively inexpensive to compute or retrievable from some existing tabulated sources. To understand which set of properties will lead to accurate prediction, thus suitable to be used for constructing the feature vectors, often requires strong prior knowledge on the dominating physics that leads to the desired physical outcomes. This also implies that such a property-oriented approach for constructing the feature vectors is highly problem specific thus not necessarily leads to a generalise scheme for constructing this type of feature vectors. As such, we shall not focus on this aspect in this section.

The overarching goal of computational materials science, as demonstrated in this monograph, is to establish the critically needed chemistry-structure-property relationship to guide future materials design, starting from the information of the materials' chemistry and atomistic structures. One of the fundamental principle that underpins this concept is that \emph{ materials with similar structures $\mathbf{R}$ should possess similar properties $p$}, and our goal is here is try to achieve a mathematical way of expressing such a concept. In the framework of kernelised supervised ML [\cref{sec:supervised_ML}], the (dis)similarities in the materials properties, which are often scalar or vectorial quantities, can be easily computed using well-known distance metrics, such as the Euclidean distance $d_{ij}=\Vert p_{i}-p_{j}\Vert_2$ (typically for scalars) and the (normalised cosine distance $d_{ij}=\sqrt{1-\bm{p}_1\cdot \bm{p}_2/(\Vert \bm{p}_1\Vert_2\cdot\Vert \bm{p}_{2}\Vert_2)}$ (typically for vectors). However, how one could similarly quantify the (dis)similarity between two atomistic structures, are generally non-trivial. Traditionally, the Root-Mean-Square-Deviation (RMSD) between the atomic positions of two structures, after they become optimally aligned through rotations,\cite{coutsias2004using,chisholm2005compack} have been commonly used for structural comparisons, but such an approach often lacks generality. For examples, method developed for comparing molecular crystals\cite{chisholm2005compack} are typically confined to crystals with identical molecule(s) in the asymmetric unit, thus not helpful for identifying the trend for different molecules that may adopt similar crystal packing structures.\cite{yang2018large}

As such, the construction of structure-oriented approach for constructing the feature vectors, or most commonly known as the structural descriptors, to facilitate structure-based ML through, most importantly, the quantification of structural similarities, is highly desired for atomistic materials ML. Over the years, a wide range of structural descriptors have been proposed and tested, including internal coordinates\cite{pulay1979systematic,echenique2006definition},  bond-order parameters,\cite{steinhardt1983bond}  atom-centred symmetry functions,\cite{behler2011atom} permutation-invariant polynomials,\cite{braams2009permutationally} Coulomb (for molecules)\cite{rupp2012fast} and Ewald (for crystal) matrix,\cite{faber2015crystal} as well as the SOAP\cite{bartok2013representing} and atomic cluster expansion (ACE) descriptor.\cite{drautz2019atomic} Most importantly, the development of SOAP (and other related\cite{drautz2019atomic,behler2011atom,shapeev2016moment}) descriptors have emerged into a unified framework\cite{willatt2019atom} for constructing \textbf{physics-inspired symmetrised atom density field representations} for molecules and materials, which has been thoroughly reviewed in a recent monograph.\cite{musil2021physics} In addition to the demonstrated examples in the previous sections on the applications of SOAP descriptors in supervised and unsupervised ML, such a unified framework provide a useful guidance to extend the SOAP descriptor to account for, such as, chemical diversity,\cite{de2016comparing,willatt2018feature}  tensorial properties,\cite{grisafi2018symmetry} as well as long-range physics,\cite{grisafi2019incorporating} that are critical in materials modelling, and have been demonstrated\cite{nigam2022unified} on how the popular scheme of  message-passing NNs for quantum chemistry\cite{gilmer2017neural} can be united under the same abstract mathematical formulation using the Dirac (bra-ket) notation from quantum physics.

Therefore, the focus of our discussion will be on the family of symmetrised atom density field representations. The rich set of existing literatures on this topic renders a mathematically detailed account here somewhat meaningless. Understanding how fundamental physical constraints, particularly symmetry, are incorporated into the SOAP descriptor, is important for the practitioners of ML/AI-driven computational materials science how these models lead to higher data efficiency, better ability for extrapolation, as well as model explainability. However, these knowledge are often embedded in lengthy mathematical derivations, which requires strong prior knowledge in topics such as tensor algebra, group and representation theories, which may not be trivial for most practitioners of materials modellings. 

Hence, following the style of the present monograph, we aim to strike here a balance between mathematical formality and intuitiveness to help readers to understand the most critical aspects behind the design of this group of structural descriptors, and building up linkages to AI models that are to be discussed in the subsequent sections.

\subsubsection{Coordinate Systems in Atomistic Modelling}\label{sec:coordinates_systems}

The starting point of any atomistic molecular/materials modelling is to provide the computer program with the structural (and chemical information) of the system being studied, and this structural information basically tells the computer program the position of each atom in the three-dimensional space $\mathbf{r}_i$ for each given atom $i$. Before we dive into the details of structural descriptor, we provide a brief account on the choices of coordinate systems that are common for atomistic modelling, and demonstrate why they are not suitable for the purposes of ML. Readers who are familiar with this topic may skip this section.

For molecular systems, the most intuitive choice to define $\mathbf{r}_i$ is to use the standard \textbf{Cartesian coordinates} system that define $\mathbf{r}_i\in\mathbb{R}^3$ as a three-dimensional vector quantity $\mathbf{r}_i=(x_i,y_i,z_i)$. The enumeration of $\mathbf{r}_i$ is thus clearly depends on the position of the origin and the orientations of the three orthogonal axes, referencing to which $\mathbf{r}_i$ is defined, thus it is non-unique, as  applying  a translational operation $\bm{t}$ to the coordinates ($\bm{r}^\prime\leftarrow \bm{r}+\bm{t}$)  and/or a rotational operation ($\bm{r}^\prime\leftarrow \hat{\bm{R}}_{\boldsymbol{\theta}}\bm{r}$, where $\hat{\bm{R}}_{\boldsymbol{\theta}}$ denotes the three-dimensional rotation operator), will lead to a new set of atomic coordinates $\{\bm{r}^\prime_i\}_{i=1}^N$ for the \textit{same} molecule. However, we know that the system properties should not change with respect to both the translation and rotation of the atomic coordinates. This shows that the Cartesian coordinates of the atoms, which are clearly dependent on the choice of the reference frame, are therefore non-unique, and not a proper choice for representing atomistic systems for ML purposes.

It is also well known that, particularly for molecular systems, performing geometry optimisations are more efficient in the curvilinear space if the system is defined in terms of its \textbf{internal coordinates},\cite{pulay1979systematic,echenique2006definition} including bond lengths ($r_{ij}$, a two-body term), bond angles ($\theta_{ijk}$, a three-body term) and the dihedrals ($\varphi_{ijkl}$, a four-body term). Physically, these coordinates are also used as the independent variables in constructing the functionals of the traditional atom-atom forcefields [\cref{sect:empirical_forcefield}], as one expect that changes in these variables lead to the most dominant changes in the systems' potential energy. In comparison, manipulating the Cartesian coordinates are more complex. Imaging changing a specific bond angle $\theta$ by a specific amount of $\Delta\theta$ would require manipulating (at least) three real numbers $(x,y,z)$ for a specific atoms in a bespoken concerted way, suggesting the Cartesian coordinates form a set of highly coupled mathematical representations of the system that are difficult to manipulate. Nevertheless, although the internal coordinates are defined with respect to the position of a single specific atom in the system, which removes the problems of frame-dependency in Cartesian coordinates thus known to be invariant with respect to the rotation of the system,\cite{duval2023hitchhiker} its enumerations, particularly for the dihedrals,\cite{echenique2006definition} are non-unique. Hence the internal coordinates by itself do not constitute a unique set of representations for atomistic ML, in general.

Finally, for crystalline materials, the positions of atoms in a crystal are typically specified with a tuple of \textbf{fractional coordinates} $\bm{f}=[f_1,f_2,f_3]$, with $f_i\in [0,1)$ that is defined with respect to its lattice vectors $\bm{L}=[\bm{l}_a,\bm{l}_b,\bm{l}_c]\in\mathbb{R}^{3\times3}$ where $\bm{l}=(l_x,l_y,l_z)\in\mathbb{R}^{3}$ that spans the parallelepiped which corresponds to the unit cell of a crystal. The transformation between the Cartesian and fraction coordinates can be usually achieved with a matrix product $\bm{R}=\bm{L}\bm{f}$. However, it is also well-known that, given a specific pattern of periodic three-dimensional arrangement of points, the definition of $\mathbf{L}$ (matrix of lattice vectors) is also non-unique.  In fact, there exists an infinite choice of possible $\mathbf{L}$, making $f$ also unsuitable to be used directly in the tasks of atomistic ML. Whilst this can be potentially overcome by normalising $\mathbf{L}$ to the \emph{reduced cell} representation that is in principle unique, the actual cell reduction algorithm\cite{kvrivy1976unified}  can be rather numerically sensitive thus not be able to assure two different but degenerate sets of $\mathbf{L}$ can always be reduced to an identical $\mathbf{L}_{\mathrm{reduced}}$.

\subsubsection{General Requirements for Structural Representations in Atomistic Machine Learning}

Based on the above discussions of the coordinate systems, we can anticipate that the first requirement for constructing meaningful structural representations is \textbf{symmetry}. More specifically, we need to remove the trivial symmetries such that the structural representation will no longer dependent on the choice of the reference system. In the context of atomistic modelling, the most relevant symmetry operations are the \emph{translation} and \emph{rotation} (one may also include mirror symmetry but in most circumstances, it can be considered as a special case of rotation), which changes the atomic positions of the materials. 

\begin{figure}[htb]
    \centering
    \includegraphics[width=0.8\linewidth]{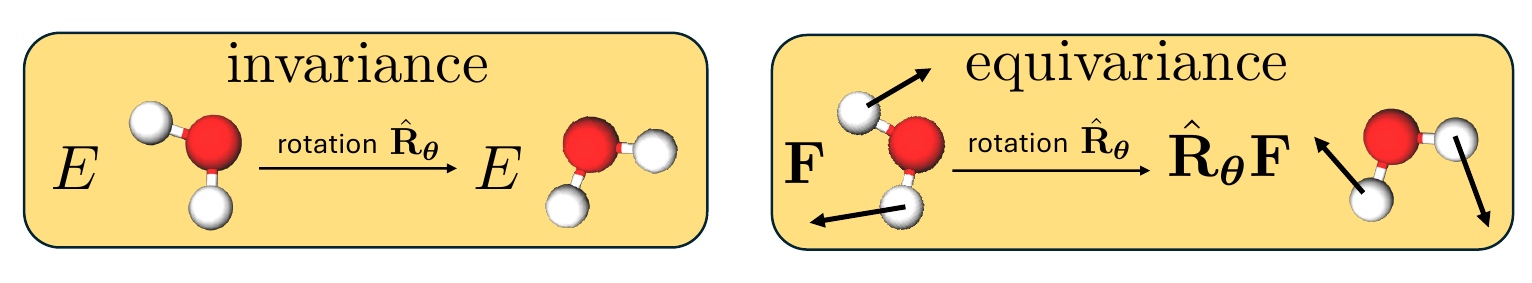}
    \caption{Illustration of two important types of symmetry transformations (using rotation as an example) that are relevant to atomistic ML/AI models for molecules and materials: (left) invariant transformation, and (right) equivariant transformation.}
    \label{fig:symmetry_transformations}
\end{figure}

More importantly, and particularly in the context of supervised ML that aims to establish a model of the form $\bm{y}=f_{\boldsymbol{\theta}}(\bm{x})$, the construction of the structural descriptors (which, we shall see later, also implies for the construction of NN models in DL), should respect two types of symmetry transformations [\cref{fig:symmetry_transformations}] that are physically quintessential: (a) the \emph{\textbf{invariant transformation}}, which typically applies to scalar properties such as energy ($E$), that remain unchanged under the symmetry operations upon the structure, and (b) the \emph{\textbf{equivariant transformation}}, which typically applies to vectorial and tensorial quantities such as forces and charge multipole moments, that change in the same way as the atomic positions upon the act of symmetry operations. As such, it is expected that any bespoken mathematical representations for the atomistic structures should ultimately preserve the equivariance (where we can treat invariance being a limiting case of equivariance).

\citet{musil2021physics} outlined three further requirements for constructing meaningful structural representations, which we shall briefly recount here. The \emph{\textbf{smoothness}} requirement originates from the fact that many physical properties are continuous and smooth function with respect to (small) changes in the atomic positions. In the context of atom density field representations, the smoothness requirement is typically fulfilled by using an atom-centred Gaussian function\cite{bartok2013representing} $\rho(\bm{r})=\mathcal{N}(\bm{r}\vert\bm{r}_i,\sigma_i)$ to represent the density instead of a delta function\cite{drautz2019atomic,shapeev2016moment} $\rho(\bm{r})=\delta(\bm{r}-\bm{r}_{i})$. The \emph{\textbf{locality and additivity}} requirement follows the designs of traditional atom-atom empirical potentials [\cref{sect:empirical_forcefield}] and underpins the design of neural-network potentials for molecular modelling since the early days\cite{behler2011atom,behler2011neural,behler2021four}, which express the total energy of a multi-atomic systems $\{A_i\}$ as the sum of contributions from individual atom (affected by its respective atomic environment\cite{bartok2013representing}) $E=\sum_{i\in A}E(A_i)$. Theoretically, such a decomposition scheme can be validated as most dominant quantum effects that contribute to the strong bondings in the chemical systems are typically short-ranged [\cref{sec:DFT}]. This also implies that the \emph{global descriptor} for a given structure is usually constructed as a concatenation of all atom-centred descriptors in the system, which is usually denoted as $\bigoplus_{i=1}^{N}\boldsymbol{\chi}_{i}$. Apparently this does not impose the \textbf{\emph{permutation invariance}} of the global descriptor upon the swapping of the atomic indices. In the context of kernel-based ML, this problem can be overcome through the designs of the similarity kernel,\cite{de2016comparing} through an approach of averaging, best-match, or optimum-transport, when comparing the global similarity between two chemical structures. Finally, an open question in the field is whether it is necessary to enforce and how to achieve the \emph{\textbf{completeness}} of the structural descriptor, such that mapping to the feature space should keep inequivalent structures in apart from each other.

\subsubsection{Physics-Inspired Symmetrised Atom Density Field Representations}
\label{sec:physics_informed_representation}

We are now posed to provide more detailed discussions on the constructions of symmetrised atom density field representations for atomistic structures, focusing primarily on the SOAP descriptor,\cite{bartok2013representing} which also provides a good fundamental understanding on the related framework such as ACE (Atom Cluster Expansion).\cite{drautz2019atomic} The latter serves as the backbone for our subsequent discussions of  the message-passing NN (MACE),\cite{batatia2022mace,batatia2025design}  which, in the past few years, we have seen the emergence of foundational models\cite{batatia2023foundation} for materials chemistry that have been trained based on such framework.  With a slightly different format to the previous sections in this monograph, we provided four \textbf{boxes} which briefly account for the key mathematical background information along the main discussions, such that the latter will not be significantly defocused by the mathematical details. We believed that these mathematical details are, nonetheless, not only necessary for the readers to attain a good understanding on the SOAP frameworks, but also subsequent discussions on modern equivariant NNs. We focus our present discussions around the construction of invariant structural descriptors, and refer interested reader to a broader context of existing literatures on the possible extensions of the SOAP framework.\cite{de2016comparing,willatt2018feature,grisafi2018symmetry,grisafi2019incorporating}

Following the general theoretical framework of atom density field representation,\cite{willatt2019atom} we shall use the \emph{bra-ket} notation inspired from quantum mechanics, where here the ket $|\mathcal{A}\rangle$ is used to annotate an abstract way of representing the elemental compositions and geometric arrangements of atoms in the structure. In the present context, this leads to the description of the \emph{chemical environment}, whereby for a given central atom $i$, it is modelled as a continuous distributions of atom-centred densities located at $j\in\mathcal{N}(i)$, that are, the indices of atoms that are the direct neighbours of the central atom:
\begin{equation}
    \label{eq:atom_env_density_field}
    \rho_i(\bm{r})=\sum_{j\in\mathcal{N}(i)}\rho(\bm{r}-\bm{r}_{ij}),
\end{equation}
such that we may think of $|\mathcal{A}\rangle=\rho_i(\bm{r})$ (which will be short-handed as $|\rho_i\rangle$ subsequently.) The neighbouring atoms $j$ may be identified, for example, based on a predefined cut-off radius $r_c$,\cite{bartok2013representing,behler2011atom} such that $j\in\mathcal{N}(i)$ if and only if $\Vert\bm{r}_{ij}\Vert_2\leq r_c$.  To ensure smoothness, this density function can be further attenuated with a smooth cut-off function $f_{\mathrm{cut}}(r)$ that decays smoothly from 1 to 0 across $r_{c}$. However, such an abstract object of $|\mathcal{A}\rangle$ does not necessarily facilitate computation. This is to be overcome in the SOAP-framework by expanding $|\mathcal{A}\rangle$ onto some basis functions $\langle \mathcal{B}|$ (the bra), which, following the bra-ket notation, can be denoted as $\bm{c}=\langle\mathcal{B}|\mathcal{A}\rangle$, with $\bm{c}$ being a vector of the expansion coefficients. This gives us a high-level general overview of the atom-centred density representation of structures, which we shall further break down our discussions by focusing on the ket- and bra- parts separately.

Using $\rho_i(\bm{r})$ alone does not lead to a good representation of a material's structure. There are two interwinding problems here. Firstly, it does not lead to an invariant representation, since it depends on the choice of $\bm{r}_i$. Although (as seen in \textbf{Box III}) geometric averaging can solve the problem of invariance, it leaves a quantity $\bar{\rho}_i$ that can only encode the information about the relative distances between atom $i$ and all of its neighbours $j$ whereas the information on the relative orientations will be lost. As such, the first step we need to take is to \emph{enrich the information content} that is encoded by the atom-centred density representation. Practically, this is achieved through performing the tensor product on $\rho_i$ (\textbf{Box I}) up to some order $\nu$, which, as shown in \textbf{Box II}, leads to $|\rho^{\otimes \nu}_i\rangle$, that leads to a translationally invariant density field, encoding $(\nu+1)$-body interactions in an atomistic structure, and is conceptually equivalent to the inclusions of bond lengths, bond angles, torsional angles, and possibly other higher-order features in traditional atom-atom forcefields [\emph{c.f.} \cref{sect:empirical_forcefield}]. For the SOAP descriptors, more specifically, it is generally limited to the case of $\nu=2$ due to the high  cost associated with the  computations of tensor products.

\begin{myquote}
\textbf{Box I: A quick primer on tensors}

\footnotesize

In this box, we shall provide readers with some intuitive understanding on tensors and why they are relevant in the field of ML/AI. 

\hspace{10pt} For readers with basic knowledges of linear algebra, they should be familiar with the three basic types of objects in linear algebra [\cref{fig:LA_objects}], which are (a) \textbf{Scalar}, that we shall denote with a lowercased italics $s$, which is a zero-dimensional object that contains a single real number, $i.e.$ $s\in\mathbb{R}$. (b) \textbf{Vector}, that we shall denote with a lowercased bold letter $\bm{v}$, is an one-dimensional object contains $n$ entries. A vector is said to be of dimensionality $n$ as $\bm{v}\in\mathbb{R}^{n}$. (c) \textbf{Matrix}, that we shall denote with an uppercased bold letter $\bm{M}$, is a two-dimensional object contains $(m\times n)$ entries, meaning a matrix is of the dimension $(m\times n)$ and $\bm{M}\in \mathbb{R}^{m\times n}$. 

\begin{minipage}[t]{\linewidth}

\centering
\includegraphics[width=0.6\linewidth]{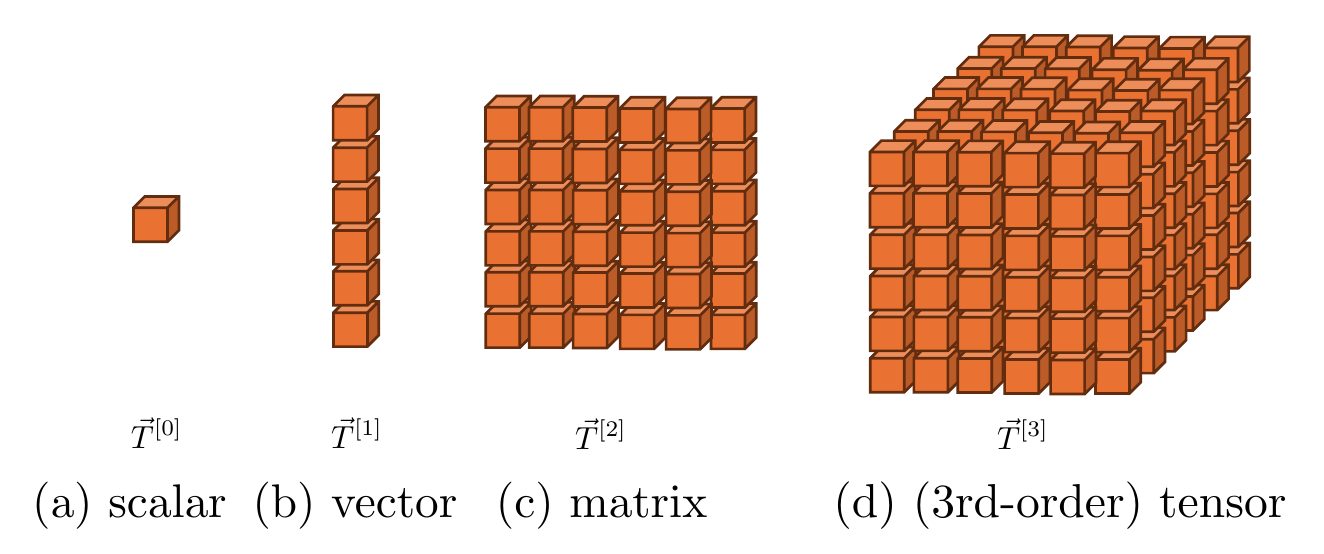}

\captionof{figure}{Schematic illustrations of the four key types of \emph{objects} that are of our subject interest, which each one can be considered as a tenor of orders 0 to 3, respectively.}
\label{fig:LA_objects}
\end{minipage}

\hspace{10pt} However, there is no reason why we should stop here, and we can form objects with higher dimensionalities. For example, stacking multiple matrices together we reach a third-order tensor [\cref{fig:LA_objects}] that can be represented by a three-dimensional cube, and we can continue to combine tensor together to form tensors of even higher orders. Following the convention of \citet{duval2023hitchhiker}, we shall denote tensor of order $c$ defined in the Cartesian space as $\vec{T}^{[c]}$,  and it is apparent to see that scalar, vector and matrix are nothing else but tensors of lower ranks of 0, 1 and 2, respectively. 

\hspace{10pt} In physics, the most common ways to express the interactions between two vectors $\bm{v}_1$, $\bm{v}_2\in\mathbb{R}^N$ are achieved by using either the dot product $a=\bm{v}_1\cdot\bm{v}_2$ or the cross product $\bm{b}=\bm{v}_1\times\bm{v}_2$, which returns a scalar $a\in\mathbb{R}$ or a vector $\bm{b}\in\mathbb{R}^{N}$ of the same dimension as the two interacting vectors. However, both these two constructs either preserves the amount of information (as for the cross-product) or significantly loss the information (as for the dot-product) that are encoded in $\bm{v}$, which, however, are often  non-desirable for learning purposes. In the latter case, maximising the information content by encoding more \emph{fine-grained} details on how components in the data interact, thus providing a larger  space to optimise the learning model, is often needed. This can be achieved by using the \textbf{tensor product} to build higher-order tensors from lower ones. For example, consider the tensor product between two three-dimensional vectors $\bm{v}_1=(v_{1x},v_{1y},v_{1z})^\top$ and $\bm{v}_2=(v_{2x},v_{2y},v_{2z})^\top$, the tensor product between them gives back a rank-2 tensor which is a $(3\times3)$ matrix:
\begin{equation}
   \label{eq:tensor_product_two_vector_3D} \bm{v}_1\otimes\bm{v}_2=\left[\begin{array}{ccc}
        v_{1x}v_{2x} & v_{1y}v_{2x} & v_{1z}v_{2x} \\
        v_{1x}v_{2y} & v_{1y}v_{2y} & v_{1z}v_{2y} \\
        v_{1x}v_{2z} & v_{1y}v_{2z} & v_{1z}v_{2z}
    \end{array}\right].
\end{equation}
This shows that the tensor product encodes the mutual interactions between all possible combinations for pairs of elements in $\bm{v}_1$ and $\bm{v}_2$. Generally, (1) the tensor product between two tensors of $c_1$ and $c_2$ gives a higher-order tensor of $c_1+c_2$, (2) one can continues performing tensor products for multiple tensors to encode higher-order information. Following the notation from \citet{willatt2019atom}, the following notation
\begin{equation}
    \label{eq:density_tensor_prod_v}
    \rho^{\otimes \nu} = \underbrace{\rho\otimes\rho\otimes\cdots\otimes\rho}_{\nu}
\end{equation}
shall be used to denote the $v$-th order tensor product of atom-centred density, which we shall provided further example to demonstrate how it encodes $(\nu+1)$-body atom-atom interactions in Box II. 

\hspace{10pt} To provide more physical context, one can refer back to the previous discussions on anharmonic vibrations in solids [\cref{sec:anharmonic_vibs}], the second- ($\boldsymbol{\Phi}_{ij}^{\alpha\beta}$), third-  ($\boldsymbol{\Phi}_{ijk}^{\alpha\beta\gamma}$) and fourth- ($\boldsymbol{\Phi}_{ijkl}^{\alpha\beta\gamma\delta}$) as in \cref{eq:higher_order_vib_Hamiltonian} are all tensors that encodes the information on how pair, triplets and quadruplets of atoms interact when they are independently displaced along a specific Cartesian direction.

\hspace{10pt} Conversely, one can also perform \textbf{tensor contraction} along specific dimension(s) of a higher-order tensor (typically through summations, which can be viewed equivalently as a process of information compression), to generate lower-order tensor. Using again the example given in \cref{eq:higher_order_vib_Hamiltonian}, the operation that involves the fourth-order tensor (which was written in the form of \emph{Einstein summation}), collapse the repeated indices in the tensor product to return a vector as a result:
\begin{equation*}
    \underbrace{\boldsymbol{\Phi}_{ijkl}^{\alpha\beta\gamma\delta}\Delta\bm{R}_l^\delta\Delta \bm{R}_{k}^{\gamma}\Delta\bm{R}_j^{\beta}}_{\mbox{\footnotesize{Einstein summation}}}\equiv\sum_{l\delta}\sum_{k\gamma}\sum_{j\beta}\boldsymbol{\Phi}_{ijkl}^{\alpha\beta\gamma\delta}\Delta\bm{R}_l^\delta\Delta \bm{R}_{k}^{\gamma}\Delta\bm{R}_j^{\beta}=\bm{F}_{i}^{\alpha,(\mathrm{anh},4)},
\end{equation*}
which effectively gives the anharmonic component for the atomic force on atom $i$ along the $\alpha$-direction originated from the fourth order vibration anharmoncity.\cite{yang2021altas}
\end{myquote}

%
%

\begin{myquote}
\textbf{Box II: $\rho^{\otimes \nu}$ encodes $(\nu+1)$-body correlations}

\footnotesize

In this box, we shall provide a simplified derivation to showcase, that  applying the tensor products on the atom-centred density fields will enable us to encode the many-body atom-atom correlations in a multiatomic systems, and how this is related to the many body interactions in traditional atom-atom forcefield as discussed in \cref{sect:empirical_forcefield}. 

\hspace{10pt} Starting from the definition of atom-centred density field \cref{eq:atom_env_density_field}, since $\rho_i^{\otimes 1}=\rho_i$, it is apparent that $\rho_i^{\otimes 1}$ encodes two-body information between $i$ and any one of its neighbours $j$. If one integrate out the dependency upon $\bm{r}$
\begin{equation}
    \label{eq:two-body-density-tensor-product-invariant}
    \mathcal{A}(\bm{r}_{ij})=\int\rho(\bm{r}-\bm{r}_{ij})\mathrm{d}\bm{r},
\end{equation}
then we essentially retrieve a quantity that is only dependent on the two-body interactions, resembling the form of the bonding term in atom-atom forcefield.

\hspace{10pt} For the second-order tensor product of $\rho_i$ [\cref{eq:density_tensor_prod_v}], we have
\begin{equation}
    \label{eq:three-body-density-tensor-product}
   \rho_i^{\otimes 2}=\sum_{j \in\mathcal{N}(i)}\sum_{k \in\mathcal{N}(i)}\rho(\bm{r}-\bm{r}_{ij})\rho(\bm{r}-\bm{r}_{ik}),
\end{equation}
which clearly encodes the information amongst a triplet of atoms $(i,j,k)$ in the system. Without going into further mathematical details, we can envisage that such an expression not only contain translational degree of freedom but also an extra one that depends on the relative orientation between the two vectors $\bm{r}_{ij}$ and $\bm{r}_{ik}$, which can be characterised by the bond-angle $\theta_{ijk}=\bm{u}_{ij}\cdot\bm{u}_{ik}/(\Vert \bm{u}_{ij}\Vert\cdot\Vert\bm{u}_{ik}\Vert)$, thus integrating out the translational part in \cref{eq:three-body-density-tensor-product} will lead us to a term $\mathcal{A}(\theta_{ijk})$. One can thus further extend this argument to demonstrate how dihedrals $\varphi_{ijkl}$ may enter in the atom-centred density field representation through $\rho^{\otimes \nu}$ with $\nu=3$, and so on.

\end{myquote}

The next step is to \emph{symmetrise} $\rho^{\otimes \nu}_i$  so it becomes invariant with respect to the rotation of the reference frame, within which the molecular structure is defined. As shown in \textbf{Box III}, for atomistic structures that lives in the space of $\mathbb{R}^3$, the formal way to introduce rotational invariance is by averaging over all copies of $\rho^{\otimes\nu}_i$ that can possibly be generated by operating the rotational operator $\hat{\bm{R}}_{\boldsymbol{\theta}}$ on $\rho^{\otimes\nu}_i$ (for $\nu$ times!), leading to the rotationally averaged atom-centred density field, that is to be denoted with $\overline{\rho^{\otimes \nu}_i}$. Such process of rotational averaging can be achieved through performing the Haar integration on $\rho^{\otimes\nu}_i$.

%
%

\begin{myquote}
\textbf{Box III: $\overline{\rho^{\otimes \nu}}$ -- rotational invariance via averaging}

\footnotesize
Physically, the integration procedures as performed in \cref{eq:two-body-density-tensor-product-invariant} is to remove the translational dependent part of the coordinate system in the definitions of the atomic positions, that eventually imposed the important \textbf{translational invariance} in $\mathcal{A}$. Such a procedure is generally referred to as the \emph{\textbf{Haar integration}}, which, in general, for any given quantity $\mathcal{A}$ (in our case, will be a tensorial quantity) where a \textbf{\textit{group action}} $\hat{S}$ upon it will equivariantly transform into a new quantity $\hat{S}\mathcal{A}$, one can obtain a  quantity that is invariant to the transformation of $\hat{S}$ by integrating over all possible $\hat{S}$ acting upon $\mathcal{A}$:
\begin{equation}
    \label{eq:Harr_integration}
    \bar{\mathcal{A}}=\int\hat{S}\mathcal{A}\mathrm{d}\hat{S},
\end{equation}
and a generalisation to the tensor product reads,
\begin{equation}
    \label{eq:Harr_integration_tensor}
    \overline{\mathcal{A}^{\otimes\nu}}=\int\underbrace{\hat{S}\mathcal{A}\otimes \hat{S}\mathcal{A}\cdots\hat{S}\mathcal{A}}_{\nu} \mathrm{d}\hat{S}.
\end{equation}

\hspace{10pt} Returning to the tensor product of the atomic density, and focusing on the term $\rho^{\otimes 1}$, one sees that, after removing the trivial symmetry of translation, the term $\mathcal{A}(\bm{u})$ [\cref{eq:two-body-density-tensor-product-invariant}] is a vector (or rank-1 tensor) property, which is \emph{not} rotationally invariant. In the three-dimensional space, the group that govern the rotation of objects is the \textbf{SO(3) rotation group} [see Box IV], with group action $\hat{\bm{R}}_{\boldsymbol{\theta}}$ that rotates objects in the space (which we shall not worry too much about its concrete form). Based on the about discussion on the Haar integration,  an invariant representation of $\rho^{\otimes 1}$ can thus be constructed as,
\begin{equation}
    \label{eq:two_body_density_rotation_invariance_haar}
    \overline{\rho^{\otimes 1}}=\int_{\mathrm{SO(3)}}\left(\hat{\bm{R}}_{\boldsymbol{\theta}}
    \rho \otimes \hat{\bm{R}}_{\boldsymbol{\theta}}
    \rho\right) \mathrm{d}\hat{\bm{R}}_{\boldsymbol{\theta}},
\end{equation}
which leads to the notation of $\overline{\rho^{\otimes\nu}}$ in the  literatures\cite{willatt2019atom,musil2021physics} that represents the \textbf{symmetrised atom centred density fields}.  As shown in \citet{willatt2019atom}, for a 3D vector, the Haar integration may be evaluated with a suitable parameterisation of the rotation, such as using the Euler angles, and to be computed as
\begin{equation}
    \label{eq:euler_Haar_integration}
    \int_0^{2\pi}\mathrm{d}\alpha\int_0^\pi\sin\beta \mathrm{d}\beta\int_0^{2\pi}\mathrm{d}\gamma \bm{u}(\alpha,\beta,\gamma),
\end{equation}
which one can anticipate that the end result is only dependent on the modulus of $\bm{u}$, \emph{i.e.} $\Vert\bm{u}\Vert_2$.
\end{myquote}

%
%

Nevertheless, explicitly performing the Haar integration to obtain $\overline{\rho^{\otimes \nu}_i}$ for every atomistic structure to be investigated will not be numerically efficient.  The mathematical trick that can be used to simplify problem is via the use of group theory (which are further explained in \textbf{Box IV}), particularly, the rotational group in three dimensions, or SO(3), and it further brings the bra-component for constructing atom-centred density field representation into the equation. Practically, this is typically achieved via three steps. Firstly, one projects $\rho_{i}(r)$ on a set of radial $\{R_n(r)\}$ and angular $\{Y_{\ell}^{m}(\hat{\bm{r}})\}$ (where $\hat{\bm{r}}$ denotes a projection of $\bm{r}$ onto a unit sphere) to obtain a set of expansion coefficients $\{c_{n\ell m}\}$ (for $n=1,2,\ldots,m_{\max}$, $\ell=0,1,2,\ldots,\ell_{\max}$ and $m=-\ell,-\ell+1,\ldots,\ell-1,\ell$) in which,
\begin{equation}
    \label{eq:soap_expansion_coeff}
    c_{n\ell m}:=\langle n\ell m|\rho_i(\bm{r})\rangle=\int R_{n}(r)Y_{\ell}^{m}(\hat{\bm{r}})\rho(\bm{r})\mathrm{d}\bm{r}.
\end{equation}
Collecting $\{c_{n\ell m}\}$ leads to a vector of expansion coefficients $\bm{c}$. The radial basis functions are typically chosen as $n$ Gaussian functions with equi-distant centres positioned in the interval $[0,r_c]$. The radial components are generally invariant and hence one only needs to focus on symmetrising the angular part in the second step.
Using the SOAP descriptor as example, $\rho^{\otimes2}_i$ can be directly obtained by performing the tensor product of $\bm{c}\otimes\bm{c}$, which gives a combination of rank-2 tensors with dimensions from $(0\times0)$ up to $(\ell_{\max}\times\ell_{\max})$ where the tensor elements can be written as $c_{n_1\ell_1 m_1,n_2\ell_2 m_2}:=\langle n_1\ell_1 m_1,n_2\ell_2 m_2 |\rho^{\otimes 2}_i\rangle=c_{n_1\ell_1 m_1}c^{*}_{n_2\ell_2 m_2}$ [\cref{eq:tensor_prod_decomp_general}]. For the na\"{i}ve SOAP descriptor,\cite{bartok2013representing} in which we are  interested in the rotationally invariant part of the tensor product, only the entries with $\ell_1=\ell_2$ are relevant, and invariance can be further imposed by summing over the magnetic quantum numbers for a given $\ell$, leading to the so-called SOAP spectrum representation:
\begin{equation}
\label{eq:SOAP_spectrum}
\chi_{n_1n_2\ell}=\sum_{m=-\ell}^{\ell}c_{n_1\ell m}c^{*}_{n_2\ell m},
\end{equation}
(or equivalently as $\langle n_1n_2\ell|\overline{\rho^{\otimes 2}_i}\rangle$),
whereby the final SOAP structural descriptor for a given centre $i$ (or $\boldsymbol{\chi}_i$) is constructed by concatenating  $\chi_{n_1n_2\ell}$ for this centre with all combinations of $(n_1,n_2,\ell)$. The global descriptor for a structure containing $N$ atoms can thus be constructed by further concatenating $\boldsymbol{\chi}_i$ as $[\boldsymbol{\chi}_1,\boldsymbol{\chi}_2,\ldots,\boldsymbol{\chi}_N]^\top$. This thus concludes our short discussions on the philosophy behind the designs of invariant SOAP structural descriptors. 
 
\begin{myquote}
    \textbf{Box IV: simplifying the computation of density-based representation with group theory}

\footnotesize
\hspace{10pt} One can envisage that, explicitly computing the integral such as \cref{eq:two_body_density_rotation_invariance_haar} to achieve rotational invariance will be numerically expensive, and applying the knowledge of group theory, in this case, the SO(3) rotational group in $\mathbb{R}^3$, will allow us to break down the rotation into its elementary building blocks, which are called the \emph{irreducible representations (irreps)}, that can be more easily manipulated independently. 

\hspace{10pt} Rather than treating the abstract object of atom-centred density $\rho$, we shall focus our discussion using more concrete examples of tensors (which will also become relevant to the subsequent discussions on equivariant NNs). For this, we start from a rank-1 tensor $\hat{T}^{[1]}\in\mathbb{R}^{3}$ defined in the Cartesian space, and tensors of higher ranks $\hat{T}^{[\nu]}$ can be constructed using tensor products as outlined in Box I. In this case, rotation of tensors of different ranks can be achieved by repeatedly applying the rotational operators, for examples:
\begin{align*}
    \rho^{\otimes 1}\sim\hat{T}^{\prime [1]}&\leftarrow \hat{\bm{R}}_{\boldsymbol{\theta}}\hat{T}^{[1]},\\
    \rho^{\otimes 2}\sim\hat{T}^{\prime [2]}&\leftarrow 
    \hat{\bm{R}}_{\boldsymbol{\theta}}\hat{\bm{R}}_{\boldsymbol{\theta}}\hat{T}^{[2]}, \\
    \rho^{\otimes 3}\sim\hat{T}^{\prime [\nu]}&\leftarrow \underbrace{\hat{\bm{R}}_{\boldsymbol{\theta}}\hat{\bm{R}}_{\boldsymbol{\theta}}\cdots\hat{\bm{R}}_{\boldsymbol{\theta}}}_{\nu}\hat{T}^{[\nu]}.
\end{align*}
Several key observations should be noted here: (1) Tensors of different types (indicated by their order $\nu$) transform differently, as characterised by the number of times that the operator $\hat{\bm{R}}_{\boldsymbol{\theta}}$ is applied. (2) Repeatedly performing the product operations to rotate higher-order tensors will not be computationally efficient. Following the pedagogical introductions in \citet{duval2023hitchhiker}, group theory gives us a mathematical tool to \emph{decompose} a higher-order tensor $\hat{T}^{[\nu]}$ into the linear combination of \emph{irreps} (thus only summation operations are required), which are the \emph{orthogonal basis} for the SO(3) rotational group. These basis are of the form of tensors themselves, which also follow their own transformation rules that depend on their ranks (dubbed as $\ell$, which is the \textit{angular momentum quantum number} in quantum mechanics), but in this case, can be achieved with a single (basis specific) Wigner D matrix, denoted as $\mathbf{D}^{\ell}(\hat{\bm{R}}_{\boldsymbol{\theta}})$).

\hspace{10pt} As a concrete example (see more detailed derivation in \citet{duval2023hitchhiker}), the decomposition of $\hat{T}^{[2]}$ (that is a tensor product between two vectors in $\mathbb{R}^3$) can be written as:
\begin{equation}
    \label{eq:rank_2_tensor_decomp}
    \vec{T}_1^{[1]}\times\vec{T}_2^{[1]}\to\vec{T}^{[2]} = \lambda_1\cdot \underbrace{\hat{T}_1^{[2]}}_{\ell=0}+\left[ \begin{array}{c}
         \lambda_2  \\
         \lambda_3 \\
         \lambda_4
    \end{array} \right] \cdot \underbrace{\left[ \begin{array}{c}
         \hat{T}_2^{[2]} \\
         \hat{T}_3^{[2]}  \\
         \hat{T}_4^{[2]} 
    \end{array} \right]}_{\ell=1} + \left[
    \begin{array}{c}
         \lambda_5  \\
         \lambda_6 \\
         \lambda_7 \\
         \lambda_8 \\
         \lambda_9
    \end{array} \right] \cdot \underbrace{\left[ \begin{array}{c}
         \hat{T}_5^{[2]} \\
         \hat{T}_6^{[2]}  \\
         \hat{T}_7^{[2]} \\
         \hat{T}_8^{[2]} \\
         \hat{T}_9^{[2]}
    \end{array} \right]}_{\ell=2}
\end{equation}
in which we use the symbol $\hat{T}^{[2]}_i$ to specify the $i$-th basis that is a rank-2 tensor. This shows that $\vec{T}^{[2]}$ can be decomposed into 9 basis of $\hat{T}^{[2]}_i$ [which is consistent with the number of elements in the matrix form of $\vec{T}^{[2]}$, see \cref{eq:tensor_product_two_vector_3D}]. These basis  can be further grouped depending on $\ell$, which leads to the sum as expansions in the 1, 3, and 5 dimensional spaces, expressed as tensor basis of ranks 0, 1 and 2, respectively, or more compactly as $1\otimes1=0\oplus1\oplus2$. In general, for the tensor products between two tensors of rank $\ell_1$ and $\ell_2$, we have:
\begin{equation}
    \label{eq:tensor_prod_decomp_general}
    \ell_1\otimes\ell_2=\bigoplus_{|\ell_1-\ell_2|}^{\ell_1+\ell_2}\hat{T}^{[\ell]}.
\end{equation}
Finally, collecting the expansion coefficients $\lambda$'s according to $\ell$ into a vector, and following the conventional notation of magnetic quantum numbers $m_\ell\in\{-\ell,-\ell+1,\cdots,\ell-1,\ell\}$, that gives $\vec{\lambda}^{(\ell)}=\left[\lambda^{(\ell)}_{-\ell},\lambda^{(\ell)}_{-\ell+1},\ldots,\lambda^{(\ell)}_{\ell-1},\lambda^{(\ell)}_{\ell}\right]^\top$, then the rotation of the rank-2 tensor product can be achieved in the following additive form, which is the result if we work with \textbf{\textit{finite dimensional representation theory}}:
\begin{equation}
    \label{eq:Wigner_D_matrix_rotate_rank_2_tensor}
    \hat{\bm{R}}_{\boldsymbol{\theta}}\hat{\bm{R}}_{\boldsymbol{\theta}}\vec{T}^{[2]} = \bm{D}^{(0)}(\hat{\bm{R}}_{\boldsymbol{\theta}})\vec{\lambda}^{(0)}\oplus \bm{D}^{(1)}(\hat{\bm{R}}_{\boldsymbol{\theta}})\vec{\lambda}^{(1)}\oplus \bm{D}^{(2)}(\hat{\bm{R}}_{\boldsymbol{\theta}})\vec{\lambda}^{(2)}.
\end{equation}

\hspace{10pt} Is this the best solution? Not necessarily! Firstly, one is yet to perform rotational average over the Wigner D-matrices to achieve rotationally invariance. Secondly, particularly for encoding interactions in NNs through tensor products, it does not seems efficient to perform tensor decomposition after every tensor product operations, as the possible number of tensor products to handle can be extremely large. Finally, one may also question whether the use of a finite basis to decompose rotation can sufficiently encode finite orientational details in molecular modelling (\emph{e.g.} in the case of continuous geometric change during MD simulations). The alternative approach, as more widely adopted in both descriptor-based ML\cite{bartok2013representing,drautz2019atomic} and equivariant neural networks\cite{duval2023hitchhiker} is to harness the \textbf{spherical harmonics} $\{Y_{\ell}^{m}\}$ as the \emph{\textbf{infinite dimensional basis set}}, to first expand the angular part of the (uncorrelated) atom-centred density $\rho(\bm{r})$ and then construct higher-order invariant and/or equivariant features from it. 

\hspace{10pt} In the subsequent discussions, we focus on the angular part of the density field $\rho(\bm{r})$, assuming it has been projected on to the surface of the unit sphere in $\mathbb{R}^3$, thus the radial part can be temporarily discarded. In this case,   $\rho(\bm{r})$ (a rank-1 vector) can be expanded in the basis of spherical harmonics as 
\begin{equation}
    \label{eq:sph_expand_density}
    \rho_i(\bm{r})=\sum_{\ell m}c_{\ell m}Y_\ell^m (\bm{r}),
\end{equation}
up to a predefined hyperparameter of $\ell_{\max}$. Similar to $\vec{\lambda}^{(\ell)}$, the expansion coefficients $c_{\ell m}$ can be grouped into tensors of different orders of $\ell$, which may be denoted as $\vec{c}^{(\ell)}$. Using the bra-ket notation from Dirac's formulations of quantum mechanics, that is also adopted in the literatures of density-based structural representations,\cite{willatt2019atom} we can compute $c_{\ell m}=\langle\ell m|\rho_i\rangle$, which projects an abstract density onto a concrete representation as denoted by the state of $|\ell m\rangle$. At the limit of the $\delta$-function-like density field, $c_{\ell m}\leftarrow \sum_j Y_{\ell m}(\bm{r}_{ij})$ (\emph{c.f.} Steinhardt order parameters\cite{steinhardt1983bond}), thus can be easily computed from the interatomic vector. In this case, collecting the expansion coefficients 
\begin{equation}
    \label{eq:expansion_coeff_vector_sph}
    \vec{\bm{c}}=\left[\vec{c}^{(0)},\vec{c}^{(1)},\ldots,\vec{c}^{(\ell_{\max})}\right]
\end{equation}
leads to a \emph{representation} that encodes the angular information of the density field, which is yet to be symmetrised. 

\begin{minipage}[t]{\linewidth}

\centering
\includegraphics[width=0.75\linewidth]{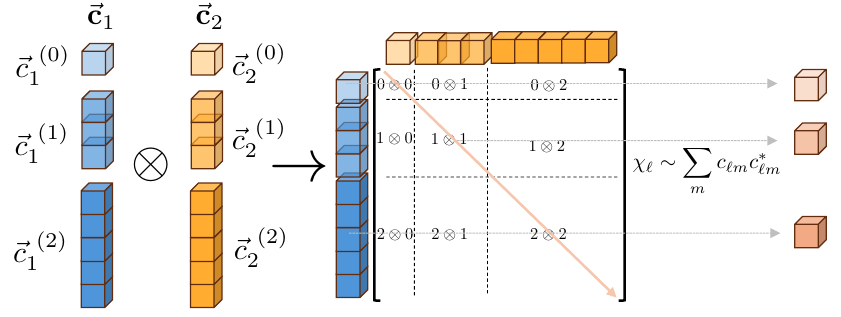}

\captionof{figure}{Illustration of the construction for the (only the angular part) of the SOAP descriptor, starting from tensor product of two list of tensors consisting the expansion coefficients (left). The tensor product leads to a range of multiple tensors, which we arrange into a two-dimensional matrix form (middle), where the final SOAP descriptor comes from extracting only the diagonal part ($\ell_1=\ell_2$) followed by summing over the magnetic quantum number $m$ for a given $\ell$ to retain invariance. }
\label{fig:soap_invariant_tensor_prod}
\end{minipage}

\vspace{10pt}

\hspace{10pt} To proceed, the second-order tensor product between atom-centred density fields can now be computed through the tensor product $\rho^{\otimes 2}\sim\vec{\bm{c}}\otimes\vec{\bm{c}}$, which in terms we need to form tensor products between all possible pairs of tensors $\vec{c}^{(\ell_1)}\otimes\vec{c}^{(\ell_2)}$ in $\vec{\bm{c}}$ with $\ell_1,\ell_2\in[0,\ell_{\max}]$, leading to a set of multiple tensors (each of the rank $\ell_1+\ell_2$), as shown in the middle section of \cref{fig:soap_invariant_tensor_prod}. As indicated in \cref{eq:tensor_prod_decomp_general}, each one of these tensors can be further decomposed into a sum of tensors with ranks from $|\ell_1-\ell_2|$ to $(\ell_1+\ell_2)$, which is usually expressed as $\ell_1\otimes\ell_2=|\ell_1-\ell_2|\oplus|\ell_1-\ell_2|+1\oplus\cdots\oplus \ell_1+\ell_2$.  Another important point, which will becomes useful later, is that each decomposed tensor do not contribute to the sum equally, but is weighted by some coefficients, known as the \textbf{\textit{Clebsh-Gordan (CG) coefficients}}. In the case of $\rho^{\otimes 2}$, we have
\begin{equation}
    \label{eq:CG_decomposition}
    \rho^{\otimes 2}_i\sim\sum_{LM}\sum_{\substack{\ell_1 m_1\\ \ell_2 m_2}}C^{LM}_{\ell_1 m_1\ell_2 m_2}c_{\ell_1 m_1}c_{\ell_2 m_2},
\end{equation}
in which the CG coefficient $C^{LM}_{\ell_1 m_1\ell_2 m_2}$ is defined based on the combinations of $(\ell_1,\ell_2)$ and $(m_1,m_2)$ that can lead to the decomposed tensor element with $L\in[|\ell_1-\ell_2|,\ldots,\ell_1+\ell_2]$ and $M\in[-L,\ldots,L]$.

\hspace{10pt} At this moment, we are only interested in retrieving the invariant components $L=0$ to construct the final SOAP descriptor,\cite{bartok2013representing} so we will only be interested in the tensor products $\vec{c}^{(\ell_1)}\otimes\vec{c}^{(\ell_2)}$ with $\ell_1=\ell_2=\ell$ and extract only the decomposed component with $L=0$. (In the SO(3) group theory language, we say there exists only one \emph{pathway} to combine $\ell$'s that gives $L$, in this case, the only relevant CG coefficient in \cref{eq:CG_decomposition} is left with $C^{0M}_{\ell m_1\ell m_2}$.) From here, it is obvious that $\vec{c}^{(\ell)}\otimes\vec{c}^{(\ell)}$ is a rank-2 tensor with the dimension of $((2\ell+1)\times(2\ell+1))$, and matrix element $c_{\ell m_1}c_{\ell m_2}$. Without going into further mathematical details, we can intuitively understand that, a practical way to extract a scalar (thus invariant) component of the tensor can be achieved by taking the trace of this rank-2 tensor, leading to the final construct of the SOAP descriptor (angular part only, rightmost of \cref{fig:soap_invariant_tensor_prod}):
\begin{equation*}
\chi_{\ell}=\sum_{m}c_{\ell m}c^{*}_{\ell m},
\end{equation*}
which here we put the complex conjugate back, as in the original formulation\cite{bartok2010gaussian} to signify that the spherical harmonic basis are complex numbers.


\end{myquote}

At this point, it is natural to think about following the above recipe to construct many-body correlations $\rho^{\otimes \nu}_i$ ($\nu>2$) that accounts for more complex multiatomic interactions that may contribute to the total energy of the system, that is in line with the expansion scheme introduced in \cref{eq:total_energy_atom_pos_expand}. A particular example is for the modelling of solid-state materials within the conceptual framework of the embedded atom scheme [as introduced in \cref{sect:empirical_forcefield}] where all the surrounding atoms around the central atom can mutually interact with each other. However, the na\"{i}ve way of the computing the tensor product and extracting the invariant components, as explained in \textbf{Box IV} become practically intractable for higher-order terms. Considering, for example $\rho^{\otimes 3}_i$, which, following \cref{eq:CG_decomposition}, we should have (following the ACE notation where $A_{n\ell m}=\langle n\ell m|\rho_{i}\rangle$):
\begin{equation}
    \label{eq:CG_decomposition_3rd_order}
    \rho^{\otimes 3}_i\sim\sum_{\substack{\ell_1 m_1\\ \ell_2 m_2\\ \ell_3 m_3}}C^{00}_{\ell_1 m_1\ell_2 m_2 \ell_3 m_3}A_{\ell_1 m_1}A_{\ell_2 m_2}A_{\ell_3 m_3}.
\end{equation}
Focusing only on the invariant part thus enforcing $\ell_1\otimes\ell_2\otimes\ell_3\to0$, one can see there are multiple pathway to select the combinations of $\ell_i$ and $m_i$ (i=1,2,3) to achieve this outcome. For example, considering the case with $\ell_1=\ell_2=\ell_3=2$, we may first combine $\ell_1\otimes\ell_2=0\otimes1\otimes2\otimes3\otimes4$, in which only $\ell_3=2$ can be combined with the $\ell_1\otimes\ell_2=2$ (which will be denoted as $L_{12}$ and correspondingly $M_{12}$) component to obtain $L=0$. However, now with $M_{12}\in[-2,2]$ and $m_3\in[-2,2]$, we have five combinations of $(M_{12},m_3)$ that can lead to the final outcomes of $M=0$, which are $(-2,2)$, $(-1,1)$, $(0,0)$, $(1,-1)$ and $(2,-2)$. One can easily envisage that this can soon become a book-keeping exercise that, the above approach, that aim to surrogate higher-order body correlations based on the uncoupled basis of $\{A\}$, is prone for errors for large $\nu$. The ACE framework, as formulated by \citet{drautz2019atomic}, is to first replace the uncoupled basis $\{A\}$ with a coupled basis $\{B^{(\nu)}\}$ that is formed by first summing over all combination of indices that fulfil the condition of $\sum_{i=1}^{\nu}m_{i}=0$, for example, in the cases of $\rho^{\otimes \nu}$ with $\nu=0,1,2$, it leads to 
\begin{subequations}
\label{eq:ACE_basis}
\begin{align}
    B_{in}^{(1)}&=A_{in00},\label{eq:ACE_1_basis}\\
    B_{in_1 n_2\ell}^{(2)} &=\sum_{m=-\ell}^{\ell}(-1)^m A_{in_1\ell m}A_{in_2\ell -m},\label{eq:ACE_2_basis}\\
    B_{\substack{in_1 n_2 n_3\\ \ell_1 \ell_2 \ell_3}}^{(3)} &= \sum_{m_1=-\ell_1}^{\ell_1}\sum_{m_2=-\ell_2}^{\ell_2}\sum_{m_3=-\ell_3}^{\ell_3}\left(\begin{array}{ccc}
        \ell_1 & \ell_2 & \ell_3\\
         m_1 & m_2 & m_3
    \end{array}\right)A_{in_1\ell_1 m_1}A_{in_2\ell_2 m_2}A_{in_3\ell_3 m_3}, \label{eq:ACE_3_basis}
\end{align}
\end{subequations}
in which the coefficients in \cref{eq:ACE_3_basis} are given in the Wigner $3j$ symbol. With this construct, the part of the total system energy in \cref{eq:total_energy_atom_pos_expand} that is contributed from the $i$-th central atom interacting with its surrounding environment (\emph{i.e.} the cluster $\boldsymbol{\sigma}$ around $i$), can be expressed in the $B$-basis without invoking the tensor product among $B$'s:
\begin{equation}
    \label{eq:ACE_atom_energy_expansion}
    E_i(\boldsymbol{\sigma})=\sum_{n}c_n^{(1)}B_{in}^{(1)}+\sum_{n_1n_2\ell}^{\prime}c_{n_1 n_2\ell}^{(2)}B_{in_1 n_2\ell}^{(2)}
    +\sum_{\substack{n_1 n_2 n_3\\ \ell_1 \ell_2 \ell_3}}^{\prime}c_{\substack{n_1 n_2 n_3\\ \ell_1 \ell_2 \ell_3}}^{(3)}B_{\substack{in_1 n_2 n_3\\ \ell_1 \ell_2 \ell_3}}^{(3)}+\cdots,
\end{equation}
where $c^{(\nu)}$ are no longer fixed expansion coefficients but learnable coefficients for fitting the ML interatomic potentials. With such a construction, it can be shown that the complexity of computing the $(\nu+1)$-body correlation features can be significantly reduced from that of $\mathcal{O}((2\ell+1)^\nu)$ for na\"{i}ve SOAP to that of $\mathcal{O}(\nu(2\ell+1)^2)$ for ACE with a given $\ell$.

\begin{figure}[htb]
    \centering
    \includegraphics[width=\linewidth]{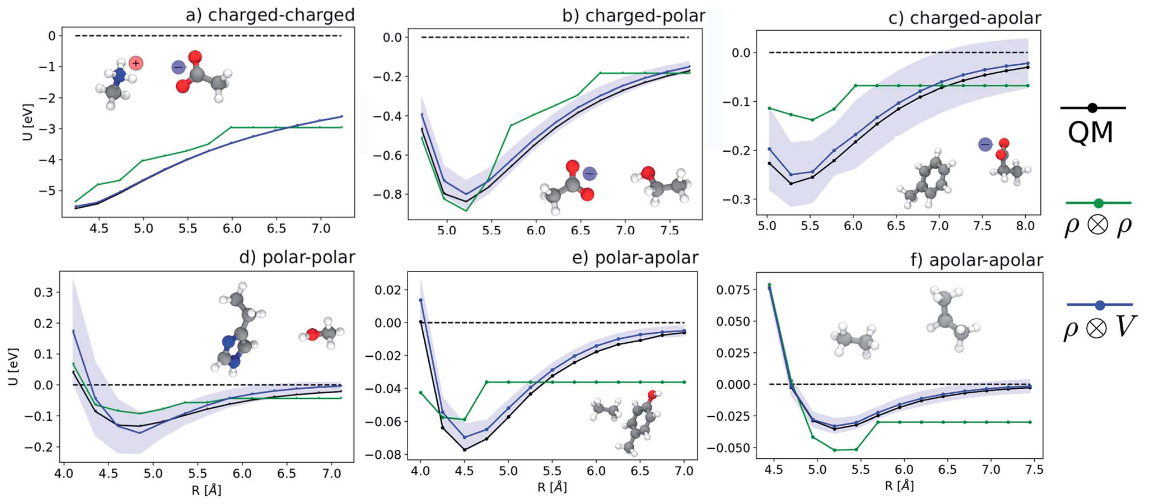}
    \caption{Median-error binding curves for molecular dimers across six different classes of intermolecular interactions. (black lines) Quantum-mechanical calculations. (green lines) Predictions with the SOAP descriptor constructed based on the $\left\vert \overline{\rho_i\otimes\rho_i}\right\rangle$ model. (blue lines) Predictions based on a $\left\vert \overline{\rho_i\otimes V_i}\right\rangle$ models that takes into account for the long-ranged interactions [adopted from \citet{grisafi2021multi}].}
    \label{fig:binding_curve_soap_longrange}
\end{figure}

\textbf{\emph{Interpretability of descriptor-based ML models}} Before we conclude this part that has discussed the physics-inspired constructions of the atom density field representations for atomistic structures for ML, we would like to briefly account on how the descriptor-based ML  generally lead to better model interpretability. At the very first place, as shown in \cref{appendix:supervised_ML} and \ref{appendix:unsupervised_ML}, the underlying ML models are usually quite simple in terms of the mathematical operations that are to be performed on the input data in order to reach the target output are generally very simple, mostly linear operations. The possible existence of the complex nonlinearity is handled by the constructions of the similar kernel matrix (or equivalently, in the context of atom density field representations, the structural descriptors). As mentioned above, such a principled model design is already based on a strong physical prior on our fundamental understandings on the structural-property relationships for materials, in which similar materials should adopt similar properties. 

On the other hand, such design also implies that, the predictive accuracy of supervised ML model or the insightfulness of the unsupervised ML model, will be predominantly determined by the quality of the descriptors that are hand-crafted based on the specific domain knowledge that underpins the specific problems to be solved at hand. As seen above, such knowledge entails the structural, physical and/or physical properties of the datasets, that may additionally include the incorporations of constraints, such as symmetries in the case of structural descriptors, to enable efficient learnings. This is critical when it comes to asking the question on which set of factors contributes strongly to specific properties for the materials, or conversely, what features might be lacking that are critically needed for improving the predictive accuracy of the model. For example, as shown in \cref{fig:binding_curve_soap_longrange} when it comes to the long-ranged noncovalent interactions between molecular fragments,\cite{grisafi2021multi} simply using the density field $\overline{\rho_i^{\otimes 2}}$ that is constructed from the neighbouring atoms identified based on a pre-defined cut-off radius $r_c$ leads to early plateauing thus no able to surrogate the full quantum-mechanical binding curve for the molecular dimers. This naturally leads to one proposal (of, possibly many others) to improve the SOAP model by augmenting $\rho_i$ with a Coulomb-type long-range potential:
\begin{equation*}
    |\rho\otimes V\rangle\sim\int \mathrm{d}\bm{x}^\prime \frac{|\rho\rangle}{|\bm{x}-\bm{x}^\prime|},
\end{equation*}
with which one is able to arrive at the ML predicted binding curves that closely follows that of the quantum-mechanical ones. Model development efforts as such can draw parallel comparison to those earlier endeavours on the development of polarisable forcefields [\cref{sect:polarisable_FF}] that both aim to capture how long-range electrostatic interactions and chemical environments interdepend on each other.

It is therefore clear, that, the descriptor-based ML represents a paradigm in which the designs and optimisations of the materials representations are, in principle, separated from the optimisations of the actual learning algorithm, which is distinctively different from the deep learning approaches to be discussed in the next section. This means that the descriptors are fixed prior to, and not further optimised, during the model learning process, which may otherwise leads to a new set of features in the model's latent space that are significantly different from the initial input, highlighting other sets of features of the training data that may not be immediately apparent on what these features really are. In this case, the use of hand-crafted descriptors becomes particularly helpful when it comes to enhancing the model's interpretability, especially many hyperparameters that control the computations of the descriptors bear strong physical meaning, such as the cut-off radius ($r_c$, controlling how far apart a central atom can see its neighbours) and the many-body order ($\nu$, controlling the maximal number of interacting neighbouring atoms), in the construction of the SOAP-type structural descriptors.

\subsection{Fundamentals of Deep Learning}\label{sec:dl_fundamentals}
\subsubsection{Motivations}

Whilst the relatively simple mathematical constructs of the ML models, that are applied together with the physics-inspired structure descriptors, provide us with high degrees of interpretability to deepen our fundamental understanding on the chemistry-structure-property relationships of materials, which have been reviewed with many examples in \cref{sec:supervised_ML} and \ref{sec:unsupervised_ML}, they also imposed significant constraints for broadening the scopes, as well as increasing the levels of physical complexities, of the problems to which these models can be applied. There are a few fundamental reasons behind these limitations. 

Firstly, physics-inspired structure descriptors handle symmetries through a `baked-in' manner via the extensive use of tensor decomposition techniques, which,  as seen in \cref{sec:physics_informed_representation}, are typically achieved through the computations of the CG coefficients whose values are generally fixed. Whilst such an approach is pragmatic for preserving the invariance and equivariance of the structural features, they do not enlarge the space for \emph{optimising} the representations. The descriptors are thus \emph{fixed} in a sense that they are chosen and optimised for one class of material system (\emph{e.g.} bulk inorganic solids) and a specific learning task (being either unsupervised or supervised for a chosen target property), which may not work well for another system and/or learning task. This is easy to understand, as descriptors constructed in this way are pre-fixed into a feature space to highlight a subset of dominating features towards the learning target, which may not be transferable to another learning task. In another word, such a fixed descriptor approach is weak in terms of generalisability and adaptabilities, often capped by the sparsity of the high-dimensional space\cite{beyer1999nearest} and the size of the kernel matrix that can be handled numerically. 

On the other hand, in neural-network-based DL approaches, the latent representations of materials and the model parameters can be dynamically and jointly optimised with respect to the prediction targets and the additional exposures to new training data (the  process that has been commonly referred to as  \textit{fine-tuning}\cite{radova2025fine,bertani2025atomic,grandel2026parameter}), due to the flexibility provided by the large hyperparameter space ($\sim10^{6}$) over which the models can be optimised with the use of huge set of training data ($\sim10^{4}-10^{6}$).  This opens the opportunities to train universal foundational models that can be generally applied across materials system(s)\cite{batatia2023foundation,feng2025universal,wood2025family} for a wide variety of downstream tasks\cite{martin2025general,zhong2025machine,elena2025machine,ho2026equivariant} with the opportunity of solving classical theoretical problems with new tools from AI,\cite{goonnheimer2025beyond} making the DFT-level universality and accuracy for atomistic materials modelling achievable at the cost of forcefields to significantly expedite the process of materials discoveries.

Secondly, most structural descriptors were constructed with strong prior assumption in locality through the usages of, the cut-off distance $r_c$ that predefines the spatial range of interactions, and the body order $\nu$ that predefines the maximal number of neighbours that a given central atom can `see'. It is clear that a wide range of physical interactions, including electrostatics, non-covalent interactions,\cite{grisafi2021multi} charged defects,\cite{freysoldt2011electrostatic} anharmonic vibrations,\cite{darby2022compressing,yang2021altas} and conformational changes in molecules,\cite{frank2022so3krates} are long-ranged and/or entailing the cooperative interactions across many atoms. As such, a pragmatic approach would be to learn the effective cut-off and body-order from the data itself. Such a mechanism is apparently lacking in the paradigm of descriptor-based ML at this stage.

Lastly, the complexity involved in mapping the atomistic structure to its descriptor $\bm{R}\to\boldsymbol{\chi}$ implies that the reverse mapping from $\boldsymbol{\chi}$ back to a plausible structure $\bm{R}$, an endeavour that is needed for the inverse generative designs of new materials, will be almost impossible with simple ML-based approaches.

In these regards, the adaptation of the DL paradigm  have provided new opportunities to enhance our capabilities in data-driven  atomistic molecular and materials modelling. Increasing interests and development efforts that are devoted to the development of DL techniques can be seen over the past two decades.\cite{behler2021four} Rather than providing a full historical account, in the present and the next section, we shall provide an account on the current state-of-arts DL architectures for materials modelling, focusing on the key design concepts that enable us in building data-efficient, stable and robust models, respecting the fundamental physical principles. A particular focus will be on how equivariance is enforced in the modern graph neural networks (GNNs),\cite{duval2023hitchhiker} that are subtlety different from the construction of equivariant structural descriptors for atomistic ML, as discussed in \cref{sec:physics_informed_representation}.

\subsubsection{Preliminaries on the Principles behind Neural Networks}

As discussed above, kernel-based ML heavily leverages on the designs and engineering of feature vectors and kernels to account for nonlinearity in the data. In contrast, NNs from the DL paradigm take a distinctively different approach, which, in this section, we shall provide the readers with some baseline understandings on the principles behind NNs to facilitate the subsequent discussions on the modern equivariant NN architectures for atomistic molecular and materials modellings. 

\begin{figure}[tbh]
    \centering
    \includegraphics[width=0.8\linewidth]{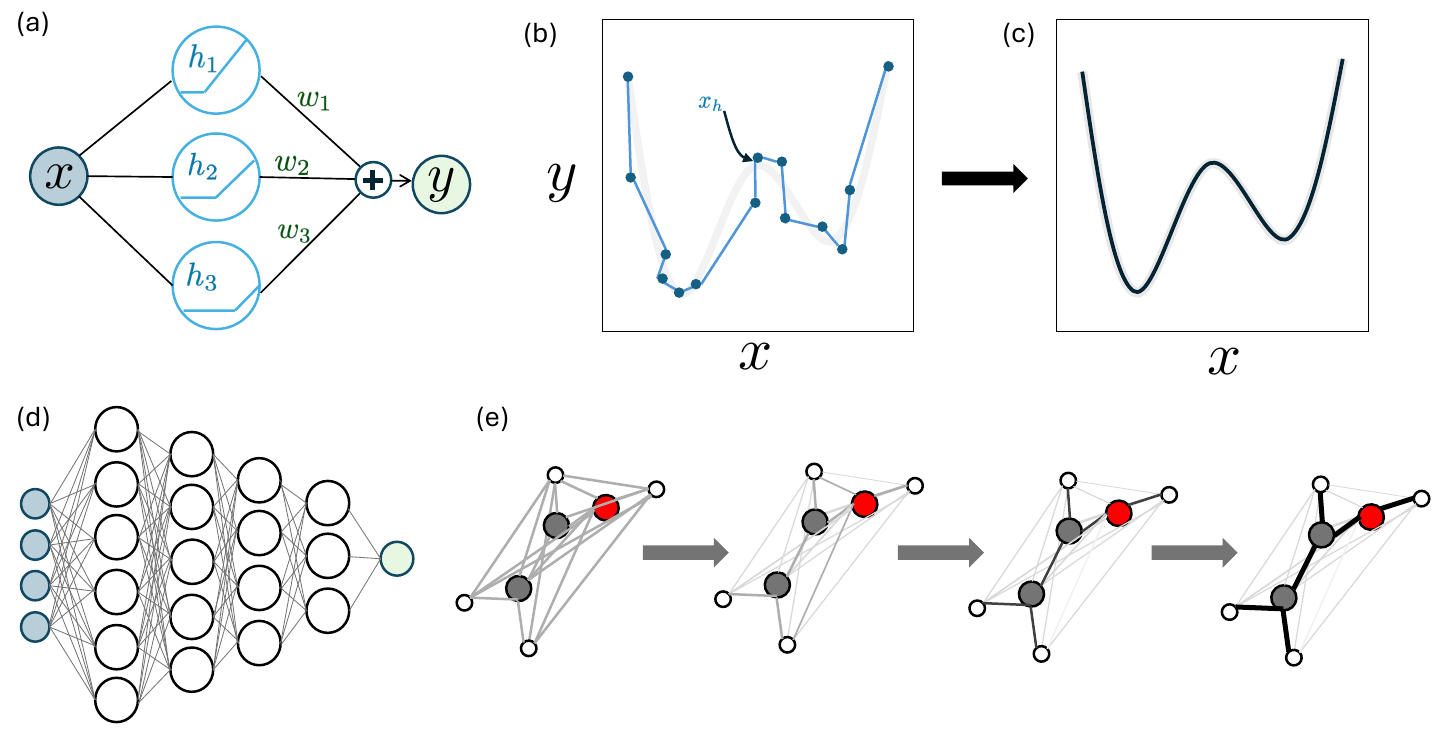}
    \caption{Illustrations on the conceptual understandings on the working principles behind neural networks and their relevance to pairwise interactions in atomistic systems. (a) A simple network that maps an input $x$ to an output $y$ with three nodes $h_i$ in the middle layer. Each node is modelled by a ReLU-type activation function, the output of which will be weighted by $w_i$, before the output of all three nodes are added together to generate the output. (b) Using simple network to surrogate a smooth function of the type of a double potential well can be seen as fitting piecewise linear functions with different gradients ($k$) that are joined at multiple `hinge points' $x_h$, where $k$ and $x_h$ are to be optimised. With sufficient number of hyperparameters, this surrogate model will be able to reproduce the nonlinear target function as shown in (c). (d) An schematic illustration of a simple, densely interconnected, deep neural networks with 4 input nodes (blue circles), 4 hidden layers (open circles)  and 1 output node (green circle). (e) Analogous example on how simple neural network is trained to understand the connectivity in a molecule, in which the training starts by putting equal weight (leftmost) on all pairs of atoms in the structure, and as training progress, the weights essentially become dominated by the connections that correspond to the existence of interatomic covalent bonds (rightmost). }
    \label{fig:simple_nn_illustrate}
\end{figure}

In its simplest form, NN tries to map simple input  with zero or minimum feature engineering, into highly non-linear functions (or features) by passing the data through multiple \textit{interconnected neurons}, each carries out some simple (often linear) mathematical transformations,\cite{prince2023understanding} that collectively allows the model to learn complex relationships between the input and output. As an example, \cref{fig:simple_nn_illustrate}(a) demonstrates the construct of a simple NN that contains a single input node that maps a simple input $x\in\mathbb{R}$ (which for simplicity, we shall assume $x\sim \mbox{Uniform}([0,1])$) into a single output node with the corresponding output of $y$. The middle layer contains 3 hidden units, each of which takes the form of a \emph{rectified linear unit (ReLU)} or otherwise known as the \emph{activation function}, that transforms the input with the following function:
\begin{equation}
    \label{eq:ReLU}
    h_i(x)=\max(0,k_ix+b_i).
\end{equation}
Suppose $k_i>0$, then it is easy to see that each neuron will act on the input and map $\forall x>x_h=-b_i/k_i$ (which we shall call $x_h$ the hinge point) to a linear segment $k_ix+b_i$. The output from each hidden unit is then subsequently weighted by a weighting parameter $w_i$ before the outcomes from all three hidden units are added together for the final outcome, \emph{i.e.} 
\begin{equation}
    \label{eq:NN_additivitiy}
    y=\sum_i w_ih_i(x).
\end{equation}

As shown in \cref{fig:simple_nn_illustrate}(b), this approach effectively breaks down a smooth non-linear function into a piecewise linear one that is joined at multiple hinge points. It is therefore easy to envisage, that by increasing the number of hidden layers and nodes in each layers to form densely connected deep NN [\cref{fig:simple_nn_illustrate}(d)], the model will have the potential to optimise a large set of hyperparameters $\{\bm{k},\bm{b},\bm{w}\}$ that allows the target functions to be accurately regressed in finer details as more piecewise linear functions were used in the fitting    [\cref{fig:simple_nn_illustrate}(c)]. This provides us with an intuitive understanding on how, by stacking multiple simple mathematical transformations in layers, NN is able to learn the complex nonlinear relationships in the the underlying data. 

Most importantly, the usage of simple activation functions $h_i(x)$ enable its derivatives to be efficiently computed analytically, where the total functional derivative of the output with respect to the input  ($\partial y/\partial x$) can now be computed through the \emph{back propagation} algorithm, enabling the gradient-based minimisations of the loss function for model training. This is typically combined with \emph{stochastic gradient descents},\cite{prince2023understanding} especially when training NN with large datasets, that introduces randomness in the gradient-descent process by using different batches of the training samples at each step, allowing the optimisation algorithm to more effectively jump out from the local minima and cross the saddle points in the landscape of the loss function.  This provides significant numerical advantages compared to, for example, the optimisations of parameters in traditional atom-atom forcefields [\cref{sect:empirical_forcefield}], in which the different forcefield terms can become highly coupled, leading to trench-shaped loss function landscape that does not lead to an optimum solution. Problems like this often require tailored designed training sets that allow different terms in the forcefields to be individually fitted, or imposing  constraints building upon prior knowledges, to regularise the overall fitting process, which can be mitigated by adapting the DL approach for developing NN potentials for atomistic modellings.

In a more physical context, when the NN is used to train an energy model that outputs the system energy $E$ with respect to the input, that is usually  the atomic positions $\bm{R}$, the back-propagation algorithm enables us to compute and learn the atomic forces via $F=-\partial E/\partial \bm{R}$ that is particularly useful for dynamic simulations. This leads to the so-called \emph{energy conserving} energy model, as opposed to those \emph{non-conserving} models in which the atomic forces are trained with a separate model from the system energies.
 
Training the NN, in which the weight parameters are optimised, reflecting how strongly one neuron influences another, can be understood by drawing parallel examples on how chemical bond emerge in atomistic systems heuristically [\cref{fig:simple_nn_illustrate}(e)]. In this analogous example, one can consider the initial state of the system being one that contains a collection of atoms, in which all pairs of atoms interact with equal strength. In terms of NN training, this corresponds to initialising the NN with equal weight across all connected neuron pairs. As time continues, electrons in space will `feel' the field established by the nuclei and gradually redistribute towards regions that minimise the total system energy, during which chemical bonds will appear between selected pairs of atoms in the system where the covalent interactions should exist. This is the same process in which weights are optimised throughout the NN, by putting higher weights on neuron interactions (or information passing pathways) that are expected to contribute most significantly in transforming the input data to the corresponding outputs.  Such concepts underpin the designs and implementations of modern equivariant GNNs for atomistic modellings, which,  we shall provide a brief account on the key features of equivariant GNNs in the following section.

\subsubsection{Equivariant Graph Neural Networks}

In mathematical chemistry, molecules have seen a long tradition in being modelled as graph-type objects, dating back to 1870s.\cite{cayley1874lvii} It is therefore, unsurprising to see chemistry being the early adaptor of the GNN model.\cite{duvenaud2015convolutional} In 2017, \citet{gilmer2017neural} further proposed the message-passing GNNs (MPNNs) for molecules, which has since became one of the most popular and mainstream DL architectures for atomistic molecular and materials modellings. Most importantly, the incorporation of symmetries through extensive use of group and representation theory in the model architecture, as opposing from earlier NN models\cite{behler2021four} which heavily reply on the use of handcrafted descriptors,  has become an important inductive bias in the model design to enhance the models' data efficiency and ability of extrapolations.\cite{fu2022forces,frank2022so3krates} In terms of fundamental architectural designs behind MPNNs, the hitchhiker's guide written by \citet{duval2023hitchhiker} has provided a good comprehensive pedagogical reference to understand the fundamental mathematical principles behind the invariant and equivariant GNNs. On a more application-oriented focus, \citet{ reiser2022graph} reviewed the development of GNNs for molecular and materials science up to 2022. 

To be inline with our previous discussions on physics-inspired density-field structure descriptors in \cref{sec:physics_informed_representation}, in this section,  we shall provide a more focused and concise account of the key concepts behind, and general approaches in incorporating the invariant and/or equivariant symmetry constraints, as well as accounting for many-body orders, in modern message-passing GNNs, that are different from the descriptor-based approaches. As a more concrete example, we will review how the MACE\cite{batatia2022mace} model architecture has up-taken a more principled approach that combines equivariant atom-centred interatomic potential\cite{drautz2019atomic} and MPNNs, that allows features up to any arbitrary order of $\nu$ to be learned with MPNN. We further assume that the primary focus here is the application of GNN for developing energy models, although equivariant GNNs see a wide range of scenarios in which it may be applied, for instances, as the backbones for generative models, to be discussed in \cref{sec:generative_AI}.

\begin{figure}[tbh]
    \centering
    \includegraphics[width=\linewidth]{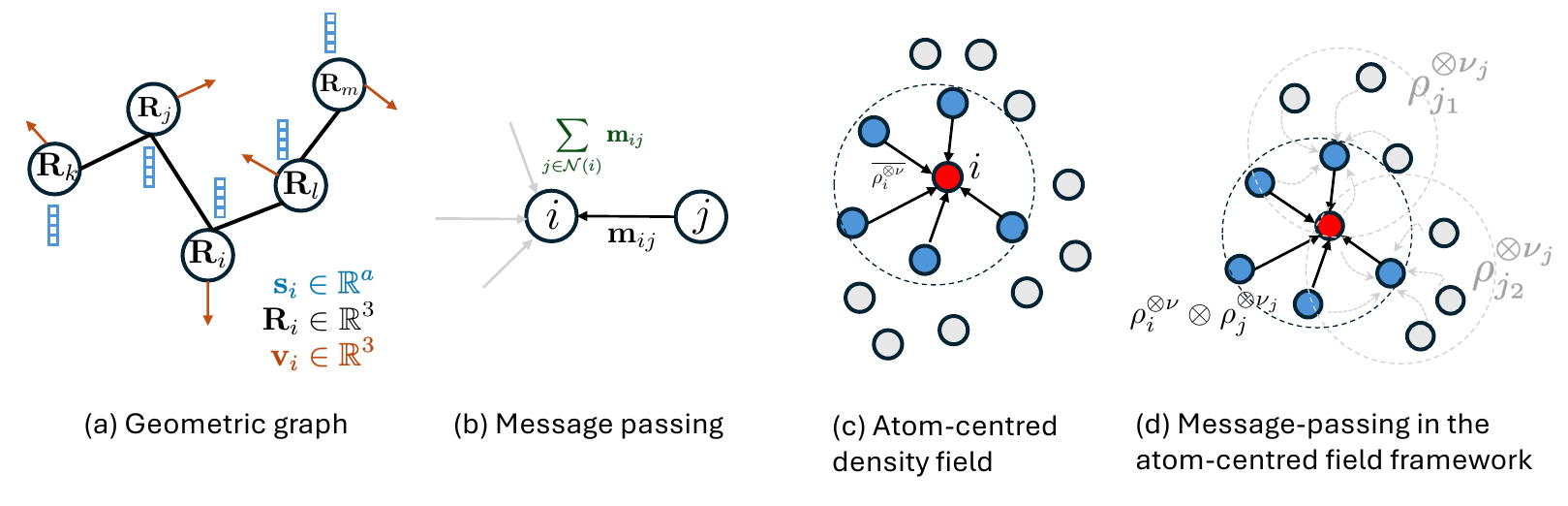}
    \caption{Core concepts behind message-passing graph neural networks. (a) Atomistic structure can be modelled as geometric graph in 3D Euclidean space\cite{duval2023hitchhiker} consists of nodes that represent atoms that are connected by edges, which represents bondings between atoms. Each node is associated with a set of mathematical attributes such as $\bm{s}_i$ which carriers chemical information constructed through N.N. embedding and transforms invariantly, and the Cartesian coordinates $\bm{R}_i$ and velocity (or forces) $\bm{v}_i$ that transform equivariantly under translation and rotation. (b) Message passing operation, where each central node $i$ learns its latent representation by receiving and aggregating the messages $\bm{m}_{ij}$ from all of its local neighbours $j\in\mathcal{N}(i)$. The way that message-passing formalism builds latent representations can be understood from the perspectives of atom-centred density field representations in which (c) in its original construct, the central atom $i$ sees the `bare' densities from each one of its neighbouring atoms $j$ within a pre-defined cut-off, whereas in the message-passing scheme (d) the centre atom interacts with the `dress' densities from the neighbours, whereby  each neighbour atom $j$ also aggregates information from its own surrounding neighbours through tensor product operation $\rho_{j_k}^{\otimes\nu_j}$ before they are interacting with the central atom $i$ through the operation of $\rho_i^{\otimes\nu}\otimes\rho_{j_k}^{\otimes\nu_j}$.
     }
    \label{fig:massage-passing}
\end{figure}

The starting point of GNN is to treat an atomistic structure of molecule/material as graphs in which atoms that are connected together by chemical bonds can be treated as \emph{nodes} that are connected by \emph{edges} in the graph representation. Two nodes are said to be connected if there exists an edge connecting the two nodes, the information of which is generally encapsulated in the connectivity matrix $\bm{A}$ that takes a matrix element of value 1 for two connected nodes and 0 otherwise. Practically, the connectivity between nodes are generally determined using the `neighbourhood' approach based on a pre-defined cut-off radius  $r_c$,  as  similar to that of \cref{eq:atom_env_density_field}, so the definition of an \emph{edge} is not necessarily confined to the existence of a covalent/ionic bond between a pair of atoms, but more align with the physical concept of `field' within the 3D space.  Furthermore, the graph object that is constructed for atomistic modelling is not a simple graph but is what generally referred to as the \emph{geometric graph} [\cref{fig:massage-passing}(a)], in which the nodes (optionally for edges as well) are further decorated with  a set of attributes that are either invariant (such as chemical identities $\bm{s}_i\in\mathbb{R}^a$ as a vectorial quantity constructed through embedding) or equivariant (such as the atomic coordinates $\bm{R}_i\in\mathbb{R}^3$ and atomic forces/velocities  $\bm{f}_i\in\mathbb{R}^3$) with respect to the translation and SO(3) rotations of the atomistic structures, in addition to the connectivity information $\bm{A}$.

With the establishment of the geometric graphs, MPNN further model the physical processes for formation of chemical bonds (or equivalently, the interactions among neighbouring atoms) as a mathematical operation of \emph{message passing} between connected nodes [\cref{fig:massage-passing}(b)]. In the message-passing paradigm, a given central node $i$ exchange message $\bm{m}_{ij}$ from every one of its connected neighbouring nodes $j$, in which we consider $\bm{m}_{ij}=f_{\boldsymbol{\varphi}}(i,j)$  as some parameterised NN function on node attributes, and the messages from all the neighbours are summed together in each round of message-passing, \emph{i.e.} $\sum_{j\in\mathcal{N}(i)}\bm{m}_{ij}$ to update the information on $i$. Conceptually, this can be considered as equivalent to constructing a two-body (one from the centre atom) atom-centred density field representation  [\cref{fig:massage-passing}(c)]. However, one must note that each neighbouring node $j$ is also connected to its own neighbours, and so on. In the simplest case, if we combine two message-passing layers, then we think of the information from the neighbours if a given atom $j$ are aggregated onto $j$ before its information is passing onto $i$, allowing $i$ to effectively `sees' atoms that are two connections away. In the atom-centred density-field framework, this effectively means that the central atom is not convoluting the `bare' density but a `dressed' one from its neighbouring environments [\cref{fig:massage-passing}(d)].\cite{nigam2022unified} In this regards, stacking multiple MP hidden layers in GNN essentially establishes a densely connected NN model [\cref{fig:simple_nn_illustrate}(a)] to mimic the interconnected atomistic systems in 3D space, whereas the \emph{receptive field} of a given central atom $i$, \emph{i.e.} how far away the chemical and structural information from distant atoms (measured in terms of number of edges), are passed onto $i$ in MPNN, can be controlled by the number of MP operations used in an MPNN.

Unlike the descriptor-based approach, where symmetry is baked into the descriptor designs, in MPNN, symmetry of the geometric graphs are preserved by carefully designing  mathematical transformations in the operations that construct, pass and update messages through diligently counting the allowed operations that preserve the invariant and/or equivariant geometry features up to some pre-defined maximum body order $\nu_{\max}$.\cite{duval2023hitchhiker}

\begingroup 
\LTcapwidth=\textwidth

\begin{longtable}{p{1.25cm}p{2.5cm}p{3cm}p{10.25cm}}
    \caption{A concise chronological list of major geometric/equivariant GNNs for molecular and materials modellings, which are predominantly developed for the application as MLFFs to accelerate PES samplings.}
    \label{tab:equivariant_GNNS}\\
    \hline\hline
     \textbf{\footnotesize{Year}}    & \textbf{\footnotesize{Model}}  & \textbf{\footnotesize{Domain}} &
      \textbf{Key features}\\
     \hline
     \endfirsthead
\caption*{Table \ref{tab:equivariant_GNNS} (Continued):A concise chronological list of major geometric/equivariant GNNs.  }\\\toprule
     \textbf{\footnotesize{Year}}    & \textbf{\footnotesize{Model}}  & \textbf{\footnotesize{Domain}} &
      \textbf{Key features}\\
     \hline
\endhead
\hline\hline
\endfoot
\hline\hline
\endlastfoot

\footnotesize{2017} &  
\footnotesize{SchNet\cite{schutt2017schnet}}&
\footnotesize{Molecules \& materials} & 
\footnotesize{Invariant features, up to 2-body order, convolution}\\

\footnotesize{2017} &  
\footnotesize{CGCNN\cite{xie2018crystal}}&
\footnotesize{Crystalline materials}& 
\footnotesize{Invariant features, up to 2-body order, convolution}\\

\footnotesize{2018} &
\footnotesize{TFN\cite{thomas2018tensor}} &
\footnotesize{Molecules} &
\footnotesize{SO(3) equivariant feature in spherical harmonic basis, up to 2-body order}\\

\footnotesize{2019} &
\footnotesize{PhysNet\cite{unke2019physnet}}&
\footnotesize{Molecules}&
\footnotesize{Invariant features, up to 2-body order with radial basis functions, convolution}\\

\footnotesize{2019} &
\footnotesize{DimeNet\cite{gasteiger2020directional}} &
\footnotesize{Molecules} &
\footnotesize{Invariant features, up to 3-body with radial and angular (spherical harmonic) basis functions, message-passing }\\

\footnotesize{2019} &
\footnotesize{Comorant\cite{anderson2019cormorant}} &
\footnotesize{Molecules} &
\footnotesize{Rotationally covariant features, up to 2-body order with spherical harmonics}\\


\footnotesize{2020} &
\footnotesize{DimeNet++\cite{gasteiger2020fast}} &
\footnotesize{Molecules} &
\footnotesize{Invariant features, up to 3-body, message-passing}\\

\footnotesize{2020} &
\footnotesize{MXMNet\cite{zhang2020molecular}} &
\footnotesize{Molecules} &
\footnotesize{Invariant features, up to 3-body, message-passing}\\

\footnotesize{2020} &
\footnotesize{LieConv\cite{finzi2020generalizing}} &
\footnotesize{Molecules} &
\footnotesize{Lie-group equivariant features, convolution}\\

\footnotesize{2020} &
\footnotesize{LieTransformer\cite{hutchinson2021lietransformer}} &
\footnotesize{Molecules} &
\footnotesize{Lie-group equivariant features, attention mechanism\cite{vaswani2017attention}}\\

\footnotesize{2020} &
\footnotesize{SE(3)-Transformer\cite{fuchs2020se}} &
\footnotesize{Molecules} &
\footnotesize{Equivariant SE(3) features, spherical harmonic basis functions, attention mechanism}\\

\footnotesize{2021} &
\footnotesize{SphereNet\cite{liu2022spherical}} &
\footnotesize{Molecules} &
\footnotesize{Invariant features, up to 4-body order, message-passing}\\

\footnotesize{2021} &
\footnotesize{GemNet\cite{gasteiger2021gemnet}} &
\footnotesize{Molecules \& materials \& catalysts} &
\footnotesize{Invariant features, up to 4-body order, directional message-passing}\\

\footnotesize{2021} &
\footnotesize{SEGNN\cite{brandstetter2021geometric}} &
\footnotesize{Molecules} &
\footnotesize{E(3) equivariant features, up to 2-body order in spherical harmonic basis, meassage-passing}\\

\footnotesize{2021} &
\footnotesize{EGNN\cite{satorras2021n}} &
\footnotesize{Molecules} &
\footnotesize{Invariant E($n$) features, up to 2-body order in Cartesian basis, message-passing}\\

\footnotesize{2021} &
\footnotesize{PaiNN\cite{schutt2021equivariant}} &
\footnotesize{Molecules} &
\footnotesize{Equivariant features, up to 3-body order in Cartesian basis, message-passing}\\

\footnotesize{2021} &
\footnotesize{NequIP\cite{batzner20223}} &
\footnotesize{Molecules \& materials} &
\footnotesize{E(3) equivariant features in spherical basis, up to 2-body order, convolution}\\

\footnotesize{2021} &
\footnotesize{SpookyNet\cite{unke2021spookynet}} &
\footnotesize{Molecules} &
\footnotesize{Invariant features, up to 2-body order, attention mechanism}\\

\footnotesize{2022} &
\footnotesize{TorchMD-Net\cite{tholke2022torchmd}} &
\footnotesize{Molecules} &
\footnotesize{Equivariant features in Cartesian basis, up to 2-body order, attention mechanism}\\

\footnotesize{2022} &
\footnotesize{EQGAT\cite{le2022equivariant}} &
\footnotesize{Molecules} &
\footnotesize{Equivariant features in Cartesian basis, up to 2-body order, message-passing and attention mechanism}\\

\footnotesize{2022} &
\footnotesize{ComENet\cite{wang2022comenet}} &
\footnotesize{Molecules} &
\footnotesize{Invariant features up to 4-body order, message-passing}\\

\footnotesize{2022} &
\footnotesize{So3krates\cite{frank2022so3krates}} &
\footnotesize{Molecules \& materials} &
\footnotesize{SO(3) equivariant features in spherical basis, up to 2-body order, message-passing with attention mechanism}\\

\footnotesize{2022} &
\footnotesize{Equiformer\cite{liao2022equiformer}} &
\footnotesize{Molecules \& catalysts} &
\footnotesize{SE(3) equivariant features in spherical basis, message-passing with attention mechanism}\\

\footnotesize{2022} &
\footnotesize{GemNet-OC\cite{gasteiger2022gemnet}} &
\footnotesize{Catalysis} &
\footnotesize{Invariant features up to 4-body order, message-passing}\\

\footnotesize{2022} &
\footnotesize{MACE\cite{batatia2022mace}} &
\footnotesize{Molecules \& materials} &
\footnotesize{Equivariant features in spherical and cluster-expansion basis, up to many-body-orders, message-passing}\\

\footnotesize{2022} &
\footnotesize{Allegro\cite{musaelian2023learning}} &
\footnotesize{Molecules \& materials} &
\footnotesize{Equivariant features in spherical basis, up to many-body-orders, message-passing}\\

\footnotesize{2022} &
\footnotesize{ClofNet\cite{du2022se}} &
\footnotesize{Molecules} &
\footnotesize{SE(3) equivariant features in Cartesian basis, message-passing}\\

\footnotesize{2023} &
\footnotesize{LEFTNet\cite{du2023new}} &
\footnotesize{Molecules} &
\footnotesize{Equivariant features in Cartesian basis, message-passing}\\

\footnotesize{2023} &
\footnotesize{ViSNet\cite{li2024long}} &
\footnotesize{Molecules} &
\footnotesize{Equivariant features in Cartesian basis, long-short-range message-passing}\\

\footnotesize{2023} &
\footnotesize{TensorNet\cite{simeon2023tensornet}} &
\footnotesize{Molecules} &
\footnotesize{Equivariant features in Cartesian basis, message-passing}\\

\footnotesize{2023} &
\footnotesize{eSCN\cite{passaro2023reducing}} &
\footnotesize{Materials \& catalysts} &
\footnotesize{Equivariant features in spherical harmonic basis, up to 2-body order, convolution}\\

\footnotesize{2023} &
\footnotesize{CHGNet\cite{deng2023chgnet}} &
\footnotesize{Materials} &
\footnotesize{Charge-aware equivariant features, convolution}\\

\footnotesize{2023} &
\footnotesize{M3GNet\cite{chen2022universal}} &
\footnotesize{Materials} &
\footnotesize{Equivariant features in Cartesian space, up to 3-body order, }\\

\footnotesize{2023} &
\footnotesize{EquiformerV2\cite{liao2024equiformerv2}} &
\footnotesize{Molecules \& materials \& catalysts} &
\footnotesize{SO(3) equivariant features, up to 2-body order, message-passing \& attention mechanism.}\\

\footnotesize{2024} &
\footnotesize{UMA\cite{wood2026family}} &
\footnotesize{Molecules \& materials \& catalysts} &
\footnotesize{Foundational model for atomistic modelling building on the eSCN equivariant model.}\\

\footnotesize{2024} &
\footnotesize{CACE\cite{cheng2024cartesian}} &
\footnotesize{Molecules \& materials}&
\footnotesize{Equivariant features in Cartesian basis as an alternative framework to MACE  to achieve higher-order correlations.}\\

\footnotesize{2025} &
\footnotesize{E$^2$GNN\cite{yang2025efficient}} &
\footnotesize{Molecules \& materials}&
\footnotesize{Equivariant features in Cartesian basis, message-passing}\\

\footnotesize{2025} &
\footnotesize{MGNN\cite{chang2025mgnn}} &
\footnotesize{Molecules \& materials}&
\footnotesize{Rotationally invariant features with moment-based representation learning, message-passing.}\\

\footnotesize{2026} &
\footnotesize{mMACE\cite{ho2026equivariant}}&
\footnotesize{Materials}&
\footnotesize{Building on the MACE framework to additionally take into account the equivariance of magnetic moments}
\end{longtable}

Most of the early implementations of MPNNs [\cref{tab:equivariant_GNNS}] have been focusing on constructing messages from scalar many-body features that transforms invariantly under symmetry operations, such as the chemical information, bond lengths (2-body), bond angles (3-body) and torsional angles (4-body), depending on the choice of $\nu_{\max}$. These messages are easy to construct, and closely resembles the use of internal coordinates [\cref{sec:coordinates_systems}] for constructing the atom-atom forcefields [\cref{sect:empirical_forcefield}]. For instance, using the node information on the geometric graph [\cref{fig:massage-passing}(a)] discussed above, a two-body invariant learnable message can be constructed by selecting tensor operations amongst the nodal and two-body features that will lead to scalar output, such as:\cite{duval2023hitchhiker}
\begin{equation}
    \label{eq:2-body-invariant-message}
    m_{ij}=f_{\boldsymbol{\varphi}_1}\left(\bm{s}_i,\Vert \bm{v}_i\Vert_2\right)+\sum_{j\in\mathcal{N}(i)}f_{\boldsymbol{\varphi}_2}(
    \bm{s}_i,
    \bm{s}_j,
    \Vert \bm{x}_{ij}\Vert_2,
    \Vert \bm{v}_j\Vert_2,
    \bm{x}_{ij}\cdot\bm{v}_i,
    \bm{x}_{ij}\cdot\bm{v}_j,
    \bm{v}_{i}\cdot\bm{v}_j),
\end{equation}
in which the interatomic vector $\bm{x}_{ij}=\bm{R}_i-\bm{R}_j$, whereas equivariant messages of a given order can be constructed by scaling the tensorial features that particular order with the scalar outputs from the learnable messages:
\begin{align}
\bm{m}_{ij}=f_{\boldsymbol{\varphi}_3}\left(\bm{s}_i,\Vert \bm{v}_i\Vert_2\right)\odot \bm{v}_i &+ \sum_{j\in\mathcal{N}(i)}f_{\boldsymbol{\varphi}_4}(
    \bm{s}_i,
    \bm{s}_j,
    \Vert \bm{x}_{ij}\Vert_2,
    \Vert \bm{v}_j\Vert_2,
    \bm{x}_{ij}\cdot\bm{v}_i,
    \bm{x}_{ij}\cdot\bm{v}_j,
    \bm{v}_{i}\cdot\bm{v}_j)\odot\bm{x}_{ij}\nonumber\\
    &+\sum_{j\in\mathcal{N}(i)}f_{\boldsymbol{\varphi}_5}(
    \bm{s}_i,
    \bm{s}_j,
    \Vert \bm{x}_{ij}\Vert_2,
    \Vert \bm{v}_j\Vert_2,
    \bm{x}_{ij}\cdot\bm{v}_i,
    \bm{x}_{ij}\cdot\bm{v}_j,
    \bm{v}_{i}\cdot\bm{v}_j)\odot\bm{v}_{ij}.
    \label{eq:2-body-equivariant-message}
\end{align}
It is thus easy to understand the inclusion of three-body features in the message would naturally include terms like the bond angles $\theta_{ijk}$, which may be constructed from dot-products $\bm{x}_{ij}\cdot\bm{x}_{ik}$, for an dihedral centred on $i$. Although this provides a physically intuitive way to understand how nodes can exchange information through the constructions of tensor products using geometric information that are native to computationalists on an practical basis, it does not leads to a pragmatic way to construct an arbitrary $\nu$-body invariant message as one needs to diligently keep track of all possible operations on different combinations of nodal features up to $\nu$ interacting nodes that can output a scalar feature.

The above problem is effectively overcome in the later variants of MPNNs [\cref{tab:equivariant_GNNS}], that uses the knowledge of tensor algebra, group and representation theory [\cref{sec:physics_informed_representation}] to construct and interact amongst tensorial quantities to construct equivariant messages from 0 up to some pre-defined maximal ranks. Drawing from our discussions on atom-centre density field representation, we can envisage that the primary goal here is to preserve and exchange angular information on the local chemical environments amongst the interacting pairs, triplets, and higher-order atomic clusters, which, as can be seen in \textbf{Box IV} that using the irreps of the SO(3) group in the spherical harmonic basis can provide the principled way to achieve our goal. 

In equivariant MPNN, this is typically started from the implementation of a feature mapping layer, that maps any vector properties, such as $\bm{x}_{ij}$ into the spherical harmonic basis up to the maximal rank of $\ell_{\max}$, giving a list of tensors $\vec{\bm{Y}}=[\vec{Y}^{(0)},\vec{Y}^{(1)},\ldots,\vec{Y}^{(\ell_{\max})}]$, where we only retain the angular information with $Y_\ell^m(\bm{x}_{ij}/\Vert \bm{x}_{ij}\Vert_2)\to Y_\ell^m$. Now the process of constructing the message, we can, in the message function $f_{\boldsymbol{\varphi}}$ to allow two list of tensors $\vec{\bm{T}}$ (which, in line with the definition, will encode the angular information between two bodies $i$ and $j$) and $\vec{\bm{S}}$ (which encodes two-body information between $i$ and $k$) to interact through tensor product, similar to the process of constructing $\rho^{\otimes 2}$ in SOAP (\cref{fig:soap_invariant_tensor_prod}), but instead of retrieving only the invariant component of the tensor products, we can retrieve tensors of all other allowed ranks as well.

\begin{figure}[tbh]
    \centering
    \includegraphics[width=\linewidth]{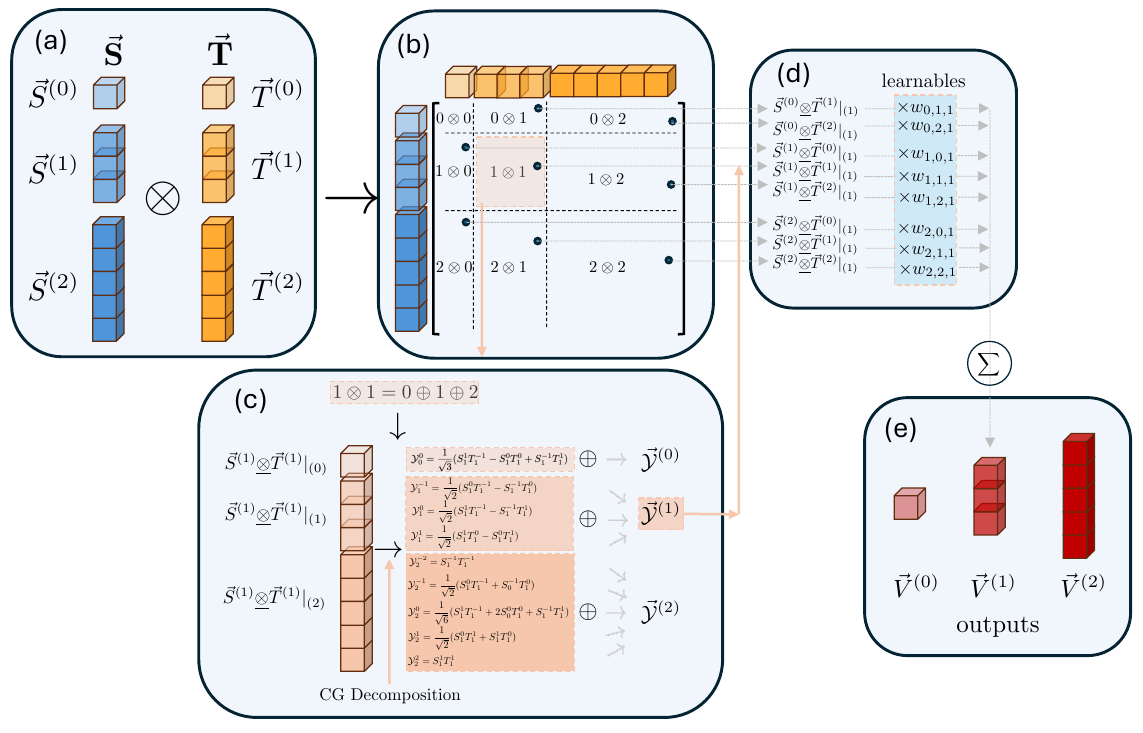}
    \caption{Preserving symmetry while enabling learnings through parameterising the tenor products in equivariant network (as implemented in, for example the \texttt{e3nn} model\cite{geiger2022e3nn}). Similar to \cref{fig:soap_invariant_tensor_prod}, performing the tensor product between two list of tensors $\vec{\bm{S}}\otimes\vec{\bm{T}}$ (a) leads to a matrix of tensors of multiple different ranks (b). For each block of tensors, one first perform a tensor decomposition [\emph{c.f.} \cref{eq:rank_2_tensor_decomp}] to decompose the tensor into a combination of basis tensors, which is illustrated in (c) that decompose $1\otimes 1$ into $0\oplus1\oplus 2$, where we denote $\left.\vec{S}^{(\ell_1)}\underline{\otimes}\vec{T}^{(\ell_2)}\right\vert_{(\ell_3)}$ as extracting the $\ell_3$-components for the tensor product $\vec{S}^{(\ell_1)}\otimes\vec{T}^{(\ell_2)}$. Such a tensor decomposition procedure leads to a well defined sets of tensors with fixed orders, which we knows how they should be transformed individually to ensure equivariance. For $\left.\vec{S}^{(\ell_1)}\underline{\otimes}\vec{T}^{(\ell_2)}\right\vert_{(\ell_3)}$ with a given $\ell_3$, one can further apply CG decomposition to express it in the basis of spherical harmonics $\vec{Y}^{(\ell_3)}$. Note that the coefficients for such a CG decomposition is fixed, which in principle do not lead to learnables for NN. However, one can assign a single scalar learnable weight to the output tensor $\left.\vec{S}^{(\ell_1)}\underline{\otimes}\vec{T}^{(\ell_2)}\right\vert_{(\ell_3)}$ as a whole, which will be denoted it as $w_{\ell_1,\ell_2,\ell_3}$. As can be seen in (d), this decomposition can be performed for each block of tensor from (b), from which we can obtain a weighted sum of all the  $\ell_3$ components of $\left.\vec{S}^{(\ell_1)}\underline{\otimes}\vec{T}^{(\ell_2)}\right\vert_{(\ell_3)}$ for all possible combinations of $\ell_1$ and $\ell_2$ that may lead to a specific decomposed $\ell_3$ component, which is illustrated for the case of $\ell_3=1$.  This  leads to the vector output $\vec{V}^{(1)}=\displaystyle{\sum_{\ell_1,\ell_2\to 1}w_{\ell_1,\ell_2,1}\left.\vec{S}^{(\ell_1)}\underline{\otimes}\vec{T}^{(\ell_2)}\right\vert_{(1)}}$, in which the 1-st order rotational equivariance of the tensor product operation $\vec{\bm{S}}\otimes\vec{\bm{T}}$ in the neuron is preserved. Repeating the process for all possible $\ell_3$ from the combination of $(\ell_1,\ell_2)$ gives the full set of equivariant output of $\vec{\bm{S}}\otimes\vec{\bm{T}}$, which is illustrated in (e).}
    \label{fig:tensor_product_equ_nn}
\end{figure}

Such a mechanism is schematically illustrated in \cref{fig:tensor_product_equ_nn}. As explained in details by \citet{duval2023hitchhiker}, the key here is through the use of tensor product decomposition to first decompose a given tensor product $\vec{S}^{(\ell_1)}\otimes\vec{T}^{(\ell_2)}$ and extract individual component from the decomposition depending on the output rank $\ell_3$, which will be denoted as $\vec{S}^{(\ell_1)}\otimes\vec{T}^{(\ell_2)} \vert_{(\ell_3)}$. Through the further use of CG decomposition, this output component can be further expressed in the spherical harmonic basis as $\vec{\mathcal{Y}}^{(\ell_3)}$, which is shown in \cref{fig:tensor_product_equ_nn}(c). Nevertheless, such decomposition is mathematically fixed in order to preserve the equivariant properties hence do not lead to learnable quantities in the MPNN framework. On the other hand, there exists multiple different combinations of $(\ell_1,\ell_2)$ from the tensor products between which an $\ell_3$ component of $\vec{\mathcal{Y}}^{(\ell_3)}$ can be extracted, and are decoupled from each other from the perspective of the SO(3) rotation group. Consequently, as shown in  \cref{fig:tensor_product_equ_nn}(d), we can assign learnable weights $w_{\ell_1,\ell_2,\ell_3}$ and combined a weighted average of $\vec{\mathcal{Y}}^{(\ell_3)}$ as the final rank-$\ell_3$ output of the tensor product of $\vec{S}^{(\ell_1)}\otimes\vec{T}^{(\ell_2)}$. Repeating such a procedure $\forall\ell_3\in[0,1,\ldots,\ell_1+\ell_2]$ will enable us to learn equivariant features across a wide range of ranks [\cref{fig:tensor_product_equ_nn}(e)], and this highlights the key difference in the way equivariance is preserved between the atom-centred density field and MPNN framework.

As mentioned above, utilising  the tensor product $\vec{S}^{(\ell_1)}\otimes\vec{T}^{(\ell_2)}$ to construct the message $\bm{m}_{ij}$ essentially only encodes the information on two-body correlations, and if we denote the (learnable) state of $i$-th node in the $t$-th MP-iteration as $\sigma_i^{(t)}$, then the message on the node $i$ constructed on the current MP-iteration, that is to be passed onto the next iteration, can be more abstractly written as $\bm{m}_i^{(t)}=\sum_{j}\bm{u}(\sigma_i^{(t)},\sigma_j^{(t)})$, that only depends on the states of a two-body pairs, whereas the \emph{effective} many-body correlations are learned through stacking multiple MP layers in the NN. The MACE\cite{batatia2022mace} framework, instead, asks the question of whether we can directly construct many-body correlated message in one-shot:
\begin{equation}
    \label{eq:mace_msg}
    \bm{m}_i^{(t)}=\sum_{j}\bm{u}_i\left(\sigma_i^{(t)},\sigma_j^{(t)}\right)+\sum_{j_1,j_2}\bm{u}_2\left(\sigma_i^{(t)},\sigma_{j_1}^{(t)},\sigma_{j_2}^{(t)}\right)+\cdots+\sum_{j_1,j_2,\ldots,j_\nu}\bm{u}_{\nu}\left(\sigma_i^{(t)},\sigma_{j_1}^{(t)},\cdots,\sigma_{j_\nu}^{(t)}\right),
\end{equation}
and correspondingly reduced the number of MP layers to enable the learning of many-body atom-atom interactions up to any predefined order of $\nu_{\max}$. From the discussions in \cref{sec:physics_informed_representation}, we know that n\"{i}vely repeating the tensor product operations for $\nu$ times does not lead to a tractable way to construct equivariant features for learning many-body correlations, and the ACE framework solved this problem through the use of the uncoupled basis of $\{\bm{A}\}$. MACE harnesses this idea, by further attaching learnable object $\bm{h}$ to the uncoupled basis as $\bm{A}^\prime=\bm{A}\bm{h}$, before they are used to constructed the coupled B-basis [\cref{eq:ACE_basis}].  Since now the B-basis of different body orders are uncorrelated, following the ideas in \cref{fig:tensor_product_equ_nn}(d), we can further assign body-order-specific learnable weights and form linear combinations to construct the message, in an alternative  to \cref{eq:mace_msg} as $\bm{m}_i^{(t)}\sim\sum_{\nu}W^{(t)}_{\nu}\bm{B}_{\nu}^{(t)}$. This constitutes the core design concept behind MACE that provides it with a unique pathway that enable the use of flexible body-order $\nu_{\max}$ as a hyperparameter in model training, rather than as a fixed model prior as in most of other GNN implementations as shown in \cref{tab:equivariant_GNNS}. 

\subsubsection{Applications of Deep Learning in Atomistic Modelling}
\label{sec:DL_applications}

Having outline the general conceptual framework behind the modern equivariant neural networks, in this section, we shall briefly review some of the exemplary applications of DL in atomistic modellings. Our focuses here will be  on the development of the energy models to showcase the fundamental conceptual difference with the energy models  developed through simple supervised ML (\cref{sec:supervised_ML}) techniques, which is followed by the learning of electronic Hamiltonian to accelerate the electronic structure calculation, bypassing the expensive steps of solving the self-consistent field (\cref{sect:practical_DFT}), thus showcasing DL models as a good \emph{universal approximator for any arbitrarily complex mathematical functions.}\cite{Hermann2020-ao}

\textbf{\emph{Energy Models}} On the highest level, the goal of developing a parameterised surrogate model $E=f_{\boldsymbol{\theta}}(\bm{R})$ to compute the energy and atomic forces from a given structure using either DL or simpler ML models are the same, that is to accelerate the sampling of PES at a high accuracy (typically at the DFT level), with the speed of empirical forcefields, such that we can perform the statistical samplings on atomic systems with larger sizes, over longer time scales, and include phenomena in which electronic contributions are important (such as chemical reactions and polar solvents) that cannot be well parameterised by simple empirical forcefields (\cref{sect:empirical_forcefield}). However, as briefly mentioned at the beginning of our discussions on DL, NN provides us with a methodology that scales better to larger systems with a more diverse range of chemical environments compared to ML. ML approaches, especially kernel ML, heavily depend on the sparse chemical environments that are used to construct the similarity kernels and train the corresponding supervised model. In this case, the kernel must be recomputed, with its corresponding size and memory storage requirements grows, as training samples containing new chemical environments are to be included, whereas the model complexity of NN is (generally) prefixed by the number of hyperparameters, which can be dynamically optimised when the model is exposed to the new training data. By combining with the advantages of GPU accelerations, it makes DL energy models more adaptable to large and complex atomistic systems. In the context of materials modelling, this  can be best exemplified by the recently developed foundation model\cite{batatia2023foundation} for materials chemistry based on the MACE\cite{batatia2022mace} framework, which, its initial version  is trained approximately 1.5M structures with 90\% of them of less than 70 atoms per unit cell, with demonstrated ability to be applied in over 30 different categories of atomistic simulations, ranging from heterogeneous catalysts, ice structures, metal organic frameworks, amorphous structures, to complex liquid-solid interfaces. More specific examples on the applications of (tailor-trained) NN models in atomistic modellings are briefly summarised as following.

Water-solid interfaces play an important role in technologies such as heterogeneous catalysts, water treatments, and semiconductor manufacturing. It is well known that the many-body cooperative effects\cite{mahadevi2016cooperativity} contribute significantly to the collective properties on the ensemble of water molecules, which leads to, for example, the   solvent-mediated concerted proton transfer, that lowers the water dissociations on \ce{TiO2} surfaces.\cite{andrade2020free} Conventional DFT calculations often treat water-surface interactions between a single water molecule and surface \emph{in vacco} under 0 K, due to the high computational costs associated with DFT, thus the important physical insights on the thermal and many-body effects cannot be obtained, and it was only in the past ten years, with the increasing computational power and GPU-accelerated DFT packages, we start to see more adaptations of AIMD surface simulations under the influences of explicit solvents.\cite{sakong2018electric} The complex and dynamic chemical environments among the water molecules and ions in the surfaces make the NN a more suitable type of architecture to train energy models for accelerating the long-time dynamic simulations of this kind\cite{andrade2020free,zeng2023mechanistic,chen2023accelerating,zhang2024modelling,elsner2025atomistic} (also refer to the recent review by \citet{wang2026machine}).

The fluctuations in the local structural and chemical environments has strong linkage with the anharmonic lattice dynamics in solids. In particular, the pseudo-Jahn-Teller effect,\cite{bersuker2013pseudo} for example, has important contributions towards the dynamic stabilities of anharmonic solids, particularly in modulating the local environments around the lone pair electrons,\cite{laurita2022chemistry} suggesting this particular type of lattice dynamics has its strong electronic origins. In anharmonic lattice vibrations, one typically see atoms traverse a wider part of the PES thus experiencing a much broader distributions of local environments, making NN a promising choice to facilitate the investigations of exotic physical phenomena that are coupled to anharmonic dynamics. Recent research examples on this include, but not limited to, liquid-like-phonon-mediated fast ion conduction in solid electrolytes,\cite{ding2025liquid} lone-pair-induced ultralow $\kappa_L$ in metal chalcogenides,\cite{minhas2026unravelling} as well as the structural dynamics of halide perovskite photovoltaic materials.\cite{mao2025correlated} A particularly interesting example is the development of the CHGNet model,\cite{deng2023chgnet} that not only provides the energy and force predictions to drive MD simulations, but also the atomic magnetic moments, enabling the study of the spin-phonon coupling effects, in which the fluctuation in the spin state is induced by the Jahn-Teller distortion around the transition metal centres. 

Finally, another category of materials system in which a diverse range of chemical environments may be found is the disordered and/or high-entropy systems, where DL energy models are expected to provide great promise not only in determining the configurational entropy contributions to their free energy stabilities across a large configurational space, but also other physical properties, such as mechanical and thermal properties,\cite{dai2020theoretical,liu2025predicting} as well as catalytic performances\cite{zhai2022high} (for example, identifying catalytic active sites on surface with each active centre being surrounded by a completely different set of neighbouring elements), which all deemed to be tasks that used to be impossible to accomplish with DFT. With the acceleration of energy and force evaluations from the GPU-accelerated DL energy models, we are no longer confined to the computation of the static properties of disordered solids by also their dynamic properties. For examples, recent studies from \citet{deng2025accelerating} has charted the stabilities and ionic conductivity landscapes in disordered ion conductors such as Li$_{3}$In$_{x}$Y$_{1-x}$Br$_{6y}$Cl$_{6-6y}$ to guide the optimisation of materials' compositions towards highest phase stabilities and ionic conductivities, and our own work\cite{yang2025accounting} took into account the vibrational free energy contribution to the configurational entropy  [\cref{eq:conf_entropy}] in the site-disordered solid electrolyte \ce{Li7La3Zr2O12} by explicitly performing MD simulations on each configuration to extract its corresponding vibrational free energy. The later work follows the generic thermodynamic framework for disordered solids that had been put forward by \citet{van2002effect} but yet remain less adopted practically due to the extremely high computational costs associated with sampling both the configurational and vibrational degree-of-freedoms in these material systems.

Before leaving this part of the discussion, we return to a more theoretical ground and noting that the large number of DL potential models proliferated within a short period of time (\cref{tab:equivariant_GNNS} for GNN only) poses an immediate challenge for us to gain deeper understanding on the strengths and limitations on how different physical priors (\emph{e.g.} the body-order and invariance/equivariance) correlates with the model performances. Emerging benchmark studies becomes particularly helpful in enhancing our understanding on these aspects. For example, \cite{fung2021benchmarking} benchmarked several GNNs (such as SchNet,\cite{schutt2017schnet}  and CGCNN\cite{xie2018crystal}) and the SOAP-descriptor-based approach in energy predictions across five groups of systems, including bulk, surfaces, metal-organic-frameworks, 2D materials and Pt clusters. It was shown that GNN shows significantly improved performance for the first four groups of the system, whereas SOAP still excels in Pt clusters, that can be attributed to the high sensitivities to the angular informations in random structures. \citet{kovacs2023evaluation} showed enhanced performance of MACE in modelling amorphous silicon, which may be attributed to its capabilities of encoding higher-body-order correlations that are essential in random structures exhibiting highly complex chemical environments.  Lastly, \citet{gong2023examining} showcased the challenge in correctly capturing the periodicity of crystalline solids, particularly for systems with large lattice constants that are beyond the receptive field of MPNN. 

\textbf{\emph{Charge Densities or Electronic Hamiltonian}} Rather than directly learning the system energies $E(\bm{R})$ from the atomistic structures $\bm{R}$, one may harness the  strength of NN in learning arbitrarily complex mathematical functions to learn the (converged) charge densities $\rho(\bm{r})$ or the electronic Hamiltonians $\hat{\mathcal{H}}$ from $\bm{R}$, that can be directly used for predicting a series of downstream physical properties (\cref{chapter:physical_properties}), including electronic and dielectric properties in additional to the system energies, based on either the KS, generalised-KS (\cref{sec:DFT}) or GW (\cref{sect:GW_approximation}) schemes. The application of equivariant NN in this case, is particularly relevant, as the orbital interactions that contribute to the elements in the  electronic Hamiltonian matrix ($\bm{H}_{i\alpha,j\beta}=\langle\phi_i^\alpha|\hat{\mathcal{H}}|\phi_j^\beta\rangle$ denoting the interaction between the $\alpha$ atomic orbital on atom $i$ and the $\beta$ atomic orbital on atom $j$) are characterised by their angular momentum quantum numbers $\ell$, which must transform equivariantly (depending on the two coupling orbitals $(\ell_1,\ell_2)$) with respect to $\bm{R}$. Here, we shall provide a high-level account (in chronological order) over the recent DL models being developed along this line to facilitate the understanding on how the direct learning of the electronic information using NN may be achieved.

\citet{Schutt2019-mu} developed the SchNOrb model, which is one of the earliest works in using NN models to predict the quantum mechanical wavefunctions for molecular systems. Starting from the molecular electronic wavefunction as a linear combination of atomic basis $|\psi_m\rangle=\sum_i c_m^i|\phi_i\rangle$, the model aims to solve the eigenvalue problem of $\bm{H}\bm{c}_{m}=\varepsilon_m\bm{S}\bm{c}_{m}$, by learning the symmetry-adapted pairwise elements of the Hamiltonian and orbital overlap ($\bm{S}$) matricies from some  learnable features that encodes the information on the interactions between pairs of atoms. By training the model with a loss function that is constructed on reconstructing $\bm{H}$, $\bm{S}$ as well as the system energy and forces, the trained model can be directly applied for investigating, for example, conformational dependent electronic properties of molecules by computing the electronic properties on-the-fly in MD simulations. 

The PauliNet developed by \citet{Hermann2020-ao} was an interesting model, targeted at solving the  electronic Schr\"{o}dinger equation, accounting for the full electron correlation effect through overcoming the challenges behind the use of quantum Monte Carlo to solve the multielectron wavefunctions (with up to 30 electrons). The multielectron wavefunction $\Psi(\bm{r}_1,\bm{r}_{2},\ldots,\bm{r}_N)$ is a complex mathematical object that lives in the space of $\mathbb{R}^{3N}$, the PauliNet seeks the inspiration from the variation principle to construct a parameterised approximate $\Psi_{\boldsymbol{\theta}}$ to the true wavefunction $\Psi$ in which the true ground-state energy of a many electron system $E_{0}\leq\min_{\Psi}E[\Psi]\leq \min_{\boldsymbol{\theta}}E[\Psi_{\boldsymbol{\theta}}]$. More specifically, starting from an HF ansatz of uncorrelated wavefunction $\{\phi\}_p$ (per Slater determinant $p$), the authors utilise the fact that the correlated wavefunction $\Psi_{\boldsymbol{\theta}}$ can be constructed from the uncorrelated ones as $\Psi_{\boldsymbol{\theta}}(\bm{r})\sim e^{J_{\boldsymbol{\theta}}(\bm{r})}\sum_p c_p\mbox{det}[\phi_{\boldsymbol{\theta}}^{\uparrow}(\bm{r})]\mbox{det}[\phi_{\boldsymbol{\theta}}^{\downarrow}(\bm{r})]$, in which 
 $J_{\boldsymbol{\theta}}(\bm{r})$ is the Jastrow factor, thus the key problem that is to be solved by the PauliNet is to predict $\{J_{\boldsymbol{\theta}},\phi_{\boldsymbol{\theta}}^{\uparrow},\phi_{\boldsymbol{\theta}}^{\downarrow}\}$  from the nuclear ($\bm{R}$) and electron ($\bm{r}^{\uparrow,\downarrow}$) coordinates, as well as the chemical information $Z$.

DeepH constitutes a family of NN models that were developed for approximating the KS/generalised-KS Hamiltonians for \emph{crystalline materials.} The initial model published by \citet{Li2022-ie} was designed to harness the `\emph{near-sightedness}' of KS-DFT to predict the electronic properties of crystalline solids that are represented by the tight-binding Hamiltonian matrix $\bm{H}_{i\alpha,j\beta}$, which is computed from DFT results using a pseudo-localised basis function as implemented in the openMX\cite{ozaki2003variationally,ozaki2004numerical} package. Upon Fourier transformation of the Hamiltonian, we obtain its corresponding representation in the reciprocal space $\bm{H}(\bm{k})$, which is used for constructing the eigenvalue problem $\bm{H}(\bm{k})\bm{c}_{n\bm{k}}=\varepsilon_{n\bm{k}}\bm{S}(\bm{k})\bm{c}_{n\bm{k}}$ to obtain the electronic band structures. By adapting the MPNN architecture, DeepH approximate the DFT Hamiltonian by learning $H_{ij}(\{\bm{R}\}_{N})$ as a function of the local chemical environment around each atom in the material. The model's capabilities in predicting the electronic properties of solids have been demonstrated on a set of large scale systems that is beyond the reach of standard DFT calculations, including: band structures of zigzag carbon and MoS$_{2}$ nanotubes, as well as the band structures of magic angle twisted vdW bilayers of 2D materials. In particular, DeepH did not preserve the rotational equivariance through the use of tensor product and CG decompositions, which becomes computationally expensive for large-scale systems. Instead, a strategy employing spherical harmonic basis set transformation was adopted.

The XDeepH\cite{Li2023-wf} extended the DeepH model by taking into account the magnetic states in condensed matter systems, and had been applied to simulate, for example, the spin-polarised electronic band structures for the Moir\'{e} bilayer of CrI$_{3}$. For spin polarised systems, the electronic Hamiltonian now has the spin-dependent blocks with a significant portion of its non-diagonal elements being non-zero to reflect the spin-spin coupling, hence the number of Hamiltonian matrix elements that needs to be learned are enlarged substantially compared to the non-spin-polarised counterpart. Incorporating the spin degree-of-freedom also implies that the model must now respect to both the Euclidean and time-reversal symmetry ($E(3)\times\{\mathcal{I},\mathcal{T}\}$, in order to allow one to construct the mapping from $(\{\bm{R}\},\{\bm{M}\})$ to $\bm{H}$, in which $\bm{M}$ is the information of atomic magnetic moments. Practically, XDeepH imposes rotational equivariance through the use of \emph{Darstellung} matricies [\emph{c.f.} \cref{eq:Wigner_D_matrix_rotate_rank_2_tensor}] and CG decompositions. The time-reversal symmetry  implies that,  a spin-dependent message of the type $\bm{h}^{\ell_1,\ell_2}_{m_1\sigma_1,m_2\sigma_2}$ should satisfy the condition of 
\begin{equation*}
\mathcal{T}\bm{h}^{\ell_1,\ell_2}_{m_1\sigma_1,m_2\sigma_2}=\begin{cases}
\left(\bm{h}^{\ell_1,\ell_2}_{m_1(-\sigma_1),m_2(-\sigma_2)}\right)^* &\text{if }  \sigma_1=\sigma_2\\
-\left(\bm{h}^{\ell_1,\ell_2}_{m_1(-\sigma_1),m_2(-\sigma_2)}\right)^* &\text{if }  \sigma_1\neq\sigma_2.
\end{cases}
\end{equation*}
In XDeepH, treating the spin symmetry in the complex domain is solved by applying the transformation of $\frac{1}{2}\otimes\frac{1}{2}=0\oplus 1$, and physically means that any neurons that implemented the time-reversal operation will flip the sign for the imaginary part of the $\ell=0$ vector and the real part of the $\ell=1$ vector while all the other numbers remain unchanged. Another extension of the DeepH model is the DeepH-Hybrid\cite{Tang2024-ek} model that was published in 2024 that aims to learn the hybrid-GGA (generalised-KS) Hamiltonian for condensed matters, which was trained on a specifically collated dataset on graphene, carbon nanotubes and MoS$_2$. The general question to be answered in this work is whether the DeepH model, that harnessed the `near-sightedness' of KS-DFT, can be adopted to model the exact exchange  [\emph{c.f.} \cref{eq:exchange_HF_DFT}] that is non-local in nature. This is validated by realising that the Coulomb integral that goes into the exact exchange is generally short-ranged, meaning that it is non-zero when there is a finite orbital overlap between two interacting electrons, thus enabling the extension of DeepH model into its DeepH-Hybrid variant.

\citet{Koker2024-vs} developed the ChargE3Net model which employed higher-order equivariant (up to $\ell=4$) GNN to predict the charge densities for solids, instead of the Hamiltonians, that can be directly used as an input to solve the KS eigenvalue problem non-self-consistently while achieving near-DFT accuracy in predicting the electronic and thermodynamic properties. The higher-order tensors were employed to ensure accurate predictions of the charge densities which exhibit high angular variations. The model was designed to predict charge densities on a set of probe points $\{\bm{g}_1,\ldots,\bm{g}_{m}\}\in\mathbb{R}^{3}$ given a set of atomic positions $\{\bm{R}_1,\ldots,\bm{R}_n\}\in\mathbb{R}^3$ and chemical information $\{Z_1,\ldots,Z_n\}\in\mathbb{R}$ as the model inputs.

The successes behind the development of DL models as approximators for electronic Hamiltonians thus have definitely opened up new opportunities for theoreticians to tackle more complex many-body problems in the domain of electronic structure theory. As a final note for this section, without going into further details, we refer the readers to more recent examples in the development of the MBGF-Net\cite{Venturella2025-fh} targeted at predicting the electron interaction self energies $\Sigma(\omega)$ for GW-type calculations [\emph{c.f.} \cref{sect:GW_approximation}] based on the results of low-level electronic structure theory calculations, as well as the N$^2$AMD\cite{Zhang2025-sk} model for accelerating the non-adiabatic AIMD simulations for studying the ion-coupled electronic excited state dynamics in photovoltaics and similar energy-harvesting materials.

\subsection{Generative AI}\label{sec:generative_AI}
\subsubsection{Motivations}

Till this point, we have discussed extensively on the theoretical approaches that enables us to determine the energetic stabilities and physical properties of materials, starting from their chemical compositions and atomistic structures, which constitute a forward mathematical mapping of $\{\mathcal{C},\bm{R}\}\xrightarrow{f}\{\mathcal{P}\}$. However, the ultimate goal in the computer-guided materials discoveries, is to inversely design new materials\cite{zunger2018inverse} that aims to propose/sample the compositions and structures of new materials that possess certain properties, or to be  mathematically formulated as to solve and develop an inverse mapping of   $\{\mathcal{P}\}\xrightarrow{f^{-1}}\{\mathcal{C},\bm{R}\}$. As NN has shown its unique capability as a universal approximator for solving the forward mapping problem $\mathcal{P}=f_{\boldsymbol{\theta}}(\mathcal{C},\bm{R})$, which, in principle, can also be trained to realise a model that approximate the inverse mapping of $(\mathcal{C},\bm{R})=f^{-1}_{\boldsymbol{\varphi}}(\mathcal{P})$. This motivates the development of generative AI methods for exploring new materials. 

Focusing further on the energetic stabilities $\mathcal{P}(\bm{R})=E(\bm{R})$ together with the data-driven and probabilistic perspective of AI models, the core problem that one would like to address in generative AI modelling can be formulated as following. Given the knowledge of a set of structures $\{\bm{R}\}$ that possess the stabilities $\{E\}$ is $p(\bm{R})$ (which, in this case, is the Boltzmann distribution), then can we directly draw new structures $\bm{R}_{\mathrm{new}}$ from $p(\bm{R})$, provided that we can somehow learn the highly complex probability distribution of $p(\bm{R})$ with a universal approximator, such as NN. Answering such a question leads to a fundamentally different, and conceptually broader, statistical sampling paradigm for exploring the materials space compared to traditional energy-based PES sampling approaches based on MD/MC [\cref{chapter:sampling_methods}]. 

Firstly, the bespoke generative model should, in principle, help to overcome the ergodicity challenge in statistical samplings that often require long time/Markov chain to discover rare minima separated by high energy barriers, thus leading to extremely low probabilities in discovering new structures. Secondly, whilst energy-based models strongly rely on the knowledge of gradient (or forces, $\nabla_{\bm{R}}E(\bm{R})$) to drive the sampling process to reach new local minima, in general, we do not necessarily know how to compute $\nabla_{\bm{R}}\mathcal{P}(\bm{R})$ with respect to any property of interest, particularly, in a broader context, the \emph{alchemical derivatives}\cite{anatole2009accurate,sheppard2010alchemical}   $\nabla_\mathcal{C}\mathcal{P}$ that informs us on how the property changes with respect to the change in the chemical compositions, that can be applied to guide our exploration of materials across the chemical space. The development of generative AI models for materials discoveries are posed to address these challenges.  

\subsubsection{Frameworks of Common Generative Models}
\label{sec:common_generative_models}

In this section, we shall provide a brief account on the architectures behind three most prominent generative models that have been applied to material and molecule discoveries in the past five years, namely, the variational autoencoders (VAE),\cite{kingma2013auto} the diffusion generative model\cite{sohl2015deep,song2019generative,song2020score,ho2020denoising} (which shares a close linkage with the hierarchical VAE\cite{sonderby2016ladder}), and finally the flow-based generative model.\cite{rezende2015variational} 

\begin{figure}[htb]
    \centering
    \includegraphics[width=0.7\linewidth]{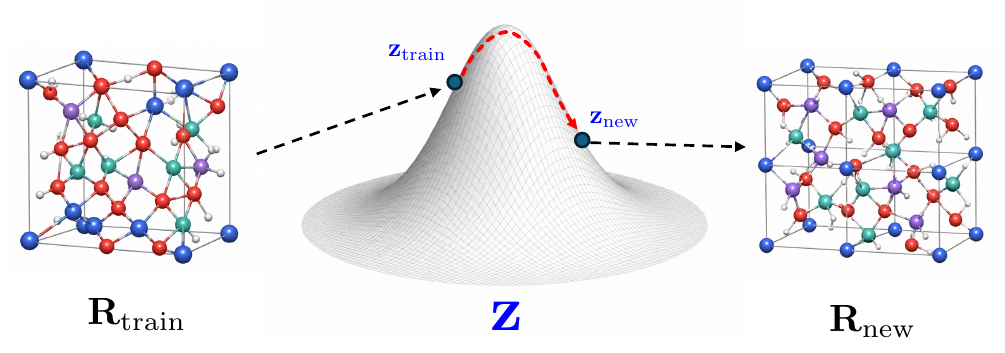}
    \caption{General design concept behind modern generative AI models for materials discovery, which the training data $\bm{R}_{\mathrm{train}}$ that follows complex multimodal probability distribution is first converted into a latent variable $\bm{z}_{\mathrm{train}}$ that follows a simple, easy-to-sample distribution in the latent space. Subsequently, sampling a new latent variable $\bm{z}_{\mathrm{new}}$ in the latent space, and use it as the input in the inverse generative step will lead to a new material structure $\bm{R}_{\mathrm{new}}$. }
    \label{fig:generative_general_concept}
\end{figure}

In a high level overview (\cref{fig:generative_general_concept}), all these three generative models aims to overcome the problem of sampling complex multimodal probability distributions of $p(\bm{R})$ by first mapping the input/training structure $\bm{R}_{\mathrm{train}}\in \mathbb{R}^{3N}$ into a \emph{latent space} variable $\bm{z}_{\mathrm{train}}$, which may or may not be of a lower dimensionality of $\bm{R}_{\mathrm{train}}$, but most importantly, follow a simple probability distribution $p(\bm{z})$ that is easy to sample from, such as $p(\bm{z})\sim\mathcal{N}(\bm{z}|0,\bm{I})$. Providing that we can achieve the forward mapping by training a model $\bm{z}=f_{\boldsymbol{\varphi}}(\bm{R})$, and conjunctively learn a model that enables the inverse mapping $\bm{R}=g_{\boldsymbol{\theta}}(\bm{z})$, then, in principle, one could easy generate new samples $\bm{R}_{\mathrm{new}}$ by first sampling a new latent variable $\bm{z}_{\mathrm{new}}\sim p(\bm{z})$ (with negligible cost) followed by the application of the inverse mapping  $\bm{R}_{\mathrm{new}}=g_{\boldsymbol{\theta}}(\bm{z}_{\mathrm{new}})$.

\begin{wrapfigure}[28]{rt}{0.5\textwidth}
  \begin{center}
    \includegraphics[width=0.5\textwidth]{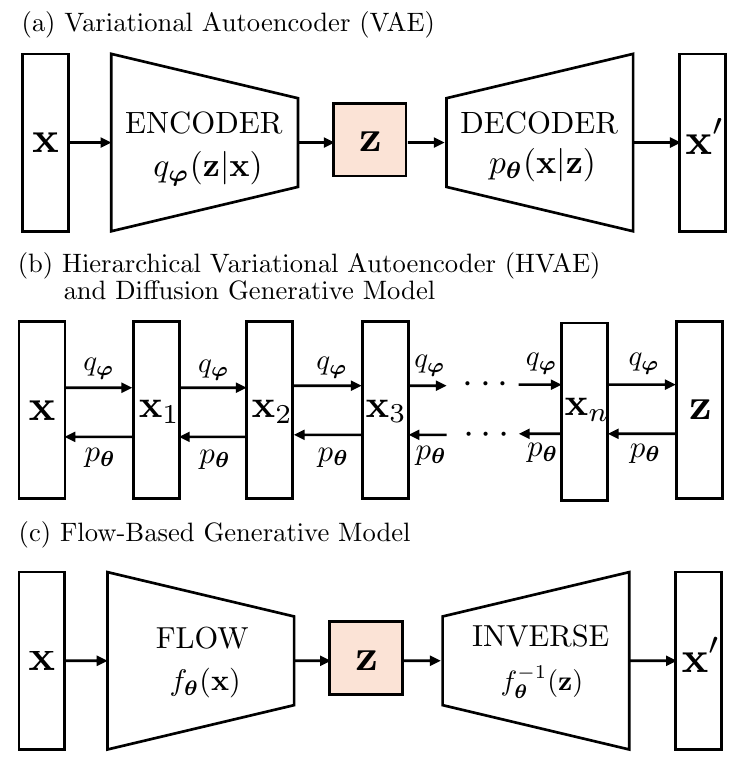}
    \caption{Schematic illustration of the architectures behind the three key generative models discussed in this work.}
    \label{fig:generative_models}
  \end{center}
\end{wrapfigure}

Without lost of generality, most modern structural generative models\cite{xie2021crystal,midgley2023se,zeni2025generative,kohler2019equivariant,klein2023equivariant}  tend to adopt equivariant NNs [\cref{sec:dl_fundamentals}] as the inductive bias to approximate the functions $\{f_{\boldsymbol{\varphi}},g_{\boldsymbol{\theta}}\}$ to achieve the efficient transformations between the real and latent space, such that we do not need to drill into the details of the NN architectures here but focusing more on the transformation process $\bm{x}\Leftrightarrow\bm{z}$ itself\footnote{In what follow, we shall be using $\bm{x}$ and $\bm{R}$ interchangeably to denote the inputs to the generative models.}. For the latter, we can further classify the transformations between $p(\bm{x})$ and $p(\bm{z})$ to be either achieved through an \emph{one-shot} (in VAE and flow-based models) or \textit{progressive} (in HVAE and diffusion models to enhance the quality of the generated models) procedure [\cref{fig:generative_models}], whereby the \textit{progressive} generative models would also enable us  \cite{song2019generative,song2020score} to apply techniques from stochastic differential equations to evolve the samples, in a way similar to the case of evolving structures on the PES with MD simulations while maintaining a desired thermodynamic ensemble through stochastically interacting with the thermo-/baro-stats (\cref{sect:molecular_dynamics}). Finally, building the foundation of probability interpretation of AI models, one could easily cast the problem of \emph{property-guided materials discovery} into the mathematical concepts of transforming between the  probabilities  of observing a specific structure/latent variable that is conditioned on the desired property, \emph{i.e.} $p(\bm{x}|\mathcal{P})\Leftrightarrow p(\bm{z}|\mathcal{P})$,\cite{ho2022classifier}  opening the implementable pathways to achieve property-driven inverse design of materials in modern atomistic materials simulations.


Conceptually, VAE [\cref{fig:generative_models}(a)] possesses a model architecture that is the easiest to understand, which largely follows the general conceptual frameworks for generative models discussed in the beginning.  Here, a forward NN is used to learn a map from the input $\bm{x}$ into the latent variable $\bm{z}$, or equivalently, the parameterised probability distribution of the latent variable conditioned on the input space $q_{\boldsymbol{\varphi}}(\bm{z}|\bm{x})$. This is to be combined with a backward NN that learns how to \emph{reconstruct} the real sample $\bm{x}$ from a given latent variable $\bm{z}$, which is in term governed by the conditional decoding distribution of $p_{\boldsymbol{\theta}}(\bm{x}|\bm{z})$. 

Denoting both model parameters collectively as $\boldsymbol{\phi}=(\boldsymbol{\varphi},\boldsymbol{\theta})$, training the generative model is similar to the training of supervised ML models (\cref{appendix:supervised_ML}) that aims to maximise the probability of observing the input samples over the parameter space, \emph{i.e.}
\begin{equation}
    \label{eq:argmax_phi_VAE}
    \hat{\boldsymbol{\phi}}=\argmax_{\boldsymbol{\phi}}p_{\boldsymbol{\phi}}(\bm{x}),
\end{equation}
whereby in VAE, $p(\bm{x})$ is obtained by marginalising the joint probability $p(\bm{x},\bm{z})$ over the latent space, \emph{i.e.} $p(\bm{x})=\int p(\bm{x},\bm{z})\mathrm{d} \bm{z}$. However, working directly with $p(\bm{x})$ is not mathematically tractable because in statistical thermodynamics, the full distribution $p(\bm{x})$ is normalised as $p=p^\prime/Z$ by the normalisation constant $Z=\int p^\prime(\bm{x})\mathrm{d}\bm{x}$ integrated over the full sample space, computing which thus requires a full configurational space sampling that satisfies the ergodic condition. This problem is commonly circumvented by taking the logarithm of the probability $\log p(\bm{x})=\log p^\prime(\bm{x})-\log Z$, which factors out a constant term of $\log Z$ that will not affect the outcome of the optimisation problem.\cite{sohl2015deep} This effectively cast the model optimisation problem of (\ref{eq:argmax_phi_VAE}) into a minimisation one as
\begin{equation}
    \label{eq:argmax_log_prob_phi_VAE}
    \hat{\boldsymbol{\phi}}=\argmin_{\boldsymbol{\phi}}[-\log p_{\boldsymbol{\phi}}(\bm{x})].
\end{equation}
The next problem to be solved is how to tackle the intractable joint probability distribution of $p(\bm{x},\bm{z})$. This is also easy to understand, as $p(\bm{z})$ is established with a trained NN model, hence it is not possible to compute $p(\bm{x},\bm{z})$ during the model training process. The trick here is to apply the Baye's rule followed by the Jansen's inequality (\cref{fig:generative_model_relations}) to derived an ELBO, which is to be used as the upper bound of the negative log-likelihood [\cref{eq:argmax_log_prob_phi_VAE}] in model training that will jointly optimised the parameters in both the encoder and decoder:\cite{luo2022understanding}
\begin{equation}
    \label{eq:VAE_ELBO}
    \mbox{ELBO}=\underbrace{ \mathbb{E}_{q_{\boldsymbol{\varphi}}(\bm{z}|\bm{x})}\left[p_{\boldsymbol{\theta}}(\bm{x}|\bm{z})\right]}_{\mbox{\scriptsize{reconstruction loss}}}
    -\underbrace{\mathcal{D}_{\mathrm{KL}}\left(  q_{\boldsymbol{\varphi}}(\bm{z}|\bm{x})||p(\bm{z}) \right)}_{\mbox{\scriptsize{prior matching}}}.
\end{equation}
Here, the reconstruction loss measures the averaged agreement between the input data and the latent variables, whereas the prior matching term measures how well the latent variables generated from the forward network follows the prior distribution of the latent variables. Minimising the ELBO is numerically achievable if we further parameterise the conditional and prior distributions with common distributions, such as Gaussians [see, for example, the derivation of the training target for ridge regression in \cref{appendix:ridge_regression}]. 

To improve the quality of the samples being generated, hierarchical VAE (HVAE) [\cref{fig:generative_models}(b)] stacks multiple VAE together that enables us to gradually transform between $\bm{x}$ and $\bm{z}$ through multiple intermediate latent variables $\bm{x}_1$, $\bm{x}_2$, $\cdots$, and $\bm{x}_n$. The key modification to the training target [\cref{eq:VAE_ELBO}] that will enable its extension to HVAE is to utilise the fact that the joint probability distributions $q_{\boldsymbol{\varphi}}(\bm{x},\bm{z})$ and $p_{\boldsymbol{\theta}}(\bm{x},\bm{z})$ now take the form of products of all intermediate transition probabilities:
\begin{equation}
\label{eq:HVAE_backward}
    p_{\boldsymbol{\theta}}(\bm{x},\bm{z})\Rightarrow p(\bm{x},\bm{x}_1,\bm{x}_2,\ldots,\bm{x}_n,\bm{z})=p(\bm{z})p_{\boldsymbol{\theta}}(\bm{x}|\bm{z})\prod_{i=1}^{n}p_{\boldsymbol{\theta}}(\bm{x}_{i-1}|\bm{x}_i),
\end{equation}
for the backward process and
\begin{equation}
\label{eq:HVAE_forward}
    q_{\boldsymbol{\varphi}}(\bm{z}|\bm{x})\Rightarrow q_{\boldsymbol{\varphi}}(\bm{z},\bm{x}_{n},\bm{x}_{n-1},\ldots,\bm{x}_1|\bm{z})=q_{\boldsymbol{\varphi}}(\bm{x}_1|\bm{x})\prod_{i=2}^{n}q_{\boldsymbol{\varphi}}(\bm{x}_i|\bm{x}_{i-1}),
\end{equation}
for the forward process. This does not necessarily provide us with an easily implementable target function to train the HAVE-based generative model, however, it leads to a good foundation for us to understand diffusion generative models.

The diffusion model simplifies HVAE by modifying the forward process (\cref{fig:generative_model_relations}), such that it no longer requires the training of a parameterised NN model, but collapse the input $x$ into  pure noise  $\bm{z}\sim\mathcal{N}(0,\bm{I})$ by sequentially adding (time-dependent) Gaussian noises $\boldsymbol{\varepsilon}\sim\mathcal{N}(0,\bm{I})$ onto the sample as:
\begin{equation}
    \label{eq:diffusion_forward}
    \bm{x}_{t+1}=\alpha_t\bm{x}_{t}+\beta_t\boldsymbol{\varepsilon}_t \quad \Leftrightarrow\quad \bm{x}_{t+1}\sim\mathcal{N}(\bm{x}|\alpha_t\bm{x}_t,\beta_t\bm{I}).
\end{equation}
Here, the time-dependent hyperparameters $\alpha_t$ and $\beta_t$  are interdependent by imposing the condition that $\alpha_t+\beta_t=1$.  This enables us to collapse the input sample $\bm{x}_0$ into pure white noise in a single step:
\begin{equation}
    \label{eq:diffusion_forward_single_step}
    \bm{x}_t=\bar{\alpha}_t\bm{x}_0+\bar{\beta}_t\bar{\varepsilon}_t \quad \Leftrightarrow\quad \bm{x}_{t}\sim\mathcal{N}(\bm{x}|\bar{\alpha}_t\bm{x}_0,\bar{\beta}_t\bm{I}),
\end{equation}
with $\bar{\varepsilon}_t\sim\mathcal{N}(0,\bm{I})$ and $\bar{\alpha}_t=\prod_{i=0}^t\alpha_i$, leading to $\bar{\beta}_t=\sqrt{1-\bar{\alpha}_t^2}$. In the practical setting, the time-dependent noise $\boldsymbol{\varepsilon}_t$ is typically drawn from a \emph{noise scheduler}, which allows small noises to be added to the sample during the initial stage when $\bm{x}_t$ is similar to the input sample $\bm{x}_0$, whereas large $\boldsymbol{\varepsilon}_t$ can be used towards the end of the forward process when $\bm{x}_t$ is close to pure white noise. To ensure stable generative process, one typically applies a large number of steps $T\sim10^3$ in both the forward and reverse process of the diffusion generative process. 

The diffusion generative process takes the inspiration from this simple forward process [\cref{eq:diffusion_forward}] and ask the question, that if we can use a model to learn the time-dependent noise $\boldsymbol{\varepsilon}_t$ that is added to $\bm{x}_{t-1}$ to generate $\bm{x}_t$, from the outcome $\bm{x}_t$, through a parameterised model, such that $\boldsymbol{\varepsilon}_t\sim\boldsymbol{\varepsilon}_{\boldsymbol{\theta}}(\bm{x}_t,t)$, then could we reverse the data-collapsing process to generate plausible new samples by sequentially removing the added noised from the collapsed sample (from $T$ to 0), \emph{i.e.}
\begin{equation}
    \label{eq:diffusion_generative_process}
\bm{x}_{t-1}=\frac{1}{\alpha}_t\big(\bm{x}_t-\beta_t \boldsymbol{\varepsilon}_{\boldsymbol{\theta}}(\bm{x}_t,t) \big).
\end{equation}
This suggests that the most straightforward way to write the loss function for training the diffusion generative model is to minimise the difference between the learnt and actual noises that are added to collapse the samples:
\begin{equation}
    \label{eq:DDPM_loss}
    \mathcal{L}(\boldsymbol{\theta})=\sum_{t=1}^{T}\frac{\beta_t^2}{\alpha_t^2}\big\Vert \boldsymbol{\varepsilon}_t -\boldsymbol{\varepsilon}_{\boldsymbol{\theta}}(\bm{x}_t,t)\big\Vert_2^2,
\end{equation}
in which the prefactors $\beta_t^2/\alpha_t^2$ are usually ignored in the actual training process, that had been shown to lead to a numerically more stable training curve.\cite{ho2020denoising} Although \cref{eq:DDPM_loss} is trivial to understand and is usually how the training of diffusion models is implemented numerically, it does not necessarily facilitate the understanding on the deep relationships between the diffusion model and HVAE,  developing other parameterisation strategies for the loss function, as establishing the relationship between the backward generation process [\cref{eq:diffusion_generative_process}] and energy-based samplings. In \cref{fig:generative_model_relations} and \cref{fig:denosing_process} we provided the reader with flowcharts that highlight these key relationships based on the work from \citet{luo2022understanding}, and summarise some of the key ideas as following.

We return to the starting point of training the generative model by minimising the negative-log-likelihood of the data distribution $-\log p(\bm{x})$ through the ELBO, that is to be constructed by plugging in \cref{eq:HVAE_backward} and \cref{eq:HVAE_forward} into \cref{eq:VAE_ELBO}. By further factorising out the intermediate transition probabilities as well as the first term $q(\bm{x}_1|\bm{x}_0)$ from the joint probability distribution $q(\bm{x}_{1\ldots T}|\bm{x}_0)$, we arrive at the following KL-divergence (denoising matching term for training the diffusion model), that can be aligned with \cref{eq:VAE_ELBO}:
\begin{equation}
    \label{eq:diffusion_KL_div}
\argmax_{\boldsymbol{\theta}}p_{\boldsymbol{\theta}}(\bm{x})\sim\argmin_{\boldsymbol{\theta}}\left\{ -\sum_{t=2}^T \mathbb{E}_{q(\bm{x}_t|\bm{x}_0)} \Big[\mathcal{D}_{\mathrm{KL}}\big( q(\bm{x}_{t-1}|\bm{x}_t,\bm{x}_0)\Vert p_{\boldsymbol{\theta}}(\bm{x}_{t-1}|\bm{x}_t) \big) \Big] \right\},
\end{equation}
which provides a clear probabilistic interpretation of the loss function as to match the backward transition probability from $\bm{x}_{t-1}$ to $\bm{x}_t$ (as achieved by the learned model $p_{\boldsymbol{\theta}}$), with the corresponding transition probability (by reversing \cref{eq:diffusion_forward} to make $\bm{x}_t$ as the subject from the forward process. Furthermore, anchoring the forward transition probability $q(\bm{x}_{t-1}|\bm{x}_t,\bm{x}_0)$ through conditioning on the input data $\bm{x}_0$ leads to a more stable training process through aligning with the probability distribution for the input data. As further shown in \cref{fig:denosing_process}, numerically implementable KL-divergence in \cref{eq:diffusion_KL_div} can be derived depending on how we parameterised the transition probability $(\bm{x}_{t-1}|\bm{x}_t,\bm{x}_0)$, from which \cref{eq:DDPM_loss} is one of the three possible outcomes (or interpretations) of these parameterisation schemes. 

The \emph{\textbf{score-matching}}\cite{song2020score} interpretation provides a more interesting pathway to reveal the close relationship between the diffusion generative process and the energy-based PES sampling from statistical thermodynamics.\cite{sohl2015deep} Instead of using the Tweedie's formula to derive such result [\cref{fig:denosing_process}]\cite{luo2022understanding}, it is more inspiring to derive the score-matching denoising process based on stochastic differential equations (SDEs). Consider the time-dependent forward process of collasing the sample that is to be described by an SDE in the limit of $\Delta t\to0$ as 
\begin{equation}
    \label{eq:diffusion_forward_SDE}
    \mathrm{d}\bm{x}=f_t(\bm{x})dt+g_t \mathrm{d}\boldsymbol{w},
\end{equation}
 with $\mathrm{d}\boldsymbol{w}\sim\mathcal{N}(0,\bm{I})$ denotes some random noises. This allows us to parameterise the forward transition probability with a conditional normal distribution: 
\begin{equation}
    p(\bm{x}_{t+\Delta t}|\bm{x}_t)=\mathcal{N}\left( \bm{x}_{t+\Delta t}; \bm{x}_t+f_t(\bm{x}_t)\Delta t,g_t^2\Delta t\bm{I}\right)
    \propto \exp\left(-\frac{\| \bm{x}_{t+\Delta t}-\bm{x}-f_t(\bm{x})\mathrm{d}t\|_2^2}{2g_t^2\Delta t}\right),\label{eq:forward_prob_SDR}
\end{equation}
From here the transition probability for the reverse generative process can be derived using the Baye's rule as
\begin{align}
  p(\bm{x}_t|\bm{x}_{t+\Delta t})&=\frac{p(\bm{x}_{t+\Delta t}|\bm{x}_t) p(\bm{x}_t)} {p(\bm{x}_{t+\Delta t})} = p(\bm{x}_{t+\Delta t}|\bm{x}_t)\exp\left[ \log p(\bm{x}_t) - \log p(\bm{x}_{t+\Delta t}) \right] \nonumber\\
  &\propto \exp \left( -\frac{\| \bm{x}_{t+\Delta t}-\bm{x}_t-f_t(\bm{x}_t)\mathrm{d}t\|_2^2}{2g_t^2\Delta t}   + \log p(\bm{x}_t) - \log p(\bm{x}_{t+\Delta t}) \right).\label{eq:reverse_prob_Bayes}
\end{align}
The missing puzzle in the above equation is how $\log p(\bm{x}_t)$ will evolve into $\log p(\bm{x}_{t+\Delta})$ in the forward process. This can be expressed with Taylor expansion:
\begin{equation}
    \label{eq:taylor_exp_probability_diffusion}
    \log p(\bm{x}_{t+\Delta t    }) \approx \log p(\bm{x}_t)+\big(\bm{x}_{t+\Delta t}-\bm{x}_t \big)\nabla_{\bm{x}_t}\log p(\bm{x}_t)+\frac{\partial}{\partial t}p(\bm{x}_t),
\end{equation}
in which $s(\bm{x},t)=\nabla_{\bm{x}_t}\log p(\bm{x}_t)$ is  the so-called \emph{score function}. Substituting \cref{eq:taylor_exp_probability_diffusion} into \cref{eq:reverse_prob_Bayes}, one can show that \cref{eq:reverse_prob_Bayes} can be further simplified into
\begin{equation}
    \label{eq:reverse_prob_Bayes_2}
    p(\bm{x}_t|\bm{x}_{t+\Delta t})\approx \exp\left(-\frac{\|\bm{x}_t- \bm{x}_{t+\Delta t}-[f_{t+\Delta t}(\bm{x}_{t+\Delta t})-g_{t+\Delta t}^2 \nabla_{\bm{x}_{t+\Delta t}} \log p(\bm{x}_{t+\Delta t})]\Delta t\|_2^2}{2g_{t+\Delta t}^2\Delta t}\right),
\end{equation}
This shows that $p(\bm{x}|\bm{x}_{t+\Delta t})$ is approximately a normal distribution with the following mean:
\begin{equation*}
    \bm{\mu}=\bm{x}_{t+\Delta t}-[f_{t+\Delta t}(\bm{x}_{t+\Delta t})-g_{t+\Delta t}^2 \nabla_{\bm{x}_{t+\Delta t}} \log p(\bm{x}_{t+\Delta t})]\Delta t,
\end{equation*}
and standard deviation of $\bm{\sigma}=g_{t+\Delta t}^2\Delta t\bm{I}$. Taking the limit of $\Delta t\to 0$, then the corresponding SDE for the reverse generative process is
\begin{equation}
    \label{eq:reverse_generation_SDE}
    \mathrm{d}\bm{x}=[f_t(\bm{x}_t)-g_t^2\nabla_{\bm{x}}\log p(\bm{x})]dt +g_t \mathrm{d}\boldsymbol{w}. 
\end{equation}
It can be further shown that the standard deviation of $\bm{x}_t$ is proportional to $\nabla_{\bm{x}}\log p(\bm{x})$, allowing the loss function [\cref{eq:DDPM_loss}] to be rewritten in the score-matching form as:
\begin{equation}
    \label{eq:DDPM_score_matching_loss}
    \mathcal{L}(\boldsymbol{\theta})=\sum_{t=1}^T\big\Vert s_{\boldsymbol{\theta}}(\bm{x}_t,t)-\nabla_{\bm{x}_t}\log p(\bm{x}_t)\big\Vert_2^2,
\end{equation}
where the NN $s_{\boldsymbol{\theta}}(\bm{x}_t,t)$ are parameterised to learn the score function. 

Following our previous discussions around \cref{fig:dummy_MC_sampling}, given the fact that, in atomistic simulations, the probability distribution for materials follows the Boltzmann distribution of $p(\bm{x})=\exp(-\beta U(\bm{x}))$, such that the corresponding score function $-\nabla_{\bm{x}}p(\bm{x})=\beta\nabla_{\bm{x}}U(\bm{x})=\beta \bm{F}$ is essentially the atomic forces, suggesting that applying \cref{eq:reverse_generation_SDE} for generating new material structures is essentially a process that first randomly placing  a set of atoms in the simulation box and subsequently letting them to aggregate into a new structure based on the interaction forces between atoms (provided that we have a good model for approximating $U$).\cite{cheng2024response} In the case of image generation, the original application for which diffusion generative models were designed for, the score function can equivalently be considered as the \emph{forces} that brings the image pixels together that is learnt from the underlying data distribution. Furthermore, based on the SDE formulation, one can further integrate existing methods of statistical samplings [\emph{c.f.} \cref{sect:molecular_dynamics}]. In the spirit of molecular dynamic samplings as a way to solve the SDE, one could seek to  incorporate additional bias terms, as well as utilising numerical methods that accelerates the solution of SDEs, to explore a wide range of alternative approaches for generating new and interesting samples (\emph{e.g.} to ensure the newly generated materials are thermodynamically stable). In the spirit of Monte Carlo sampling, the trained generative model $q_{\boldsymbol{\theta}}(\bm{x}^\prime|\bm{x})$ can be used to guide he Metropolis-type MC moves that could lead to higher probability of guiding the sampling to generate thermodynamically stable structures. Theoretically, to incorporate the generative guidance in MC sampling, one could adapt the following modification to the standard Metropolis acceptance criterion [\cref{eq:metropolis}] into:
\begin{equation}
    \label{eq:diffusion_metropolis}
    p(\bm{R}^{\mathrm{new}}|\bm{R})=\min\left[1,\frac{\pi(\bm{R}^{\mathrm{new}})q_{\boldsymbol{\theta}}(\bm{R}|\bm{R}^{\mathrm{new}}) }{\pi(\bm{R})q_{\boldsymbol{\theta}}(\bm{R}^{\mathrm{new}}|\bm{R}) } \right],
\end{equation}
which is an essential change to ensure the sampling process is still Markovian.

It can be understood from the discussion above,  that the diffusion generative model is designed to achieve a  transformation between the sample $\bm{x}_0$ and the latent variable $\bm{x}_T$ through a process that is \emph{continuous}. The flow-based generative model  [\cref{fig:generative_models}(c)] adapts the fundamental model architecture of a VAE, while preserving the idea of continuity by treating the forward and reverse processes as ones similar to the flow of fluids through a bottleneck region, during which the total volume of the fluid is preserved. Unlike VAE that trains two separate NNs for the encoder and decoder, respectively, the flow model trains an \emph{invertible NN approximator} for the forward process $\bm{z}=f_{\boldsymbol{\theta}}(\bm{x})$, such that the reverse generative process can be more easily achieved with $\bm{x}=f_{\boldsymbol{\theta}}^{-1}(\bm{z})$. Given that $f_{\boldsymbol{\theta}}(\bm{x})$ is a continuous and differentiable function, then the data distribution (conditioned on the model parameters $\boldsymbol{\theta}$ is related to the probability distribution of the latent variables (\emph{e.g.}  normal distribution) through the following volume-preserving transformation:
\begin{equation}
    \label{eq:volume_preserving_norm}
    p(\bm{x}|\boldsymbol{\theta})=\left\vert\frac{\partial f_{\boldsymbol{\theta}}(\bm{z})}{d\bm{z}}\right\vert^{-1}p(\bm{z}),
\end{equation}
(where $\bm{J}_{\boldsymbol{\theta}}(\bm{z})=\left\vert\frac{\partial f_{\boldsymbol{\theta}}(\bm{z})}{d\bm{z}}\right\vert$ is the Jacobian of $f_{\boldsymbol{\theta}}(\bm{z})$.) such that the optimum model parameters are given by $\hat{\boldsymbol{\theta}}=\argmax_{\boldsymbol{\theta}}p(\bm{x}|\boldsymbol{\theta})$ by computing \cref{eq:volume_preserving_norm} via sampling $\bm{z}\sim\mathcal{N}(0,\bm{I})$.   The key design consideration here is therefore how to construct a NN approximator $f_{\boldsymbol{\theta}}$ that is invertible, as standard NN are generally non-invertible. \cite{prince2023understanding} The flow-based model achieve this goal by adopting the following trick. Instead of working on the input $\bm{x}$ as a whole, it first splits the input into two parts $\bm{x}=(\bm{x}_a,\bm{x}_b)$, and leave one part to be unchanged when it is passed through the NN. This  can be achieved with a following network design:
\begin{equation}
    \label{eq:invertible_NN}f_{\boldsymbol{\theta}}(\bm{x})=
    \begin{cases}
        \bm{y}_a=\bm{x}_a\\
        \bm{y}_b=\bm{x}_b\odot\exp(s(\bm{x_a}))+t(\bm{x}_a),
    \end{cases}
\end{equation}
in which $s$ and $t$ are NN that are not required to be invertible. Such a design leads to the following simple form of the inverse function:
\begin{equation}
    \label{eq:invertible_NN_inverse}
    f_{\boldsymbol{\theta}}^{-1}(\bm{y})=
    \begin{cases}
        \bm{x}_a=\bm{y}_a\\
        \bm{x}_b=(\bm{y}_b-t(\bm{y}_a))\odot \exp(-s(\bm{y}_a)).
    \end{cases}
\end{equation}

The above discussions on three representative deep generative models (\cref{fig:generative_models}) show a common feature behind their general design frameworks, that is the generative model itself plays very little focus on the actual architectural designs of the NN that is to be used in the generative model. The main focus for each generative model is on the design of mathematical transformations between between the distributions of the real data, which is unknown and difficult to sample,  and their corresponding latent representations, that are designed to be easy to sample (which is often referred to as a particular \textit{parameterisation} of the general model). One may further appreciate that, all different transformations of the probability distributions being proposed are grounded by certain mathematical foundations of the probability distributions, such as the Bayes' rule  [\cref{eq:reverse_prob_Bayes} for deriving the score-matching results] or volume conservation [\cref{eq:volume_preserving_norm} for flow-matching]. NN in this case, predominantly plays the role of a universal approximator, that in theory, will enable us to determine the unknown parameters which are necessary to complete these transformation. The exact form of the NN is largely left as an implementation details depending on the problem at hand. In this sense, the general design frameworks behind different deep generative frameworks are rather straightforward to understand.

The main challenges in implementing the generative models for the discoveries of new molecules and materials at the atomistic level lie in the actual designs of the NNs, as well as the incorporation of necessary modifications to the transformations shown above, in order for the model to respect the geometry and symmetry information, and different prior distributions (\cref{tab:discrete_prob_distributions} and \cref{tab:continuous_prob_distributions}) that should be followed by the chemical and  structural components in the systems that we are trying to model. For examples, the discrete distribution on $\mathbb{Z}^{+}$ for the occurrence of different chemical elements in the materials, the uniform distribution of bond angles on a unit circle $\theta\in[0,2\pi)$, the fractional coordinates for atoms in a crystal $f_i\in[0,1)$ (which may be described by a periodic distribution), and so on. As such, in what follows where we chronologically review some of the prominent generative models that have been developed for atomistic modellings, we aim to provide readers with some general understandings on the different approaches that the chemical and structural informations are represented and transformed mathematically to achieve the generative targets, based upon which future model developments may draw inspirations.

\subsubsection{Applications of VAE-Based Generative Models in Atomistic Modelling}
\label{sec:generative_VAE_examples}

\paragraph{\textbf{Generation of Molecules}}

We begin by reviewing the development of VAE for the tasks of molecular structure generations. \citet{gomez2018automatic}  presented one of the seminal work on generative model that is applied for molecular generative task, before the message-passing GNN\cite{gilmer2017neural} become popular and as a mainstreamed architecture in the field of atomistic modelling. Here, a VAE is trained to encode the SMILES string representations for molecules into a continuous numerical latent space representation. As such, new latent representations generated through sampling the latent space can be reconstructed into a new SMILES string through the decoder, which can be further converted into molecular structures with traditional chemoinformatic approaches. Since the SMILES strings only represent the chemical and connectivity information in a molecule with no explicit 3D geometric information, it does not require the model to implement symmetry constraint as the inductive bias to ensure stable generations.

Instead of using the string representation of molecules, \citet{jin2018junction} developed the junction-tree variational autoencoders (JTVAE) for molecular generations that applied the GNN architectures to enable molecular structure generations using information that is directly learned from the three-dimensional atomistic structures of molecules. This can help to overcome the critical problem of similar chemical structures having highly dissimilar SMILE strings, which prevents the effective learning of the latent representations over a continuous manifold with VAE. To align the model learning with the general chemical intuitions in which large (organic) molecules are built by connecting functional groups together,  JTVAE learns and generates molecular graph structures in a two-phased approach by encoding a molecule into a two-part latent representation $\bm{z}=[\bm{z}_\mathcal{T},\bm{z}_\mathcal{G}]$. 

Here, $\bm{z}_\mathcal{T}$ is the latent representation of a \emph{tree} structure (\emph{i.e.} without loops, yellow box in \cref{fig:junction-tree-net}), in which each leaf of the tree represents a chemical substructure (or cluster) and $\bm{z}_\mathcal{T}$ essentially captures the connectivities among the clusters. At the same time $\bm{z}_\mathcal{G}$ encodes the detailed atomistic structures of each cluster (which can contain loops, such as in the case of a phenyl ring, as represented by the blue box in \cref{fig:junction-tree-net}). Consequentially, in additional to the concept of atom graph $\mathcal{G}$ for representing individual cluster that is the same as in the conventional GNN for structural learning,\cite{duval2023hitchhiker} the concept of a directional tree-graph $\mathcal{T}_{\mathcal{G}}$ over interconnecting cluster $\{\mathcal{C}_1,\mathcal{C}_2,\ldots,\mathcal{C}_n\}$ is also introduced, such that the full atom-graph for the entire molecule $\hat{\mathcal{G}}=\bigcup_i\mathcal{C}_i$. Under this foundation, the encoder is trying to learn the conditional probabilities of $q(\bm{z}_\mathcal{T}|\mathcal{T})$ and $q(\bm{z}_\mathcal{G}|\mathcal{G})$  to established the joint latent space of $\bm{z}$. Molecule generation is then achieved via a two-staged approach, where a structural scaffold $\mathcal{T}_{\mathrm{new}}$ is first generated by sampling the latent variable of $\bm{z}_\mathcal{T}$ that is to be fed into the decoder $p(\mathcal{T}|\bm{z}_\mathcal{T})$, then the detailed atomistic structures will be reconstructed by sampling $\bm{z}_{\mathcal{G}}$ to reconstruct individual substructures that is conditioned on the newly generated junction-tree $\mathcal{T}_{\mathrm{new}}$ based on $p(\mathcal{G}|\bm{z}_\mathcal{G},\mathcal{T}_{\mathrm{new}})$. Box V provides the readers with more detailed explanation on the implementation details of JTVAE.

\begin{myquote}

\textbf{Box V: architectural details of the junction-tree variational autoencoder for molecular generation}

\footnotesize

\textbf{(i) Molecular graph encoder through atom-level message-passing}

\hspace{10pt} At the top level, a standard MPGNN is implemented that map a molecular graph $\mathcal{G}=\{\bm{x}_{u},\bm{x}_{uv}\}$ into its latent space representation $\bm{h}_{\mathcal{G}}$, in which $\bm{x}_{u}$ and $\bm{x}_{uv}$ are the nodes (atoms) and edges (bonds) that constitute the molecular graph. More specifically, after $t$-th MP iterations, the message that is being passed from the atomic node $u$ to $v$ (blue box in \cref{fig:junction-tree-net}) is constructed with the following NN-based MP operation:
\begin{equation*}
    \boldsymbol{\nu}_{uv}^{(t)}=\tau\left( w_1^g\bm{x}_u+w_2^g \bm{x}_{uv} +w_3^g \sum_{w\in\mathcal{N}(u)\backslash v}\boldsymbol{\nu}_{wu}^{(t-1)}\right),
\end{equation*}
\begin{wrapfigure}[34]{rt}{0.5\textwidth}
  \begin{center}
    \includegraphics[width=0.3\textwidth]{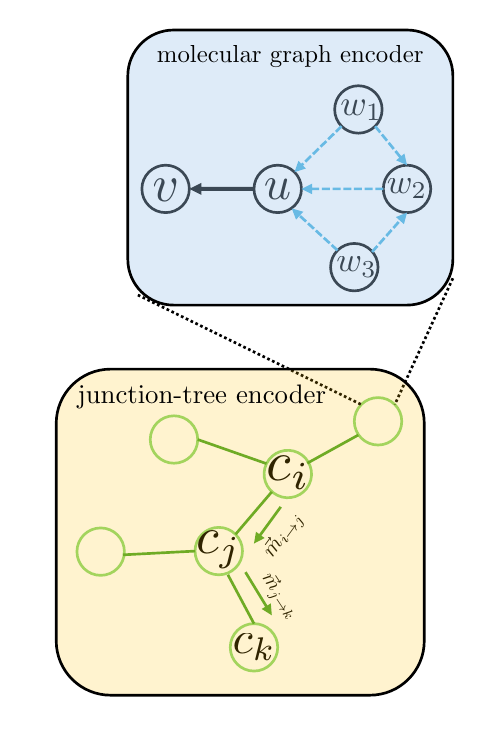}
    \caption{Schematic illustrations on the hierarchical encoders in the junction-tree variational autoencoder\cite{jin2018junction} for molecular generation, in which a molecular graph encoder (blue box) is first deployed to encode (neighbouring) atoms into clusters, and a junction-tree encoder is then deployed to exchange information amongst clusters to build up the entire structure of the molecule.}
    \label{fig:junction-tree-net}
  \end{center}
\end{wrapfigure}
which is initialised with $\boldsymbol{\nu}_{wu}^{(0)}=0$. Essentially the last term in this operation progressively aggregates the information from all the neighbouring nodes $w$ (that are not $v$)  around $u$ to the central node $u$, before it is used to update the information encoding the bond that connects $u$ and $v$ (blue dashed arrows in the blue box of \cref{fig:junction-tree-net}). After $T$ MP operations, the message on a given node $u$ from all of its incoming bonds are aggregated into the final \emph{per-atom} representation:
\begin{equation*}
    \bm{h}_u=\tau\left( u_1^g\bm{x}_u + u_2^g\sum_{v\in\mathcal{N}(u)} \boldsymbol{\nu}^{(T)}_{vu}\right),
\end{equation*}
from which the global latent representation for the entire molecular graph $\mathcal{G}$ is averaged over all per-atom representations:
\begin{equation*}
    \bm{z}_{\mathcal{G}}=\frac{1}{\Vert V\Vert}\sum_{u\in V}\bm{h}_{u}.
\end{equation*}
This is similar to the construction of an averaged structural similarity kernel from atom-centred structural descriptors,\cite{de2016comparing} which was described in \cref{sec:physics_informed_representation}. The advantage of constructing the global latent representation of molecular structures is that in ensure molecules with different number of atoms are mapped into a latent representations with fixed dimensionality. However, it has essentially discarded the ordering information behind atomic connectivities, which makes it difficult to regenerate molecules based on $\bm{Z}_{\mathcal{G}}$ alone. It should be also easy to envisage that the way messages $\boldsymbol{\nu}^{(T)}_{vu}$ are aggregated does not depends on the topology of the molecular graph, so it should automatically be able to handle atoms that are interconnected in closed loop structures after some number of steps of MP.

\textbf{(ii) Junction-tree encoder through cluster-level message-passing}

\hspace{10pt} Here, the model treats the molecule as  clusters of atoms forming a tree-like graph structure without loop, which initiate itself from a single root that is sequentially branched into various leaves. (Whether all molecules confine with this directional tree-like structure, however, is questionable.) In this sense, the MP architecture should respect this directional tree-like structure.  

\hspace{10pt} To implement the junction-tree MP, we start from two interconnected clusters $\mathcal{C}_{i}$ and $\mathcal{C}_j$ in the junction tree $\mathcal{T}$ (yellow box of \cref{fig:junction-tree-net}), and we construct a \emph{directed message} $\bm{m}_{i\to j}$ that is computed from the label of cluster $\mathcal{C}_i$ and all messages entering $i$ from all of its neighbouring clusters other than $\mathcal{C}_j$:
\begin{equation*}
    \bm{m}_{i\to j}=\tau_{\mathrm{tree}}\left(\bm{x}_{i},\Big\{ \bm{m}_{k\to i} \Big\}_{k\in\mathcal{N}(i)\backslash j} \right).
\end{equation*}

\hspace{10pt} Here, the directional information is preserved by not explicitly aggregating the messages from all messages coming from nodes $\mathcal{C}_{k}$ into node $\mathcal{C}_j$ \textit{before} it is passed into $\tau_{\mathrm{tree}}$. More explicitly,   $\tau_{\mathrm{tree}}$ implements the following operations of MP:
\begin{enumerate}
    \item $\bm{s}_{ij}=\sum_{k\in\mathcal{N}(i)\backslash j} \bm{m}_{k\to i}$, which aggregate all messages coming into $\mathcal{C}_i$ from all neighbouring clusters $\mathcal{C}_k$ except $\mathcal{C}_j$.
    \item $\bm{z}_{ij}=\sigma\left(w^{z} \bm{x}_i+u^z \bm{s}_{ij}+\bm{b}^{z}\right)$, which is a gate operation that update the bonding information between  $\mathcal{C}_i$ and $\mathcal{C}_j$ based on the information on cluster $\mathcal{C}_i$  and $\bm{s}_{ij}$, the incoming information on $\mathcal{C}_i$. This is a standard message construction operation, which essentially gathers the information on how much  previously pooled bonding information affects the next one.
    \item $\bm{r}_{ki}=\sigma\left(w^r \bm{x}_i + u^r \bm{m}_{k\to i} +\bm{b}^r \right)$ for all $k\in\mathcal{N}(i)\backslash j$, which is a separate gate that update the information coming from each connected branch $k\to i$.
    \item $\bm{\tilde{m}}_{ij}=\tanh\left(w\bm{x}_i+u\displaystyle{\sum_{k\in\mathcal{N}(i)\backslash j}} \bm{r}_{ki}\odot  \bm{m}_{k\to i}\right)$, which gives a candidate message in which the weighting factor $\bm{r}_{ki}$ informs how much a particular branch $k\to i$ should influence the new message, and finally,
    \item $\bm{m}_{i\to j}=(\mathds{1}-\bm{z}_{ij})\odot \bm{s}_{ij}+\bm{z}_{ij}\odot\bm{\tilde{m}}_{ij}$.
\end{enumerate}
Similar to the all-atom MP operations outlined in (i), the junction-tree MP leas to the following per-cluster latent representation by gathering all the incoming messages to a specific cluster $\mathcal{C}_i$ as:
\begin{equation*}
    \bm{h}_i=\sigma \left(w^{o}\bm{x}_i+ \sum_{k\in\mathcal{N}(i)} u^{o} \bm{m}_{k\to i} \right),
\end{equation*}
and one simply takes the learned representation of the root node $z_{\mathcal{T}}=\bm{h}_0$ as the latent representation of the tree. More specifically, the junction-tree encoder implemented the MP operations along two directions. For MP from the root to the leaves, it will allow the leaves to receive instruction on whether to further branch out, whereas the information passed from the leaves to the tree enables the root to keep track on the global structures of the newly generated molecules. These are more specifically discussed as follows.

\textbf{(iii) Tree decoder that grows tree through message-passing}

\hspace{10pt} Taking the analogy from the MP operations in the tree encoding step, in the generative process, the model should decide whether one should generate a new node $\mathcal{C}_j$ from the current one $\mathcal{C}_i$ based on: (a) the features on $\mathcal{C}_i$, and (b) all other nodes $\mathcal{C}_k$ that had been traversed that lead to $\mathcal{C}_i$. Casting this mathematically, it reads
\begin{equation*}
    \bm{h}_{i\to j}=\tau_{\mathrm{tree}}\left(\bm{x}_i,\Big\{ \bm{h}_{k\to i}\Big\}\mbox{ for }(k,i)\mbox{ that has been traversed with } k\neq j\right).
\end{equation*}
 Practically, this message constructed is used by the decoder, upon reaching the tree node $\mathcal{C}_i$, to make a topological decision on whether to make a new node, or to backtrack:
\begin{equation*}
    p_j =\sigma\left( u^d\cdot \tau\left( w_1^d \bm{x}_i+w_2^d \bm{z}_{\mathcal{T}} +w_3^d \sum_k \bm{h}_{k\to i} \right) \right),
\end{equation*}
in which $\bm{z}_T$ is a new latent variable that is sampled in the sub-latent space for the junction-tree representations of molecule, this ensures that the generation process follows the conditional distribution $p(\mathcal{T}|\bm{z}_\mathcal{T})$ for the decoder. If a new leaf node $\mathcal{C}_j$ is created, then we can predict the actual cluster label for $\mathcal{C}_j$, which tells us what functional group should go into this particular cluster. Again, this will be conditioned upon $\bm{z}_T$  as:
\begin{equation*}
    q_{j}=\mbox{Softmax}\left(u^{l}\tau\left(w_1^l \bm{z}_{\mathcal{T}}+w_2^l \bm{h}_{i\to j} \right)\right).
\end{equation*}

\textbf{(iv) Graph decoder using tree-conditioned message-passing to assemble clusters}

\hspace{10pt} At this stage, all necessary components to build a new molecules would have been generated, and one is left with the final problem of identifying which pairs of atoms should be connected in order to join all the clusters together to form the final atomistic structure of the molecule. This is the reason why we need another full-atom MP scheme to update the bonding information. Following the notations used in the molecular graph encoder, we need a MP operation to update the bond connectivity between a candidate pair of atoms $(u,v)$, each coming from a different cluster that is connected between each other, mathematically, this reads
\begin{equation*}
    \boldsymbol{\mu}_{u\to v}^{(t)}= \tau\left(w_1^{\alpha} \bm{x}_u+ w_2^\alpha \bm{x}_{uv}+ w_3^\alpha \tilde{\boldsymbol{\mu}}_{u\to v}^{(t-1)}\right),
\end{equation*}
in which 
\begin{equation*}
    \tilde{\boldsymbol{\mu}}_{u\to v}^{(t-1)}=\sum_{w\in\mathcal{N}(u)\backslash v}\boldsymbol{\mu}_{u\to v}^{(t-1)}+\mathds{1}[\alpha_u\neq\alpha_v]\hat{\bm{m}}_{\alpha_u\to\alpha_v}.
\end{equation*}
Here, $\alpha_u$ bears the information on which cluster atom $u$ comes from, such that $\mathds{1}[\alpha_u\neq\alpha_v]\hat{\bm{m}}_{\alpha_u\to\alpha_v}$ can be considered as a penalty term that enforcing two interconnecting atoms must each comes from a different cluster. One may see that a potential bottleneck in this approach is that one must compute the message for all possible pairs of $(\alpha_u,\alpha_v)$ to optimise the connectivity of the fragments. 

\hspace{10 pt} Finally, one needs to assess the quality of the structure that is being generated by the model, which the authors proposed the score $f(\bm{G}_i)=\bm{h}_{G_i}^\top \bm{z}_{G}$ that measures the 
the alignment between the latent space representation $\bm{z}_{G}$ that is being sampled and the latent space representation (mapped by the all-atom encoder) for the molecules generated from the decoders. 
\end{myquote}

As one can see in the above discussions, a key aspect behind the designs of JTVAE for molecular structure generation is to progressively build the molecular framework (the tree) in a cluster-by-cluster manner. This progressive approach is similarly applied in the VAE model developed by  \citet{liu2018constrained} for molecular generation, in which, instead of trying to map a single $\bm{z}$ into a full atomistic molecular structure with the decoder, the model was designed to generate molecule atom-by-atom. At each generative step, the model first selects an atom of `current focus', and then decide whether one should add a new bond and atom, or the generative step should stop (through encountering a STOP signal). Practically, the generative model is designed to output a score $s_{vu}$ that can be used for determining the likelihood to generate a new bond between nodes $u$ and $v$, which is normalised amongst all possible nodes $\{u\}$ that might from a bond with $v$ as $p_{vu}=M_{vu}\exp(s_{vu})/\sum_w M_{vw}\exp(s_{vw})$, in which $M_{vw}$ allows to further implement a hard constraint for bond formation by defining $M_{vu}=1$ if a connection between $u$ and $v$ is allowed or zero otherwise. 

One of the key advantages to generate structures is the enabling of more fine-grained control on molecular structure generations, which leads to higher chance of generating valid structures. However, it is at the sacrifice on increasing model complexity, as can be seen from the discussions above. Nevertheless, one-shot approach to predict the global structures of molecules does exist, For example, in the same year as the VAE model proposed by \citet{liu2018constrained}, \citet{simonovsky2018graphvae} published the graph-VAE model for small molecular graph generation, which treats the entire molecular structure as a single graph $\mathcal{G}$, such that the VAE was designed to learn both the forward and backward conditional probability distributions between a graph and its corresponding latent space representations $p_{\boldsymbol{\theta}}(\mathcal{G}|\bm{z})$ and $q_{\boldsymbol{\varphi}}(\bm{z}|\mathcal{G})$. On a more practical level, the problem of structure generation become further simplified into the problems of predicting the adjacent matrix $\tilde{\bm{A}}\in[0,1]^{k\times k}$ for a molecule containing $k$ atoms, which defines the atomic connectivities. 

As a last, and more recent example of (progressive) molecule generations with the VAE framework, we examine the NEVAE model put forward by \citet{samanta2020nevae} Similar to the JTVAE model, rather than mapping an entire molecule $\{\bm{C},\bm{R}\}$ (chemical identity and atomic position vectors) into a single laten variable $\bm{z}$, NEVAE tries to map each atom in a molecule into a separate latent space variable as $\bm{z}=\{\bm{z}_1,\bm{z}_2,\cdots, \bm{z}_n\}$. The encoder then takes the inspirations from MPGNN that aggregates information up to $k$-th connected atoms in distant into the message $\bm{h}_i$. More specifically, the model implements this aggregation operation to be a symmetric one, such that the aggregated information does not depend on the order of atomic connectivities. This can be achieved with, for example, a summation operation as
\begin{equation*}
    \begin{cases}
        \phi(\bm{h}_i) \to\bm{z}_i\\
        \bm{h}_{i}^{(k)}=f_k\left( \bm{h}_i^{(k-1)},\mbox{AGG}\left\{  g_k\left(\bm{h}_j^{(k-1)},\bm{e}_{ij}\right): j\in \mathcal{N}(i)\right\} \right).
    \end{cases}
\end{equation*}
Similarly, the decoder in the VAE generates molecular structures in stages. Here, starting from the number of atoms $n\sim\mbox{Poisson}(\lambda_n)$, the model first sample the per-atom latent variables $\bm{z}_i\sim\mathcal{N}(0,\bm{I})$, from which the corresponding atom type is determined based on a learned conditional probability distribution of $p(x_i=C|\bm{z}_i)$. Subsequently, the latent variables $\bm{z}_i$ are used for sampling the number of bonds from $m\sim\mbox{Poisson}(\lambda_{E}(\bm{z}))$ such that the model knows how many edge-generation decisions it needs to make. From here, the model then builds an edge between any two nodes $i$ and $j$ based on a computed probability for edge formation that is similar to \citet{liu2018constrained} The model then determines the bond type for each connection and finishes by sampling the 3D coordinates for each atom. 

\paragraph{\textbf{Generation of Crystalline Solids}}
Moving on to the designs and implementations of VAE-based generative models for crystalline materials. The model put forward by \citet{noh2019inverse} was one of the earliest proof-of-concept generative models for  inorganic crystals, which had already demonstrated performances that were comparable to traditional genetic algorithms. As mentioned by the authors, the key advantage in VAE-based crystal structure prediction compared to genetic algorithm is that the former is physically grounded by the probability distributions of the \emph{thermodynamically stable} crystal structures, which should naturally enhance the successful rates in generating feasible inorganic crystal structure. Being a model that was developed pre-dating the uprise of equivariant NNs, the model drew analogy of structure generation to that of image generations by mapping the positions of atoms inside a cubic box into an atom-centred Gaussian distributions inside a pixelated 3D grid, which is learnt by the model and also acts as the target generative outcome for the backward decoder.

\citet{xie2021crystal} proposed the crystal diffusion variational autoencoder (CDVAE) that combined the frameworks of VAE and diffusion models for generating periodic crystal structures. Here, the encoder of the VAE tries to learn a parameterised conditional distribution $q_{\boldsymbol{\phi}}(\bm{z}|\bm{M})$ of the latent representation $\bm{z}$ for a given crystal structure $\bm{M}$ that is collectively represented by $(\bm{A},\bm{X},\bm{L})$, which denotes the atomic number, atomic coordinates, and the lattice parameters of the crystal, respectively. To generate a new sample of the crystal structure, one first samples a new variable $\bm{z}^\prime$ in the latent space, which is then put into the decoder to generate a \emph{coarse} initial structure of $\tilde{\bm{M}}_0=(\tilde{\bm{A}}_0,\tilde{\bm{X}}_0,\hat{\bm{L}})$. Then the mode uses the score function to progressively refine the chemical identity and the atomic positions with a diffusion model: $(\tilde{\bm{A}}_0,\tilde{\bm{X}}_0)\to\cdots\to(\tilde{\bm{A}}_\mathrm{final},\tilde{\bm{X}}_\mathrm{final})$. Note that, if we do not wish to refine the chemical identities, but only the positions of the atoms in the crystal that is being generated, the score function essentially resembles an atom-atom forcefield (as discussed in \cref{sec:common_generative_models}), meaning that the refining step is equivalent to the use of an energy function to optimise the newly generated structure to a nearby local energy minimum. In this sense, this particular VAE framework can be conceptually understood as a simulated annealing process that first samples a structure on a high-temperature PES to overcome the problem of crossing high energy barriers in traditional MD/MC-based approaches [\cref{sect:MonteCarlo} and \cref{sect:molecular_dynamics}], followed by quenching the structure into a stable one. To provide an example of property-guided structure generation, \citet{luo2024deep} extended the CDVAE\cite{xie2021crystal} model for generating high-pressure crystal structures, which is achieved by further augmenting the vectorial representation of crystal structure with an additional term that encodes the information of pressure $P$. The later was realised by mapping $P$ onto a basis constructed using equidistant Gaussian functions.

The Wyckoff positions for atoms  are unique symmetry-embedded encoding of atomic positions in a crystal structure, for this reason, the WyCryst model proposed by \citet{zhu2024wycryst} explored the usage of the  Wyckoff positions for representing and generating candidate inorganic crystal structure. Different from the CDVAE model, WyCryst seeks to represent a given crystal structure as a tuple of $\bm{W}_i=(\bm{X}_i,\bm{S}_i,\bm{F}_i)$, which correspond to the Wyckoff positions, one-hot-encoded space group and chemical element representations, respectively. This allows the VAE model developed to learn the statistical distribution (as a function of $\bm{W}$) for all inorganic crystals in the training data. One could identify multiple advantages in using the Wyckoff representations for the tasks of inorganic crystal structures. For examples, the Wyckoff positions naturally define a constrained space $\bm{X}_i\in[0,1)^3$ that is beneficial for enhancing the learning and inference stabilities of the model. Utilising the Wyckoff positions also reduces the dimensionality of the problems to be tackled, as one only needs to generate atoms in a set of unique symmetry-inequivalent Wyckoff sites, and complete the whole crystal by using the symmetry operations (which is also called the \emph{generators} that are embedded in $\bm{S}_i$) to generate the other symmetry-related Wyckoff positions. This also overcomes the problem of permutation invariance in structure generations, when only a single site amongst a set of symmetry-related Wyckoff positions is needed for model learning and inference. The complexity of the problem can be further reduced  by realising the fact that many special Wyckoff positions are fixed (such as $(0,0,0)$ and $(\frac{1}{2},\frac{1}{2},\frac{1}{2})$), or contain fewer independent parameters (such as $(x,\frac{1}{4},\frac{1}{4})$ with $x\in[0,1)$). A possible extension of Wyckoff-position-based generative models for inorganic crystal could be through the explorations of packing different `packing-motifs' (such as octahedra and tetrahedra), that are represented by a group of related Wyckoff positions, to develop crystal generative models with greater physical and structural interpretability. 

Finally, to highlight the cross-disciplinary nature in the development of AI for scientific discovery, we discuss the model developed by \citet{chen2025transformer} which utilised the transformer architecture\cite{vaswani2017attention} from the large language model for the task of crystal structure prediction. Here, instead of using GNN, the transformer is applied to encode the latent representation $\bm{z}$ for a given crystal structure. This is inspired by the fact that, the distributions of the neighbouring coordination environments in an inorganic crystal structure are highly correlated thus non-uniform, in which certain pairs of neighbouring atoms (\emph{e.g.} a cation surrounded by anions) are more likely to occur compared to the other pairs, which is highly similar to the appearance of word sequences in sentences. Here, each local environment around a given centre is first encoded into a set of SO(3) equivariant features up to a prescribed order $n$ as $\bm{h}_i=[\bm{h}_i^{(0)},\bm{h}_i^{(1)},\cdots,\bm{h}_i^{(n)}]$. Then the transformer learns the attention coefficient $\alpha_{ij}\propto \bm{h}_i\cdot\bm{h}_j$ amongst all pairs of chemical environments, which provides the information on the likelihood that a pair of atomic environments may occur as neighbour to each other. Most importantly, the structural equivariance (and hence the equivariance in the attention coefficient thus computed) can be strictly imposed if the attention coefficients are computed by following the group representation theory (see \cref{sec:physics_informed_representation}). Furthermore, the model developed a dataset-adaptive encoding scheme for better encoding the interatomic distances $r_{ij}$  into their respective latent space representation, for example, the authors had chosen to map $r_{ij}$ for the \texttt{MP-20} training set into equidistant Gaussians, whereas for the \texttt{Perov-5} set (which are structurally more uniform), $r_{ij}$ is mapped into a set of Bessel functions, which are itself oscillatory in nature that better reflects the periodicity of the crystal structures. 

\subsubsection{Applications of Diffusion Generative Models in Atomistic Modelling}
\label{sec:generative_diffusion_examples}

Since around 2020, diffusion generative model\cite{ho2020denoising} has become one of the dominant and rapidly-developing deep generative model, particularly for its demonstrated capability to generate images with higher qualities compared to the GAN (Generative Adversarial Network) model.\cite{dhariwal2021diffusion} In the field of atomistic molecule and material generations, the development of diffusion model has been a vibrant multidisciplinary field in model development, requiring the practitioners in this particular area to possess a wide range of knowledges to further advance the field. These include: (a) Probability theory, particularly in terms of the Bayesian statistics [\cref{sec:prob_theory_fundamentals}], which plays an important role in determining the appropriate probability distributions that the system should follow, such that the forward and reverse diffusion processes are implemented in a way that these distributions are respected. (b) Choosing the appropriate way to represent the atomistic structures as the learning and generative targets. (c) Fundamentals of group and representation theories [\cref{sec:physics_informed_representation}] that enables the NN models to learn the noises while respecting the invariant and equivariant symmetries of the systems to be modelled. (d) Stochastic processes, that is grounded by the SDE formulation of the diffusion model [\cref{eq:diffusion_forward_SDE} and \cref{eq:reverse_generation_SDE}] which enables one to explore \cite{song2020score,song2020denoising} the mathematical techniques in solving differential equations to accelerate the sampling of new materials, as well as designing a diverse range of data-dependent forward collapsing process of the input data.\cite{cornet2024equivariant} (e) Differential geometry, which enables one to design the coordinate-specific transformation of atomistic structures on the high-dimensional manifold, and derive the corresponding score functions for model training. Finally, (f) Statistical mechanics, which provides the conceptual bridge to derive the physical interpretation of the SDE with the conventional energy-based PES samplings [\cref{chapter:sampling_methods}].

In this regard, the development of diffusion generative models is largely the process of exploring and incorporating different mathematical techniques that have been developed for the subfields listed above, to improve the model efficiency, stability and capabilities to extrapolate beyond the known materials space. These requirements may be much beyond the knowledge base for general materials modellers, which poses a significant challenges for them to develop deep understanding on the key features in the models that have been developed and why they may be applicable to specific categories of targeted materials systems. Hence, the purpose of this section is aiming at providing the community with an overview on how the above listed aspects are intertwined together in the designs and implementations of diffusive generative models for molecules and crystalline materials through a chronological review of the diffusion generative models proposed over the recent years. In particular, we aim to highlight how the model design are strongly influenced by the design choices on how the structure and chemistry are represented as the training and generative targets for the diffusive generative models.

\paragraph{\textbf{Generation of Molecules}}
 Following the same sequence as the discussions around VAE models, we begin here by reviewing different implementations of the diffusion generative models that have been developed for the task of molecular structure generations. \citet{niu2020permutation} presented one of the earliest diffusion models for generating graph objects, that is applicable for molecules when being treated as graphs, in which the molecular generation task is being cast here as one that samples different adjacency matrices $\bm{A}$ for representing the molecular graphs with the diffusion generative model. In this regard, a key considerate when constructing the NN model to approximate the score function $s_{\boldsymbol{\theta}}(\bm{A})=\nabla_{\bm{A}}\log p_{\boldsymbol{\theta}}(\bm{A})$ is the \emph{permutation invariance of graph} (also \emph{c.f.} \citet{simonovsky2018graphvae}), meaning that with $\Pi$ being the set of row/column permutation operations on $\bm{A}$, then: (a) $\forall \pi_i,\pi_j\in\Pi$ and $\pi_i\neq\pi_j$, $s_{\boldsymbol{\theta}}(\pi_1(\bm{A}))=s_{\boldsymbol{\theta}}(\pi_2(\bm{A}))$, which can be achieved with NN operations that only operates element- or edge-wise, or not sensitive to the atomic ordering, such as the aggregation operations, as well as explicitly constructing the score function to be symmetric thus permutation invariant \emph{e.g.} $s=\left(s+s^\top\right)/2$. Furthermore, the author pointed out an important geometric interpretation of the score function $s(\pi(\bm{A}))$, which is the gradient along a specific transformation pathway from a noised starting point $\bm{A}=\bm{0}$ to $\pi(\bm{A})$,  hence the final log-probability distributions over graphs $\log p(\bm{A})$ can be obtained by integrating $\langle s(\bm{X}),d\bm{X}\rangle_F$ (with $\langle\cdot\rangle_F$ being the  Frobenius inner product)  over all possible transformation pathways in the diffusion process. Conceptually, this resembles the construct of \emph{averaged} structural similarity kernel\cite{de2016comparing} in atomistic machine learning (\cref{sec:feature_eng}) that removes the problem of permutational matching of atom pairs. In this case, with the NN-modelled score function $s_{\boldsymbol{\theta}}(\bm{A})$ being permutational invariant will also ensure $p(\bm{A})$ is also permutational invariant.  (b) One must also ensure that in the reverse diffusion process that samples the adjacent matrix does not repetitively samples adjacent matrices that are the permuted copies of each other. For this, the author proposed that with learnt $s_{\boldsymbol{\theta}}(\bm{A})$, the sampling process can start with $\tilde{\bm{A}}^{(T)}$ with $\tilde{\bm{A}}_{ij}^{(T)}\sim\mathcal{N}(0,1)$ and $\tilde{\bm{A}}_{ij}^{(T)}=\tilde{\bm{A}}_{ji}$, followed by repretitive sampling using Langevin dynamic as $\tilde{\bm{A}}_{ij}^{(t-1)}=\tilde{\bm{A}}_{ij}^{(t)}+\frac{\alpha_t}{2}s_{\boldsymbol{\theta}}(\tilde{\bm{A}}^{(t)},\sigma_t)+\sqrt{\alpha_t}\bm{Z}_t$, whereby $\bm{Z}_t$ is a symmetric Gaussian noise that is added to prevent the sampling from converging prematurely to regions with high probability. Readers may already appreciate from this early work that, even though the underlying object being transformed (the adjacency matrix $\bm{A}$ with $\bm{A}_{ij}\in[0,1]$) may be quite simple, the  details to be considered in designing mathematical transformations of this object through diffusion model while ensuring permutational invariance is already very involved.

To progress one step further from a binary encoding of atomic connectivities, \citet{shi2021learning} addressed the problem of generating the 3D coordinates $\bm{R}$ for molecules from a given molecular graph $\mathcal{G}$ via the interatomic distances, in which case the model needs to learn a score function with respect to $\bm{R}$ that is conditioned on $\mathcal{G}$, \emph{i.e.} $s_{\boldsymbol{\theta}}(\bm{R}|\mathcal{G})=\nabla_{\bm{R}}p(\bm{R}|\mathcal{G})$. In particular, this model utilise the chain rule to reach an explicit expression of the score function with respect to the atomic coordinates and interatomic distances as $s_{\boldsymbol{\theta}}(\bm{R}_i)=\sum_{j\in\mathcal{N}(i)}s_{\boldsymbol{\theta}}(d_{ij})(\bm{R}_i-\bm{R}_j)/d_{ij}$, which is a quantity that is rotational and translational invariant up to 2-body, and can be learnt with GNN models that implement invariant operations at this body-order (\cref{tab:equivariant_GNNS}). With this learnt score function, the model can then achieve the sampling of atomic positions, starting from a random geometry, and progressively denoised following the Langevin dynamics.

\citet{jo2022score} further consider the molecular representation as a full graph $\mathcal{G}=(\bm{X},\bm{A})$ in which $\bm{X}\in\mathbb{R}^{N\times F}$ is the node feature vector for $N$ nodes each with a dimension of $F$, and $\bm{A}\in \mathbb{R}^{N\times N}$ is the \emph{weighted} adjacency matrix. The main uniqueness behind this work, is that, rather than simply just adding the Gaussian noise to collapse the samples, it evolves $\bm{X}$ and $\bm{A}$ separately in the forward process by designing the according SDEs [\emph{c.f.} \cref{eq:diffusion_forward_SDE}]:
\begin{equation}
    \label{eq:JLH_forward_SDE}
    \begin{cases}
        \mathrm{d}\bm{X}_t=f_{1,t}(\bm{X}_t)dt+g_{1,t}\mathrm{d}\bm{w}_{1,t}\\
        \mathrm{d}\bm{A}_t=f_{2,t}(\bm{A}_t)dt+g_{2,t}\mathrm{d}\bm{w}_{2,t},
    \end{cases}
\end{equation}
in  which $\mathrm{d}\bm{w}\in\mathcal{N}(0,\bm{I})$. Hence this generative model extended the previous two by allowing both the chemistry and the structure of the molecule to be changed in the diffusion process. The main advantage of using SDEs for the forward process is that it can better ensure the finally collapsed samples are well centred around a predefined mean (typically zero), such that the original sample can be gradually collapsed by following a noise with well defined variance to ease the subsequent samplings. This may be achieved, for example, by designing the appropriate drift functions $f$ for different parts of molecular representations. Following \cref{eq:JLH_forward_SDE}, the model needs to train two score functions 
\begin{equation*}
    \begin{cases}
        s_{\boldsymbol{\theta},t}(\bm{X}_t,\bm{A}_t) = \nabla_{\bm{X}_t}\log p_t(\bm{X}_t,\bm{A}_t)\\
         s_{\boldsymbol{\varphi},t}(\bm{X}_t,\bm{A}_t) = \nabla_{\bm{A}_t}\log p_t(\bm{X}_t,\bm{A}_t)
    \end{cases}
\end{equation*}
that are to be used in the generating process with the reverse SDE [\emph{c.f.} \cref{eq:reverse_generation_SDE}]:
\begin{equation*}
    \begin{cases}
        \mathrm{d}\bm{X}_t = \left[f_{1,t}(\bm{X}_t)-g_{1,t}^2s_{\boldsymbol{\theta},t}(\bm{X}_t,\bm{A}_t)\right]\mathrm{d}t +g_{1,t}\mathrm{d}\bm{w}_{1,t}\\
        \mathrm{d}\bm{A}_t = \left[f_{2,t}(\bm{A}_t)-g_{2,t}^2s_{\boldsymbol{\varphi},t}(\bm{X}_t,\bm{A}_t)\right]\mathrm{d}t +g_{2,t}\mathrm{d}\bm{w}_{1,t}.
    \end{cases}
\end{equation*}
Since both score functions are modelled as the joint probability distribution that is dependent on the adjacency matrix $\bm{A}$, one can appreciate that the above discussion on implementing NN approximator that preserves the \emph{permutation invariance} is also a necessary model design constraint here.

\citet{wu2022diffusion} also harnessed the flexibility of SDE in the design of diffusion generative model, which they considered the drift-term $f(\bm{x})\sim p_t(\bm{x}_t|\bm{x}_0)$ as determined by a time-dependent probability distribution that is conditioned upon the training samples $\bm{x}_0$. This enables them to design the drift term by including an additional \emph{informative bias force} that is force-field based to evolve samples into physically and statistically meaningful distribution, $\emph{i.e.}$
\begin{equation*}
    p_t(\bm{x}_t|\bm{x}_0)=\underbrace{\sigma_t^2\frac{\bm{x}_0-\bm{x}_t}{\beta_t-\beta_0}}_{\mbox{\scriptsize{Brownian term}}}+\underbrace{\sigma_t\left(-\nabla_{\bm{x}_t}E(\bm{x}_t)\right)}_{\mbox{\scriptsize{Informative bias force}}}.
\end{equation*}
As demonstrated in the paper, a possible choice of $E(\bm{x})$ is the traditional atom-atom forcefield [\emph{c.f} \cref{sect:empirical_forcefield}], but this does not exclude the use of modern NN forcefields [\cref{tab:equivariant_GNNS}] in which \emph{per-atom} gradients are available. The model training process then adopt the score-matching process, in which the conditional distribution $p_t(\bm{x}_t|\bm{x}_0)$ becomes the learning target for the score function. 

\citet{xu2022geodiff} designed the \textsc{GeoDiff} model for molecular conformation generation. Conceptually, the authors treat both the forward and backward diffusion processes as the Markov chains with the corresponding transition probabilities $q(\bm{x}_{t}|\bm{x}_{t-1})$ and $p(\bm{x}_{t-1}|\bm{x}_t)$ being treated as the Markov transition kernels. Based on this formulation, the model focuses on designing transition kernels that are SE(3)-equivariant, meaning that  $\forall\hat{\mathcal{T}}_g\in$SE(3), it follows $p(\bm{x}_{t-1}|\bm{x}_t)=p\left(\hat{\mathcal{T}}_g\left(\bm{x}_{t-1}\right)|\hat{\mathcal{T}}_g\left(\bm{x}_t\right)\right)$. This is achieved by designing the kernel as 
\begin{equation*}
    \hat{\rho}\left(\frac{\bm{x}_{t-1}-\mu_{\boldsymbol{\theta}}(\mathcal{G},\bm{x}_t,t)}{\sigma_t}\right),
\end{equation*}
which is naturally conditioned on a given molecular graph $\mathcal{G}$, with the molecular conformations $\bm{x}_{t}$ and $\bm{x}_{t-1}$ centred around their respective centres-of-masses to remove the translational degree-of-freedom, while the rotational equivariance is further imposed by using an equivariant Graph Field Network\cite{satorras2021n} to parameterise the mean as 
\begin{equation*}
    \mu_{\boldsymbol{\theta}}(\mathcal{G},\bm{x}_t,t)=\frac{1}{\alpha_t}\left(\bm{x}_t -\frac{\beta_t}{\sqrt{1-\bar{\alpha}_t}}\varepsilon_{\boldsymbol{\theta}}(\mathcal{G},\bm{x}_t,t) \right).
\end{equation*}
Up to this stage, the model still resemble a standard diffusion generative model. The main innovation of this work lies in how the authors formulated the model training processes. In the first variant, the model derived the training loss following the minimisation of ELBO which essentially cast the training process into the minimisation of $\left\Vert \boldsymbol{\varepsilon}_t -\varepsilon_{\boldsymbol{\theta}}(\bm{x}_t,t) \right\Vert_2^2$, in which $\boldsymbol{\varepsilon}_t\sim \frac{1}{\sqrt{1-\bar{\alpha}_t}}\left(\bm{x}_t-\sqrt{\bar{\alpha}_t}\bm{x}_0\right)$ that are computed by taking the differences of two molecular conformations. The model solve this problem by first overlaying the centres-of-masses between the two conformations followed by the solution of the optimum rotational matrix using the Kabsch algorithm to align them before the differences are taken. This nonetheless introduces additional computational costs in each step of the reverse diffusion process, and may lead to the problem of the noise $\boldsymbol{\varepsilon}_t$ no longer following a white noise as the Kabsch algorithm may not always find the global minimum for the rotation matrix, leading to instabilities in the reverse generation process.  In the alternative variant to mitigate the alignment problem, the author followed the score-matching framework used by \citet{shi2021learning}, which allows the time-dependent noise to be computed based on the differences in pair-wise interatomic differences $\boldsymbol{\varepsilon}_t\sim\frac{1}{\sqrt{1-\bar{\alpha}_t}}[\partial_{\bm{x}_t}\bm{d}_t\cdot\left( \bm{d}_t-\sqrt{\bar{\alpha}_t}\bm{d}_0 \right)]$ that can be approximated with a two-body invariant GNN.
We note that this alignment problem is usually overcome in later model by learning and directly using $\boldsymbol{\varepsilon}_{\boldsymbol{\theta}}$ modelled by equivariant NN.

The model developed by \citet{hoogeboom2022equivariant} in 2022 is a landmark $E(3)$-equivariant model that jointly diffuse the atomic coordinates (continuous features) and atomic types (categorial features), building on the standard diffusion denoising probabilistic model (DDPM)\cite{ho2020denoising} and the Markov transition formalism by \citet{xu2022geodiff}. In particular, the model harness the two-body invariant EGNN model\cite{satorras2021n}, together with the inspirations of the score functions from \citet{shi2021learning}, to learn the noises that are dependent on both the coordinates $\bm{x}_i\in\mathbb{R}^3$ and nodal features $\bm{h}_i\in\mathbb{R}^d$ through the following the following message constructions:
\begin{equation}\label{eq:Hoogeboom_MP}
    \begin{cases}
        \bm{m}_{ij}=\phi_{e}\left(\bm{h}_i,\bm{h}_j,\bm{d}_{ij},a_{ij}\right),\\
        \bm{h}_{j}^{(\ell+1)}=\phi_{h}\left( \bm{h}_i^{(\ell)},\sum_{ij}\tilde{\bm{e}}_{ij}\bm{m}_{ij}\right),\\
        \displaystyle{\bm{x}_i^{(\ell+1)}=\bm{x}_i^{(t)}+\sum_{j\neq i}\frac{\bm{x}_i^{(\ell)}-\bm{x}_j^{(\ell)}}{d_{ij}+1}\phi_{x}\left(\bm{h}_i^{(\ell)},\bm{h}_j^{(\ell)},d^2_{ij},a_{ij}\right)}.
    \end{cases}
\end{equation}
In which $\ell$ is the index of the MP layer.
With this model, for each input structure $\bm{x}_t$, the model is able to directly learn the time-dependent noise $\boldsymbol{\varepsilon}_{\boldsymbol{\theta}}(\bm{x}_t,t)$ that is invariant with respect to  the ro-translations of $\bm{x}_t$. In addition, the intermediate diffusion steps jointly diffuse both the coordinates $\bm{x}_{t}$ and the feature $\bm{h}_{t}$ that samples both the molecular chemistry and structures together, for which the Markov transition probability\cite{xu2022geodiff} needs to be extended to reflect the joint probability distributions of the structure- and chemistry-dependent noises. By explicitly parameterising this joint distribution as $\mathcal{N}\left(\left .\bm{z}_t^{(\bm{x})}\right\vert \bar{\alpha}_t\bm{x}_0,\bar{\sigma}_t^2\bm{I}\right)\cdot \mathcal{N}\left(\left .\bm{z}_t^{(\bm{h})}\right\vert \bar{\alpha}_t\bm{h}_0,\bar{\sigma}_t^2\bm{I}\right)$, the model is trained to predict structure- and chemistry-dependent noises that can be concatenate together to assemble a total time-dependent noise $\boldsymbol{\varepsilon}_t=(\boldsymbol{\varepsilon}_t^{(\bm{x})},\boldsymbol{\varepsilon}_t^{(\bm{h})})^\top$. This allows one to sample molecular structure in the denoising process as $(\bm{x}_{t-1},\bm{h}_{t-1})^\top=\frac{1}{\alpha_t}[(\bm{x}_{t},\bm{h}_{t})^\top-\boldsymbol{\varepsilon}_t\sigma_t]$ with the initial state of $(\bm{x}_T,\bm{h}_T)$ drawn from their respective probability distributions that should be followed. Subsequently, \citet{igashov2024equivariant} extended this model for the conditional design of molecular linkers, whereas \citet{schneuing2024structure} further tailored the model for structure-based drug designs with equivariant diffusion.

Similarly, \citet{huang2023mdm}  adopted the SchNet\cite{schutt2017schnet} [\emph{c.f.} \cref{tab:equivariant_GNNS}] to enable the learning of equivariant score $s_{\boldsymbol{\theta}}(\bm{x}_t)$ that leads to a similar two-body form of the score function as \citet{shi2021learning} to achieve the generation of new molecular structure. An additional noise term is incorporated into the revserse sampling process to enhance the diversification of the samples being generated by the diffusion model

As discussed in \cref{sec:descriptor_motivation},  choosing the appropriate coordinates system plays an important role in determining the sampling efficiency. In this regard,  \citet{jing2022torsional}  considered the diffusion generative process for conformational flexible molecules in the space of torsional angle that operates on a hypertorus $\mathbb{T}^{m}$ of $m$ torsional angles. Correspondingly, the distribution of torsional angles $\boldsymbol{\tau}\in[0,2\pi)^m$ follow  that of a wrapped normal distribution that should be respected in the diffusion generative process (\emph{c.f.} \cref{tab:continuous_prob_distributions}), as they do not live in the usual Cartesian space of $\mathbb{R}^{m}$. In this case, the diffusion process on $\mathbb{T}^{m}$ can be efficiently implemented as $\boldsymbol{\tau}_{t+1}=\left(\alpha_t\boldsymbol{\tau}_t+\beta_t\boldsymbol{\varepsilon}_t\right)\mbox{mod 2}\pi$, with $\boldsymbol{\varepsilon}_t\sim\mathcal{N}(0,\bm{I})$, that properly ensures $\forall\boldsymbol{\tau}_t\in[0,2\pi)^m$.
However, the author also noted that the knowledge of a complete set of $\{\tau_i\}_{i=1}^{m}$  does not directly lead to a new molecular structure. To address this problem, the model adopted an extrinsic-to-intrinsic score model that first receives the current molecule as a 3D point cloud, then predict and assign a torsion angle for each rotatable bond. Physically, the as-proposed model can be considered equivalent to the establishment of a torsion-angle-dependent energy model for sampling the molecular conformation,\cite{cheng2024response} or a Boltzmann machine\cite{noe2019boltzmann} that draws samples of torsion angles from the distributions of which that are seen in the stable molecules.

\citet{vignac2022digress} developed the \textsc{DiGRESS} model, which can be considered as a diffusion-based editing model, that aims to generate graph objects $\mathcal{G}$ with edges and nodes that both follow discrete categorial distributions. This leads to the Markov transitions that  edits the graphs by successively adding or deleting nodes and edges, or mathematically expressed as $q(\bm{x}_{i,t}|\bm{x}_{i,t-1})=\mbox{Pr}_{x,t}(x=i)$ and $q(\bm{e}_{ij,t}|\bm{e}_{ij,t-1})=\mbox{Pr}_{E,t}(E=ij)$, which can be use to train a model $p_{\boldsymbol{\theta}}(\mathcal{G}_{t-1}|\mathcal{G}_t)$ through minimising a loss function based on cross-entropy $\mathcal{L}=\sum_i\mbox{CE}(\bm{x}_i,\hat{\bm{x}}_{i})+\sum_{ij}\mbox{CE}(\bm{e}_{ij},\hat{\bm{e}}_{ij})$. 
With the trained transition kernel, the reverse generation  process $\mathcal{G}_T\to\cdots\to\mathcal{G}_0$ can be proceeded by successively applying the graph editing operation. One can easily envisage that, similar to \citet{niu2020permutation}, enforcing the permutation invariance would be the key design constraint in designing the NN that learns the Markov transition kernels. Subsequently, the same group of authors\cite{vignac2023midi} further advanced the \textsc{DiGRESS} model that leads to the \textsc{MiDi} model for the task of molecular generation by simultaneously generating both the 2D connectivity and 3D structures, with the additional capabilities of generating the formal atomic charges. The model is designed based on representing the molecular graph as $\mathcal{G}=(\bm{X},\bm{C},\bm{R},\bm{Y})$, which correspond to the vectors representing the atom types, formal charges, and matrices representing the atomic coordinates and bond types, respectively. Hence, the probability distribution over the molecular graphs now follows a mixture of continuous and discrete distributions, which is reflected in the modified training loss $\mathcal{L}=\lambda_{\bm{R}}\Vert \hat{\bm{R}}-\bm{R}\Vert_2^2 +\lambda_{\bm{C}}\mbox{CE}(\hat{\bm{C}},\bm{C})+\lambda_{\bm{X}}\mbox{CE}(\hat{\bm{X}},\bm{X})+\lambda_{\bm{Y}}\mbox{CE}(\hat{\bm{Y}},\bm{Y})$, in which the part contributed from $\bm{R}$ follows that of the standard diffusion model on continuous variables whose distribution is parameterised as the normal distribution.\cite{ho2020denoising} Such design of the training loss naturally takes into account the interdependencies among the four variables that constitutes the full representation of $\mathcal{G}$, or equivalently, the 3D atomistic structures of the molecules. Implementation-wise, the model adapted a modified EGNN\cite{satorras2021n} for predicting the Markov transition probabilities, enforcing two-body equivariance in the model training and inference.
 
The MuDiff model proposed by \citet{hua2024mudiff} shares similar design concept as \textsc{MiDi} whereby the model also tries to generate both the 2D graph and the 3D molecular structures simultaneously, except they adopt a molecular representation in the form of $\mathcal{M}=(\bm{H},\bm{E},\bm{X})$, which correspond to the atomic features, edge features and the matrix of the Cartesian coordinates, respectively. Consequentially, this also leads to a loss function that is established based on a mixed continuous (for $\bm{H}$ and $\bm{X}$) and discrete categorial function (for $\bm{H}$), except now the continuous variable follows the standard denoising models in establishing the training loss, such that $\mathcal{L}(t)=\Vert \boldsymbol{\varepsilon}_{\bm{H},t}- \hat{\boldsymbol{\varepsilon}}_{\bm{H},t}\Vert_2^2 +\Vert \boldsymbol{\varepsilon}_{\bm{X},t}- \hat{\boldsymbol{\varepsilon}}_{\bm{X},t}\Vert_2^2+\mbox{CE}(\bm{E}_t,\hat{\bm{E}}_t)$, and the collapsing of the input sample follows the following dynamics:
\begin{equation*}
    \begin{cases}
        \bm{H}_t=\alpha_t \bm{H}_{t-1}+\sigma_t \boldsymbol{\varepsilon}_{\bm{H},t}\quad \boldsymbol{\varepsilon}_{\bm{H},t}\in\mathcal{N}_{\bm{H}}(0,\bm{I})\in\mathbb{R}^{n\times d}\qquad &\mbox{(atom feature noising)}\\
        \bm{X}_t=\alpha_t \bm{X}_{t-1}+\sigma_t \boldsymbol{\varepsilon}_{\bm{X},t}\quad \boldsymbol{\varepsilon}_{\bm{X},t}\in\mathcal{N}_{\bm{X}}(0,\bm{I})\in\mathbb{R}^{n\times 3}\qquad &\mbox{(coordinates noising)}\\
        \bm{E}_{t}=\bm{E}_{t-1}\bar{\bm{Q}}_t\quad \bar{\bm{Q}}_t=\alpha_t\bm{I}+(1-\alpha_t)\mathds{1}_b\mathds{1}_b^t/b\in\mathbb{R}^{b\times b}\qquad &\mbox{(edge noising)}.
    \end{cases}
\end{equation*}
The  sampling processes essentially reverse the above forward processes, by first sampling $\bm{H}$, $\bm{E}$ and $\bm{X}$ from their respective priors, then progressively denoise the atom features and coordinates with the learned $\boldsymbol{\varepsilon}_{\bm{H},\boldsymbol{\theta}}$, $\boldsymbol{\varepsilon}_{\bm{X},\boldsymbol{\theta}}$, and evolves the edges according to the transition kernel $p(\bm{E}_{t-1}|\bm{E}_t)$. In particular, the model for learning the noises is build based on the transformer\cite{vaswani2017attention} architecture, which works on both the 2D graphs $\mathcal{M}=(\bm{H},\bm{E})$ with an invariant GNN and 3D graphs $\mathcal{M}=(\bm{H},\bm{X})$ with an equivariant GNN. Here, the attention mechanism allows both the 2D graph (invariant features) and 3D structure (equivariant features) to interact with each other, enabling both to be updated during the learning and generating processes. 

Instead of directly diffusing the (graph representation of) molecular structures, \citet{xu2023geometric} applied a VAE to first map the molecular graph $\mathcal{G}=(\bm{H},\bm{X})$ into the latent space representation $(\bm{z}_{h},\bm{z}_x)^\top$ for the (invariant scalar) features encoding the chemical identities  and the (equivariant vectorial) features encoding the atomic coordinates.  Subsequent diffusion process will enable the sampling of new latent representations of $(\bm{z}_{h}^\prime,\bm{z}_x^\prime)^\top$ that will be mapped back to a new molecular structure with the decoders from VAE. Both the encoding and decoding are achieved by using the EGNN\cite{satorras2021n} model to ensure the invariance and equivariance are preserved in the mapping. The standard diffusion process is applied to jointly collapse $\bm{z}_{h}$ and $\bm{z}_x$ as $( \bm{z}_h^{(t+1)} , \bm{z}_x^{(t+1)})^\top=( \bm{z}_h^{(t)} , \bm{z}_x^{(t)})^\top+\sigma_t\boldsymbol{\varepsilon}_t$, with the time-dependent noise $\boldsymbol{\varepsilon}_t=(\boldsymbol{\varepsilon}_h^{(t)},\boldsymbol{\varepsilon}_x^{(t)})^\top$ that can be learned by another from another equivariant network $\boldsymbol{\varepsilon}_{\boldsymbol{\theta }}(\bm{z}_h^{(t)} , \bm{z}_x^{(t)})$ based on the current state of the latent variable along the diffusion pathway. In particular, the EGNN network is modified to ensure that the ro-translational transformations on  $\bm{z}_{x}$ does not change the noise that is to be learned. As the latent space is generally a simpler and smoother geometric manifold compared with the raw molecular structure, such that diffusion in the latent space can facilitate the direct generation of stable 3D molecular structures. Similar approach of diffusing graphs in the latent space has also been demonstrated in the work by \citet{you2024latent}


As the last and latest example, \citet{cornet2024equivariant} adopted comparable ideas as \citet{wu2022diffusion}, which focused on the design of the forward diffusion process in the generative model. Here, the authors define the forward process implicitly with a \emph{learnable transformation} of the injected noise $\boldsymbol{\varphi}$, time $t$ and the data point $\bm{x}$, such that the standard forward diffusion [\cref{eq:diffusion_forward_single_step}] is cast into the the problem of $\bm{z}_t=F_{\boldsymbol{\varphi}}(\boldsymbol{\varepsilon},t,\bm{x})$. Under this construct, the forward process becomes \emph{data dependent} in which $F_{\boldsymbol{\varphi}}$ knows how each input structure $\bm{x}$ should be transformed specifically at any given time $t$, providing the model with more flexibility in constructing the latent space through the forward process. Correspondingly, the time-dependent evolution of $\bm{z}_t$ is governed by the ODE of $\partial_t\bm{z}_t=\partial_t F_{\boldsymbol{\varphi}}(\boldsymbol{\varepsilon},t,\bm{x})$. However, such an ODE cannot be solved, as we do not actually know $\boldsymbol{\varepsilon}$, which is supposed to be learnt by a model, that is also dependent upon $\bm{z}_t$. Borrowing the ideas from the flow model, we can mitigate this problem by enforcing the model $F_{\boldsymbol{\varphi}}$ to be an invertible one, such that $\boldsymbol{\varepsilon}_t$ becomes inferrable from $\bm{z}_t$ as $\boldsymbol{\varepsilon}_t=F^{-1}_{\boldsymbol{\varphi}}(\bm{z}_t,t,\bm{x})$, this leads to the following forward SDE:
\begin{equation}
    \label{eq:forward_ODE_cornet}
    \frac{\partial\bm{z}_t}{\partial t}=f_{\boldsymbol{\varphi}}(\bm{z}_t,t,\bm{x})\quad\Rightarrow\quad f^F_{\boldsymbol{\varphi}}(\bm{z}_t,t,\bm{x})=\left.\frac{F_{\boldsymbol{\varphi}}(\boldsymbol{\varepsilon},t,\bm{x})}{\partial t}\right\vert_{\boldsymbol{\varepsilon}_t=F^{-1}_{\boldsymbol{\varphi}}(\bm{z}_t,t,\bm{x})}.
\end{equation}

However, we still need to align \cref{eq:forward_ODE_cornet} with the forward SDE [\cref{eq:diffusion_forward_SDE}], which combines the drift term (in this case, given by $f_{\boldsymbol{\varphi}}(\bm{z}_t,t,\bm{x})$) and being stochastic in nature, and is needed because we need a well-defined reverse SDE for the sampling process. This means that we want an equation of the form:
\begin{equation}
    \label{eq:forward_SDE_cornet}
    \mbox{d}\bm{z}_t=f^F_{\boldsymbol{\varphi}}(\bm{z}_t,t,\bm{x})\mbox{d}t+g_{\boldsymbol{\varphi}}(t)\mathrm{d}\boldsymbol{\omega},\qquad\mbox{d}\omega\sim\mathcal{N}(0,\bm{I}).
\end{equation}
The main question then becomes how should one design $f^F_{\boldsymbol{\varphi}}$ such that the SDE [\cref{eq:forward_SDE_cornet}] and ODE [\cref{eq:forward_ODE_cornet}] lead to the same time-dependent conditional distribution of $q_{\boldsymbol{\varphi}}(\bm{z}_t|\bm{x})$? The author addressed the problem by adapting an approach that is similar to the one used by \citet{ni2026straight}, and shows that $f_{\boldsymbol{\varphi}}$ and $f^F_{\boldsymbol{\varphi}}$ can be transformed into each other through the score of the conditional distribution:
\begin{equation*}
    f^F_{\boldsymbol{\varphi}}(\bm{z}_t,t,\bm{x})=
    f_{\boldsymbol{\varphi}}(\bm{z}_t,t,\bm{x})
    +\frac{g^2_{\boldsymbol{\varphi}}(t)}{2}\nabla_{\bm{z}_t}\log q_{\boldsymbol{\varphi}}(\bm{z}_t|\bm{x}).
\end{equation*}
This theoretical results thus enables us to flexibly explore the design space of the forward diffusion process via  $f_{\boldsymbol{\varphi}}$ while not sacrificing the techniques of solving SDEs for reverse sampling. This may be beneficial, for example, to preserve certain rigid motifs that often occurs as the repetitive building blocks in the molecular structures from fully collapsing into individual noised atoms, thus increasing the probability of reassembling molecules with higher chemical validity. Nevertheless, the inclusion of a data-dependent forward diffusion process $f_{\boldsymbol{\varphi}}(\bm{z}_t,t,\bm{x})$ inevitably leads to a more complex training process, for which a \emph{joint-reverse-drift} training protocol was designed by the authors. In brief, the training starts from drawing a sample of $\{\bm{x},t,\boldsymbol{\varepsilon}\}$, a latent variable is generated as $\bm{z}_t=F_{\boldsymbol{\varphi}}(\bm{x},t,\boldsymbol{\varepsilon})$. With this, we can (1) compute the drift term directly as $f_{\boldsymbol{\varphi}}(\bm{z}_t,t,\bm{x})$, and (2) regenerate $\bm{x}^\prime$ from $\bm{z}_t$ by denoising the later using the trained noise model, that shall be denoted as $\bm{x}^\prime=\bm{x}_{\boldsymbol{\theta}}(\bm{z}_t,t)$, from which we can also compute the corresponding drift as $\hat{f}_{\boldsymbol{\varphi}}\left(\bm{z}_t,t,\bm{x}_{\boldsymbol{\theta}}(\bm{z}_t,t) \right)$. As such, the diffusion generative model can be trained by solving the joint minimisation problem of $\argmin_{(\boldsymbol{\varphi},\boldsymbol{\theta})}\Vert f_{\boldsymbol{\varphi}}-\hat{f}_{\boldsymbol{\varphi}}(\bm{x}_{\boldsymbol{\theta}}(\bm{z}_t,t))\Vert_2^2$. One can appreciate from the above discussion that the proposed framework is rather general and not tight to a specific representation of (the molecular structure) $\bm{x}$, this may be explored for a wider range of generative tasks that are relevant to molecular and materials modellings.

\paragraph{\textbf{Generation of Crystalline Solids}}
We now proceed to further examine the development of diffusion generative models that are applied to the problem of generative designs of periodic crystals, in order to demonstrate how crystal symmetries and periodicity are incorporated into the diffusion models as the \emph{inductive bias} to enhance the model stability and efficiency. 

\citet{jiao2023crystal} presented the seminal \textsc{DiffCSP} model, which a joint equivariant diffusion process that allows both the lattice vectors and the fractional coordinates to evolve concurrently in the generative process. The model design begins with the representation of crystal structure as $\mathcal{C}=(\bm{A},\bm{F},\bm{L})$, in which $\bm{A}=(\bm{a}_1,\bm{a}_2,\ldots,\bm{a}_N)\in\mathbb{R}^{k\times N}$ are the one-hot representations of the chemical identities, $\bm{F}=(\bm{f}_1,\bm{f}_2,\ldots,\bm{f}_N)\in [0,1)^{3\times N}$ with $\bm{f}_i\in[0,1)^3$ representing the fractional coordinates of atoms and $\bm{L}=(\boldsymbol{\ell}_1,\boldsymbol{\ell}_2,\boldsymbol{\ell}_3)\in(0,+\infty)^{3\times 3}$  is the matrix of lattice vectors. With this representation, the task to be solved by \textsc{DiffCSP} is essentially the prediction of the most likely crystal structure for a material with a specific chemistry, or mathematically coined as learning the conditional probability of $p(\bm{L},\bm{F}|\bm{A})$. 

The \textsc{DiffCSP} model is designed and implemented by enforcing three critical invariances that are related to the problem of crystal structure predictions: (1) The \emph{permutation invariance} of the atomic ordering, such that $p(\bm{L},\hat{\mathcal{P}}\bm{F}|\hat{\mathcal{P}}\bm{A})=p(\bm{L},\bm{F}|\bm{A})$, with $\hat{\mathcal{P}}$ being the permutation operator. Similar to the models discussed above,\cite{simonovsky2018graphvae,niu2020permutation} \textsc{DiffCSP} enforces permutation invariance in noise learning with GNN operations that are not sensitive to the atom ordering in graph constructions and updates. (2) $O(3)$ \textit{rotational invariance} with respect to the rotation of the crystal lattice, whereby given a rotation operator $\hat{\mathcal{Q}}\in\mathbb{R}^{3\times3}$ and $\hat{\mathcal{Q}}\hat{\mathcal{Q}}^\top=\bm{I}$, it requires $p(\hat{Q}\bm{L},\bm{F}|\bm{A})=p(\bm{L},\bm{F}|\bm{A})$. This translates specifically to the design of a lattice noise model that is rotational invariant $\boldsymbol{\varepsilon}_\bm{L}=\hat{\mathcal{Q}}\boldsymbol{\varepsilon}_\bm{L}$, $\forall\hat{\mathcal{Q}}\hat{\mathcal{Q}}^\top=\bm{I}$. This is achieved in \textsc{DiffCSP} through constructing lattice-dependent messages using invariant such as $\bm{L}^\top\bm{L}$. Finally, (3) \emph{periodic translational invariance} which requires $p(\bm{L},\bm{F}+\bm{t}|\bm{A})=p(\bm{L},\bm{F}|\bm{A})$, $\forall \bm{t}\in\mathbb{N}^{3\times1}$. This requires the diffusion of fractional coordinates satisfies the wrapped normal distributions [\cref{tab:continuous_prob_distributions}], which the model achieves this by using a Fourier-transform embedding of the type $f_{\mathrm{FT}}(\bm{f}_i-\bm{f}_j)$ to construct the coordinate-dependent messages. More specifically, the model implemented the joint diffusion process by enabling the fractional-coordinate-noise to be dependent on the current state of the lattice, \emph{i.e.} $\boldsymbol{\varepsilon}_\bm{F}\sim\boldsymbol{\phi}(\bm{L}_{t},\bm{F}_{t},\bm{A},t)$, which is used to denoise $\bm{f}_i^{(0)}\sim\mathrm{Uniform}(0,1)^3$ towards the generative outcome $\bm{f}_i^{(T)}$. With this, the training of the denoising model can be simply achieved by minimising the loss function in the form of $\mathcal{L}=\sum_t\Vert \boldsymbol{\varepsilon}_\bm{L}^{(t)}-\hat{\boldsymbol{\varepsilon}}_\bm{L}^{(t)}\Vert_2^2+\Vert \boldsymbol{\varepsilon}_\bm{F}^{(t)}-\hat{\boldsymbol{\varepsilon}}_\bm{F}^{(t)}\Vert_2^2$. 

This foundational model construct enables the authors to adapt two further extensions of the \textsc{DiffCSP} model in the same paper.\cite{jiao2023crystal} In the first example, which can be considered as a crystal-equivalent to the molecular generative model by \citet{hoogeboom2022equivariant}, the authors formulated the \emph{ab initio crystal generations} that allows the chemical identities $\bm{A}$ to be jointly diffuse with $\bm{L}$ and $\bm{F}$, enabling both the chemical and crystal structure spaces to be explored simultaneously. Secondly, by harnessing  Langevin dynamics, the authors demonstrated how the reverse generation process can be biased towards more stable structures by incorporating the energy-guidance through the drift term $\nabla_{\bm{L},\bm{F}}E$ using energy models with accessible lattice and coordinates derivatives.

In their follow up work, \citet{jiao2024space} presented the \textsc{DiffCSP}$++$ model. Instead of freely diffusing the lattice and fractional coordinates, the model seeks to impose constraints on both the crystal systems and space group symmetries to generate crystal structures in a given space group $\bm{G}$, hence the problem is cast into one that generate crystal structures based on the conditional probability $p(\mathcal{C}|\bm{G})$, and the main innovations of the paper lies in how the space group constraints are decomposed and implemented in the diffusion model. To follow the \textit{constraints on the lattice parameters (crystal systems)}, the authors adopted an approach based on \emph{polar decomposition}, which is not often seen and used in standard atomistic materials modelling. Here, we seek to express $\bm{L}=\bm{Q}\exp(\bm{S})$, in which $\bm{S}$ is a symmetric $O(3)$-invariant matrix in $\mathbb{R}^{3\times3}$, whereas rotations and reflections only change $\bm{Q}$. Most importantly, $\bm{S}$ can be decomposed into a set of six-dimensional matrix basis as $\bm{S}=\sum_{i=1}^6 k_i\bm{B}_i$, where, depending on the specific crystal family, only a subset of $\{k_i\}$ may take any arbitrary values with the others remain fixed. Hence, by only diffusing the unconstrained $k$'s will enable us to sample the lattices in the appropriate crystal family throughout the diffusion process, effectively overcoming the problem of imposing constraints on $\bm{L}$ directly in which many matrix elements are coupled together. The site-symmetry constraints imposed by specific space group, on the other hand, is realised with the use of Wyckoff positions and generators. Instead of generating all fractional coordinates independently, they are grouped into the Wyckoff classifications, such that only a unique Wyckoff position will be diffused and the rest generated with the space group symmetry operations. 
Similar to the WyCryst model,\cite{zhu2024wycryst} this significantly reduces the dimensionality that is dealt by the generative tasks.

Along the same line, \citet{levy2025symmcd} implemented the \textsc{SymmCD} which also tries to harness crystal symmetry in generating crystal structures, as one of the major problems faced by the \textsc{DiffCSP} model is the high tendency for the model-generated crystal structures ending up in $P1$ symmetry when all the fractional coordinates were allowed to diffuse independently. Here, \textsc{SymmCD} tries to generate the asymmetric unit by representing it with $(\bm{X},\bm{A},\bm{S})$ being the fractional coordinates, element embedding and the associated site symmetries, respectively.   The site symmetries are the symmetry operations in a space group that maps a given Wyckoff position into another that belongs to the same Wyckoff group, which is usually denoted by a lower-cased letter. In \textsc{SymmCD}, the site symmetries $\bm{S}$ are generated rather than prefixed according to $\bm{G}$ in the \textsc{DiffCSP}++, thus allowing the model to explore a wider range of site symmetries that are adaptable by specific chemical element, rather than being fixed to a specific $\bm{G}$. Mathematically, $\bm{S}$ is a site-symmetry operation which combines 15 possible crystallographic symmetry axes and 13 possible local operations on it, thus \textsc{SymmCD} models $\bm{S}$ as $\{0,1\}^{15\times 16}$ and diffuse it according to a discrete binary distribution.

\citet{yang2024scalable} proposed the \textsc{UniMat} model, aiming to achieve the generation of inorganic crystals with high fidelity in the chemical space across the full periodic table. This is achieved predominantly through a new representation of crystal structures, which is composed of: (1) fractional coordinates of the atoms $\bm{f}_i\in[0,1)^{3\times3}$, (2) the lattice parameters $(a,b,c,\alpha,\beta,\gamma)$, (3) elements in the periodic table $\bm{M}\in\mathbb{R}^{L\times H\times W\times C}$, in which $H=9$, $W=18$ are the numbers of periods and groups, $L$ is the maximum number of atoms per element, with $C=3$ corresponds to the dimension of Cartesian space $\mathbb{R}^3$. In addition, a \emph{null} location is included for positioning the atoms. The diffusion process starts from $\bm{x}_T$ that contains a a set of randomised elements placed in the random positions within a unit cell. The denoising model is trained to learn how to (a) remove atoms that were not supposed to be in the crystal into the null space, and (b) gradually refine the positions of the allowed atoms back to their original positions from the random start.

\citet{kelvinius2025wyckoffdiff} proposed the \textsc{WyckoffDiff} model for generating inorganic crystals that also takes into account the space group symmetry. Here, instead of generating the Wyckoff positions for each element that is going to be present in the crystal the model tries to sample the \emph{occupation} of each Wyckoff position, \emph{i.e.} which element and how many corresponding atoms is occupying the given set of Wyckoff positions. This leads to the following representation of the crystal structure $\mathcal{M}=(s,\bm{z},\bm{z}_\infty)$, in which $s\in\mathbb{Z}$ represents the space group, and the Wyckoff positions in $s$ are further divided into $\mathcal{L}(s)=\mathcal{L}_0(s)\bigcup\mathcal{L}_{\infty}(s)$ with $\mathcal{L}_0(s)$ denote the set of Wyckoff positions with no positional degree-of-freedom and $\mathcal{L}_{\infty}(s)$ are those with at lease one positional degree-of-freedom. Then, $\forall i\in\mathcal{L}_0(s)$, $z_{0,i}=j$ with $\bm{z}_{0}\in\mathbb{N}^{N_a}$, specifying the $i$-th position is occupied by element $j$, where as $\forall i\in\mathcal{L}_{\infty}(s)$, $z_{\infty,ij}=n$ with $\bm{z}_\infty\in\mathbb{N}^{N_a\times N_w}$, representing the number of atom of type $j$ that is occupying the $i$-th Wyckoff position. The entire diffusion process then works on the discrete categorial distribution. Starting from a given $s$, the model applies the diffusion generative process to determine $\bm{z}_0$ and $\bm{z}_\infty$ that are conditioned upon $s$. After all Wyckoff positions have their occupancies determined, the model then assign a compatible lattice parameters, with which the full crystal structure can be generated.

\citet{zeni2025generative} developed the \textsc{MatterGen} model that is probably one of the most well-known diffusion-based materials generative model, which was implemented with many baseline features that resemble those of the \textsc{DiffCSP},\cite{jiao2023crystal} hence we will use the same notation as the \textsc{DiffCSP} model in the discussion. In addition to the translational and rotational invariances that are already recognised by the  \textsc{DiffCSP} model, \textsc{MatterGen} further considered the (1) \emph{invariance with respect to the cell choice} as $p(\bm{A},\bm{F},\bm{L})=p(\bm{A},\hat{\bm{C}}^{-1}\bm{F},\bm{L}\hat{\bm{C}})$ with $\hat{\bm{C}}\in\mathbb{Z}^{3\times3}$ and $\mathrm{det}\hat{\bm{C}}=1$, as well as (2) \emph{supercell invariance}, that can be mathematically expressed as $p(\bm{A},\bm{F},\bm{L})=p(\oplus_{i=0}^{\mathrm{det}(\hat{\bm{C}})}\bm{A}, \oplus_{i=0}^{\mathrm{det}(\hat{\bm{C}})}\hat{\bm{C}}^{-1}\left(\bm{F}+\bm{k}_i\mathds{1}_n^\top \right) , \bm{L}\hat{\bm{C}})$, with $\hat{\bm{C}}\in\mathbb{Z}^{3\times 3}$ (diagonal) and $\bm{k}_i\in\mathbb{N}^3$  being the cell repetition vector. Under these constraints, the generation model designs a joint diffusion process as a Markov chain, and factorises the joint distributions into their individual components as $q\left( \bm{A}_{t+1},\bm{F}_{t+1},\bm{L}_{t+1}|\bm{A}_t,\bm{F}_t,\bm{L}_t\right)= q(\bm{A}_{t+1}|\bm{A}_t) q(\bm{F}_{t+1}|\bm{F}_t) q(\bm{L}_{t+1}|\bm{L}_t)$, with which: (1) The atom type is diffused according to a transition kernel $q(\bm{A}_{t+1}|\bm{A}_t)$ parameterised with a discrete categorial distribution. (2) The diffusion of the fractional coordinates was parameterised to follow a wrapped normal distribution $q(\bm{F}_{t+1}|\bm{F}_t)\sim \mathcal{N}_w(\bm{F};\bm{F}_t,\sigma^2\bm{I})$. Although one could also optionally choose to diffuse the Cartesian coordinates, the conversion between the fractional and Cartesian coordinates is dependent on the current state of the lattice $\bm{L}_t$, which also needs to be sampled in the diffusion process, making it unsuitable to adopt the above factorisation scheme of the joint distribution, hence this approach was not adopted in the \textsc{MatterGen}. Finally, (3) diffusion of the lattice according to the transition kernel $q(\bm{L}_{t+1}|\bm{L}_t)$ was implemented with a \emph{fixed-rotation diffusion} scheme, in which a canonical rotational alignment is used throughout the diffusion process. In addition, the noise being added/removed to $\bm{L}_t$ was also forced to be diagonally symmetric. These invariant constraints are jointly implemented through an SE(3)-equivariant NN based on the GemNet architecture.\cite{gasteiger2021gemnet} Finally, the model training loss can be designed based on this factorisation and parameterisation scheme, to derive an implementable loss based on the score-matching\cite{jo2022score} formalism.

\citet{luo2024deep} presented the cond-\textsc{CDVAE} that is based on the CDVAE model\cite{xie2018crystal} to achieve the conditional generation of crystalline materials. As an example, the paper considered the generation of crystal structure with a given chemical composition at a specific pressure, as given by the conditional probability distribution of $p(\mathcal{C}|\bm{A},P)$. The physical implication behind this model is easy to understand from the point of basic thermodynamics, as pressure is an additional degree-of-freedom on the phase diagram that shifts the equilibrium distribution of stable phases at a given temperature. The model achieved conditional generation by augmenting the laten variable $\bm{z}_t$ with an additional information of $\bm{c}=(\bm{A},P)$ that are jointly fed into the decoder to generate new crystal structures with specific compositions at the designated pressure.

As a final example, we consider the \textsc{DYNACRYS} model developed by \citet{jin2026dynacrys} Instead of generating crystal structures with a given chemistry in a specific space group, the model also treat the space group as a diffusable variable that is jointly evolved with both the Wyckoff position and the chemical elements. In particular, the diffusion of the space group is not a random process but is designed to follow a tree-structure of space groups that is established based on the group-subgroup relationship, mathematically encoded by an adjacent matrix $\bm{A}$. This ensures that the diffusion of space group is biased towards neighbouring symmetries that are structurally meaningful. With a space group $\bm{G}_t$ being generated at each time step of the diffusion model, the corresponding Wyckoff rows for this particular  $\bm{G}_t$ will be generated. This is performed with a process being coined as \textit{symbolic generation} in the paper. Specifically, one define the Wyckoff rows $R_{G}=\{r_1,r_2,\ldots,r_{n}\}$ as the available set of Wyckoff positions for $\bm{G}$ that can be retrieved from the International Handbook of Crystallography, except the number of the occupied orbits $K$ and the elements $Z$ assigned to each row, \emph{i.e.} $\{(Z_i,r_i)\}_{i=1}^{K}$, which is to be learnt by the model. Probabilistically, the generative task to be solved by \textsc{DYNACRYS} is then to find the most probable occupation pattern of Wyckoff positions conditioned on $\bm{G}_t$ and the current chemical state that is `legal'. This leads to a \emph{protostructure} $P=\{\bm{G}_t,\bm{Z}_t,\bm{r}_t\}$ for the crystal to be generated.

Lastly, the model harness the diffusion generative framework from \textsc{DiffCSP}++\cite{jiao2024space} to generate the lattice parameters and the atomic coordinates for $P$. More specifically: (a) For generating the lattice parameters, the model starts from the lattice vector matrix $\bm{L}\in\bm{R}^{3\times3}$, from which a symmetric matrix $\bm{S}=\log(\sqrt{\bm{L}\bm{L}^\top})$ is constructed with the  matrix elements rearranged into $\bm{b}\in\mathbb{R}^6$, forming the object to be diffused with the standard diffusion generative models. Constraints in the lattice system can be imposed by projecting $\bm{b}_t$ onto its subspace that is allows by the crystal family corresponding to $\bm{G}$. The final lattice vectors can be retrieved from $\bm{b}_T$ by inverting the process of constructing the symmetric matrix as $\exp[2\bm{S}(\bm{b}_T)]$. (b) For generating the atomic coordinates,  \textsc{DYNACRYS} followed the wrapped normal distribution to perform the diffusion of coordinates on a torus $\mathbb{T}^3$, with a learnt noise model that is conditioned upon $P$, to generate the fractional coordinates that are compatible with crystal protostructure.

\subsubsection{Applications of Flow-Based Generative Models in Atomistic Modelling}
\label{sec:generative_flow_examples}

We now proceed to provide a brief survey of the recently developed flow-based  models for the atomistic generative samplings for molecules and crystalline solids. As one shall appreciate from the designs of these  normalising flow models, that, whilst many of the key concepts that we have highlighted at the beginning of  \cref{sec:generative_diffusion_examples} still applies, there is a strong focus on ensuring the \emph{continuity} is enforced in the high-dimensional manifold on which the generative targets will evolve (more specifically, \emph{flow}), such that the probability distributions of the sample and latent variables can be continuously differentiated along the sampling pathway. 

\paragraph{\textbf{Generation of Molecules}} Along a very similar trail as the development of other generative models for molecules, we see the initial proposals of flow-based model were primarily focusing on the generations of 2D molecular graphs, which is subsequently evolved to the concurrent generations of both the 2D graphs and 3D coordinates for molecular structures.

\citet{madhawa2019graphnvp} proposed the first invertible, normalising-flow-based molecular graph generative model (GraphNVP) that simultaneously generate both the adjacency matrix and node attributes in an \emph{one-hot} generation procedure. Here, a molecular graph adapts a discrete-type representation of $\mathcal{G}=\{\bm{A},\bm{X}\}$, in which $\bm{A}\in\{0,1\}^{N\times N\times R}$ is the adjacent matrix and $\bm{X}\in\{0,1\}^{N\times M}$ is the node attribute matrix, where $N$ is the number of nodes, $M$ is the number of types of the nodes and $R$ is the number of types of the bonds (\emph{e.g.} single, double and triple bonds in a molecule). The generative task is then formulated as the development of an invertible model $f_{\boldsymbol{\theta}}$, such that one can map a molecular graph directly to a simple latent space $\bm{z}=f_{\boldsymbol{\theta}}(\mathcal{G})$, and achieve the reverse graph generation from a new latent variable through direct inversion  $\mathcal{G}^\prime=f_{\boldsymbol{\theta}}^{-1}(\bm{z}^\prime)$. A key modification to such a representation, which we shall see repetitively in subsequent models, is the \emph{dequantisation of the graph representations} to allow the application of density models to discrete variables. This is achieved by adding a small noise drawn from the uniform distribution [\cref{tab:continuous_prob_distributions}] to the elements of $\bm{A}$ and $\bm{X}$. The invertible model  $f_{\boldsymbol{\theta}}$ is implemented with the RealNVP\cite{dinh2016density}-type affine coupling layer [\cref{eq:invertible_NN} and \cref{eq:invertible_NN_inverse}]. More specifically, for matrix type input, such as the adjacency matrix $\bm{A}$, if $f_{\boldsymbol{\theta}}$ is to update the information on a given node $i\in N$, then the information from all other $(N-1)$ nodes is kept unchanged. In addition, note that what chemical elements are allowed on a specific node of the graph depends on the connectivity of that node. Hence, in the forward flow, the model first generate the latent representation of the adjacent matrix $\bm{z}_{\bm{A}}$, conditioned upon which $\bm{z}_{\bm{X}}$, the latent representation for the node attributes will be generated, and such a two-staged process is also applied in the reverse sampling. 

\citet{liu2019graph} introduced the graph normalising flow, which also seeks to establish an invertible mapping between the input graph and latent representation space. Unlike the GraphNVP, the model works on a graph representation $\mathcal{G}=(\bm{H},\boldsymbol{\Omega})$ that is designed to be continuous in nature. Here. $\bm{H}=\left(\bm{h}^{(1)},\bm{h}^{(2)},\ldots,\bm{h}^{(N)}\right)\in\mathbb{R}^{N\times d_n}$ are the feature matrix for $N$ nodes, and $\boldsymbol{\Omega}\in \mathbb{R}^{ N\times N\times (d_e+1) }$ is the edge feature tensor with the first channel being the connectivity matrix in $\mathbb{R}^{N\times N}$. More specifically, the authors implemented the GRevNet,  a MPNN that is designed to be invertible. Here, (1) the attention mechanism\cite{vaswani2017attention} is harnessed in the message construction that allows messages from different edges in the graph are appropriately weighted before they are aggregated into the next stage, and (2) modified the standard MP operation, which operates on the entire message in one go, into one that operates on a changeable and an invariant part of the message, that is accompanied with an intermediate step to generate an intermediate node feature $\bm{H}_{t+\scriptscriptstyle\frac{1}{2}}$. The constraint of volume conservation in normalising-flow is more explicitly considered in the design of the multi-staged message-passing operations, whereby \cref{eq:volume_preserving_norm} is cast into an autoregressive form [\emph{c.f.} \cref{eq:HVAE_forward}] such that the probability distributions over the input graph (after being properly \emph{dequantised}) and final node representations are preserved:
\begin{equation*}
    p(\mathcal{G})=P(\bm{H}_T)\prod_{t=1}^T\mathrm{det}\left\vert \frac{\partial \bm{H}_t}{\partial \bm{H}_{t-1}}  \right\vert.
\end{equation*}
To train this invertible model for subsequent sampling, a permutation invariant graph encoder is first trained to map a set of node features $\bm{H}$ and adjacency matrix $\bm{A}\in\{0,1\}^{N\times N/2}$ into a set of node embedding $\bm{X}\in\mathbb{R}^{N\times k}$. The decoder (and equivalently, the sampling process) then takes an embedding $\hat{\bm{X}}$ and reconstruct it (in \emph{one-shot} that contains $T$ steps of MP operations) to a set of edge probabilities $\hat{\bm{A}}\in [0,1]^{N\times N/2}$ where $\hat{\bm{A}}_{ij}=[1+\exp (C \Vert \hat{\bm{X}}_i-\hat{\bm{X}}_j \Vert_2^2) ]^{-1}$ with trainable parameter $C$. Then the model can be trained by minimising the cross-entropy between $\bm{A}$ and $\hat{\bm{A}}$.

\citet{shi2020graphaf}  developed the GraphAF model, which may be considered as a hybrid between a normalising-flow and a diffusion generative model for generating molecular graphs. Similar to  \citet{liu2019graph}, GraphAF represents a molecular graph as $\mathcal{G}=(\bm{A},\bm{X})$, in which $\bm{A}\in\{0,1\}^{n\times n\times(b+1)}$ that is a tensor containing the connectivity between two given nodes $i$ and $j$ as well as its corresponding one-hot embedding vector, where as $\bm{X}\in\{0,1\}^{n\times d}$ that is a one-hot embedding of the atom type for each node. Both representations are dequantised in a similar scheme as \citet{madhawa2019graphnvp} The actual graph generation process is autoregressive in nature, which resemble many similarity to that of the junction-tree algorithm from  \citet{jin2018junction} (\emph{c.f} \textbf{Box V}) by growing the full graph in a node-by-node, rather than a one-shot, fashion. Here, the model first generates the type of a new atom based on the state of the current sub-graph, according to a learned conditional distribution of $p_{\bm{X}}(\bm{X}_i|\mathcal{G}_i)$. Then, the model decides the connectivities to this newly generated atom based on the connectivities of the existing sub-graph with $p_{\bm{A}}(\bm{A}_{ij}|\mathcal{G}_{i},\bm{X}_i,\bm{A}_{i,1:j-1})$, $\forall j<i$. This process is repeated until a full molecular graph containing $N$ nodes is generated, thus the corresponding probability distributions of the generated molecular graph follows the autoregressive form:
\begin{equation}
\label{eq:graphAF_autoregressive}
    p(\mathcal{G})\sim\prod_{i=1}^N p_{\bm{X}}(\bm{X}_i|\mathcal{G}_i) \prod_{i=1}^N p_{\bm{A}}(\bm{A}_{ij}|\mathcal{G}_{i},\bm{X}_i,\bm{A}_{i,1:j-1}),
\end{equation}
in which if we adopt the same approach in formulating the diffusion generative model by parameterising $p_{\bm{X}}$ and $p_{\bm{A}}$ as Gaussians with learnable means and variances, then a numerically implementable loss function can be further derived. Such a parameterisation will also enable us to establish the link between autoregressive generations with the construct of normalising-flow model. More specifically, let us write the conditional distributions in \cref{eq:graphAF_autoregressive} as
\begin{equation}
    \label{eq:graphAF_parameterisation}
    \begin{cases}
        p_{\bm{X}}\sim\mathcal{N}(\bm{X}_i|\boldsymbol{\mu}_{i}^X, (\alpha_i^X)^2 ) \\
        p_{\bm{A}}\sim \mathcal{N}(\bm{A}_{ij}|\boldsymbol{\mu}_{ij}^{A},(\alpha_{ij}^A)^2)
    \end{cases}\quad \Rightarrow \quad
    \begin{cases}
        \bm{x}_i=\boldsymbol{\varepsilon}_i\odot\alpha_i^X+\boldsymbol{\mu}_i^X\mbox{ with } \boldsymbol{\varepsilon}_i\in\mathbb{R}^{d}\\
        \bm{A}_{ij}= \boldsymbol{\varepsilon}_{ij}\odot\alpha_i^A+\boldsymbol{\mu}_{ij}^A\mbox{ with } \boldsymbol{\varepsilon}_{ij}\in\mathbb{R}^{b+1}.
    \end{cases}
\end{equation}
Then, the process of reconstructing the initial starting graph is a process to remove $\boldsymbol{\varepsilon}$ (as in the diffusion model), which can be achieved by reversing the processes in \cref{eq:graphAF_parameterisation}, providing that the variances are learned by an invertible network:
\begin{equation*}
    \begin{cases}
        \boldsymbol{\varepsilon}_i=\left(\bm{x}_i-\boldsymbol{\mu}_i^X\right)\odot\frac{1}{\alpha_i^X}\\
        \boldsymbol{\varepsilon}_{ij}=\left(\bm{A}_{ij}-\boldsymbol{\mu}_{ij}^A\right)\odot\frac{1}{\alpha_{ij}^A}
    \end{cases}.
\end{equation*}

In another comparable model, the MolGrow proposed by \citet{kuznetsov2021molgrow} also followed the similar designs of the junction-tree VAE\cite{jin2018junction} to generate molecular structures hierarchically with the normalising-flow model. Here, the molecular graph is represented as $\mathcal{G}=(\bm{V},\bm{E})$, in which $\bm{V}\in\mathbb{R}^{N\times d_v}$ is the node feature matrix for $N$ nodes in the graph, and $\bm{E}\in\mathbb{R}^{N\times N\times d_e}$ that is the edge feature tenor, with its elements dequantised while enforcing $\bm{E}_{ij}=\bm{E}_{ji}$ to ensure the representation corresponds to an undirected graph. To achieve a hierarchical learning and generation process, the model starts with $\mathcal{G}$ with $N$ nodes, that is progressively merged $\mathcal{G}^{(0)}\to\mathcal{G}^{(1)}\to\cdots\mathcal{G}^{(L)}$, by successively merging two adjacent nodes at each step, and such a sequence is reversed in the generation process. To merge two neighbouring nodes,  the two node features and the features of the edge connecting the two nodes:
\begin{equation*}    \bm{V}_{i}^{k+1}=\mbox{cat}\left(\bm{V}_{2i}^k,\bm{V}_{2i+1}^k,\bm{E}_{2i,2i+1}^k \right),
\end{equation*}
and all edge features originated from the two nodes to be merged are concatenated to  a new edge feature:
\begin{equation*}
\bm{E}_{ij}^{k+1}=\mbox{cat}\left(\bm{E}_{2i,2j}^k,\bm{E}_{2i=1,2j}^k,\bm{E}_{2i,2j+1}^k,\bm{E}_{2i+1,2j+1}^k \right).
\end{equation*}
In contrast to the junction tree algorithm\cite{jin2018junction}, which learns the latent representation $\bm{z}$ solely from the final node $\mathcal{G}^{(L)}$, latent variables are produced after each node concatenation operations, which corresponds to the transformation in the form of $(\bm{V}^k,\bm{E}^k)\to(\bm{V}^{k+1},\bm{E}^{k+1},\bm{z}_k)$ such that the forward process learns  multileveled set of latent variables $\{\bm{z}_i\}_{i]0}^L$. Whilst $\bm{z}_L$ enables the control of the global molecular graph structure, the access to the remaining latent variables enables us to gain more fine-grained control on the lower-level structures during the generation process. Similar to the model by \cite{liu2019graph}, the data-to-latent-space mapping is implemented with the invertible RealNVP\cite{dinh2016density} that is augmented by the attention layer, making each transformation to be aware of the global graph structure at each step. In line with \cref{eq:graphAF_autoregressive}, both the forward and reverse transformations can be abstracted into the following hierarchical processes:
\begin{equation*}
    \begin{cases}
        \bm{z}=f_k\circ \cdots f_2\circ f_1(\bm{x})\\
        \bm{x}=f_1^{-1}\circ \cdots\circ f_{k-1}^{-1}\circ f_k^{-1}(\bm{z}).
    \end{cases}
\end{equation*}
Denoting $f_\ell(\cdot)=h_{\ell}$, then the model is trained by maximising the likelihood of:
\begin{equation*}
    \log p(\mathcal{G})=\log p(\bm{z}_0,\ldots,\bm{z}_L)+\sum_\ell \log\left\vert \mbox{det} \frac{\partial h_\ell}{\partial h_{\ell-1}}\right\vert,
\end{equation*}
which follows the construction of the normalising-flow model. Based such a construct for the forward mapping process, reverse sampling essentially starts from a set of randomly sampled latent representations of $\{\hat{\bm{z}}_{0},\hat{\bm{z}}_1,\ldots,\hat{\bm{z}}_L\}\in\mathcal{N}(0,\bm{I})$, from which a top single-node graph $\hat{\mathcal{G}}^{(L)}=(\hat{\bm{V}}^{(L)},\hat{\bm{E}}^{(L)})=f^{-1}(\hat{\bm{z}}_L)$ is first generated. Then, at each successive level, the currently generated molecular graph, together with the corresponding latent map for the next step, is injected to update the graph as $f^{-1}(\hat{\mathcal{G}}^{(k+1)},\hat{\bm{z}}_k)\to\hat{\mathcal{G}}^{(k)}$m and the process is repeated until all $2^L$ nodes were generated. Finally, a requantisation step is applied to the fully generated graph to represent the molecular structure.

\citet{zang2020moflow} proposed the MoFlow model, in which the molecular graph representation from the GraphAF\cite{shi2020graphaf} model was followed, but adopted an one-shot generation approach to generate the molecular graph $\mathcal{G}$. Similar to \citet{madhawa2019graphnvp}, MoFlow also implemented two invertible NN, such that, starting from a random latent variable of $\hat{\bm{z}}_{\bm{A}}$ and $\hat{\bm{z}}_{\bm{X}}$, the first NN was trained to enable the mapping of $\hat{\bm{A}}=f_{\boldsymbol{\theta}}^{-1}(\hat{\bm{z}}_{\bm{A}})$, which is then served to predicted the chemistry of molecule conditioned on the given connectivity $\bm{\hat{X}}=f_{\boldsymbol{\varphi}}^{-1}(\hat{z}_{\bm{X}},\hat{\bm{A}})$. From here, the final molecular graph $\hat{\mathcal{G}}=(\hat{\bm{A}},\hat{\bm{X}})$ can be assembled, and the validity of which can be further checked in a \emph{post-hoc} manner.

Moving forward from generating 2D molecular graphs only, \citet{Garcia_e_n_generative} designed the $E(n)$ equivariant normalising flow model to directly generate both the atomic features and the 3D coordinates with continuous-time normalising flow, while respecting the translation, rotation, reflection and node-permutation symmetries.  
As such, the model represents a molecule directly as $\mathcal{M}=\{\bm{x},\bm{h}\}$, in which $\bm{x}\in\mathbb{R}^{M\times 3}$, which is the matrix collecting the Cartesian coordinates in $\mathbb{R}^3$ for $M$ atoms, and $\bm{h}\in \mathbb{R}^{M\times n_h}$ that contains the invariant node feature, each lies in the space of $\mathbb{R}^{n_h}$. The model is then designed by considering the evolution of a molecule from its initial state of $(\bm{x}(0),\bm{h}(0))$ to the final latent state of $(\bm{z}_{\bm{x}}=\bm{x}(1),\bm{z}_{\bm{h}}=\bm{h}(1))$, as one the follows the dynamics that is defined by the ODE:
\begin{equation}
    \label{eq:E_n_flow_ODE}
    \frac{\partial\bm{x}(t)}{\partial t},\frac{\partial\bm{z}(t)}{\partial t}=\phi_{\boldsymbol{\theta}}(\bm{x}(t),\bm{h}(t)),
\end{equation}
in which the vector field $\phi_{\boldsymbol{\theta}}(\bm{x}(t),\bm{h}(t))$ is modelled by an $E(n)$ equivariant GNN based on the EGNN\cite{satorras2021n} architecture developed by the same group, that was also adopted in the equivariant diffusion model developed by \citet{hoogeboom2022equivariant} In addition, the translational invariance is imposed by removing the centre-of-mass drift in $\bm{x}(t)$ in each time step along the trajectory. To train the NN approximator $\phi_{\boldsymbol{\theta}}$, one solves the ODE [\cref{eq:E_n_flow_ODE}], which gives the following solution that links the probability distributions between the data and latent distributions:
\begin{equation}
    \label{eq:E_n_flow_solution}
    \log p_{\mathcal{M}}(\bm{x},\bm{h})= \log p_\bm{z}(\bm{z}_{\bm{x}},\bm{z}_{\bm{h}})+\int_0^1 \mathrm{Tr} \left[  J_{\phi_{\boldsymbol{\theta}}} (\bm{x}(t),\bm{h}(t)) \right]dt,
\end{equation}
in which the second term maintains the volume conservation through the Jacobian matrix for the ODE as $J_{\phi_{\boldsymbol{\theta}}}=\partial \phi_{\boldsymbol{\theta}}/\partial(\bm{x},\bm{h}) $. Without going into further mathematical details, \cref{eq:E_n_flow_solution} essentially defines the upper bound for maximising the log-likelihood of the data distribution, which is the target function to be optimised during the model training to attain the optimum model hyperparameters $\hat{\boldsymbol{\theta}}_{\mathrm{opt}}$.

As a last and more interesting example that directly connects normalising-flow with MD samplings [\cref{sect:molecular_dynamics}], we consider the \textsc{MDGEN} model developed by  \citet{jing2024generative}, which, instead of generating one particular molecular structure at a time, the model treats an entire MD trajectory containing $T$ frames $\bm{X}_{1:T}=[\bm{X}_1,\bm{X}_2,\ldots,\bm{X}_T]\in\mathbb{R}^{T\times 3N}$ to be the model's generative target, where each MD frame is initially represented in the Cartesian coordinates for all $N$ atoms. Focusing on generating peptides and proteins, the model adopted the residue-based representation from AlphaFold\cite{jumper2021highly} to remove the redundancies in the Cartesian representation, which divide each molecule into $L$ residue, and represents each residue at time $t$ as $g_t^\ell$ that is encoded by the rigid-body pose and internal torsion angle, collectively denoted as $(\bm{R},\boldsymbol{\tau})$. Such a construct allows one to reframe the representation of the MD trajectory, which is now being divided into $K$ sets of \emph{key frames} $\{\bm{g}_{t_1},\ldots,\bm{g}_{t_K}\}$, and the rest of the frames are defined as a change in the pose with respect to a specific key frame $\Delta g^\ell_{t,t_i}=(\Delta \bm{R}^\ell_{t,t_i},\Delta\boldsymbol{\tau}^\ell_{t,t_i})$. To generate an MD trajectory, the key frames are then supplied to the model as the conditional information, with respect to which the trajectory can be generated by predicting $\Delta g^\ell_{t,t_i}$ for all the missing frames. The flow model is then designed to learn the transition probability of the molecular trajectory along a path defined by the flow-ODE $\partial_\mathcal{T}\bm{Z}_{\mathcal{T}}=v_{\boldsymbol{\theta}}(\bm{Z}_{\mathcal{T}},\mathcal{T}|c)$ conditioned on the given chemical information $c$. Here, we use the symbol $\mathcal{T}\in[0,1]$ to denote the time along the flow-trajectory that is different from the time $t$ in the MD trajectory, and the initial condition from which the reverse sampling starts is generated by sampling in $\mathcal{N}(0,1)^{T\times 3N}$ for all $T$ frames in the trajectory.

\paragraph{\textbf{Generation of Crystalline Solids}} Now we proceed to discuss examples of flow-based generative models that are applied to the generations of crystalline inorganic solids. Compared to VAE and diffusion models, the adaptation of the flow models in this particular area comes at a relative later stage. As one shall see in the detailed discussions below, the main challenge behind this is how to identify and operate on the complex and continuous geometric manifolds in which the representations of a crystal structures live, particularly ones that explicitly takes into account the crystal periodicity  that these representations ought to follow.

\citet{miller2024flowmm} developed the FlowMM model for the \emph{ab initio} generation of crystal structures, in which the chemical compositions, fractional coordinates and the lattice constants are jointly flowed to generate new material compositions and corresponding crystal structures. Unlike the diffusion  models, the developments of which are fundamentally grounded by the (time-dependent) evolution of the \emph{probability distributions} that each component should follow, here, one seeks to directly evolve each component on the geometric manifold that it belongs to. This is achieved by following the tangents to the time-dependent state on the surfaces of these manifold. As such, the authors pointed out that these components do not lie on a simple Euclidean manifold, but a \emph{Riemannian} one, and is how the flow model should be designed accordingly, which we discuss in more details in \textbf{Box VI}.

\begin{myquote}
\textbf{Box VI: design details of the Riemannian Flow Matching model for \emph{ab initio} crystal structure generation }

\footnotesize

Here, we provide the readers with a more detailed account behind the implementation of Riemannian Flow Matching model for the tasks of \emph{ab initio} generations of crystal structures.

\hspace{10pt} \textbf{Representations of crystal structures} The flow model begins by adopting the representation of a crystal structure in the form of $\mathcal{C}=(\bm{A},\bm{F},\bm{L})$, which, on a high level, is similar to many of the diffusion-based generative models discussed above,\cite{jiao2023crystal,jiao2024space,zeni2025generative} but with subtle differences. More specifically, 
\begin{itemize}
    \item $\bm{A}$ contains the information of atom types that is encoded in $\{-1,1\}^{d_{N_a}}$, with ${d_{N_a}}=\lceil \log_2 h\rceil$, with $h$ being the number of possible elements. This means that, instead of the commonly applied one-hot vectors for encoding chemical elements, each element here is assigned with with a $d_{N_a}$-dimensional vector with entries in $\{-1,1\}$. This enables one to work on manifold with a lower dimensionality. To be more explicit, $\bm{a}\in\{-1,1\}\in\mathbb{R}$ is able to represent 2 elements, $\bm{a}\in\{-1,1\}^2\in\mathbb{R}^2$ leads to 4 possible combinations of the two bits thus can represent up to 4 elements. Hence, $\bm{A}\in\{-1,1\}^{d_{N_a}}$ is capable of representing up to $h=2^{N_a}$ different elements.
    \item $\bm{F}$ is the matrix containing the fractional coordinates  of each atoms in the crystal, or $\bm{f}_i\in[0,1)^3$, such that $\bm{f}_i\in\mathbb{T}^3$ lives in a 3D torus.
    \item $\bm{L}$ which is the lattice parameters $\{a,b,c,\alpha,\beta,\gamma\}$ for a \emph{Niggli-reduced} cell, such that $a,b,c>0$ and $60^\circ\leq\alpha,\beta,\gamma\leq 120^\circ$.
\end{itemize}

\hspace{10pt} \textbf{The flow manifold} Such choice of crystal representation leads to $\mathcal{C}$ become living in the manifold of $\mathcal{M}=\mathbb{R}^{d_{N_a}}\times\mathbb{T}^{3n}\times\mathbb{R}^{d_\ell}$, that combines the manifolds for elements, fractional coordinates, and the lattice parameters, respectively. Then the goal of the Riemannian flow is to learn a time-dependent vector field on this manifold, in the form of $v_{\boldsymbol{\theta}}(m,t)\in \mathcal{T}_m\mathcal{M}$, which is the tangent space of $\mathcal{M}$ at a given point $m\in\mathcal{M}$, such that the samples shall evolves on this manifold according to an ODE:
\begin{equation}
    \label{eq:riemannian_flow_ODE}
    \frac{\partial m_t}{\partial t}=v_{\boldsymbol{\theta}}(m,t).
\end{equation}
Consequently, reverse sampling can be achieved by starting from a tuple of $(\bm{A}_0,\bm{F}_0,\bm{L}_0)$ that are sampled from their respective prior distributions, and integrate the the ODE towards its final state  $(\bm{A}_1,\bm{F}_1,\bm{L}_1)$.

\hspace{10pt} \textbf{Training of the NN parameterisation of the vector field} As seen in some of the model discussed above, the inspiration here to to treat $v_{\boldsymbol{\theta}}(m,t)$ as one that defines the transition probability from a training crystal of $\mathcal{C}_{1}$ to a state $\mathcal{C}_t$ with $t\in [0,1)$, which will be denoted as $u_t(\mathcal{C}_t|\mathcal{C}_1)$. Therefore, conditioned upon $\mathcal{C}_1$, the goal to train the Riemannian flow model is to minimise the following loss:
\begin{equation}
\label{eq:riemannian_flow_lost}
\mathcal{L}(\boldsymbol{\theta})=\mathbb{E}_{t,\mathcal{C}_1,\mathcal{C}_t}\left\Vert v_{\boldsymbol{\theta}}(\mathcal{C}_t,t) - u_t(\mathcal{C}_t|\mathcal{C}_1)\right\Vert_2^2,
\end{equation}
and the job left here is to factorise and parameterise this loss function to a computable form based on the geometric manifold $\mathcal{M}$. More specifically,

\begin{itemize}
    \item For the flow of chemical identities $\bm{A}$, the transition path is essentially a linear interpolations between $\bm{a}_0$ and $\bm{a}_1$ that follows $\bm{a}_t=(1-t)\bm{a}_0+t\bm{a}_1$, which has a tangent, or a transition probability that is $\bm{a}_1-\bm{a}_0$ to be matched in the model training.
    \item For the lattice parameters $\bm{L}$, one parameterises the cell lengths to follow a log-normal distributions such that only positive cell lengths are allowed, and the cell angles that follow the sigmoidal distributions are mapped to a logit function first so they become properly restrained within $[60^\circ,120^\circ]$ [see \cref{tab:continuous_prob_distributions}]. As such, the lattice parameters can also be treated as flowing linearly in the Euclidean space with a transition probability of $\bm{l}_1-\bm{l}_0$.
    \item For the fractional coordinates, which needs to be flowed on a torus $\mathbb{T}^3$, the trick of \emph{torus logarithmic mapping} is applied to derive the transition probability. The mechanisms behind work as follows. For any given component $i$ of a given atom, $f_i\in[0,1)$, and in order to represent the distance between two fractional coordinates, FlowMM first express this distance as an angular displacement of $\omega_i=2\pi(f_{1,i}-f_{0,i})$. A standard trigonometric trick is then applied to recovered the shortest signed angular distance as $\mathrm{atan2}(\sin\omega_i,\cos\omega_i)$, such that it is automatically wrapped within the interval of $(-\pi,\pi]$, and dividing this by $2\pi$ gives back the distance in terms of the fractional coordinates, resulting in the following torus logarithmic mapping:
    \begin{equation}
        \label{eq:torus_log_map}
        \mathrm{Log}_{f_{0,1}}(f_{1,i})=\frac{1}{2\pi}\mathrm{atan2}\left[ \sin\left( 2\pi(f_{1,i}-f_{0,i})  \right) , \cos\left( 2\pi(f_{1,i}-f_{0,i})  \right) \right].
    \end{equation}
    \cref{eq:torus_log_map} essentially defines the transition probability $u_t^F(\bm{F}_t|\bm{F}_1)$, that goes into the the corresponding loss for the fractional coordinates. 
\end{itemize}
\end{myquote}

\citet{luo2025crystalflow} developed  CrystalFlow, which combined the normalising flow with a symmetry-aware equivariant NN to gradually transport a simple distribution to one that of realistic materials. Here, it adapted the similar form of crystal representation $\mathcal{C}$ as in FlowMM,\cite{miller2024flowmm} except now: (1) for the lattice vectors $\bm{L}$, it adapted the polar decomposition approach as in DiffCSP++\cite{jiao2024space} into a set of rotationally invariant basis of $\bm{k}\in\mathbb{R}^6$ that will be transported by the flow model following a continuous linear trajectory, and (2) a continuous atom-type embedding of $\bm{A}\in\mathbb{R}^{a\times N}$ for all $N$ atoms are used to describe the atom type on each lattice position. Similar to FlowMM, CrystalFlow also trains a vector field model $v_{\boldsymbol{\theta}}(\mathcal{C}_t,\bm{y})$ to describe the transition probability, in which $\bm{y}$ encodes conditional information such as the chemical composition and/or pressures, and leads to a linear combinations of losses in all three components of the crystal representation, respectively. 

Finally, as the last example, \citet{wu2026dmflow} considered the design of normalising flow model for the generation of \emph{disordered} crystal structures (DMFlow), an important system that has strong implication in terms of PES sampling [\cref{sect:sampling_examples}], but had not been encountered in the previously discussed crystal generative models. To be more aligned with the above discussion on the FlowMM model, \textbf{Box VII} provides a more detailed account on how disorders are represented and mathematically manipulated in DMFlow.

\begin{myquote}
    \textbf{Box VII: design details of the DMFlow model for disordered crystal structure generations}

\footnotesize

\hspace{10pt} The DMFlow model was designed primarily to account for the presence of both \emph{substitutional} and \emph{positional} disorders in a crystal structure probabilistically, while preserving some of the approaches that had been developed in the FlowMM model for the transportation of perfect crystals on a continuous manifold. The keys lie in how the model represents both types of disorders in the crystal structures.

\hspace{10pt} \textbf{Representations of the disorders} Starting from the highest level, DMFlow seeks to represent a crystal containing disorders as $\mathcal{C}=(\bm{L},\bm{S},\bm{F},\bm{W},\bm{F}^\prime)$, in which $\bm{L}$ is the matrix of lattice vectors and the rest are matrices containing informations of site positions and corresponding occupancies. More specifically, a given tuple of $(\bm{s}_i,\bm{f}_i,\bm{w}_i,\bm{f}_i)$ is a representation of \emph{a single lattice site} that may contain the substitutional and/or positional disorders. More specifically, the model represents
\begin{itemize}
    \item Substitutional disorder at site $i$ by an occupancy vector
    \begin{equation*}
        \bm{s}_i=[s_{i,0},s_{i,1},\ldots,s_{i,D-1}]^\top\in\Delta^{D-1},
    \end{equation*}
    that represents the occupancy of a specific elements (among $D$ possible choices) at site $i$, which changes the standard site representations $(\bm{a}_i,\bm{f}_i)$ (\emph{e.g.} in the FlowMM model) into $(\bm{s}_i,\bm{f}_i)$ for a site that may contain substitutional disorder, and naturally $\bm{s}_i\to\bm{a}_i$ for one that contains no substitutional disorder.
    \item Positional disorder (in case of a binary positional disorder) at the same site is to be encoded by introducing a secondary position $\bm{f}^\prime$ that is associated with the primary position $\bm{f}$, with the corresponding site occupancy vector $\bm{w}_i=(w_{i,0},w_{i,1})\in[0,1]^2$, with $w_{i,0}+w_{i,1}=1$.
\end{itemize}

\hspace{10pt} \textbf{Model designs in the framework of Riemannian Flow} The flow model now needs to consider all five components in $\mathcal{C}$ that are to be transported jointly, from which the training loss can be derived as a linear combination of the losses for each one of the five components.

\hspace{10pt} More specifically, we need to design how the disorders are transported along both the forward and reverse trajectories. Consider an arbitrary disorder vector $\boldsymbol{\mu}=(\mu_1,\mu_2,\ldots,\mu_D)\in[0,1]^D$ with $\mu_k\geq 0$ $\forall \mu_k\in\boldsymbol{\mu}$ and $\sum_k\mu_k=1$. As such, performing flow na\"{i}vely on the Euclidean space would easily violate this normalisation constraint, and this is different from the flow of fractional coordinates, in which each component is restrained in $[0,1)$ independently from the other components. To overcome this problem, the model considers two points $\boldsymbol{\mu},\boldsymbol{\nu}\in [0,1]^D$ whose distance in between being considered by the Fisher-Rao distance:
\begin{equation*}
    d_{\mathrm{FR}}(\boldsymbol{\mu},\boldsymbol{\nu})=2\mathrm{arccos}\left(\sum_{k=1}^D\sqrt{\mu_k\nu_k}\right).
\end{equation*}
However, this metric becomes problematic when $\mu_k\to0$. This is overcome through a spherical parameterisation of $\pi(\boldsymbol{\mu})=\sqrt{\boldsymbol{\mu}}=(\sqrt{\mu_1},\sqrt{\mu_2},\ldots,\sqrt{\mu_D})$ which automatically satisfies the constraint of $\sum_k\pi(\mu_k)^2=1$, such that $\pi(\boldsymbol{\mu})\in\mathbb{S}_{+}^{D-1}$ is a probability vector ties on the positive orthant of a unit sphere, with the corresponding distance defined by:
\begin{equation*}
    d_{\mathbb{S}}=\mathrm{arccos}\left( \sum_{k=1}^D\sqrt{\mu_k\nu_k}\right).
\end{equation*}
This then enables the model to follow what has been done in FlowMM, that construct a torus logarithmic map to give the tangent vector for training as $\log_{\pi(\boldsymbol{\mu}_0)}[\pi(\boldsymbol{\mu}_1)]$ to transport the site occupancy vectors.

\hspace{10pt} \textbf{Modification of GNN architecture to account for the interaction between positional disorders} This becomes another unique contribution from this particular paper. Consider the case in which we have a set of two position disordered sites: $(\bm{f}_i,\bm{f}^\prime_i)$ with weights $(w_i,w_i^\prime)$ for site $i$, and $(\bm{f}_j,\bm{f}^\prime_j)$ with weights $(w_j,w_j^\prime)$ for site $j$. Because in a standard GNN, one must construct an edge between $i$ and $j$, instead of using the average positions between $\bm{f}$ and $\bm{f}^\prime$ to represent a given node, the edge was constructed by considering all four possible pairing between $(\bm{f}_i,\bm{f}^\prime_i)$ and $(\bm{f}_j,\bm{f}^\prime_j)$. Changing the notation to $\bm{f}_{i,a}$ with $a\in\{0,1\}$ such that $\bm{f}_{i,0}=\bm{f}$ and $\bm{f}_{i,1}=\bm{f}^\prime$, then edge construction and update between two positional disordered sites is realised through
\begin{equation*}
    \bm{e}_{ij}=\bigoplus_{a,b\in\{0,1\}}w_{ia}w_{jb}\Phi(\bm{f}_{i,a},\bm{f}_{j,b},\bm{L}).
\end{equation*}

\end{myquote}




\section{Conclusions and Future Outlooks}
\label{sec:conclusion_outlook}
With the advancements in computational power, modern scientific programming practices, community-driven open-source and data-sharing initiatives, computational materials modelling is now stepping away from simply computing the properties for a small set of structures for ones with some prescribed compositions, to a paradigm in which high-throughput computations and statistical tools (broadly encompassing both PES samplings and ML/AI techniques) are applied jointly to holistically investigate the structure-energy-property relationships for a given material system. In particular, this is empowered by the developments and applications of ML and AI tools that help us accelerate the energy and property computations, enhance our understanding of the computational results in high dimensional space, and ultimately, to be used to suggest new research directions, specifically manifests itself through the proposal of new, stable material candidates with targeted properties.

These integrated approaches have impact significantly how atomistic modelling is driving the development of materials in the new era. For examples, it has enabled the establishment of holistic structure-energy-property relationships to understand how atomistic structures translate into the physical properties of materials. It promotes interdisciplinary researches, in which mathematical and statistical tools are developed in a material-specific context to help us derive agnostic, unbiased and physically informed materials representations and associated algorithms, to facilitate the establishment of these understandings, that are subsequently captured to build models to accelerate the sampling and predictions of materials properties. 

These advancements have played a critical role in enabling us to tackle much complex problems in materials physics and chemistry that traditional methods suffer, predominantly limited by the approximations for the corresponding theories were adopted, in order to convert these theories into practically workable computational tools. We have already seen promising developments in breaking these limitations across multiple fronts, such as in the investigation of electrocatalytic surfaces under explicit solvents,\cite{Qin2023CO2Electroreduction,Sakong2018ElectricDoubleLayer,Andrade2020ProtonTransfer} electro- and magneto-caloric effects for energy conversions,\cite{graf2021unified} phonon-mediated superconductivity beyond harmonic phonons,\cite{zappacosta2025nonlinear} to name just a few.

By providing a pedagogical review of the established methodologies, by balancing the mathematical formality, theoretical rigorous, and intuitive explanations, we hope that the present monograph can facilitate the computational materials practitioners to further: (a) enhance their understanding on the working principles, capabilities, and limitations behind the existing methodologies in a conceptually unified framework, (b) build competency in the applications of the appropriate methods for solving problems in a more physically informed way, and (c) equip themselves to capture the emerging capabilities of modern AI tools for tackling challenging problems in materials physics and chemistry.

Looking forward,  there are still plenty of open questions to be answered, which are providing the momentum for the continuous developments of  atomistic molecular and materials modellings. In the final section of this monograph, some of these open questions that we believe are most relevant to the topics being covered here  shall be raised. One could easily appreciate that, many of these questions are cross-cutting problems that, each one of them  correlate multiple topics that have been discussed here. 

\textit{\textbf{What are the roles of transitional methods, such as empirical atom-atom potentials and DFT, in materials modelling, with respect to the development of more generalised, broadly applicable AI models?}} Modern deep-learning potentials, for examples, multiple rounds of MP operations in GNN-based potentials, are usually computationally demanding, and require GPU accelerations to make them competitive with non-accelerated empirical atom-atom potentials.\cite{leimeroth2025machine} This makes classical forcefields still possess competitive advantages when it comes to atomistically simulating very large systems close to a mesoscopic scale. Most importantly, even with foundational energy models,\cite{batatia2023foundation} our pursuit of scientific discoveries meaning that more than often we will be dealing with problems that are out-of-distributions in the training set of the foundation models. For examples, in the studies of anharmonic solids,\cite{yang2026mace} fast-ion conductors,\cite{ding2025liquid} where the ions frequently visit lattice positions/chemical environments that are not the equilibrium positions presented in the training data, structure and dynamics of magnetic materials where foundational models for these systems are still at the emerging state,\cite{ho2026equivariant} as well as the modelling of complex interfacial phenomena in which the charge, spin and orbital interactions interwind that leads to unconventional physical properties. Systems such as these still require detailed and reliable electronic structure theory calculations, to facilitate the modelling of these systems and data generations and validations for training future AI models. 

Most importantly, the continuous developments of the electronic structure theory, which is heavily grounded by the fundamental physical principles, rather than data representations, still play a pivotal role in facilitating the understanding of exotic physical phenomena in material systems. At the same time, we are yet to discover more routes to incorporate these fundamental knowledge in materials physics and chemistry into future AI models. Hence, it is still very immature to think AI is taking over the traditional tools in atomistic modellings.  

\textit{\textbf{How to know if one have sampled enough?}} This is probably one of the most persistent and challenging problem to answer in all investigations that involve PES sampling. Certainly this is a very problem-specific question, meaning to a large extent, it depends on the nature of the problems that we are trying to solve. If one is trying to determining the free energies of high-entropy alloys, in which we know the material does not confine into a specific form of atomic arrangements, the convergence of the statistical sampling can be gauged by monitoring the evolution of the probability distributions of the energies of the local minima being sampled. This may be considered as a relatively \emph{loose criterion}. On the other hand, in the case such as organic crystal structure predictions,\cite{case2016convergence}  where there is a strong interest in locating the global minimum (a \emph{strong} criterion), then one may have to keep running the sampling until the same global minimum has been repetitively visited for multiple times. 

The generally lack of a robust check for the convergence also affects our capabilities in achieving predictive computations of certain materials properties through statistical sampling, such as the computation of lattice thermal conductivities in real space with Green-Kubo method [\cref{tab:properties_from_MD}].\cite{knoop2023ab} In materials where long wavelength $|\bm{q}|\to 0$ phonons (that needs to be modelled with an indefinitely large supercell) can be  strongly scattered  to suppress $\kappa_L$, simply using larger and larger supercells in the calculation may never converge $\kappa_L$ or converge it to a wrong value, and the Green-Kubo method itself does not necessarily provide us with a diagnostic that we are dealing with such a kind of materials.  

\textit{\textbf{Can generative modelling help us to sample better?}} It is therefore natural to ask, whether the development of generative models can help us to sample better. We are still yet to see more evidences that will allow us to answer this questions in a more objective way. Nevertheless, do not forget that, on the very fundamental level, in method such as the score-matching diffusion model,\cite{song2019generative} the generative process is still largely \emph{guided} by a forcefield\cite{cheng2024response} that are trained predominantly on the equilibrium distributions of stable materials. Whether this imposes a limit on our capability in exploring the chemistry and structural landscape of the materials, is probably worth further investigations. Answering this question, in the context of this monograph, would help us to unleash the greater promises of the generative models that we are yet to capitalise. 

\emph{\textbf{What are the future perspectives and promises behind the development of DFT and higher-level electronic structure methods for materials modelling?}} The 2022 ``DFT exchange'' paper,\cite{dftexchange} has gathered the leading theoreticians behind the developments of DFT and related theoretical chemistry methods to share their views behind the current status and future trends in DFT from a theoretical perspective. The present monograph is partially inspired by this unique publication, and trying to focus on a more practical side on the applications of DFT. In this aspect, some of the ongoing challenges in electronic structure theory that are relevant to the atomistic materials modellings include (but are not necessarily limited to) the followings:

(a) \emph{What are the perspectives on the development of more accurate exchange-correlation functionals?} Perhaps of the most interests to materials scientists would be the perspective of having a universally good functional to achieve reliable predictions on the electronic band gaps across a wide range of materials, by overcoming the self-interaction errors but at a computational cost that is comparable to GGA. Tackling this problem would certainly depending on the viewpoints of the DFT developers on whether one should achieve a universal functionals that works for all systems, or one should tailor-design them  for different categories of systems with their unique physics.\cite{mardirossian2016accurate} This is not a judgement that we tend to make or suggest here.

(b) \emph{What are the roles of the experimental data for fitting and benchmarking density functionals?} 
In the development and testing of new density functionals, the role of experimental data cannot be neglected, as the ultimate goal of atomistic modelling is for us to guide the experimental discoveries of materials. Nevertheless, one should always bear in mind that, (i) We should not expect a theory or model to be able to fully account all the aspects that may influence the experimental outcomes, as a complex model itself may not lead to valuable and explainable insights. (ii) The experimental data itself, even carefully curated, often need to be corrected, such as the thermal effects,\cite{otero2012benchmark} before they can be used to benchmark the theory, in which those effects are not accounted for. In addition, these experimental data should ideally be collated and corrected with a consistent set of settings, which  impose additional challenges on the experimental designs. 

Certainly different material systems and measurement conditions may require different correction scheme, and whether the contributions from different factors, such as defects, can always be isolated, will remain a highly material-dependent problem. Hence we expect the development on the aspect of experimental data may be lying something in the middle. The same argument  probably also applies to as to what extent future (foundational) AI models for materials can be trained on experimental data.

(c) \emph{AIMD only account for dynamics of nuclei under thermal excitation, how about electrons?} Thermal excitations not only affects the dynamics of nuclei, the electronic subsystems are also excited and can play a critical role in determining a variety of excited state behaviours in molecules and materials. However, such effect is not accounted in ground-state DFA. Recent developments on workable implementations of ensemble DFAs\cite{valone1980one} provides the community a workable route to tackle this problem, and we are yet to see more exciting and practical applications of such methods in molecular and materials modellings.

(d) \emph{What are the future directions of modelling the response behaviours under external stimuli?} Fundamentally, this question is linked to the excited state properties of materials, which have not been extensively discussed in the monograph. Amongst all the possible response properties of materials under external fields, materials scientists would probably argued that the modelling of magnetic responses in materials is properly one of the most interesting aspects to be looked at. The theoretical challenges behind the formulation of DFT in a magnetic field are still nevertheless enormous,\cite{DFT_exchange} so we are still to see where the future promises in this aspect are lying.

(e) \emph{How about the many-body methods?} As we have already demonstrated here, GW method has become increasingly important and popular in modelling materials properties in which many-body effect plays a critical role, such as the optical properties of 2D materials. A more accurate, faster and scalable GW methods, probably with the help of AI models,\cite{hou2024unsupervised,hou2025data} that are more integrated with computational workflow managements,\cite{ganose2025atomate2} would be important to make these methods more accessible to materials modelling practitioners. To make this happens would also largely depend on one final question raised as following.

(f) \emph{How can we better integrate atomistic  materials  modelling into the existing educational program of chemical and materials science?}  Successfully achieving this will be critical to drive the future developments of this field. A critical challenge is likely to be how the educators can convey and deliver the difficult concepts in physics, highly abstract ideas in mathematics and practical skills in computational programming in an approachable way, such that the students can slowly build up their confidences in mastering the many faceted fundamentals that underpins atomistic materials modellings, which will ultimately be translated into algorithms and research software to deliver meaningful and valuable outcomes. 

\textbf{\emph{What are the future promises in the application of atomistic modelling in understanding and predicting materials' properties?}} With the increase in computational power, the development of new materials theory and integrated workflow managements for high-throughput computation, together with better education and advocation of good atomistic modelling practices, we should see an increasing adaptation and integration of the existing and future tools in solving many-body problems in materials science beyond a mean-field or lower-order perturbative approaches. This will significantly wider the scope over the materials systems and problems we can tackle. Many of these had been encountered in the previous discussions, such as the thermal properties of highly anharmonic solids, superconductivities mediated by anharmonic phonons, and many others. Advances in this front will also enable us to curate more highly specialised and accurate dataset for training ML/AI models that cover a wider range of materials properties.

We now move on to raise some more specific questions related to the development of ML/AL in materials science.

\textbf{\emph{How should we treat the `zoo' of AI models for materials modelling? }} DFT practitioners are often said to be challenged by a `zoo'\cite{goerigk2017look} of exchange-correlation functionals that have been developed over many decades, each one being developed as some approximation to the complex quantum behaviours of electrons, that are relevant in the chemical systems (\emph{either broadly or specifically to some specific ones}). Their successes and pitfalls are identified through extensive benchmarks across a wide chemical space.

As one can see in \cref{tab:equivariant_GNNS} and \cref{sec:generative_AI}, the number of AI models developed for atomistic molecular and materials modelling is also rapidly increasing over the past few years, with many of them not only being developed by chemists and materials scientists, but also increasingly by mathematicians and computer scientists who are not traditionally working in this field. In response to this, there is an increasing focus in benchmarking the performance of different AI models against a consistent training and validation set to facilitate the community in the selection of best performing models for materials modelling,\cite{leimeroth2025machine,eastman2026accuracy,yu2024systematic,focassio2025performance,deng2025systematic,riebesell2025framework} which is definitely  much needed information. However, a more deeper scientific questions behind these benchmarking is that whether some numerical performance indicators (such as prediction errors), though simple to compare, may also lead to a similar problem of `\emph{choice by popularity}' with the use of B3LYP\cite{kruse2012standard} in DFT practices, as more users are converged into the `\emph{best performing}' models without the desire to properly understand why these models perform well. We envisage that more analyses, tools and frameworks that are formulated and developed to facilitate the community in understanding how subtle differences behind model designs can lead to the differences in model performances will definitely be very beneficial.

\textbf{\emph{Are foundational models for materials modelling truly `foundational'?}} Needless to say, the concept of foundational models are borrowed from the Large Language Models,\cite{vaswani2017attention} that lead to successful applications such as ChatGPT. It is foundational in a sense that it is trained on all the available text-based resources so it can infer to almost all range of available human knowledge. In a sense that it is a universal model. 

However, it is very obvious that in materials physics and chemistry, we do not yet have a universal theory (or an approximation of which), that is applicable to all application scenario for all existing and to-be-discovered materials. Hence, the word foundational model when applied to this specific scientific disciplinary may lead to some incorrect perceptions on where we are currently standing on.  At the time of writing this monograph, the problems that are common to all ML/AI models are still highly relevant to foundational models, being (i) the ability for the models to extrapolate is highly correlated with the data on which the models are being trained on,\cite{yang2026mace,fu2022forces} and (ii) the models' accuracies are also highly correlated with the underlying theory (\emph{e.g.} the exchange-correlation functionals\cite{shi2026influence}) that are adopted to generate the training data. Hence, one still needs to be cautious when using the foundational models, as one would approach the problem of choosing appropriate DFAs to use in a similar way. Nevertheless, foundational models have indeed provided us a faster alternative to DFT for pre-screening the material space to facilitate the subsequent, more accurate downstream  tasks in atomistic modellings.

\textbf{\emph{To what extent the trained AI models can help us discovering new knowledge in materials science at the atomistic level?}}
NN are often considered as another type of `black-box' method that maps an input to some desired output, whereas how and what has been learnt by the models remain largely under-explored. Models such as GNN\cite{duval2023hitchhiker} and attention layers,\cite{vaswani2017attention} in principle, are trying to encode atom-atom interactions into a simpler latent space that may facilitate the understanding of which interactions are important, for example, in determining materials stabilities. For examples, \citet{hafizi2026towards} analysed the attention matrices learnt by the NN model and tried to correlate them with the stabilities of organic crystal structures.
On the other hand, one may also argue to what extent the information encoded in these matrices are meaningful as they may be just  an ansatz corresponding to a plausible local minimum on a high dimensional landscape of the loss function. We believe that the continuous developments in the `\emph{Explainable AI}'\cite{minh2022explainable} will offer us with more opportunities in solving this problem. Another promising direction is behind the development of \emph{neural operators},\cite{azizzadenesheli2024neural} which aims at better incorporating existing physical knowledges, such as electrostatics, electron densities and wavefunctions, into the model design to enhance their capabilities in solving more physically relevant tasks, this could simultaneously help to make the model becoming more explainable. 


\textbf{\emph{Will the AI models eventually have the capability to discover equivariance and the high-level (such as the graph structures) of materials themselves, rather being enforced as an inductive bias?}} The final questions that we believe is worth asking, is on the importance of enforcing symmetry in the design of AI models for materials modelling, which, in this monograph, we have covered extensively on how symmetry can be incorporated in both ML/AI models to enhance models' data efficiency and abilities to extrapolate. However, the counter argument behind this is that, in a truly `data-driven' paradigm, the model should be able to discover these symmetries and enforcing them, without explicit data augmentations,\cite{duval2023hitchhiker} which can be noted in some recent research being devoted to address such a problem.\cite{domina2026unconstrained,bigi2026pushing,langer2024probing} 

We believe that the ongoing theoretical advances across all fields related to atomistic materials simulations in the coming years will continuous to provide us with exciting discoveries related to the above questions. 

\section*{Appendix}

\appendix
\renewcommand{\thesubsection}{\Alph{section}.\arabic{subsection}}
\renewcommand{\thesubsubsection}{\Alph{section}.\arabic{subsection}.\arabic{subsubsection}}

\renewcommand{\thefigure}{A\arabic{figure}}
\setcounter{figure}{0} 


\section{Supervised Machine Learning Models}
\label{appendix:supervised_ML}
\subsection{Linear Regression Models}\label{appendix:linear_regression}

In this section, we will provide detailed explanations on a few prominent linear regression ML models, including Ordinary Least Squares (OLS), Ridge Regression,\cite{hoerl1970ridge} LASSO (Least Absolute Shrinkage Operator) Regression\cite{tibshirani1996regression}  and Logistic Regressions.\cite{cox1958regression} Our focus here is to provide a comprehensive understanding of these models from the perspective of Bayesian statistics, and how this leads to the common forms of the loss functions used in training these models.

\subsubsection{Ordinary Least Squares Regression}\label{appendix:ols}

For the OLS regression, we consider the problem of fitting a linear model $y=\bm{x}^{\top}\boldsymbol{\beta}$, with $\boldsymbol{\beta}$ being the model parameter, to a set of data points $\{(\bm{x}_i,y_i)\}_{i=1}^{N}$, with $\bm{x}_i \in \mathbb{R}^d$ being a set of features and $y_i \in \mathbb{R}$ being the corresponding target variables, which, for simplicity, we assume to be scalars. In the probabilistic framework, instead of thinking the target variables take their deterministic values at all $y_i$, we think of them as being normally distributed around the target values with a common variance $\sigma^2$, such at $y_i\sim q_i(y)=\mathcal{N}(y|y_i,\sigma^2)$. In this csae, the goal of the OLS regression is to arrive at a model distribution $p_i(y)=\mathcal{N}(y|\bm{x}^{\top}\boldsymbol{\beta},\sigma^2)$ conditioned on the model parameters $\boldsymbol{\beta}$ that can best approximate the target distribution $q_i(y)$. 

Under this view, the training of the OLS regression model can be formulated as the process of minimising the KL divergence between the model distribution $p_i(y)$ and the target distribution $q_i(y)$, which, for two normal distributions, can be computed in closed-form as
\begin{equation}
\label{eq:ols_kl_div}
\mathcal{D}_{\mathrm{KL}}(q_i\Vert p_i) = \mathcal{D}_{\mathrm{KL}}(\mathcal{N}(\mu_q,\sigma^2)\Vert \mathcal{N}(\mu_p,\sigma^2)) = \frac{(\mu_q-\mu_p)^2}{2\sigma^2},
\end{equation}
where $\mu_q$ and $\mu_p$ are the means of the two normal distributions, which, in our case, are $y_i$ and $\bm{x}_i^{\top}\boldsymbol{\beta}$, respectively. Therefore, the training of the OLS regression model can be formulated as minimising the following loss function:
\begin{equation}
\label{eq:ols_loss}
\hat{\boldsymbol{\beta}} = \argmin_{\boldsymbol{\beta}} \sum_{i=1}^{N} \mathcal{D}_{\mathrm{KL}}(q_i\Vert p_i) =  \argmin_{\boldsymbol{\beta}}   \frac{1}{2\sigma^2}\sum_{i=1}^{N} \left\Vert y_i-\bm{x}_i^{\top}\boldsymbol{\beta}\right\Vert_2^2,
\end{equation}  
in which the constant term $\frac{1}{2\sigma^2}$ can be ignored during the optimization process, leading to the commonly used L2-loss function for OLS regression. Practially, the above minimisation problem can be solved in closed-form, by first expanding the loss function as (in which we concatenate the data points into a data matrix $\bm{X}\in \mathbb{R}^{N\times d}$ and a target vector $\bm{y}\in \mathbb{R}^N$):
\begin{equation}
    \label{eq:ols_loss_expand}
    \mathcal{L}(\boldsymbol{\beta})=(\bm{y}-\bm{X}^\top\boldsymbol{\beta})^\top (\bm{y}-\bm{X}^\top\boldsymbol{\beta}) = \boldsymbol{\beta}^\top \bm{X}\bm{X}^\top \boldsymbol{\beta} - 2\bm{y}^\top \bm{X}^\top \boldsymbol{\beta} + \bm{y}^\top \bm{y},
\end{equation}
and we take the gradeint of the loss function with respect to $\boldsymbol{\beta}$  using the fact that $\nabla_{\boldsymbol{\beta}} (\boldsymbol{\beta}^\top \bm{A} \boldsymbol{\beta}) = (\bm{A}+\bm{A}^\top)\boldsymbol{\beta}=2\bm{A}\boldsymbol{\beta}$ for $\bm{A}=\bm{A}^\top$ and $\nabla_{\boldsymbol{\beta}} (\boldsymbol{\beta}^\top\bm{a} ) = \bm{a}$ for $\bm{a}$ being a vector, which gives us 
\begin{equation}
    \label{eq:ols_loss_derivative}
    \nabla_{\boldsymbol{\beta}}\mathcal{L}(\boldsymbol{\beta}) = 2\bm{X}\bm{X}^\top \boldsymbol{\beta} - 2\bm{X}^\top\bm{y}.
\end{equation}
Setting the above gradient to zero, we obtain $\bm{X}^\top\bm{X}\boldsymbol{\beta}=\bm{X}^\top\bm{y}$, where $\bm{X}^\top\bm{X}$ is the covariance matrix of the data, and $\bm{X}^\top\bm{y}$ is the covariance between the data and the target variables. The OLS problem will have the following closed-form solution \emph{when the covariance matrix $\bm{X}^\top\bm{X}$ is invertible:}
\begin{equation}
    \label{eq:ols_solution}
    \hat{\boldsymbol{\beta}}=(\bm{X}^\top\bm{X})^{-1}\bm{X}^\top\bm{y}.
\end{equation}

\subsubsection{Ridge Regression}\label{appendix:ridge_regression}
The most straightforward way to extend the OLS regression is to add a regularisation term to the loss function, to avoid the problem of non-invertibility of the covariance matrix $\bm{X}^\top\bm{X}$, which can happen when the data is high-dimensional or when there are multicollinearity among the features. This leads to the following analytical expression for the optimum model parameters for the Ridge regression model:
\begin{equation}
    \label{eq:ridge_solution}
    \hat{\boldsymbol{\beta}}=(\bm{X}^\top\bm{X}+\lambda \bm{I})^{-1}\bm{X}^\top\bm{y},
\end{equation}  
where $\lambda$ is the regularisation parameter that controls the strength of the regularisation, and $\bm{I}$ is the identity matrix. In what follows, we shall provide two derivations for the above solution, one from the perspective of directly aligning the optimised model parameter distribution with a target prior, and one from the perspective of \emph{variational inference}.\cite{blei2017variational}

\paragraph{Alignment with a Target Prior}
This viewpoint allows us to view the training of the linear ridge regression model from the perspective of optimising the probability distribution of the model parameters $\boldsymbol{\beta}$ with respect to a target prior distribution, from which a posterior distribution of the model parameters can be obtained, where the mean of the posterior distribution corresponds to the optimal  parameters for the linear model $y=\bm{x}^\top\boldsymbol{\beta}$. 

We begin with a \emph{\textbf{prior distribution}} over the model parameters $\boldsymbol{\beta}$, which represents our guess on the range in which the model parameters should reside, and is commonly chosen to be a zero-mean Gaussian distribution with a diagonal covariance matrix to allow easy sampling:
\begin{equation}
    \label{eq:linear_ridge_param_prior}
    \boldsymbol{\beta}\sim \mathcal{N}(\bm{0},\tau^2\bm{I})\quad \Longleftrightarrow \quad p(\boldsymbol{\beta})\propto\exp\left(-\frac{\Vert\boldsymbol{\beta}\Vert_2^2}{2\tau^2}\right).
\end{equation}
Based on this prior distribution, we can write down the \textbf{\emph{likelihood}} of observing the output $y$ with given input $\bm{x}$ and model parameters $\boldsymbol{\beta}$ as a conditional nornal distribution:
\begin{equation}    \label{eq:linear_ridge_likelihood}
    p(y|\bm{x},\boldsymbol{\beta}) = \mathcal{N}(y|\bm{x},\boldsymbol{\beta},\sigma^2) \quad \Longleftrightarrow \quad p(y|\bm{x},\boldsymbol{\beta})\propto \exp\left(-\frac{\left\Vert y-\bm{x}^\top\boldsymbol{\beta}\right\Vert_2^2}{2\sigma^2}\right).        
\end{equation}
Using the Baye's theorem [\cref{eq:bayes_rule}], we can obtain the \textbf{\emph{posterior distribution}} of the model parameters after seeing some training data $\{(\bm{x}_i,y_i)\}_{i=1}^{N}$ as
\begin{equation}    \label{eq:linear_ridge_param_posterior}
    p(\boldsymbol{\beta}|\bm{X},\bm{y}) \sim p(\bm{y}|\bm{X},\boldsymbol{\beta})p(\boldsymbol{\beta})   \propto \exp\left(-\frac{1}{2\sigma^2}\left\Vert \bm{y}-\bm{X}_i^\top\boldsymbol{\beta}\right\Vert_2^2 - \frac{\Vert\boldsymbol{\beta}\Vert_2^2}{2\tau^2}\right),
\end{equation}
which is a simple product of two Gaussian distributions, and thus is also a Gaussian distribution. Primarily, the goal of training the linear ridge regression model is to find the optimal model parameters $\hat{\boldsymbol{\beta}}$ that maximise the posterior distribution $p(\boldsymbol{\beta}|\bm{X},\bm{y})$. Practically, this is often casted into a minimisation problem by taking the negative logarithm of the posterior distribution:
\begin{equation}
    \label{eq:log_likelihood_linear_ridge}
    \mathcal{L}(\boldsymbol{\beta})=-\log p(\boldsymbol{\beta}|\bm{X},\bm{y}) = \frac{1}{2\sigma^2} \left\Vert \bm{y}-\bm{X}_i^\top\boldsymbol{\beta}\right\Vert_2^2 + \frac{\Vert\boldsymbol{\beta}\Vert_2^2}{2\tau^2}+\mbox{const},
\end{equation}
in which the constant term comes from the renormalisation constant of the posterior distribution, and can be ignored during the optimization process, so that we arrive at the commonly used loss function for training the linear ridge regression model:
\begin{equation}\label{eq:linear_ridge_loss}
    \hat{\boldsymbol{\beta}} = \argmin_{\boldsymbol{\beta}} \mathcal{L}(\boldsymbol{\beta}) = \argmin_{\boldsymbol{\beta}} \left[\left\Vert \bm{y}-\bm{X}_i^\top\boldsymbol{\beta}\right\Vert_2^2 + \lambda\Vert\boldsymbol{\beta}\Vert_2^2 \right],       
\end{equation}
where $\lambda=\sigma^2/\tau^2$ is the regularisation parameter that controls the strength of the regularisation. By taking $\nabla_{\boldsymbol{\beta}}\mathcal{L}(\boldsymbol{\beta})=-2\bm{X}^\top\bm{y}+2(\bm{X}^\top\bm{X}+\lambda\bm{I})\boldsymbol{\beta}=0$, we can obtain the closed-form solution for the optimal model parameters as shown in \cref{eq:ridge_solution}. 

\paragraph{Variational Inference for Linear Ridge Regression}

The above discussions of the linear ridge regression model gives a more rigid derivations for the loss function \cref{eq:ridge_solution} of the linear ridge regression model. However, it only leads to a point estimate of the model parameters $\hat{\boldsymbol{\beta}}$, without providing any information on the uncertainty of the model parameters. This problem can be solved by using the variational inference framework, which allows us to obtain the full posterior distribution of the model parameters $p(\boldsymbol{\beta}|\bm{X},\bm{y})$ instead of just a point estimate. Let us assume that our prior distribution for the model parameter $\boldsymbol{\beta}$ follows some distribution that can be written as $q(\boldsymbol{\beta})$, and we shall explore how the optimum posterior distribution $p(\boldsymbol{\beta}|\bm{X},\bm{y})$ is affected by the choice of the prior distribution $q(\boldsymbol{\beta})$.

In the variational inference framework, we start by noting that what allows us to estimate the posterior distribution $p(\boldsymbol{\beta}|\bm{X},\bm{y})$ is to fitst invoke the Bayes' theorem by writing:
\begin{equation}
    \label{eq:bayes_linear_variational_inf}
p(\boldsymbol{\beta}|\bm{X},\bm{y})=\frac{p(\bm{y},\boldsymbol{\beta}|\bm{X})}{p(\bm{y}|\bm{X})},
\end{equation}
where $p(\bm{y},\boldsymbol{\beta}|\bm{X})$ is the joint distribution of the data and the model parameters, and $p(\bm{y}|\bm{X})$ is what we called the \emph{evidence}, because it directly tells how the obaservable should distribute upon a given input $\bm{X}$. However, the evidence $p(\bm{y}|\bm{X})$ is often intractable to compute because it is formally given by the marginalisation of the joint distribution over the model parameters $\boldsymbol{\beta}$,
\begin{equation}
    \label{eq:bayes_linear_variational_inf_evidence_margnialisation}
p(\bm{y}|\bm{X}) = \int p(\bm{y},\boldsymbol{\beta}|\bm{X}) \mathrm{d}\boldsymbol{\beta}.
\end{equation}
One can approximately thinks that this joint distribution along the $\boldsymbol{\beta}$-dimension is a delta function that is non-zero only at the optimal model parameters $\hat{\boldsymbol{\beta}}$, so unless the ridge model has been trained, which is what we are aiming for, the evidence \cref{eq:bayes_linear_variational_inf_evidence_margnialisation} remains unsolved. To solve this problem, we can introduce a variational distribution $q(\boldsymbol{\beta})$ to approximate the true posterior distribution $p(\boldsymbol{\beta}|\bm{X},\bm{y})$. This allows us to sample $\boldsymbol{\beta}$ from the variational distribution $q(\boldsymbol{\beta})$ and use these samples to estimate the evidence $p(\bm{y}|\bm{X})$ by using the following identity:
\begin{equation}    \label{eq:bayes_linear_variational_inf_evidence_estimation}
p(\bm{y}|\bm{X}) = \int q(\boldsymbol{\beta}) \frac{p(\bm{y},\boldsymbol{\beta}|\bm{X})}{q(\boldsymbol{\beta})} \mathrm{d}\boldsymbol{\beta} = \mathbb{E}_{ q(\boldsymbol{\beta})}\left[\frac{p(\bm{y},\boldsymbol{\beta}|\bm{X})}{q(\boldsymbol{\beta})}\right],
\end{equation}
taking the logarithm of both sides, we can obtain the following expression for the log-evidence:
\begin{equation}    
\label{eq:bayes_linear_variational_inf_log_evidence}
\log p(\bm{y}|\bm{X}) = \log \mathbb{E}_{ q(\boldsymbol{\beta})}\left[\frac{p(\bm{y},\boldsymbol{\beta}|\bm{X})}{q(\boldsymbol{\beta})}\right].   
\end{equation}
This allows us to apply the Jensen's inequality to obtain  the \textbf{evidence lower bound} (ELBO) for the log-evidence:
\begin{equation}
    \label{eq:bayes_linear_variational_inf_ELBO}
\log p(\bm{y}|\bm{X}) \geq \mathbb{E}_{q(\boldsymbol{\beta})}\left[\log\frac{p(\bm{y},\boldsymbol{\beta}|\bm{X})}{q(\boldsymbol{\beta})}\right] = \underbrace{\mathbb{E}_{q(\boldsymbol{\beta})}[\log p(\bm{y},\boldsymbol{\beta}|\bm{X})] - \mathbb{E}_{q(\boldsymbol{\beta})}[\log q(\boldsymbol{\beta})]}_{\mathrm{ELBO}(\boldsymbol{\beta})}.
\end{equation}
Comparing the above ELBO with the log-evidence  with the logarithm of $p(\bm{y}|\bm{X})$ from the Bayes' theorem [\cref{eq:bayes_linear_variational_inf}]:
\begin{equation*}
    \log p(\bm{y}|\bm{X}) = \log p(\bm{y},\boldsymbol{\beta}|\bm{X}) - \log p(\boldsymbol{\beta}|\bm{X},\bm{y}),
\end{equation*}
in which we further introduce the variational distribution $q(\boldsymbol{\beta})$ and take the expectation with respect to $q(\boldsymbol{\beta})$ on both sides, we can obtain the following expression for the log-evidence:
\begin{equation}
    \label{eq:bayes_linear_variational_inf_expectation_variational_dist}
\mathbb{E}_{q(\boldsymbol{\beta})}[\log p(\bm{y}|\bm{X})] = 
\underbrace{\mathbb{E}_{q(\boldsymbol{\beta})}\Big[\log p(\bm{y},\boldsymbol{\beta}|\bm{X}) - \log q(\boldsymbol{\beta}) \Big]}_{\mathrm{ELBO}(\boldsymbol{\beta})} + \underbrace{
\mathbb{E}_{q(\boldsymbol{\beta})}\Big[\log q(\boldsymbol{\beta}) - \log p(\boldsymbol{\beta}|\bm{X},\bm{y})\Big]}_{\mathcal{D}_{\mathrm{KL}}(q(\boldsymbol{\beta})\Vert p(\boldsymbol{\beta}|\bm{X},\bm{y}))}.
\end{equation}
At this stage we can write the conditional distribution on the left hand side equivalently as $p(\bm{y}|\bm{X}) = p(\bm{y}|\bm{X},\boldsymbol{\beta})$ because the evidence is independent of the model parameters $\boldsymbol{\beta}$, such that the above expressions leads to the interpretation of the model trainin as a process of maximising the evidence:
\begin{equation}
    \label{eq:bayes_linear_variational_evidence_maximisation}
\argmax_{\boldsymbol{\beta}} \mathbb{E}_{q(\boldsymbol{\beta})}[\log p(\bm{y}|\bm{X},\boldsymbol{\beta})] = \argmax_{\boldsymbol{\beta}} \mathbb{E}_{q(\boldsymbol{\beta})} \Big[\mathrm{ELBO}(\boldsymbol{\beta}) + \mathcal{D}_{\mathrm{KL}}(q(\boldsymbol{\beta})\Vert p(\boldsymbol{\beta}|\bm{X},\bm{y}))\Big].
\end{equation}      
To recast the above maximisation problem into a minimisation problem, we can take the negative of the above expression, and making the negsative of ELBO as the subject of the minimisation, we can obtain the following expression for the loss function for training the linear ridge regression model:
\begin{equation}
    \label{eq:bayes_linear_variational_inf_loss}        
\hat{\boldsymbol{\beta}} = \argmin_{\boldsymbol{\beta}} \mathbb{E}_{q(\boldsymbol{\beta})}[-\mathrm{ELBO}(\boldsymbol{\beta})] = \argmin_{\boldsymbol{\beta}} \mathbb{E}_{q(\boldsymbol{\beta})}\Big[-\log p(\bm{y}|\bm{X},\boldsymbol{\beta}) + \mathcal{D}_{\mathrm{KL}}(q(\boldsymbol{\beta})\Vert p(\boldsymbol{\beta}|\bm{X},\bm{y})) \Big].
\end{equation}

Here, the first term on the right hand side of \cref{eq:bayes_linear_variational_inf_loss} leads directly to the L2-loss term in the loss function as in \cref{eq:ols_loss} by assuming a Gaussian likelihood for the data distribution. The second term on the right hand side of \cref{eq:bayes_linear_variational_inf_loss} leads to the regularisation term in the loss function and its expression depends on the choice of the variational distribution $q(\boldsymbol{\beta})$. In the most simplest and general form, we assume that the variational distribution $q(\boldsymbol{\beta})$ is a Gaussian distribution $q(\boldsymbol{\beta})\sim\mathcal{N}(\bm{m}_{\boldsymbol{\beta}},\bm{s}_{\boldsymbol{\beta}})$, which takes into account the uncertainty of the model parameters $\boldsymbol{\beta}$ by introducing a covariance matrix $\bm{s}_{\boldsymbol{\beta}}$, and has a mean $\bm{m}_{\boldsymbol{\beta}}$ that corresponds to the optimal model parameters. Further assuming that the true posterior distribution $p(\boldsymbol{\beta}|\bm{X},\bm{y})$ is also a Gaussian distribution $p(\boldsymbol{\beta}|\bm{X},\bm{y})\sim \mathcal{N}(\bm{0},\tau^2\bm{I})$, we can obtain the following closed-form expression for the KL divergence between the variational distribution and the true posterior distribution:
\begin{equation}
    \label{eq:bayesian_linear_KL_div}
    \mathcal{D}_{\mathrm{KL}}(q\Vert p)= \frac{1}{2}\left[\frac{\mathrm{tr}(\bm{s}_{\boldsymbol{\beta}})}{\tau^2} + \frac{\bm{m}_{\boldsymbol{\beta}}^\top \bm{m}_{\boldsymbol{\beta}}}{\tau^2} - d + \log \frac{\tau^{2d}}{\det(\bm{s}_{\boldsymbol{\beta}})} \right],
\end{equation}
in which $d$ is the dimension of the model parameters $\boldsymbol{\beta}$. For ridge regression, we have a single prior belief on the model parameters $\boldsymbol{\beta}$, such that $q(\boldsymbol{\beta})=\delta(\boldsymbol{\beta}-\bm{m}_{\boldsymbol{\beta}})$, with $\bm{s}_{\boldsymbol{\beta}}\to\bm{0}$. Although this will blow up the KL divergence to infinity due to the log term, the key point is that the KL divergence will be dominated by the term $\frac{\bm{m}_{\boldsymbol{\beta}}^\top \bm{m}_{\boldsymbol{\beta}}}{\tau^2}$, which leads to the L2 regularisation term ($\Vert\bm{m}_{\boldsymbol{\beta}}\Vert_2^2/2\tau^2$) in the loss function \cref{eq:linear_ridge_loss}. This is how the regularisation term in the loss function for training the linear ridge regression model can be derived from the perspective of variational inference. The more interesting case is when we start working with the Gaussian variational distribution $q(\boldsymbol{\beta})\sim\mathcal{N}(\bm{m}_{\boldsymbol{\beta}},\bm{s}_{\boldsymbol{\beta}})$, in which we no longer have a single point estimate for the model parameters $\boldsymbol{\beta}$, but instead have a distribution over the model parameters that captures the uncertainty of the model parameters. It can be shown that solving the optimisation problem in \cref{eq:bayes_linear_variational_inf_loss} gives the following analytical solutions:
\begin{equation}
    \label{eq:bayes_linear_variational_best_solutions}
\bm{S}_{\boldsymbol{\beta}}^{-1} = \frac{1}{\sigma^2}\bm{X}^\top\bm{X} + \frac{1}{\tau^2}\bm{I},\quad \bm{m}_{\boldsymbol{\beta}} = \bm{S}_{\boldsymbol{\beta}}\left(\frac{1}{\sigma^2}\bm{X}^\top\bm{y}\right),
\end{equation}
which one can show that  $\bm{m}_{\boldsymbol{\beta}}$ is of the same forms as  \cref{eq:ridge_solution}. The above solutions allow us to extimate the uncertainties when making predictions with the linear ridge regression model, which is not possible with the OLS regression model. For a given input $\bm{x}_{*}$, the predictive distribution of the output $y_*$ is given by the following Gaussian distribution:
\begin{equation}    \label{eq:bayes_linear_variational_inf_predictive_dist}
p(y_*|\bm{x}_*,\bm{X},\bm{y}) = \int p(y_*|\bm{x}_*,\boldsymbol{\beta})p(\boldsymbol{\beta}|\bm{X},\bm{y}) \mathrm{d}\boldsymbol{\beta} = \mathcal{N}(y_*|\bm{x}_*^\top\bm{m}_{\boldsymbol{\beta}},\bm{x}_*^\top\bm{S}_{\boldsymbol{\beta}}\bm{x}_*+\sigma^2),
\end{equation}
in which the mean of the predictive distribution is given by $\bm{x}_*^\top\bm{m}_{\boldsymbol{\beta}}$, and the variance of the predictive distribution is given by $\bm{x}_*^\top\bm{S}_{\boldsymbol{\beta}}\bm{x}_*+\sigma^2$, which captures the uncertainty of the prediction.

\subsubsection{Logistic Regression}\label{appendix:logistic_regression}
The above discussions on the Bayesian interpretation of the linear ridge regression model allows us to easily extend to other linear ML models. In the case of the logistic regression model, the output of the models follows the logistic distribution [see \cref{tab:continuous_prob_distributions}] $p(y|\bm{x},\boldsymbol{\beta}) = [1+\exp(-\bm{x}^\top\boldsymbol{\beta})]^{-1}$. In this case, we only need to modify the first term in the loss function \cref{eq:bayes_linear_variational_inf_loss} by replacing the Gaussian likelihood with the logistic likelihood, which leads to the following loss function for training the logistic regression model:
\begin{equation}
    \label{eq:logistic_reg_loss}
    \hat{\boldsymbol{\beta}}=\argmin_{\boldsymbol{\beta}} \left[-\sum_{i}\log p(y_i\vert \bm{x}_i,\boldsymbol{\beta})+\frac{1}{2\tau^2}\Vert\boldsymbol{\beta}\Vert_2^2\right].
\end{equation}

It can be seen that the logistic regression model essentially provides two possible outputs for the target variable $y\in\{0,1\}$, such that for any given output $y_i$, its conditional distribution $p(y_i|\bm{x}_i,\boldsymbol{\beta})$ is a Bernoulli distribution, which can be written as $p(y_i|\bm{x}_i,\boldsymbol{\beta}) = p_i^{y_i}(1-p_i)^{1-y_i}$ where $p_i=p(y_i=1|\bm{x}_i,\boldsymbol{\beta})$. 
With this, for a i.i.d. dataset $\{(\bm{x}_i,y_i)\}_{i=1}^{N}$, the likelihood of observing the data can be written as $p(\bm{y}|\bm{X},\boldsymbol{\beta}) = \prod_{i=1}^{N} p(y_i|\bm{x}_i,\boldsymbol{\beta}) = \prod_{i=1}^{N} p_i^{y_i}(1-p_i)^{1-y_i}$, which leads to the following expression for the likelihood term in the loss function \cref{eq:logistic_reg_loss}:
\begin{equation}    \label{eq:logistic_reg_loss_likelihood}
    \sum_{i}\log p(y_i\vert \bm{x}_i,\boldsymbol{\beta}) = \sum_{i=1}^{N} \left[y_i\log p_i + (1-y_i)\log (1-p_i)\right],
\end{equation}
which is the commonly used log-likelihood term for training the logistic regression model. The above expression can be further simplified by if we take $y_i\in\{-1,1\}$ instead of $y_i\in\{0,1\}$. In this case, the loss function for training the logistic regression model can be written as 
\begin{equation}    \label{eq:logistic_reg_loss_simplified}
    \hat{\boldsymbol{\beta}}=\argmin_{\boldsymbol{\beta}} \left[\sum_{i=1}^{N} \log\left(1+\exp(-y_i\bm{x}_i^\top\boldsymbol{\beta})\right)+\frac{1}{2\tau^2}\Vert\boldsymbol{\beta}\Vert_2^2\right].
\end{equation}

\subsubsection{LASSO Regression}\label{appendix:lasso_regression}
In this model, instead of training the L2 norm of the model parameters $\boldsymbol{\beta}$, we take the L1 norm of the model parameters $\boldsymbol{\beta}$ as the regularisation term, which leads to the following loss function for training the LASSO regression model:
\begin{equation}    \label{eq:lasso_loss}
    \hat{\boldsymbol{\beta}} = \argmin_{\boldsymbol{\beta}} \mathcal{L}(\boldsymbol{\beta}) = \argmin_{\boldsymbol{\beta}} \left[\left\Vert \bm{y}-\bm{X}^\top\boldsymbol{\beta}\right\Vert_2^2 + \lambda\Vert\boldsymbol{\beta}\Vert_1^2 \right],       
\end{equation}
where $\Vert\boldsymbol{\beta}\Vert_1=$$\sum_{j=1}^{d} \Vert \beta_j\Vert$ is the L1 norm of the model parameters $\boldsymbol{\beta}$. Compared to the L2 regularisation term in the linear ridge regression model, the L1 regularisation term in the LASSO regression model has the effect of encouraging sparsity in the model parameters $\boldsymbol{\beta}$, which is particularly beneficial for feature selections and enhancing the interpretability of the model. The LASSO regression model can be derived from the perspective of variational inference by assuming a Laplace distribution for the prior distribution of the model parameters $\boldsymbol{\beta}\sim\exp\left(-\frac{1}{\tau}\Vert\beta_i\Vert\right)$, which naturally leads to the penalty term $\log p(\boldsymbol{\beta})\sim\frac{1}{\tau}\sum_j \Vert\beta_j\Vert$.

\subsubsection{Multimodal Bayesian Linear Regression}\label{appendix:multimodal_bayesian_regression}

The introduction of the variational distribution $q(\boldsymbol{\beta})$ in the variational inference framework provides us with great flexibility in developing more agile model that is able to capture the complex distributions in the plausible model parameters $\boldsymbol{\beta}$, which is not possible with the standard linear ridge regression model. An interesting and genernal case is to consider the situation where $q(\boldsymbol{\beta})$ is best fitted by Gaussian mixture model (GMM) that reflects the existence of multiple regions in the parameter space where plausible model parameters $\boldsymbol{\beta}$ reside [\cref{fig:multimodal_gaussian_parameter_space}]. In GMM, the variational distribution $q(\boldsymbol{\beta})$ is given by a weighted sum of multiple Gaussian distributions:
\begin{equation}    \label{eq:multimodal_gaussian_variational_dist}
    q(\boldsymbol   {\beta}) = \sum_{k=1}^{K} \pi_k \mathcal{N}(\boldsymbol{\beta}|\bm{m}_k,\bm{s}_k), 
\end{equation}  
where $\pi_k$ is the weight of the $k$-th Gaussian component, which satisfies the normalisation constraint of $\sum_{k=1}^{K} \pi_k=1$, and $\mathcal{N}(\boldsymbol{\beta}|\bm{m}_k,\bm{s}_k)$ is the $k$-th Gaussian component with mean $\bm{m}_k$ and covariance $\bm{s}_k$.

\begin{figure}[h]
    \centering
    \includegraphics[width=0.8\textwidth]{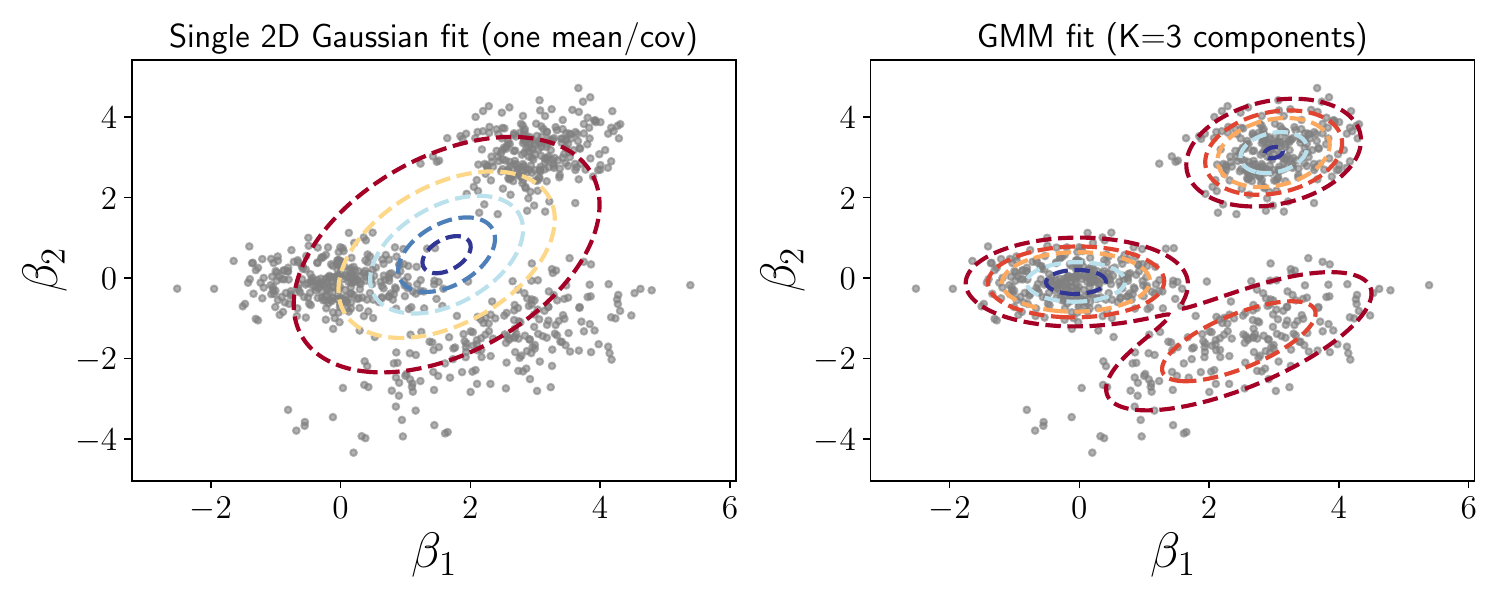}
    \caption{Schematic illustration of the distribution of the model parameters $\boldsymbol{\beta}$ in the parameter space: (Left) showing a unimodal Gaussian distribution, which corresponds to the standard linear ridge regression model. The unimodal distribution over the parameter space leads to a single optimal set of model parameters $\hat{\boldsymbol{\beta}}$ which is positioned in a region with lowest density of the plausible model parameters. (Right) showing a multimodal Gaussian distribution, which corresponds to the multimodal Bayesian linear regression model. The multimodal distribution over the parameter space leads to multiple optimal sets of model parameters, which are positioned in regions with high density of the plausible model parameters.}
    \label{fig:multimodal_gaussian_parameter_space}
\end{figure}

When training the multimodal Bayesian linear regression model, we work directly by maximising the ELBO [\cref{eq:bayes_linear_variational_inf_ELBO}] with respect to the parameters of the variational distribution $q(\boldsymbol{\beta})$.  For the first term in the ELBO, subsituting \cref{eq:multimodal_gaussian_variational_dist} into \cref{eq:bayes_linear_variational_inf_ELBO}, we obtain:
\begin{equation}
    \label{eq:multivariant_gaussian_ELBO_first_term}
    \mathbb{E}_{q(\boldsymbol{\beta})}\left[\log p(\bm{y},\boldsymbol{\beta}\vert\bm{X})\right]=\sum_{k=1}^{K} \pi_k \mathbb{E}_{\mathcal{N}(\boldsymbol{\beta}|\bm{m}_k,\bm{s}_k)}\left[\log p(\bm{y},\boldsymbol{\beta}\vert\bm{X})\right],
\end{equation}
which is a weighted sum of Gaussian distributions that can be analytically evalauted. For the second (entropy) term in the ELBO, we have:
\begin{equation}
    \label{eq:multivariant_gaussian_ELBO_second_term}
    \mathbb{E}_{q(\boldsymbol{\beta})}\left[\log q(\boldsymbol{\beta})\right] = \mathbb{E}_{q(\boldsymbol{\beta})}\left[\log \sum_{k=1}^{K} \pi_k \mathcal{N}(\boldsymbol{\beta}|\bm{m}_k,\bm{s}_k)\right],
\end{equation}
which, unfortunately, does not have a closed-form solution. As such, practically, when maximising the ELBO for training the multimodal Bayesian linear regression model, some types of statistical samplings of $\boldsymbol{\beta}$ from $q(\boldsymbol{\beta})$ (e.g. Monte Carlo) are often needed to obtain an estimate of the expectation value in \cref{eq:multivariant_gaussian_ELBO_second_term} to jointly optimise $\{\pi_k\}$ and $\boldsymbol{\beta}$ for the model in an iterative manner,\cite{murphy2012machine} which we shall not go into the details here for brevity.

\subsection{Nonlinear Regression Models}
\label{appendix:nonlinear_regression}

In this section, we shall provide some brief account of the key technical details for three mostly used nonlinear ML models in materials and molecular simulations, namely, the Kernel Ridge Regression (KRR), the Support Vector Machine (SVM),\cite{cortes1995support} and the Gaussian Process (GP) Regression.\cite{williams2006gaussian}

   \subsubsection{Kernel Ridge Regression}\label{appendix:krr}

KRR provides us with a convenient bridge between linear and nonlinear regressions. In KRR, rather than directly learning a linear function of the input features $y=\bm{X}^\top\boldsymbol{\beta}$, which places a strong prior assumption that the property $y$ behaves linearly with respect to the input features $\bm{X}$, KRR employs the so-called \emph{kernel trick} that explores the similarities among the input data through the \emph{similarity kernel} $\bm{K}$, which is a symmetric and positive-definite matrix with the $(i,j)$-th entry $\bm{K}_{ij}=k_{\boldsymbol{\varphi}}(\bm{x}_i,\bm{x}_j)$ measuring the similarity between the $i$-th and $j$-th data points. To keep the dicussion general, $k_{\boldsymbol{\varphi}}(\cdot,\cdot)$ is a kernel function, which may or may not be parameterized by a set of parameters $\boldsymbol{\varphi}$. The kernel function may take a highly nonlinear form, [such as the popular radial basis function (RBF) kernel, which is defined as $k_{\boldsymbol{\varphi}}(\bm{x}_i,\bm{x}_j)=\exp(-\varphi \|\bm{x}_i-\bm{x}_j\|^2)$, see \cref{fig:RBF_kernel_1D}], and in most ML models, the kernel is precomputed and fixed during the training process. Thus the kernel hyperparameters $\boldsymbol{\varphi}$ are typically tuned through a separate optimisation process, such as grid search.

\begin{figure}[htb]
    \centering
    \includegraphics[width=0.7\textwidth]{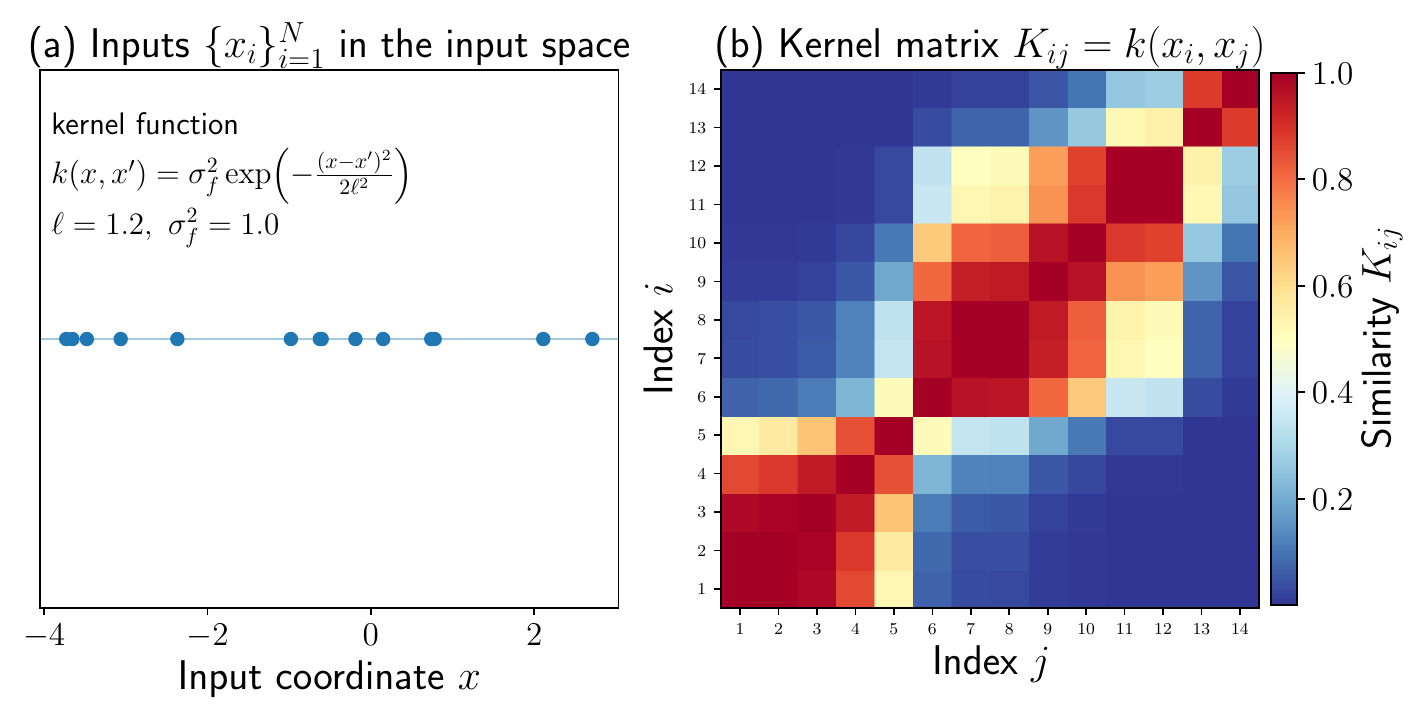}
    \caption{Schematic illustration of the usage of a RBF kernel that maps a series of one-dimensional inputs (a) to a corresponding similarity kernel (b), where closely poisitoned points in the input space are mapped to high similarity values (red), while distantly positioned points are mapped to low similarity values (blue).}
    \label{fig:RBF_kernel_1D}
\end{figure}

With the kernel matrix $\bm{K}$, the KRR model can be expressed as $y=\bm{K}^\top\boldsymbol{\alpha}$, where $\boldsymbol{\alpha}$ is the vector of regression coefficients. In line with the loss function for the ridge regression [\cref{eq:linear_ridge_loss}], the training of KRR is formulated as the following its corresponding loss function:
\begin{equation}
    \label{eq:KRR_loss}
    \hat{\boldsymbol{\alpha}}=\argmin_{\boldsymbol{\alpha}}\Big\{ \|\bm{K}^\top\boldsymbol{\alpha}-\bm{y}\|^2+\lambda \boldsymbol{\alpha}^\top \bm{K} \boldsymbol{\alpha}\Big\},
\end{equation}
which has a closed-form solution as follows:
\begin{equation}
    \label{eq:KRR_optimum_solution}
    \hat{\boldsymbol{\alpha}}=(\bm{K}+\lambda \bm{I})^{-1} \bm{y}.
\end{equation}

   \subsubsection{Support Vector Machine}
\label{appendix:SVM}

In the na\"ive KRR, all the data points that are used for constructing the kernel matrix are treated equally and included in the model training. In another word, this kernelised data representation is \emph{dense}. However, in the presence of outliers, the KRR model can be significantly biased by the outliers and failed to capture the true underlying trend of the data [see \cref{fig:SVR_illustration}]. The support vector regression (SVR) is a variant of the SVM, that is designed for regression problems, provides an active mechanism to detect the data points that are more likely to be outside the model fitting boundaries, providing more \emph{support} to the data points that are more likely to be within the model fitting boundaries.

\begin{figure}[htb]
    \centering
    \includegraphics[width=0.5\textwidth]{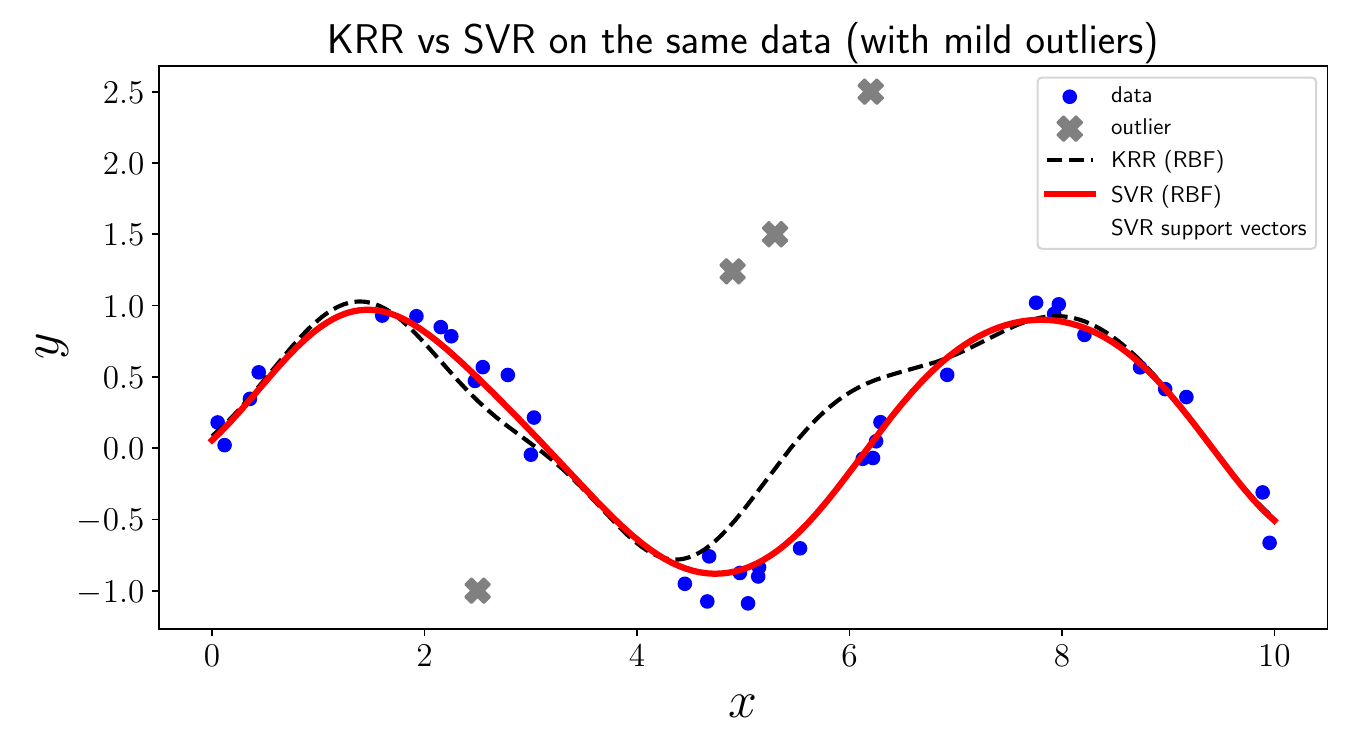}
    \caption{Comparison of  Kernel Ridge Regression (KRR) and Support Vector Regression (SVR) for fitting a 1D dataset with mild outliers. It can be observed in the region $x\in[4,7]$ that KRR is significantly affected by the outliers while SVR is able to follow the trend of the data where majority of the data points are located.}
    \label{fig:SVR_illustration}
\end{figure}

SVR achieves this by introducing a margin of tolerance ($\varepsilon$) around the true data points, which is incorporated into the first term of the target loss function \cref{eq:KRR_loss} as follows:
\begin{equation}
    \label{eq:SVR_loss}
    \mathcal{L}_{\varepsilon}(\boldsymbol{\alpha}) = 
        \begin{cases}
            \Vert \bm{K}^\top \boldsymbol{\alpha} - \bm{y} \Vert^2 &\text{if} \Vert \bm{K}^\top \boldsymbol{\alpha} - \bm{y} \Vert < \varepsilon,\\
             0 & \text{otherwise}
        \end{cases}
\end{equation}
This leads to the following optimization problem for SVR:
\begin{equation}
    \label{eq:SVR_optimization}
    \argmin_{\boldsymbol{\alpha}} \big\{ \mathcal{L}_{\varepsilon}(\boldsymbol{\alpha}) + \lambda \boldsymbol{\alpha}^\top \bm{K} \boldsymbol{\alpha}\big\}.
\end{equation}

To apply the SVM machinary to the classification problem, we start from the loss function for logistic regression [\cref{eq:logistic_reg_loss_simplified}] by first rewriting the term after the summation sign in a more compact form as follows:
\begin{equation}
    \label{eq:logistic_reg_loss_compact}
    \ell(y_i,f_i)=\log\big(1 + \exp(-m_i)\big),
\end{equation}
in which we define an auxiliary variable $m_i = y_i(\bm{x}_i^\top \boldsymbol{\alpha})$. Then it can be observed that the loss function $\ell(y_i,f_i)$ is a smooth monotonically decreasing function of $m_i$, which goes to zero as $m_i\to\infty$ and 1 as $m_i\to0$, such that $\ell(y_i,f_i) \in (0,1)$. To follow the idea behind SVR by excluding the outlying data points from the training of classification model, we need to correct the problem in the limit of $m_i\to\infty$. This can be achieved by replacing the logistic loss function with the hinge loss function as follows:
\begin{equation}
    \label{eq:SVM_classification_loss}
     \argmin_{\boldsymbol{\alpha}} \left\{ \sum_{i=1}^N \max\left(0, 1 - y_i\left(\bm{x}_i^\top \boldsymbol{\alpha}\right)\right) + \lambda \boldsymbol{\alpha}^\top \bm{K} \boldsymbol{\alpha} \right\}.
\end{equation}
   \subsubsection{Gaussian Process}
\label{appendix:GP}

Linear models [\cref{appendix:linear_regression}] are fundamentally constrained by their linearity, in which the only degree of freedom comes from the parameters to be optimised. The introduction of kernels in KRR [\cref{appendix:krr}] partially alleviated this problem through the use of kernel functions that implicitly encode non-linear relationships among the features, but a linear relationship between the features and the target variable is still assumed. In this regard, GP regression is the most flexible non-parametric Bayesian regression model, which move from optimisating the probability distribution over the model parameters to a regime of optimising the probability distribution over the space of functions $\{f(\bm{x})\}$ to arrive at the best surrogate model to fit the underlying data [\cref{fig:GP_illustration}].

\begin{figure}[htb]
    \centering
    \includegraphics[width=\textwidth]{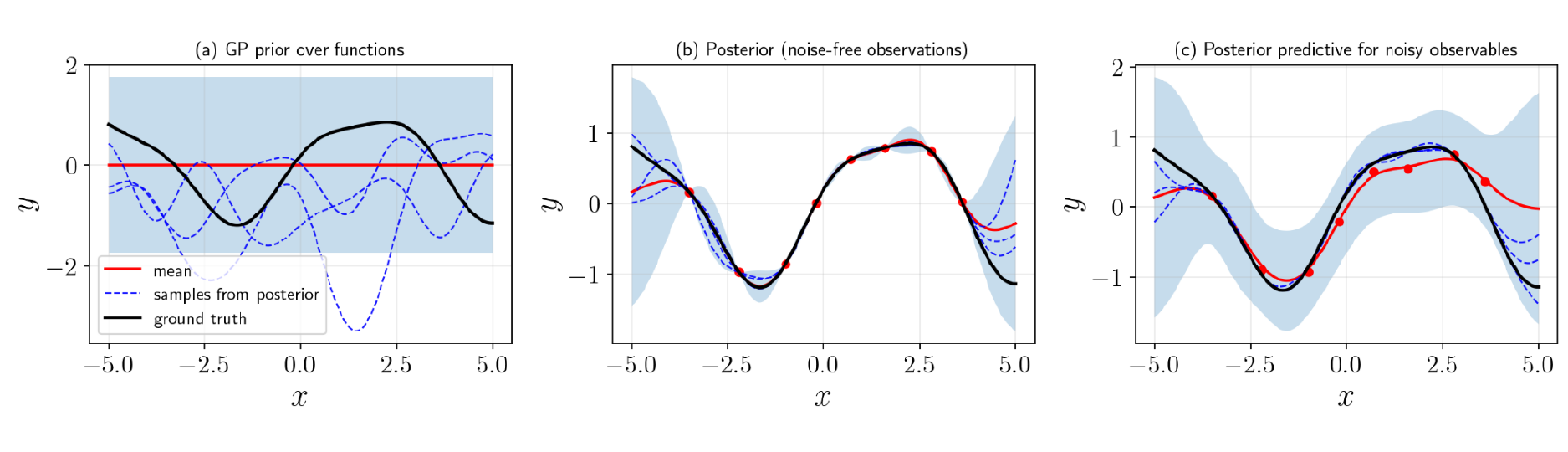}
    \caption{Graphical illustration of the GP regression process to fit a set of data points (red dots) that are drawn from an underlying ground truth function (solid black curve). (a) The prior distribution over functional space that is defined by the mean (red solid curve) and the covariance (shaded area) of the GP, with a few samples drawn from the prior distribution (dashed blue curves). The posterior distribution over the functional space after observing the data points after training the GP model with noise-free (b) and noisy (c) data points. For noise-free data points, the posterior mean of the function passes exactly through the data points with small prediction uncertainties (i.e. small shaded area) around the data points, which is a constraint that is relaxed when fitting the  noisy data points.}
    \label{fig:GP_illustration}
\end{figure}

In this case, the prior distribution in the GP regression is no longer one that is defined over the parameter space [\cref{eq:linear_ridge_param_prior}] but over the function space. To start, for a set of data $\{(\bm{x}_i,y_i)\}_{i=1}^N$, $\bm{x}\in\mathbb{R}^d$ and $y\in\mathbb{R}$, we define the prior distribution over the function space as a Gaussian Process from which plausible functions can be sampled, which is fully specified by a mean function $m(\cdot)$ and a covariance function (or kernel) $k(\cdot,\cdot)$ as follows:
\begin{equation}
    \label{eq:GP_function_prior}
    f(\cdot)\sim\mathcal{GP}\big(m(\cdot),k(\cdot,\cdot) \big).
\end{equation}
Practically, a given function is represented as a finite-dimensional vector $\bm{f}=[f(\bm{x}_1),f(\bm{x}_2)\ldots,f(\bm{x}_N)]^\top\in\mathbb{R}^N$ evaluated at the training data points, which is drawn from a multivariate Gaussian distribution $\bm{f}\sim\mathcal{N}(\bm{m},\bm{K})$, in which: (a) The mean vector $\bm{m}$ has the entries $m_i=m(\bm{x}_i)$, which is some function of the input $\bm{x}_i$ but is often set to zero for simplicity after centering $\{y_i\}_{i=1}^N$ around zero. (b) The kernel matrix $\bm{K}$ has the entries $K_{ij}=k_{\boldsymbol{\varphi}}(\bm{x}_i,\bm{x}_j)$, which is positively defined and encodes some prior belief on how strongly the function values at $\bm{x}_i$ and $\bm{x}_j$ are correlated with each other (i.e. whether we expect $f(\bm{x})$ to be a widly oscillating or a smooth function). The kernel function is often parameterised by a set of hyperparameters $\boldsymbol{\varphi}$, which can be optimised by maximising the marginal likelihood of the data, or may need to be separatedly optimised, depending on the form of the kernel function $k_{\boldsymbol{\varphi}}$.

One of the key advatanges of GP is that it encourages the incorporation of the information of noises in the target variable $y$, that can be concurrently accounted during the model training. In this case, we first rewrite $\{(\bm{x}_i,y_i)\}_{i=1}^N$, $\bm{x}\in\mathbb{R}^d$ as $\{\bm{X},\bm{y}\}$ with $\bm{X}=(\bm{x}_1,\bm{x}_2\ldots,\bm{x}_N)^\top\in\mathbb{R}^{N\times d}$ and $\bm{y}=(y_1,y_2\ldots,y_N)^\top\in\mathbb{R}^N$, from which we can replace $\bm{y}$ by its noised version $\bm{y}^\prime=\bm{y}+\varepsilon\bm{I}$, where we assumed that the noise $\varepsilon$ is a single scalar drawn from a Gaussian distribution $\varepsilon\sim\mathcal{N}(0,\sigma^2)$ with zero mean and variance $\sigma^2$ (a hyperparameter to be optimised). Although there is nothing stopping us from assuming a more complex sample-dependent noise model by having $\boldsymbol{\sigma}^2=\mathrm{diag}\{\sigma^2_1,\sigma^2_2,\cdots,\sigma^2_N\}$.

From here, the primary goal of optimising the GP model is to maximise the probability distributions of the target observables conditioned on the input data and model parameters, \emph{i.e.} $p(\bm{y}^\prime\vert \bm{X},\boldsymbol{\varphi},\sigma^2)$. To align with the viewpoint of optinmising the distribution over the function space, we can use the fact that the conditional distribution $p(\bm{y}^\prime\vert \bm{X},\boldsymbol{\varphi},\sigma^2)$ can be obtained by margionalised over all possible functions whose distribution is conditioned on the data and model parameters, such that:
\begin{equation}
    \label{eq:GP_marginal_likelihood}
    p(\bm{y}^\prime\vert \bm{X},\boldsymbol{\varphi},\sigma^2)=\int p(\bm{y}\vert\bm{f},\sigma^2)p(\bm{f}\vert\bm{X},\boldsymbol{\varphi})\mathrm{d}\bm{f},
\end{equation}
in which $p(\bm{f}\vert\bm{X},\boldsymbol{\varphi})\sim\mathcal{N}(\bm{m},\bm{K}_{\boldsymbol{\varphi},\sigma})$, where $\bm{K}_{\boldsymbol{\varphi},\sigma}=\bm{K}_{\boldsymbol{\varphi}}+\sigma^2\bm{I}$. Without elaborating on the details of the derivation,\cite{williams2006gaussian} the logarithm of the marginal likelihood, which is the objective function to be maximised, can be expressed as:
\begin{equation}    \label{eq:GP_log_marginal_likelihood}
    \mathcal{L}(\boldsymbol{\varphi},\sigma^2)=-\frac{1}{2}\bm{y}^{\prime\top}\bm{K}_{\boldsymbol{\varphi},\sigma}^{-1}\bm{y}^\prime-\frac{1}{2}\log\vert\bm{K}_{\boldsymbol{\varphi},\sigma}\vert-\frac{N}{2}\log 2\pi.
\end{equation}  
The first term in \cref{eq:GP_log_marginal_likelihood} is the data fit term, which encourages the model to fit the data well, while the second term is the complexity penalty term, which penalises models that are too complex (i.e. with a large kernel matrix determinant). The third term is a constant that does not depend on the model parameters and can be ignored during optimisation. 

To optimise $\mathcal{L}(\boldsymbol{\varphi},\sigma^2)$ defined in \cref{eq:GP_log_marginal_likelihood}, one typically first apply the Cholesky decomposition to $\bm{K}_{\boldsymbol{\varphi},\sigma}$ to obtain $\bm{L}\bm{L}^\top=\bm{K}_{\boldsymbol{\varphi},\sigma}$, which allows us to compute the inverse and determinant of $\bm{K}_{\boldsymbol{\varphi},\sigma}$ efficiently. More specficially, the inverse can be computed by first solving the eigenvalue problem of $\bm{L}\bm{v}=\bm{y}^\prime$ from which $\bm{L}^\top\boldsymbol{\alpha}=\bm{v}$ or $\boldsymbol{\alpha}=\bm{K}_{\boldsymbol{\varphi},\sigma}^{-1}\bm{y}^\prime$, and the determinant can be computed as $\vert\bm{K}_{\boldsymbol{\varphi},\sigma}\vert=\vert\bm{L}\vert^2=\prod_{i=1}^N L_{ii}^2$, such that $\log\vert\bm{K}_{\boldsymbol{\varphi},\sigma}\vert=2\sum_{i=1}^N \log L_{ii}$. With these efficient computations, the gradient of $\mathcal{L}(\boldsymbol{\varphi},\sigma^2)$ with respect to the model parameters can be computed as:
\begin{equation}    \label{eq:GP_log_marginal_likelihood_gradient}
    \frac{\partial\mathcal{L}}{\partial\boldsymbol{\varphi}}=\frac{1}{2}\mathrm{tr}\bigg(\big(\boldsymbol{\alpha}\boldsymbol{\alpha}^\top-\bm{K}_{\boldsymbol{\varphi},\sigma}^{-1}\big)\frac{\partial\bm{K}_{\boldsymbol{\varphi},\sigma}}{\partial\boldsymbol{\varphi}}\bigg),\quad\frac{\partial\mathbf{K}_{\boldsymbol{\varphi},\sigma}}{\partial\sigma^2}=\bm{I}.
\end{equation}
An interesting point to note is that, with the optimised kernel, we have $\boldsymbol{\alpha}=\hat{\bm{K}}^{-1}\bm{y}$, which is the same as the solution of KRR [\cref{eq:KRR_optimum_solution}] with $\lambda=0$. This is because, with the optimised kernel, the GP regression can be viewed as a KRR with a kernel that is optimised to fit the data well, which is equivalent to a KRR with a fixed kernel and zero regularisation. However, it is important to note that the optimisation of the kernel in GP regression is not equivalent to the optimisation of the regularisation parameter in KRR, as they are two different aspects of the model.

To make the predictions with the trained GP model, we start from a set of training data $\{(\bm{x}_i,\bm{y}_i)\}_{i=1}^{N}$ and a set of predictive inputs $\{\bm{x}_*\}_{i=1}^M$. From here, we have the mean of the distribution of the predictive function values $m(\bm{X})=\bm{m}\in\mathbb{R}^N$ and $m(\bm{X_*})=\bm{m}_*\in\mathbb{R}^M$. Correspondingly, we have three sets of kernel matrices: $\hat{\bm{K}}\in\mathbb{R}^{N\times N}$, $\hat{\bm{K}}_{*}\in\mathbb{R}^{N\times M}$ and $\hat{\bm{K}}_{**}\in\mathbb{R}^{M\times M}$, which are computed by evaluating the kernel function amongst the training data points, between the training data points and the predictive inputs, and amongst the predictive inputs, respectively. These ingredients leads to the following joint probability distribution of the training and predictive function values:
\begin{equation}    \label{eq:GP_predictive_distribution}
    \begin{bmatrix}\bm{f}\\\bm{f}_*
    \end{bmatrix}\sim\mathcal{N}\bigg(
    \begin{bmatrix}\bm{m}\\\bm{m}_*
    \end{bmatrix},
    \begin{bmatrix}\hat{\bm{K}}&\hat{\bm{K}}_*\\\hat{\bm{K}}_*^\top&\hat{\bm{K}}_{**}
    \end{bmatrix}\bigg).
\end{equation}
By properly centering the data thus making $\bm{m}=\bm{m}^*=0$, this gives us the probability distribution of the predictive function values $\bm{f}_*$ that pass through the predictive inputs $\{\bm{x}_*\}_{i=1}^M$ as follows:
\begin{equation}    \label{eq:GP_predictive_distribution_final}
    p(\bm{f}_*\vert\bm{X},\bm{y},\boldsymbol{\varphi},\sigma^2)\sim\mathcal{N}\Big(\hat{\bm{K}}_*^\top(\hat{\bm{K}}+\sigma^2\bm{I})^{-1}\bm{y},\hat{\bm{K}}_{**}-\hat{\bm{K}}_*^\top(\hat{\bm{K}}+\sigma^2\bm{I})^{-1}\hat{\bm{K}}_*\Big).
\end{equation}
and correspondinly the point prediction at $\bm{x}_*$ and its associated uncertainty can be obtained by the mean and variance of the predictive distribution as:
\begin{equation}    \label{eq:GP_predictive_single_point}
    p(f_*|\bm{x}_*,\bm{X},\bm{y},\boldsymbol{\varphi},\sigma^2)\sim\mathcal{N}\Big(\hat{\bm{k}}_*^\top(\hat{\bm{k}}+\sigma^2\bm{I})^{-1}\bm{y},k_{**}-\hat{\bm{k}}_*^\top(\hat{\bm{k}}+\sigma^2\bm{I})^{-1}\hat{\bm{k}}_*\Big),
\end{equation}
in which $\hat{\bm{k}}_*\in\mathbb{R}^N$ and $k_{**}\in\mathbb{R}$ are the kernel vector and scalar obtained by evaluating the kernel function between the predictive input $\bm{x}_*$ and the training data points, and between the predictive input $\bm{x}_*$ and itself, respectively.

Before concluding this section, it is worth noting that the kernelised formulation of the GP regression suffers the same bottleneck in dealing with the outliers as the KRR model. In the spirit of SVM, different approaches to remedify this problem have been proposed, such as the use of a Huber-type loss function\cite{huber1992robust} to approximate $p(\bm{y}|\bm{f},\sigma^2)$ in \cref{eq:GP_marginal_likelihood} to down weight the outlying points during model training.\cite{algikar2023robust} 

\section{Unsupervised Machine Learning Models}
\label{appendix:unsupervised_ML}
\subsection{Density Estimation}
\label{appendix:density_estimation}

Given the strong connection between ML/AI methodologies and the theory of probability, estimating the probability distribution function  of a given dataset is of foundamental importance to understand the underlying data structure, as well as to perform various downstream tasks. In this section, we will briefly outline the basis behind the parameteric and non-parametric density estimation methods for the density estimation tasks.

\subsubsection{Parametric Density Estimation}
\label{appendix:parametric_density_estimation}

The parametric model assumes that the underlying data distribution can be modeled by the probability distribution function with a fixed parameterised functional form $p(\bm{x})\sim p_{\boldsymbol{\theta}}(\bm{x})$. This is usually the case when one has a strong physical knowledge on the prior distribution of the data, such as the normal distribution [\cref{eq:particle_velocity_distribution}] that is used to sample the initial particle velocities in the MD simulations under a given temperature $T$. Generally speaking, one can assume any well-known probability distribution functions as outlined in \cref{tab:discrete_prob_distributions} and \cref{tab:continuous_prob_distributions}. Nevertheless, the most common choice is to model the data distribution with a mixture of Gaussian distributions, which leads to the GMM [\cref{fig:multimodal_gaussian_parameter_space}]:
\begin{equation}
  \label{eq:GMM_density_estimate}
  p(\bm{x})=\sum_{k=1}^{K}\pi_k \mathcal{N}(\boldsymbol{\mu}_k,\boldsymbol{\sigma}_k).
\end{equation}
Denoting $\boldsymbol{\theta}_k=(\boldsymbol{\mu}_k,\boldsymbol{\sigma}_k)$, then the optimal parameter $\hat{\boldsymbol{\theta}}$ can be obtained through maximising the log likelihood function $\hat{\boldsymbol{\theta}}=\argmax_{\boldsymbol{\theta}}\sum_{k=1}^{K}\log p_{\boldsymbol{\theta}_k}(\bm{x})$, which, however, does not have a general solution for $K>1$ (for $K=1$, the optimum solution is given by the mean and the standard deviation of the dataset $\{\bm{x}_i\}_{i=1}^{M}$). As one can envisage, the fundamental challenge for the GMM is to determine the optimal number of Gaussian components $K$ to use, which requires some good prior knowledge on the data distribution.

\subsubsection{Non-parametric Density Estimation}
\label{appendix:non_parametric_density_estimation}

In contrary to the parametric model, the non-parametric density estimation does not assume any specific functional form for the underlying probability distribution function for the data, but trying to directly estimate the probability distribution function from the data itself. Notably, the most common choice of the non-parametric density estimation is the kernel density estimation (KDE), which, in its most general form, the probability distribution function can be expressed as
\begin{equation}
  \label{eq:KDE_density_estimate}
  p(\bm{x})=\frac{1}{nh}\sum_{i=1}^{n} k_{\boldsymbol{\theta}}\left(\frac{\bm{x}-\bm{x}_i}{h}\right),
\end{equation}
in which $k_{\boldsymbol{\theta}}$ is a kernel function [as discussed in \cref{appendix:krr}], and $h$ is the so-called bandwith parameter that controls the sharpness/smoothness of the probability distribution function. Since the kernel function typically has a high (low) value for closely (distant) spaced data points, it is intuitively to understand why this is also suitable for density estimation, as the probability distribution function will be higher in the regions where the data points are more densely distributed, and \emph{vice versa}. Again, the Gaussian kernel is probably the most common (default) choice for KDE, whereas the bandwidth $h$ critically controls the levels of the details that can be captured by the KDE-estimated probability distribution function. A common default choice for the bandwidth is the Silverman's rule of thumb,\cite{silverman1986density} which is given by $h=1.06\hat{\sigma}n^{-1/5}$, where $\hat{\sigma}$ is the standard deviation of the dataset $\{\bm{x}_i\}_{i=1}^{M}$. However, this default choice may not be optimal for all datasets, and it is often necessary to perform some parameter searches to find the optimal bandwidth that can best capture the underlying data distribution.

\subsection{Signal Decomposition Analysis} 
\label{appendix:signal_decomposition_analysis}

Signal decomposition analysis, such as the principal component analysis (PCA) and its kernelized version (KPCA), are widely used unsupervised machine learning methods for dimensionality reduction and feature extractions, which try to find a small set of `basis' that correspond to the directions of maximum variance in the data, upon which the data can be projected to obtain a low-dimensional representation.

\subsubsection{Principal Component Analysis}
\label{appendix:pca}

\begin{wrapfigure}[22]{r}{0.3\textwidth}
  \begin{center}
    \includegraphics[width=0.3\textwidth]{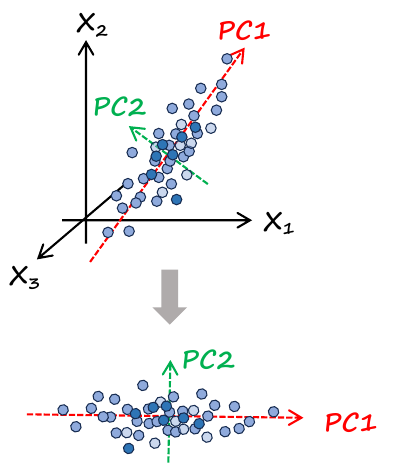}
    \caption{Schematic illustration on the principal component analysis.}
    \label{fig:PCA_illustration}
  \end{center}
\end{wrapfigure}

Given a set of high-dimensional data points $\bm{X}=\{\bm{x}_i\}_{i=1}^M$ with $\bm{x}_i \in \mathbb{R}^N$ that lie in the $N$-dimensional space,  PCA\cite{jolliffe2016principal} [\cref{fig:PCA_illustration}] aims to find a set of orthogonal directions $\{\bm{v}_i\}_{i=1}^{N}$, \emph{i.e.}, the \emph{principal components}, such that along $\bm{v_1}$. the data exhibits the largest variances, followed by $\bm{v_2}$, $\bm{v_3}$, and so on.  

If we take the first $k$ $(k<N)$ dominant principal components, we can define a matrix of the principal components $\bm{V}_k=\left[\bm{v}_1,\bm{v}_2,\cdots,\bm{v}_k\right]^\top$, such that a projection of the original high-dimensional data onto the low-dimensional subspace can be obtained by the dot product of $\bm{Z}=\bm{X}\bm{W}_k$. 

To compute the principal components, we first compute the mean $\bar{\bm{x}}$ from each data point, and then compute the covariance matrix $\bm{\Sigma}=\frac{1}{M}\bm{X}^\top\bm{X}=\frac{1}{M}\sum_{i=1}^M (\bm{x}_i-\bar{\bm{x}})(\bm{x}_i-\bar{\bm{x}})^\top$. By solving the eigenvalue problem $\bm{\Sigma}\bm{v}=\lambda\bm{v}$, we can obtain the eigenvalues $\{\lambda_i\}_{i=1}^{N}$ and the corresponding eigenvectors $\{\bm{v}_i\}_{i=1}^{M}$. By sorting the eigenvectors in the descending order of their corresponding eigenvalues, the first $k$ eigenvectors with the largest eigenvalues can be selected as the principal components. Typically, the eigenvalues $\{\lambda_i\}_{i=1}^{N}$ will show a large discontinuous drop from $\lambda_k$ to $\lambda_{k+1}$,\cite{jolliffe2016principal} this can be used as a criterion to determine the optimal number of principal components $k$ to retain, which also corresponds to the intrinsic dimensionality of the data.

To show that $\bm{v}_1$ is the direction of the largest variance, we reformulate the PCA as an optimisation problem, in which we see for a unit vector $\bm{v}$ to maximise the variance of the data being projected onto it, \emph{i.e.}, $\max_{\bm{v}} \mathrm{Var}[\bm{X}\bm{v}]$, subected to the normalisation constraint of $\Vert\bm{v}\Vert_2^2=1$.  This constrained maximisation problem can be solved by the method of Lagrangian multiplier, which for this problem, read as $\mathcal{L}(\bm{v},\lambda)=\mathrm{Var}[\bm{v}^\top\bm{X}\bm{v}]-\lambda(\bm{v}^\top\bm{v}-1)$. The optimum soltion can be obtained from the equation $\nabla_{\bm{v}}\mathcal{L}(\bm{v},\lambda)=0$, which leads to the eigenvalue problem $\bm{\Sigma}\bm{v}=\lambda\bm{v}$, where $\lambda$ is the Lagrangian multiplier. The solution of this eigenvalue problem gives us the principal components $\{\bm{v}_i\}_{i=1}^{N}$ and their corresponding variances $\{\lambda_i\}_{i=1}^{N}$, which are the eigenvalues of the covariance matrix $\bm{\Sigma}$.

\subsubsection{Kernel Principal Component Analysis}
\label{appendix:kpca}

\begin{figure}[htb]
  \centering
  \includegraphics[width=\textwidth]{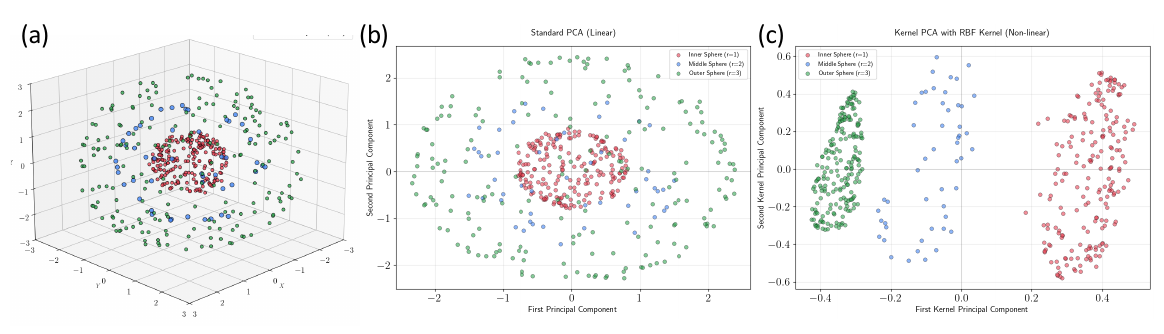}
  \caption{Dimensionality reduction analysis performed on a set of 3D concentrically distributed data points [shown in (a)], in which the results obtained from PCA (b) and KPCA (c) are compared. The KPCA analysis was performed using a radial basis function (RBF) kernel, which is defined as $k(\bm{x}_i,\bm{x}_j)=\exp(-\gamma\Vert\bm{x}_i-\bm{x}_j\Vert^2)$, where $\gamma$ is a hyperparameter that controls the width of the kernel. }
  \label{fig:PCA_KPCA_compare}
\end{figure}

Similar to the KRR [\cref{appendix:krr}] that is a kernelised version of the ridge regression [\cref{appendix:ridge_regression}] which enable the model to capture highly nonlinear relationship amongst the input data and the target variable, the KPCA\cite{scholkopf1998nonlinear}
 is a kernelised version of the PCA, in which the covariance matrix $\bm{\Sigma}$ is replaced by a kernel matrix $\bm{K}$. As dicussed above, the matrix elements of the kernel may be parameterised, that can be written as $k_{\boldsymbol{\varphi}}(\bm{x}_i,\bm{x}_j)$. When the input are feature vectors, such as structural descriptors that are built from the atomistic coordinates, themselves can also be parameterised, thus the full matrix element can be written as $k_{\boldsymbol{\varphi}}(\mathcal{F}_{\boldsymbol{\varphi}^\prime}(\bm{x}_i),\mathcal{F}_{\boldsymbol{\varphi}^\prime}(\bm{x}_j))$. Under such construction, the principal components obtained from the KPCA reflect the directions of maximum variance in the feature space, which is a high-dimensional space that is implicitly defined by the kernel, and if exits, the feature mapping function. As shown in \cref{fig:PCA_KPCA_compare}, depends on the choice of the kernel, which, in this case, a Gaussian radial basis function (RBF) kernel, the KPCA put more focus on the local structures in the data, where densly distributed data points in the high-dimensional space are more likely to be projected onto the same region within the low-dimensional space, with some tradoff in reproducing the global structure of the original data, which is more faithfully captured by the PCA.
 
\subsection{Mitidimensional Scaling}
\label{appendix:mds}

\begin{figure}[htb]
  \centering
  \includegraphics[width=0.8\textwidth]{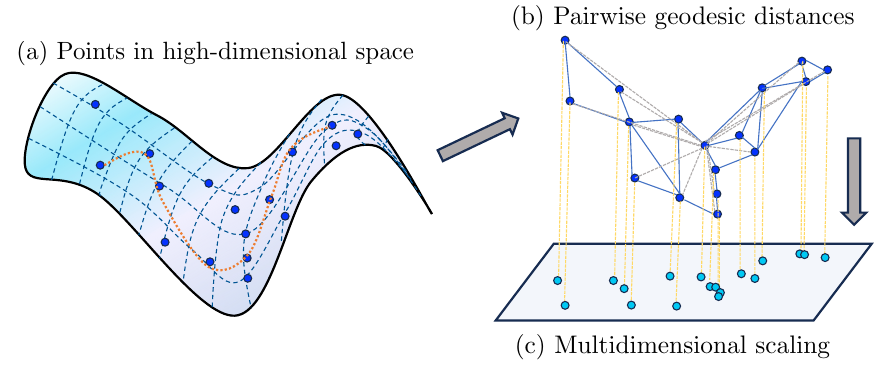}
  \caption{Schematic illustration on the dimensionality reduction analysis performed by the multidimensional scaling (MDS). Start from a set of data points in the high-dimensional space (a), the MDS first computes the pairwise distance matrix $\bm{D}$ (b), then the data is projected onto the low-dimensional space (c) by preserving the pairwise distances among all the points through minimising the cost function which measures the descrepency between the pairwise distances in the high- and low-dimensional spaces.}
  \label{fig:mds_illustration}
\end{figure}

Instead of focusing on extracting the highest-variant patterns among the entire dataset, sometimes, we want to put more focus on the local structure of the data, particularly the similarities and dissimilarities within each pair of data in the entire dataset. For examples, when analysing the structures sampled from an MD simulation spanned by different time intervals $\Delta t$, to understand the degrees of structural fluctuations at different time scales. For this purpose, the multidimensional scaling (MDS)\cite{borg2005modern} offers an approach to project high-dimensional data onto a low-dimensional representation by preserving the pairwise distances among all the data points [\cref{fig:mds_illustration}].

Again, starting from high-dimensional data points $\bm{X}=\{\bm{x}_i\}_{i=1}^M$ with $\bm{x}_i \in \mathbb{R}^N$, we can construct a pairwise distance matrix $\bm{R}\in\mathbb{R}^{M\times M}$, where the element $R_{ij}$ represents the distance between the data points $\bm{x}_i$ and $\bm{x}_j$. Based on the above fundamental concept behind MDS, the low-dimensional representation $\bm{Z}=\{\bm{z}_i\}_{i=1}^M$ with $\bm{z}_i \in \mathbb{R}^k$ $(k<N)$ can be obtained by minimising the cost function, which measures the descrepancy between the pairwise distances in the high- and low-dimensional spaces, \emph{i.e.}, 
\begin{equation}
  \label{eq:MDS_minimisation}
  \min_{\bm{Z}} \sum_{i<j} (R_{ij}-r_{ij})^2,
\end{equation}
 where $r_{ij}$ is the distance between the data points $\bm{z}_i$ and $\bm{z}_j$ in $\mathbb{R}^{k}$. By minimising this cost function, we can obtain a low-dimensional representation of the data that preserves the pairwise distances among all the data points as much as possible, which can be used for visualisation and further analysis. 


\subsubsection{ISOMAP}
\label{appendix:isomap}

The ISOMAP\cite{tenenbaum2000global} represents one of the earliest variants of the MDS algorithm, which aims to closely map the pairwise geodesic distances among the data points in the high-dimensional manifold into the low-dimensional projection. There are three key steps behine the ISOMAP algorithm. Firstly, clusters of points in the high-dimensional space are identified based on the Euclidean distances to identify the local structures in the data. Secondly, the connecting clusters, \emph{e.g.} those that are identified by the $k$-nearest neighbour (KNN) algorithm, are furtehr connected together based on the shortest geodesic paths that are traced on the high-dimensional manifold surface. Finally, the low-dimesional representation of the data is generated by applying the classical MDS algorithms on the geodesic distances traced on the high-dimensional manifold. 

\begin{figure}[htb]
  \centering
  \includegraphics[width=0.9\textwidth]{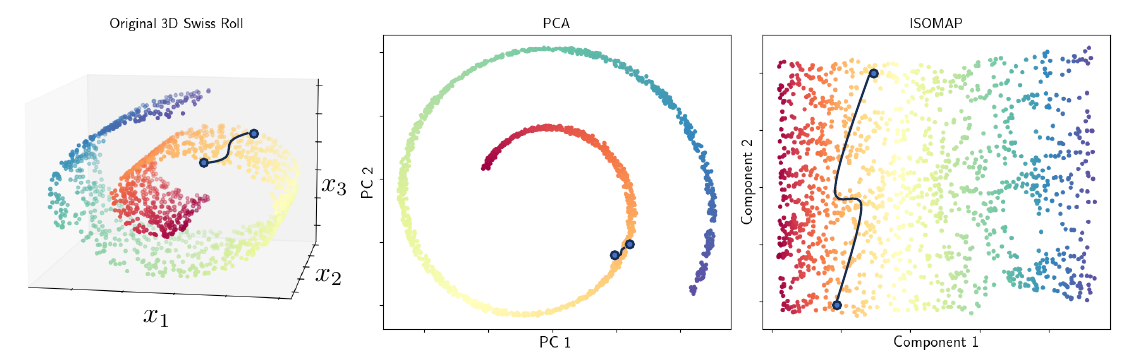}
  \caption{Comparsion of the low-dimensional representation of the 3D Swiss roll data (left panel) obtained from the PCA (middle panel) and ISOMAP (right panel) analyses.}
  \label{fig:PCA_ISOMAP_compare}
\end{figure}

The 3D Swiss roll data is commonly used to demonstrate the differences between PCA and ISOMAP in generating the low-dimensional representation of the complex data.\cite{glielmo2021unsupervised} As also shown in \cref{fig:PCA_ISOMAP_compare}, PCA is able to capture the dominant variance of the data that occurs along the $x_1$-direction in the 3D space, but fails to reproduce the geodesic relationships along the $x_2$-directions, making the two blue points appear close to each other in the low-dimensional space. On the other hand, the ISOMAP algorithm effectively `unrolled' the original Swiss roll data, whereby the spatial relationships along the curved 3D manifold Swiss roll surface have been well preserved in the low-dimensional representation, where the two blue points are correctly projected far away from each other. 

The ISOMAP algorithm provided an attractive approach for analysing complex data, such as the free-energy landscapes for protein foldings and conformational changes in large molecules, that are simulated with MD. However, na\"{i}ve ISOMAP becomes computationally expensive, and often intractable, when both the number of data points $M$ and the dimensionality of the data $N$ are large, which is often the case for long-time (bio)molecular simuations. \cite{das2006low} proposed the landmark-based ISOMAP to overcome such a bottleneck. Here, a small set of landmark points are selected from the original data, and the geodesic distances are only computed between the landmark points and the rest of the data points. In this case, when optimising the low-dimensional representation of the data, one only needs to preserve the geodesic distances between the landmark points and the rest of the data points, which significantly reduces the computational cost of the ISOMAP algorithm. Such a landmark-based approach has also been applied in other variants of the unsupervised machine learning algorithms, such as the  KPCA\cite{zeng2023mechanistic} and sketch-map\cite{ceriotti2011simplifying} discussed below, to reduced the overall computational cost for analysing large datasets. Even with such a landmark-based approach, the ISOMAP algorithm can still be computationally expensive when applied to analyse large datasets. In the original study published in 2006,\cite{das2006low} analysis perofmed on the folding trajectory of a small protein cost $\sim$500 CPU hours.
 
\subsubsection{Diffusion Map}
\label{appendix:diffusion_map}

The diffusion map algorithm was first proposed by \citet{coifman2006diffusion} and was reviewed in the work by \citet{ferguson2011nonlinear} as an unsupervised ML method for analysing the molecular simulation data. The inspiration behind the diffusion map is by realising the fact that the dynamics of a complex system, such as the conformational changes in macromolecules, are governed by a small number of slow modes, whereas the remaining fast degrees of freedom couple to them effectively as stochastic noise. Furthermore, the evolution of the system is a Markovian process, which depends only on the instaneous state, not its previous history. Hence, it is critical that we can extract these fundamental underlying slow modes, to facilitate the understanding of the system dynamics,  parameterising the meaningful dynamic models to characterise the free energy surfaces, as well as provide robust classification of the metastable and stable configurations. 

Starting from a set of high-dimensional samples $\bm{X}=\{\bm{x}_i\}_{i=1}^M$ with $\bm{x}_i \in \mathbb{R}^N$, we first construct a Gaussian kernel matrix $\bm{K}$, where the matrix element is computed with a parameterised Gaussian function $k_\sigma(i,j)=\exp(-\Vert \bm{x}_i-\bm{x}_j\Vert_2^2/2\sigma^2)$. This can be considered as a soft-thresholding operation to retain information in which the pairwise distance between a given pair of data is of the similar order of $\sigma$, which allows us to synthesise distinct embeddings of disconnected sub-regions of the global energy landscape. This can be achived by first computing the corresponding normalised right-stochastic Markov transition matrix $\bm{P}$ can be obtained by normalising the kernel matrix $\bm{K}$ as $\bm{P}_{ij}=\bm{K}_{ij}/\sum_{i=1}^M K_{ij}$, and its corresponding eigenvalues $\{\lambda_i\}_{i=1}^M$ and eigenvectors $\{\bm{\psi}_i\}_{i=1}^M$ in the descending order of the eigenvalues. From here, the diffusion map embedding is defined as the mapping of the $i$-th snapshot onto the $i$-th component of each of the top $k$ non-trivial eigenvectors of the $\bm{P}$ matrix as $\bm{z}_i=[\psi_1(i),\psi_2(i),\cdots,\psi_k(i)]^\top$. Essentially, the diffusion distance, which is related to the Markov transition matrix, provides an indication on how easy the system can diffuse between two configurations $\bm{x}_i$ and $\bm{x}_j$, which is then reflected from the diffusion map embedding.

\subsubsection{\textit{t}-Distributed Stochastic Neighbour Embedding}
\label{appendix:t_sne}

The basic idea behind the $t$-distributed stochastic neighbour embedding ($t$-SNE)\cite{van2008visualizing} shares some similarity with KPCA using a Gaussian RBF kernel, in which the dimensionality reduction focuses on preserving the local spatial relationship among data that lie closely to each other in the high-dimensional space, whereas the relationship among the distant data points is down weighted/rejected. It is formulated based on a probabilistic model that describes the similarity between data points in the high-dimensional space and the low-dimensional space, in which we can define a probability for any point $\bm{x}_j$ to be a neighbouring point to $\bm{x}_i$ in the high-dimensional space as
\begin{equation}
  \label{eq:prob_high_d_distance}
  p(j|i)=\frac{\exp(-\Vert\bm{x}_i-\bm{x}_j\Vert_2^2/2\sigma_i^2)}{\sum_{k\neq j}\exp(-\Vert\bm{x}_i-\bm{x}_k\Vert_2^2/2\sigma_i^2)},
\end{equation}
in which $\sigma_i$ is a hyperparameter that controls the how closes two points may be considered as neighbours. Similarly, we can define a probability for any point $\bm{z}_j$ to be a neighbouring point to $\bm{z}_i$ in the low-dimensional space as:
\begin{equation}
  \label{eq:prob_low_d_distance}
  q(j|i)=\frac{(1+\Vert\bm{z}_i-\bm{z}_j\Vert_2^2)^{-1}}{\sum_{k\neq j}(1+\Vert\bm{z}_i-\bm{z}_k\Vert_2^2)^{-1}},
\end{equation}
where the Student's $t$-distribution with one degree of freedom is used to compute the similarity between data points in the low-dimensional space, which has heavier tails than the Gaussian distribution, thus can better capture the local structure of the data. The low-dimensional representation $\bm{Z}=\{\bm{z}_i\}_{i=1}^M$ can be obtained by minimising the Kullback-Leibler divergence between the two probability distributions $\mathcal{D}_{\mathrm{KL}}(p\Vert q)$, which is typically solved by gradient optimisations.

\subsubsection{Sketch-Map}
\label{appendix:sketch_map}

Generally, the distance metric used to construct the pairwise distance $R_{ij}$ (in high-dimensional space, or $r_{ij}$ in the low-dimensional space) are all normalised such that $R_{ij}\in[0,1]$. As such, ISOMAP is a procedure that focuses on preseving the information $\forall R_{ij}\in[0,1]$, whereas diffusion map and $t$-SNE are more focused on preserving the local structure of the data $\forall R_{ij}\in[0,\sigma]$.  The general idea behind the sketch-map\cite{ceriotti2011simplifying} is to provide a more flexible approach to preserve the pairwise distance information on length-scales that are particularly interesting in the high-dimensional space, which recasts the cost function in \cref{eq:MDS_minimisation} as
\begin{equation}
  \label{eq:sketch_map_cost_function}
  \min_{\bm{Z}} \sum_{i<j} \left[s_{R}(R_{ij})-s_{r}(r_{ij})\right]^2,
\end{equation}    
with a non-linear sigmoid function parameterised by $\sigma$, $a$, and $b$ defined as
\begin{equation}
  \label{eq:sketch_map_sigmoid_function}
  s(x)=1-\left(1+(2^{a/b}-1)(x/\sigma)^a\right)^{-b/a}.
\end{equation}

\subsection{Clustering Analysis}
\label{appendix:clustering_analysis}

In this section, we will briefly introduce the K-mean clustering algorithm,\cite{mcqueen1967some} which will provide some fundamental understanding on the process of clustering analysis. This will be followed by a brief introduction on the density-based spatial clustering of applications with noise (DBSCAN)\cite{ester1996density} and its hierarchical variant,\cite{campello2015hierarchical} which are most widely used in molecular and materials modellings.

\subsubsection{K-Means Clustering}
\label{appendix:k_means_clustering}

\begin{figure}[htb]
  \centering
  \includegraphics[width=0.8\textwidth]{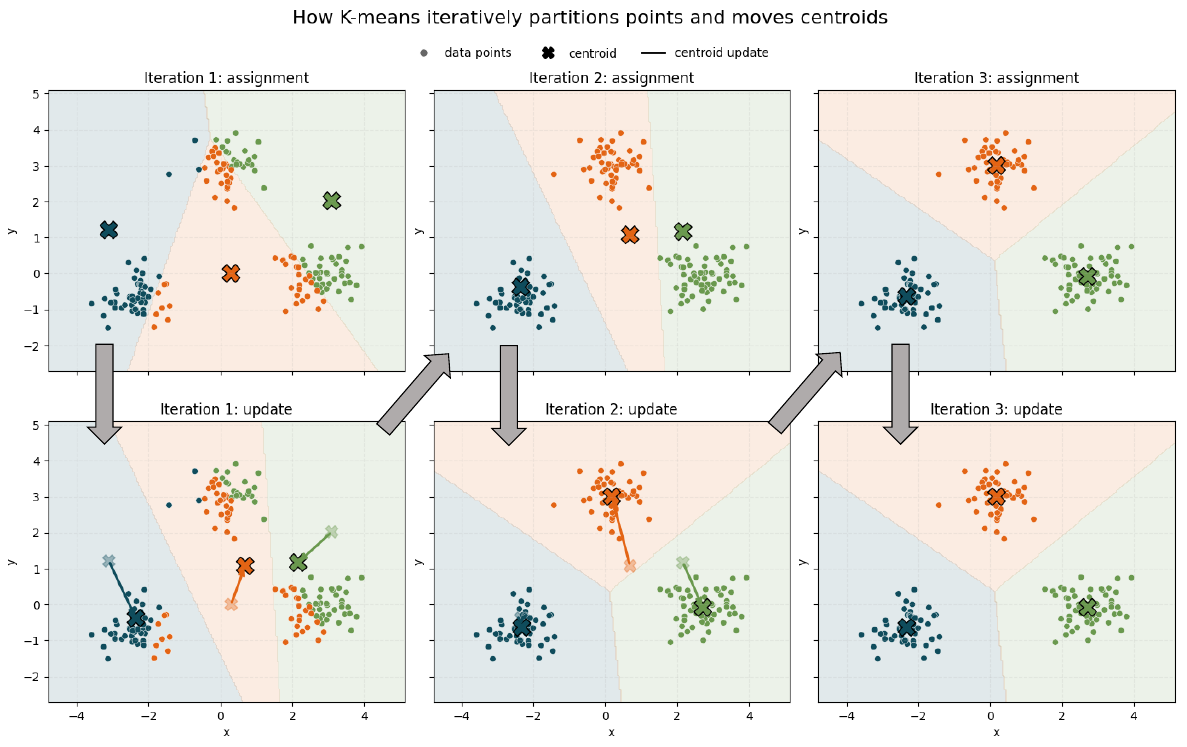}
  \caption{Schematic illustration on the process of K-means clustering analysis.}
  \label{fig:K_means_clustering}
\end{figure}

The basic process behind the K-means clustering algorithm is schematically illustrated in \cref{fig:K_means_clustering} for a set of 2D data points $\{\bm{x}_i\}_{i=1}^{M}$ ($\bm{x}_{i}\in\mathbb{R}^2$). The key hyperparameter for the algorithm is the number of clusters $K$, which is typically assigned based on some prior knowledge on the data. The algorithm starts with the random assignment of the positions for $K$ cluster centroids $\{\bm{c}_k\}_{k=1}^K$ ($\bm{c}_{k}\in\mathbb{R}^2$),  each data point is then assigned to the cluster with the closest centroid, which can be mathematically expressed as $\mathrm{argmin}_k \Vert\bm{x}_i-\bm{c}_k\Vert_2^2$. After this, the mean position of the data points within each cluster is computed as $\bar{\bm{c}}_k=\frac{1}{N_k}\sum_{i=1}^{N_k} \bm{x}_i$ for all $N_k$ data points that belong to the $k$-th cluster. This allows the cluster centroids to be updated to $\bar{\bm{c}}_k$, at which the algorithm cycle repeats again from the assignment of data points to clusters, until all the cluster centroids converge to the converged positions.

\subsubsection{DBSCAN Clustering}
\label{appendix:DBSCAN_clustering}

\begin{figure}[htb]
  \centering
  \includegraphics[width=0.4\textwidth]{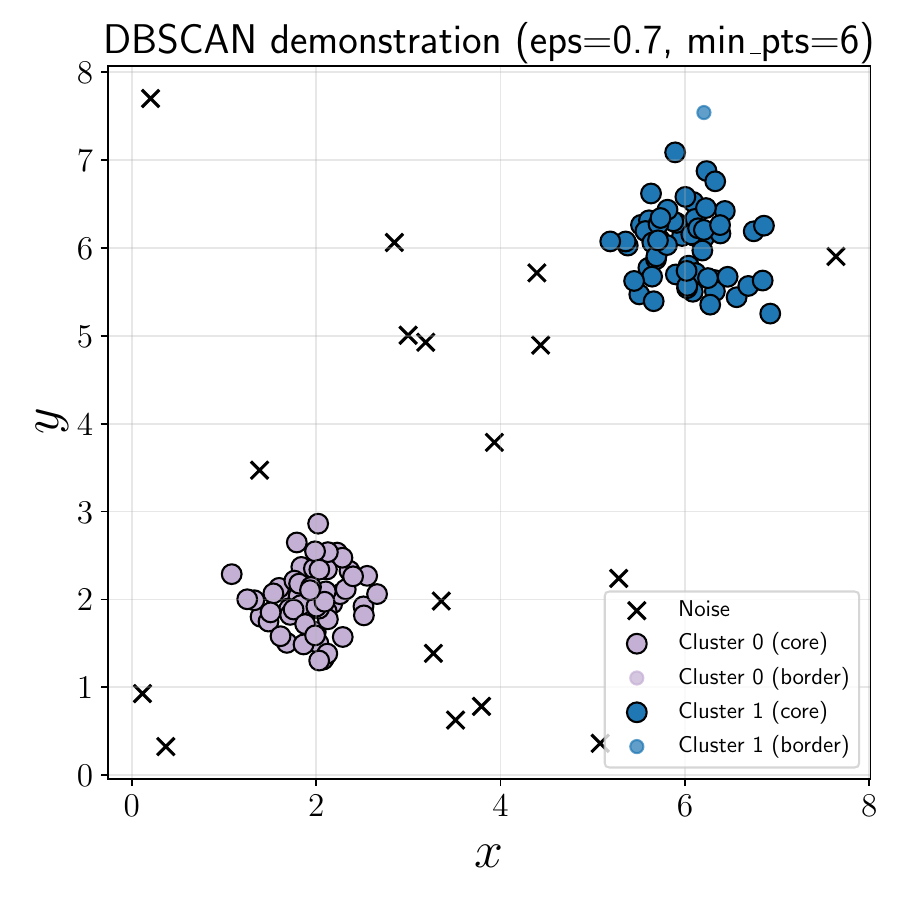}
  \caption{Schematic illustration on the DBSCAN clustering analysis of 2D data, in which points aggregated in high-density regions are classified into the `core' points, points that are close to the `core' points are classified into the `boarder' points, and the remaining points are classified as `noise'.}
  \label{fig:DBSCAN_clustering}
\end{figure}

Instead of assigning points into clusters based on the distances to the cluster centroids, the DBSCAN clustering algorithm (\cref{fig:DBSCAN_clustering}) adopts a density-based approach to classify points into clusters. The algorithm requirws the pre-definitions of parameters of $\varepsilon$, which defines the radius within which to find the neighbouring points for a given centre, and \texttt{minpts}, which defines the minimum number of points required to form a dense region (or a cluster). Starting from a given data point $\bm{x}_i$, if there are at least \texttt{minpts} data points within the radius of $\varepsilon$ from $\bm{x}_i$, then $\bm{x}_i$ is classified as a `core' point, and all the data points within the radius of $\varepsilon$ from $\bm{x}_i$ are classified as `boarder' points. If there are less than \texttt{minpts} data points within the radius of $\varepsilon$ from $\bm{x}_i$, then $\bm{x}_i$ is classified as a `noise' point. The process is repeated for all the data points until all the points are classified into either `core', `boarder', or `noise' points. The clusters can be formed by connecting the `core' points together, and including the `boarder' points that are close to the `core' points, while excluding the `noise' points. 

In its hierachical variant, the HDBSCAN\cite{campello2015hierarchical} algorithm, it remove the specific requirement for predefining the $\varepsilon$ parameter. Instead, it perform clustering across \textit{many different levels} of $\varepsilon$ and build a hierarchical tree of clusters based on the denisties. From this hierarchical tree, the most stable clusters can be identified based on the stability of the clusters across different levels of $\varepsilon$, which can be quantified by the persistence of the clusters across different levels of $\varepsilon$.

\section{Details for Generating \cref{fig:ridge_regression_prob_dist_loss_landscape}}
\label{appendix:details_ridge_regression_loss_landscape}

Here, we provide the technical details of generating \cref{fig:ridge_regression_prob_dist_loss_landscape} in the main text, which exmplifies how the probability distribution of the hyperparameters changes at different points on the landscape of the loss function for model training, following the underlying example given in \citet{prince2023understanding}.

The simple ridge regression model is applied to fit the data generated from the \emph{Gabor model} function that maps an 1D input to an 1D output, which is defined as follows:
\begin{equation}
    \label{eq:gabor_model}
 y=f(x|\phi_0,\phi_1)=\sin\left(\phi_0+0.06\phi x\right)\cdot\exp\left( -\frac{(\phi_0 + 0.06\phi_1 x)^2}{32.0}\right),
\end{equation}
in which $\phi_0$ and $\phi_1$ are the hyperparameters of the Gabor model, that we shorthanded it with a vectorial quantity $\boldsymbol{\phi}=(\phi_0,\phi_1)^\top$. We generated 40 testing samples by first randomly sampled 40 $x$ values in the interval $[-15,15]$ and then computed the corresponding $y$ values using \cref{eq:gabor_model} with additional noises $\epsilon\sim\mathrm{Uniform}(-0.2,0.2)$ added.

The landscape of the loss function is generated by computing the values of the loss function over a uniform 2D grid of $(\phi_0,\phi_1)$. We take the loss function for the ridge regression:
\begin{equation*}
    \label{eq:gabor_ridge_loss}
    \mathcal{L}(\boldsymbol{\phi})=\sum_{i=1}^{N}\left\Vert y_i-f(x_i|\boldsymbol{\phi} )\right\Vert_2^2+\lambda\left\Vert \boldsymbol{\phi} \right\Vert_2^2,
\end{equation*}
in which we set the regularisation strength $\lambda=10^{-4}$. 

Unfortunately, computing $\mathcal{L}(\boldsymbol{\phi})$ over the $\boldsymbol{\phi}$-grid does not give us the direct access to the probability distribution of the hyperparameters $p(\boldsymbol{\phi})$, which is what we want to show in \cref{fig:ridge_regression_prob_dist_loss_landscape}. This problem is overcomed by realising the fact that the local curvature of the loss function at a given point $\hat{\boldsymbol{\phi}}$ can be used to obtain a Boltzmann-like distribution of the hyperparameters around $\hat{\boldsymbol{\phi}}$ as $p(\hat{\boldsymbol{\phi}})\propto \exp(-\beta \hat{\boldsymbol{\phi}})$, which is approximately a Gaussian distribution $\mathcal{N}(\hat{\boldsymbol{\phi}},\boldsymbol{\Sigma})$ with mean $\hat{\boldsymbol{\phi}}$ and a covariance matrix $\boldsymbol{\Sigma}\sim(\beta\bm{H})^{-1}$, if we perform a second-order Taylor expansion of the loss function around $\hat{\boldsymbol{\phi}}$:
\begin{equation*}
    \label{eq:taylor_expansion_loss}
    \mathcal{L}(\boldsymbol{\phi})\approx \mathcal{L}(\hat{\boldsymbol{\phi}})+\frac{1}{2}(\boldsymbol{\phi}-\hat{\boldsymbol{\phi}})^\top \bm{H} (\boldsymbol{\phi}-\hat{\boldsymbol{\phi}}),
\end{equation*}
in which $\bm{H}$ is the Hessian matrix of the loss function at $\hat{\boldsymbol{\phi}}$, which can be computed using a finite difference method around the vicinity of $\hat{\boldsymbol{\phi}}$. 

Nevertheless, for an arbitrary point $\hat{\boldsymbol{\phi}}$ on the loss landscape, the Hessian matrix $\bm{H}$ may not be positively definitive and hence cannot be directly inverted. To solve this problem, we perform an eigen-decomposition of $\bm{H}$ and then set the negative eigenvalues to a small positive value of $10^{-3}$ to recompute the modified Hessian matrix $\tilde{\bm{H}}$ to make it become positively definitive, from which we can compute the covariance matrix $\boldsymbol{\Sigma}\sim(\beta\tilde{\bm{H}})^{-1}$ and visualise the probability distribution $\mathcal{N}(\hat{\boldsymbol{\phi}},\boldsymbol{\Sigma})$ as shown in \cref{fig:ridge_regression_prob_dist_loss_landscape}. Here, the temperature $\beta$ is set to $10^{3}$ to make the probability distribution more concentrated around $\hat{\boldsymbol{\phi}}$ for better visualisation.

\section{Relationships amongst Different Generative Models}
\label{appendix:generative_model_relations}
\begin{figure}[htb]
    \centering
    \includegraphics[width=\linewidth]{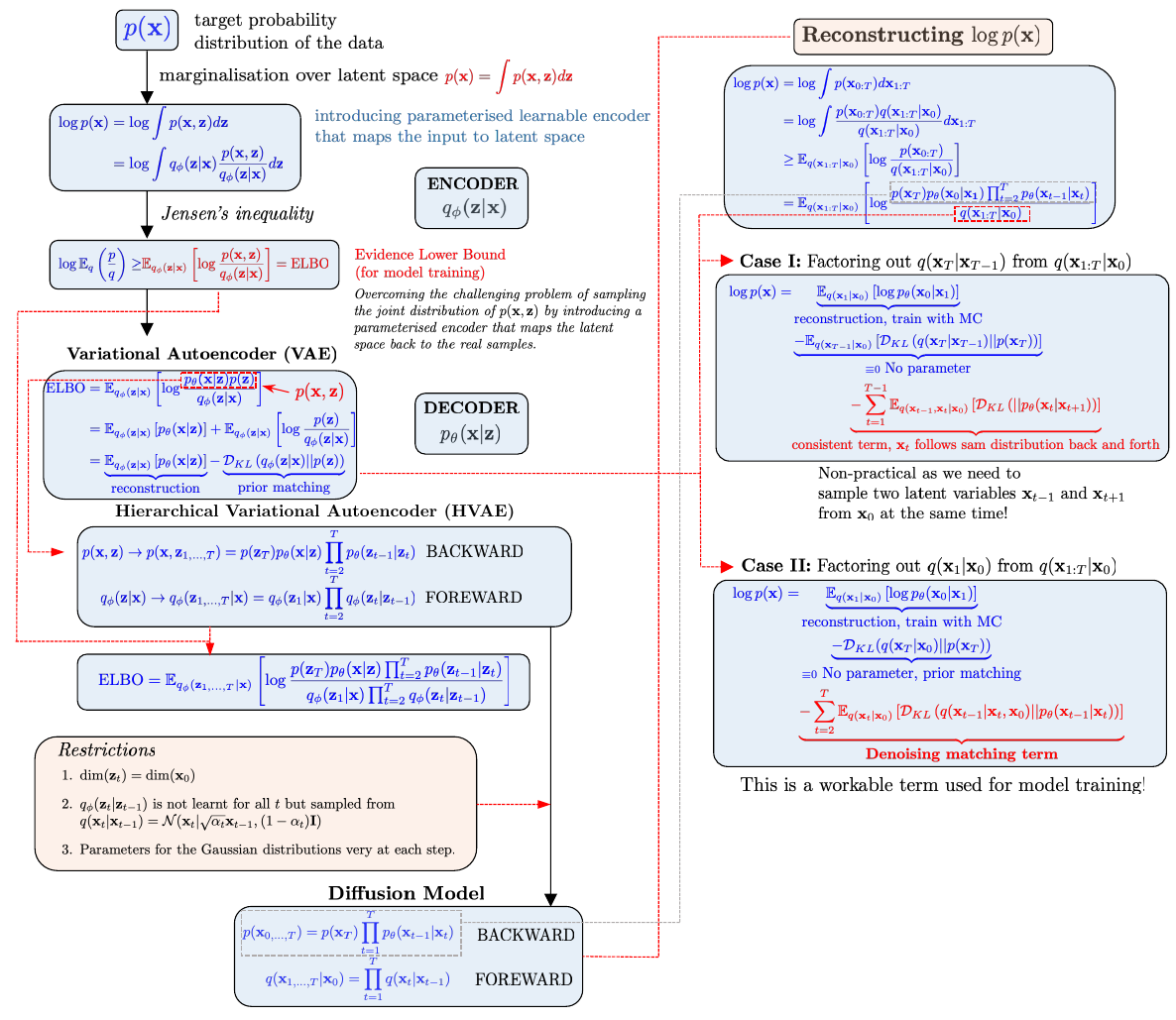}
    \caption{Relationship between variational autoencoder and the diffusion generative model, that is bridged via the hierarchical variational autoencoder, showcasing the key mathematical equations that are used to reconstruct the probability distribution of the target data $p(\bm{x})$, which are discussed in details by \citet{luo2022understanding}. Different variants and interpretations of the denoising term, that is used for training the diffusion generative models are further showcased in \cref{fig:denosing_process}.}
    \label{fig:generative_model_relations}
\end{figure}

\begin{figure}[htb]
    \centering
    \includegraphics[width=\linewidth]{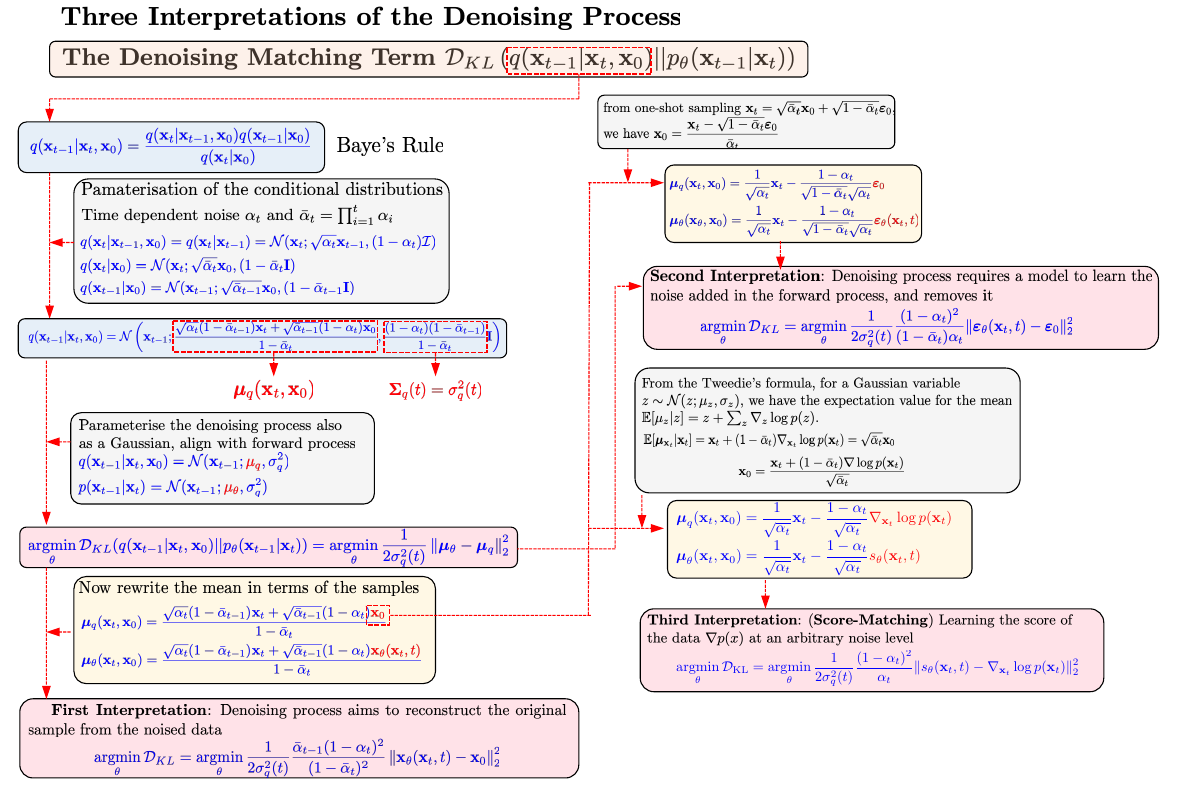}
    \caption{Relationship amongst three different ways to parameterise the transition probability in the reverse generative process for the diffusion generative model, from which computable expressions for the KL divergence between the forward and reverse process can be derived and interpreted for model training.}
    \label{fig:denosing_process}
\end{figure}

\section*{Acknowledgment}
I would like to thank Tim Gould for providing a Pythonic version of the GMTKN database in preparation for \cref{fig:DFA_benchmarks}.

\newpage 
\small

\bibliography{ref.bib}

\end{document}